\documentclass[pdftex,11pt]{article}
\usepackage[pdftex]{graphicx}
\usepackage[pdftex,                
bookmarks=true,
bookmarksnumbered=true,
hypertexnames=false,
breaklinks=true,
linkbordercolor={0 0 1}]{hyperref} 
\usepackage{tabularx}    
\usepackage{import}      
\usepackage{amsmath, amssymb} 
\usepackage{pdfpages}
\usepackage[24hr,mmddyyyy]{datetime}
\usepackage{hyphenat}			
\usepackage{lineno}
\usepackage{booktabs}			
\usepackage{textcmds}	
\usepackage{xspace}
\usepackage{longtable}	
\usepackage{bm}		
\usepackage{afterpage}

\usepackage[margin=0.25in,font=small,labelfont=bf,labelsep=colon]{caption}

\usepackage[backend=bibtex8,doi=false,style=numeric,sorting=none]{biblatex}

\newcommand{\citep}[1]{\cite{#1}}

\usepackage[bottom]{footmisc}

\def\bfseries{\fontseries \bfdefault \selectfont \boldmath}

\usepackage[outermarks]{titlesec}
\titleformat{\section}
  {\normalfont\Large\bfseries}{\thesection}{1em}{}
\titleformat{\subsection}
  {\normalfont\large\bfseries}{\thesubsection}{1em}{}
\titleformat{\subsubsection}
  {\normalfont\normalsize\bfseries}{\thesubsubsection}{1em}{}
\titleformat{\paragraph}
  {\normalfont\normalsize\bfseries}{\theparagraph}{1em}{}
\titleformat{\subparagraph}
  {\normalfont\normalsize\bfseries}{\thesubparagraph}{1em}{}

\titlespacing*{\section}      {0pt}{3.5ex plus 1ex minus .2ex} {2.3ex plus .2ex}
\titlespacing*{\subsection}   {0pt}{3.25ex plus 1ex minus .2ex}{1.5ex plus .2ex}
\titlespacing*{\subsubsection}{0pt}{3.25ex plus 1ex minus .2ex}{1.5ex plus .2ex}
\titlespacing*{\paragraph}    {0pt}{3.25ex plus 1ex minus .2ex}{1.5ex plus .2ex}
\titlespacing*{\subparagraph} {0pt}{3.25ex plus 1ex minus .2ex}{1.5ex plus .2ex}

\newcommand{\ihmpc}{{\ensuremath{h{\rm\,Mpc}^{-1}}}\xspace}

\newcommand{\lya}{{Ly-$\alpha$}\xspace}
\newcommand{\simlt}{\lower.5ex\hbox{$\; \buildrel < \over \sim \;$}}
\newcommand{\simgt}{\lower.5ex\hbox{$\; \buildrel > \over \sim \;$}}

\newcommand{\ihMpc}{\ihmpc}

\newcommand{\micron}{{\mbox{$\mu$m}}\xspace}

\newcommand{\celsius}{{\mbox{$^\circ$C}}\xspace}
\newcommand{\degree}{{\mbox{$^\circ$}}\xspace}

\renewcommand\thetable{\arabic{section}.\arabic{subsection}.\arabic{table}}
\renewcommand\theequation{\arabic{section}.\arabic{subsection}.\arabic{equation}}

\newcommand{\lyaf}{{Ly-$\alpha$\ forest}\xspace}

\newcommand{\ie}{{i.e.}\xspace}
\newcommand{\eg}{{e.g.}\xspace}
\newcommand{\etc}{{etc.}\xspace}

\newcommand{\otwo}{{[O\,II]}\xspace}
\newcommand{\othree}{{[O\,III]}\xspace}

\def\pvm#1{
}

\begin{document}

\renewcommand{\thefigure}{\thesection .\arabic{figure}}
\renewcommand{\thetable}{\thesection .\arabic{table}}
\renewcommand{\theequation}{\thesection .\arabic{equation}}




\begin{titlepage}
\begin{center}
\begin{Large}
{The DESI Experiment Part II: Instrument Design\\[0.1in]}
\end{Large}
\end{center}

\begin{center}
DESI Collaboration: 
Amir Aghamousa$^{73}$,
Jessica Aguilar$^{76}$,
Steve Ahlen$^{85}$,
Shadab Alam$^{41,59}$,
Lori E. Allen$^{81}$,
Carlos Allende Prieto$^{64}$,
James Annis$^{52}$,
Stephen Bailey$^{76}$,
Christophe Balland$^{88}$,
Otger Ballester$^{57}$
% ...

\end{center}

\begin{center}
{\it (Affiliations can be found after the references)}
\end{center}

\begin{abstract}

DESI (Dark Energy Spectropic Instrument) is a Stage IV ground-based dark energy experiment that will study
baryon acoustic oscillations and the growth of structure through
redshift-space distortions with a wide-area galaxy and quasar
redshift survey.  The DESI instrument is a robotically-actuated, fiber-fed spectrograph
capable of taking up to 5,000 simultaneous spectra over a wavelength
range from 360 nm to 980 nm.  The fibers feed ten three-arm
spectrographs with resolution $R= \lambda/\Delta\lambda$ between 2000 and 5500,
depending on wavelength.  The DESI instrument will be used to conduct a five-year survey
designed to cover 14,000 deg$^2$.  This powerful instrument will be installed
at prime focus on the 4-m Mayall telescope in Kitt Peak, Arizona,
along with a new optical corrector, which will provide a three-degree
diameter field of view.  The DESI collaboration will also deliver a
spectroscopic pipeline and data management system to reduce and
archive all data for eventual public use. 

\end{abstract}

\end{titlepage}

\cleardoublepage


\setcounter{secnumdepth}{5}
\setcounter{tocdepth}{5}	

\pagenumbering{roman}		
\tableofcontents
\cleardoublepage

\pagestyle{headings}		
\pagenumbering{arabic}		



\section{Overview}
\label{sec:InstrOverview}
\subsection{DESI Scope}
\begin{figure}[!hb]
\centering
\includegraphics[width=5in]{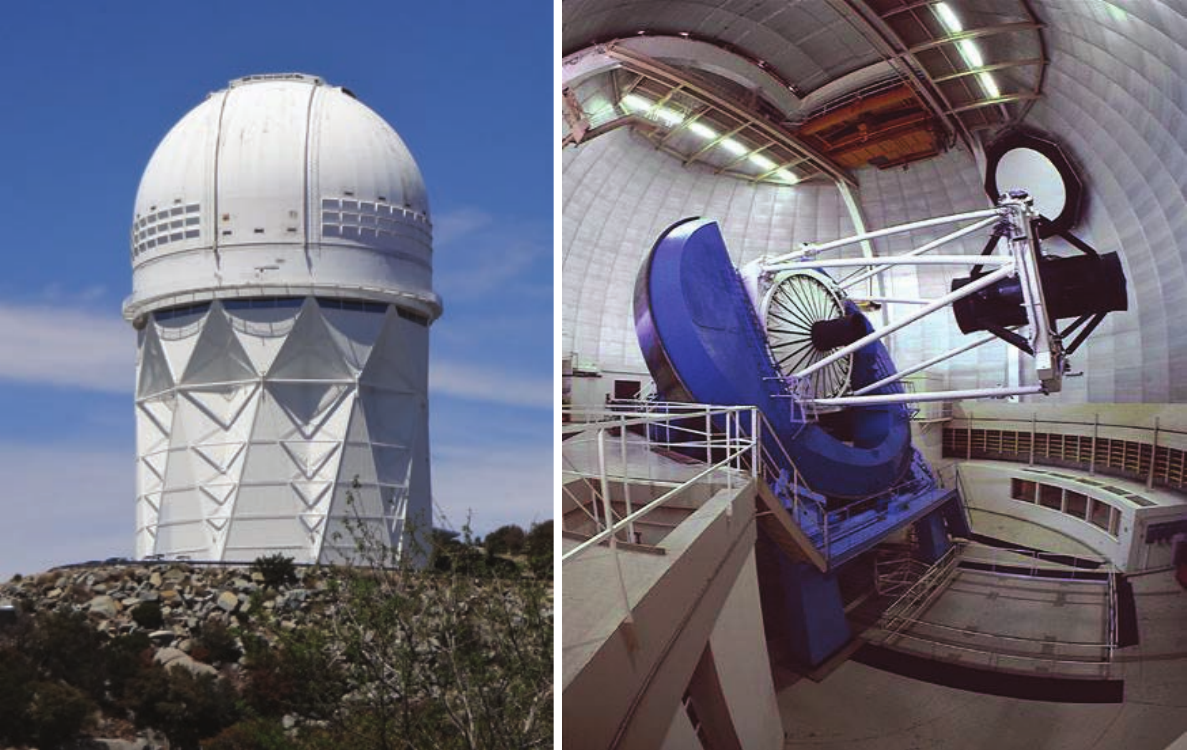}
\caption{On the left is the Mayall 4m telescope dome structure.  On the right is the telescope.  The black cylinder is the MOSAIC corrector that will be replaced as part of the DESI project to instrument a 3 degree diameter field of view.}
\label{fig:Mayall}
\end{figure}

DESI, the first Stage IV dark energy experiment, will use the
redshifts of 25 million galaxies to probe the nature of dark energy
and test General Relativity.  
DESI builds on the successful Stage-III BOSS redshift
survey, which has established BAO as a precision technique for dark
energy exploration.  DESI will make an order-of-magnitude advance over
BOSS in volume observed and galaxy redshifts measured 
by using 5,000 robotically controlled fiber-positioners to feed a
collection of spectrographs covering 360 nm to 980 nm.  Newly designed
optics for the National Optical Astronomy Observatory's 4-m Mayall telescope at Kitt Peak, Arizona, will
provide an 8-square-degree field of view.  The telescope dome and telescope with the existing MOSAIC corrector are shown in Figure~\ref{fig:Mayall}.

DESI is an important component of the DOE Cosmic Frontier program,
meeting the need for a wide-field spectroscopic survey identified in
the 2011 ``Rocky-III'' dark energy community planning report. It
complements the imaging surveys Dark Energy Survey (DES, operating
2013-2018) and the Large Synoptic Survey Telescope (LSST, planned
start early in the next decade).  In addition to providing Stage-IV
constraints on dark energy, DESI will provide new measurements that
can constrain theories of modified gravity and inflation, and will
provide cutting-edge limits on the sum of neutrino masses.  The DESI
collaboration will also deliver a spectroscopic pipeline and data
management system to reduce and archive all data for eventual public
use.

The DESI collaboration is planning a five-year spectroscopic survey
covering 14,000 deg$^2$, targeting three classes of galaxies
identified from imaging data.  We will measure luminous red galaxies
(LRGs) up to $z=1.0$, extending the BOSS LRG survey in both redshift
and survey area.  To probe to even higher redshift, DESI will target
bright \otwo emission-line galaxies (ELGs) up to $z=1.7$.  Quasars
will be targeted both as direct tracers of the underlying dark matter
distribution and, at higher redshifts ($ 2.1 < z < 3.5$), for the
\lyaf~absorption features in their spectra, which will be used to
track the distribution of neutral hydrogen.  More than 20 million
galaxy and quasar redshifts will be obtained to measure the BAO
feature and determine the matter power spectrum, including redshift
space distortions.  In this document we consider primarily this
baseline survey.  We also provide numbers for a reduced survey
spanning 9,000 deg$^2$, which is still sufficient to meet the
requirements of a Stage-IV project.
  
DESI will obtain more than 30 independent distance-vs-redshift
measurements, a powerful probe of the nature of dark energy.  This can
be quantified with the Dark Energy Task Force figure of merit (DETF
FoM~\cite{DETF06}), which measures the combined precision on the present day dark energy
equation of state, $w_0$, and the rate of its evolution with
redshift, $w_a$.  DESI galaxy BAO measurements achieve a DETF FoM of
125, more than a factor of three better than the DETF FoM of all
Stage-III galaxy BAO measurements combined.  The FoM increases to 152
with the inclusion of \lyaf BAO, and 326 including galaxy broadband
power spectrum to $k=0.1$ \ihMpc.  DESI thus satisfies the DETF
criteria for a Stage-IV experiment, generally taken as a factor of three increase
over Stage-III.  Moreover, the FoM grows to 756
when the galaxy broadband power spectrum data out to $k<0.2$ \ihMpc
are included.

DESI's science reach will extend beyond dark energy.  It will measure the
sum of neutrino masses with an uncertainty of 0.017 eV (for $k_{\rm
  max} < 0.2$ \ihMpc), sufficient to make a direct non-zero
detection of the sum of the neutrino masses at 3-$\sigma$
significance. If the neutrino masses are minimal, and the hierarchy is
normal, the inverted hierarchy will be ruled out at 99\% CL.  DESI
will also place significant constraints on theories of modified
gravity by measuring redshift space distortions, and on theories of
inflation by measuring the spectral index $n_s$ and its running with
wavenumber, $\alpha_s$.

DESI will provide an unprecedented multi-object spectroscopic
capability for the U.S. using an existing NSF telescope
facility. Many other science objectives can be addressed with the DESI
wide-field survey dataset,  bright time survey, and piggy-back
observation programs.  Much as with SDSS, a rich variety of projects
will be enabled with the legacy data from the DESI survey.

DESI will overlap with the DES and LSST survey areas, which are
primarily in the southern hemisphere but which will have equatorial
and northern ecliptic regions.  DESI will be a pathfinder instrument
for the massive spectroscopic follow-up required for future large area
imaging surveys such as LSST.

This Technical Design Report summarizes the DESI scientific
goals, the instrument and optical design, the target selection, the
integration and test plan, and the data management plan.  The
companion Science Requirements Document provides information that
guides the design.  The DESI construction management plan is presented
in the accompanying Project Execution Plan.  Likewise, project cost
and schedule are available in relevant Project Office documents.

\subsection{DESI Subsystems}
The DESI subsystems comprise

\begin{enumerate}  
\item Prime focus corrector optics.  
\item Focal plane with robotic fiber positioners. 
\item Optical fiber system.
\item Spectrographs.
\item Real-time control and data acquisition system.
\item Data management system.   
\end{enumerate}

Each subsystem has a specification, including performance
requirements.  Tests will be developed to demonstrate that the
subsystems meet these requirements prior to shipment to the Mayall.
After delivery to the Mayall, a subset of the pre-ship tests will be
repeated to verify survival of shipping along with general system
function.

\begin{figure}[p]
\centering
\includegraphics[width=0.95\textheight,angle=90]{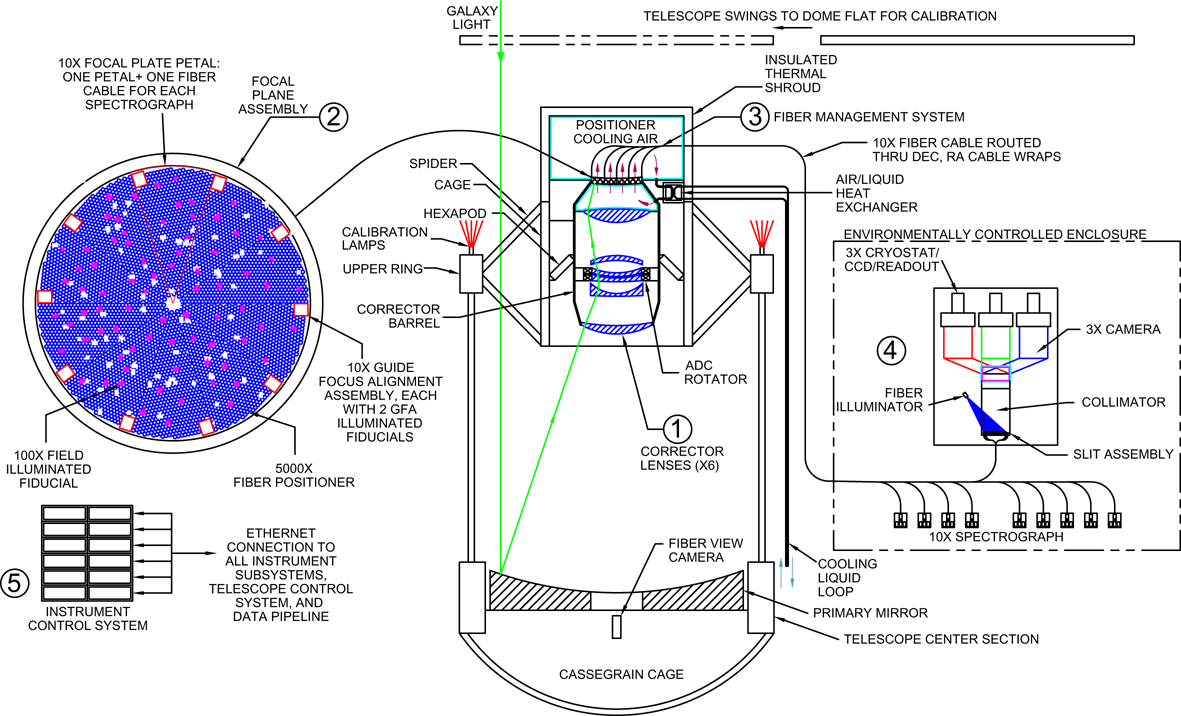}
\caption{DESI block diagram.}
\label{fig:block-diagram}
\end{figure}

\begin{figure}[tb]
\centering
\includegraphics[width=\textwidth]{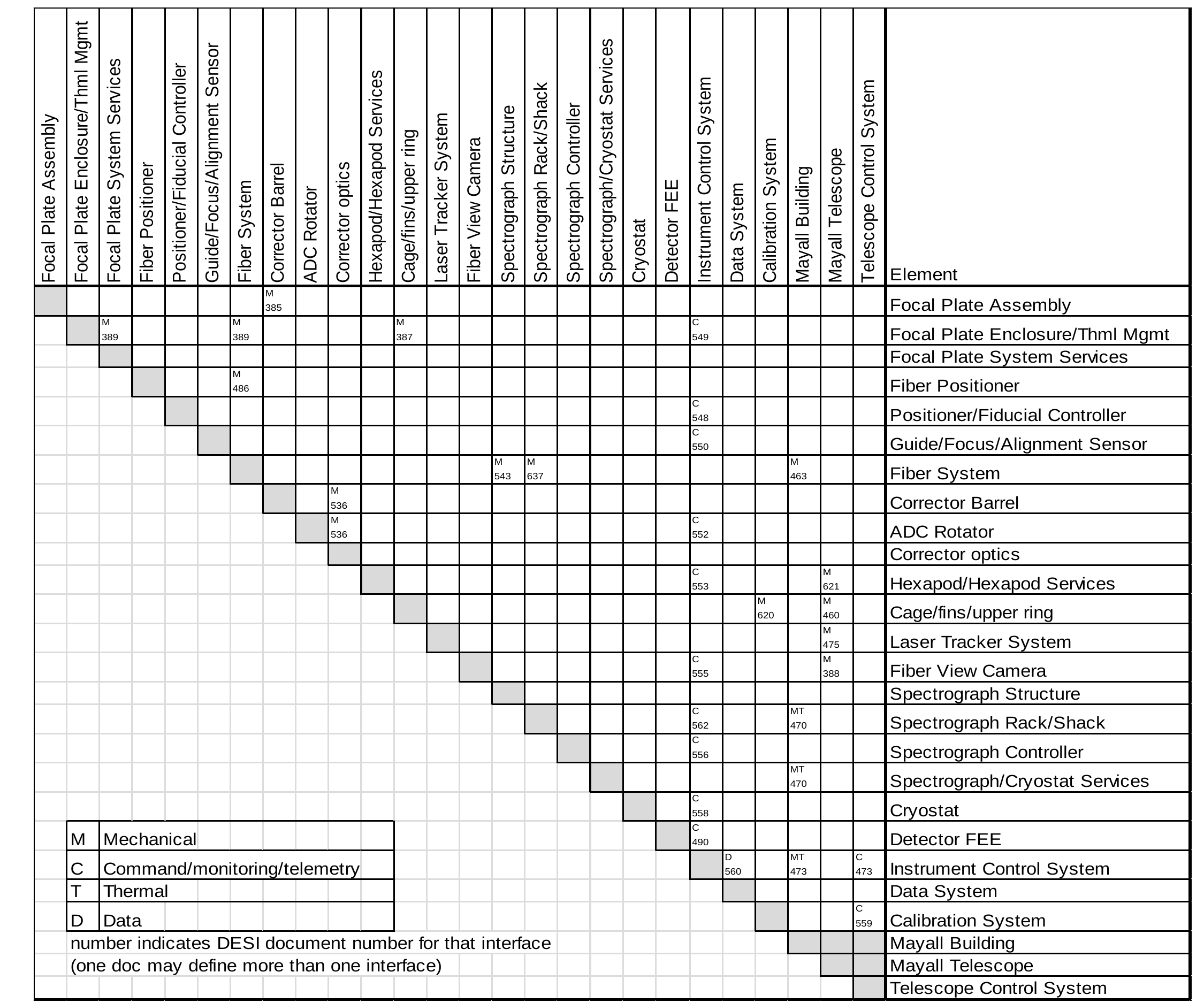}
\caption{High-level DESI Interface Matrix'}
\label{fig:interfacematrix}
\end{figure}

A critical aspect of each element of the instrument is its interface
or interfaces with other elements.  The system block diagram
(Figure~\ref{fig:block-diagram}. The number of illuminated fiducial fibers has been increased since this diagram was created:  See Figure~\ref{fig:FiducialLayout}.) is used to identify interfaces
between elements from different suppliers.  Figure~\ref{fig:interfacematrix} shows a high-level DESI Interface Matrix (or
N-squared matrix), illustrating the locations and types of interfaces
between the major elements of the DESI instrument, as integrated with
the Mayall Telescope.  
Each of these interfaces will be managed with
an Interface Control Document (ICD).

\subsection{DESI Design Fulfills Requirements}

The design of the DESI instrument is set by the key science project and
operational requirements defined in the DESI Level 1 through Level 3 Requirements Document (SRD)~\cite{SRD13}.
Instrument subsystems impacted by these requirements are  outlined below.

\begin{itemize}
  \setlength{\itemsep}{1pt}
  \setlength{\parskip}{0pt}
  \setlength{\parsep}{0pt}

\item[A.] Field of view (effective) not less than 7 square degrees [Corrector optics, Focal plane].
\item[B.] S/N $>7$ for the detection of an \otwo flux of $8\times10^{-17}$~erg/s/cm$^2$ in ELGs [Fiber optics, Spectrographs].
\item[C.] A baseline 14,000 deg$^2$ survey, with a minimum-requirements of 9,000 deg$^2$ [Data management, Corrector optics].
\item[D.] 30 million cosmological targets composed of LRGs, ELGs, QSO tracers, and 
\lyaf\ QSOs [Data management, Corrector optics].
\item[E.] Survey duration of 5 years for a 14,000 deg$^2$ baseline survey not including commissioning and validation [Data management, Real-time control].
\item[F.] Fiber density of $< 700$ per square degree [Focal plane].
\item[G.] Spectral range of 360--980 nm [Spectrographs].
\item[H.] Spectroscopic resolution ($\Delta\lambda/\lambda$) and precision better than  $< 0.0005(1+z)$, \ie,   for redshift error (precision and accuracy) 
$R>$1500 for $\lambda >$360 nm and longer, 
$R>$3000 for $\lambda >$555 nm, 
and $R>$4000 for $\lambda >$656~nm [Spectrographs].
\end{itemize}

\subsubsection{Corrector Optics}
The corrector design with four large, fixed fused silica elements and two rotating borosilicate lenses provides a linear field of view of 3.2$^\circ$, corresponding to a diameter of about 812~mm at prime focus and an area of 8 square degrees [SRD requirement A].  Combined with a reference exposure time at zenith of 1000 seconds, this is sufficient to cover the 14,000 square degree footprint approximately five times, enough to guarantee approximately 25 million successful spectra from a  larger target list, with multiple exposures of certain targets [SRD requirements C,D].  

\subsubsection{Focal plane}
The 5,000 positioners instrument 7.5 square degrees of  the 8 available [SRD requirement A], giving a fiber density of 667/square degree [SRD requirement F]. The focal plane also hosts field fiducials and Guide, Focus, and Alignment (GFA) sensors, located at the periphery of the instrumented focal plane. The positioners are arranged in a hexagonal pattern with a 10.4~mm pitch between centers.  Each positioner has two rotational degrees of freedom allowing it to reach any point within a 6 mm radius.  The focal plane is divided into ten pie-slice-shaped petals.

\subsubsection{Optical Fiber System}
Optical fibers run 49.5 m from each positioner to the spectrographs.  The core diameter of the fiber is 107~\micron.  The fiber system throughput varies from 0.499 at 360 nm to 0.838 at 980 nm [SRD requirements B,G].  

\subsubsection{Spectrographs}
Two dichroics split the output of the optical fibers into three channels. The throughput for the dichroics is greater than 95\% throughout the full wavelength range, except for the dichroic transition intervals (747-772 nm) and (566-593 nm), where it still exceeds 90\%. Volume phase holographic gratings provide dispersion in the three channels, [360-593 =nm], [566-772 nm], and [747-980 nm] [SRD requirement G], with resolutions greater than 2000, 3200, and 4100 respectively [SRD requirement H].  The blue arm of the spectrographs use CCDs from Imaging Technologies Lab, while the red and NIR channels use LBNL CCDs.  All will be 4k x 4k with 15~\micron pixels [SRD requirement B].





\subsubsection{Real-Time Control and DAQ}
DESI real-time control and DAQ components such as the dynamic exposure time
calculator, the realtime data quality assessment, and complex algorithms to convert on-sky
target coordinates to fiber positions on the focal plane are based on the SDSS-III/BOSS
online system.  The time between exposures will be less than 120 seconds [SRD requirement E].  


\subsubsection{Data Management}
An initial list of targets will be assembled from imaging completed prior to the survey, with criteria designed to select LRGs, ELGs, and QSOs.  The 14,000 square degree footprint will be covered by about 10,000 tiles, providing average coverage of each spot of  $10,000 \times 7.5 {\rm sq.\ deg.}/14,000 {\rm sq. deg.} \approx 5.3$ times [E].  

\subsection{Integration and Performance}
\label{sec:performance}

  A block diagram of DESI at the
Mayall telescope is shown in Figure~\ref{fig:block-diagram}.  The
Mayall is a 4-m telescope capable of supporting a 3.2-degree diameter
field of view, which DESI achieves with a new prime focus optic, lens
barrel, support cage with hexapod adjusters, and a new ring attachment
to the top of the telescope.

Careful management of all the elements that contribute to overall
light throughput is a systems engineering activity.  There are 70+
DESI elements that affect throughput presently being tracked
against current estimates (DESI-0347).  These instrument performance
parameters are combined with sky, atmospheric, galaxy, and survey
parameters to simulate and predict DESI performance (see
Section~\ref{sec:desimodel}).  These in turn drive the exposure time range described in FDR Part I, Section~4.4.2.  Rollup values from the instrument
throughput tracking spreadsheet are shown in
Table~\ref{tab:throughput} and we briefly describe the items below.
Note that all values are field-weighted averages.

\begin{table}[!htb]
\centering
\caption{DESI throughput current best estimates rollup (DESI-0347). }
\label{tab:throughput}
\small
\begin{tabular}{lcccccccccccccc}

\hline 
~ & \multicolumn{7}{c}{Wavelength (nm)} \\ 
~ & 360 & 450 & 550 & 650 & 750 & 850 & 980  \\
 \hline 
Blur (\micron 1-D sigma) 	&	15.6	&	14.3	&	13.4	&	13.9	&	14.8	&	15.4	&	15.6	\\	
Lateral Errors (\micron) 	&	13.4	&	11.7	&	11.0	&	12.3	&	13.0	&	12.7	&	12.9	\\	
Telescope  to Fiber Input Throughput (\%)	&	56.5	&	71.8	&	74.2	&	67.1	&	68.4	&	70.4	&	69.9	\\	
Fiber System Throughput (\%)	&	49.7	&	70.0	&	78.0	&	81.1	&	82.4	&	82.7	&	83.6	\\	
Spectrograph Throughput (\%)	&	39.4	&	73.1	&	68.5	&	74.5	&	67.4	&	77.5	&	55.3	\\	
\hline
Instrument Throughput (\%)	&	11.1	&	36.8	&	39.6	&	40.6	&	38.0	&	45.2	&	32.4	\\	
\hline
& & & & & & & \\
& & & & & & & \\

\end{tabular}
\end{table}

{\it Blur.}  Includes contributions to spot size at the fiber tip from intrinsic blur of the corrector optics prescription, manufacturing and alignment errors in the corrector, alignment errors of the corrector relative to the mirror, and defocus due to gravity, manufacturing and alignment errors.  Note that the blur tabulated below does not include atmospheric effects or telescope jitter or tracking errors.  
These effects are accounted for in the simulations modeling the delivered image quality.

{\it Lateral errors.}  Summarizes inaccuracies in placing the target centroid onto the center of the optical fiber, including effects such as fiber positioner inaccuracies, fiber view camera inaccuracies, uncompensated atmospheric refraction, focal plane thermal and gravity distortions, metrology errors between guiders and fibers, and misalignments of the corrector relative to the mirror.  Note that the lateral errors tabulated below do not include atmospheric effects, telescope jitter, or telescope tracking errors.  
Again, these effects are accounted for in the simulations modeling the delivered image quality.

{\it Telescope to Fiber Input Throughput.}  Accounts for losses due to primary mirror reflectivity, corrector lens absorption and anti-reflectance coatings, corrector vignetting, and scattering in the corrector optics.

{\it Fiber System.}  Includes losses due to fiber bulk absorption, reflections at entrance and exit, and focal ratio degradation (FRD).  FRD effects include losses due to fiber construction, end polishing, bends/twists, and tilt of the chief ray relative to the fiber normal. 

{\it Spectrograph.} Spectrograph losses, including collimator reflectivity, dichroic effectiveness, grating efficiency, camera glass absorption and detector quantum efficiency are considered in the spectrograph row.

{\it Instrument throughput.} Sky throughput is a measure of the net throughput of the instrument from light at the aperture to electronic signals from the detectors.  It is the product of the telescope-to-fiber-input, fiber system, and spectrograph throughputs. It does not include the atmosphere, obscurations, or galaxy effects, nor does it include losses due to blur or lateral errors, which are separately reported.

\clearpage

\section{Corrector}
\setcounter{equation}{0}\setcounter{figure}{0}\setcounter{table}{0}
\label{sec:echo}
The DESI multi-object spectrograph uses the existing primary mirror at the Mayall telescope with a newly developed prime focus corrector that allows $\sim$5,000 robotically positioned fibers to be arranged over 8~deg$^2$. 

%

\subsection{Corrector Optics Design}
\label{sec:corrdesign}

\begin{figure}[hb]
\centering 
\includegraphics[width=.8\textwidth]{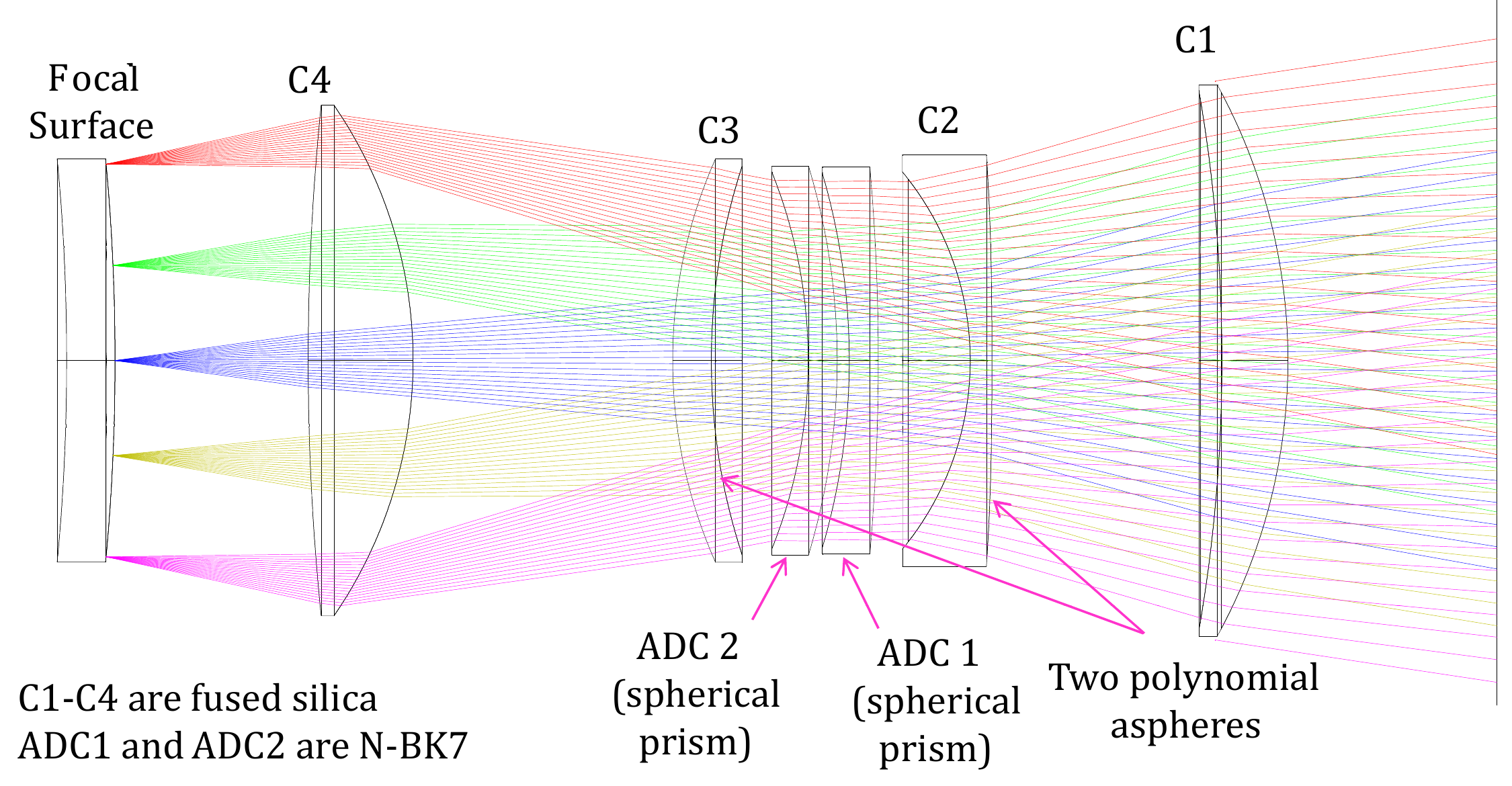}
\caption{Optical layout of Echo 22.}
\label{fig:corF1} 
\end{figure}

Optical requirements as well as predicted performance of the  
baseline DESI Prime Focus Corrector design (Echo 22) (DESI-0329) are listed in Table~\ref{tab:corT2}. The design includes four large fused silica elements.  The  two smaller lenses have aspheres on one surface each; all other silica surfaces are spherical.  The lens specifications are within the capabilities of multiple vendors.  An atmospheric dispersion compensator (ADC) is included in the design, in order to meet blur requirements at off-zenith angles. This compensator includes two large wedged borosilicate lenses; all surfaces are spherical. 
The layout of the optical corrector elements is shown in Figure~\ref{fig:corF1}.

\begin{table}[!t]
\centering
\caption{Prime Focus Corrector Requirements (DESI-0478).  Blurs are FWHM, maximums are for any field position, and means are field averaged.}
\footnotesize
\newcolumntype{R}[1]{>{\raggedright\arraybackslash}m{#1}}
\begin{tabularx}{0.9\textwidth}{@{}R{0.8in}|R{2.0in}|R{2.5in}} \hline 
Requirement & Value & Note \\ \hline 
Optical design & Compatible with Mayall 4 m primary mirror & Mayall telescope selected for DESI. \\ \hline

Wavelength band & 360--980~nm &  Required for z range of \lya\ QSOs and ELGs \\ \hline 
Design residual blur (arcsec) & zenith:\newline \hspace{10pt}360--450~nm: $<$0.4 mean, 0.60 max \newline \hspace{3pt}450--980~nm: $<$0.4  mean, 0.50  max \newline 60~deg from zenith: \newline \hspace{3pt}360--450~nm: 0.4 mean, 0.75 max \newline \hspace{3pt}450--980~nm: 0.4 mean, 0.6 max & 
360--450 nm: required for QSO/LRG system throughput  \newline 450--980~nm: required for LRG/ELG system throughput \\ \hline 
Field of view & $>3^\circ$ diametral &  Required for Poisson statistics of targets. \\ \hline 
f/\# variation & f/3.7 to f/4.3 & f/\# number variation is allowed, but must be constrained. \\ \hline
Focal surface diameter & Sufficiently large to contain $\geq 5,000$ fiber positioners & Verified by the focal plate layout simulator. \\ \hline 
Focal surface curvature & Radius of curvature (convex), greater than 3000~mm  & Focal plate may be curved, with aspheric shape, but maximum curvature is constrained for actuator packaging.  \\ \hline 
Aspheric departure slope & $<$30~mrad & Manufacturability; Aspheres are permissible on one surface of smaller, non-ADC elements.  \\ \hline 
Chief ray deviation & $<0.5^\circ$ average, $<1.0^\circ$ maximum & Required for system throughput. This is to be met at all field points, and thus all f/ratios.  \\ \hline
Glass throughput of combined elements & 
$>75\%$ at 360 nm \newline
$>88\%$ at 375 nm \newline
$>94\%$ at 400 nm \newline
$>95\%$ for $>450$ nm  & Required for system throughput. \\ \hline
Vignetting & $0\%$ up to 1.45\degree; $<1\%$ elsewhere & Required for system throughput. \\ \hline
Ghosting & $<1\%$ of observation region contaminated by ghosts brighter than 0.01$\times$ Night sky. & Focal and pupil ghosts must not contaminate a large fraction of the survey region. \\ \hline
Glass mass & $<900$ kg & To ensure barrel and mount the lenses and telescope can support the corrector \\ \hline
Lens diameter & $\leq 1175$ mm & Required for obscuration \\ 

\hline 
\end{tabularx}
\label{tab:corT2}
\end{table}

\begin{table}[!ht]
\centering
\caption{DESI Corrector Optical Design Summary.  Refer to DESI-0329 for details about the optical design.}
\begin{tabularx}{0.7\textwidth}{l|l} 
\hline
Parrameter & Value \\
\hline
Primary Mirror diameter &	3.8 meters circular \\ 
\hline
Field of View	& 3.2 degrees circular \\
	& 811.8 mm at the focal plane \\
\hline
F/\#	& 3.68  on-axis \\
	& 3.86 average over FOV \\
\hline
Plate scale	& 243.0 mm/deg on-axis \\
	& 254.8 mm/deg, average over FOV \\
\hline
\end{tabularx}
\label{tab:corrparams}
\end{table}

The Mayall telescope primary mirror (M1) is 3.8~m diameter, has a radius of curvature of 21.3~m, and a  focal length of 10.7~m.  This physical structure sets the scale for a corrector with a linear $3.2^\circ$~FOV, corresponding to a diameter of $\sim$812~mm at prime focus.  
Table~\ref{tab:corrparams} is a summary of the DESI corrector optical design parameters.
Because optical elements are placed between the prime focus and M1, optical elements are as large as 1140~mm in diameter.  Table~\ref{tab:corT3} shows manufacturing details of the six corrector elements, including mass, materials and dimensions.  The large size of these elements reduces the pool of potential vendors,
but DESI has successfully contracted for the production of all lenses with reliable vendors.
The blanks for lenses C1 and C4 are provided by Corning, and are polished by L3 Brashear.  
The blanks for lenses C2 and C3 are provided by Ohara, and are polished by Arizona Optical Systems.  
The blanks for the ADC lenses are provided one each by Schott and Ohara, and are
polished by Rayleigh Optical Corporation.

\begin{table}[tb]
\centering
\caption{Echo 22 corrector lenses.  Total mass of corrector glass is 864~kg.}
\begin{tabularx}{0.85\textwidth}{c|c|c|c|c|c|c|c} \hline 
Lens & Diam & Mass & Material & Index & Vertex & Edge  & Total Lens \\  
     &          &        &          &       & Thick  & Thick & Thick  \\  
     & (mm) & (kg) &          &       &  (mm)  &  (mm) & (mm) \\ \hline
C1 & 1140 & 201 & Silica &      $1.46 $ & 136.4 & 39.9 & 186.0 \\ 
C2 & 850 & 152 & Silica &        $1.46 $ & 45 & 204.7 & 216.5 \\ 
ADC1 & 800 & 102 & N-BK7 &  $1.52 $ & 60 & 102.1 & 119.7 \\ 
ADC2 & 804 & 89 & N-BK7 &   $1.52 $ & 60 & 80.7 & 140.0 \\
C3 & 834 & 84 & Silica &         $1.46 $ & 80 & 59.9 & 148.3 \\ 
C4 & 1034 & 237 & Silica &     $1.46 $ & 217 & 34.9 & 216.9 \\ 
Focal & 811.8 &  & Al 6061 & --- &  &  &  \\ \hline 
\end{tabularx}
\label{tab:corT3}
\end{table}

Vendors recommended that aspheric departure slope should be limited to values less than 30~mrad (\micron/mm), to allow interferometric testing with better than Nyquist fringe sampling.  The peak aspheric slope departures of lenses C2 and C3 are $<$15 and $<$11, respectively.

\subsubsection{Atmospheric Dispersion Compensator}

\begin{figure}[p]
\centering
\includegraphics[height=3in]{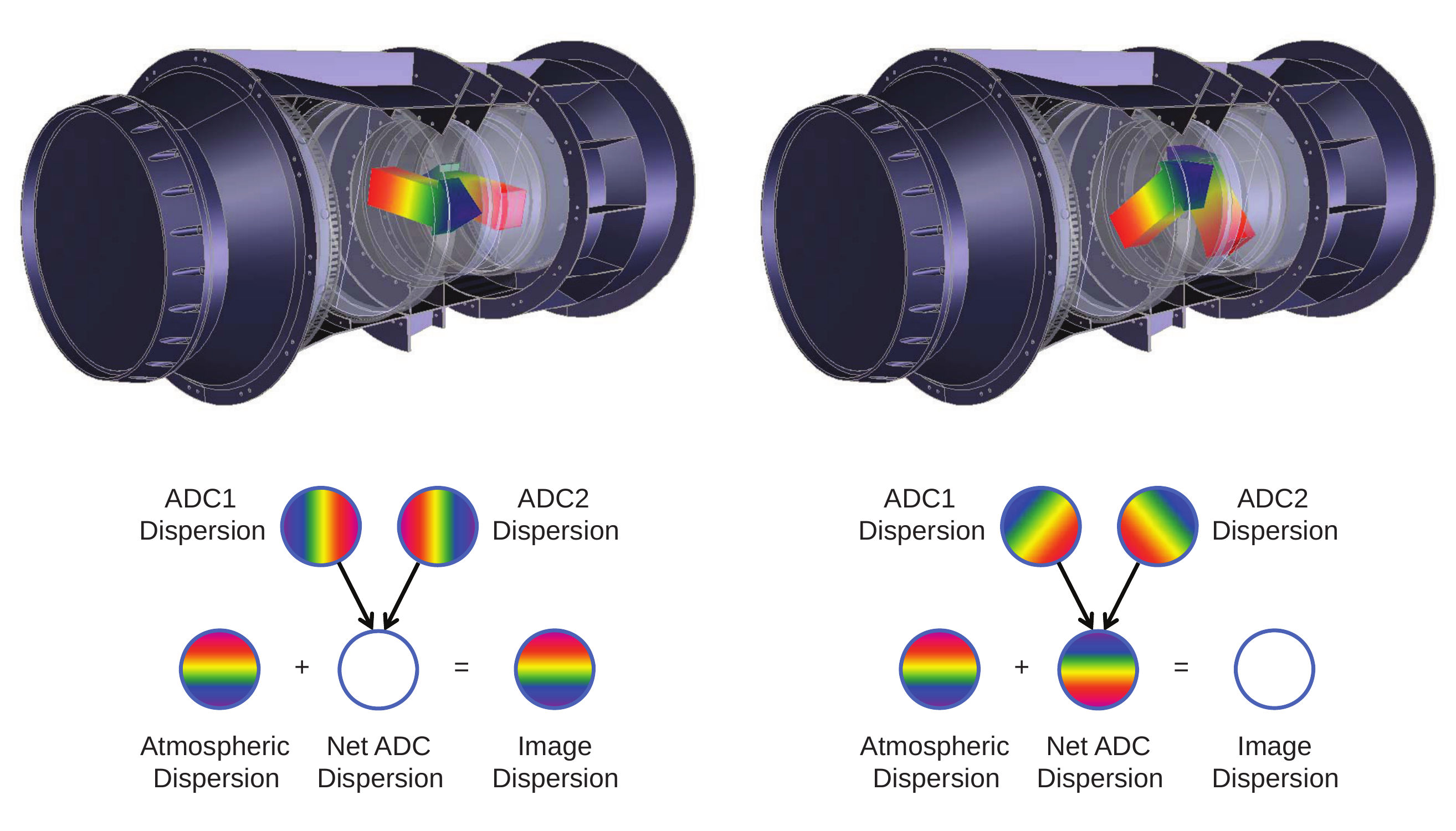}
\caption{ The atmospheric dispersion compensator consists of two oppositely oriented prisms 
(figures at left) which produce leftward and rightward dispersions that exactly cancel each other. Prism pairs can be rotated in opposite directions to produce a net dispersion opposite to that of the atmospheric contribution. Left and right components still cancel (figures at right), but allow a 
variable net dispersion that can be selected to negate atmospheric dispersion. }
\label{fig:ADC}
\vspace{0.25in}
\centering
\includegraphics[height=3in]{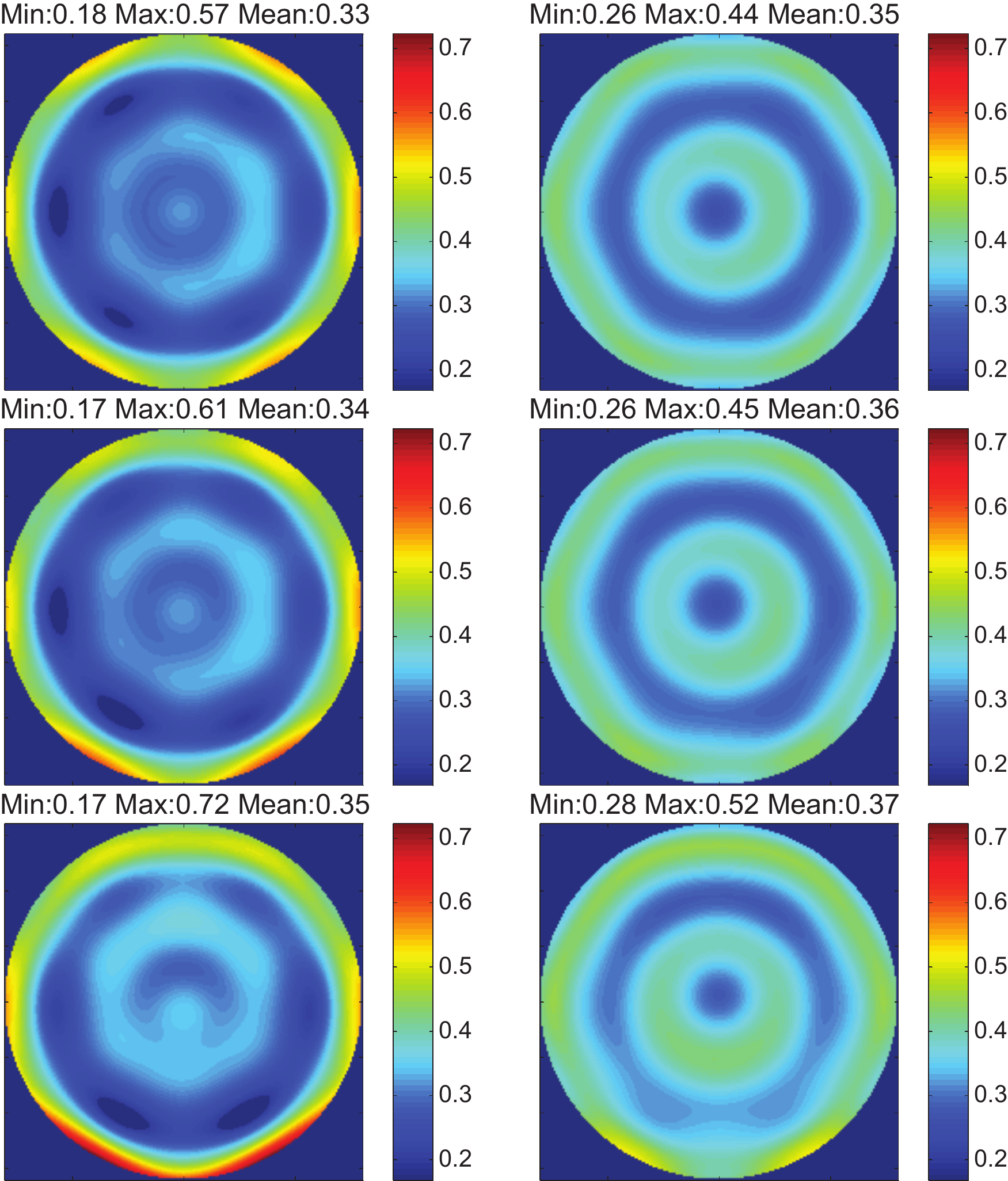}
\caption{Geometric blur is shown as a function of zenith angle. Top figures: zenith. Middle figures: $30^\circ$ zenith angle. Bottom figures: $60^\circ$ zenith angle. Numbers above figures are FWHM blur in arcsec.}
\label{fig:blur}
\end{figure}

The ADC includes two monolithic N-BK7 elements, each with two spherical surfaces and an internal wedge angle.  ADC elements are monolithic, and have no bonded joints.  Although designs exist which use fused silica for all lens elements, curved N-BK7 ADC elements allow better control of lateral color on the blue end of the DESI band. The lens elements have a slight ($\sim0.25^\circ$) internal wedge angle, which leads to predictable lateral color.  The net dispersion magnitude and direction is set by rotating the ADC elements (Figure~\ref{fig:ADC}) in opposite directions, along the corrector optical axis.  Thin ring Kaydon$^\circledR$~bearings (similar to those used on the One Degree Imager for WIYN) are employed to support the ADC lens cells, and allow independent rotation of the ADC elements.  When oriented in opposite directions, the wedges introduce minimal chromatic aberrations.  Opposite direction rotations introduce increasing dispersion, while identical rotation of the elements changes the direction of dispersion (required for an equatorial mount telescope).

Due to the wedge angle of the spherical ADC elements, the system has no optical symmetries.  Blur distributions cannot be represented as simple radial distributions, but must be quantized over the entire focal surface.  In the figures below, the full width, half maximum (FWHM) of the geometric blur for the perfect system is plotted.  This includes design residual phase error, but does not include estimates of system performance with manufacturing errors.  Figure~\ref{fig:blur} shows blur at zenith, $30^\circ$, and $60^\circ$ after correction by the ADC.

\subsubsection{Material Optical Properties}

The lenses of Echo 22 are large, but not of unprecedented size.  Larger diameter lenses are currently envisioned for the LSST camera and a 1.5~m diameter fused silica conformal window was produced by Heraeus and Corning for the ABL project.  
In the case of DESI, capable vendors with a history of successful deliveries have been
chosen from a limited pool to provide the fused silica and N-BK7 blanks. 
The main question is the impact of homogeneity grades on delivered image quality.  
Optical modeling of the DESI design suggests that variation in the refractive index of the raw material mainly affects wavefront distortion in low-order Zernike modes, \ie, variations up to about three cycles per lens diameter.  

A statistical (Monte Carlo) study of the effects of various homogeneity grades was performed, and contributions to the blur budget investigated. 
Cost of glass increases with improved homogeneity, but this cost may be balanced against polishing cost, as transmissive multi-pass optical testing allows the effects of homogeneity to be polished out, but at greater polishing cost.  Results of this study, in conjunction with programmatic considerations, have allowed procurement of appropriate-value optical substrates for DESI.

\subsubsection{Lens Tolerancing and Alignment}

Preliminary manufacturing tolerance studies have been performed (DESI-0335) on the individual lens elements.  
The studies show that given melt data from the fused silica and N-BK7 suppliers, bulk index of
refraction variations are easily compensated by small changes to the lens spacings.

Instead of specifying tight tolerances on the lens thickness and radii of curvature, a compensation scheme is employed that allows lens spacings to be adjusted based on as-built measurements of lens thickness and radii.  A Monte Carlo inverse analysis was performed to determine the sensitivity of blur performance on compensated lens parameters.  
Given this, manufacturing tolerances can be specified using values that are readily met.

Lenses are held in lens cells, which are in turn held by the barrel assembly.  Errors in the positions of the lenses result from a stackup of tolerances including lens mount, cell mounts, corrector barrel thermal and mechanical distortion.  Allowable limits for decenter and tilt were established by Monte Carlo simulation.  These tolerances assume 5-DOF compensation by the hexapod and perfect knowledge of the misalignment (wavefront errors are budgeted elsewhere). 
This budget is parceled out to the various contributing components in DESI-0335.

\subsubsection {Focal Surface}

While maintenance of the focal array assembly is not required, it is possible that the assembly will be required to be removed after installation at some point.  For that reason, a repeatable mount (pins or semi-kinematic mounts) is being considered for the focal plate interface to the corrector barrel adapter.  Tolerances for the focal plate mount, relative to the optical axis, are $\pm$150~\micron~lateral and $\pm$75~\micron~tip/tilt (across the diameter).

\begin{figure}[p]
\centering
\includegraphics[width=0.8\textwidth]{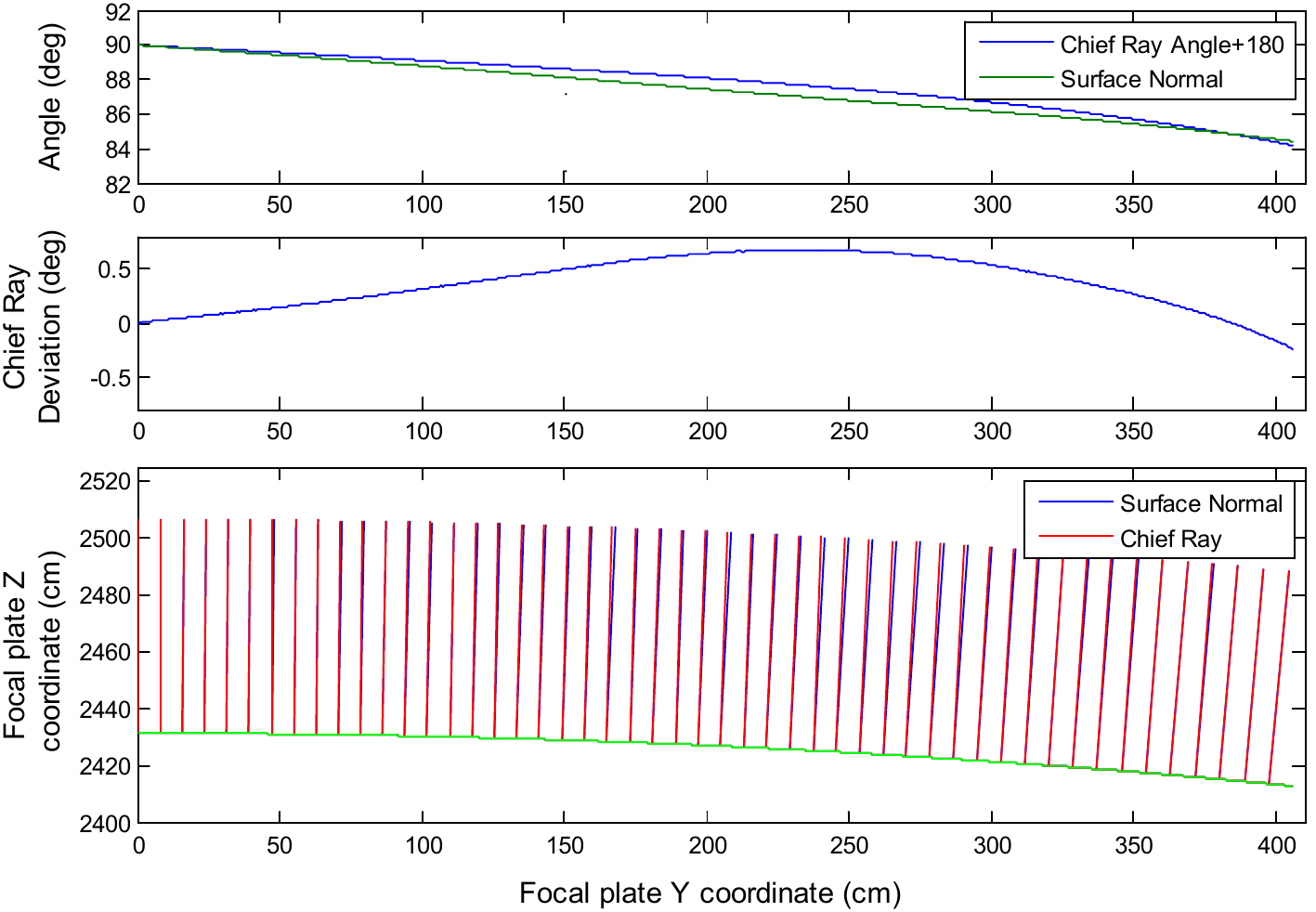}
\caption{ Chief rays are generally not normal to the local surface.  The upper figure shows the absolute angle of the chief ray and local surface normal, while the middle figure shows the difference between the two.  The bottom figure shows schematically direction of the ray deviation.}
\label{fig:chiefray}
\vspace{0.15in}
\includegraphics[width=0.8\textwidth]{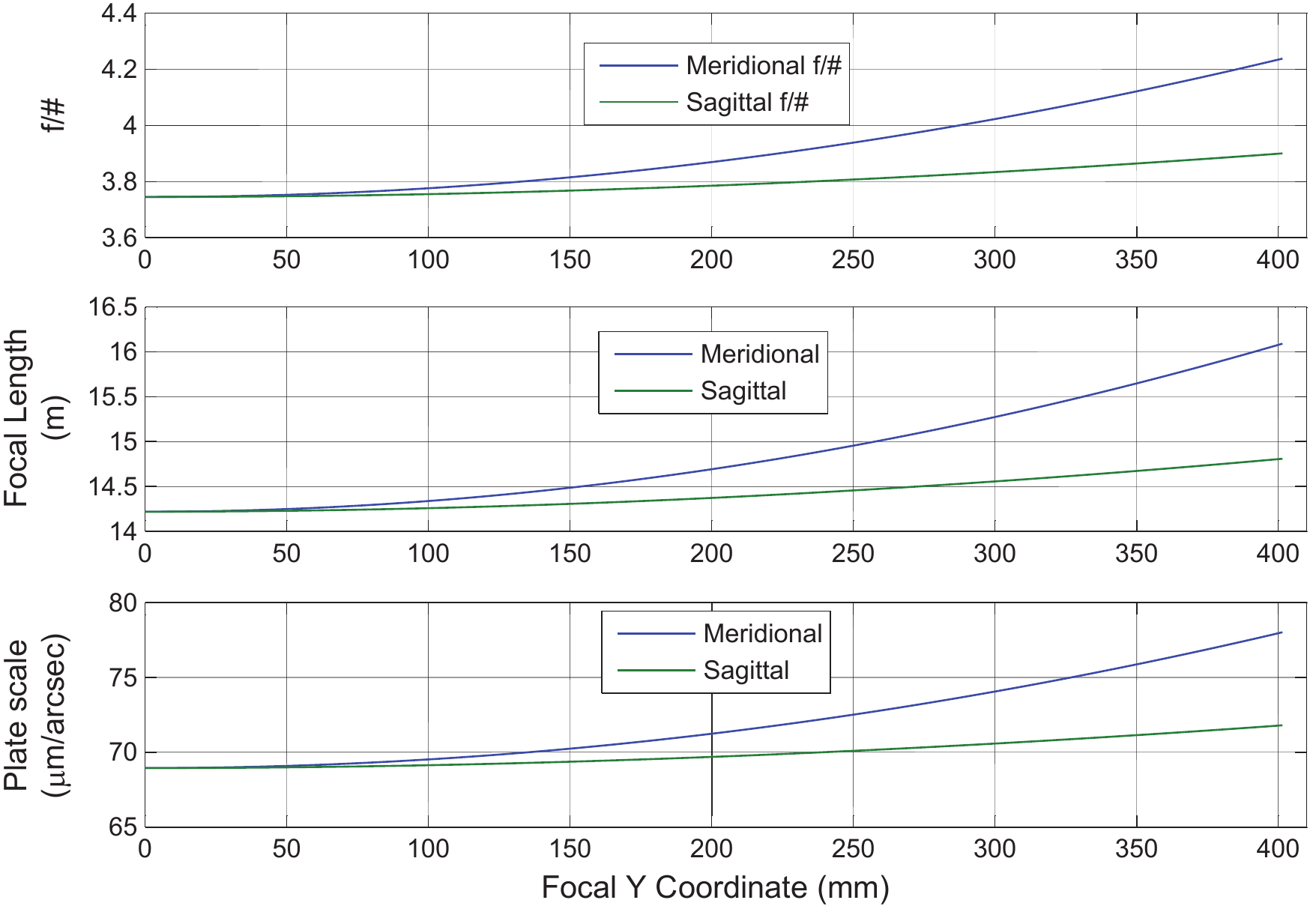}
\caption{Anamorphic pupil distortion of Echo 22.}
\label{fig:anamorphic}
\end{figure}

The design of Echo 22 included constraints on the maximum tilt of the chief ray.  The chief ray should be aligned with the fiber axis for maximum injection efficiency.  Generally in an optical system, the chief ray will not be normal to the local surface.  The relatively large field of view of this optical corrector exacerbates the issue, and the chief ray must be constrained to a manageable level in the optimization merit function.  Because the actuator mounting holes are drilled by a 5-axis CNC machine, holes can be tilted to match the chief ray direction.  This technique, and effects on actuator packaging are discussed in the focal plate section of this report.  Figure~\ref{fig:chiefray} shows the residual chief ray tilt as a function of field angle.

Figure~\ref{fig:anamorphic} shows the anamorphic (non-symmetric) distortion of the f/\# cone that is present as a function of focal plate coordinate.  Note, the cone narrows and the plate scale increases
in the meridional plane, which means the circular fiber subtends less sky in that direction.  Sagittal and meridional f/\#'s are plotted and averaged.

\subsubsection{Ghosts}

\begin{figure}[tb]
\centering 
\includegraphics[width=\textwidth]{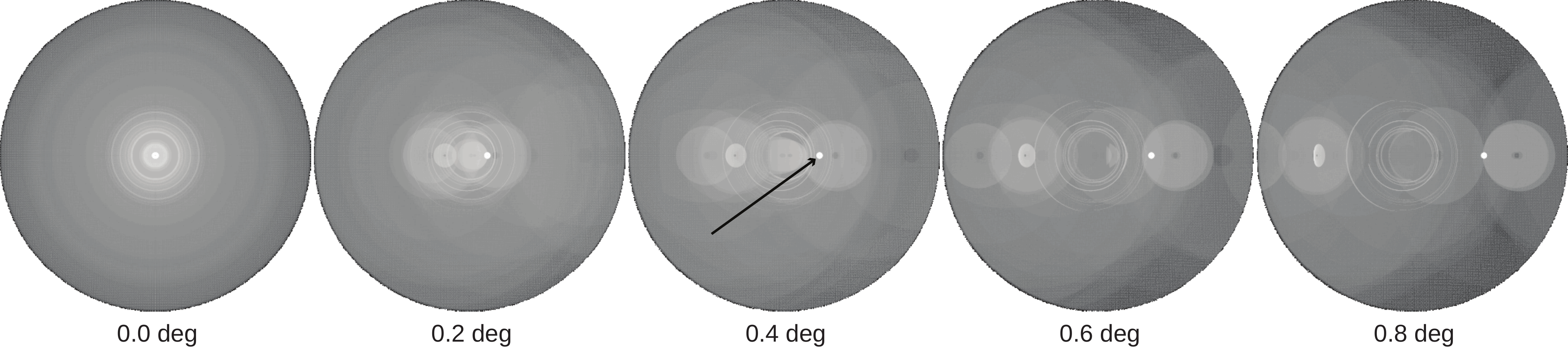} 
\caption{Images showing the complete set of two-bounce focal ghosts present at the Echo 22 focal surface for five field points.  The arrow points toward the brightest focal ghost feature, a reflection between the two ADC lenses.  In a 14,000 degree survey, reflections of the 1575 brightest stars along this ghost path contaminates 0.036\% of the sky area with irradiance greater than 1\% of NIR sky background. }
\label{fig:corF3}
\end{figure}

Ghost reflections between the optical surfaces of the corrector, including lenses, M1, and focal array sensors are unavoidable.  Management of the degrading effects of these reflections involves non-sequential ray tracing to determine the peak irradiance of ghost reflections, and quantization of the brightness of these ghost features on the focal surface for a given stellar input.  Focal, pupil and sensor reflection ghosts are 
detailed in DESI-0358, which analyzes possible double reflections amongst all surfaces in the corrector.  The brightest ghost results from a two-bounce reflection between the ADC lens elements.  This feature has a diameter of $\sim$13~mm, or about 0.024 deg$^2$ on the sky.  According to Hanuschik \cite{Hanuschik2003}, near-infrared sky background flux is $1 \times 10{}^{-17}$ erg/cm${}^{2}$/s/(sq arcsec)/\micron, or 400 photons/m${}^{2}$/s/(arcsec$^2$)/\micron at 0.8~\micron.  With an assumed reflectivity of 2\% per surface, the 1575 brightest stars in the {\it Hipparcos2} I-band catalog will have a ghost reflection greater than 1\% of the NIR sky background.  Aggregate area of these ghosts is 4.8 deg$^2$, or 0.036\% of the 14,000 deg$^2$ survey.  The ghosts are shown at three field angles in Fig~\ref{fig:corF3}. 

\subsubsection{Corrector Lens Acquisition and Polishing}

\begin{figure}[!b]
\centering 
\includegraphics[width=\textwidth]{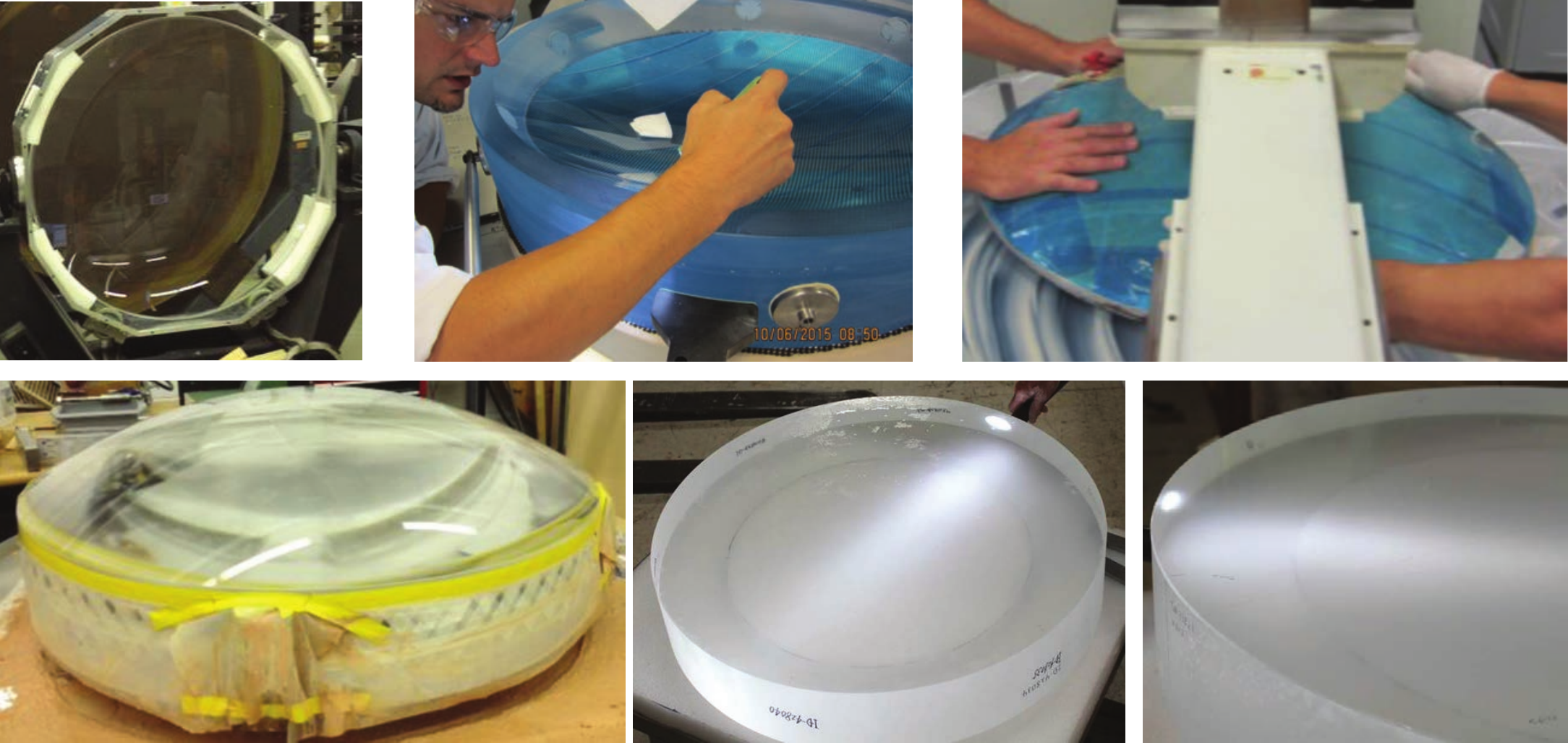} 
\caption{The six corrector lenses are show in various phases of fabrication. From top left to bottom right: C1 C1 in transmission test, C2 concave (spherical) side being checked during final figuring, C3 convex surface being polished, C4 ready for final figuring, ADC1 blank, and ADC2 being machined.}
\label{fig:corrlenses}
\end{figure}

The DESI project has purchased the optical glass blanks and provided them to the chosen optical fabrication shop. Contracts have been issued to the optical fabricators with detailed instructions on final figure and surface quality specifications.  Here we outline the main specifications for the blank procurement and the optical fabrication to meet blur requirements (see DESI-0338).  The six lenses are shown in Figure~\ref{fig:corrlenses}.

As described above, correction lens elements C1-C4 are made of fused silica, and atmospheric dispersion correctors ADC1-ADC2 are made of N-BK7.  The materials for C1-C2 and ADC1-ADC2 are required to be homogeneous to 3 ppm and C3 and C4 to 4 ppm.  Inclusions are expected to be less than 0.1 mm${}^2$ in cross section and less than 0.28 mm in size (0.15 mm for ADC1-ADC2).
Inhomogeneity introduces optical aberrations in much the same manner as a lens figure error by introducing a non-uniform optical path difference across the beam.  Effects of inhomogeneity were modeled as random Zernike modes over limited bands,
with magnitudes scaled to the maximum thickness of the lens, and specified inhomogeneity, in order to determine the effects of random inhomogeneity distributions in the lens substrate material.
The magnitudes were then flowed into the specifications for all of the lens blanks.  For the four of six lens blanks already procured, the actual measured homogeneity is analyzed in the Zemax corrector model and found to be sufficiently small enough to meet blur requirements.

All non-optical surfaces, edges, chamfers and flat faces are to be fine ground to a surface equivalent to a 600-mesh SiC loose abrasive surface texture, 1.5 to 2.5~\micron~roughness amplitude. All optical surfaces are to be ground to a better than 320-mesh SiC loose abrasive surface texture, 8~\micron~roughness amplitude. There shall be no hairline, conchoidal or other fractures or gouges in any surface.

The axis of rotation of the figure of revolution should pass through the mechanical axis of the lens with an error not exceeding 0.1~mm. The tilt between the two lens faces surfaces should not be more than 0.1~mm (across the diameter) and the run out between the two surfaces should not be more than 0.1~mm.

The diameters are to be ground to size, but the lens faces require additional material left on for the optical fabricator to fine grind and refine the surface prior to polishing. The amount of material to remain is to be left to the optical fabricator but it is not more than 1~mm nor less than 0.5~mm per face. 

The optical surfaces shall be polished to a 2 nm Ra surface texture or better. The Ra is to be measured at 50~mm intervals from center to edge using a Wyko RST 500 or similar type of non-contact surface profiler. Test results are to be given over a 1~mm${}^2$ area of surface. The scratch dig allowance shall be over the entire clear aperture. The allowance will vary from optic to optic. Surface form and tolerance on radius of curvature will be stated for each optical surface. Tilt, decenter and wedge will be stated for each optical element. Clear instructions are given on the marking of the maximum wedge angle position, as it is very important to know this for the alignment of the complete system. A detailed test procedure including any ray trace models for each optical surface has been supplied by each respective vendor. 

The lenses will be coated to maximize system throughput. Based on the experiences of DECam, the lenses will be coated with hard multilayer coatings.

\label{sec:corrector}
\subsection{Corrector Mechanical Design}

\begin{table}[!t]
\centering
\caption{Alignment and deflection requirements on the Corrector lenses and Focal Plane,  
weight requirements on Barrel/Cage and Cells and assembly/disassembly repeatability
requirements.}
\label{tab:lens_place}

\addtolength{\tabcolsep}{-2pt}
\newcolumntype{C}{>{\centering\arraybackslash}X}
\newcolumntype{R}[1]{>{\raggedright\arraybackslash}m{#1}}

\begin{tabularx}{\textwidth}{R{2in}CCCCCC}

\hline
 & \multicolumn{2}{c}{Total tolerances} & \multicolumn{2}{c}{Static tolerances} 
& \multicolumn{2}{c}{Dynamic tolerances} \\
\hline
 Element & Lateral & Tilt & Lateral & Tilt & Lateral & Tilt  \\
         & ($\pm\micron$) & ($\pm\mu$rad) & ($\pm\micron$) & ($\pm\mu$rad) 
& ($\pm\micron$) & ($\pm\mu$rad) \\
\hline
 C1         & 200 & 123 &  80 &  49 &  58 &  59 \\
 C2         &  75 & 176 &  55 &  83 &  20 &  60 \\
 ADC1       & 200 & 250 &  50 &  88 &  81 & 123 \\
 ADC2       & 200 & 175 &  50 &  87 &  81 &  79 \\
 C3         & 100 & 180 &  45 &  84 &  51 &  68 \\
 C4         & 200 & 105 &  70 &  66 &  62 &  38 \\
Focal Plane & 150 &  92 &  20 &  25 &  40 &  25 \\
\hline \hline
Barrel material & \multicolumn{6}{c}{Carbon Steel} \\
Telescope top end mass & \multicolumn{6}{c}{10,700 Kg} \\
Mass of outer ring/spiders/cage/ & \multicolumn{6}{c}{$\leq$ 7,000 Kg} \\
 /hexapod/barrel & \multicolumn{6}{c}{} \\
Mass of cells/base rings/spacers & \multicolumn{6}{c}{$\leq$ 560 Kg} \\
Barrel assembly/disassembly repeatability & \multicolumn{6}{c}{$\pm$5~\micron per flange} \\
\hline \hline
Flange flatness & \multicolumn{6}{c}{ 15~\micron} \\
Flange parallelism & \multicolumn{6}{c}{ 15~\micron} \\
\hline

\end{tabularx}
\end{table}

The six lenses of the DESI optical corrector require mounting into their 
respective lens cells, which then are mounted into the barrel of the camera.  
The four lenses C1-C4 will be mounted in individual cells with flexure 
systems to provide essentially athermal performance.  The two atmospheric 
dispersion corrective lenses ADC1--ADC2 will be housed in rotating cells.  

The total decentering and tip/tilt specifications for the lenses and
Focal Plane are outlined in the first part of Table \ref{tab:lens_place} (DESI-0612).
The static tolerance, columns 
4 and 5 in the table, refer to the maximum errors introduced in the process of aligning the lenses in the barrel. These performance requirements apply over a range of Zenith angles
from 0 to 60\degree. The static and dynamic tolerances in the table are subdivided into allocations for the barrel and the cell subassemblies, this is detailed in DESI-0335.

The second part of Table \ref{tab:lens_place} contains the barrel material, weight 
limits for the barrel and the cells, and the assembly/disassembly repeatability.
The barrel and cage mass specifications are from DESI-0617.
Table \ref{tab:corrmech} shows the specification for the cage and the hexapod that are
relevant for this writeup.  These specifications are also  from document DESI-0617.

\begin{table}[!t]
\begin{center}
\caption{Relevant Cage and Hexapod requirements (DESI-0617).}
\label{tab:corrmech}
\begin{tabular}{lc}
\hline
 Item & Current Design \\
\hline \hline
Cage outer diameter & 1.8 m \\
\hline
Hexapod resolution: & \\
\hspace{3mm} Lateral motion & 15~\micron \\
\hspace{3mm} Axial motion &  10~\micron \\
\hspace{3mm} Tip, tilt and twist &  2 arcsec \\
\hspace{3mm} Roll &  3 arcsec \\
Hexapod range: & \\
\hspace{3mm} Lateral range & $\pm$8 mm \\
\hspace{3mm} Axial range &  $\pm 10 $ mm \\
\hspace{3mm} Tip, tilt range &  $\pm$250 arcsec \\
\hline
\end{tabular}
\end{center}
\end{table}

\subsubsection{Lens Cells}\label{sec:lenscells}

The lens cell provides the interface between the lens and the barrel of the wide field corrector.  Lenses can be mounted to the cell in a variety of ways.  One of the simplest is to use an elastomeric mount where the lens to cell interface is an RTV (room temperature vulcanized) rubber-like material. RTV rubbers have much higher coefficient of thermal expansion (CTE) values than typical glass and metals, with common materials ranging from 200 to 300 ppm/$\celsius$. RTV rubbers are essentially incompressible, and exhibit varying effective stiffness and CTE, depending on boundary conditions (constrained and free surfaces determine where the rubber can expand under load). Elastomeric mounts can be made to be essentially athermal with a suitable choice of elastomer thickness (\ie, there is little induced stress in the components with temperature change). This solution has been adopted by several of the current large wide field correctors on telescopes such as the Blanco, CFHT, LBT and MMT. 

\begin{figure}[tb]
\centering
\includegraphics[width=2.5in]{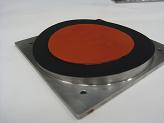}  \hspace{.25in} \includegraphics[width=2.5in]{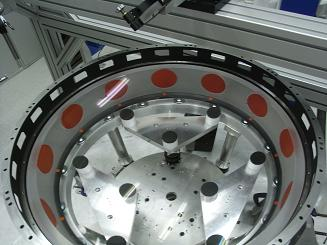}
\caption{Mounting of the DECam optics. Shown at left is a pad and insert, and at right the radial and axial pads attached to the lens.}
\label{fig:UCL1}
\end{figure}

The design chosen for  DESI  is to use discrete pads of RTV for both radial and axial support (Figure~\ref{fig:UCL1}). This method, which was adopted by both the MMT and DECam projects, has the advantage that the pads can be manufactured to high dimensional accuracy and consistency before mounting.  The pads can be connected to inserts, made of the same material as the cells, which are inserted through holes in the side of the lens cell and screwed into place. The pads themselves are either glued to the lens with a thin layer of RTV (radial pads) or just pressed against the lens (axial pads).  For each case the RTV will exhibit different properties, and this is taken into account in the design. 

The cell material is an important choice in the design, since it affects the thickness of the radial pads, as discussed below. The radial pad thickness should not be too great, as this means the lens will not be held stiffly under a changing gravity vector.  The thickness also should not be too small (less than $\sim$1 mm) as this poses problems for attachment to the inserts.  

Thermal stability is a key issue in the design. Ideally the lens cell would be made of the same material as the barrel, so that the expansion and contraction with temperature change is matched. The choice of the barrel material is based on cost, ease of production and weight. In the DESI case the preferred choice of barrel material is steel.  Steel, however, has a large coefficient of temperature expansion (CTE), $\sim$12 ppm/$\celsius$.  This is greatly different from that of fused silica, with a CTE of 0.5 ppm/$\celsius$, and somewhat different from that of N-BK7 with a CTE of 7.1 ppm/$\celsius$. If steel were used for the cell as well as the barrel, a strictly athermal cell design for a fused silica lens might require a thick layer of RTV ($\sim$9 mm) to compensate for the difference in CTE (though a thinner layer could be used as long as the thermal stresses stayed within certain allowances).  This thickness would make it difficult to meet the system decentering tolerances. 

An alternative material for the fused silica lens cell is nickel-iron alloy. This can have a low CTE, more closely matching the fused Silica CTE, allowing a thinner layer of RTV to be used in an athermal design. Invar is a well known nickel-iron alloy (36\% nickel) with CTE of $\sim$1 ppm/$\celsius$. However this choice of material for DESI fused silica lens actually leads to very thin radial pads ($<$1 mm). An alternative is to use a nickel-iron alloy with a higher percentage of nickel.  For DECam and the MMT designs, an alloy of nickel-iron with 38\% nickel was chosen, giving a CTE of $\sim$3 ppm/$\celsius$ which led to radial pad thicknesses of 1.4 to 2 mm.  However, if Ni/Fe is used for the cell there will be a thermal mismatch between the cell and the barrel if the barrel is made of steel. To overcome this, a series of thin flexures can be used to connect the lens cell to the barrel. These flexures then compensate for the differential thermal expansion. 

\paragraph{Lens Cell Design} 

\begin{figure}[!t]
\centering
\begin{minipage}[b]{0.45\linewidth}
\includegraphics[width=2.5in]{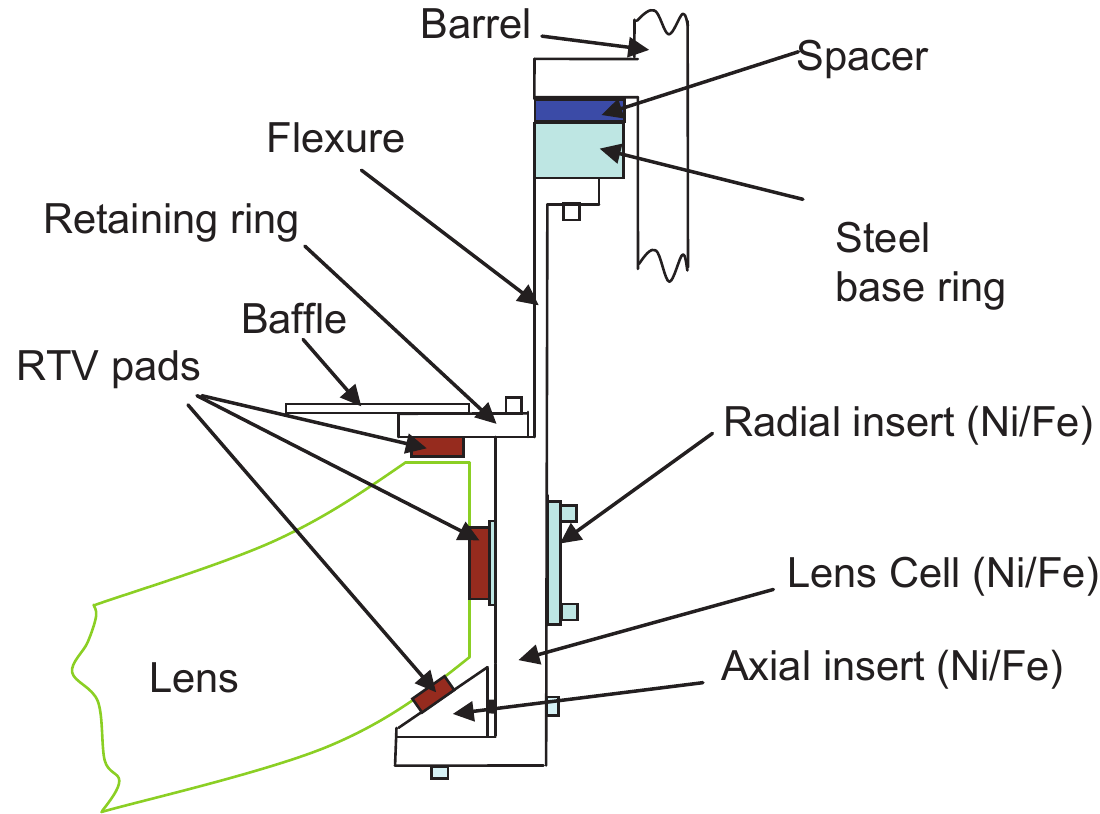}
\caption{Schematic of fused silica lens cell design.}
\label{fig:UCL3}
\end{minipage}
\hspace{0.25in}
\begin{minipage}[b]{0.45\linewidth}
\includegraphics[width=2.5in]{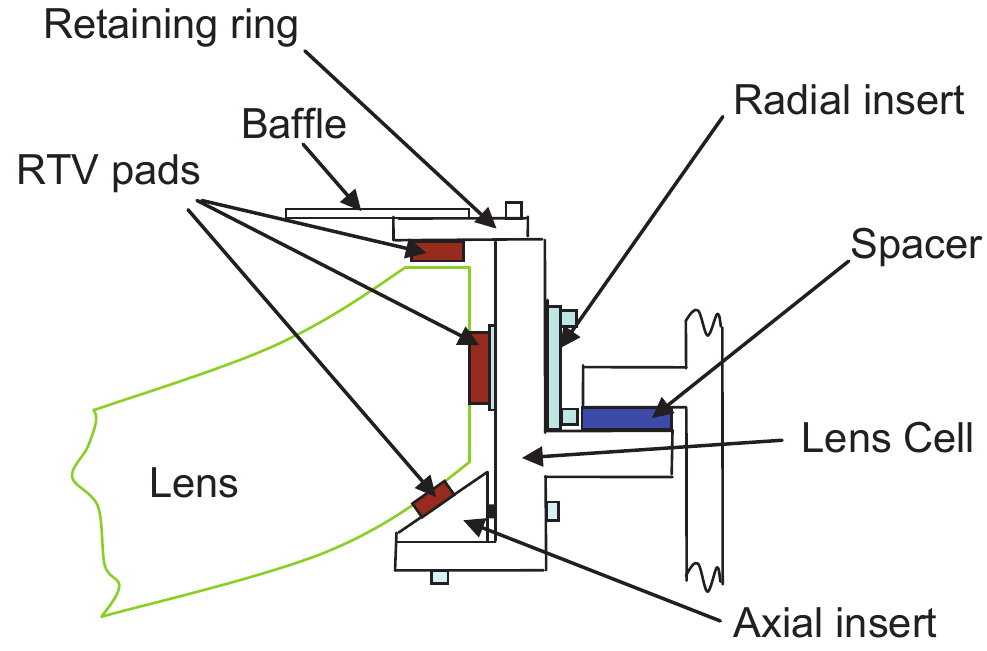}
\caption{Schematic of N-BK7 lens cell design.}
\label{fig:UCL4}
\end{minipage}
\end{figure}

A baseline design for the cells for the fused silica lenses is shown in Figure~\ref{fig:UCL3}.  It is similar to the one used for the DECam lenses with a few modifications.                       
Both the axial and radial pads are mounted on inserts. These inserts can be precision machined and accurately positioned to meet the tight tolerances on the lens mating surfaces derived from the optical system alignment tolerances. The axial pad would be glued to the axial insert but not to the lens, whilst the radial pads would be glued both to the inserts and the lenses.  A retaining ring and baffles would also be included; the retaining ring pads do not touch the lens so as not to over-constrain it. The Young's modulus of the pad is affected by both its S-factor and by whether the pad is glued either on one side or on both sides \cite{Fata04}.  The allowable stresses in the lens is governed by the stress induced birefringence with a limit adopted of $<3$~MPa. The target for the max cell stress is to be a factor of 2 below the yield strength (228 MPa). For C1, due to its position above the primary mirror, a safety factor of 4 would be required but this can be reduced to a factor of 2 for the stresses if additional backup safety constraints are installed on the cell (it is planned to install wire safety straps on cells connected from the main body to the steel ring part of the cell).

The cell design for the C1 and C4 cells is slightly different to that of the C2 and C3 cells. The C1 and C4 cells have narrow slots (5mm width) between the flexures whilst the C2 and C3 cells have a more crenelated design with the gap between flexures being the same size as the flexure as in the designs for DECam. The thinner gaps for C1 and C4 allow for a stronger support for these heavier lenses and easier sealing of the cells to restrict airflow and dust ingress.

Due to the higher CTE of N-BK7 a steel cell design is proposed, yielding an athermal radial pad thickness of $\sim$3.5 mm (Figure~\ref{fig:UCL4}).  This can meet the dynamic lens decentering tolerances.
As both the cell and barrel structure are steel, the design does not need flexures between the two.  
Again the axial and radial pads are mounted on inserts.

\paragraph{Radial and Axial Pad Design}

The axial and radial RTV pads provide the interface between the metal cell and the lens. The optimum radial pad thickness for an athermal behavior can be calculated from equations given in Doyle, Michels \& Genberg \cite{athermalbonds} and the thicknesses determined for each lens and cell are shown in the last column of Table~\ref{tab:UCL2}. 

The axial pads have two functions; one is to take up any irregularity of the lens pad surfaces (and hence lens cell surface) and ensure an even spread of the loading on the lens; the other is to be stiff enough to hold the lens so that a change in gravity vector does not produce too much tilt or decentering of the lens. (Of course the radial pads also provide a measure of resistance to tilt and decenter.)  For DECam the axial pads for the C1 lens, which is directly comparable in size to the DESI lenses, were $10\times10\times1$~mm in size (S-factor of 2.5).

\begin{table}[!b]
\centering
\caption{ Lens masses and pad parameters.}
\begin{tabular}{c|c|c|c|c|c} 
\hline 
Lens & Mass & Axial Pad Size & S  & No.  & Radial Pad Size\\ 
     &  kg  & w, l, t    & Factor & Axial and &  w, l, t \\ 
     &      &    (mm)    &        & Radial Pads &     (mm)  \\ 
\hline 
C1    & 201   & 10, 10, 1    & 2.5 & 48  &  32,32,2.17 \\ 
C2    & 152   & 10, 10, 1    & 2.5 & 24  &  80,80,1.62 \\ 
C3    & 84    & 10, 10, 1.5  & 1.67 & 24 &  45,45,1.60 \\
C4    & 237     & 10, 10, 1    & 2.5 & 48  &  25,25,2.00 \\ 
ADC1  & 102   & 10, 10, 1    & 2.5 & 24  &  60,60,3.50 \\ 
ADC2  & 89     & 10, 10, 1.5  & 1.67 & 24 &  50,50,3.55 \\ 
\hline
\end{tabular}
\label{tab:UCL2}
\end{table}

\paragraph{FEA Analysis} 

A finite element analysis (FEA) has been made of the lens/cell designs to investigate performance. The cells and barrel are to be assembled at 20\celsius and the survivability temperature range for the corrector is -20\celsius to +60\celsius (the latter is accommodate shipping). For survey operations the lens and cells need to meet the alignment requirements over zenith angles of 0-60\degree. 
 
\begin{figure}[ht]
\centering
\begin{minipage}[t]{0.44\linewidth}
\includegraphics[width=\textwidth]{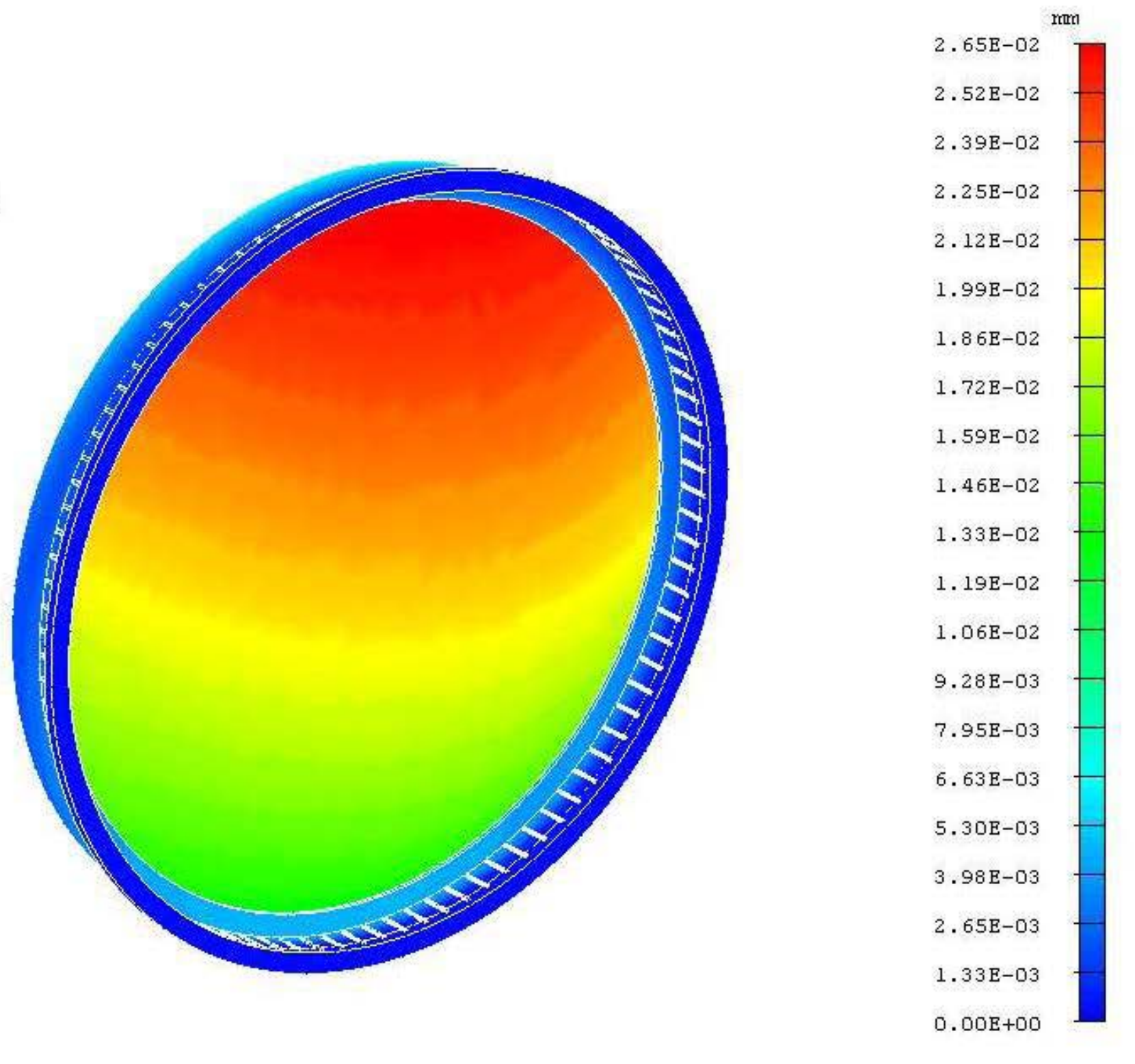}
\caption{FEA model showing tilt of C4 lens and cell at a 60\degree Zenith angle.}
\label{fig:UCL_FEA1}
\end{minipage}
\hspace{0.1\linewidth}
\begin{minipage}[t]{0.44\linewidth}
\includegraphics[width=\textwidth]{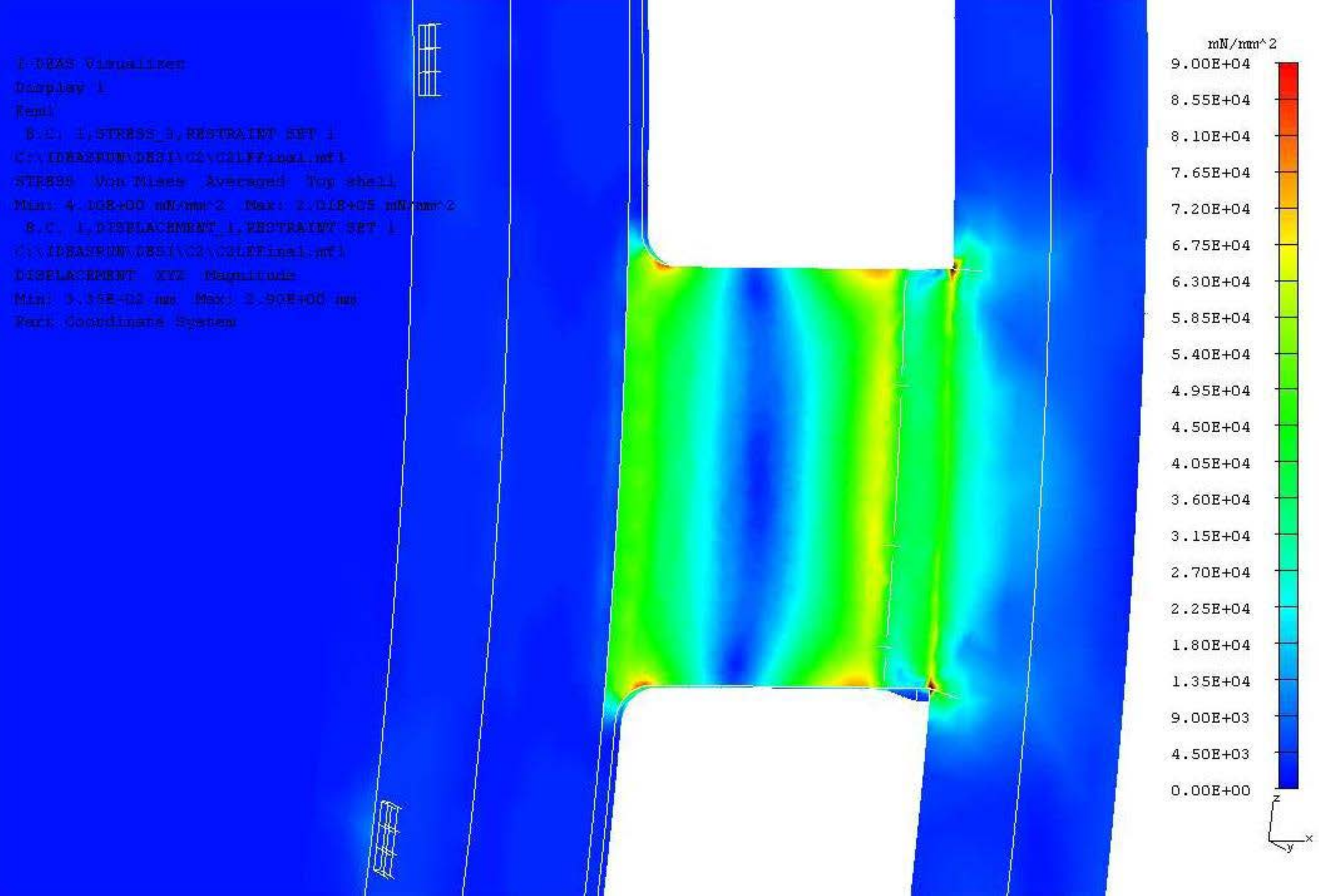}
\caption{FEA model showing stresses in C2 flexure with a temperature change of 40\celsius.}
\label{fig:UCL_FEA2}
\end{minipage}
\end{figure}

The FEA models used designs that were slightly simplified for ease of modeling.  The radial pads are modeled as attached to a continuous cell and the axial pads are modeled as effectively stuck to the lens as well as the cell.  This affects the value of the Young's modulus, which is accounted for by adjusting the material properties of the RTV.  
In all designs the cell wall thicknesses were 20 mm, the flexure lengths for the Ni/Fe cells were 42~mm, and the flexure thicknesses were 2~mm. Figures \ref{fig:UCL_FEA1} and \ref{fig:UCL_FEA2} show examples of the FEA modelling results.  Table~\ref{tab:UCL5} summarizes the predicted max stresses, tilts and decenters. These are small contributions to the overall budget in the requirements Table~\ref{tab:lens_place}.
 
\begin{table}
\centering
\caption{Predicted lens cells stresses, lens tilt (across diameter) and lens and cell decenter tilt and decenters for 30\degree and 60\degree from zenith.}
\begin{tabular}{c|cc|cc}
\hline 
\\ 
Lens & Max stress cell & Max stress cell & Lens tilt & Lens  and cell decenter \\
 &  -40\celsius change & 90\degree Zenith angle & 60\degree Zenith angle & 60\degree Zenith Angle \\
 & MPa & MPa &   $\mu $m & $\mu $m  \\ 

\hline 
  C1       & 91 & 5.8  &12.7 & 6.0 \\
  C2       & 88 & 8.3  & 0.3 &2.8 \\
  C3       & 76 & 4.1  & 5.2 & 2.3 \\
  C4       & 80 & 5.0  &15.6 &10.5\\
  ADC1   & $<$1.0 & 0.43 & 2.1 &1.0 \\
  ADC2   & $<$1.0 & 0.21 & 6.9 &2.0 \\
\hline 
\end{tabular}
\label{tab:UCL5}
\end{table}

\subsubsection{Barrel}\label{sec:barrel}
The details of the final design of the barrel for the Echo 22 optics have been
outlined in the document ``Final design of the DESI corrector barrel for the
Echo 22 optics'' \cite{barrel_cdesign}.  Here we will summarize the main points of 
the design.

Three main concepts were followed in the design of the barrel:

\begin{itemize}

\item  
Use a single shell design as a way of using mass effectively 
(see DESI-0330) and to minimize primary mirror obscuration.

\item
Minimize the number of barrel sections. The barrel has 3 
sections (plus the Focal Plate Adaptor) in order to be able to insert 
one lens or ADC from either end of each section.

\item
Connect the hexapod through a single flange to the barrel.  This reduces 
primary mirror obscuration and allows a cost, schedule and risk reduction 
by using essentially the same hexapod that were used in DES.

\end{itemize}

\begin{figure}[!h]
\begin{center}
 \includegraphics[height=.30\textheight]{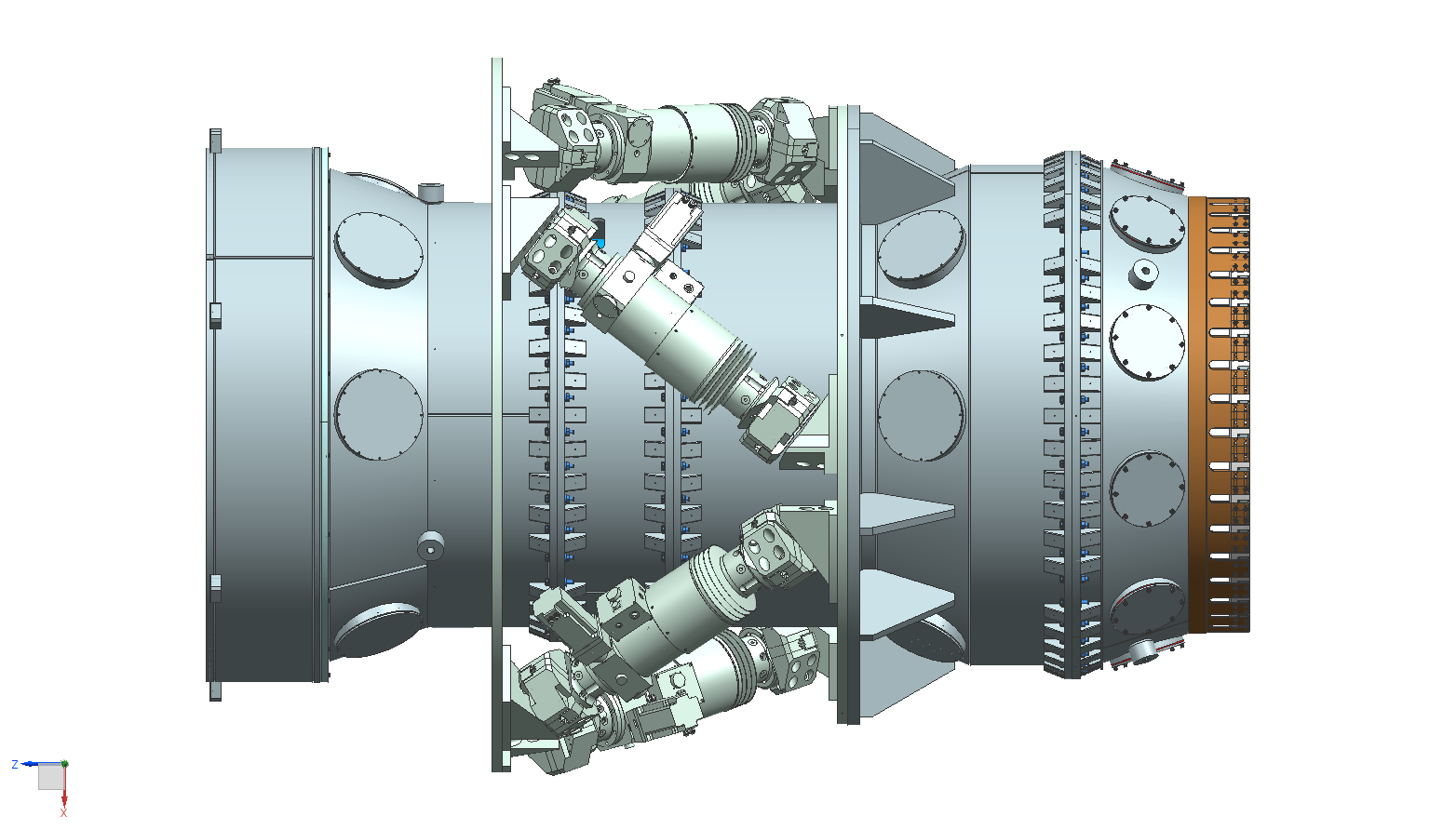}
  \caption{This picture shows the DESI corrector barrel with the DECam 
hexapod attached to the barrel.  The hexapod configuration for the DESI barrel is very similar to the one for DECam.}
\label{fig:hex_oneside}
\end{center}

\begin{center}
  \includegraphics[width=0.90\textwidth]{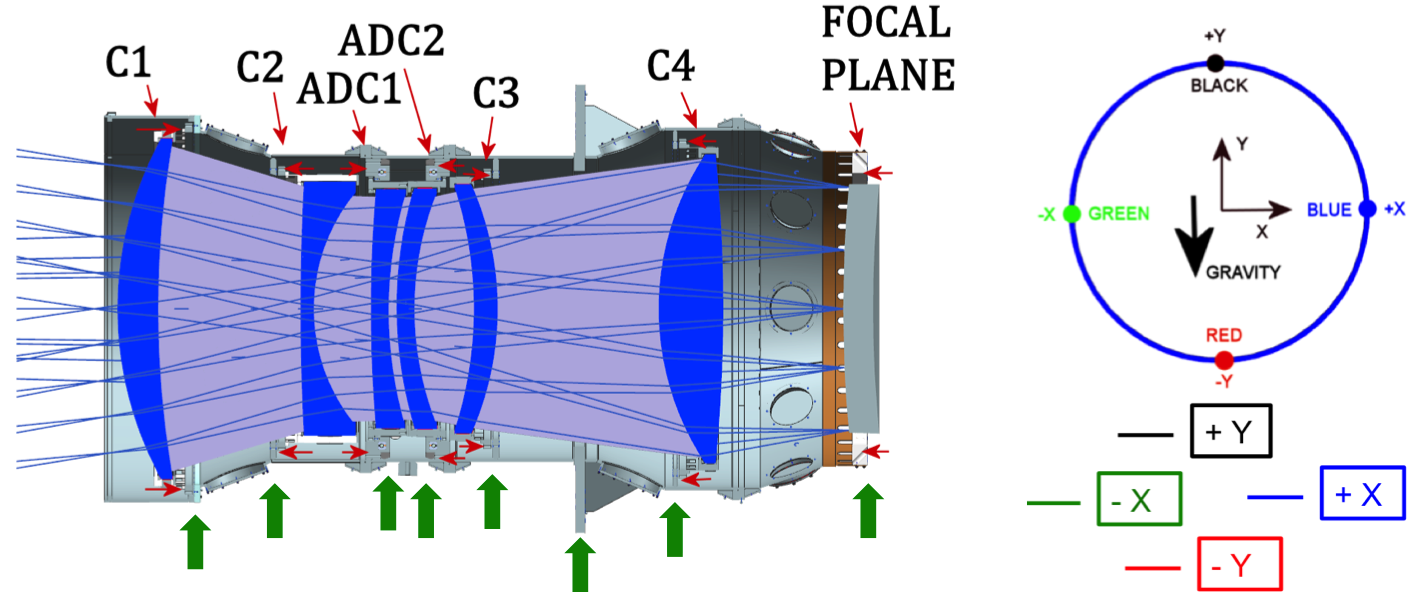}
  \caption{Section view of the final barrel design showing how the lenses are
positioned inside the barrel.  The red arrows show the direction in which the 
lenses attach to the barrel.  The light purple volume is the keep out volume
for the light created by joining the clear apertures for each lens given in 
Table \ref{tab:clear_appertures}.
For the FEA analysis deflections were tracked at eight different planes located where
the green arrows point to the barrel.  In each plane four points were tracked
as shown in the picture on the right hand side.  The points are color
coded as black $+Y$, blue $+X$, red $-Y$ and green $-X$.}
\label{fig:FEA_points}
\end{center}
\end{figure}

Figures \ref{fig:hex_oneside} and \ref{fig:FEA_points} shows the barrel design for the Echo 22 optics. The order of the lenses from left to right is C1, C2, ADC1, ADC2, C3,
and C4.  Figure \ref{fig:hex_oneside} also shows how the hexapods attach to the large diameter flange in the middle the barrel.  The last flange on the right is where the focal plane assembly attaches to the barrel.  The circular holes with covers on the barrel sides are access ports.

\paragraph{Barrel Design}\label{sec:prel_barrel_design}

The final design of the corrector barrel meets all the DESI requirements 
and takes into consideration all the Interface Control Documents. Some of the requirements are listed on Tables \ref{tab:lens_place} and \ref{tab:corrmech}, a full compliance matrix is given in Appendix A of the barrel final design report \cite{barrel_cdesign}.
The barrel design provides a minimum number of flanges and sections, a homogeneous shell thickness, and the interfaces for the hexapod movable flange, the lens cells, the ADC rotation mechanisms and the focal plane assembly.

\begin{figure}[thb]
\begin{center}
  \includegraphics[height=.45\textwidth]{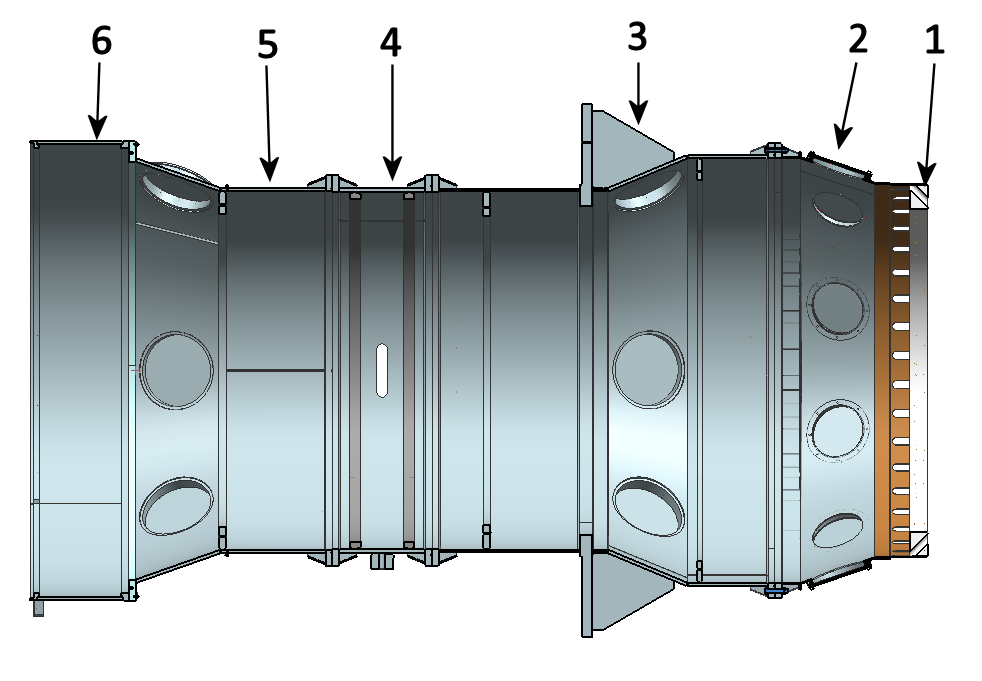}
  \caption{Barrel sections: (1) FPD ring, (2) FPD section, (3) AFTER section, 
(4) MIDDLE section, (5) FRONT section, and (6) SHROUD section.}
\label{fig:barrel_g1}
\end{center}
\end{figure}

As shown in Figure \ref{fig:barrel_g1}, the barrel consists of four main sections.
From left to right the sections are: the shroud (SHROUD), the front section (FRONT),
the middle section (MIDDLE), the after section (AFTER) and the focal plane 
adaptor (FPD).  The barrel includes flanges to support the C1, C2, ADC1, ADC2, C3 
and C4 lens cells and the FPD.  

The AFTER section is a thick shell which provides connection flanges on 
both ends and support rings for the cells.  The shell provides six ports 
with an inner bore of 180 mm used to provide access to measure lens spacings and 
lens cleaning if necessary. Each port has a cover bolted with a set of M6 screws 
and a gasket to seal the inner environment of the barrel.  The downstream flange 
is the interface connection with the FPD section and the upstream end provides the 
connection flange with the MIDDLE section. The AFTER section is welded to the 
Hexapod Mounting Plate (HMP) which is the connection plate to the moving plate 
of the Hexapod system. The inner bore includes two thick rings welded to the 
shell and used to support the C3 and C4 lens cells.

The MIDDLE section is the shortest section of the barrel and provides 
connection flanges on both ends, similar to the AFTER section, and support 
rings for the ADC rotators. The downstream flange is used to connect the 
MIDDLE section to the AFTER section while the upstream flange connects the 
MIDDLE section to the FRONT section. Two thick rings are welded to the inner 
bore to support the ADC1 and ADC2 bearings and rotating mechanism. The overall length of the MIDDLE section covers both ADC lenses. Two plates with threaded holes are provided on the shell to accommodate the driving mechanism of the ADCs.
   
The FRONT section consists of two welded pieces. The Upstream piece has a 
conical shape, the large aperture of this cone is connected to the SHROUD and 
the small aperture is welded to the cylindrical piece of the FRONT section.  
This cylindrical piece provides the connection with the MIDDLE section through 
a bolted flange. In addition there are six ports in the conical section, these 
ports are similar in design to the access ports used in the AFTER section. 
A thick ring is welded to the end of the large cone aperture and it supports 
the C1 lens cell through a set of M6 threaded holes. Another ring welded at 
the intersection between the conical and cylindrical pieces supports the C2
lens cell.

The SHROUD section extends 38 mm beyond the C1 cell with the purpose to protect 
the C1 lens.  The upstream side of the SHROUD includes a set of tapped holes 
used to install a protective cover.  The other side is attached to the large 
side of the FRONT section cone through a set of bolts and it is align to it 
through a set of pins. A set of ears are welded to the outer shell of the SHROUD
to hold a set of laser retroreflectors that will be use to locate the barrel 
with respect to the primary mirror during the installation phase.

All sections, except the SHROUD, provide two machined surfaces on the outer surface 
of the end flanges to be used as reference during the process of installing the 
cells plus lenses in the barrel at UCL.  Each section 
also includes three rows of tooling balls that will be used as reference during 
the barrel and cell alignment at Fermilab, and a set of lifting points used to 
support the section from its center-of-gravity when needed. A pinning system, 
designed to align each barrel section relative to each other with high 
accuracy is also included in the matting flanges.  
The mating flanges for all barrel sections, except the SHROUD, include an 
O-ring system used to seal the inner environment of the barrel.

The barrel shell is designed taking in consideration the installation of the 
cells, the ADC rotating mechanism and the ADC transmission system. On the 
inner bore the radial contact width for the cell support ring is 30mm for 
C1 to C4, and 13.5mm for ADC1 and ADC2. 
As shown in Figure \ref{fig:FEA_points} the barrel shape is design to accommodate
the light flow through the barrel and at the same time minimize the barrel
diameter.  This done in order to fit the barrel and the hexapods inside a cage
with a maximum outer diameter requirement of 1.8 meters.
The light purple volume in Figure \ref{fig:FEA_points} is the keep out volume
for the light; this volume was created by joining the clear apertures for each 
lens given in Table \ref{tab:clear_appertures}.

\begin{table}[htb]
\begin{center}
\caption{Clear apertures for each of the barrel lenses and ADCs.} 
\label{tab:clear_appertures}
\begin{tabular}{l c c}
\hline
 Lens & Surface &  Clear aperture  \\
      &         & (diameter in mm) \\
\hline
\hline
 C1   & s1 &  1110 \\
      & s2 &  1086 \\
\hline
 C2   & s1 &   820 \\
      & s2 &   750 \\
\hline
 ADC1 & s1 &   750 \\
      & s2 &   746 \\
\hline
 ADC2 & s1 &   750 \\
      & s2 &   750 \\
\hline
 C3   & s1 &   780 \\
      & s2 &   804 \\
\hline
 C4   & s1 &  1004 \\
      & s2 &  1004 \\
\hline
\end{tabular}
\end{center}
\end{table}

The bolted joints between sections are designed with high stiffness 
using a single set of bolts. To maintain equal spacing between bolts,  
flanges with different diameters have different number of bolts, 
as indicated in Table \ref{tab:bolts}.  Also to minimize barrel deflections
and to make the flanges stiff enough for high precision machining, the
flanges are reinforced with gussets.

\begin{table}[htb]
\caption{Bolt sizes, number of bolts and flange thicknesses for different flanges.} 
\begin{center}
\label{tab:bolts}
\begin{tabular}{l c c c}
\hline
 Connecting flange & Bolt size & No. of Bolts & Joint Thickness \\
 &  &  & (mm) \\
\hline
FRONT to MIDDLE      & Class 10.9, M10x1.5 & 40 & 37  \\
MIDDLE to AFTER      & Class 10.9, M10x1.5 & 40 & 37  \\
AFTER Section to FPD & Class 10.9, M10x1.5 & 50 & 38  \\
\hline
\end{tabular}
\end{center}
\end{table}

A set of tapped holes is provided on the inner flanges of each barrel section 
to support the spacers and the lens cells base rings. Each cells has a specific 
number of holes as specified by the Interface Control Document between the 
lens cells and the corrector barrel documents.  For a summary see
Table \ref{tab:lens_par}, for the full ICD document see Reference 
(\cite{bib:cell-barrel}).

\begin{table}[htb]
\caption{Relevant parameters for attaching the lens cells to the barrel and the
ADC lenses to the ADC rotator mechanism.}
\begin{center}
\label{tab:lens_par}
\begin{tabular}{ccccccc}
\hline
\hline
 Lens & Number of bolts & Bolt hole & Bolt hole & Screw & 
\multicolumn{2}{c}{Max/Min base ring} \\
     & (equally spaced) & PCD (mm) & size (mm) & type & 
\multicolumn{2}{c}{diameter (mm)} \\
\hline
C1  &  96  &  1224  &  7  &  M6x1  &  \multicolumn{2}{c}{1242/1182} \\
C2  &  48  &  934   &  7  &  M6x1  &  \multicolumn{2}{c}{952/892}   \\
C3  &  48  &  918   &  7  &  M6x1  &  \multicolumn{2}{c}{936/876}   \\
C4  &  96  &  1118  &  7  &  M6x1  &  \multicolumn{2}{c}{1136/1076} \\
\hline
\hline
 Lens & Number of bolts & Bolt hole & Bolt hole & Screw & 
\multicolumn{2}{c}{Maximum diameter (mm)} \\
\cline{6-7}
     & (equally spaced) & PCD (mm) & size (mm) & type & 
 cell flange & cell body \\
\hline
ADC1  &  24  &  868  &  7  &  M6x1  &  882  &  855 \\
ADC2  &  24  &  868  &  7  &  M6x1  &  882  &  855 \\
\hline
\end{tabular}
\end{center}
\end{table}

The main barrel weights and materials for each section are given in Table \ref{tab:section_weights}.  According to the DESI specs (see DESI-0617) the total mass of the Upper Ring, Spiders, Cage, Hexapods, Corrector Barrel and associated fasteners is not to exceed 7000 kg.  The weight of the barrel is 1229 kg.  Our best estimate of the other 
elements are: ADC rotator (220), hexapods (936 kg), cage rails (492 kg), cage hexapod 
ring (679 kg), cage end ring (314 kg), cage fins (351 kg), brackets and pins 
(145 kg), top ring (1437 kg).  With a 20\% safety margin in all the weights except 
the barrel we get a total of 6720 Kg.  So we are within the requirements.

\begin{table}[htb]
\begin{center}
\caption{Barrel Section Weights and Materials} 
\label{tab:section_weights}
\begin{tabular}{clrc}
\hline
 & Section & Weight & Material\\
 &  & (Kg) & \\
\hline
1 & FPD ring        &   25 & Aluminum 6061-T6 \\
2 & FPD             &   82 &        AISI 4130 \\
3 & AFTER Section   &  651 &         ASTM-A36 \\
4 & MIDDLE Section  &  136 &         ASTM-A36 \\
5 & FRONT Section   &  227 &         ASTM-A36 \\
6 & SHROUD          &   97 &         ASTM-A36 \\
7 & Hardware        &   11 &       Class 10.9 \\
 & Total Mass & 1229 \\
\hline
\end{tabular}
\end{center}
\end{table}

The hexapod is attached to the barrel through the Hexapod Mounting Plate (HMP).  
This mounting plate is located at about $150 \pm 100$ mm downstream of the 
center-of-gravity (CG) of the barrel.  The CG of the barrel was placed
a bit away from the HMP and toward the hexapod fixed flange to reduce the barrel 
rotation induced by the hexapod actuators finite stiffness and to increase the
space allowed for the ADC rotator mechanisms.  The weights and distances of the
lenses, ADCs, the FPD and the barrel that were used in the calculation of the 
CG are given in Table~\ref{tab:CoG}.

\begin{table}[htb]
\begin{center}
\caption{Cells, Focal Plane and Barrel weights and center of gravity.} 
\label{tab:CoG}
\begin{tabular}{lcc}
\hline
 & Mass & CoG Distance from (0,0,0) \\
 & (Kg) & [mm] \\
\hline
 C1     &  280 & 2294 \\
 C2     &  271 & 1750 \\
 ADC1   &  243 & 1523 \\
 ADC2   &  223 & 1414 \\
 C3     &  148 & 1228 \\
 C4     &  308 &  484 \\
 FP     &  870 & -82  \\
 Barrel & 1229 & 1222 \\
\hline
\end{tabular}
\end{center}
\end{table}

\begin{figure}[!h]
\begin{center}
  \includegraphics[height=2.8in]{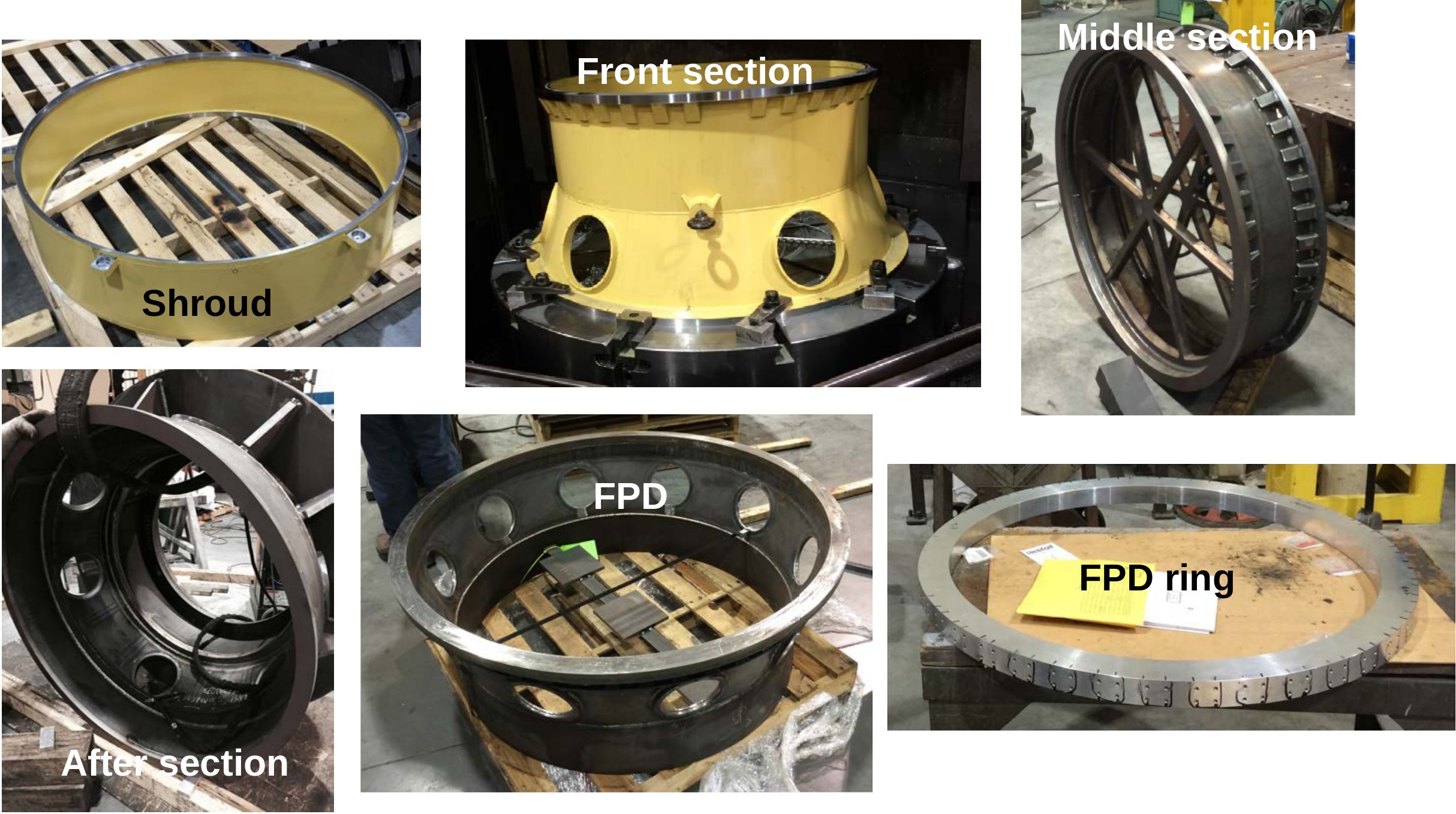}
  \caption{Barrel pieces as being fabricated during February 2016.  Counterclockwise from the top left plot we see the Shroud, the Front section, the Middle section, the After section, the FPD and the FPD Aluminum ring.}
\label{fig:barrel_parts}
\end{center}
\end{figure}

Fabrication of the barrel pieces has started.  Figure \ref{fig:barrel_parts} shows the
Shroud, the Front section, the Middle section, the After section, the FPD and the
FPD Aluminum ring as were being fabricated in February of 2016.

\paragraph{The interface between the barrel and the Focal Plane Assembly}\label{sec:FPA}

The Focal Plane Adapter assembly consists of three main components: the FPD 
section, the FPD ring and the FPD housing. The FPD assembly is in the LBNL 
scope of work, see references \cite{bib:focal_plate}, \cite{bib:ICD_FP-barrel} 
and \cite{bib:coordinates}.

\begin{figure}[!thb]
\begin{center}
  \includegraphics[height=2.5in]{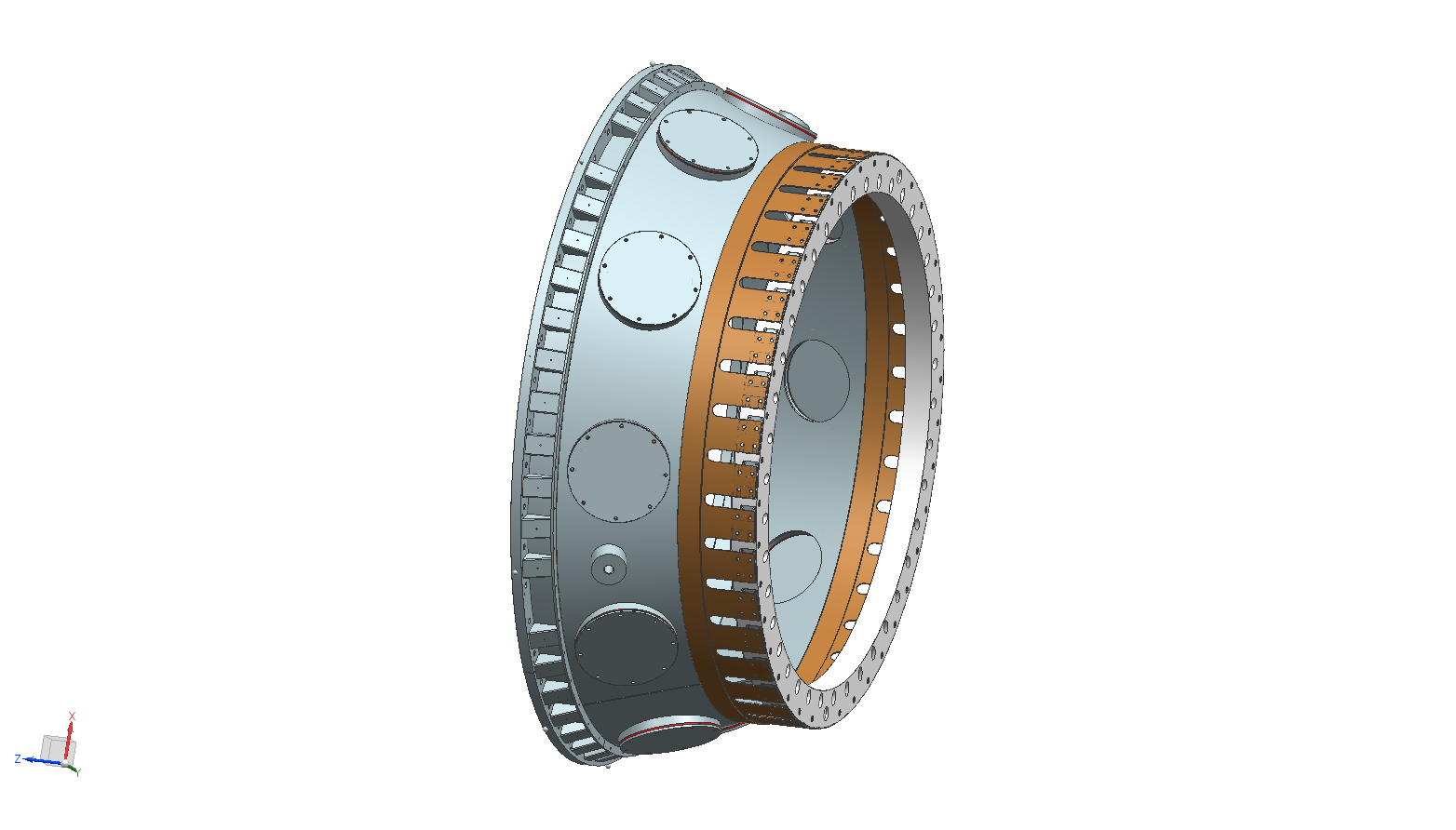}
  \caption{This picture shows the Focal Plane Adaptor or FPD.  The left side
attaches to the barrel and the right side to the Focal Plane Assembly.}
\label{fig:fpd}
\end{center}
\end{figure}

The FPD section is a 1/8 of an inch thick shell which on one end provides a 
large aperture bolted to the AFTER section.  On the other end the FPD is 
made with multiple free flexures which provide bolted lap joints as interfaces 
to the FPD ring as shown in Figure \ref{fig:fpd}. The main reason of 
the multiple free flexures is to match the different CTEs of the barrel 
(steel) and the Focal plane system (Aluminum).

The FPD ring is a thick aluminum ring used as connection with the FPD section 
and the FPD housing. It has on the outer surface tapped holes used to bolt down 
the flexures joints and the radial shims while the other side has through holes 
used to attach to the FPD housing.  LBNL did the structural design of the FPD assembly 
and provided the 3D model to Fermilab along with the ICD. Fermilab created the 
production model and drawings taking in considerations the LBNL design. 
Fermilab will align the FPD relative to the barrel coordinate system, install 
enough tooling balls on the FPD so the coordinate system can be reproduced and 
ship the FPD to LBNL. The FPD ring will be aligned together with the FPD 
section during the manufacturing process then LBNL will align the Focal Plane 
Assembly relative to the FPD.  The LBNL FPD housing is a second ring which 
surrounds the Focal Plane and it is secured to the FPD ring through preloaded 
screws.  The LBNL Focal Plane Assembly is aligned and attached to the downstream 
end of the AFTER section through bolted flanges and an O-ring to reduce 
the dust flow from the external environment. 

\paragraph{The FEA analysis}\label{sec:FEA}

In this section we will describe the results of the FEA analysis.
For deflection purposes, the deflections were measured in different planes as 
shown by the green arrows in Figure \ref{fig:FEA_points}.  These planes 
correspond to the mounting flanges for the C1, C2, C3 and C4 lenses, the ADC1 and 
ADC2 rotators, the Hexapod Mounting Plate (HMP) and the FPD.  In each plane we 
tracked the deflections at four points.  The points are shown in Figure 
\ref{fig:FEA_points} in the plot on the right hand side.  The points
are color coded as black $+Y$, blue $+X$, red $-Y$ and green $-X$.
This code was used in all the plots in this Section.  The plane tilt was calculated as the axial displacement of the top point minus the lower one divided by the diameter.
For the out-of-roundness checks we plotted the transverse shape of the
C2, C4 and the FPD flanges.  Even though the C1 flange is further away from the 
HMP, and we would therefore expect a larger out-of-roundness, we see that the 
out-of-roundness for the C1 flange is smaller than the out-of-roundness of the C2 
flange.  This is due to the fact that the C1 flange needs to be much wider because
the SHROUD attaches to this flange.  The values of the barrel out-of-roundness are
all within specification, for the details of these studies see Reference \cite{barrel_cdesign}.

\paragraph{FEA results for the final barrel design} \label{sec:rflanges}

In this section we will describe the FEA analysis for the final barrel
design.  We will show the FEA deflections results for the barrel in the horizontal position, which corresponds to pointing the telescope at 90 degrees from zenith.  
Deflections scale as $\sin\alpha$, where $\alpha$ is the zenith angle.  So, for example, to get the deflections at 60$^\circ$ from zenith we have to multiply the deflections 
by $\sin60^\circ=0.866$.  As the telescope rotates around the polar axis the 
hexapods will change their positions with respect to gravity.  So to check the 
stability of the deflections with respect to telescope rotations we ran FEA 
analysis for 0, 40 and 80 degrees of rotation around the barrel axis.  Since the 
hexapod geometry is such that it looks the same after a 120$^\circ$ rotation 
around the axis of the barrel, we considered that three angles between 0 and
120 degrees would be enough to characterize the deflections for different telescope
orientations.  For completeness we also run a case with the barrel in the vertical
position, or with the telescope at zenith.

\begin{figure}[htb]
\begin{center}
  \includegraphics[width=.95\textwidth]{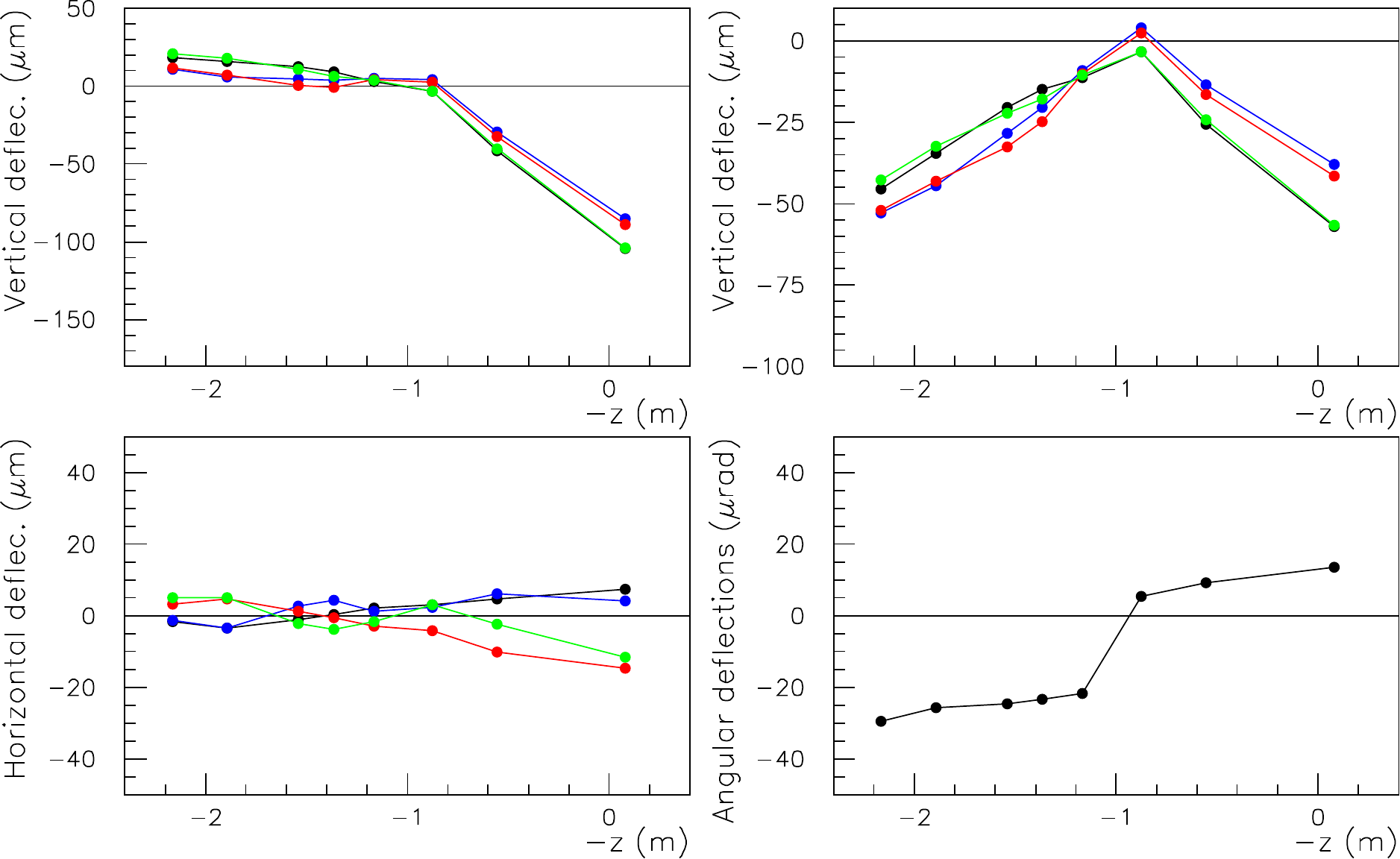}
  \caption{Deflections for the final barrel design with the telescope
in the horizontal position and a 0$^\circ$ rotation around the barrel axis.  The plots 
show the vertical deflections produced by the FEA calculations after subtracting 
the sag due to the hexapods (top left), the same deflections after an overall 
rotation of the barrel to make the deflections at both ends equal (top right), 
the horizontal deformations (middle left), and the angular deflections
after the barrel has been rotated to make deflections at both ends equal (middle right).  
The bottom left figure shows the FEA vertical deflections.  The bottom right figure shows 
the 3D model of the barrel.}
\label{fig:case1}
\end{center}
\end{figure}

The top left plot in Figure \ref{fig:case1} shows the vertical deflections 
calculated in the FEA analysis for the final barrel design with the telescope 
at 90$^\circ$ from zenith and a zero degree rotation around the axis of the barrel.  
The overall sag produced by the hexapods was subtracted from these deflections, 
that is why average deflections are zero at the Hexapod Mounting Plate.
The values for this ``hexapod'' translation are listed in Table \ref{tab:comp_FEA}.
The same deflections after an overall rotation of the barrel to make the 
deflections at both ends equal are shown in the top right plot of 
Figure \ref{fig:case1}.
The horizontal deformations are shown in the middle left plot of the same figure.  
Finally the angular deflections after the barrel has been rotated to make deflections
at both ends equal are shown in the middle right plot of Figure \ref{fig:case1}.

The spread in the deflections seen by the spread in the different color lines is 
related to the stiffness of the HMP.  The stiffness of this flange is essentially 
what keeps the barrel round.  The forces exerted by each hexapod actuator are very 
high, reaching about 2/3 of the weight of the barrel \cite{bib:hexapods}.  We 
designed the HMP to minimize the total weight of the barrel and to keep the 
out-of-roundness deformations below the required level of 50 $\mu$m.

The FEA results for the other barrel positions can be found in Reference \cite{barrel_cdesign}.

\begin{figure}[thb]
\begin{center}
  \includegraphics[width=0.95\textwidth]{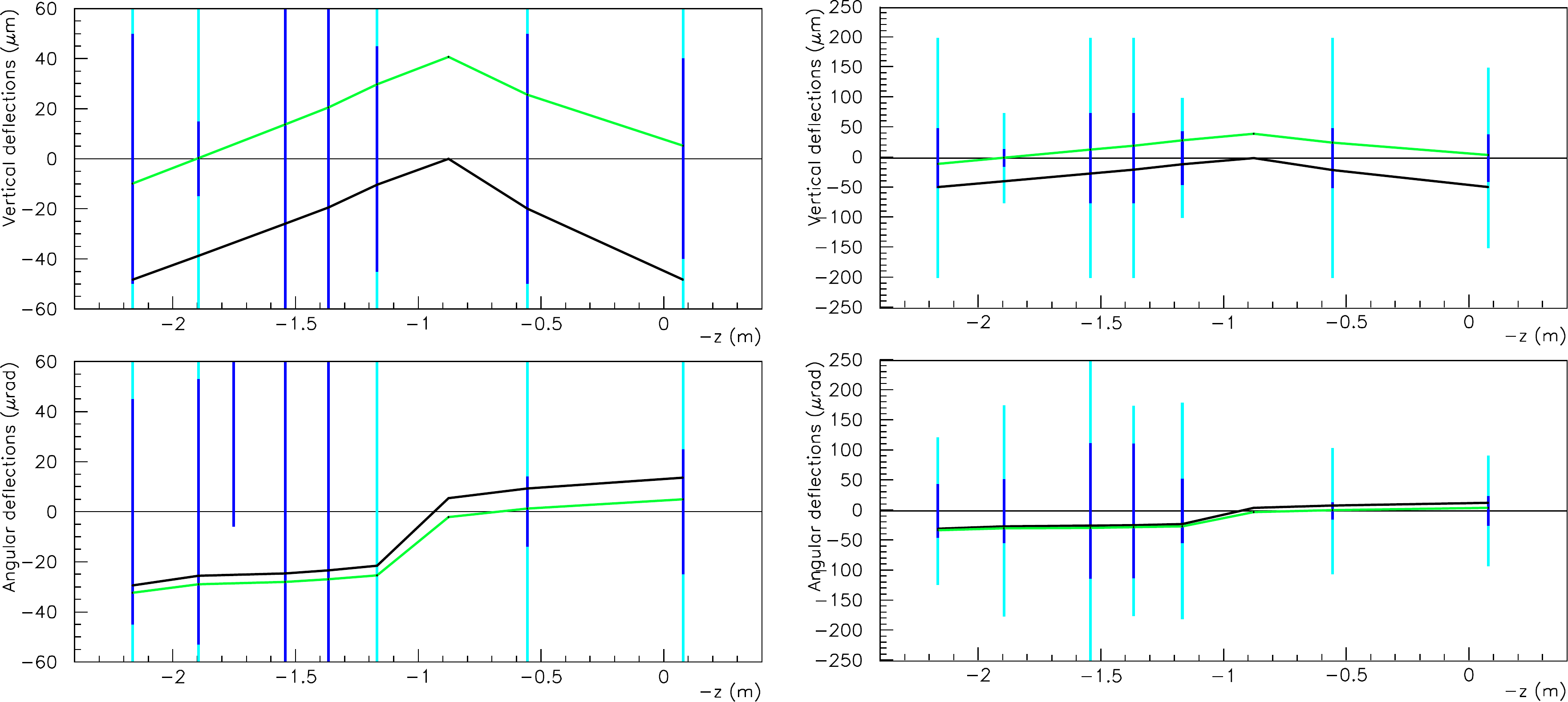}
  \caption{Comparison between requirements and the FEA analysis for the final barrel 
design with the barrel in the horizontal position and a 0$^\circ$ rotation around the 
barrel axis.  The full lens lateral and tilt requirements are shown in light blue.  
The lateral and tilt requirements due to barrel deflections are shown in blue.  The top 
plots show the vertical deflections and the bottom plots show the angular deflections.  
The right hand side plots are the same as the left hand ones but on a  different scale.  
The black lines show the average FEA results, the green line shows the deflections 
after an over all rotation of the barrel as if it would be performed with the hexapods.}
\label{fig:comp_1}
\end{center}
\end{figure}

\paragraph{Comparing the FEA deflection results with the DESI requirements}
\label{sec:comparison}

In this section we will compare the FEA results of the previous section
with the deflection requirements.  The comparison of the FEA results with the specifications for the barrel final design with the telescope pointing at 90$^\circ$ from zenith and with a 0$^\circ$ rotation around the barrel axis is shown in Figure \ref{fig:comp_1}.  

In this figure the full lens lateral decentering and tilt requirements are 
displayed in light blue.  The lateral and tilt requirements for the barrel due to barrel deflections are shown in blue.  Then the barrel satisfies specs if deflections are 
inside the blue lines.  The right hand plots are the same as the left hand 
ones but on a different scale.  The top plots show the vertical deflections the 
lower ones the angular deflections.  The black lines show the FEA results, that is 
the average of the deformations shown in the top right plot of 
Figure \ref{fig:case1} for the vertical deflections and the deformations shown in the middle right plot of the same figures for the angular deflections.  
We see that these black lines often lie outside
of the blue lines.  This is not a problem as the full camera can be rotated
using the hexapods.  This can be done because the optics only cares about
deflections of the lenses relative to each other and to the focal plane.
The green lines show the deflections after performing an over all rotation of the 
barrel as it would be carried out with the hexapods.  We see that in this case
all green lines satisfy the specifications.  The values of the deflections
given by these green lines were entered in Table \ref{tab:comp_FEA}, together with the 
same deflections for two more rotation angles.

\begin{table}[htb]
\begin{center}
\caption{Tables showing the comparison between the requirements and the FEA results.  
The FEA results shown in the last three column were performed with the barrel in the 
horizontal position.  The last two rows are the ``hexapod'' translations and rotations 
that were performed to eliminate the overall translation and rotation of the barrel 
as a unit.}
\label{tab:comp_FEA}
\begin{tabular}{ccccc}
\hline
\hline
 Element & Decenter & \multicolumn{3}{c}{Rotation around the barrel axis} \\
\cline{3-5}
 & $\pm\mu$m &  0$^\circ$ &  40$^\circ$ &  80$^\circ$ \\
\hline
 C1    &  50 &    -9.9 &    -9.8 &    -9.7 \\
 C2    &  15 &     0.3 &     0.2 &     0.4 \\
 ADC1  &  75 &    13.7 &    13.5 &    13.6 \\
 ADC2  &  75 &    20.4 &    20.2 &    20.4 \\
 C3    &  45 &    29.7 &    29.8 &    29.6 \\
 C4    &  50 &    25.5 &    25.5 &    25.7 \\
 FP    &  40 &     5.3 &     6.4 &     5.5 \\
C3-HMP &  25 &   -11.9 &   -12.9 &   -10.2 \\
\hline
\hline
 Element & Tilt & \multicolumn{3}{c}{Rotation around the barrel axis} \\
\cline{3-5}
 & $\pm\mu$rad &  0$^\circ$ &  40$^\circ$ &  80$^\circ$ \\
\hline
 C1    &  45 &   -32.3 &   -31.8 &   -31.9 \\
 C2    &  53 &   -28.9 &   -28.4 &   -28.5 \\
 ADC1  & 113 &   -28.1 &   -27.7 &   -27.8 \\
 ADC2  & 112 &   -27.0 &   -26.6 &   -26.7 \\
 C3    &  54 &   -25.5 &   -24.7 &   -25.1 \\
 C4    &  14 &     1.3 &     0.8 &     0.9 \\
 FP    &  25 &     5.0 &     4.6 &     0.0 \\
C3-HMP &  50 &   -27.0 &   -40.6 &   -31.2 \\
\hline
\hline
\multicolumn{2}{l}{HD trans. in $\mu$m (\% range)} &480.8 (6.0\%)&474.2 (5.9\%)&464.8 (5.8\%)\\
\multicolumn{2}{l}{HD rotat. in arcsec (\% range)} & 11.3 (4.5\%)& 10.5 (4.2\%)& 9.5 (3.8\%)\\
\hline
\end{tabular}
\end{center}
\end{table}

\noindent
Table \ref{tab:comp_FEA} shows a compilation of the FEA deflections shown in the 
previous plots.  The first column in the table shows the lenses or the focal 
plane where the deflections are given.  The second column shows the lateral 
deflections and tilt specifications for the barrel deflections.  Columns 3 to 5 
show the average vertical deflections and angular deflections for: a) the final barrel design for the 
telescope in the horizontal position and a 0$^\circ$ rotation around the axis of 
the barrel, b) same as in (a) but with a 40$^\circ$ rotation around the axis of 
the barrel, and c) same as in (a) but with an 80$^\circ$ rotation around the 
axis of the barrel.  Also shown in the table are the ``hexapod'' translations and 
rotations that were performed to eliminate the overall translation and rotation 
of the barrel, or the optics, as a unit.  We see that all requirements are satisfied.

\subsubsection{Lens Cells and Barrel Alignment}\label{sec:alignment}

Alignment of the six lenses of the DESI optical corrector will be achieved using mechanical and optical methods. The alignment of this corrector will be in keeping with the methods used for DECam corrector, which we will frequently refer to in describing the methods proposed for aligning the DESI corrector.
 Recall that corrector lenses C1--C4 will be mounted in individual cells with flexure systems, and atmospheric dispersion corrector lenses ADC1--ADC2 will be housed in rotating cells. The distances between optical elements will be controlled by steel spacers between the barrel shell segments, which can be ground to the required length.

\paragraph{Alignment of Lenses to Cells} 

The cells and the lenses will have to move relative to each other to facilitate central alignment.  For adjustment in the $x$ and $y$ directions, a translation stage can be made from two sets of NSK Ltd. linear rails and bearings set at 90 degrees to each other mounted on flat plates. Lens movement in $x$ and $y$ can be accomplished with the aid of spring-loaded adjustment screws. The maximum travel required is of order $\pm$10 mm. Adjustment to the cell's translation can be made using push and pull screws mounted around the cell's periphery.  These screws also can act as clamping bolts.

Moreover, the cells and the lenses will have to be aligned in tip and tilt relative to each other, and to balance any wedge angle that may have been introduced during the optical manufacture.  On DECam, the fine adjustment mechanism consisted of LJ750 ThorLab jacks (capacity 90 kg). Each individual lens will require a different set of whiffletree plates for uniform support whilst being aligned. Nylon pads are to be mounted at the corners of the whiffletree plates and machined to the curvature of the optic. Viton$^\circledR$~of 1 mm thickness will be glued onto the nylon pads to prevent damaging the lens surface.
Handling fixtures will be required for the DESI lenses.  These are expected to be similar to the DECam lens fixtures, which held the lens between two nylon rings that support the edges and face of the glass. These rings were bolted together using studding. Two aluminum plates were bolted to the nylon rings and held apart by aluminum posts.  Later, after the lenses are mounted into their cells, the cells themselves can be used as lens handling fixtures.
The measuring equipment required for the DESI lenses will be similar to that used for DECam alignment.  Rotational symmetry and tip tilt will be measured using Sylvac digital dial gauge indicators mounted on adjustable stands. These indicators are repeatable to 1~\micron.  Large bore gauges will be used to measure the distance between lens elements, along with a Faro Gage probe arm. Optical alignment measurements will be made using lasers, interferometers, CCD cameras and reference mirrors. A Micro-Epsilon laser sensor will be used to measure the placement of the RTV pads.

The alignment procedure is expected to be similar to that used for DECam.  The first step is to align the lens with respect to a Rotary Precision Instruments air bearing table (10 tonnes capacity, sub-micron run out accuracy), to which is bolted a 1500 mm diameter aluminum plate.

\begin{figure}[!ht]
\centering
\begin{minipage}[b]{0.45\linewidth}
\includegraphics[width=\textwidth]{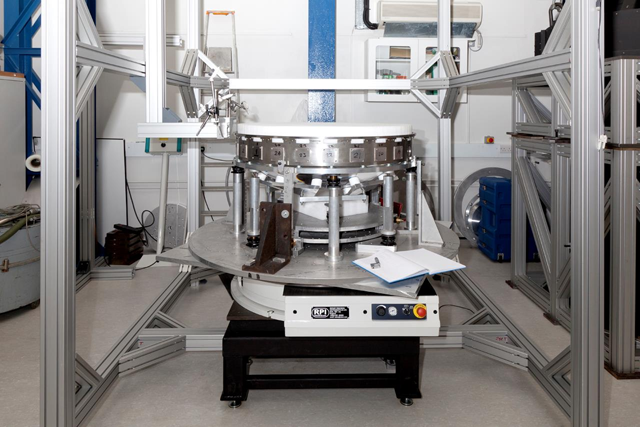}
\caption{DECam C1 lens alignment setup.}
\label{fig:DECamC1}
\end{minipage}
\hspace{0.25in}
\begin{minipage}[b]{0.45\linewidth}
\includegraphics[width=\textwidth]{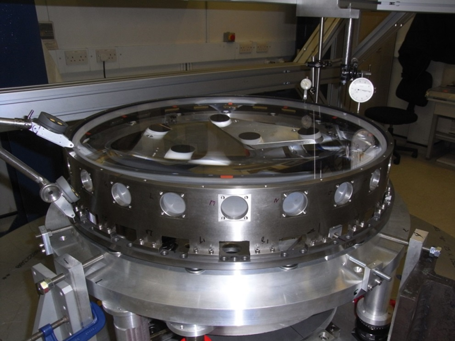}
\caption{DECam C4 Lens and cell alignment setup.}
\label{fig:DECamC4}
\end{minipage}
\end{figure}

Figure~\ref{fig:DECamC1} shows the alignment setup for DECam lens C1. The lens is lifted using the handling fixture and soft slings. With the translation stage and bottle jacks bolted in position on the rotary table, the lens is lowered onto the support flanges of the jacks, which incorporate scratch-free load spreaders conforming to the optical surface. Thus supported, the lens may roughly be aligned simply by measuring from the translation stage to its rim. In an iterative process, by rotating the table and measuring the run out of the edge of the lens with the dial gauge indicator, adjustment to the lens position can be made using the translation stage and the Thorlab jacks.

Next the cell is raised into position and mated with the lens. Hydraulic jacks accomplish long distance travel and Thorlab jacks accomplish fine movement, monitored by dial gauges. As the cell nears its contact position with the lens, smaller and smaller increments (on the order of 20~\micron) are taken until the lens is seated on the RTV pads. As this is accomplished, the radial positions of the lens and cell are monitored to maintain true alignment. Figure~\ref{fig:DECamC4}  shows the DECam C4 lens and cell alignment setup.

\paragraph{Alignment of Cells to Barrel} 

The procedure for aligning DESI cells to this barrel (DESI-0337) are similar to that employed on DECam. Using a large co-ordinate measuring machine, the FNAL-constructed barrel elements and 
UCL-fabricated lens cells will be carefully measured. Using data derived from these measurements, the cells are set at their optimal position for centering the lenses. The cells are drilled and doweled to facilitate their removal and subsequent replacement after the lenses are mounted in them. The accuracy of centering and realignment on the dowels is of order $\pm10~\micron$.

\begin{figure}
\centering
\includegraphics[height=2in]{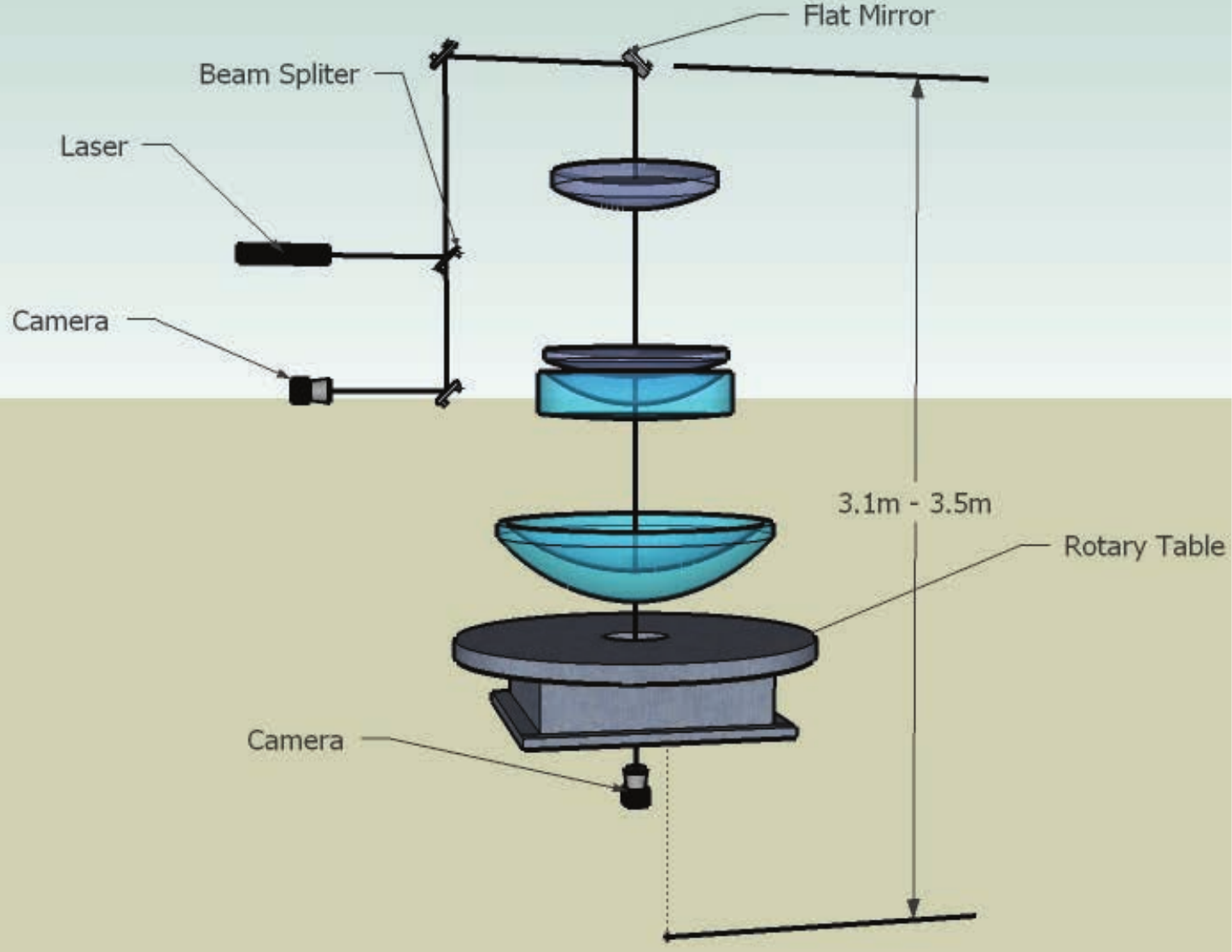}
\caption{Laser alignment schematic.}
\label{fig:laseralign}
\end{figure}

Doweled cells and barrel sections are shipped to UCL and the cells are aligned with their barrel sections.  First, the main body and cells are aligned, then the cone and the main body and optics are united. Finally the two sections are combined. An optical laser alignment provides pencil beams used to check for any run out or tilt of the lenses. Elaborate checks are required to ensure that the laser beam runs true. The schematic in Figure~\ref{fig:laseralign} shows how the beams from the system are measured with a CCD camera. A crosscheck is made by mechanically measuring each lens as mounted into the barrel.

%
%

\subsubsection{Hexapod}\label{sec:hexapods}

The hexapod system consists of six actuators, a mechanical joint with two 
degrees of freedom at both ends of each actuator, one stationary plate, 
a motion plate, and the control system. The system also comes with one 
spare actuator, the fixtures needed to change an actuator, and all 
documentation. The stationary plate bolts to the prime focus cage and 
the motion plate bolts to the barrel. 
The six actuators form three triangles. The triangle apex is located where two 
actuators meet at the motion plate. The actuators at the base of each triangle 
are attached  to the stationary plate that is fixed to the cage.  The hexapod 
effectively provides a three-point support to the motion plate.

The hexapod controller is equipped with manual controls to completely 
control hexapod motion in stand-alone, manual operation mode. The hexapod 
controller also has a remote operation mode. In remote operation mode, 
the controller is set up to move the hexapod according to the arbitrary 
position command it receives from a DESI control system. It is also setup 
to send hexapod operating status information to the same system in either 
the manual or the remote operating mode.

\subsubsection{Prime Focus Cage}\label{sec:cage}
\begin{figure}[!t]
\begin{center}
  \includegraphics[height=2.5in] {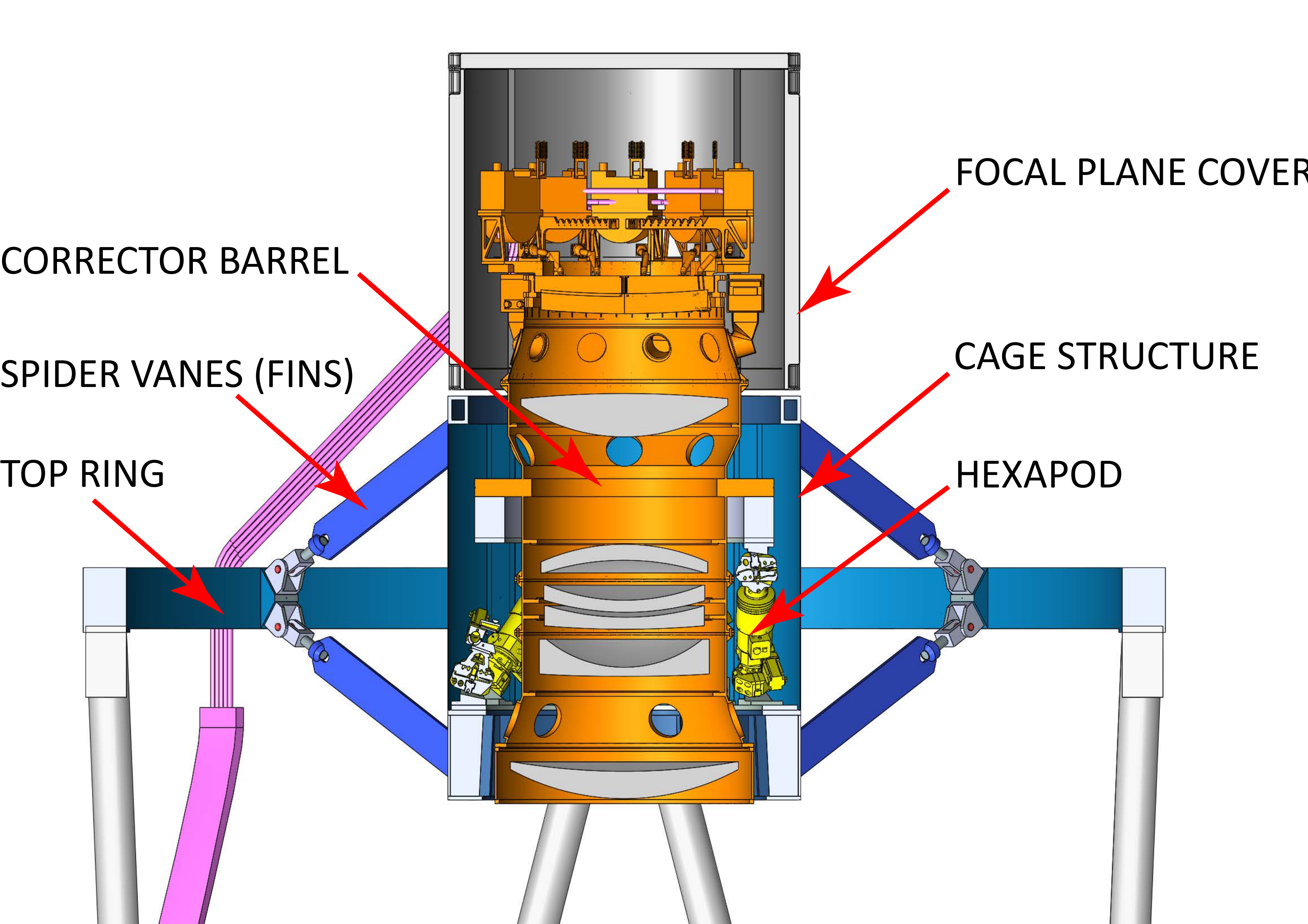}
  \caption{This figure shows the hexapod, the cage, the spider vanes (fins) and the
telescope top end ring.}
\label{fig:top_end}
\end{center}
\end{figure}

The Prime Focus Cage (PF Cage) is an assembly consisting of two end rings, 
a hexapod support structure and a cage shroud. The PF Cage will be attached to 
the telescope outer ring through a set of four fins as shown in Figure
\ref{fig:top_end}.  All elements in the cage are made of low carbon steel.

\paragraph{End Rings} 

The cage consists of a set of two end rings that interface with the ends of 
the four spider vanes. The outside circumferential dimensions of these rings match
the original cage ring dimensions. 
These rings will be made out of steel.  The rings will be attached to the fins with bolts and pins.
Mating or attachment points between 
rings and fins will be left unpainted and protected with a light coating 
of grease.  All other surfaces are painted with Aeroglaze$^\circledR$ Z306 flat black paint for
corrosion control. 
FEAs will be conducted to determine the appropriate ring cross-sectional 
configuration to best achieve the allowable assembled deflection as the 
cage moves through a varying gravitational loading.

\paragraph{Hexapod Structure} 

The hexapod structure is designed to provide support and attachment point for 
the barrel hexapod and to transfer the load to the cage. 

The hexapod support structure will be fabricated from steel and 
coated with Aeroglaze Z306 paint. Mating faces will be left unpainted and 
protected with a light coating of grease.

%

%

\paragraph{Material and Coatings} 

The end rings, hexapod support structure, and covers are fabricated from 
ASTM A36 structural steel. The cage cover sections will provide 
wind and thermal radiation protection for the camera/corrector.  
Most surfaces are painted with an optical black paint as specified in the 
stray light analysis.  The primer is Lord Corporation Aeroglaze 9947 and the top coat is
Aeroglaze Z306 (flat black and low out gassing).
Mating contact surfaces for the rings and fins are not painted. These 
surfaces are protected from corrosion by a light coat of grease.

\subsubsection{Telescope Top End}\label{sec:topend}

The telescope top end consists of the telescope outer ring and the fins that 
attach the cage to that ring.  
Since DESI is not required to support operation of Cassegrain instruments,
the inner flip ring in the original Mayall design is no longer
required.

\paragraph{Spider Vanes} 

The spider vanes are of the same design as the fins in use on the Mayall 
telescope.  The inboard end of the vanes are connected to the cage rails; 
the outboard ends are bolted to a bracket mount on the outer ring.  The 
connection between each vane and the outer ring is the same connection type 
used in the Mayall telescope.  Material for construction for the vanes is
ASTM A36 structural steel.  The current DESI design does not have the middle
vanes used originally in the Mayall telescope as these were found to over-constrain the mounting system.

\paragraph{Outer Ring} 

The current two-ring system that attaches to the top of the telescope Serrurier truss
will be replaced with a single fixed ring, saving 2250~kg in mass.

\subsubsection{Prime Focus System Integration}
As described above the lens cells are installed in the barrel assembly at University College London.  This is then broken down into three barrel sections and shipped to the telescope.  The barrel cage, hexapod system, fins, and upper ring telescope are shipped from FNAL to the telescope.   At the telescope, the barrel section are reassembled and the completed barrel installed in cage-hexapod structure.  The upper telescope ring and fins are install on the telescope, and then the barrel-cage assembly is installed.


\label{sec:straylight}
\subsection{Stray Light}

\begin{figure}[!hbt]
\begin{center}
  \includegraphics[height=2in] {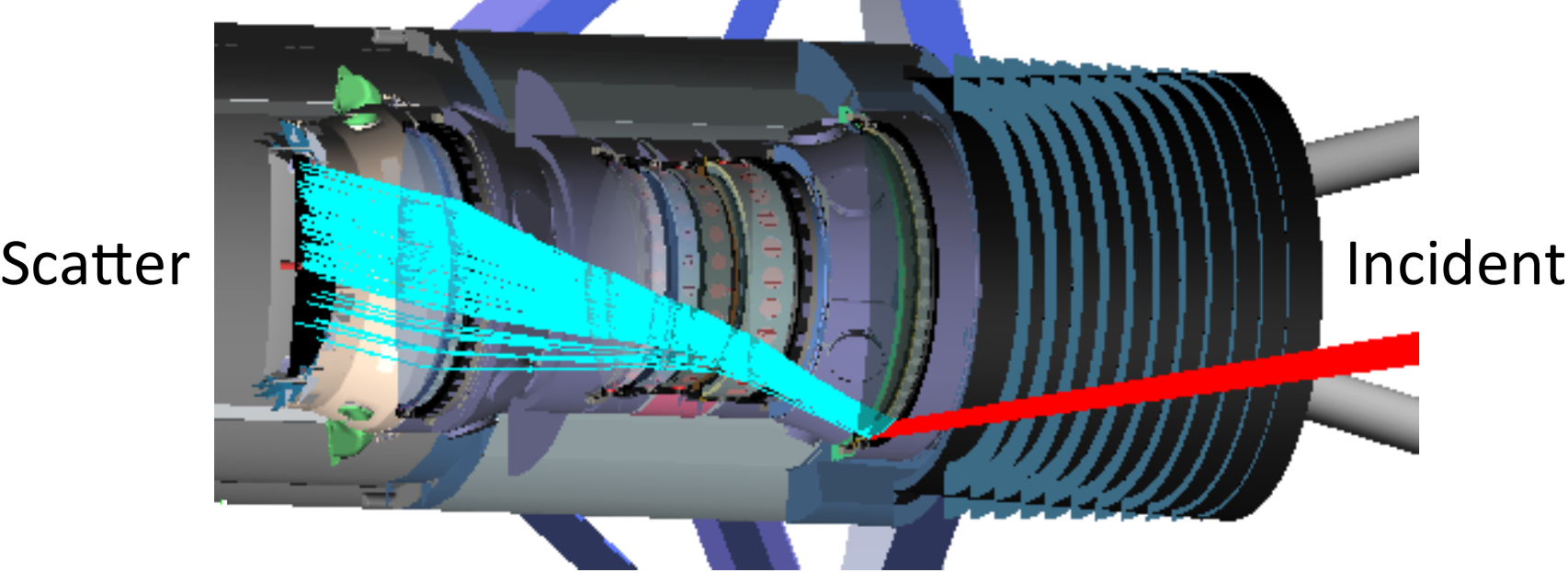}
  \caption{Stray light path found in the C1 cell region.}
\label{fig:straylight}
\end{center}
\end{figure}

Photon Engineering (Tucson AZ) performed a stray light analysis of the DESI front-end system: the corrector, Mayall telescope and the dome.  Their final report is posted to DESI-1618.  The results of this analysis have led to modifications to the design of the lens cells and barrel, as well as plans for painting sections of the telescope.
An example is the stray light source in the C1 cell shown in Figure~\ref{fig:straylight}.  The stray light source illustrated is being mitigated by scalloping and texturing the scattering surface identified on the C1 cell.

\clearpage

\section{Focal Plane System}
\setcounter{equation}{0}\setcounter{figure}{0}\setcounter{table}{0}
\label{sec:Instr_Focal_Plane}

The Focal Plane System (FPS) consists of three functional systems at optical focus:

\begin{enumerate}
  \setlength{\itemsep}{1pt}
  \setlength{\parskip}{0pt}
  \setlength{\parsep}{0pt}
	\item Fiber Positioners each carry an individual science fiber to a unique target position for each observation.
	\item Field Fiducials provide point light sources as references throughout the field for the fiber view camera.
	\item Guide, Focus, and Alignment (GFA) sensors measure the telescope pointing as well as focus and tip/tilt of the focal surface.
\end{enumerate}

\begin{figure}[!b]
\centering
\includegraphics[width=\textwidth]{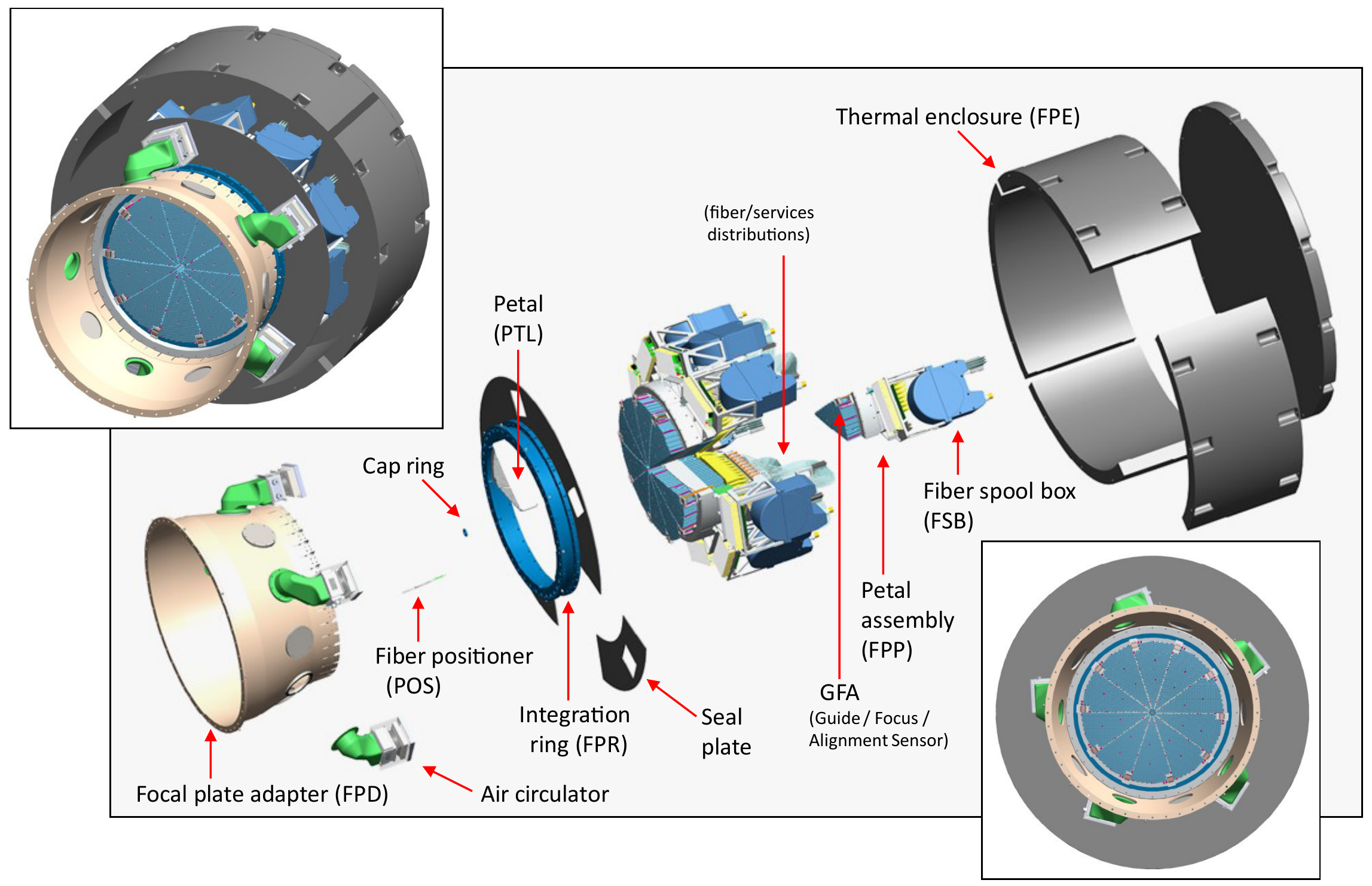}
\caption{Illustration of the Focal Plane System. The Fiber View Camera, located at the primary mirror, is not shown.}
\label{fig:focal_plane_system}
\end{figure}

These are supported mechanically in a Focal Plate Assembly, which attaches to the corrector barrel. The FPS also provides thermal insulation to block heat from escaping into dome air, thermal control to remove heat from the system, and service harnessing.
An illustration of the focal plane system is shown in Figure~\ref{fig:focal_plane_system}.

The FPS also interfaces to several external systems:

\begin{enumerate}
  \setlength{\itemsep}{1pt}
  \setlength{\parskip}{0pt}
  \setlength{\parsep}{0pt}
	\item Ferrulized fiber ends mechanically interface to the fiber positioners.
	\item Fiber spool boxes mechanically interface to the focal plate assembly.
	\item Corrector barrel mechanically interfaces to the focal plate assembly.
	\item Cage assembly mechanically interfaces to the FPS thermal enclosure (FPE).
	\item Instrument control system functionally interfaces to the fiber positioners, field fiducials, GFA, and thermal control system.
	\item Fiber View Camera measures locations of the illuminated points provided by fiber positioners, field fiducials, and GFA.
	\item Facility heat exchanger provides ambient temperature coolant to manifold on the FPE.
\end{enumerate}

Requirements on the Focal Plate design are listed in Table~\ref{tab:fpreqs}.

\begin{table}[!t]
\centering
\caption{Focal plate key requirements.$^\dagger$ Also see DESI-0455.}
\footnotesize
\addtolength{\tabcolsep}{-2pt}
\newcolumntype{C}{>{\centering\arraybackslash}X}
\newcolumntype{R}[1]{>{\raggedright\arraybackslash}m{#1}}

\begin{tabularx}{\textwidth}{R{1.5in}R{1.2in}R{2.4in}C}
	\hline
	Item & Value & Rationale   &  Current Design \\
	\hline
	Number of fiber positioners supported within field of view & 5,000 & Science throughput.   &   5,000 \\ \hline
	Number of GFAs supported within field of view & 10 & Provides sufficient sensor area when using CCD 230-42 sensors.   & 10  \\ \hline
	Number of illuminated fiducials within field of view & $\geq$ 60 & Kinematics of fitting the field when viewed through fiber view camera.   &  120 \\ \hline
	Number of fiber management units supported & 10 & Logical distribution to 10 spectrographs.   &   10  \\ \hline
	GFA sensor radial position of active area	& $\leq$ 408 mm &  Ensure sufficient bright guide stars and sufficient consistency of mode shapes.  & 407.2  \\  \hline
	Z deflection when at zenith & $\leq$ 30~\micron max (rel)~\linebreak~$\leq$ 15~\micron rms (rel) & Maintain figure of focal surface at worst-case gravity orientation.   &   22.6~\micron~\linebreak~9.8~\micron \\
	& $\leq$ 100~\micron max (abs) & Do not unnecessarily consume hexapod focus range.   &  29.0~\micron \\ \hline
	Lateral deflection when oriented at 90\degree & $\leq$ 10~\micron max (rel)~\linebreak~$\leq$ 5~\micron rms (rel) & Maintain relative (x,y) positions of fiducials and fiber positioners.   &   2.0~\micron~\linebreak~1.0~\micron \\ 
	& $\leq$ 100~\micron \newline absolute & Do not unnecessarily consume hexapod lateral range.   &  12.2~\micron   \\ \hline
	Radial deformation due to $\Delta$T during an observation & $\leq$ 10~$\micron/\celsius$ max & Maintain (x,y) positions during an observation.   &   2.5~$\micron/\celsius$ \\ \hline
	Normal deformation due to plate $\Delta$T during an observation& $\leq$ 5~$\micron/\celsius$ &	Maintain focus during momentary local heat loads.   &   0.5~$\micron/\celsius$ \\ \hline
	Plate $\Delta$T during an observation & $\leq$ 1~\celsius & Thermal-mechanical stability.   &   0.7~\celsius max \\ \hline
	Total focal plate assembly mass & $\leq$ 870~kg & Ensure corrector barrel stiff and strong enough to support focal plane system.   &  674.5~kg \\ \hline
	Heat load per observation & $\leq$ 1000~kJ & Ensure practical sizing of cooling system (1 kW of cooling for a 1000 second observation).   &  516~kJ  \\
	\hline
	\multicolumn{4}{l}{\footnotesize $^\dagger$Positioner mounting requirements are handled separately within the positioner requirements table.}
\end{tabularx}

\label{tab:fpreqs}
\end{table}


\subsection{Focal Plane Mechanical Design}{\label{FPS}}
\label{sec:focal_plane_mechanical} 

The focal plate supports the fiber positioners, Guide-Focus-Alignment sensors (GFA), and field fiducials. The focal \emph{surface} is a radially symmetric dome, defined by the corrector optical prescription; it is a definite geometric shape in space with respect to the corrector barrel. The focal surface is physically constructed of the science fiber tips and GFA active areas. The focal \emph{plate} holds these items in position on the (virtual) focal \emph{surface}.

The focal surface is 86.5~mm ahead of the front face of the focal plate. The ``front face'' is not a continuous smooth surface, but rather consists of 514 individual precision spot faces at which fiber positioners and fiducials mount (see Section \ref{sec:positioner_mounting}). The focal plate itself is 82~mm thick at the edge, increasing to 103.5~mm at the apex of the dome. Behind the focal plate are supports for fibers and electrical services. Cooling service hardware attaches outside the perimeter.

The Focal Plate Assembly is built up of 10 radially symmetric sectors which are held in place by an Integration Ring around the perimeter. These sectors (or ``petals'') are machined wedge-shaped objects, each supporting 500 fiber positioners, 10 field fiducials, and a GFA. On the rear side of each petal is attached a fiber guiding structure and a fiber spool box. These transmit the 500 individual fibers into a single cable, which is then routed down and off the telescope.

The integration ring attaches to the Focal Plate Adapter (FPD) via an FPD Ring. The FPD attaches to the rear-most flange of the corrector barrel.

The petals have precision mating faces to the integration ring, and are each bolted tightly to it. At the center of the assembly, two small Cap Rings transfer load between the petal tips.

\begin{figure}[!tb]
\centering
\includegraphics[width=\textwidth]{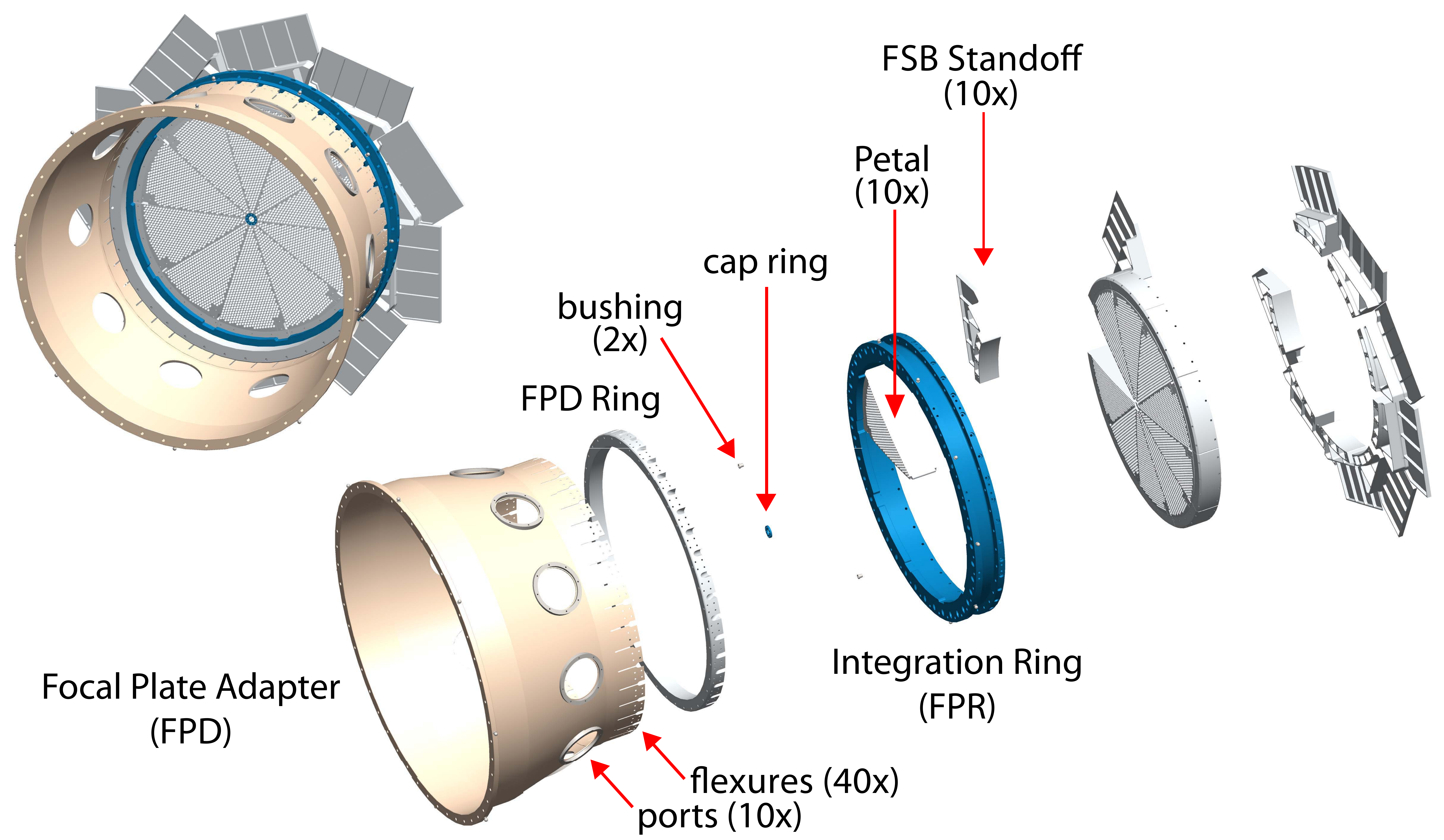}
\caption{Focal plate assembly support structures. The focal plate adapter (FPD) and FPD Ring are permanently connected to one another. They form the FPD assembly, which is what connects the corrector barrel to the focal plate integration ring (FPR). Flexuring between FPD and FPD ring accommodates CTE mismatch between focal plate and barrel structure. The FPR supports the 10 petals. Each petal supports a standoff plate, where its fiber spool box (FSB) and control electronics are mounted.}
\label{fig:focal_plate_structural_supports}
\end{figure}

The main support structures of the focal plate assembly are shown in Figure~\ref{fig:focal_plate_structural_supports}. To summarize, the Focal Plane System (FPS) consists of:

\begin{itemize}
  \setlength{\itemsep}{1pt}
  \setlength{\parskip}{0pt}

	\item Focal Plate Assembly
	\begin{itemize}
		\item Focal Plate Adapter Assembly
		\begin{itemize}
			\item Focal Plate Adapter
			\item Focal Plate Adapter Ring
		\end{itemize}
		\item Focal Plate Integration Ring 
		\item 10x Focal Plate Petal Assembly 
		\begin{itemize}
			\item Focal Plate Petal 
			\item 500x Fiber Positioner 
			\item 6-7x Field Illuminated Fiducial 
			\item 1x Guide / Focus / Alignment sensor (GFA)
			\begin{itemize}
				\item 3x GFA Illuminated Fiducial
			\end{itemize}
			\item 500x Positioner Fiber Assembly **
			\item Fiber Spool Box  **
			\item Fiber Focal Plane Guide **
		\end{itemize}
	\end{itemize}
	\item Focal Plane Enclosure
	\begin{itemize}
		\item Focal Plane Thermal Shroud 
		\item Thermal control equipment, such as heat exchanger, \etc
		\item Any electronics not on focal plane assembly, such as CPU, HV, \etc
	\end{itemize}
	\item Fiber View Camera 
\end{itemize}
Items marked ** above are elements of the Fiber System (Section~\ref{sec:Instr_Fibers}), which are closely intertwined with the Focal Plane System.

Vibration is not expected to be a problem for the focal plate structure. The design is driven by achieving enough stiffness to control gravity-induced deflection, a body force. Therefore the modal stiffness is inherently high. For the flexured aluminum petals configuration, the finite element model had a first mode at 83.4 Hz.  
Differential motion between petals is also not expected to be a problem. FEA results for petal-to-petal differentials due to thermal expansion were  $<$ 1~\micron, and due to gravity orientation were 2~\micron. See DESI-0453.

Hexapod motions of the barrel impart reaction loads to the petals via the fiber cables. Measurements of fiber cable stiffness were made in summer 2014. From initial results, the magnitude of these loads will not contribute sensibly to the petal deflections. These loads will be included in the final FEA analysis.

\subsection{Full-scale Demonstrator Petal}

\begin{figure}[!bth]
\centering
\includegraphics[width=\textwidth]{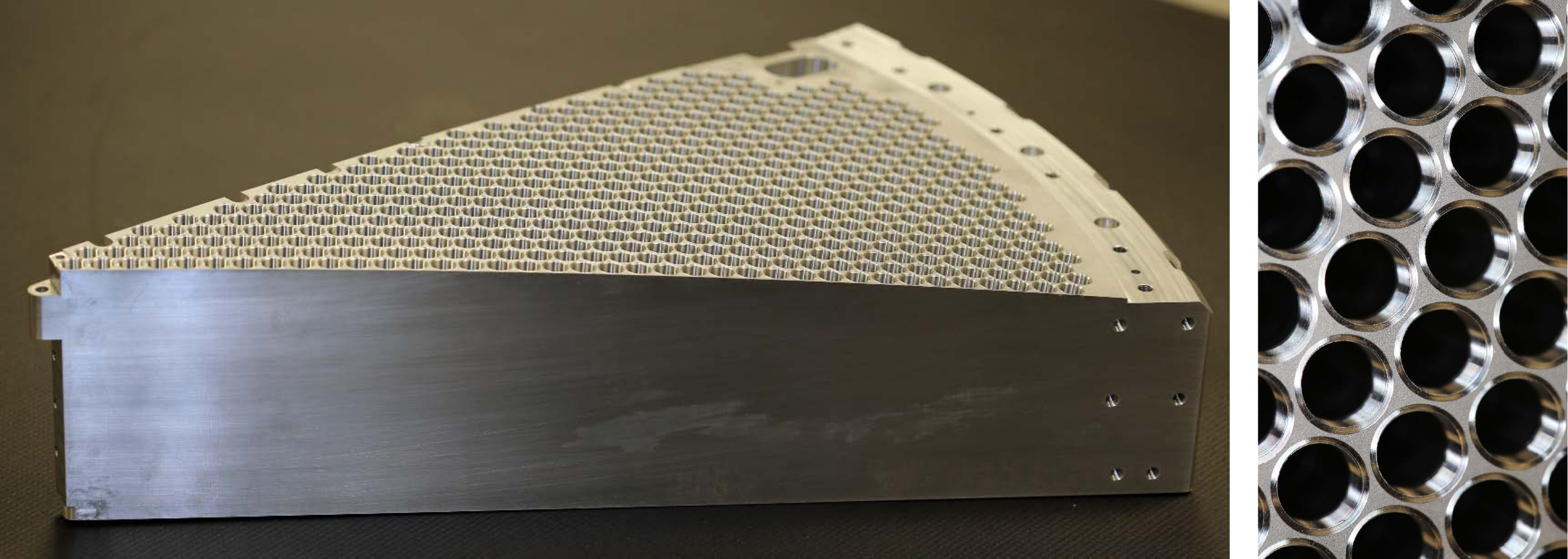}
\caption{First positioner focal plane petal on left.  Hole detail is shown on right.}
\label{fig:demonstrator_petal}
\end{figure}

A full-scale, fully-detailed prototype petal was produced and surveyed.
A photo of the petal is shown in Figure~\ref{fig:demonstrator_petal}.
This prototype is the culmination of several years of background work and
development, including multiple iterations with vendors to
understand manufacturing constraints and cost and performance drivers.

Additionally, 
subscale manufacturing samples to verify positioner mounting
interfaces and per-hole machining costs / complexity were produced. These prototypes
and tests are all very close to an expected final design.

Results from a metrology survey of the machined demonstrator petal are given in Table \ref{table:fp_demo_survey}. Mounting hole angle is the most critical tolerance for throughput, and  was held. The prototype did not meet the tolerances specified on hole diameter (worst case exceeded tolerance by 6~\micron) and hole location (worst case exceeded by 6~\micron). Additional test parts are being machined to assess whether these tolerances are fundamentally too tight to hold, or were just machined incorrectly in this one instance. Relaxing the tolerance on either hole diameter or location would consume some reserve margin on positioner close-packing. Relaxing the tolerance on the hole diameter would additionally affect positioner mounting angle error, and thus throughput.  Neither is considered to be particularly problematic, as the errors here are relatively small.

\begin{table}[!bth]
	\centering
	\caption{Hole machining accuracy on the demonstrator focal plate petal. The contributing group contracted to have 70 holes surveyed.}
	\label{table:fp_demo_survey}
	\begin{tabular}{l | c | c | c} \hline
		Hole dimension & Tolerance & Max abs deviation & rms deviation \\ \hline
		Angle (deg)& 0.045 & 0.024 & 0.010 \\
		Diameter (\micron) & 15 & 21 & 16 \\
		XY location (\micron) & 60 & 66 & 34 \\ \hline
	\end{tabular}
\end{table}

\subsection{Fiber Positioners}		
\label{sec:Instr_Fiber_Positioners}

\label{fiber_positioner_implementation}

The focal surface is composed of 5,000 fiber tips. Each fiber is uniquely positioned by an individual robot, allowing rapid reconfiguration of the entire field for each observation.

\subsubsection{Positioner Requirements} 

The basic requirement of the fiber positioner is that it consistently locate its fiber on target with accuracy $<5~\micron$.  All 5000 fibers need to be in position in less than 120 seconds between exposures. In case this activity cannot be performed in parallel with other between-exposure activities, we require the positioner reconfiguration to be complete in 45 seconds. To close the loop on positioning, after each move the fibers are back illuminated and their actual positions are measured by the Fiber View Camera (see  Section~\ref{sec:Instr_FVC}). Final accuracy may be achieved iteratively, so long as the time budget for the total number of iterations is not exceeded.

Positioners must be sufficiently robust to operate consistently for the lifetime of the project, have low mass, have low power consumption, and a sufficiently small mechanical envelope to fit 5,000 into the focal plane in non-vignetted light.

Table~\ref{tab:positioner_requirements} lists fiber positioner requirements which have been developed to meet the science needs. These have been consistently met by multiple prototypes.

\begin{table}[phtb]
	\centering
	\caption{Fiber positioner requirements and key design points. Refer to DESI-0455 for the requirement rationales.}
	\label{tab:positioner_requirements}
	\newcolumntype{C}[1]{>{\centering\arraybackslash}m{#1}}
	\newcolumntype{R}[1]{>{\raggedright\arraybackslash}m{#1}}
	\begin{tabularx}{\textwidth}{R{2.0in} R{1.0in} R{3.0in}} \hline
		Item & Requirement & Value \\ \hline \hline
		gross moves (blind) & $\le$ 100 $\micron$ & 25-30 $\micron$ max error \\ \hline
		after correction move & $\le$ 5 $\micron$ & 1-2 $\micron$ rms error (after a single correction) \\ \hline
		lifetime moves & $\ge$ 372,000 & tested $>$ 400,000 with no degradation \\ \hline
		power peak & $\le$ 3 W & 1.38 W at 4`V, 2.84 W at 6~V (while moving, 250 $\mu$~s sample) \\ \hline
		power average & \emph{(secondary)} & 1.06 W at 4~V, 2.22 W at 6~V (while moving, 100 ms sample) \\ \hline
		energy per observation & $<$ 30 J & 3-11 J at 4~V, 5-21 J at 6~V \\ \hline
		focal ratio degradation (rms) & 0.2$^\circ$ rms & 0.15$^\circ$ max $\Delta$ FWHM at f/3.75 over full patrol disk in repeated tests \\ \hline
		focal ratio degradation (max) & 0.4$^\circ$ max & 0.14$^\circ$ additional offset after 400,000 moves of positioner \\ \hline
		dynamic defocus error & $\le$ 30 $\micron$ & 2.7 $\micron$ rms measured over full patrol disk \\ \hline
		defocus mounting error & $\le$ 20 $\micron$ & 15 $\micron$ anticipated (set by butting against one stop in tooling jig) \\ \hline
		dynamic tilt error & $\le$ 0.05$^\circ$ & 0.011$^\circ$ rms measured over full patrol disk \\ \hline
		max tilt error & $\le$ 0.1$^\circ$ & 0.047$^\circ$ max measured dynamic + 0.05$^\circ$ mounting (estimated) \\ \hline
		operational temperature & -10$^\circ$C to +30$^\circ$C & tested -10$^\circ$C to +42$^\circ$C \\ \hline
		survival temperature & -20$^\circ$C to +60$^\circ$C & $<$ -30$^\circ$C during operational test several times, testing to +60$^\circ$C in Fall 2015 \\ \hline
		total reconfiguration time & $\le$ 45 sec & $<$ 24 sec to reconfigure an array of 500 positioners \\ \hline
		shaft speed & \emph{(secondary)} & 178 - 234$^\circ$/sec (depends on motor vendor gear ratio) \\ \hline
		mass & $\le$ 50 g & 30 g \\ \hline
		fiber handling radius & $\ge$ 50 mm & $\ge$ 65 mm \\ \hline
		fiber routing & \emph{(secondary)} & 68 mm free from ferrule to central axis, then gentle path through \\ \hline
		hard travel limits & required & both directions, both axes, ok to ram at full speed/power \\ \hline
		positive retention & required & M8.7 thread (like a sparkplug) \\ \hline
		mating references & \emph{(secondary)} & precision flange and cylinder on central bearing \\ \hline
		thermal expansion & \emph{(secondary)} & 6061 aluminum; matches focal plate and GFA / fiducial supports \\ \hline
		electrical connector & \emph{n/a} & 5 pin Samtec, mates one way \\ \hline
		communication & \emph{n/a} & standard CAN bus \\ \hline
		power supply & \emph{n/a} & +5 VDC nominal (4--12 V accepted) \\ \hline		                        
	\end{tabularx}
\end{table}

\subsubsection{Positioner Interfaces}

This section describes interfacing of fiber positioners to the focal plane system. This includes the subjects of  mechanical envelopes and patterning of positioners, mounting, mass and materials,  power and signal services, and thermal budget. Implementation of positioners is discussed separately, in Section~\ref{sec:positioner_design_overview}.

\paragraph{Mechanical Envelopes and Patterning} 

The packing efficiency of fiber positioners on the focal surface must be sufficient to allow 5,000 within the unvignetted field. To pack positioners together as closely as possible without interferences, detailed knowledge of the positioner mechanical envelope is necessary.

\begin{figure}[!tb]
\centering
\includegraphics[height=3in]{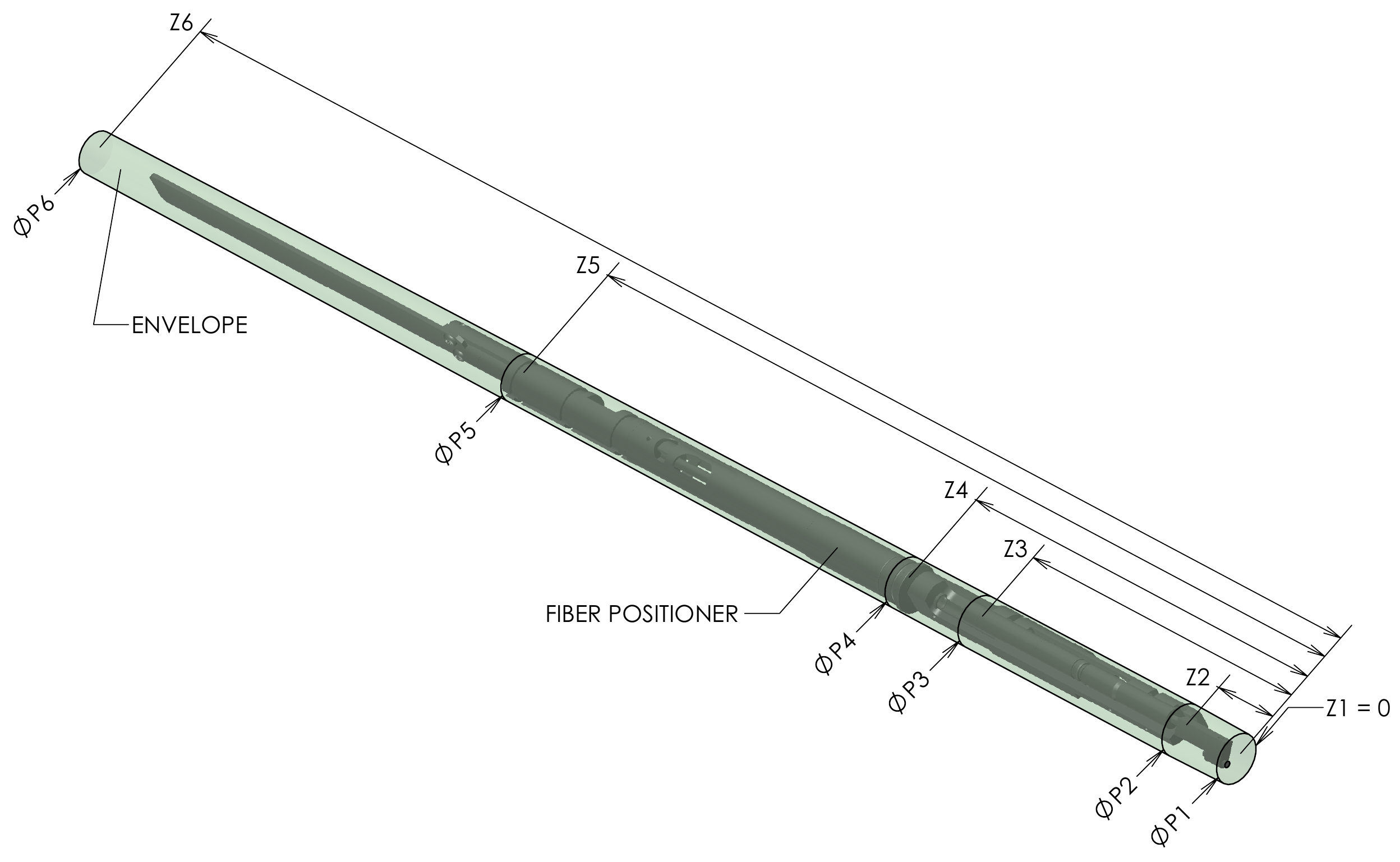}
\caption{Mechanical envelope of fiber positioner, defined by several cones. Diameters $p_k$ include dynamic and static tolerance stackups taken at axial positions $z_k$. The fiber tip is at $z_1$.}
\label{fig:positioner_in_mechanical_envelope}
\end{figure}

We define the envelope as a rotationally symmetric volume, swept about the central axis of the positioner. This volume consists of a series of cones, which together form the total mechanical envelope for the positioner. This is illustrated in Figure \ref{fig:positioner_in_mechanical_envelope}.

At several key locations along the central axis, minimum pitch values (minimum center-to-center distance between neighboring positioners) are calculated. These define the diameters of the conic envelope. The minimum pitches are calculated as a Monte Carlo sum of nominal positioner dimensions, manufacturing tolerances, and the dynamic envelopes swept out during re-positioning.

Perfect hexagonal close packing of the fiber positioners is not possible on the curved focal surface. An algorithm for efficient packing was written to maximize the number of positioners within the unvignetted field, without neighbor collisions. The algorithm is:

\begin{enumerate}
  \setlength{\itemsep}{1pt}
  \setlength{\parskip}{0pt}
  \setlength{\parsep}{0pt}
\item Collect minimum allowed pitches $p_k$ at several axial distances $z_k$ from the focal surface.
\item Call the focal surface $S_1$, and generate the several projected surfaces $S_k$ below it at distances $z_k$.
\item Project a flat hexagonal pattern (which is slightly too tight) onto $S_1$.
\item Project the flat pattern to the surfaces $S_k$.
\item Iteratively:
\begin{enumerate}
\item Select positioner $i$ of the array.
\item For positioner $i$, select the worst interfering neighbor $j$, considering all surfaces.

\item Nudge $i$ slightly away from $j$.
\item Increment $i$.
\end{enumerate}
\item Convergence is when for all surfaces $S_k$, all center-to-center pitches $\ge p_k$.
\end{enumerate}

\begin{figure}[tb]
\centering
\includegraphics[height=3in]{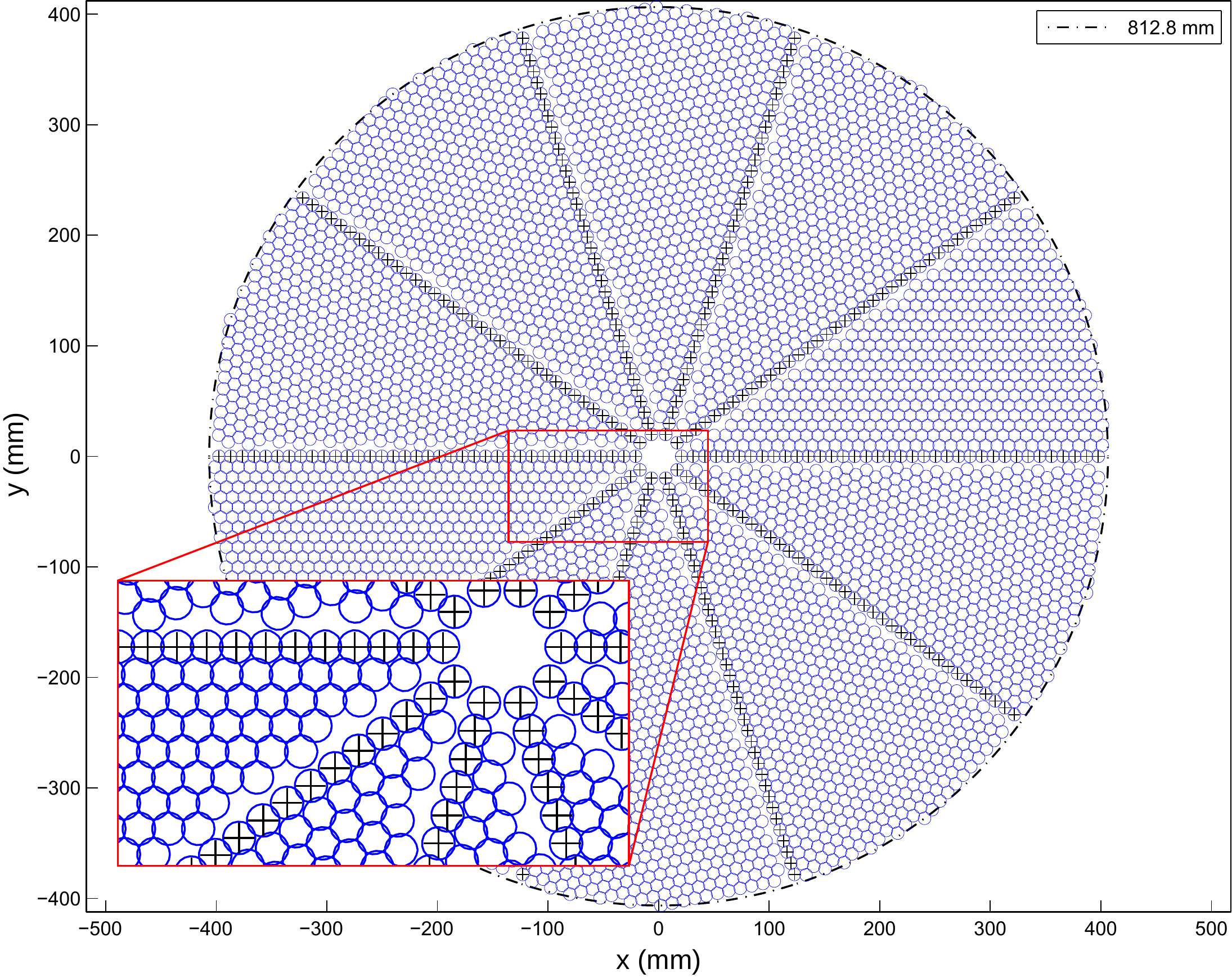}
\caption{Typical output of fiber positioner patterning algorithm. Closeup view near center of the field is inset. ``+'' signs indicate locations where additional special constraints have been applied.}
\label{fig:typical_hole_patterning_output}
\end{figure}

A top view plotting the typical output from the algorithm is shown in Figure~\ref{fig:typical_hole_patterning_output}. The field is composed of 10 identical wedge patterns, arrayed about the central axis. Additional constraints were simple to include in the algorithm. For instance, in Figure~\ref{fig:typical_hole_patterning_output}, locations marked with a ``+'' sign have additionally been constrained to lie along straight radial lines. This ensures a clean separation between wedges.

\paragraph{Mounting}
\label{sec:positioner_mounting}  

\begin{figure}[tb]
\centering
\includegraphics[height=1.75in]{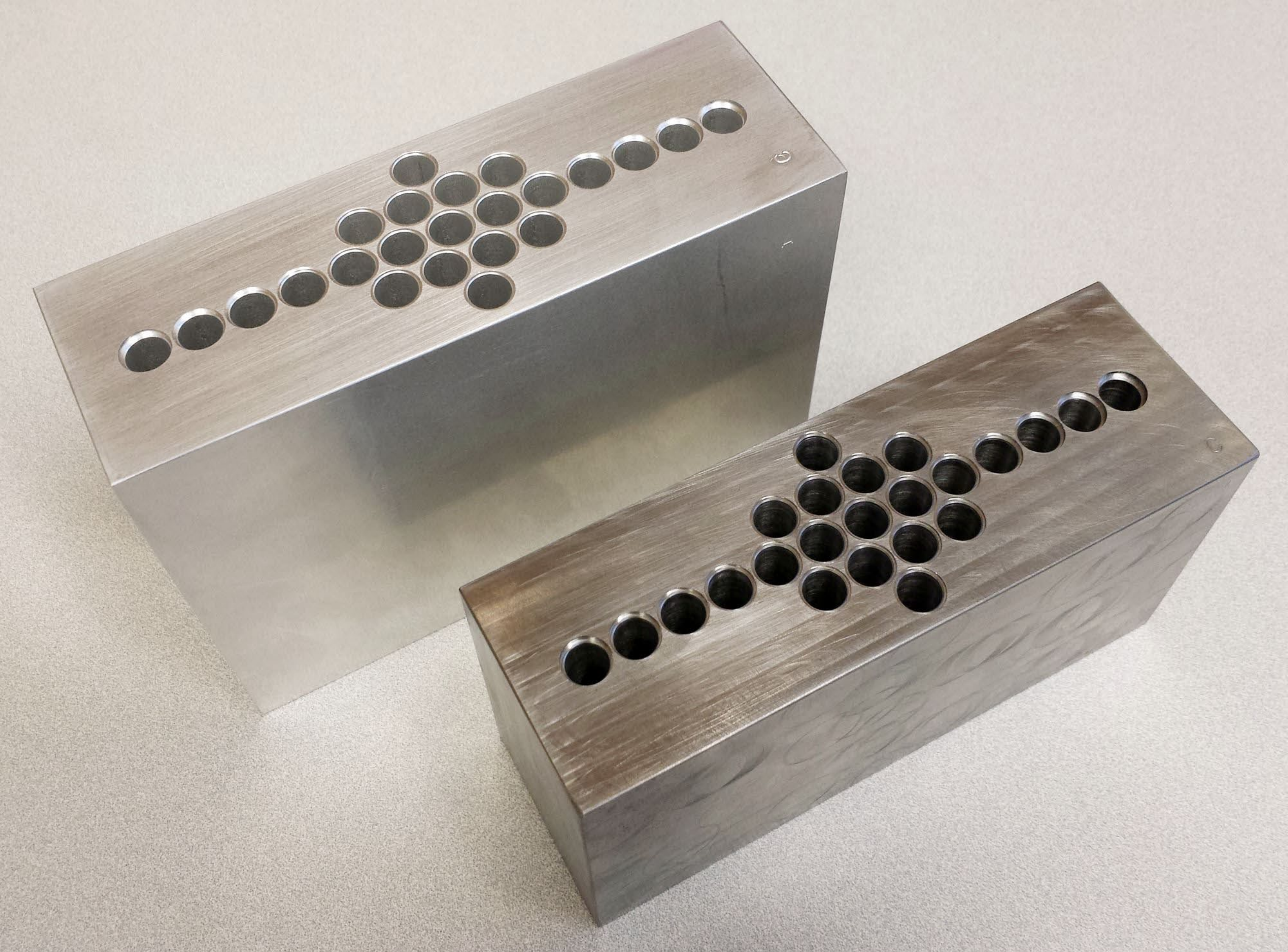} \hspace{0.25in}
\includegraphics[height=1.75in]{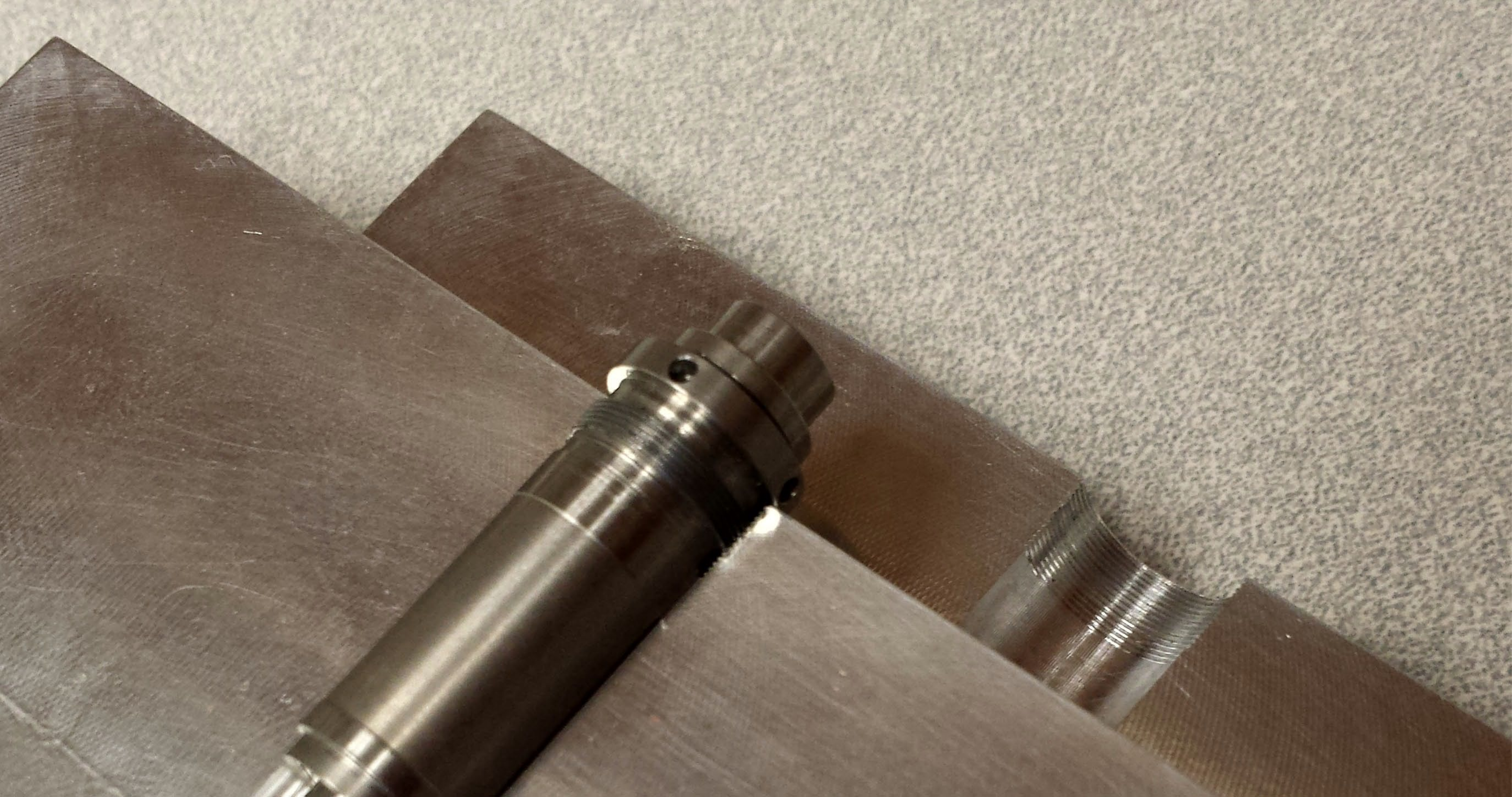}
\caption{Left: Drill test blocks for positioner interface in 6061-T6 aluminum (rear) and 4140 alloy steel (front). Hole sizes, spacings, and tilts were machined at correct scales.  Right: Section of a machining sample of the ``sparkplug'' style positioner retention feature. A typical precision bearing cartridge from an eccentric axis positioner is shown mated in the threads and registering below to the precision bore.}
\label{fig:positioner_if_machining_samples}
\end{figure}

Fiber positioners will be mounted in a hole plate, discussed in Section~\ref{sec:focal_plane_mechanical}. At each hole the plate provides a precision bore for tip/tilt and lateral positioning, and a precision spotface for transverse (focus) positioning. Machining samples have been made and surveyed in multiple materials (6061-T6 aluminum, 4140 alloy steel, and CE Al-Si alloy) by multiple vendors to confirm tolerances. Two such samples are shown in Figure~\ref{fig:positioner_if_machining_samples}(left). Positioners are positively mechanically locked by a threaded
 ``sparkplug'' style feature shown in Figure~\ref{fig:positioner_if_machining_samples}(right), which has been tested and is attractive for its simplicity and machinability.

%
%

\paragraph{Mass and Materials} 

The focal plate can stably and accurately support significant loads, allowing some freedom in the mass of positioners. The plate is designed to support positioners up to 50~g each, totaling 250~kg for the system of 5,000. The plate mass inherently provides significant thermal capacity, such that the brief heat loads generated by moving positioners between observations are readily absorbed without significantly affecting plate geometry.

\paragraph{Fiber, Power and Signal Services} 

\begin{figure}[!tb]
\centering
\includegraphics[width=0.9\textwidth]{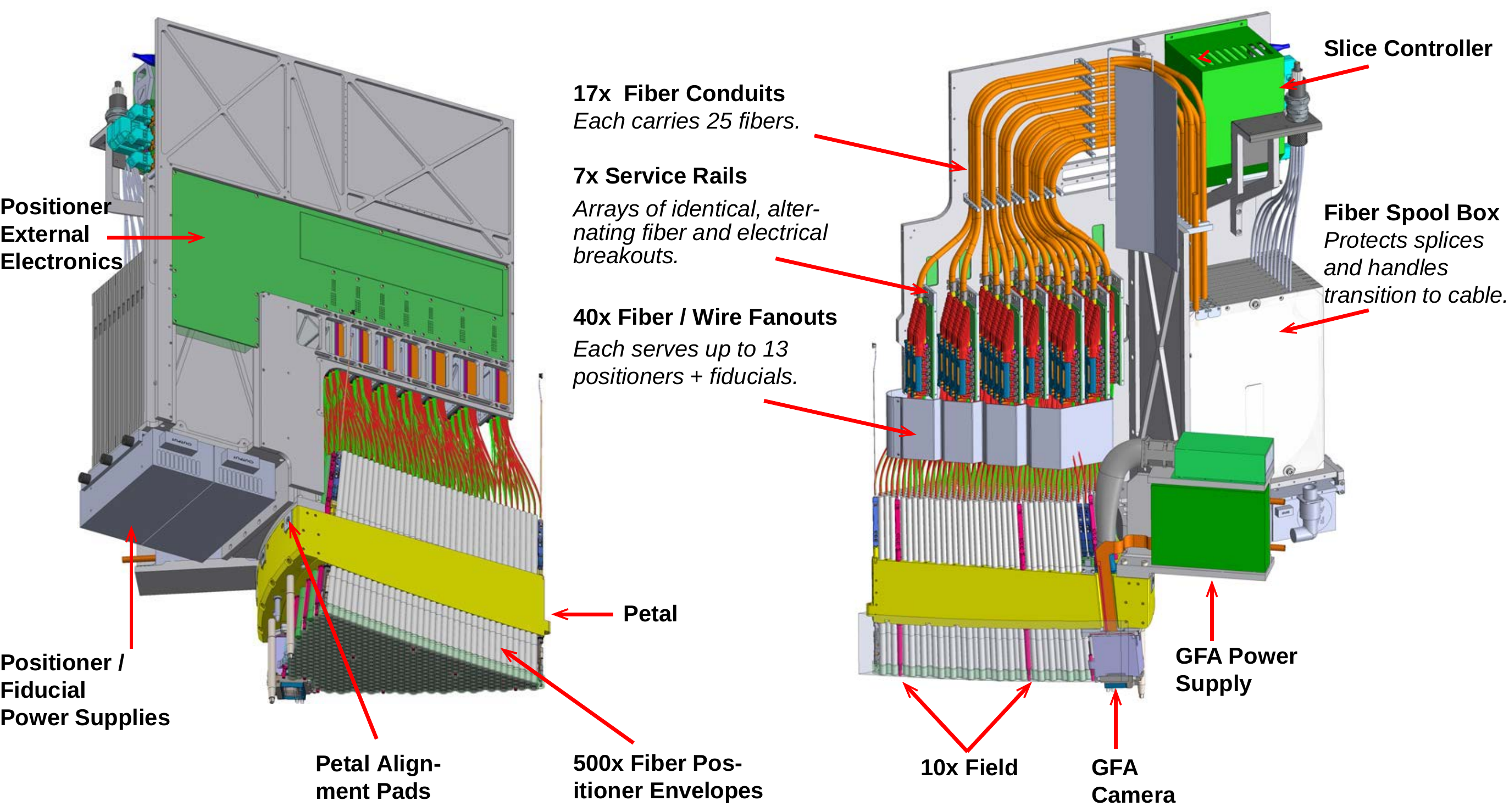}
\caption{Illustration of completed petal assembly with fiber positioners, fiducials, GFA, and fiber spool box. This is a single unit ready for loading into the integration ring.}
\label{fig:petal_assembly}
\end{figure}

In addition to distribution of science fibers, power and signal must be routed to the individual positioners. These 5,000 sets of fibers and wires are of narrow gauge and thus take up little total volume; to avoid a nest of complexity, however, their routing requires a consistent and organized approach. The first level of this organization occurs naturally at the division into 10 petals, which immediately and logically reduces the complexity to 500 sets of services. From this point we dispatch fibers into 14 furcations of 36 each. Each furcation then divides into 3 smaller groups, where each group services 12 individual positioners (really 4 of these 42 groups will service only 11 positioners). This is illustrated in Figure~\ref{fig:petal_assembly}.

The electrical services logically follow the fibers' distribution. On each petal, the 42 groups (14 furcations x 3 groups) are serviced in 7 transverse rows. Each row has a printed circuit board with individual electrical connectors to the positioners in their groups of 12. These boards plug into a main service board by a robust connector with backup screws.

The PCB on each row is mechanically supported by a stiff transverse aluminum rail. The positioner connector groups are placed in alternation with fiber fanout hardware, also mounted to the rail.

In summary, this service organization scheme reduces the potential complexity of fiber and electrical distribution from 5000 to only 12 positioners at a time. Moreover, every group of 12 positioners is serviced in a standard and repeatable way, by identical electrical and fiber support hardware.


\paragraph{Thermal budget} 

Due to the heat capacitance of the focal plate structure, the system is insensitive to the particular step profile of the heat outputs of the fiber positioners. These are relatively momentary compared to the length of the observation. The momentary heat impulse during re-positioning is smoothed by the plate capacitance and gradually removed over the course of the observation. Control electronics are shut down during observation such that positioner power consumption is effectively zero when not moving. A total of 30~J each is allotted per observation cycle. The cooling system will be sized to these allotments, though it is noted that current positioner prototypes have total energy usage $<$~11~J per reconfiguration.

\subsubsection{Positioner Mechanical Design}
\label{sec:positioner_design_overview}

The DESI fiber positioner has eccentric axis kinematics and is sized for a 10.4 mm center-to-center pitch. Actuation is by two $\phi$4~mm DC brushless gearmotors. Each positioner has one electronics board attached to the rear. This board accepts a DC voltage for power and CAN messages for communications and then drives the two motors. Power is shut off when the positioner is not operating.  The positioner mounts to the focal plate petal by screwing it into a threaded hole like a sparkplug.

\begin{figure}[!ht]
	\centering
	\includegraphics[width=0.95\textwidth]{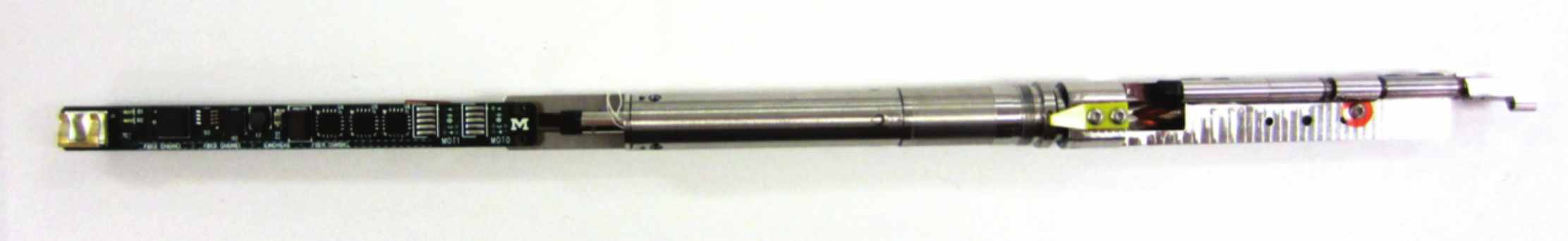}
	\caption{Prototype of the DESI 10.4~mm pitch fiber positioner.}
	\label{fig:proto10mm_pos}
\end{figure}

\begin{figure}[!ht]
	\centering
	\includegraphics[width=0.95\textwidth]{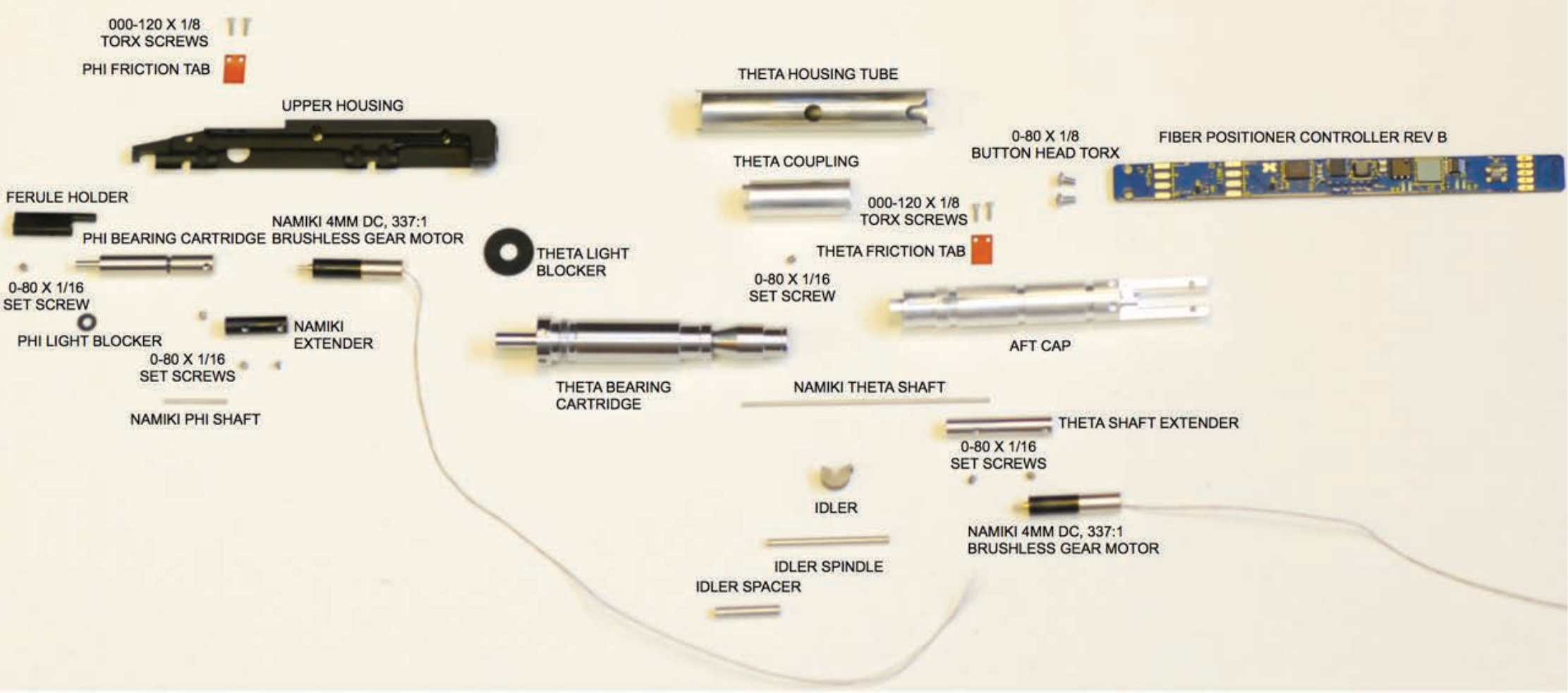}
	\caption{The components of the DESI 10.4~mm pitch fiber positioner. The design is based on commercial off the shelf $\phi$4~mm DC brushless gearmotors.}
	\label{fig:proto10mm_dino_dig}
\end{figure}

\begin{figure}[!ht]
	\centering
	\includegraphics[width=0.95\textwidth]{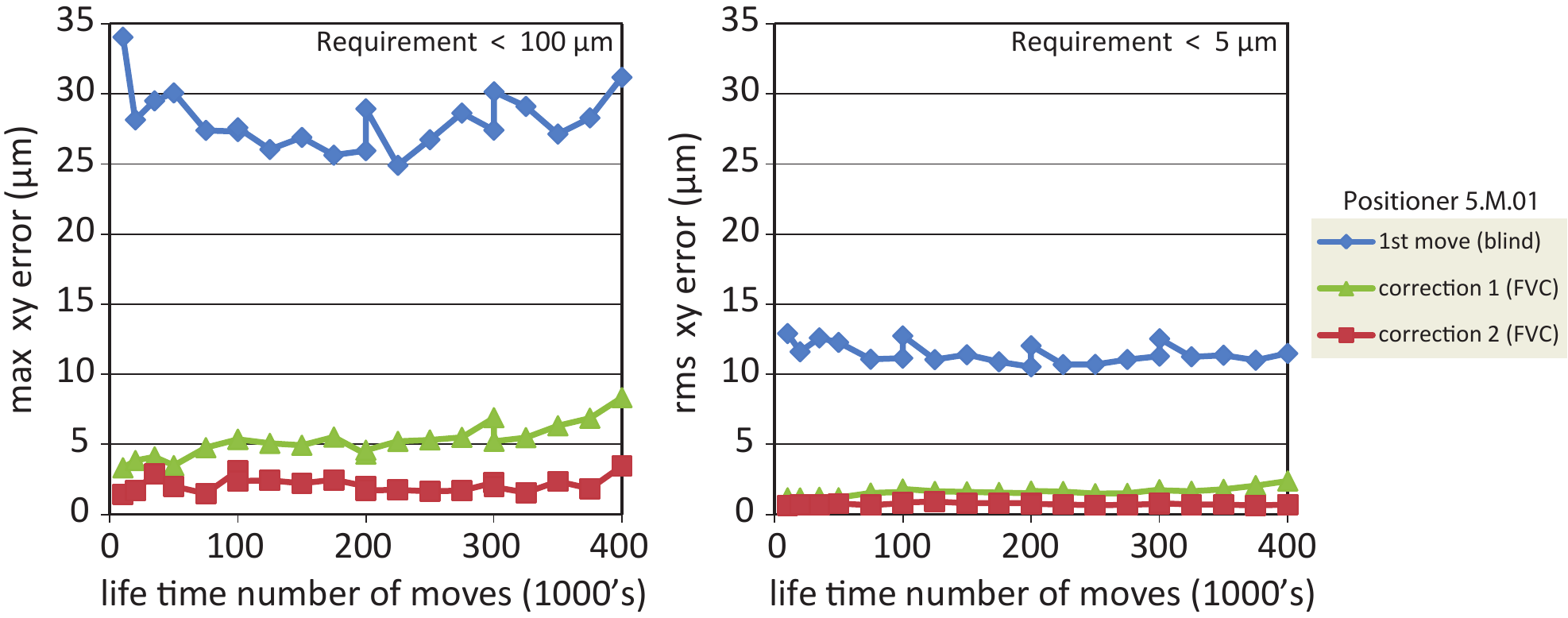}
	\caption{Positioning accuracy of prototype over a 400,000 move test. (The anticipated lifetime number of moves on the DESI instrument is 186,000.) The test is fully automated with a fiber view camera. The test takes place with continuous duty for over a month. The positioner met all requirements with comfortable margin throughout. Over the 400,000 move test, we observe a small increase in 1st correction max positioning error, which we believe is due to micron-scale wear on the internal gears. Two prototypes with motors from two different vendors were tested to the 400,000 move lifetime during prototype development. More positioners will be tested as they become available in Fall 2015, with a minimum of two more complete life tests before the first engineering model round of production.}
	\label{fig:lifetime_summary_5M01}
\end{figure}

Prototypes are fast (180$^\circ$/sec), accurate (1st move error: 10-13 \micron rms, 25-30 \micron max; 1st correction error: 1-2 \micron rms, 4-6 \micron max), and robust (tested to 400k reconfigurations with no degradation in performance). One of the prototypes is shown in Figure~\ref{fig:proto10mm_pos} and the individual components are shown in Figure~\ref{fig:proto10mm_dino_dig}. Plots of positioning accuracy measured in a typical prototype are shown in Figure~\ref{fig:lifetime_summary_5M01}.

\begin{figure}[!t]
	\centering
	\includegraphics[height=2.5in]{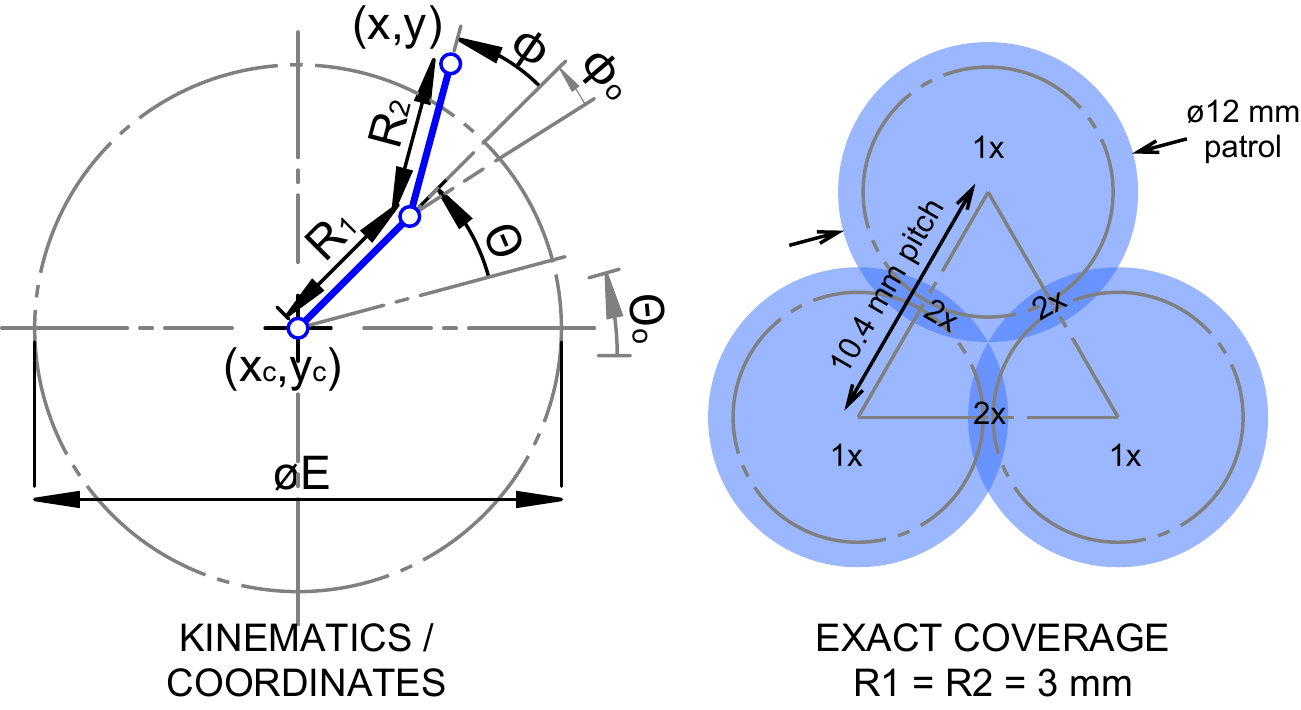}
	\caption{Left: Eccentric axis (``$\theta$-$\phi$'') kinematics and coordinate systems. Whenever R2 is retracted within the dashed circle E, the positioner is guaranteed free rotation about $\theta$ without obstruction from its neighbors. Right: Patrol coverage regions for neighboring positioners. Regions labeled ``1x'' indicate coverage by a single positioner and ``2x'' by two positioners.}
	\label{fig:eccentric_axis_patrol_area}
\end{figure}

The positioner moves the fiber about a planar disk region. There are two rotational degrees of freedom, a central axis and an eccentric axis. These axes are parallel to each other. Often these axes are referred to as $\theta$ and $\phi$. The kinematics and coverage are illustrated in Figure~\ref{fig:eccentric_axis_patrol_area}.  

Each of the two motors directly drives an axis. There are no custom gears or off-axis transmissions needed. The motors can be purchased from any of several vendors. Motors come from the factory with an integral gear head that has a reduction ratio of 256:1 to 337:1, depending on the particular vendor.

We control stiffness, run-out, and parallelism of the two rotational axes by two bearing cartridges (one for the $\theta$-axis and one for the $\phi$-axis). Each cartridge is a hollow cylindrical assembly containing two ball bearings and an internal shaft. The cartridges are mounted in-line with their respective motors' output shafts. The cartridges are purchased as integrated units from a bearing manufacturer.

The positioner has mechanical hard stops in both directions of both axes. Either axis can be driven at full speed and power into either hard stop repeatedly and without damage. During all testing we regularly ram hardstops and have not observed any discernible damage. More tests are planned to attempt to quantify if any damage modes can be identified.

The fiber ferrule is held by a ferrule holder arm. The arm secures the 1.25 mm DESI ferrule against a precise cylindrical mating surface. The ferrule is contacted by a nylon-tipped set screw under controlled preload, with a small quantity of adhesive at the threads and tip to lubricate the contact point during insertion and then to guarantee it cannot later loosen.

\begin{figure}[!tb]
	\centering
	\includegraphics[height=2in]{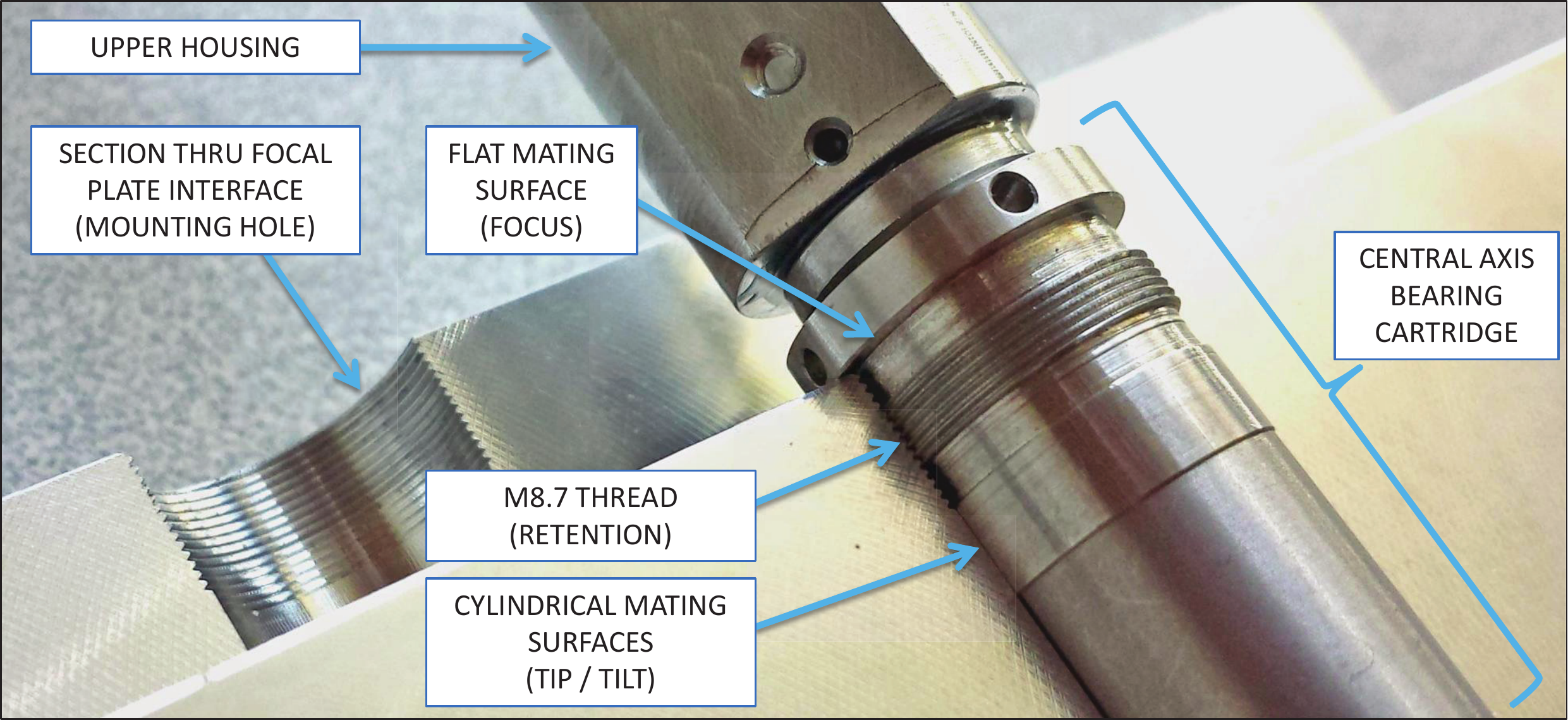}
	\caption{Positioner mechanical interface to focal plate. During insertion, the precision cylinder (g6 tolerance) of the bearing cartridge engages the precision hole (H7 tolerance) in advance of the threads. This keeps the threads well-aligned so that they engage smoothly every time.}
	\label{fig:sparkplug_interface}
\end{figure}

All the mating features for precision mounting and retention in the focal plate petal are machined into the central axis bearing cartridge. A cylindrical datum constrains tip / tilt and xy position. A flat flange datum defines focus. A thread provides positive and firm mechanical retention. These identical features are used on the assembly jigs that locate the eccentric axis bearing and the ferrule holder for each positioner. The features are illustrated in Figure~\ref{fig:sparkplug_interface}.

\subsubsection{Positioner Electrical Design}

The two motors are driven by one electronics driver board attached to the aft end of the positioner. There is no encoder on either motor. The motors each have 3 coils and are driven by pulse width modulation of the current applied to its coils. The shaft position is controlled open-loop by rotating the sum magnetic field of the 3 coils.

For large motions, the magnetic field, and thus the motor shaft, is ramped up rapidly to a rotation rate of 10,000 rpm. This is very fast in practical terms: $\sim$180$^\circ$/sec at the output shaft, depending on exact gear ratio. When re-positioning, the final $\sim$50 \micron of the move is done at low speed (15 rpm) and very fine resolution (0.1$^\circ$ at the motor shaft, $\sim$0.0003$^\circ$ at the output shaft). After positioning a shaft, the coils are completely de-energized and the position remains stable. This has been demonstrated by taking repeated centroids with the fiber view camera, before and after de-energizing the coils.

Electrical connection to each positioner is by a single connector with 5 wires. The connector is electrically, mechanically, and environmentally robust (vendor-tested to EIA-364-20, -21, -23, -09C, -28, -27, -04, -32, -17, -31, -36) and is keyed so it can only be mated the correct way. Two lines of the five are for +DC power and ground. Two lines are for the CAN bus. The fifth line is for a sync signal so that multiple positioners' motions can be easily coordinated.

Each positioner is uniquely addressed by ID number, and can be physically placed anywhere on the CAN bus. Groups of positioners are connected in parallel, and in any order, to simple 2-line power and signal rails. This decoupling is advantageous at all levels of testing, integration, and maintenance.

\begin{figure}[!ht]
	\centering
	\includegraphics[width=\textwidth]{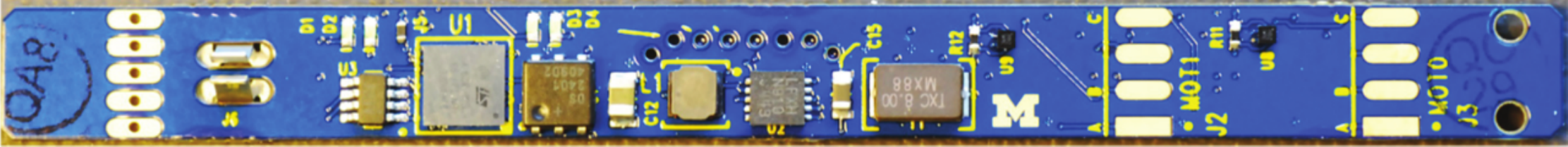}
	\caption{Electronics driver board for the fiber positioner, which was laid out at University of Michigan. On the left side is the spring clip which retains the fiber. On the right the two holes are where the board screws to the aft end of the positioner. The board is 8~mm wide and has been carefully designed so that all components, connectors, and fiber pass easily through the focal plate interface.}
	\label{fig:um_positioner_board}
\end{figure}

The PCB is a 0.8~mm thick 6-layer board, with commercial components, to achieve the necessary small size. The low-level firmware is compiled C code. The firmware is compact ($<$ 6K bytes exclusive of tables) and has been thoroughly debugged and tested, including motor control and CAN communication.
Figure~\ref{fig:um_positioner_board} shows the prototype board.

\subsubsection{Positioner Software}

The array of positioners is controlled by a Linux computer, which communicates with each positioner by ID number over the CAN bus. The controller has software which keeps track of each positioner's status and shaft angles, and sends out move commands. The software to run a single positioner has been thoroughly debugged and tested. The external software interface is a single object written in the Matlab language, with a complete set of external methods provided. All calibrations and parameters are set by a simple key/value list. The parameter values can be granularly set uniquely for each positioner, can be updated in real-time, and generally are typed and structured for ease of logging and simplicity of database configuration control. Control of an array of multiple positioners is established by making multiple instances of the object with the appropriate ID numbers.

One level of abstraction higher, there is software to coordinate the array of positioners. In particular, this software includes an anti-collision algorithm, since positioners can reach into their neighbors' patrol envelopes. This algorithm has been coded and debugged in detailed simulation. The simulation was run for $10^5$ random configurations. Because of the 180$^\circ$/sec shaft speed, the maximum time to reconfigure the array including all anti-collision moves is 4 seconds, and the average is 3.6 seconds. The algorithm will be tested on real hardware on the first array of positioners, the engineering model to be built in early 2016. It is noted that the ProtoDESI plate (being built simultaneously) has positioners spaced specifically not to collide, since any software bugs in the anti-collision code will have low technical impact at this stage, but may take up schedule time to iron out, and have nothing to do with the technical goals of ProtoDESI. Instead, over 6 months of time are scheduled for anti-collision and FVC loop-closing with the fully-populated engineering model petal, which anyway has relevant features such as asphere geometry, non-uniform spacing, non-uniform central axis angles, petal edge no-go zones, and GFA camera proximity, none of which ProtoDESI would emulate.

\subsubsection{Fiber Interface}

\begin{figure}[!h]
\centering
\includegraphics[width=0.9\textwidth]{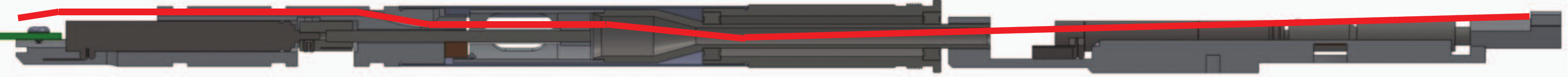}
\caption{Fiber route through the fiber positioner is illustrated by a thick red line. Ferrule holder is at the right, exit at electronics board is on the left.}
\label{fig:fiber_route_thru_ecc_axis_positioner}
\end{figure}

The positioner accepts the ferrulized and polished science fiber and provides a mechanically safe path through its internal mechanism. The fiber path is illustrated in Figure~\ref{fig:fiber_route_thru_ecc_axis_positioner}. The fiber is installed in the positioner, and the positioner is then inserted into the focal plate from the corrector side. A precision jig sets the placement of the ferrule tip with respect to those mechanical datums which contact the focal plate.

The integration of fibers into positioners is discussed further in Section~\ref{sec:Instr_Fibers}.

\subsection{Guide-Focus-Alignment}
\label{sec:Instr_Guide_Focus}

\label{gfa_section}

The guide, focus and alignment (GFA) system meets the following DESI requirements (see DESI-0526):

\begin{enumerate}
  \setlength{\itemsep}{1pt}
  \setlength{\parskip}{0pt}
  \setlength{\parsep}{0pt}
\item Provide guide signals to the telescope at 1 Hz to a precision of $<0.03$ arcsecond and latency of $<0.5$~second.
\item Monitor camera focus and alignment from at least three locations with approximately azimuthal symmetry.
\item Process and produce results from at least 50 defocused star images within 5 seconds.
\item Meet system environmental requirements from DESI-0583.
\item Determine accurate telescope pointing after slew within 20 seconds.
\item Determine camera defocus relative to M1 with better than 30~\micron accuracy.
\item Determine camera tip/tilt relative to M1 with better than 10~\micron accuracy.
\item Camera must operate in both full frame and region of interest (ROI) mode. 
\item Determine the current telescope pointing within 20 seconds after telescope slew.
\item Guide cameras will deliver empty sky images for seeing and transparency determination.
\item GFA system must operate during all conditions in which the telescope can operate.
\item GFA system can operate in event that an overly bright star is in the field of view.
\item The guide cameras can be requested to deliver full frame images at any time.
\end{enumerate}

Additionally, two design guidelines constrain the GFA design. These are
\begin{enumerate}
\item The GFA cameras will operate at ambient temperature to minimize local heating near the focal surface.
\item The GFA camera footprint will be approximately the footprint of the e2v CCD230-42 to minimize the number of science fibers displaced.
\end{enumerate}

The GFA system contains ten nearly identical cameras divided into two groups: focus and alignment cameras and guide cameras. Each camera contains a single CCD sensor and is mounted and operated as a standalone instrument. Each camera provides all readout modes used in any GFA camera. The cameras vary only in optical filter mounted above the CCD. An essentially single development stream minimizes risk, development time and cost. Each camera operates without a shutter in frame transfer mode and is sealed with a broad range (red) optical filter. Each camera contains all CCD readout electronics and all camera control and requires only DC power and a gigabit Ethernet connection. USB connectors may also be provided for in lab testing.

\subsubsection{Sensor Selection}

The GFA system is required to operate at ambient temperature to minimize the heat load in and near the DESI optical path. This leads to a sensor requirement of excellent dark current performance. This, combined with the required sensor area, reduces the number of candidate sensors. 
The e2v CCD230-42 is baselined since it has excellent ambient temperature dark current performance and read noise. This sensor has excellent manufacturer support and several commercial astronomical cameras have incorporated
it (\eg, Apogee Alta F and FLI Proline).
The sensor provides four channel readout and can operate in frame transfer mode removing the need for an optical shutter. The frame transfer operation uses 50\% of the sensor area leaving $9.4~\textrm{cm}^2$ active area. The DESI plate scale varies with radius so that radial and azimuthal plate scales are slightly different. At the edge of the focal plane where the GFA camera is located the plate scale is $\sim$70~\micron/arcsec. The active area of each sensor provides $\sim$29~arcmin${}^2$ of area on the sky.

The GFA system carries a risk that operating the sensor at ambient temperatures falls outside of the ``normal'' operating procedures for a CCD.  In order to mitigate this risk, an FLI Proline ML23042 camera was obtained and tested. While the dark current proved stable and repeatable it was found to be significantly higher than predicted in the datasheet.  e2v confirmed a mistaken scaling equation. Eventually e2v replied that the given data sheet formula of 
At 20~\celsius the dark current was found to be 140~e${}^-$/s/pixel ($\sigma=11.8~e^-$).  This is to be compared to a read noise of 20~$e{}^-$/pixel in the FLI camera.
Only on a rare  summer night would we expect the sensor to approach 28(82)~\celsius(F) producing a dark current of 400~$e{}^-$/pixel.  
%

\subsubsection{Packaging}

\begin{figure}[!ht]
\centering
\includegraphics[height=5in]{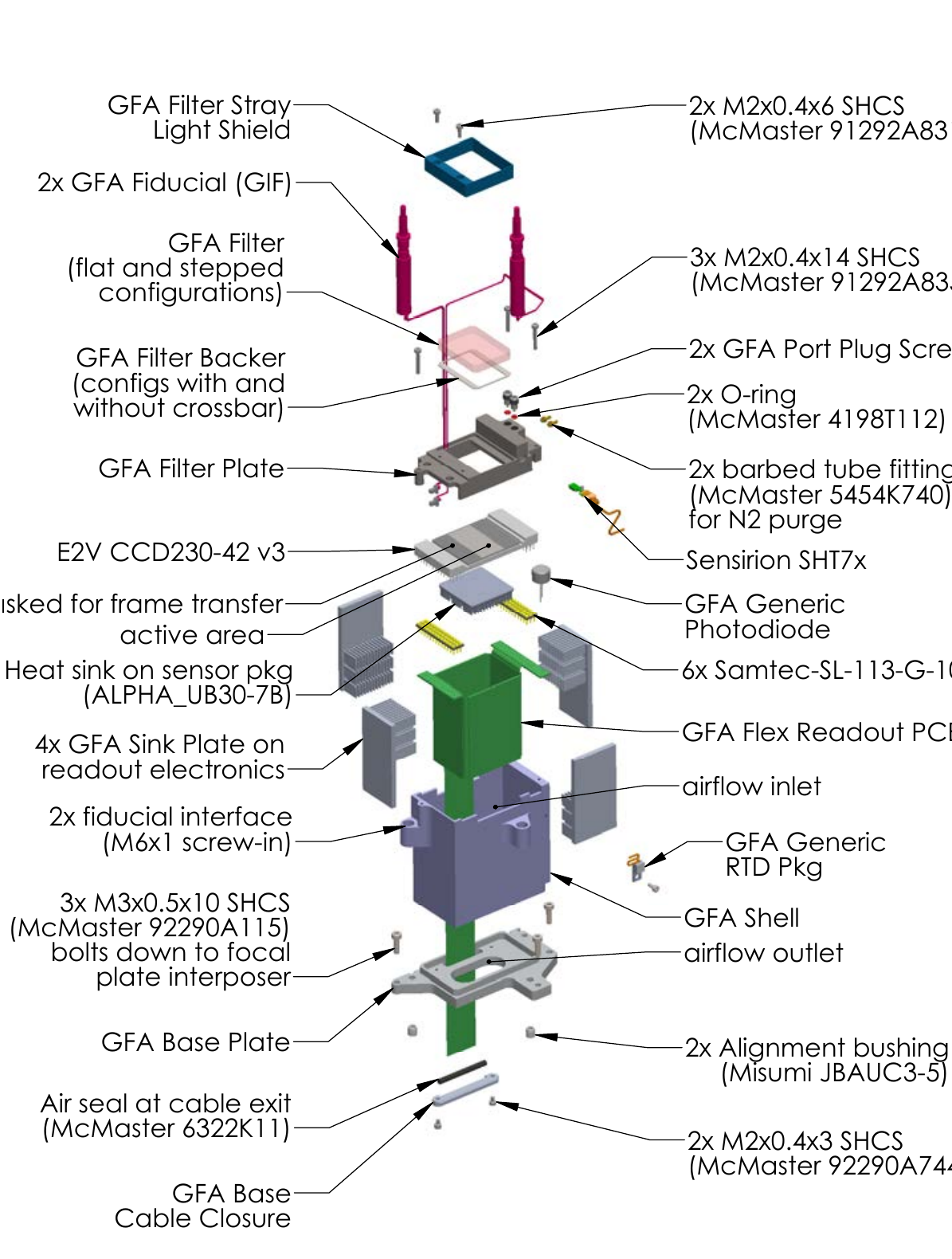}
\caption{Detailed concept design for the GFA camera.}
\label{fig:GFA_exploded}
\end{figure}

\begin{figure}[!t]
\centering
\includegraphics[height=2in]{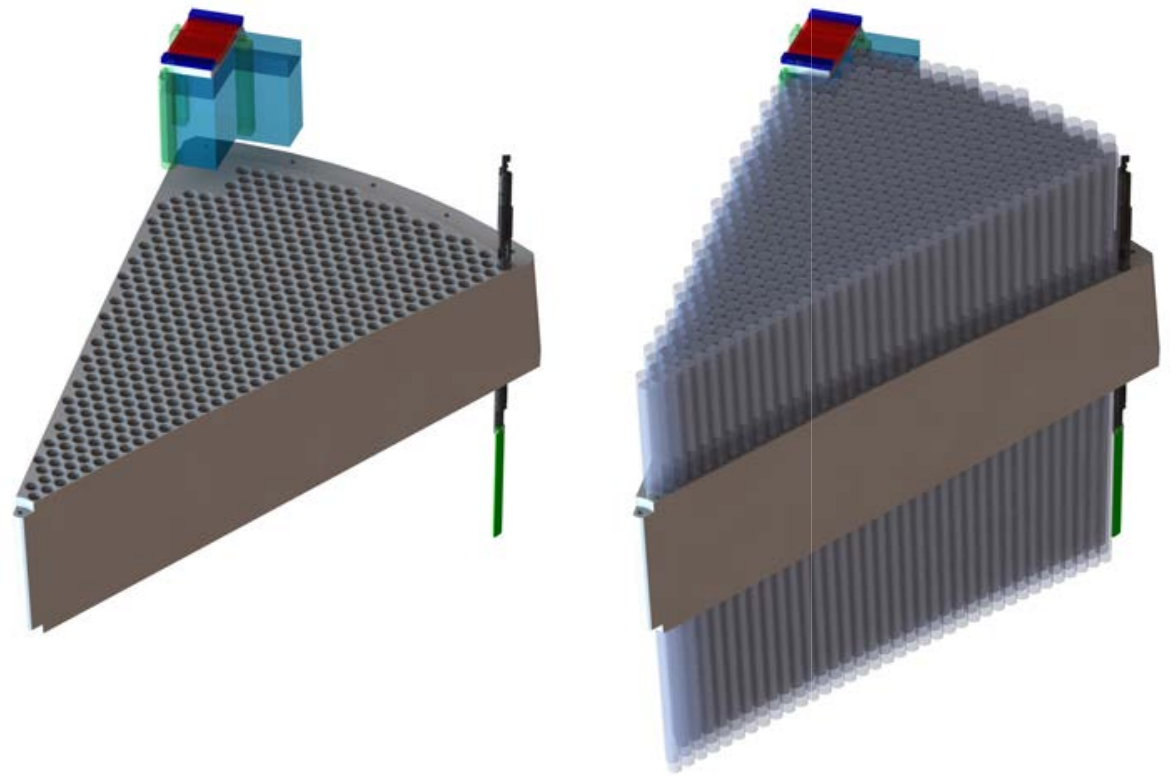}
\caption{GFA mechanical envelope is shown in relation to a focal plate petal. The envelope includes an e2v CCD230-42 package, masked for frame transfer, with the unmasked active area in unvignetted light. The GFA envelope includes generous packaging room for illuminated fiducials (transparent green) and electronics (transparent blue). A fiber positioner is shown at one corner of the petal, and in the right view positioner envelopes illustrate the full packing.}
\label{fig:gfa_nesting}
\end{figure}

The ten GFA cameras distributed around the focal surface with one camera per focal plane petal can be seen in the bottom right hand side of Figure \ref{fig:focal_plane_system}. 
A detailed concept design for the camera has been produced and is illustrated in Figure~\ref{fig:GFA_exploded}. In order to minimize the footprint of the GFA camera, all readout electronics are located on boards running perpendicular to the sensor surface. Figure \ref{fig:gfa_nesting} shows a single petal and the location of the GFA camera on that petal. In-focus sensors distributed evenly around the perimeter of the focal surface provide a cross check for any atmospheric or thermally induced changes in the magnification of the system. Having defocused sensors distributed evenly around the perimeter of the focal plane is required to fully sample the wavefront.

\subsubsection{Electronics}

\begin{figure}[!b]
\centering
\includegraphics[height=2in]{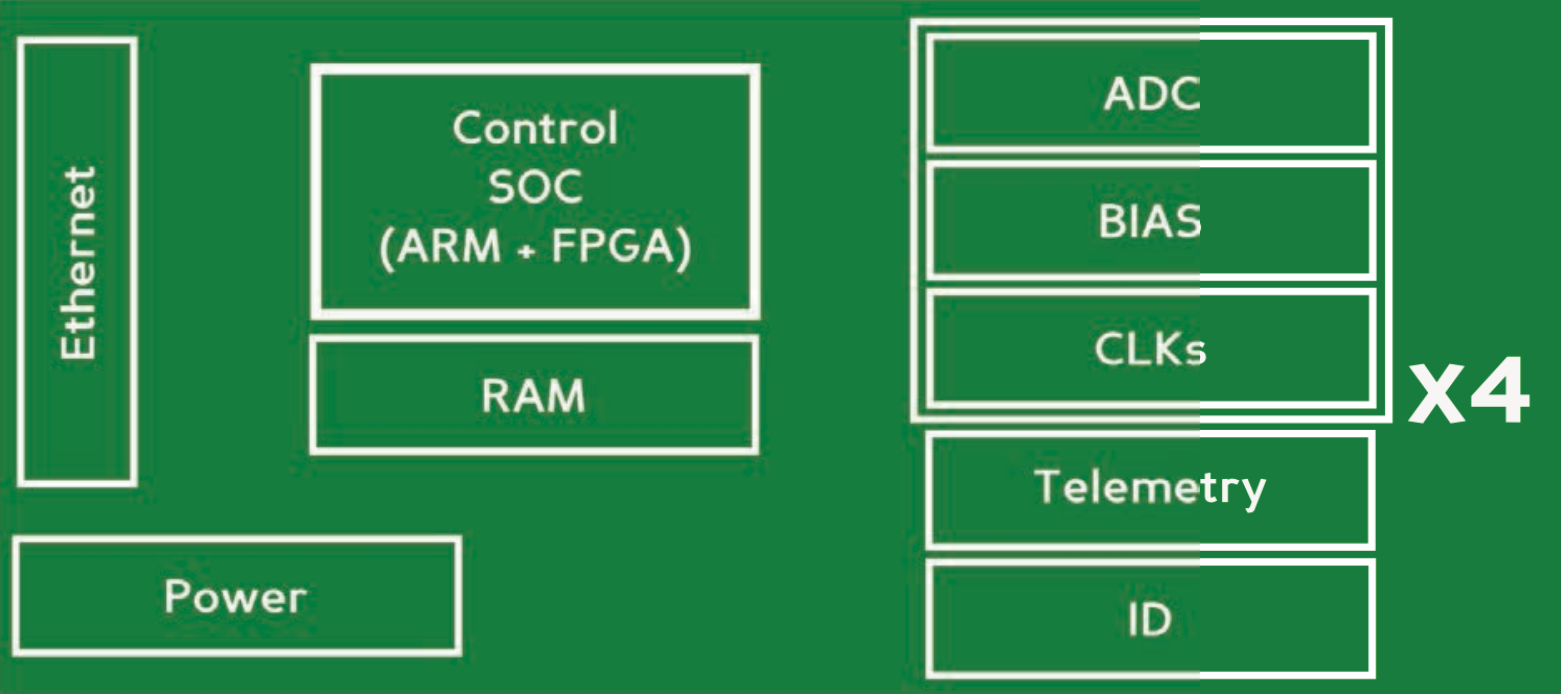} 
\caption{
All systems will be coordinated by a Xilinx System on Chip (Zynq) which will host both the low level firmware in vhdl and also a Linux embedded system.}
\label{fig:electronicslayout1}
\end{figure}

Readout of the CCD will be fully digital. This means that the output signal from each CCD amplifier will be digitized at a high rate (100 Msps) and then a digital correlated double sampling applied.
The electronics is in charge of generating the clocks to shift the charge on the CCD synchronously to the ADC and the firmware which implements the correlated double sampling. 
The voltage bias and clock values will be fixed on CCD characterization to a value, allowing a reduction of the required space to implement it.
Each CCD controller will have an Ethernet interface, a power module, a telemetry module and 4 modules to operate with each of the CCD channels as shown in Figures~\ref{fig:electronicslayout1}.
The main idea is that the system will only require 2 cables: Ethernet and power. Main power supply will be external to the system.

\subsubsection{Illuminated Fiducials}

The science fiber tips are mapped to sky positions by the Fiber View Camera and the GFA system. The acquisition step will determine the absolute pointing of the GFA sensors on the sky. Illuminated fiducials mounted near these sensors will be correlated to the position of all of the other illuminated fiducials and then the fibers located within the fiber actuators. Mechanical survey and on sky calibrations will be required to complete the guide sensor to science fiber calibration.

\subsubsection{Acquisition and Guiding}

DESI requires absolute pointing on the sky in order to deliver light into the science fibers. For each spectrographic exposure, the telescope will slew to a new location and all guide sensors will be read out in full frame mode and an astrometric solution computed. Stars found in these images will be compared to a star catalog to determine the absolute pointing of the telescope. This catalog needs to be matched absolutely to the target lists and should be created from the same data set as the targets or from an extremely good astrometry catalog such as {\it GAIA}.

Once the absolute pointing is determined the guide sensors will switch to region-of-interest (ROI) mode and start guiding the telescope to the correct absolute position on the sky and maintain that pointing during the observation.  Photometric and image quality information from stars will also be used to determine observing conditions to be used in a dynamic exposure time calculation.
To provide consistent guiding on all parts of the sky the system contains ``in focus'' optical sensors covering a total of $\sim$180 arcmin${}^2$. Six of ten GFA cameras will be used for this purpose with each camera covering  $\sim$29.3 arcmin${}^2$. In the guiding configuration, the camera is equipped with a flat optical (R-band) filter (Figure~\ref{fig:GFAfilters}a). With this area, at least 10 guide stars ($15<R<19$) will be available for more than 99\% of the survey footprint. Additionally, this area is sufficient for determining the current telescope pointing within $\sim$20 seconds assuming that the Mayall provides $\pm$10 arcminute pointing accuracy after a large slew.

The delivered optical PSF are determined from the guide signals and provided to the focus and alignment system as well as the telescope telemetry. This information will be used to help adjust the exposure times for the science fibers. 

Additionally, the 4 channel ROI readout mode of the cameras provide 1 star/object and 3 empty ROIs. These empty ROI frames are also delivered to any interested client for determining changes in total atmospheric and instrumental transmission.

\subsubsection{Focus and Alignment}

\begin{figure}[htb]
\centering
\includegraphics[width=0.9\textwidth]{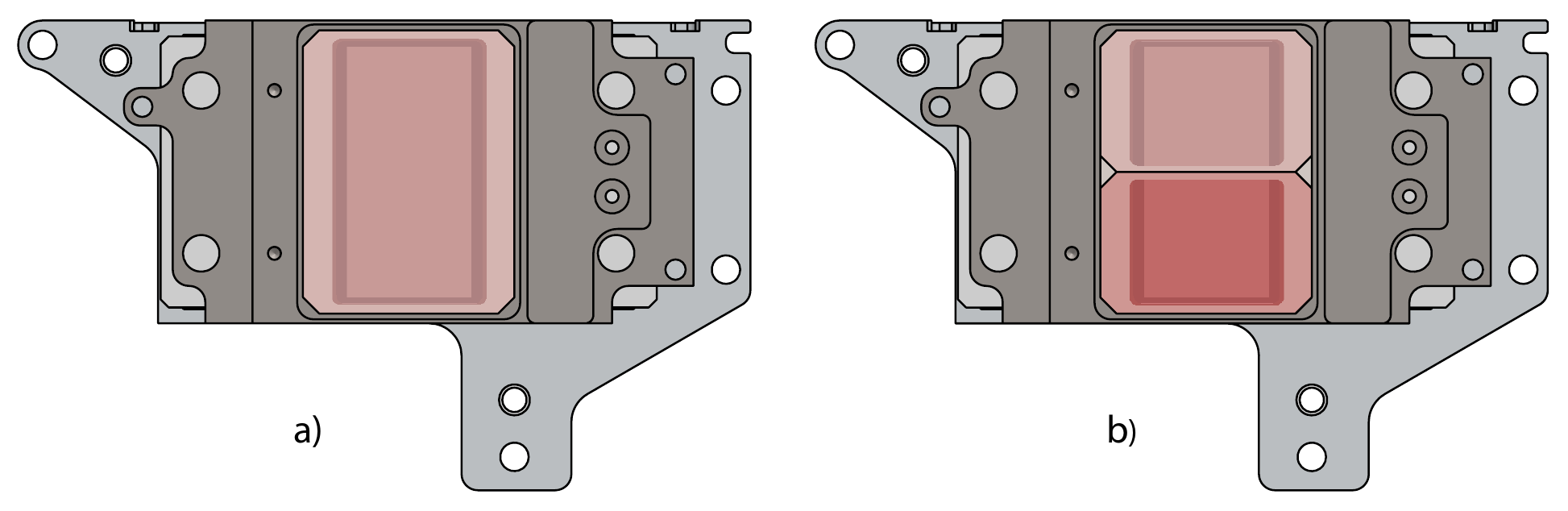} 
\caption{Top views of a GFA camera showing different  filter configurations for a) guide/acquisition and b) focus/alignment.}
\label{fig:GFAfilters}
\end{figure}

Four of the ten identical cameras will be used for focus and alignment.  In this configuration the cameras have a dual thickness optical filter  (Figure~\ref{fig:GFAfilters}b). They are otherwise identical to guide cameras. This dual thickness filter provides intra and extra focal images of stars.
One half of the acquired image contains stars nominally $\sim1.5$~mm below focus and the other half is $\sim1.5$~mm above focus. These combined images provide all needed information to correct  focus, tip/tilt and decenters with the corrector hexapod.

\begin{figure}[!t]
\centering
\includegraphics[height=2in]{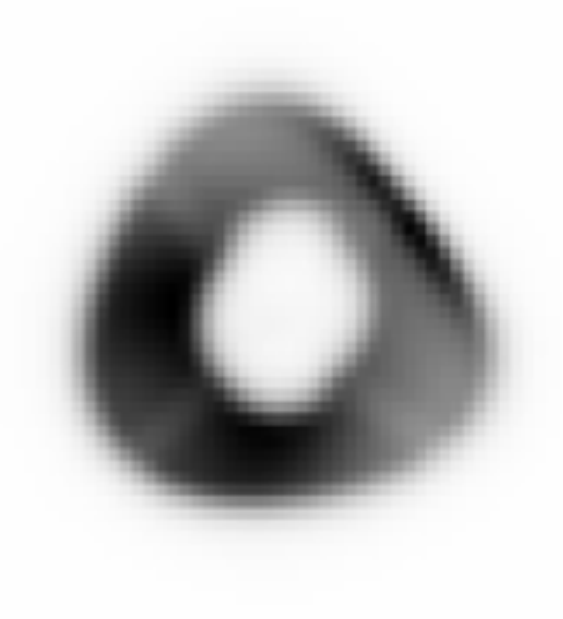} \includegraphics[height=2in]{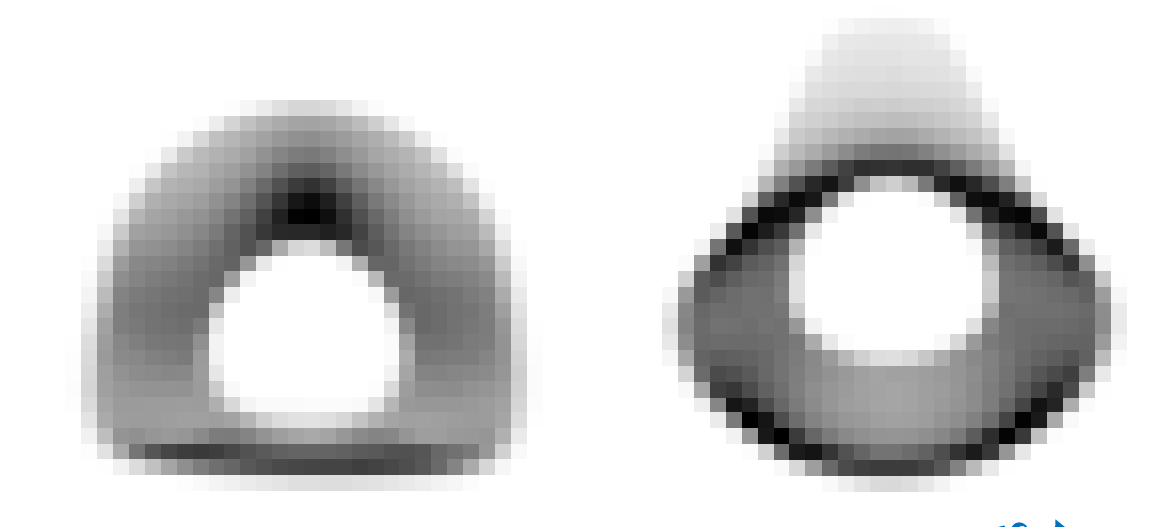} 
\caption{Left: Image of an ``out of focus'' star taken by the Dark Energy Camera. Center and right: simulated DESI intra and extra focal donuts, respectively.}
\label{fig:donutpic}
\end{figure}

The design of this system is based on the very successful system used in the Dark Energy Camera (DECam).
Out of focus stars will appear like the pupil plane, which in the case of a prime focus instrument is an annulus or ``donut''. 
 An example donut image taken from DECam is shown in Figure \ref{fig:donutpic} left.
Simulations of the optical system can generate wavefront error maps so that any observed collection of donuts can be used to determine the misalignments of the DESI focal surface. It is important to have donut samples at various locations all around the focal surface.  Intra and extra focal donut images for DESI from a Zemax simulation and shown in \ref{fig:donutpic} right.

\begin{figure}[htb]
\centering
\includegraphics[width=0.9\textwidth]{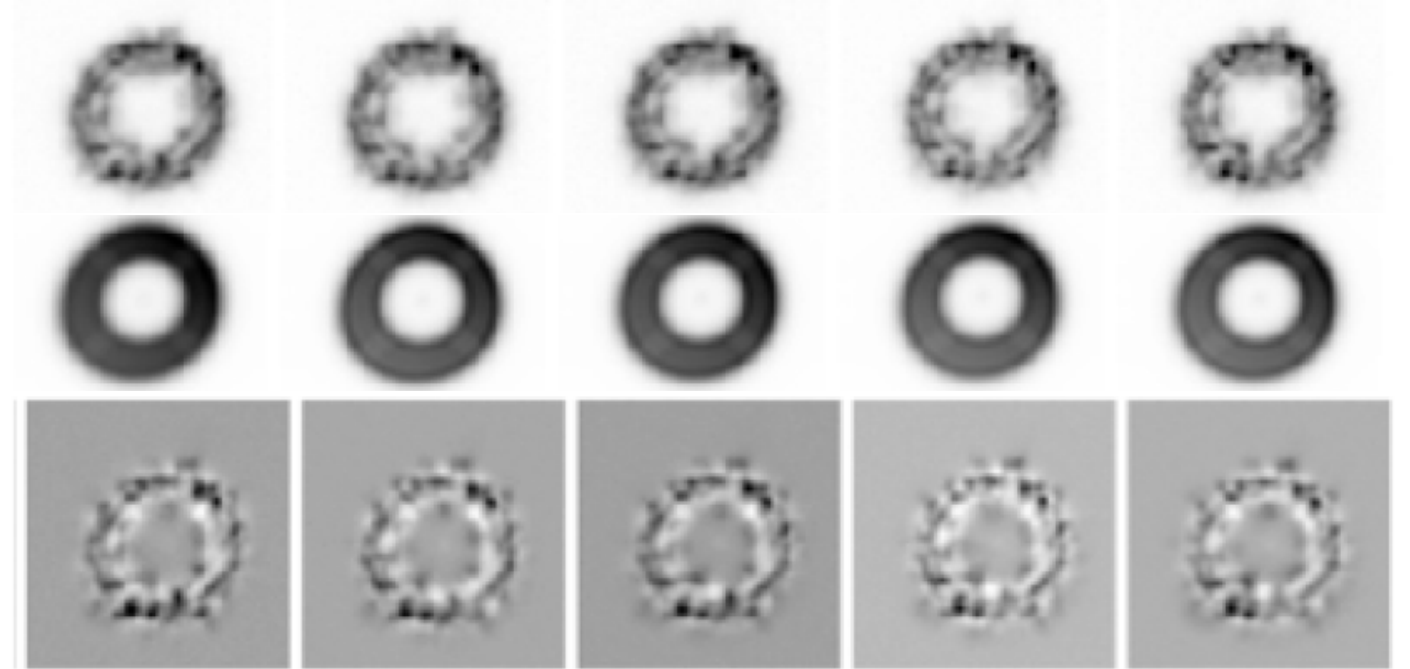}
\caption{\label{fig:donutfit} Sample donuts (top), fitted model (center), and image minus model (bottom).}
\end{figure}

The focus and alignment algorithms are based on the proven techniques used in the DECam \cite{DESfocusalignment} active optics system (AOS). The AOS currently fits one donut per CPU core in less than 2 seconds. A forward modeled $\chi^2$ fit is performed comparing the observed out of focus images (64$\times$64 pixel stamps) to a modeled expected image by varying leading order aberrations. The model includes a uniform sky background, pixelization, and a Kolmogorov kernel to model atmospheric seeing. The derivative of the $\chi^2$ with respect to the Zernike coefficients is calculated in closed form using the method described in Fienup~\cite{fienup}. The $\chi^2$ is minimized using the MIGRAD algorithm in the MINUIT~\cite{minuit} package, which implements a Davidon-Fletcher-Powell variable-metric method. Figure~\ref{fig:donutfit} shows a sample of fitted donuts from DECam showing the observed image, fitted model and model predicted image.

\begin{figure}[htb]
\centering\centering
\includegraphics[width=0.9\textwidth]{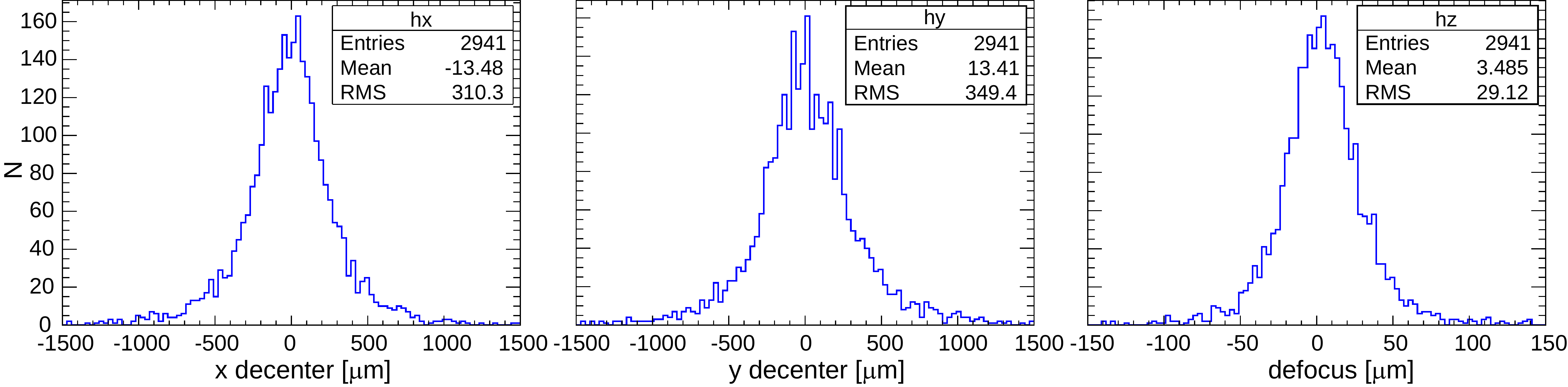}
\caption{\label{fig:dodx} Performance of DECam $x-y$ decentering and defocus from September 2013.}
\end{figure}

 Figure~\ref{fig:dodx} shows the performance of the DECam AOS for all available exposures for September 2013. Performance of 300~\micron rms in decenter and 30~\micron rms in defocus is shown. Exposures where AOS could not accurately determine the wavefront are omitted.

\subsubsection{Stellar Density}

\begin{figure}[tbh]
\centering
\begin{tabular}{c}
\includegraphics[width=0.4\textwidth]{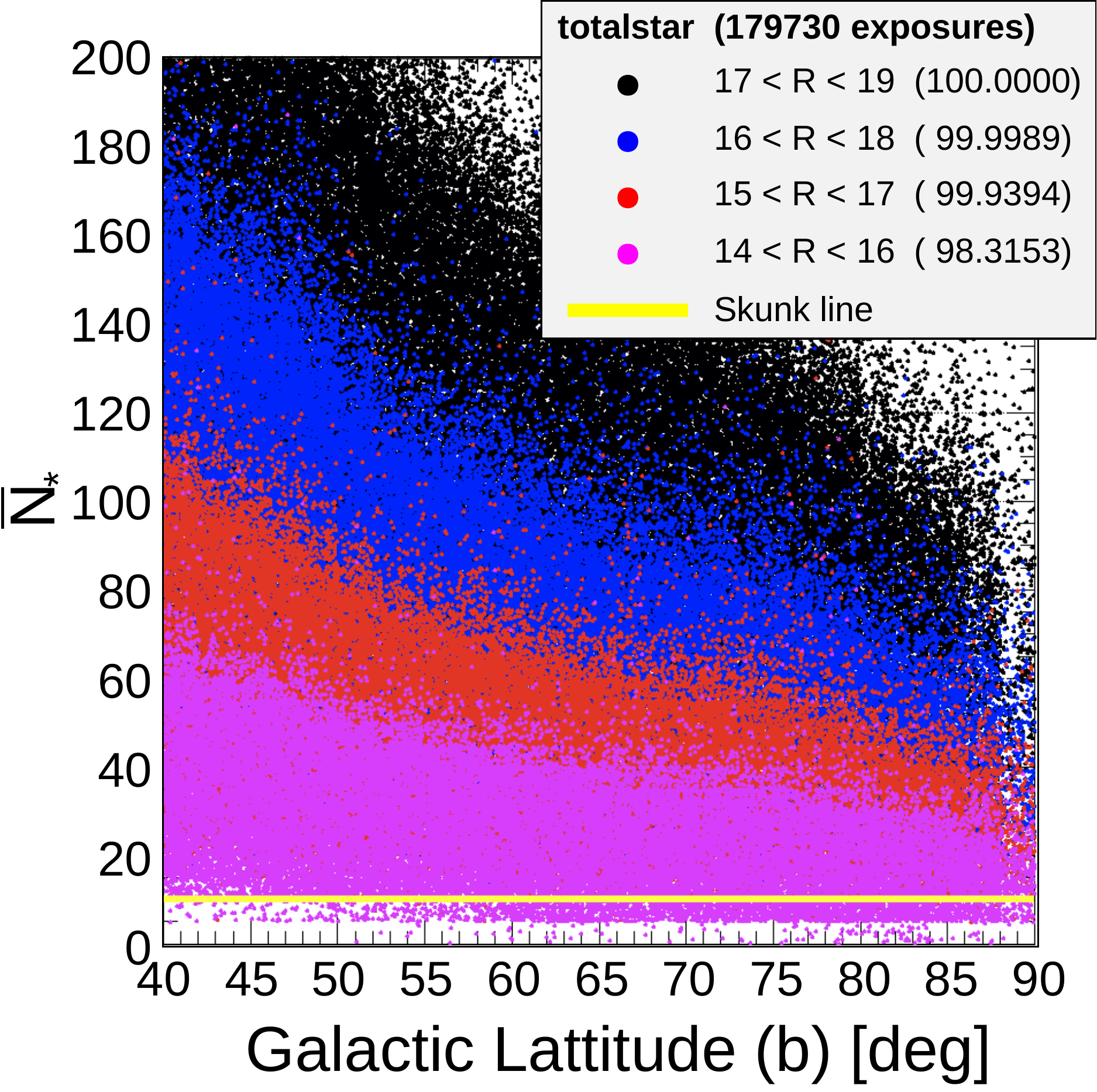} 
\includegraphics[width=0.4\textwidth]{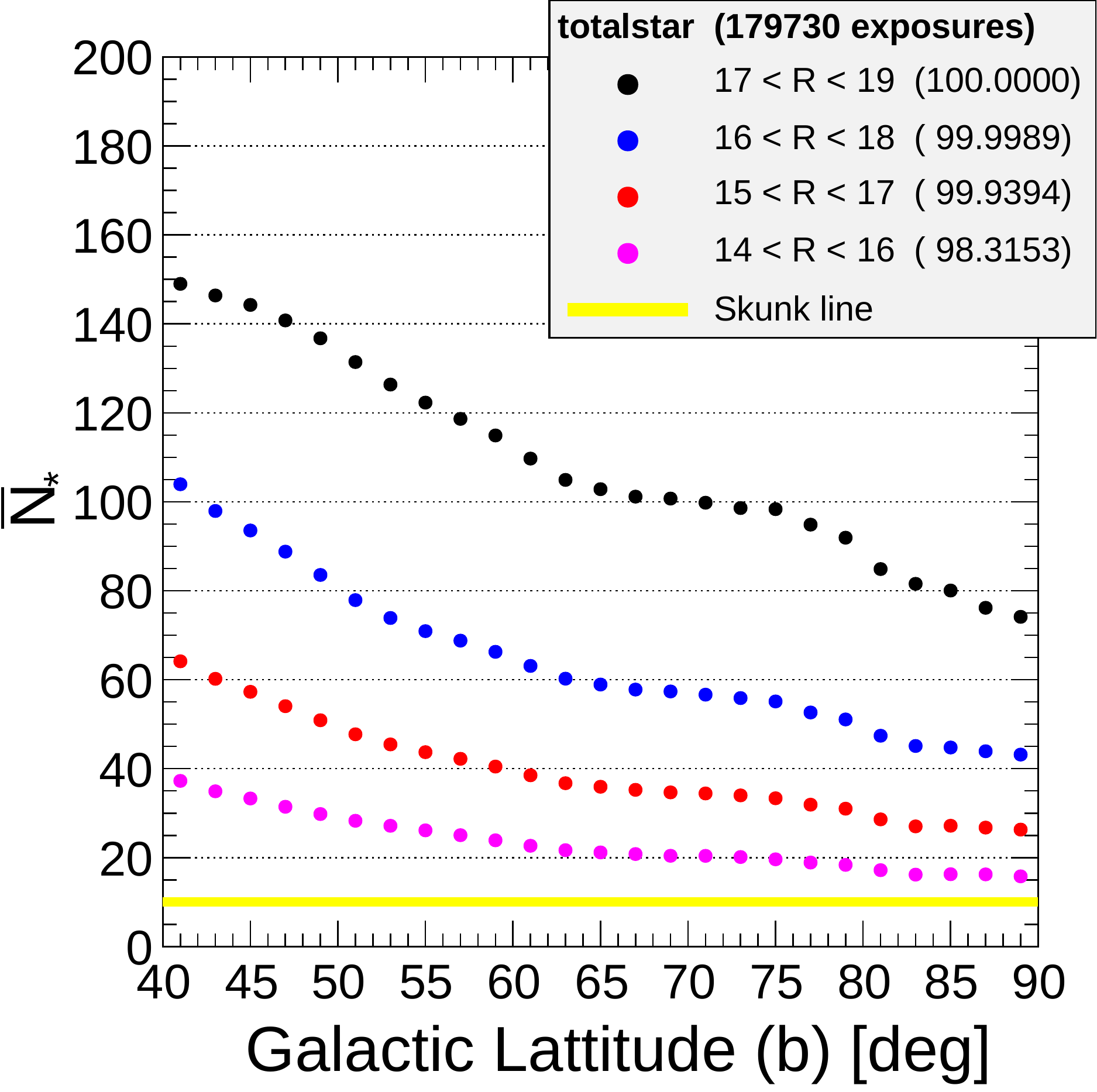}
\end{tabular}
\caption{Total number of stars seen in six DESI GFA sensors using random pointings and the NOMAD catalog.}
\label{fig:avestarfig}
\end{figure}

The GFA sensors will use stars to provide the guide signal and to measure focus and alignment of the DESI instrument. In order to determine the required area of sensors, simulations were run using both the NOMAD catalog~\cite{nomad} and tabular information from~\cite{bs2,bs1}. Figure~\ref{fig:avestarfig} shows the total and average number of stars found vs galactic latitude using the NOMAD catalog. To account for sensor dynamic range, stars were grouped into bins covering 2 magnitude bins and counted in random one degree regions of the sky. Signal-to-noise studies indicate that using magnitudes of $15<\textrm{R}<17$ and ten stars slightly exceeds our required pointing accuracy \cite{spiegfa}. The legend in figure also shows the percentage of fields that had a total of ten guide stars available. It is important to note that this simulation did not include rejection of stars with a nearby neighbor. Additionally, while the simulation used the correct sensor locations and sky area, the sensor was simulated as a circle on the sky (DESI-0341).

\subsubsection{Signal-to-Noise Studies via Simulation}

In order to better understand the effects of higher dark current and potentially higher read noise, a full simulation of guiding signal-to-noise was developed. The contents of this simulation are as follows:
\begin{itemize}
\item{A random night over the coming two years was selected. Worst case: summer and winter nights had same length.}
\item{A random telescope zenith and azimuth angle were selected: zenith $>$45\degree.}
\item{Sky brightness of the full moon always used. Worst case.}
\item{Temperature selected from Mayall historical record using full temperature range including daytime highs. Worst case. Sensor assumed to be 5~\celsius above ambient.}
\item{Delivered PSF with mean of 1.0 arcseconds and a range from 0.8 to 2.0 arcseconds was selected randomly}.
\item{For each of the 4 channels on each of 6 guide GFA cameras the brightest star in the range $15<R<19$ was selected. Used Bachall 1986 formulation for speed.}
\item{The SNR of each of the (up to 24) stars was calculated.}
\end{itemize}

\begin{figure}[tbh]
\centering
\includegraphics[height=2.5in]{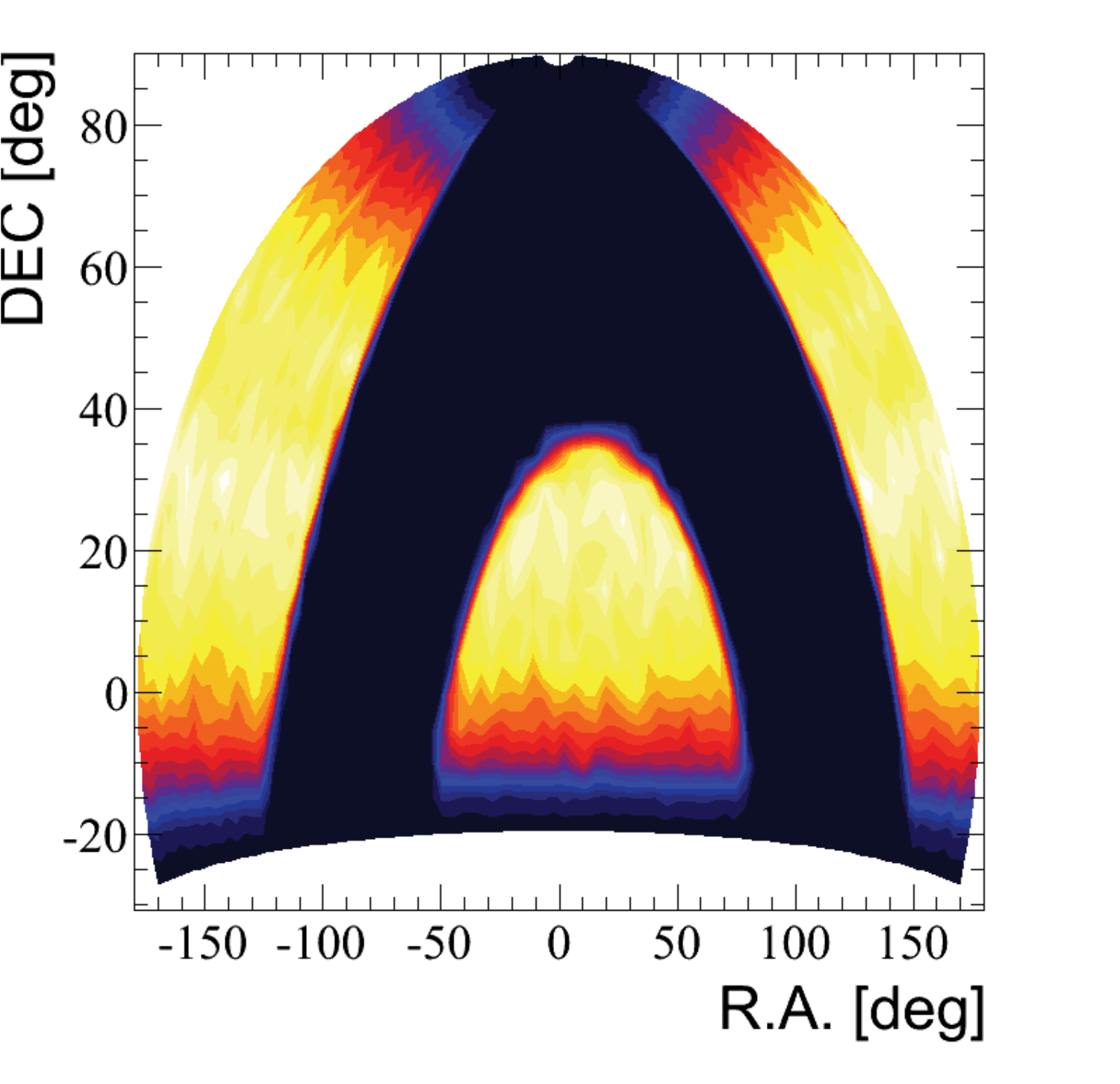} \includegraphics[height=2.5in]{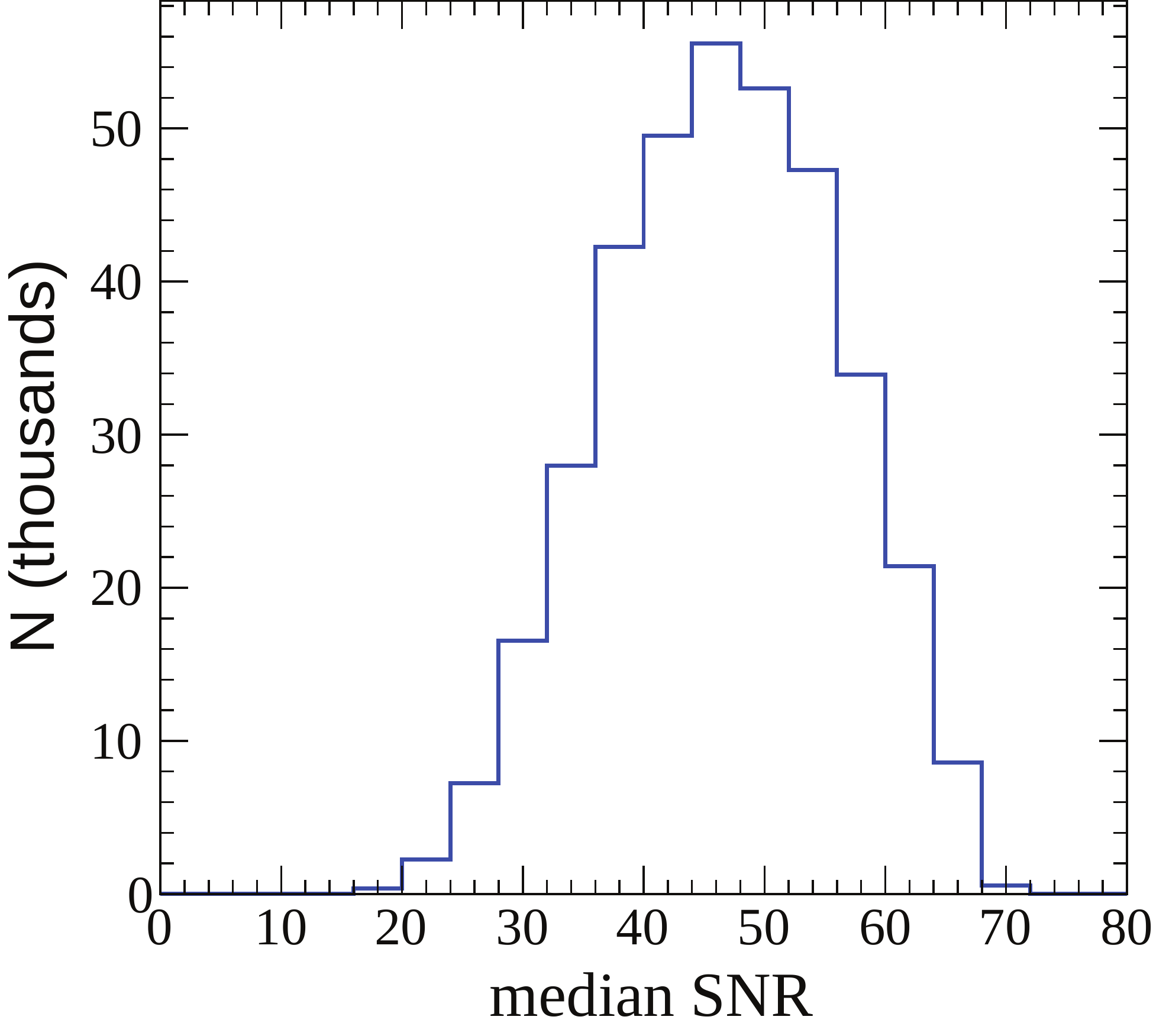}
\caption{Left: Sky coverage for guiding SNR study. Right: Simulated median signal-to-noise for guide stars over a wide range of conditions.}
\label{fig:skycover}
\end{figure}

Figure \ref{fig:skycover} left shows the sky coverage of the simulation. 
Figure \ref{fig:skycover} right shows the median signal-to-noise (here the 12th best of the up to 24 available guide stars). The mean SNR of the median guide star is more than 40 and only rarely does SNR fall below 20. Using
\begin{equation}
\sigma_{\textrm{centroid}} = \frac{\sigma_{\textrm{seeing}}}{\textrm{SNR}}
\end{equation}
we can expect to easily meet the required guide accuracy using even a single median guide star. Multiple guide stars are required, however, to account for systematic changes in focal plate change, atmospheric dispersion, \etc\  Nominally 10 guide stars will be used, increasing centroid accuracy by $\sim$3.

%

\subsubsection{Commissioning}

The GFA sensor is the only imaging system in DESI and will be vital during commissioning and engineering time for DESI.

A new pointing map will need to be created for the Mayall telescope. This is done by scanning the sky (15 degree steps in azimuth and zenith angles) and determining flexing of the telescope so that corrections for this can be made automatically by the telescope control system from the produced lookup table. Additionally, at each location the nominal hexapod settings required to maintain mechanical alignment are determined and stored in a separate lookup table.


\subsection{Fiducial Illuminators}	
\label{sec:Instr_Fid}

Illuminated fiducials are point sources distributed throughout the focal plate field. These fiducials are illuminated during Fiber View Camera (FVC) measurements (see Section~\ref{sec:Instr_FVC}). Field fiducials are attached to the focal plate petals in place of positioners. Similar fiducials are affixed to each GFA mechanical assembly (see Section~\ref{sec:Instr_Guide_Focus}). Each fiducial has precision features for mechanical survey. The mechanical datums are each measured carefully with respect to the optical source points, and with respect to focal plate datums. These are surveyed with respect to the GFA sensor active area and with respect to focal plate datums.

Once the installed field fiducials have been surveyed with respect to the focal plate datums, we have a complete set of survey data. These data are completely constrained for all the functional items on the focal surface, as measured at the survey temperature and orientation. With this set of known relations, FVC measurements of the illuminated focal plate fiducials, GFA fiducials, and back-illuminated science fibers can map accurately to physical locations, establishing the focal plate scale.

\begin{figure}[!b]
\centering
\includegraphics[height=2in]{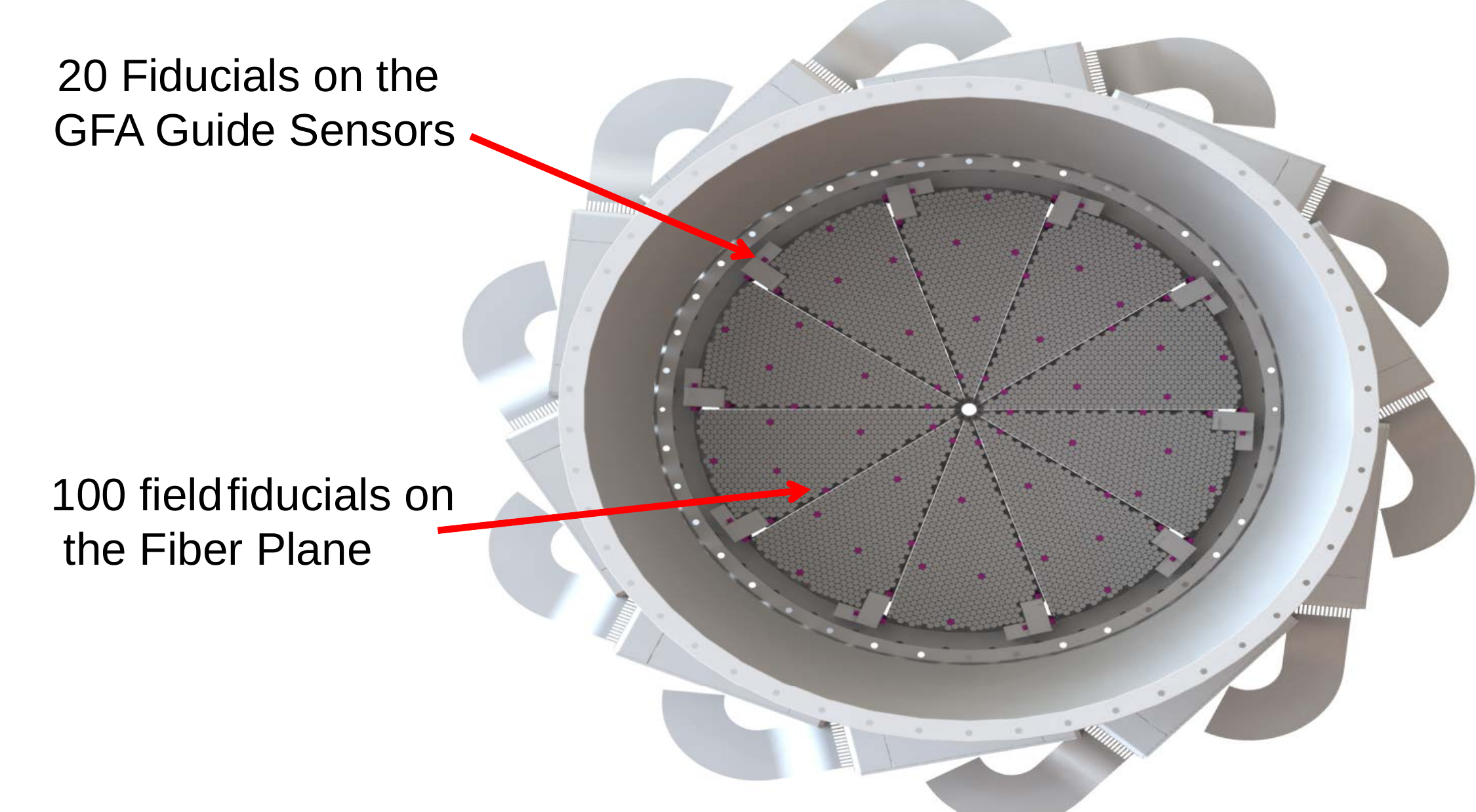}
\caption{ Illuminated Fiducials on the Focal plane (red dots).}
\label{fig:FiducialLayout}
\end{figure}

A study (DESI-0485) assessed kinematics of the field as viewed by the FVC through the corrector. The study finds that accurate and robust interpolations are made with 4 field fiducials distributed radially through roughly the center of each petal, and 6 more field fiducials at the perimeter of each petal, a total of 100 field fiducials.
Each GFA has 2 fiducials to track the guide sensor coordinates relative to the field fiducials.
Thus a total of 100 focal plane fiducials and 20 GFA fiducials will be measured on each FVC exposure. The location of these fiducials are shown as red dots on Figure~\ref{fig:FiducialLayout}. The fiducial fibers will of course be illuminated only when the positioner fibers are back illuminated during an FVC exposure, and not during  science exposures.  

\begin{figure}[!t]
\centering
\includegraphics[height=1.25in]{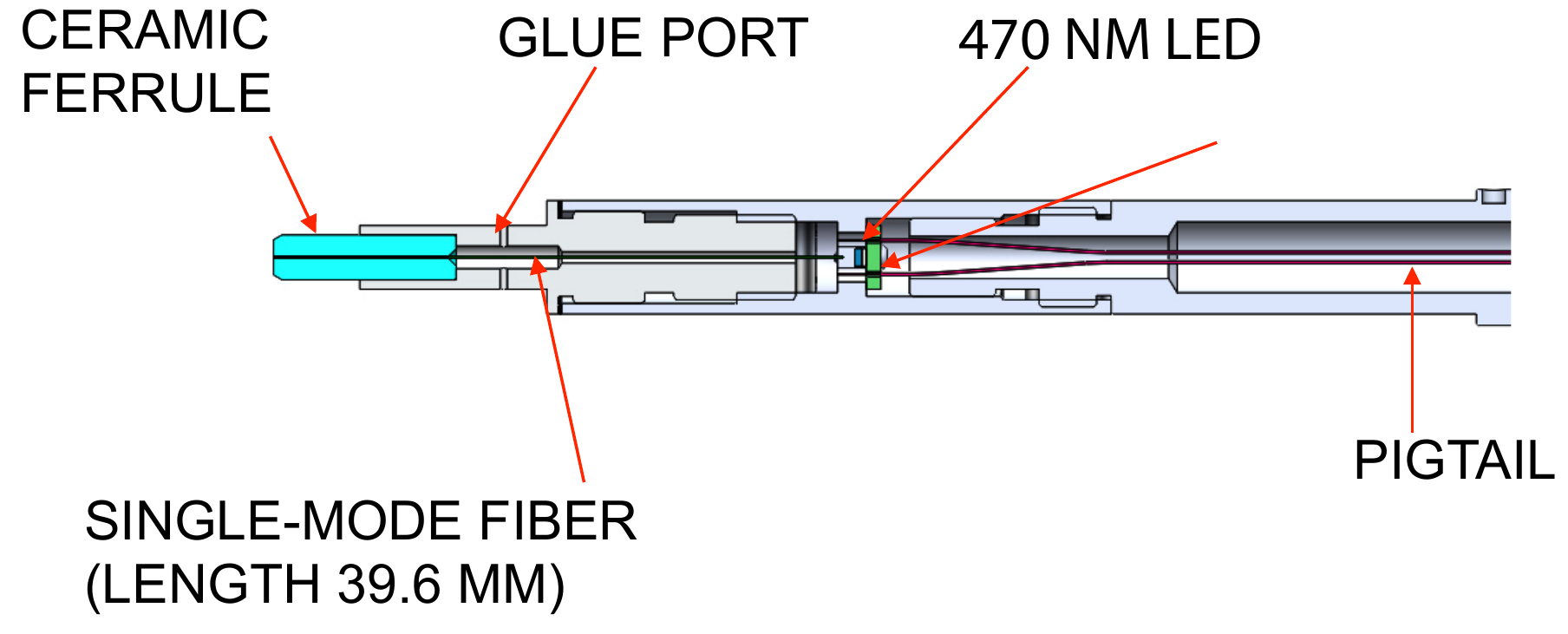}
\caption{Illuminates fiducial cross section.}
\label{fig:Fiducial}
\end{figure}

\begin{figure}[!b]
\centering
\begin{minipage}[b]{0.45\textwidth}
\centering
\includegraphics[height=2.5in]{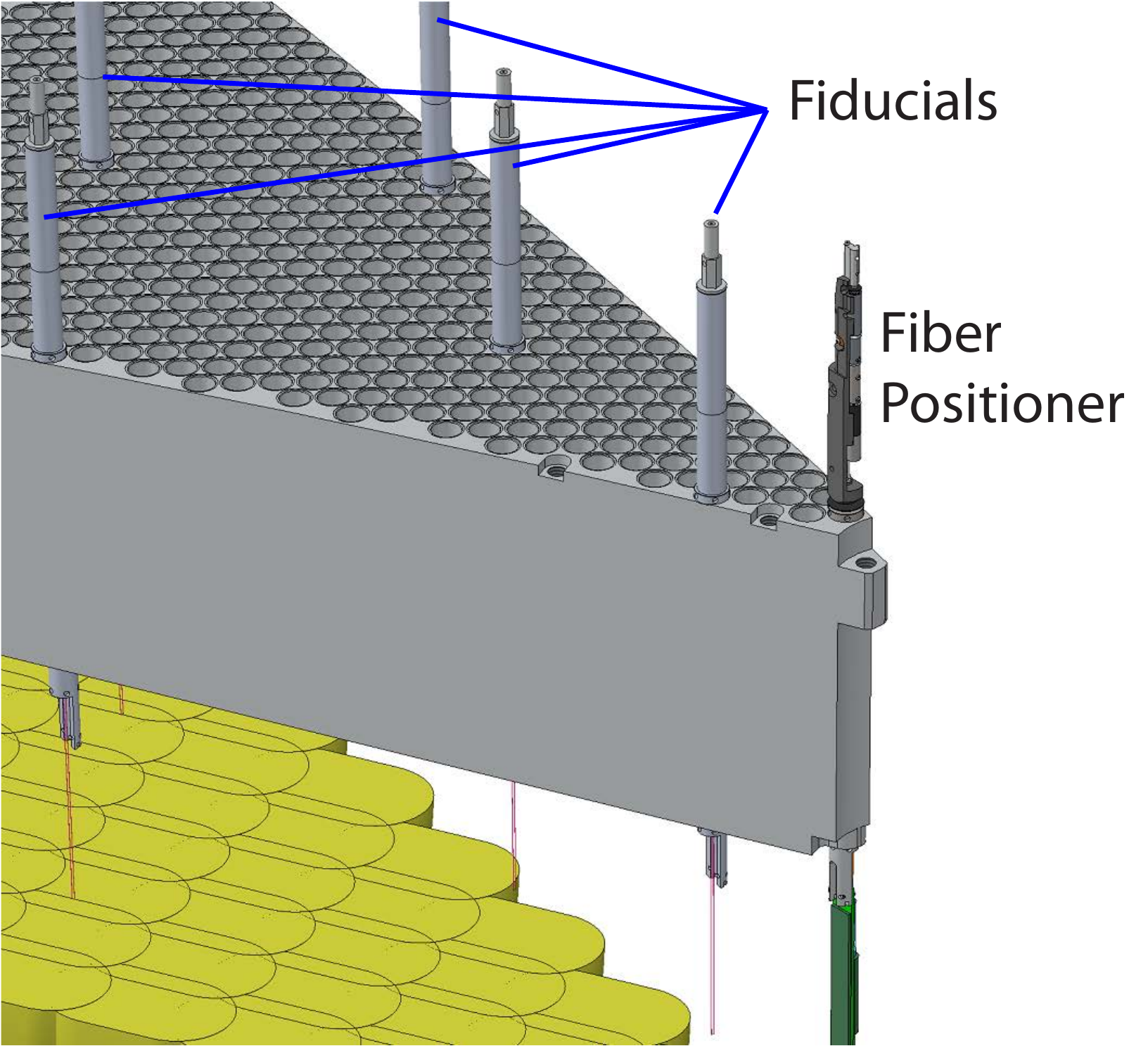}
\caption{One fiber positioner and several illuminated field fiducial on a focal plate petal.}
\label{fig:FieldFiducial}
\end{minipage}
\begin{minipage}[b]{0.45\textwidth}
\centering
\includegraphics[height=2.5in]{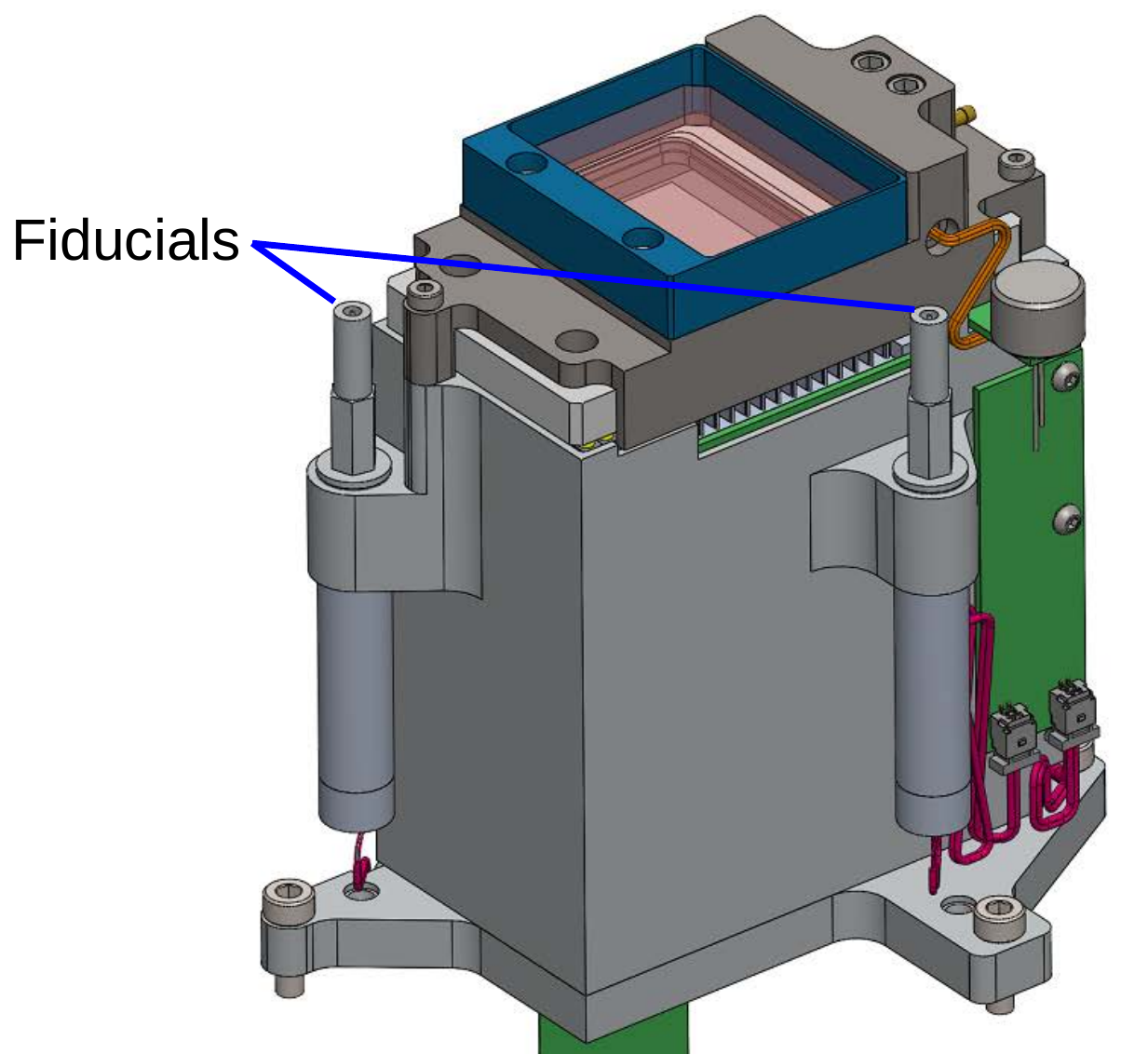}
\caption{ Location of the fiducials on the GFA enclosure.}
\label{fig:GFAfiducial}
\end{minipage}
\end{figure}

The focal plane fiducials  have the shape of long cylinders (Figure~\ref{fig:Fiducial}) the same diameter and length as the fiber positioners and they are threaded into holes on the focal plane just like the fiber positioners,  with their tips coplanar with the tips of the positioners (Figure~\ref{fig:FieldFiducial}). Each fiducial contains 
 a $\sim$40 mm long single mode fiber, 5 \micron in diameter, illuminated by an LED of 470 nm wavelength. The tip of the fiber is in a ceramic ferrule with the fiber positioned in the center with a 1 \micron precision. After installation in the fiber plane the position of the fiducial fibers will be measured to a few \micron accuracy by a measuring machine contacting the outside diameter of the ceramic ferrules. 

The GFA fiducials will be mounted on two sides of the GFA enclosure as shown in Figure~\ref{fig:GFAfiducial}. The tips of these fiducials will be identical to the tips of the focal plane fiducials with a single mode fiber illuminated by an LED and centered in a ceramic ferrule to facilitate precise location measurement.

\subsection{Fiber View Camera}	
\label{sec:Instr_FVC}

The purpose of the Fiber View Camera (FVC) is to take an image of the fiber plane after the robotic mechanism has positioned them and calculate the offset of the actual position of the fiber from its desired position to better then the 5 $\mu $m precision. These offsets are fed to the computer controlling the fiber positioners so that the positions of the fibers can be corrected. The estimate from simulations of this process is that several iterations will suffice to get all of the fibers positioned to the required positions. The performance requirements on the camera are listed in Table~\ref{tab:FVC_t1} (DESI-0589).

\begin{table}[!t]
\centering
\caption{Fiber View Camera requirements.}
\newcolumntype{R}[1]{>{\raggedright\arraybackslash}m{#1}}
\begin{tabularx}{\textwidth}{R{1.5in} R{1.0in} R{3.5in}} \hline
Item & Requirement & Rationale \\ \hline \hline
Measurement accuracy & $<$ 3~\micron rms & FVC is used to position fibers, and must be able to locate the illuminated fibers to this level of accuracy. \\  \hline
Nearest neighbor resolution & 1 mm & Minimizes planning restrictions on fiber location. \\ \hline
Integrate, read out, compute centroids & $\le 5$ seconds & FVC must image and centroid fiber tips quickly for multiple fiber positioning commands. \\ \hline
\end{tabularx}
\label{tab:FVC_t1}
\end{table}

The FVC measures the locations of the 5000 fiber positioners, the 100 field fiducial fibers, and the 20 GFA fiducials. These items are back-illuminated between observations (when the spectrograph shutters are closed), and the FVC surveys them all simultaneously in one field, looking at them through the corrector. This closes the loop unambiguously for positioning of fibers on every observation, as well as providing continuous feedback of variations in plate scale.

\subsubsection{Fiber View Camera Design}

\begin{figure}[!htb]
\centering
\includegraphics[height=2.5in]{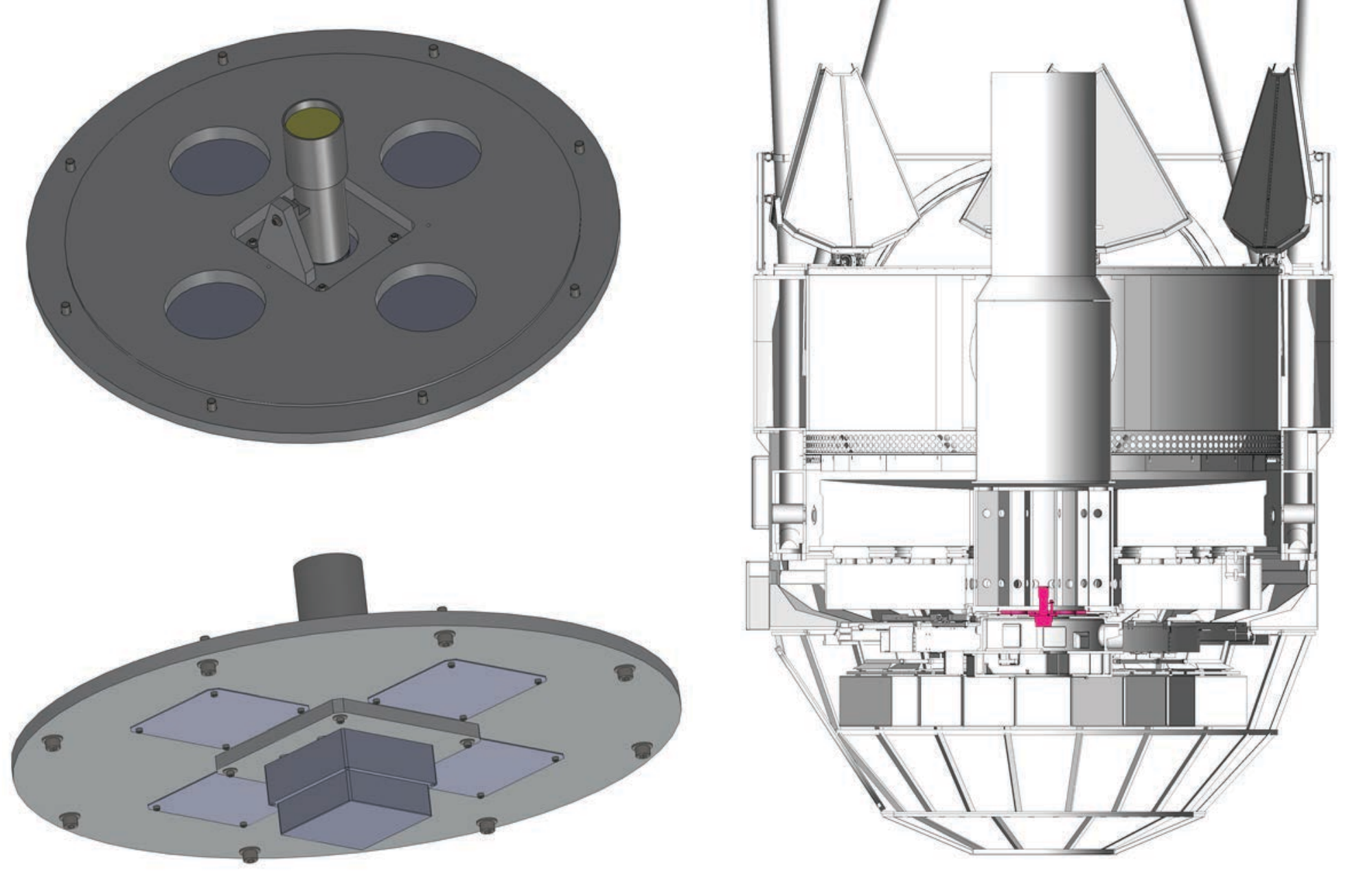}
\caption{Location of the Fiber View Camera in the Mayall Telescope.}
\label{fig:FVC_f1}
\end{figure}

In the design of the FVC a fiber plane of $\sim1$~m diameter was assumed. The camera is to be located on the axis of the telescope below the center of the primary mirror, looking up at the fiber plane through the hole in the primary mirror, about 12.25 m below the fiber plane, as shown in Figure~\ref{fig:FVC_f1}. To fit the image of the fiber plane on a $\sim40$~mm CCD requires a demagnification of $\sim25$, which can be accomplished by a 600 mm focal length camera lens. At this demagnification the 5~\micron~precision on the fiber plane translates to a 0.2~\micron~precision on the CCD. With a pixel size of 6~\micron~by 6~\micron~this means a centroid precision of 1/30 of a pixel or 33 millipixels. To satisfy these design requirements the following camera components have been chosen:

\begin{enumerate}
  \setlength{\itemsep}{1pt}
  \setlength{\parskip}{0pt}
  \setlength{\parsep}{0pt}

\item  A Kodak KAF50100 CCD with $6132{\times}8176$ pixels 6~\micron~by 6~\micron~each. This is a monochrome CCD with a 1 second readout time, with a read noise of 10 electrons. The dark current is 15 e/pixel/sec at 25~\celsius, so the camera can be operated at room temperature.

\item  A Canon EF 600 mm f/4 lens. With a Canon lens mount the lens focusing and f
   stop selection can be done under remote computer control.

\item  CCD control and readout electronics suitable for operating this CCD, utilizing a  Proline PL50100 controller available from Finger Lake Instruments.
\end{enumerate}

The fibers will be back illuminated by monochromatic LEDs in the spectrometer enclosures at a level that will provide 25,000 electrons per fiber image on the CCD.  At this level of illumination the dark current and the read noise are negligible.

Another important consideration to be taken into account in the design is the fact that the images with the camera will have to be taken through the telescope corrector optics. This configuration has been carefully simulated with the detailed design of the corrector optics and the FVC camera using the BEAM4 ray tracing program. The conclusion of this study was that the distortions introduced by the optics can be adequately corrected for by a third order polynomial transformation.  There will be 100 fixed fiducial fibers on the focal plane with precisely measured positions that will be used to calculate the coefficients of the transformation with every image taken, tracking thermal effects, position drifts and other possible perturbations. An additional 20 fiducials (2 per GFA) locate the guide/focus cameras.

\subsubsection{Development Program}

\begin{figure}[!t]
\centering
\includegraphics[height=2.0in]{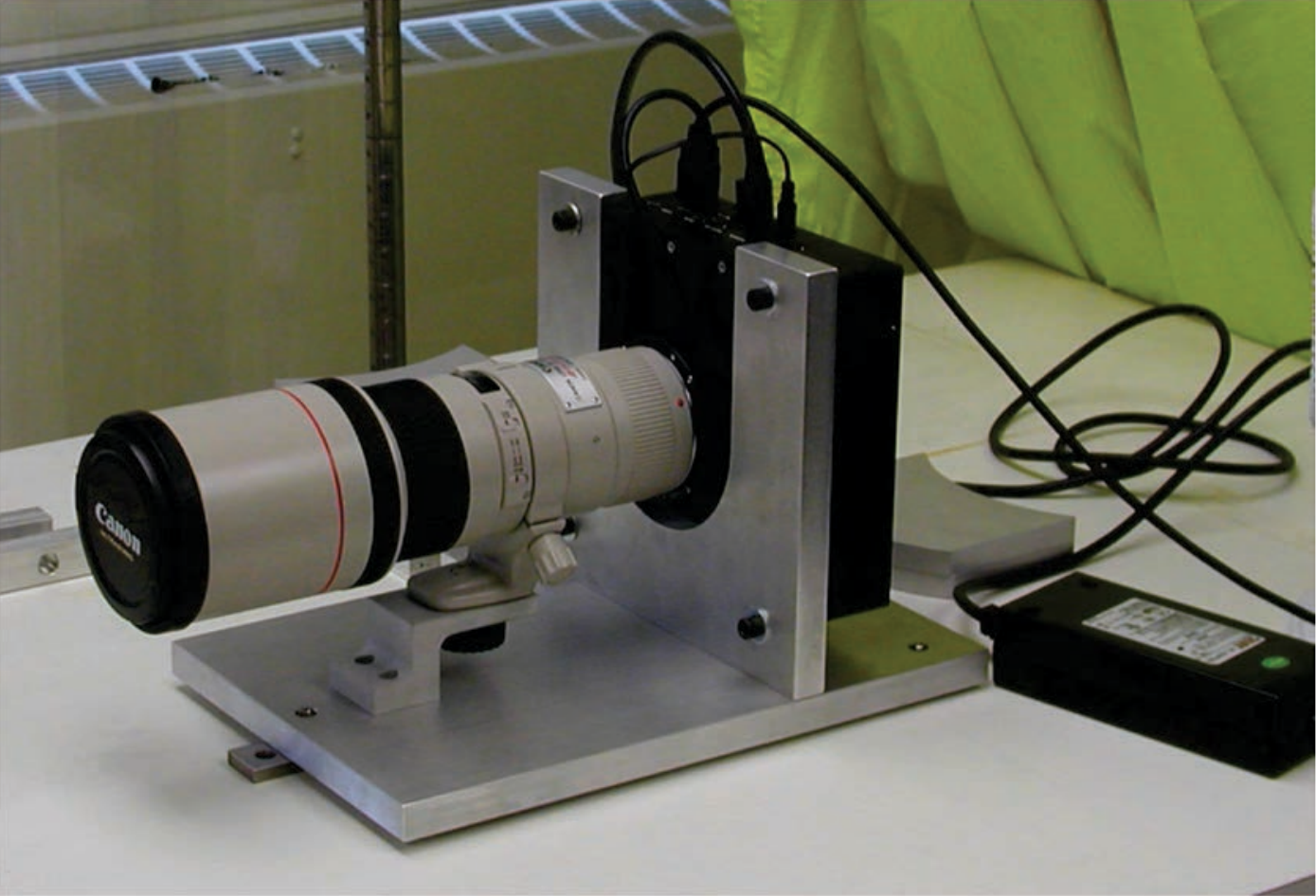}
\caption{The prototype Fiber View Camera.}
\label{fig:FVC_f2}
\end{figure}

\begin{figure}[!ht]
\centering
\includegraphics[height=2.25in]{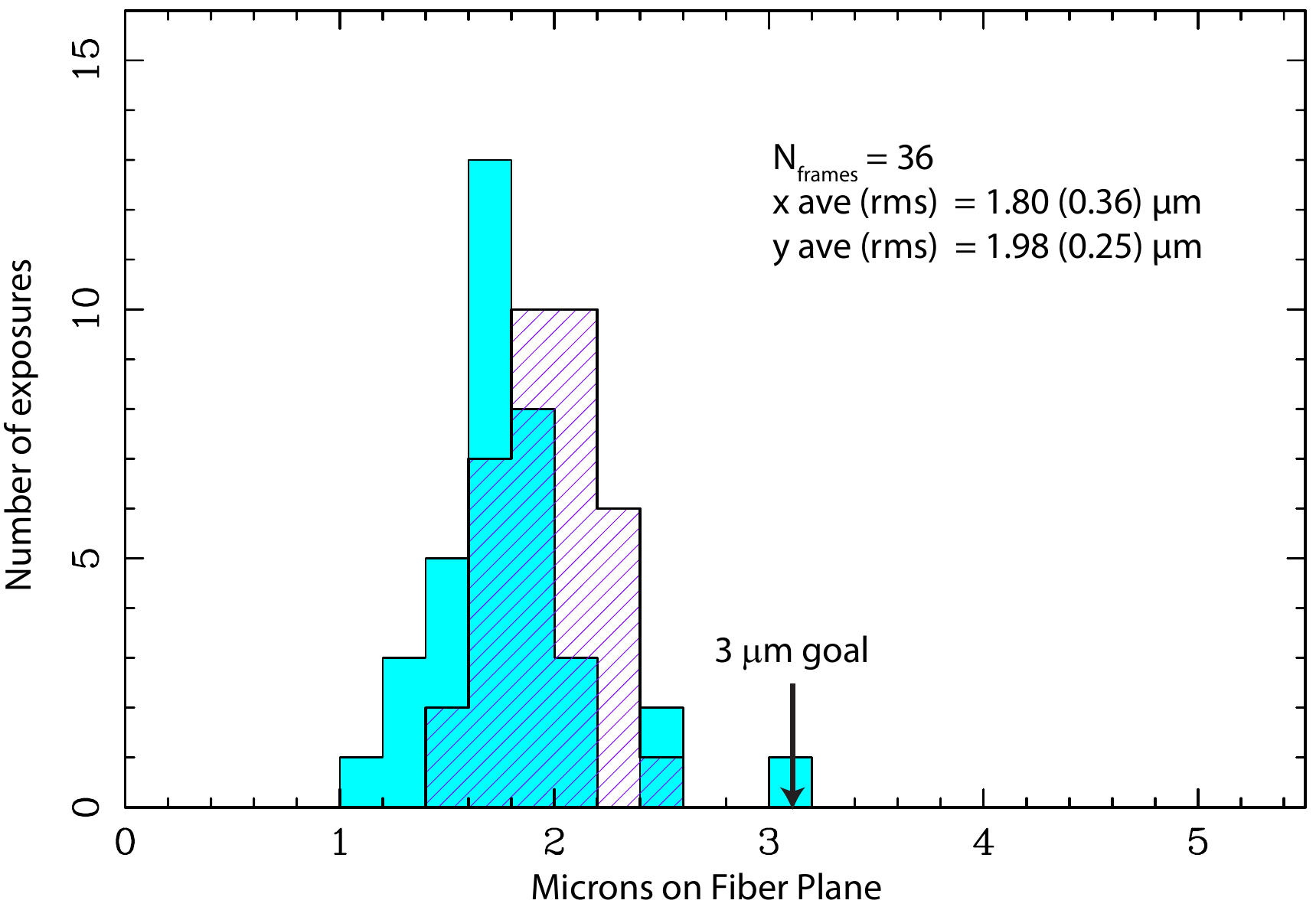}
\caption{Distribution of the rms deviations of the FVC reconstructed fiber positions relative to their OGP measured positions.  Blue is x distribution; red hash is y distribution.}
\label{fig:FVC_f3}
\end{figure}

A development program has been carried out to demonstrate the precision that can be obtained with the camera. A prototype camera has been constructed with the design described above, as shown in Figure~\ref{fig:FVC_f2}.  A prototype of a 450~mm by 600~mm section of the fiber plane has been constructed with 64  120-\micron-diameter fibers embedded in it. The fibers were back illuminated by monochromatic LEDs. The positions of the fibers were measured with an OGP (Optical Gauging Products, Rochester, N.Y.) Model Avant ZIP400 measuring machine with 1.5~\micron precision. The fiber plane was then 
placed 8.9 meters from the prototype camera. Images with a variety of illumination levels, camera lens f-stops and exposure times were taken. The reconstructed images on the CCD were transformed to the fiber plane using a third order polynomial transformation and compared with the true positions of the fibers from the OGP measuring machine.  The rms of the deviations of the reconstructed positions and the OGP measured positions is shown in Figure~\ref{fig:FVC_f3}. The precision achieved is better than the 3~\micron~rms requirement.

A number of studies have been carried out to investigate the relevant parameters of the camera, with the following results:

\begin{enumerate}
\item  The best resolution was obtained when the camera was slightly defocused to produce an image on the CCD between 1.5 and 2.0 pixels wide. Narrower images did not have sufficient flux in pixels neighboring the central pixel to allow for precise interpolation to obtain the centroid position.

\item  The most reliable good resolution was obtained with the lens f-stop set at f/16. At this aperture the image was diffraction limited with a diffraction peak 1.6 pixels wide, which gives the best resolution in a robust way.  At this f-stop the camera has a large depth of field, making it insensitive to slight shifts in camera or fiber plane motions due to flexing of the telescope support structure or thermal effects.

\item  The best resolution was obtained with between 10,000 and 40,000 electrons per image on the CCD. Below 10,000 electrons statistics started to limit the resolution, and above 40,000 saturation effects started to set in.  The design therefore calls for an illumination to produce 25,000 electrons per image, in the middle of this range.
\end{enumerate}

\subsubsection{Camera Control System}

 Ethernet communication will be used to control and read out the camera and the control software has been developed in the R\&D program. Software has also been developed and tested in the R\&D phase  to generate  centroids  from the  FVC images.  This  software  has been carefully optimized to carry out the analysis required to centroid the 5000 positioner  fibers in somewhat under 1 second. The time for one iteration with the Fiber View Camera, consisting of 1 second exposure time, 1 second readout, and 1 second centroiding software time, satisfies the 5 sec requirement.

\subsection{Thermal Management}

The primary purposes of the FPS thermal management system are:

\begin{itemize}
  \setlength{\itemsep}{1pt}
  \setlength{\parskip}{0pt}
  \setlength{\parsep}{0pt}
	\item{Prevent significant temperature changes of the focal plate during an observation. This ensures stability of the fiber and GFA locations.}
	\item{Insulate the dome air from FPS heat, sufficiently removing FPS heat throughout the night.}
\end{itemize}

Additionally the system will provide local cooling to GFA that operates at ambient temperature. Also, the thermal management system includes small circulation fans to ensure no significant density gradients in the air volume between the C4 corrector lens and the focal plate.

\begin{figure}[!t]
\centering
\includegraphics[width=.9\textwidth]{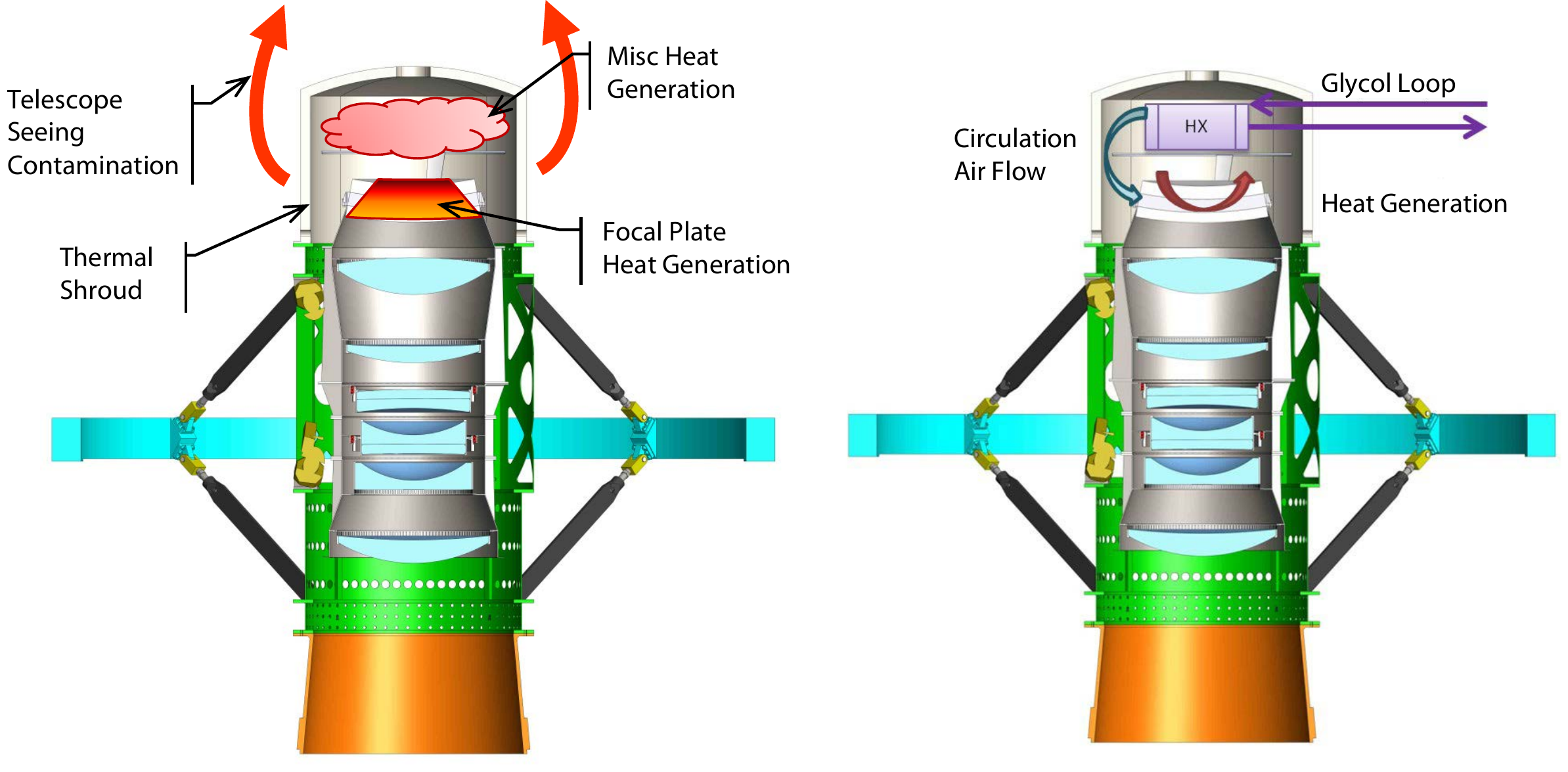}
\caption{Left: heat production in the focal plane system. Right: thermal management.}
\label{fig:fp_thermal_management_system}
\end{figure}

It is estimated that the focal plane system will consume $\sim$516~kJ per observation cycle (DESI-0995). The thermal enclosure is a 75~mm thick insulator ($<1\frac{W}{m\cdot K}$). Heat exchanger plates are located within the thermal enclosure, mounted with flush thermal contact to the petal structures. The exchangers will remove the generated heat at a sufficient rate to prevent $\Delta$T $>$1\celsius, which could contaminate telescope seeing. A coolant line will carry the heat from the exchanger and off the telescope. The coolant will track ambient dome temperature.
The thermal management system is illustrated in Figure~\ref{fig:fp_thermal_management_system}.

Neglecting entirely the heat removal by natural convection and conduction, it is still expected that the thermal capacitance of the support plate and positioners themselves will be sufficient to keep the plate temperature change well under $\Delta $T$ <$0.5~K during the course of an observation. Over a $\phi$0.82~m aluminum focal plate, $\Delta$T = 0.5~K corresponds to a worst case radial motion $<$5~\micron~of the outermost fibers and GFAs with respect to the optical axis.

We additionally anticipate between observations circulating some air from the space between C4 and the focal plate to the rear, to ensure achieving good mixing and getting warmer air to the heat exchangers.
In summary thermal management is expected to be fairly straightforward. A full-scale test is planned early on to ensure all the thermal management components have been well-sized.

\subsection{Alignment, Integration, and Test}

\begin{figure}[!htb]
\centering
\includegraphics[width=0.9\textwidth]{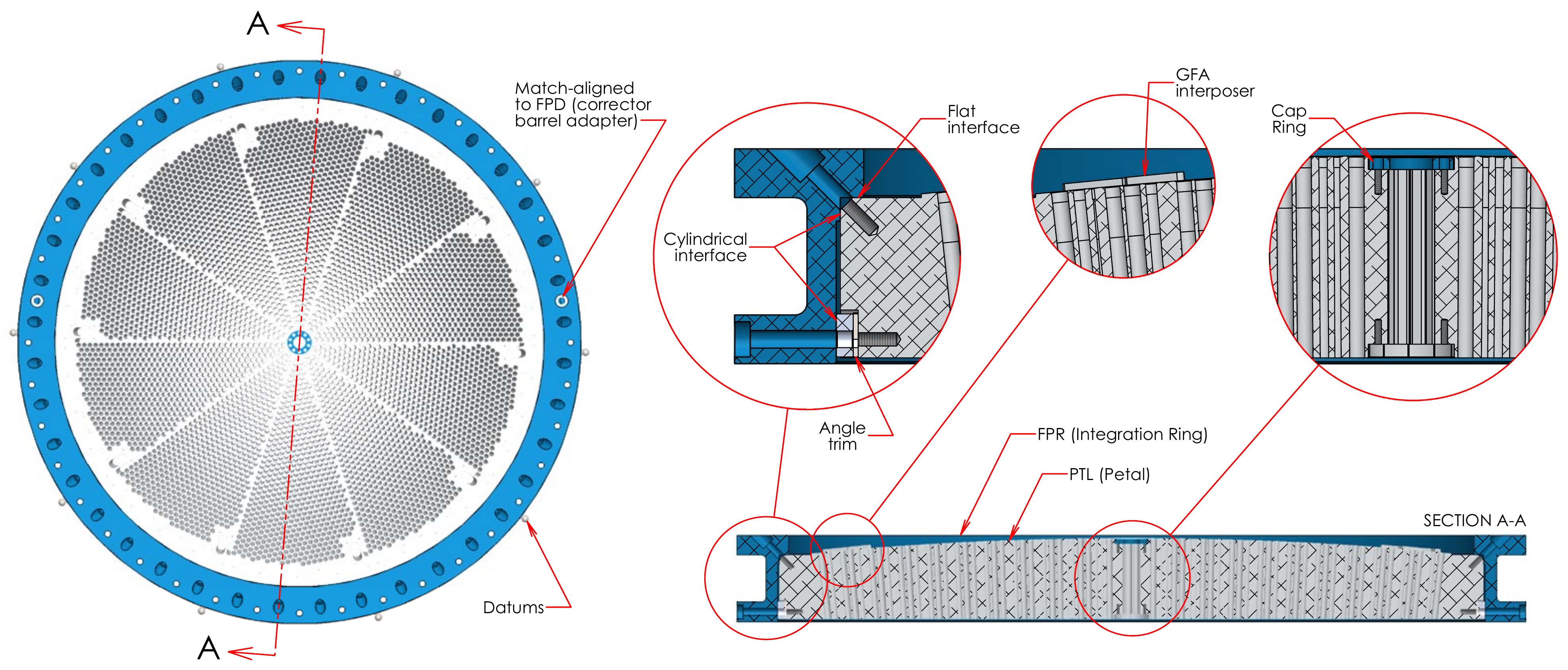}
\caption{Schematic of focal plate petal alignment to integration ring.}
\label{fig:fp_petal_alignment}
\end{figure}

The key alignment step for the focal plate assembly is bolting the petals to the integration ring. A schematic of this is illustrated in Figure~\ref{fig:fp_petal_alignment}. Each petal references along its outer radius and front face to the integration ring. Radial and $45^\circ$ bolts ensure tight mating of these surfaces.

Before integration of positioners, the petals are introduced to the ring for a dry fit. Standard machining tolerances will get the petal tilt within 0.05\degree of nominal. Petal datum balls are surveyed with respect to integration ring datums. As necessary, each petal is then trimmed in place to an angular resolution of 0.002\degree, by replacement of gauge spacers at the radial bolts. The small center cap ring (attaches to wedge tips on the front surface) is face machined appropriately at each of the 10 pads where it bolts to the wedges. After it has been attached, the compliant rear cap is attached, and the assembly is re-surveyed.

With the 10 petals aligned, the GFA interposer plates are aligned. One GFA interposer is bolted to each petal. Each interposer has two pin bushings which have been located on the same fixture, so that all 10 interposers present an identical mounting interface to the 10 GFA assemblies. Standard machining tolerances will put the interposers within 100~\micron of nominal. The interposers are surveyed in place on each petal; then they are trimmed in place to a resolution of 2.5~\micron and re-surveyed to confirm.

Subsequently, the petals are disassembled from the ring and integration of positioners commences. A key benefit of having the focal plate sectored in 10 petals is factorizing positioner loading. Each petal may independently have all its positioners loaded at one station (a process which may take some 2--3 weeks per petal), and then be taken to a completely separate station(s) for fiber dressing and splicing (another 2--3 weeks).  Petal field fiducials  are also installed. Meanwhile the positioner loading station is free for repeating the process with a fresh petal. Subsequently the latest fully-loaded petal can go to survey. 

A fully loaded petal assembly is illustrated in Figure~\ref{fig:petal_assembly}. When all the petals are fully loaded, they are re-introduced to the integration ring for final bolt-up and survey check.


There is an option to mount the GFAs to their respective petals at yet another independent station, prior to integration ring bolt-up; or after having bolted in the 10 loaded petals, to mount and survey all 10 GFAs at the same time. The second option is likely preferable, as it maximizes the decoupling in schedule of GFA production from positioner loading.

The assembled focal plate with positioners, their optical fibers and electronics, GFAs, fiber spool boxes, and coils of fiber optic cables are shipped as a unit to the telescope; the focal plane enclosure is shipped separately.

\subsection{ProtoDESI}

\subsubsection{Introduction}

ProtoDESI is the planned small-scale installation of prototype DESI elements at the Mayall scheduled for installation and commissioning at the Mayall telescope during summer 2016. ProtoDESI is a risk reduction tool for the full-scale DESI project (hereafter referred to as `DESI').

ProtoDESI will include a focal plane populated with prototype DESI fiber positioners, Field Illuminated Fiducials (FIF), and a Guide Focus Alignment (GFA) sensor module. The GFA has two fiducials (GIF) mounted directly to it. The focal plane will be mounted to the existing 0.8 degree FOV, f/3.2 refractive corrector at the prime focus (PF) of the Mayall telescope.  Short optical fibers will be routed from the positioners to the Fiber Photometry Camera (FPC), a commercial camera that will image the fiber tips.  It will also be located in the PF cage. A fiber back-illumination system will be included in the FPC.  The Fiber View Camera (FVC) will be mounted to the lower surface of the primary mirror cell.  A subset of the DESI Instrument Control System (ICS) will control the ProtoDESI subsystems, communicate with the Telescope Control System (TCS), and collect instrument monitoring data.

\subsubsection{Purpose}

There are two primary goals for ProtoDESI; both to reduce the risk related to the technical challenge of placing the fiber tips on targets and keeping them there with sufficient accuracy for the duration of a DESI exposure.  First, the GFA-ICS-TCS control loop will be tested to see if the system can place guide stars on requested GFA pixels and that they then remain stable in position over the period of an exposure.  Once this is established, the second goal is to test the GFA-positioner-FVC control loop.  Target objects will be identified relative to guide stars.  The positioners will be placed on the targets relative to assumed guide star pixel locations on the GFA by iteratively using the FVC, fiducial illuminators and fiber back-illuminator to generate corrective delta motions. The GFA-TCS will then place the guide stars on those pixels' locations.  The FPC will then measure the intensities of the positioners' fibers; the positioners can be dithered to look for intensity changes indicating how well the fibers were initially positioned on target centers.

\subsubsection{Additional Benefits}

In order to effectively use the field fiducials to account for system variations for accurate fiber positioning, metrology will be done on the assembled focal plane system.  This effort will relate the field fiducials (FIFs) to the fiducials on the GFA (GIFs), and therefore to the pixels of the GFA.  This will provide early validation of the metrology method required for the more complex DESI focal plate.

\subsubsection{Performance Goal}

The goal for the performance of ProtoDESI is to operate on the Mayall telescope for a 6 hour period per  night, successfully placing celestial targets on $>$75\% of its targeted fibers.  Exposures will be 20-30 minutes and there will be $<$15 minutes between exposures.  Completely automated observation is not required; manual interactions and inputs are allowed.  The long inter-exposure time allows for the sequence to be less parallelized than required for DESI and allows for less maturity of the software and hardware.  There is no specific throughput or signal-to-noise goal or requirement.

\subsubsection{ProtoDESI Elements}

\begin{figure}[!htb]
\centering
\includegraphics[height=3in]{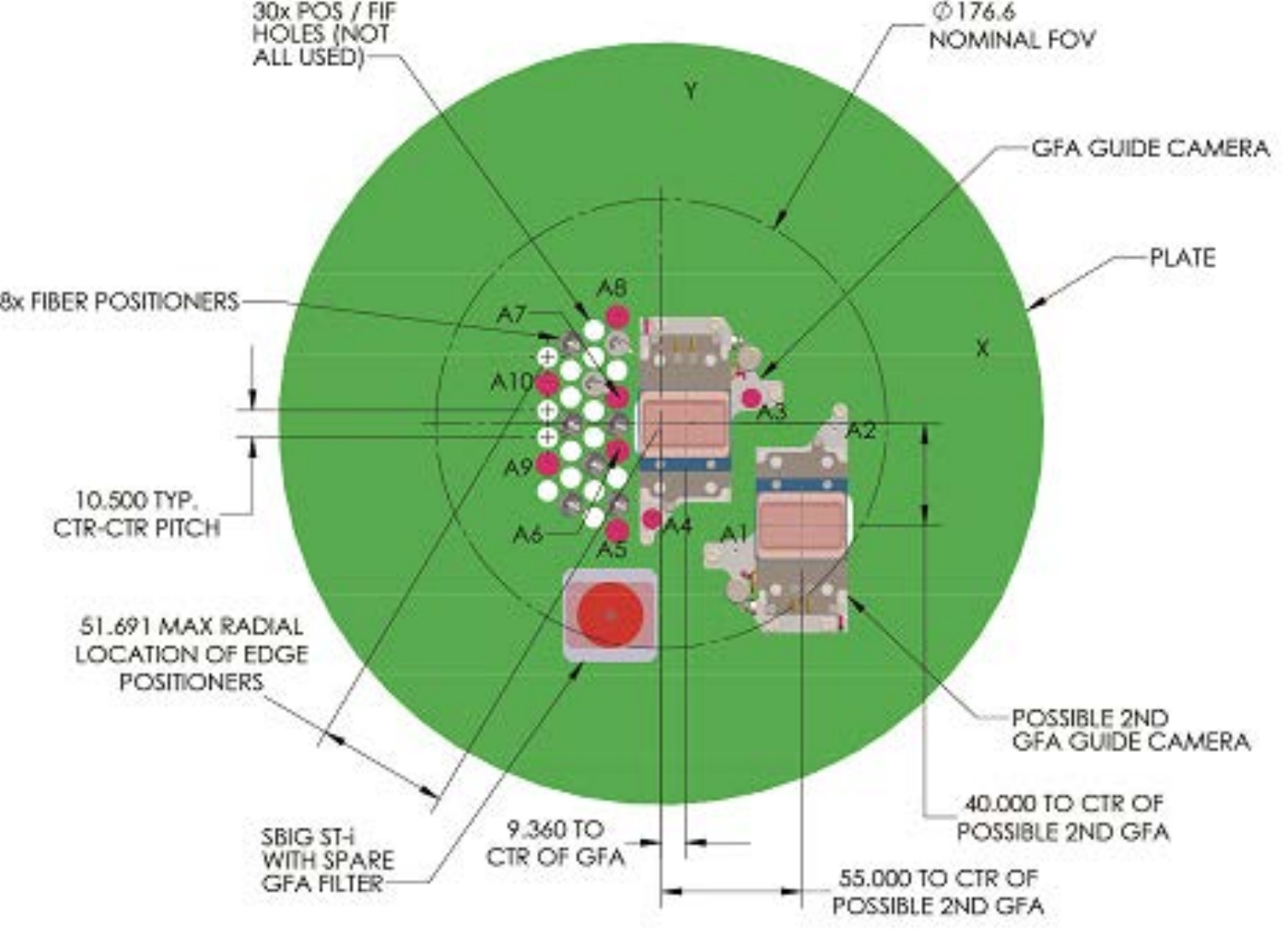}
\caption{ProtoDESI focal plate layout.}
\label{fig:protoD}
\end{figure}

Figure~\ref{fig:protoD} illustrates the layout of the elements installed on the ProtoDESI focal plate.

\paragraph{Focal plate}

The focal plate will be an aluminum part with provisions to mount two GFAs (though only one is needed for ProtoDESI). Adjacent to the GFA there will be approximately 30 standard, threaded, DESI positioner-interface holes, not all of which are planned to be populated. The front of the focal plate will be flat, with the positioner holes perpendicular to that surface. Positioner focal position is set by having all the positioner flanges contact directly to the flat surface (unlike full DESI, where each positioner flange contacts a uniquely-oriented spot face on each hole). The positioner holes will be on a hex pattern, spaced 10.5~mm apart, similar to the spacing on DESI.

\paragraph{Positioners}

Approximately 8-10 prototype positioners will be installed in the focal plate, clustered near the center of the focal plane.  Each positioner will have a short length of fiber with a ferrule at both ends installed.  After integrating positioners into the focal plate, all the free-end ferrules will be integrated in an array block so they all face the Fiber Photometry Camera.

\paragraph{Positioner Collision Avoidance}

Positioner collision avoidance is beyond the baseline scope of ProtoDESI.  The focal plate will be populated with positioners such that no positioners are adjacent.  The densest positioner packing would be a hex pattern where each positioner is surrounded by six holes that are either unoccupied or have a fiducial illuminator.  We note that the focal plane can be used later for lab testing of collision avoidance.

\paragraph{Fiber Photometry Camera}

A commercial camera behind the focal plate will simultaneously image all fiber ends to roughly measure target fluxes, to determine positioner placement accuracy.  It is possible this camera can be read out in real time for feedback as to how light throughput (and therefore fiber position accuracy) is being maintained.  Small dithers of the positioners or the telescope can be used to explore how well fibers are placed on target centers.  This camera is to be present for ProtoDESI but not for DESI.

\paragraph{Fiber Back-Illumination System}

A system to back-illuminate the fibers will be installed with the FPC. The FVC can then survey fiber positions simultaneously with the FIFs and GIFs, to iterate and confirm fiber positioner placement.

\paragraph{Field Illuminated Fiducials}

Approximately six prototype FIFs will be installed in positioner mounting holes, in a pattern to be determined by field distortion analysis of the prime focus optics. FIFs are sized such that they cannot be contacted by an adjacent fiber positioner, so there is no need for anti-collision software for this case.

\paragraph{Guide/Focus/Alignment Sensor Module}

Precision, repeatable mounting features will be provided for two GFAs with the active areas within the 176~mm diameter field of view.  ProtoDESI is baselined to use only one GFA, but the other position offers the option to install another one if available, or is a backup mounting position in case of damage to the other. The primary GFA will be located to provide significant sensor coverage at the boresight. The GFA will be mounted outside the patrol range of any adjacent fiber positioners, so there is no need for anti-collision software for this case.

\paragraph{Metrology}

In order to effectively use the field fiducials to account for system variations for accurate fiber positioning, metrology will be done on the assembled focal plane system.  This effort will relate the field fiducials to the fiducials on the GFA, and therefore to the pixels of the GFA before shipment.

\paragraph{Ancillary Hardware}

Attached to the focal plate will be any required fiber and wiring management hardware.  We anticipate CAN signals will be distributed to positioners and fiducials via either 1 or 2 small breakout boards.  The breakout boards would have the same connectors and electrical features as we anticipate for the petals in full DESI.  The breakout boards then connect to a master board.  The master board supplies DC power and is what communicates to the Ethernet network.  These local control electronics and power supplies or other ancillary focal plane systems will all be located inside the PF cage.  

\begin{figure}[!htb]
\centering
\includegraphics[height=2.5in]{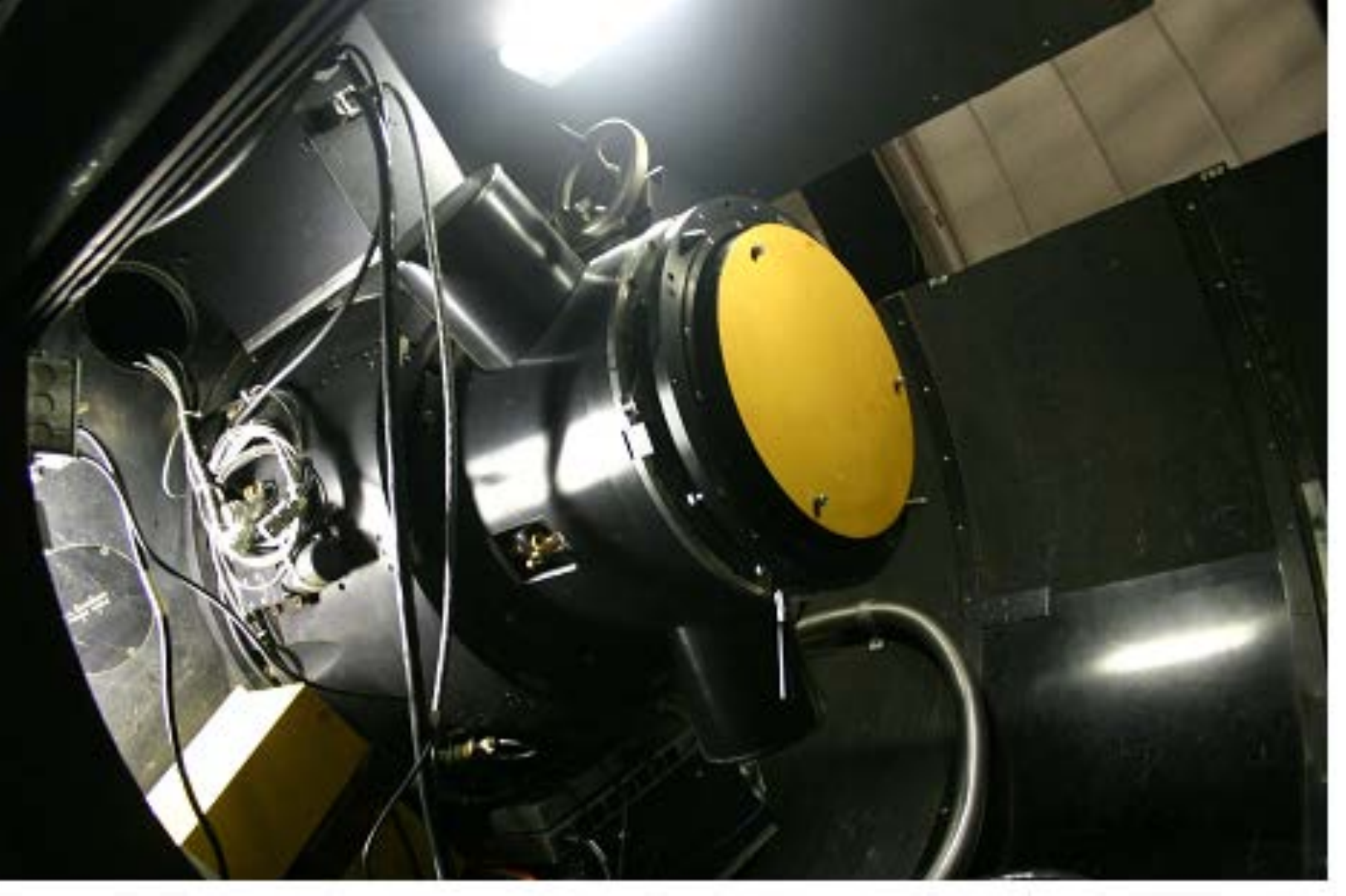}
\caption{PF cage viewed from below with telescope oriented for MOSAIC install.  Photo taken through hole that MOSAIC comes up through.  Yellow protective cover over 4th lens of corrector.  Interface plane \& alignment pins surround that. Power strip and Ethernet switch visible below corrector.  Craning cutout in cage visible in upper right.}
\label{fig:MOSAIC}
\end{figure}

\paragraph{Integration}

The ProtoDESI focal plate system will be integrated to the PF cage with the telescope pointed south at low elevation, in generally the same way MOSAIC installation and removal are currently done.  The MOSAIC installation and removal procedures are described in detail in DESI-0647.  In short, with the telescope pointed south as shown in Figure~\ref{fig:MOSAIC}, the dome jib crane will lower its hook through the slot on top of the PF cage and lift the ProtoDESI focal plane system from its cart to a height a little greater than the lowest position of the scissor lift.  Then the scissor lift will be moved into place below ProtoDESI, and the installation personnel will get on the lift (with safety harnesses).

The scissor lift and crane will be raised together, so the personnel can guide the focal plane assembly.  Then the focal plane assembly will be moved into its position and fastened to the interface ring.

Finally, the Ethernet and 120~V power cables will be plugged into the PF cage outlets.

\paragraph{Fiber View Camera}

The ProtoDESI fiber view camera hardware,  identical to the DESI FVC hardware (with the possible exception of the use of a 400~mm focal length lens rather 600~mm),  will be mounted to the lower surface of the primary mirror cell in place of the Mayall Guider and Cassegrain Instrument Rotator assemblies.  

The computing and software needs for the ProtoDESI FVC installation are less than for DESI since there is no requirement on inter-exposure timing.

Power, data, and ambient cooling are all expected to be more than sufficient in the Cassegrain cage for the FVC.

\paragraph{Instrument Control System}

The ICS will be an early release of the DESI ICS supporting only those elements needed to operate ProtoDESI and communicate with the TCS.  It also will not have the fully-developed user interfaces DESI will have.  It will have interfaces with the TCS, FVC, FPC, and positioner controller.

The ICS will require computers located in the existing Mayall computer room, but will be substantially smaller in the number of cores and amount of storage planned for DESI.  Power and cooling resources in the computer room are expected to be sufficient for ProtoDESI needs.

\subsubsection{Mayall}

\paragraph{Telescope Control System}

The TCS modernization will be completed in 2015 and will have been tested both with the MOSAIC instrument and with the ProtoDESI ICS at least several months before the complete delivery of the ProtoDESI elements.

\paragraph{Electrical Power}

As described elsewhere in this document, existing electrical capacity is expected to be sufficient for ProtoDESI.

\paragraph{Control Room Annex}

The commissioning and operation of ProtoDESI will require more control room space than is currently available at the Mayall.  A lounge area on the U floor of the Mayall will be converted to a Control Room Annex to accommodate 10 or more people.

\paragraph{Data}

Data bandwidth is not expected to require expansion for ProtoDESI.

\paragraph{Mechanical/Architectural}

Aside from the modifications required to implement the Control Room Annex, it is expected that ProtoDESI will require minimal mechanical alterations to the Mayall facility.  The focal plane system will be designed compatible with the existing envelope and cutout in the PF cage, and it should be readily balanced using existing hardware.

\subsubsection{Targeting and Observing Plan}

An observing plan will be developed for ProtoDESI based on available targets and final installation schedule.   ProtoDESI will require target locations to be identified and translated into coordinates for the fiber positioners.   Guide stars and guiding analysis code will be required for all fields to be observed.  The fiber targets will likely be a mix of galaxies and stars.


\clearpage

\section{Fiber System}
\setcounter{equation}{0}\setcounter{figure}{0}\setcounter{table}{0}
\label{sec:Instr_Fibers}

\subsection{Overview}

\begin{figure}[tbh]
\centering
\includegraphics[width=.9\textwidth]{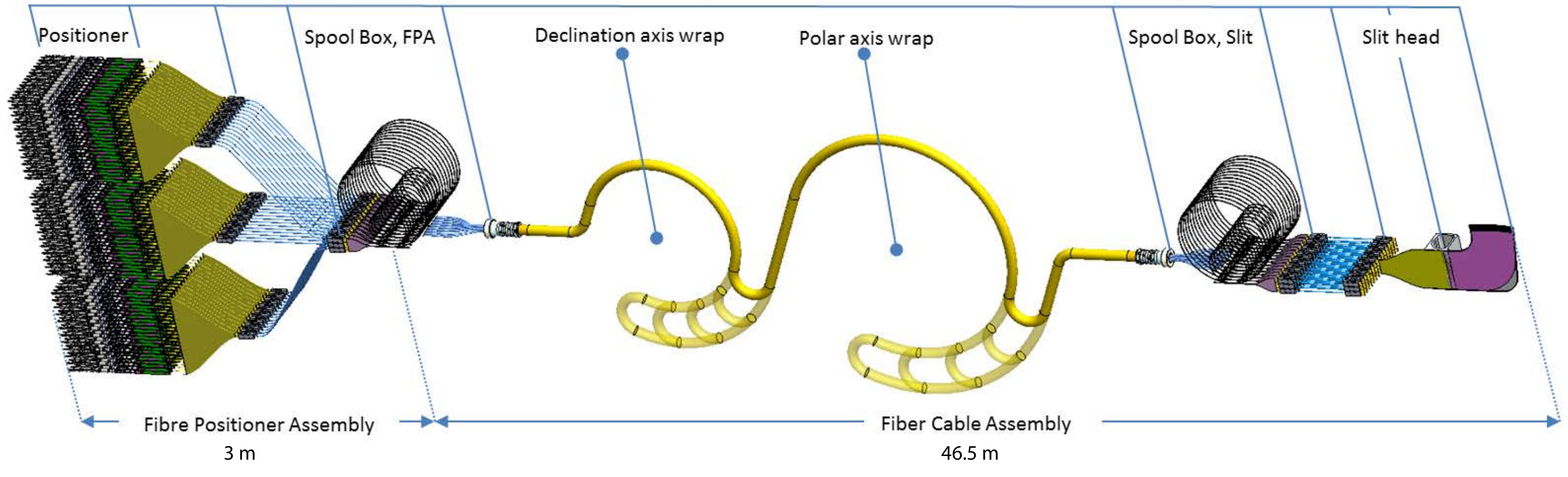}
\caption{The fiber system schematic.}
\label{fig:fs_schematic}
\end{figure}

The function of the fiber system is to transmit the light from 5,000 science-targets imaged at the telescope prime focus to the spectrograph optical input. Figure~\ref{fig:fs_schematic} shows a schematic diagram of the fiber system. Each optical fiber, with a  107~\micron diameter core, is terminated in a ferrule that is then mounted in a positioner arm (Section ~\ref{sec:PFA}).
The fibers are collected behind the focal plane into ten groups of 500. Each group spans 49.5~m from the focal plane to the spectrograph room via guides, supports, strain-relieving spool boxes, and ruggedized cables (Section ~\ref{sec:cable}). Each group of 500 fibers is then precisely arranged and terminated into a linear arc that provides the optical entrance-slit illumination (Section ~\ref{sec:slit}) to an associated spectrograph. The fiber run includes a fusion-splice connection near the focal plane to allow for positioner installation and to facilitate fabrication, integration and testing flow. The major deliverables for the fiber system are the 10 ruggedized fiber cables, each with a 500-fiber slit assembly on one end and 500 individual ferrules on the other. Declination and polar bearing cable wraps and cable and fiber routing support elements are other key deliverables.

The performance requirements for the Fiber System are summarized in Table~\ref{tab:fiberq}. Further discussion of the parameters is included with the relevant sub-system description below.

\begin{table}[hp]
\centering
\caption {Fiber System Requirements. Refer to DESI-0581 for the requirement rationales.} 
\label{tab:fiberq} 
\newcolumntype{R}[1]{>{\raggedright\arraybackslash}m{#1}}
 \begin{tabularx}{\textwidth}{ R{1.8in} R{1.3in} R{3.0in} }
\hline
Title & Value & Statement \\
\hline
Fiber core diameter& 107~\micron $\pm$ 3 & Optimum fiber core diameter is 1.46” (DESI-304 figure 1) and plate scale calculated at the edge (406 mm radius) of the focal plane (DESI-329) as 73.45 \micron/arcsec \\
\hline
Bulk throughput of FS (Neglecting FRD losses) at BOL
 & $\lambda$(~\micron) \hspace{0.0in} Efficiency
 &Required in order to meet systems engineering throughput budget (DESI-0347)\\ 
\cline{2-2}
 &  0.360\hspace{0.5in}0.555&\\ 
 &  0.375\hspace{0.5in}0.655 &\\ 
& 0.400\hspace{0.5in}0.705 & \\ 
 &0.500\hspace{0.5in}0.848 &\\ 
&0.600\hspace{0.5in}0.891 &\\ 
&0.700\hspace{0.5in}0.903 &\\ 
  &0.800\hspace{0.5in}0.914 &\\ 
&0.900\hspace{0.5in}0.929 & \\ 
&0.980\hspace{0.5in}0.937 & \\ 
\hline
FRD-induced throughput of Fiber System.& $\geq90$~\% from 360nm - 980 nm & Minimize throughput loss due to FRD (DESI-0347). (Main contributors: Fiber end polishing/cleaving/bonding, Fiber splice induced  FRD, Actuator induced FRD, Cable and guide bend induced FRD) \\
\hline
Number of Science fibers&$\geq$5000&SRD states that the fiber density shall be $\geq$700 per square degree in order to meet the combined target density requirements for all targets. \\ 
\hline
Fiber positioner flexure lifetime&376,000 positioner moves&DESI-0584: 185K positioner reconfigurations over a 5 year cycler (tested to 2x lifetime) \\ 
\hline
PFA mapping traceability&PFA to slit&It is imperative to know which spectra was obtained from which object.\\ 
\hline
Fiber Cable flexure lifetime performance	&169K cycles&DESI-0584: 1 bend per observation plus 4 additional bends per night (tested to 2x lifetime) \\ 
\hline
Number of fibers per slit&$\geq$500&Baseline Design. There will be 10 spectrographs, each with 500 fibers\\ 
\hline
Spectral stability of fiber output PSF.&The fiber output as convolved by the spectrograph optics shall not change the FWHM in the wavelength dispersion direction by more than 3\% between observations.&Spectral stability of fiber output PSF.\\
\hline
 \end{tabularx}
\end{table}


\subsection{Optical Fibers}

\begin{figure}[tbh]
\centering
$\begin{array}{cc}
\includegraphics[height=2.25in]{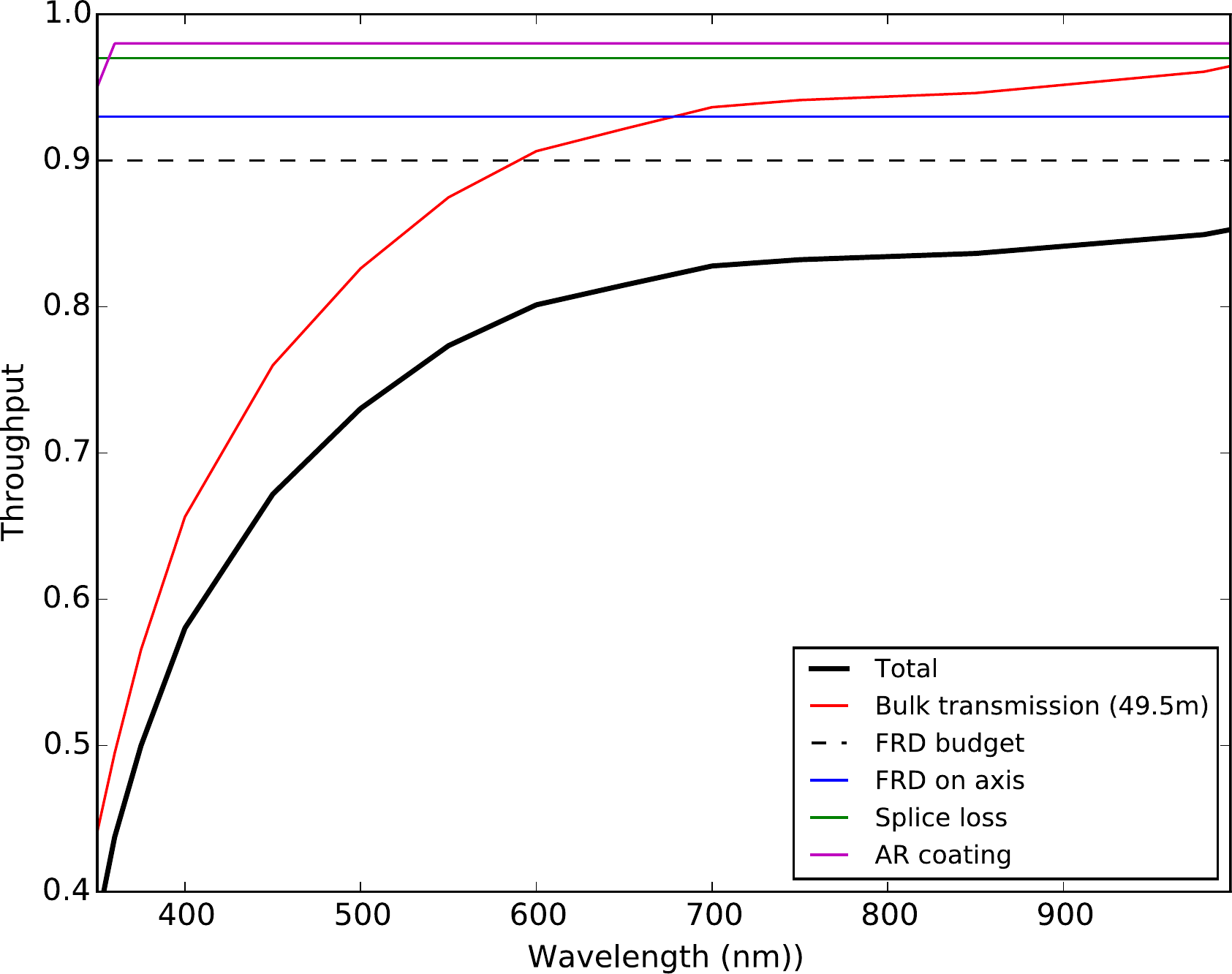} &
\includegraphics[height=2.25in]{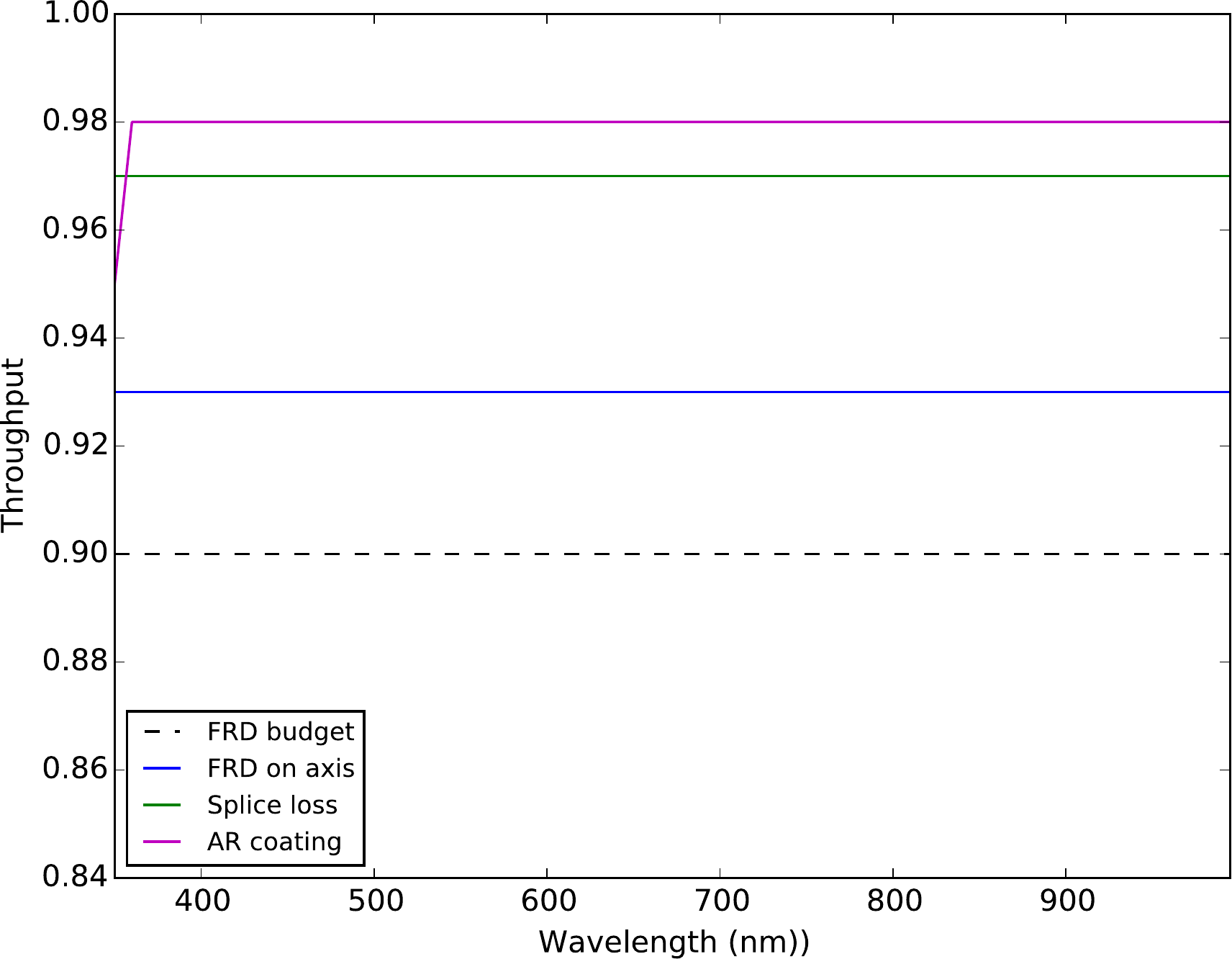}\\
\mbox{(a) }&\mbox{(b)}\\
\end{array}$
\caption[Predicted fiber throughput]{The predicted fiber throughput from 360~nm to 980~nm for 49.5~m of FBP fiber is shown in panel (a) and panel (b) shows the contributions that can be controlled. FRD, splice connection, and AR losses have been presumed constant with wavelength. }
\label{fig:fib_tp}
\end{figure}

Throughput losses cause increased exposure times, limiting the survey rate. Therefore, control of fiber losses is a critical issue in the DESI design. The fiber system throughput to the spectrograph is mainly affected by bulk transmission losses in the glass, by losses at the fiber ends due to termination imperfections and surface reflections, by geometric misalignments at the fiber entrance and exit, and by angular diffusion caused by focal ratio degradation (FRD).
Figure~\ref{fig:fib_tp} shows the budgeted total fiber throughput as well as its constituent parts, which are a combination of both measurements and predictions, as described in more detail in the following sections.

It is convenient to consider the bulk properties of the fibers separately from the issue of FRD.

\subsubsection{Bulk Fiber Losses} 

Fiber bulk attenuation is dictated by the fiber glass type and its spectral transmittance, the fiber length, and the ability of the fiber's limiting numerical aperture (NA) to fully accept the input f/\# from the corrector. We plan to use Polymicro FBP, a low-OH step-index fused silica fiber, NA = 0.22, that has been well characterized to have excellent band transmittance without the complex and potentially confusing spectral absorption features found in high-OH, UV enhanced type fibers. The fiber core size (107~\micron diameter) is selected to optimize science target flux collection versus background. The fiber cladding size (150~\micron diameter) is chosen to eliminate significant evanescent surface loss. The fiber size is bracketed by and scaled from existing commercial products so it can be purchased as a custom draw lot under the same fabrication process.

Fiber routing options have been studied to minimize the total fiber length while maintaining a plausible method of integration with the telescope tube and pivot wraps. The length of the fiber is estimated to be 49.5~m, so the fiber system will use about 200~km of optical fiber. The predicted bulk fiber throughput for 38~m FBP fiber is shown by the red line in panel (a) of Figure~\ref{fig:fib_tp}. The black
line corresponds to the total throughput when including the contributions from FRD, splice loss, and antireflection coatings.

\subsubsection{Focal Ratio Degradation} 


Light incident on a fiber at a single angle exits the fiber in a cone with finite angular width. Consequently, an incident cone of radiation exits the fiber over a larger included angle (or smaller focal ratio) that could distribute light beyond the spectrograph entrance pupil. This focal ratio degradation (FRD) can be due to modal diffusion within the fiber caused by manufacturing imperfection and mechanical stress induced by termination, bonding, or mechanical force. FRD may be characterized in two ways:

{\em Cone (or solid angle) test} measure a fiber's angular diffusion by illuminating the fiber with a filled cone of light with constant surface brightness within a specified angle. To better simulate the illumination from the telescope, a scaled obscuration is added to simulate obstruction by the telescope's secondary mirror. This test directly describes how the fibers will behave in an instrument, specifically
how much energy will be enclosed within some f/\# for this illumination geometry.

{\em Ring (or collimated) test} measures FRD by injecting a collimated beam (from a low-power laser) into the fiber. The beam is azimuthally-scrambled to form a ring which is recorded by an imaging detector. The FRD is characterized by the thickness of the ring in the radial direction. Although, less direct than the cone test, it is very simple to do and interpret the results. 

\begin{figure}[tb]
\centering
\includegraphics[height=3in]{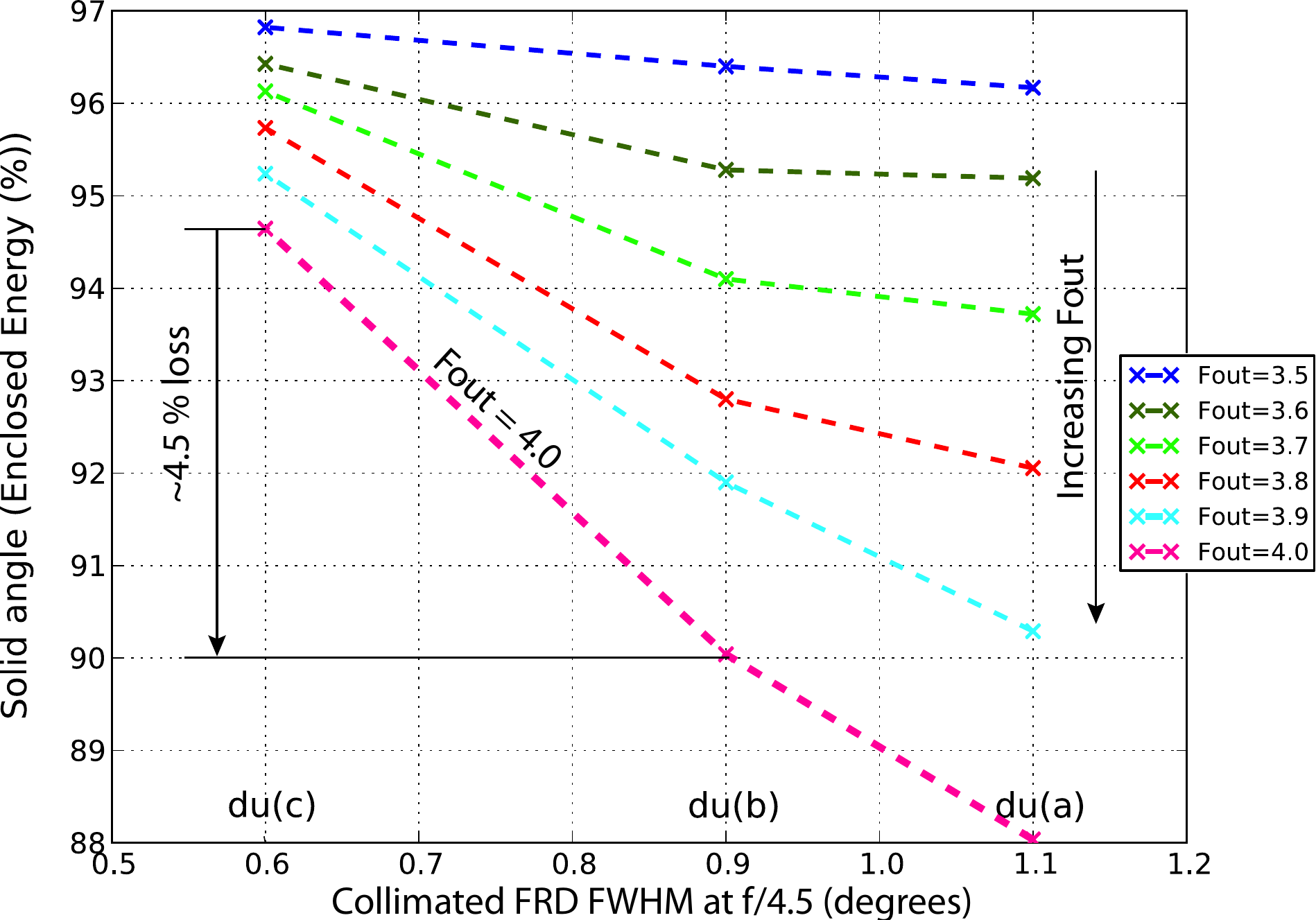}
\caption[Relationship between collimated FRD and enclosed energy from a solid angle test]
{The effect of fiber collimated FRD on encircled energy measured for three different $\sim$40~m cables, denoted by du(a) du(b) and du(c), with polished ceramic ferrules on each end. 
}
\label{fig:coll_vs_solid}
\end{figure}

It has been shown, that these methods are equivalent (\ie, give the same result) except at very small injection angles~\cite{ASmith2013MNRAS}. Determining collimated FRD is straightforward because the injection angle is obtained from the same image from which the ring thickness is measured. 
The relationship between the collimated FRD and the solid angle FRD is established by measurement and modeling as is shown in Figure~\ref{fig:coll_vs_solid}.  The horizontal axis shows the FRD (FWHM) of each fiber from a collimated FRD test at an angle corresponding to f/4.5. The vertical axis shows the enclosed energy within different output focal ratios from a solid angle FRD test using an input focal ratio of f/4.5 with a scaled obscuration. These data establish a relationship between the collimated and solid-angle FRD and establish the throughput tolerance to fiber collimated FRD. For example, there will be a throughput loss due to FRD of 4.5\% if the fiber collimated FRD is increased by 0.3 degrees FWHM (for an f/4.5 input, an f/4.0 output, and a starting FWHM of 0.6 degrees
 
The FBP fiber selected for DESI has excellent FRD performance due to its extremely uniform cladding thickness and a thermally stress-relieved polyimide jacket. Low OH fibers from alternate vendors exist (\eg, CeramOptec Optran$^\circledR$) although studies show that the FBP fiber offers the best performance overall in terms of transmission and FRD performance~\cite{FBPfibers}.

\begin{figure}[tb]
\centering
\includegraphics[height=3in]{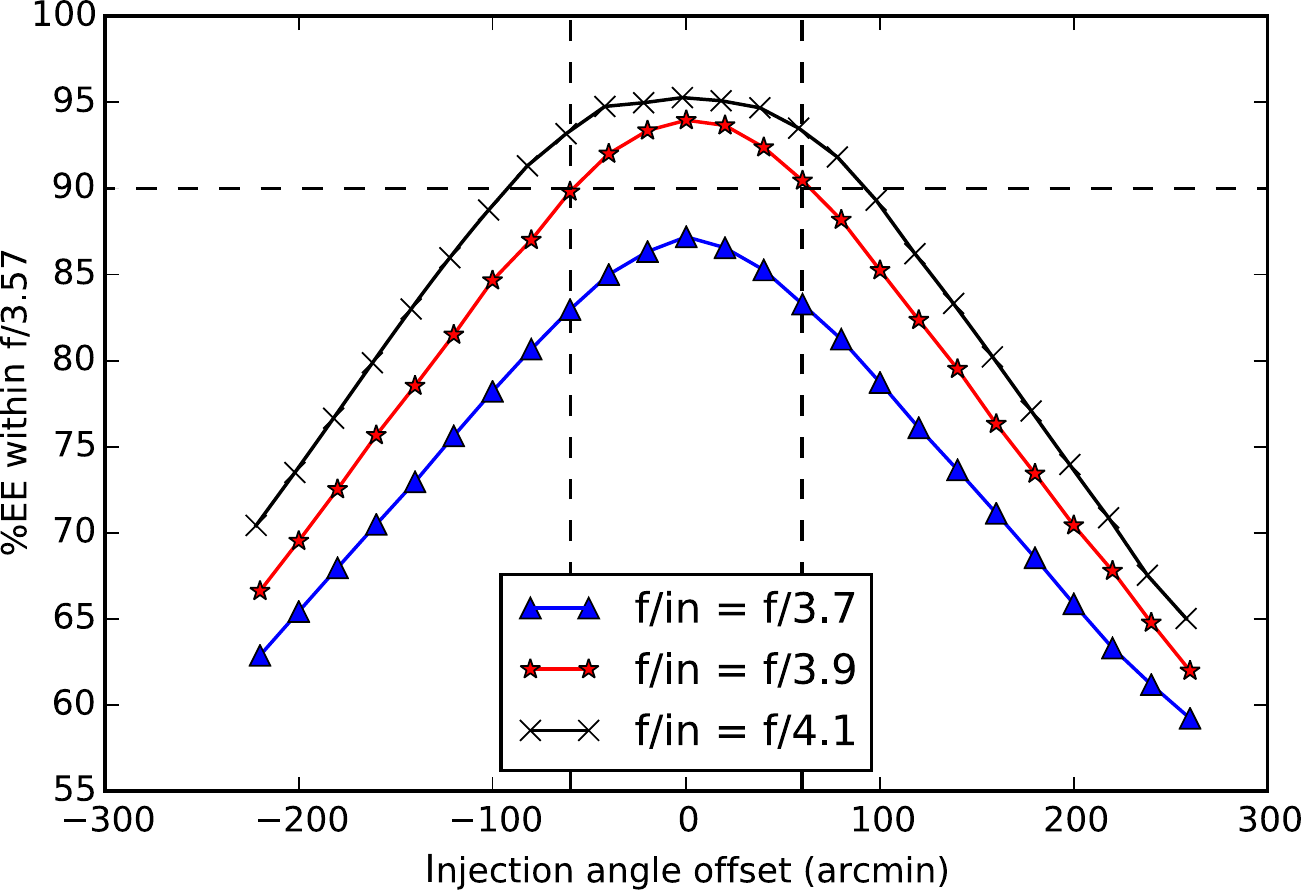}
\caption[FRD at f/3.9 for a prototype fiber assembly]
{ The effect of FRD and flux injection angle on throughput 
is derived from these measurements of the percentage of the energy contained within f/3.57 which is the spectrograph acceptance speed.
The figure shows results for a test fiber illuminated with a telescope simulator 
at different focal speeds of f/3.7, f/3.9, f/4.1 (colored lines) 
and with chief-ray injection angle tilts of $\pm$240~arcmin. The relevant specification is as follows.
The 90\% efficiency requirement will be captured 
for the average DESI injection cone of f/3.9 (red line)
at the DESI collimator acceptance speed of f/3.57
if the injection angles are kept below $\sim\pm$60~arcmin (vertical dashed lines).
Chief ray deviation shall not exceed 1.0\degree max, 0.5\degree field-weighted average 
(DESI-0478).
The test fiber was a 38~m, 120~\micron diameter core FBP fiber 
with ceramic polished ferrules and a collimated beam FRD of 1~deg FWHM at f/3.7. 
%
%
}
\label{fig:fib_FRD}
\end{figure}

In order to determine the actual fiber FRD performance, we have measured solid angle FRD as a function of input f/\# for multiple fibers. The illumination geometry, and fiber termination method were found to have significant impact on the results. We have 38~m lengths of 120~\micron~core FBP fiber that were terminated both with polished ceramic ferrules and cleaved fibers bonded into glass ferrules. The results of these measurements (see Figure~\ref{fig:fib_FRD}) were used together with corrector, positioner and ferrule injection angle values to establish that the FRD throughput loss due to angular misalignment between the input beam and the optical axis of the fiber will not exceed 10~\% given the corrector injection f/\#'s and the spectrograph's acceptance f/\#. The contribution to the throughput budget from FRD is shown  in Figure~\ref{fig:fib_tp}.

In addition to FRD increases due to injection angle offset, we have also tested the potential for FRD changes due to flexure of the fiber both in positioners and in the fiber cable as the telescope tracks over the observation lifetime. Significant fiber motions are caused by rotational type actuators and primarily constitute bending. Our tests showed no significant changes in fiber angular output distribution over 90,000~bends of 50~mm radius, although measurable throughput losses were seen at bend radii less than 50~mm. 

\begin{figure}{htb}
 \centering
$\begin{array}{cc}
\includegraphics[width=.30\textwidth]{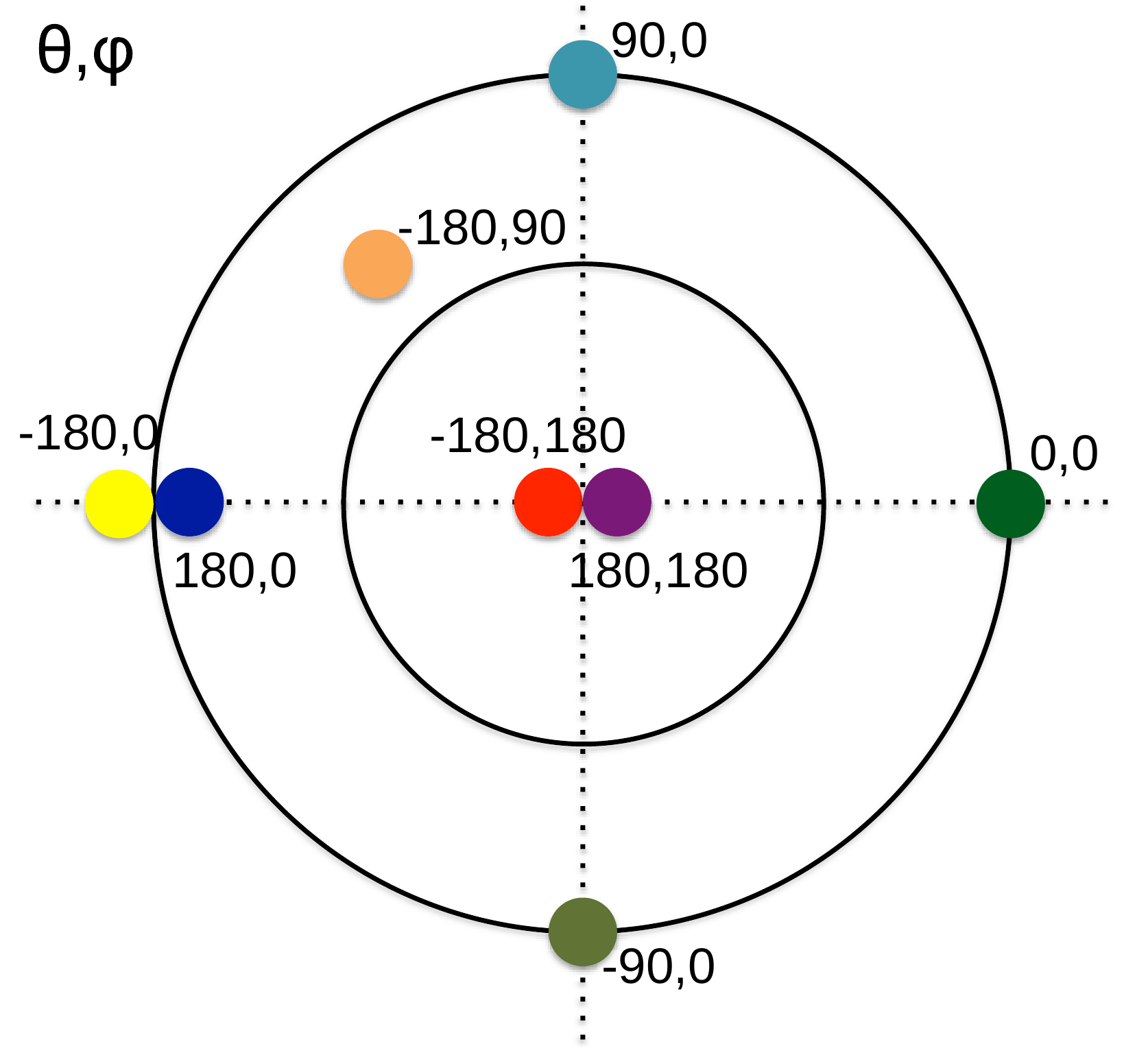}&\includegraphics[width=.60\textwidth]{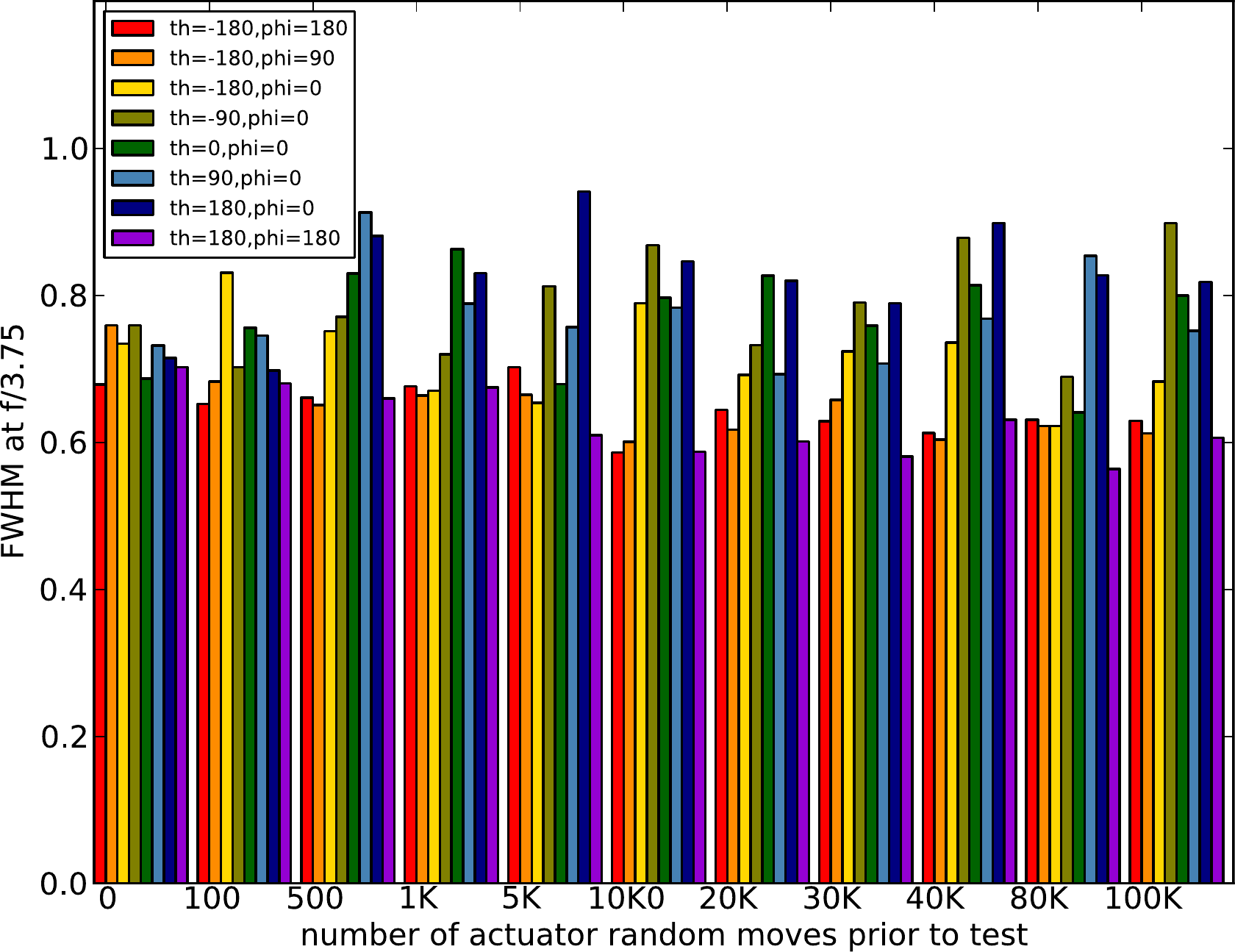}\\
\end{array}$
\caption[Lifetime FRD induced by bending and twisting in a $\theta-\phi$ actuator]{Each bar in the right panel shows the range of FRD over various actuator positions ($\theta=-180^\circ$ to $+180^\circ$, $\phi=0^\circ$ (most amount of bending) to $180^\circ$ (no bending)), see left sketch, as a function of
lifetime (\ie, number of actuator random moves prior to test). Initial tests with a bare fiber showed that in some extreme rotational actuator geometries FRD was increased. However, using the PFAs described in section \ref{sec:PFA}, the polyimide sleeve protects the bare fiber and the average FRD over the range of motion meets all requirements. To date, this performance is maintained to 100k cycles with continued testing underway.}
\label{fig:FRDactuator}
\end{figure}

Tests were conducted to see if actuation over the positioner patrol disk impacted FRD. Ceramic ferrule terminated fibers were attached to development versions of rotational and spine type positioners. Initial tests with optically terminated bare fiber indicated that the FRD performance is affected in some extreme actuation geometries for rotational positioners. Although the fraction of the patrol disk impacted was small, modifications were made to the PFA in order to minimize stress to the fiber. These modifications, which included a polyimide strain relief tube that prevents the fiber from directly touching the actuator, greatly improved the FRD performance of the fiber in the actuator over extreme actuation positions. These results in addition to lifetime tests are shown in Figure~\ref{fig:FRDactuator}.

The fibers will also be flexed in their bundled run assemblies as the telescope slews over the sky. Bending will be constrained to radii $\geq$~50~mm using guides belts, rails and soft clamps. Stress propagation to the fiber ends caused by friction-induced winding over many motion cycles will be mitigated by using a cable with spiral wind construction and low friction sleeves over the sub-bundles. A fiber cable assembly mock-up has been exercised to ensure that the telescope motion will not be affected due to the stiffness of the fiber cables.

\subsubsection{Near Field Stability} \label{sec:NFS}
During the course of measuring the FRD performance of the test fibers in the positioners, it was discovered that the near field intensity distribution is affected in addition to the far field performance. The near field intensity distribution is the distribution of flux at the exit face of an optical fiber and represents a combined intensity distribution of optical propagation modes. The near field intensity distribution is subject to variations due to many factors such as fiber run geometry, illumination angle, optical wavelength, and bandwidth. In contrast, the ``far field'' pattern, an expression of the FRD, is the flux distribution at large distances from the fiber face caused by the angular distribution of the output. Because the fiber end is directly imaged to the spectrograph sensor, the final point spread function is a combination of the near field pattern, spectrograph optical aberrations, and sensor pixelization. If the near field pattern varies at some significant level, the instrument's spectral response can be altered. Therefore, the form and stability of the near field pattern averaged over the data integration time is important for data comparisons such as calibrations.

\begin{figure}[hbt]
\centering
\includegraphics[width=\textwidth]{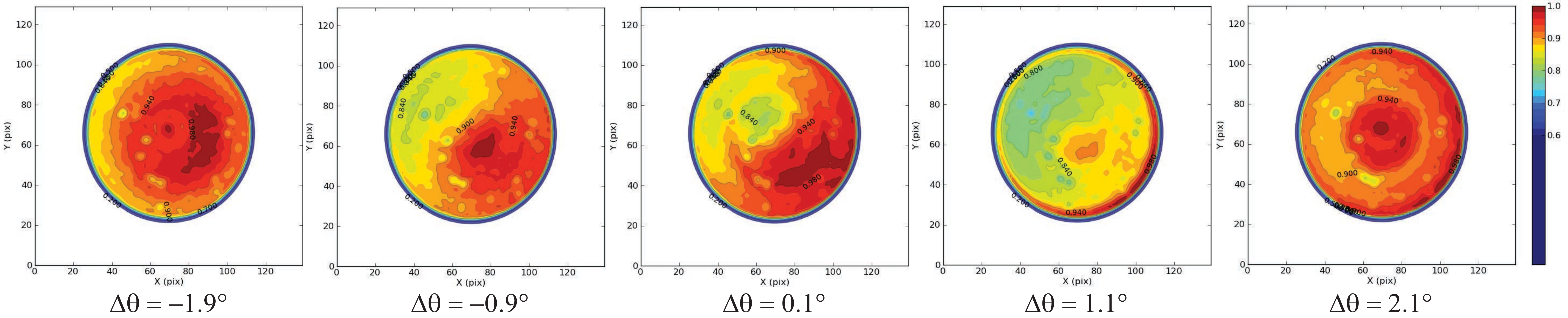}
\caption{Near field intensity distribution as the input angle is offset from on-axis.}
\label{fig:near_field}
\end{figure}

Our measurements of the near field for a representative fiber show that the near field pattern does indeed vary with illumination geometry and fiber end finishing.
Figure \ref{fig:near_field} shows how the near field intensity changes as a result of input injection angle as the fiber was held in a tilting spine actuator. Similar tests were performed for a rotating axis positioner and it was determined that if the fiber had suitable strain relief, the near field intensity remained stable. We believe that the bimodal pattern observed in the near field is a result of stress due to the hard epoxy that is required to hold the fiber during polishing. Preliminary data shows that this problem is reduced if the fiber is cleaved and bonded into a glass ferrule with a relatively soft glue.

The ultimate spectrograph response has been predicted using an end-to-end model that convolves these measured near-field patterns, ray-traced aberration functions, modeled aperture diffraction, and sensor pixelization.  Although in extreme cases, near-field inhomogeneity could cause significant departures in spectral response that would impact science investigations, the actual consequence depends on dynamical issues such as the exposure time, tracking geometry and calibration strategy, which are still under investigation. We are continuing with the near-field measurements and instrument simulations to  establish  the magnitude of the effect, and its mitigation.

\subsection{Positioner Fiber Assembly } \label{sec:PFA}

\begin{figure}[!b]
\centering
\includegraphics[height=2.5in]{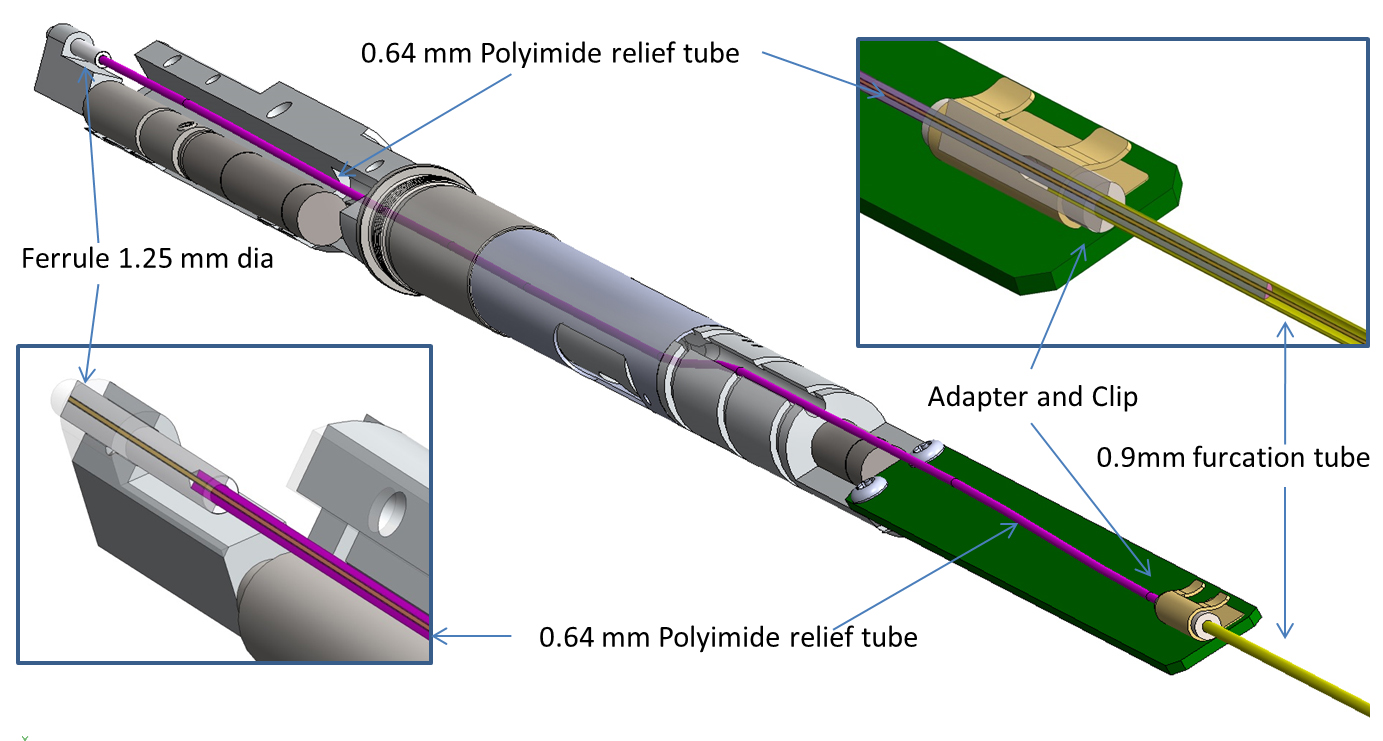}
\caption{Positioner Fiber Assembly.}
\label{fig:FPA}
\end{figure}

A 3~m long Positioner Fiber Assembly originates at the focal plane with an optically-terminated fiber connected to the positioner arm ( Figure~\ref{fig:FPA}) . The assembly begins at the focal plane area and terminates in a Fiber Spool Box where it will be connected to the long run of fiber bound for a spectrograph. The fiber input termination will be made by bonding a cleaved fiber into a close-fitting glass ferrule. A supplementary strain-relief sleeve of polyimide protects the fiber starting at the ferrule and passing through the positioner. 

This sleeve must allow unimpeded movement of the actuator and have adequate wear resistance to the repeated movement within the actuator's fiber guide channel. The strain relief tube transitions to fluoro-polymer (Hytrel$^\circledR$) furcation tubing (jacketing) at a steel adapter, also used as a tubing anchor point at the positioner's terminus. 

The fiber termination process has been subject to a substantial development effort. Two main methods of fiber termination were explored: polished fibers and cleaved fibers. The polishing development effort involved combinations of the ferrule and strain relief tube's material, geometry, and adhesive type, application and curing together with polishing conditions to optimize the FRD, end polish quality, optical axial alignment and process yield. The best results have been obtained with flat-surface polishing of hard-epoxy bonded ceramic ferrules using a minimum application of glue. 

The cleaving effort was divided into a number of stages. First, it was demonstrated that fibers could be cleaved provide excelled FRD performance. Next a glue and ferrule combination together with glue application method had to be found that did not degrade the cleaved fibers performance. Finally a method of aligning the end of the fiber with the end of the ferrule had to be established.

\begin{figure}[!t]
\centering
$\begin{array}{cc}
\includegraphics[height=2.25in]{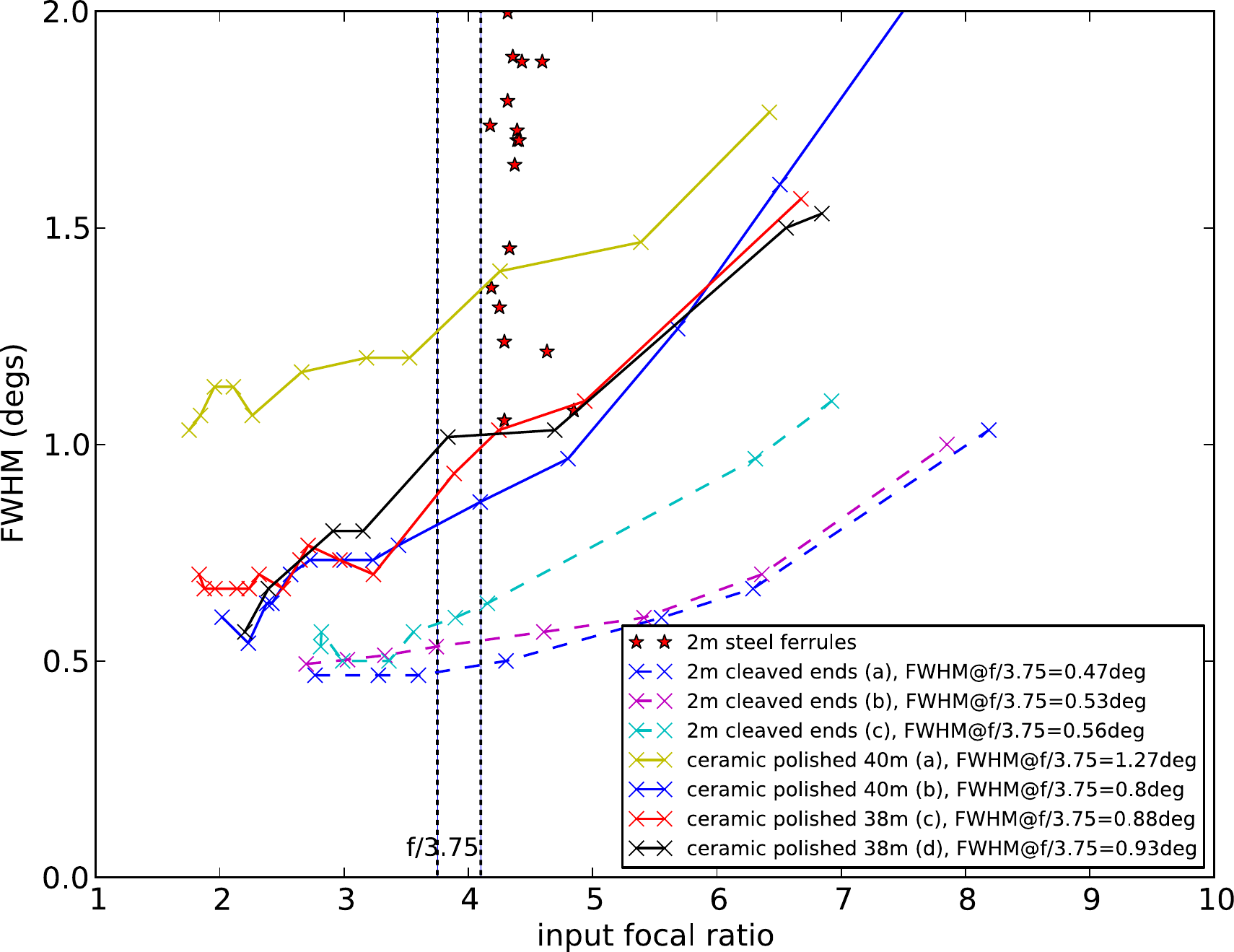}
\hspace{0.25in}
\includegraphics[height=2.25in]{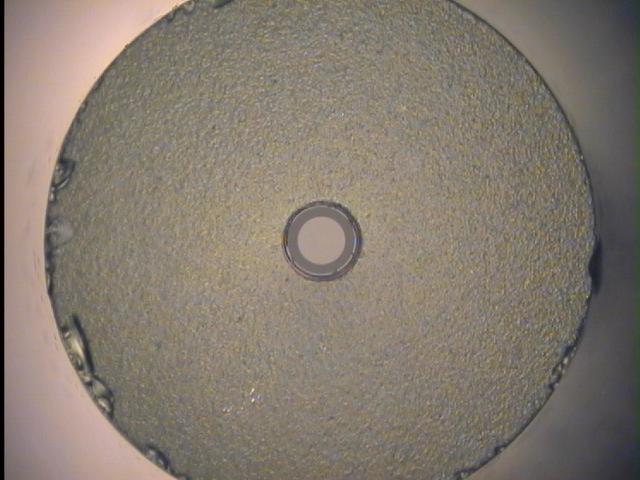}\\
\makebox[0.45\textwidth][c]{(a)} \makebox[0.45\textwidth][c]{(b)} 
\end{array}$
\caption[Collimated FRD for fibers with polished ceramic and steel ferrules] {Collimated FRD for different methods of fiber termination. The horizontal axis shows the input focal ratio and the vertical axis shows the collimated FRD (FWHM, deg). Fibers finished with steel ferrules are shown as stars, fibers with ceramic ferrules are shown with a solid line, fibers that have been cleaved are shown with a dashed line. The vertical lines bound the f/\# input range, f/3.75 to f/4.25, found at the DESI focal plane. }
\label{fig:coll_FRD}
\end{figure}

Figure~\ref{fig:coll_FRD} shows the results of collimated FRD tests of different generations of PFA fibers including both cleaved and polished ends. Panel b of this figure shows the end finish achievable with a cleaved fiber.

The ferrule will be coupled to the actuator using fixtures that precisely registers the fiber-entry axial position to the focal plane surface and that will not induce FRD from mechanical stress over the prime-focus thermal range.

The 5,000 Positioner Fiber Assembly units will be fabricated as 2,500 joined pairs, \ie, as 6~m long units with each end being optically terminated. Having science-quality ferrules on each end allows for 100\% verification of FRD and ferrule optical axis performance. Such verification appears necessary as it is expected that $\leq5\%$ of terminated fibers will require rework to meet performance standards.

A broad band anti-reflection (AR) coating will be applied to the ferrule ends to reduce reflective losses from $\sim$4\% to $\leq$1.5~\%. A low temperature coating process will be used for compatibility with the ferrule bonding. Ion-assisted-deposition will be used to densify the coating film for environmental stability. The short length of clean, high-vacuum compatible materials used in the position fiber assembly facilitates AR coating by minimizing the coating chamber out-gassing load and mechanical volume required to support the fiber assembly in the coating chamber.

After coating performance and spot checking, the joined assembly pairs will be cut and delivered for positioner integration using shipping containers that protect the terminated fiber ends. The fiber can be back-illuminated through the cut end for actuator performance verification. The free fiber end will be threaded through focal plane fiber guides to a Spool Box after the positioner is integrated to the focal plane. Optical performance testing (Collimated FRD and optical axis alignment) can be conducted on the final integrated positioner assembly by using simple collimated illumination of the free fiber end after high-quality cleaving.

\subsection{Fiber Management} \label{sec:cable}

\subsubsection{Top-End Guides}

Optical fibers originating at each positioner will be gathered into fiber bundles and supported behind the focal plane. Each fiber is mechanically constrained to its positioner via a steel adapter that is fixed only to a surrounding protective 0.8~mm diameter Hytrel$^\circledR$ furcation tube. The fiber itself is anchored only at the actuator ferrule and is free to move within the Hytrel tube. The furcation tubes are gathered and anchored into commercial, molded-plastic fiber manifolds (by Miniflex); wherein fibers are collected into sub-bundles within a polymer tube (\eg, M2fX) that exhibits desirable mechanical properties such as flexibility, toughness, crush and extension resistance. 

\begin{figure}[!b]
\centering
\includegraphics[height=3in]{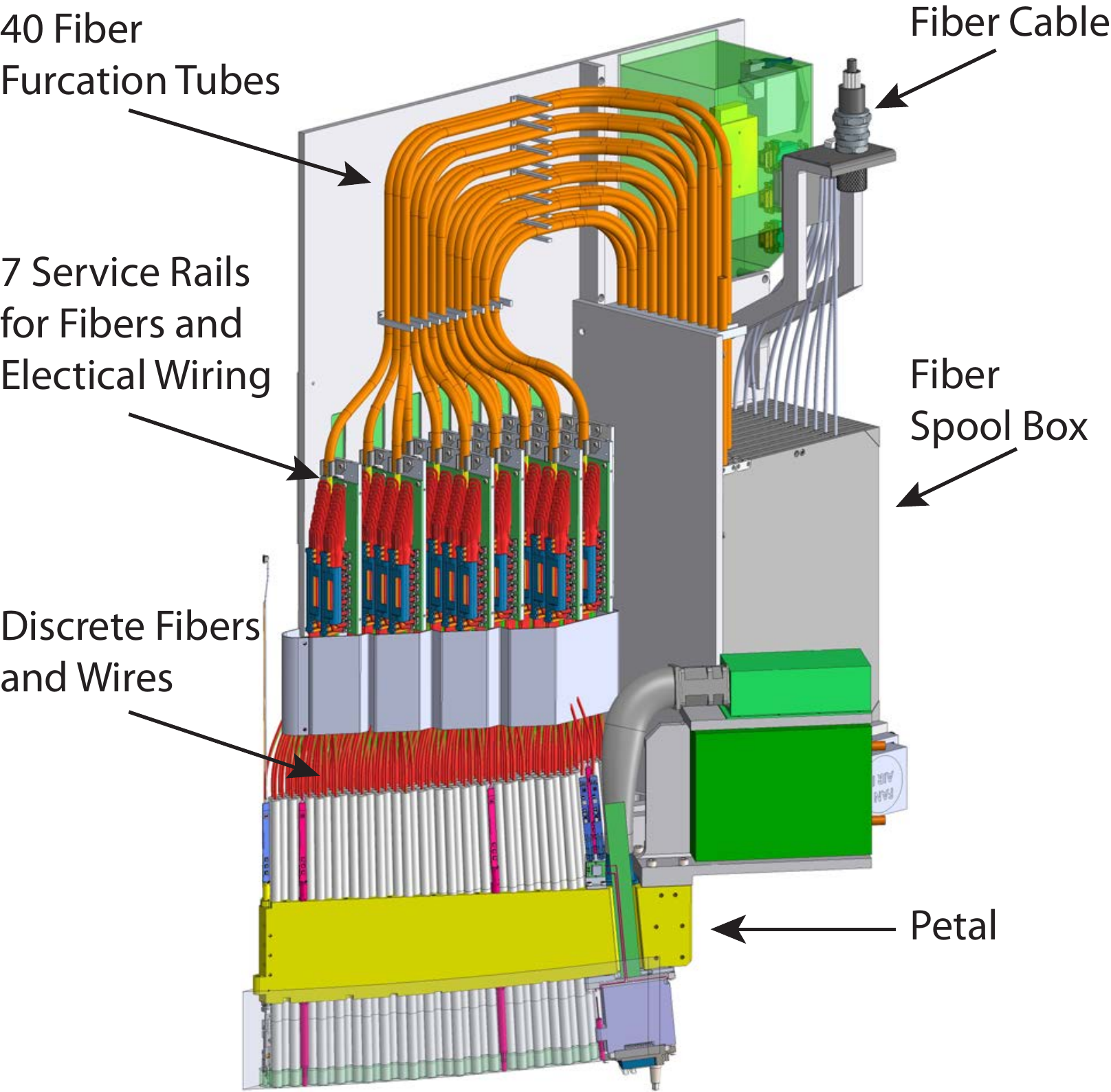}
\caption{The focal plane is modularized via 10 petals, each holding 500 fibers. The fiber cables that contain the fibers are routed to the spool boxes in a way that allows each petal to be individually installed and removed if necessary. Fibers between the spool box and the focal plane are managed via focal plane hardware in order to ensure that the minimum bend radius requirement is not violated. } 
\label{fig:focal_fib_guide}
\end{figure}

The manifolds and tubes are held by a support structure fixed to each focal plane segment. The routing tubes are collected into a single Spool Box wherein they are terminated at a support manifold. The Spool Box is also supported by ancillary structure to the same focal plane segment as its fibers so that each complete focal plane segment can be entirely and independently integrated and handled. The Focal Plane fiber guide and management system is shown in Figure~\ref{fig:focal_fib_guide}.

Two variations of Fiber Spool Boxes are planned, located at each end of the Fiber Cable. Both Spool Boxes are used for fiber routing management and contain free loops of fiber that act as a length reservoir for stress free connection to the positioners and spectrograph slit assemblies by compensating for fiber length differences due to manufacture and routing. The spools also isolate tensile stress and longitudinal fiber movement that can arise from flexing motions or thermal excursions within the main fiber cable.

Each fiber routing tube (M2fx) is directed into an associated box divider where its fibers can be looped in turns. The box includes a mechanical constraint that enforces the life-rated 50~mm minimum bend radius during fiber handling. After transit through the spool box, the fibers are again collected into M2fx routing tubes that are wound through a Fiber Cable. A total of ten spool boxes with their attached fiber cables service the full focal plane. The spool boxes are designed for layered access to the routing divider sections so that an orderly fiber integration process can be followed. 

The hexapod, which holds the prime focus corrector, flexes laterally by up to $\pm$5~mm and this will be included in the requirements of fiber lengths between the spool box and the focal plane guides.

\subsubsection{Fiber Cables}

Ten Fiber Cables, each containing over 500 fibers, will be built following under-sea cable construction techniques as developed by Durham University for astronomical telescope application.

\begin{figure}[!t]
\centering
\includegraphics[height=3in]{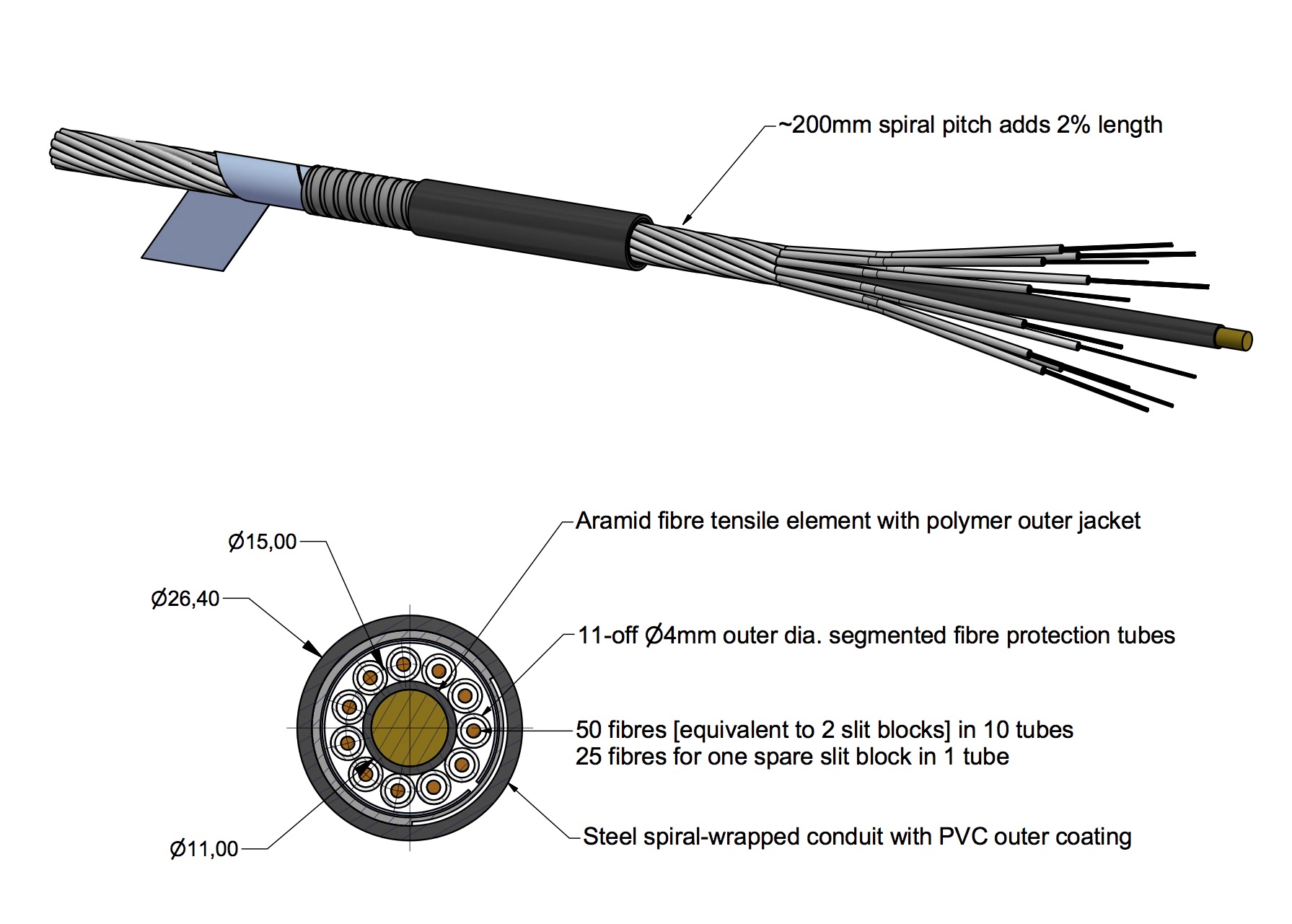}
\caption{Proposed Fiber Cable cross section.}
\label{fig:cable}
\end{figure}

Figure~\ref{fig:cable} shows a cross section through the cable. At the cable center is a strong Aramid-yarn tensile element that limits length excursions due to axial loads or thermal changes. The tensile element's diameter is built up with a polymer coating so that the M2fx routing tubes can be spiral-wound in a uniform radial packing. The spiral winding avoids cumulative tension at the end terminations that can arise from differential length strain when bending the cable. Each spiral tube is loosely filled ($\sim$60\%) with fibers so that friction between individual fibers is minimized. The tubes are pre-loaded with fibers by the tube vendor prior to cable winding. Customized cable end fittings allow fixture rotation about one end of the cable while anchoring the M2fx tubes. The helical cable is wrapped with a protective ribbon of polymer tape and a hygroscopic gel layer that maintains a dry environment within the cable volume. Each of the ten primary cables will have a rugged outer protection layer consisting of an industry-standard 25~mm diameter PVC clad steel spiral wrap (\eg, Adapataflex) conduit. The mass is estimated as 0.65 kg/m$^2$ (see also DESI-0570 for the fiber length requirement).

The Fiber Cable will include 6\% spare fibers to allow for damage mitigation during fabrication, integration and maintenance episodes. The spare fibers will terminate in the Spool Boxes at each end of the Fiber Cable so that they may be drawn from the boxes to a positioner and slit replacement location and fusion spliced. 

\begin{figure}[!b]
\centering
\includegraphics[width=.9\textwidth]{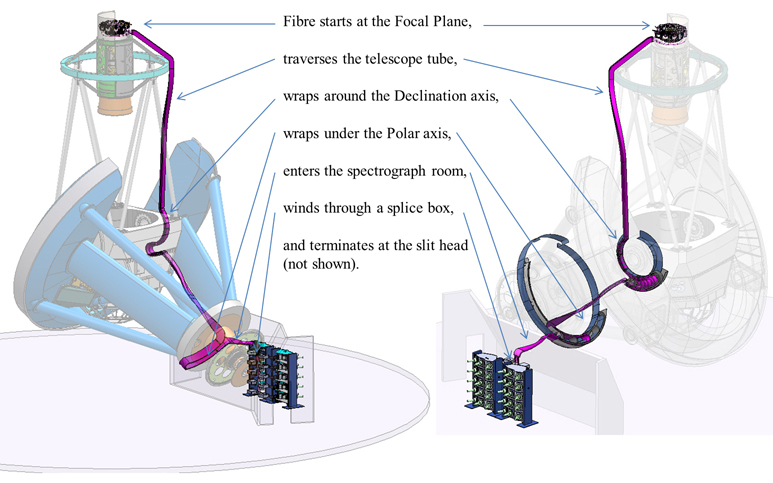}
\caption[Fiber cable routing]{The routing of the Fiber Cable starts at the Focal Plane Spool Box, traverses the telescope tube and pivots, enters the spectrograph room, and terminates at the spectrograph spool box. Two routing options are shown with the one on the left preferred.}
\label{fig:routing}
\end{figure}

The Fiber Cable runs across the prime-focus corrector support vanes, stacked to avoid obscuration, and then down the telescope truss structure toward the primary mirror cell. The cable routing to the spectrograph room has been developed based on fabrication and field test of scale and full size telescope bearing fiber-cable pivots, with particular attention to minimizing the route length and to the ease of cable assembly integration to the telescope. The cables are routed through a declination bearing pivot using guides and link belts (Igus$^\circledR$) to route to and through a polar bearing pivot shown in Figure~\ref{fig:routing}. From there it is routed to the spectrograph room where each cable terminates in a spool box that feeds a spectrograph slit assembly.

The cable guides limit twist and enforce a minimum bend radii of 200~mm that is established by the cable construction. Two 25~m lengths of mechanical sample cable were fabricated on a modified cable winding machine at Durham's Net Park facility. The sample cable was divided into ten cables, each 5~m long with complete termination connectors. The samples were used to 
 measure the spring stiffness and hysteresis of the fiber cables when integrated to the telescope bearing pivots, and to confirm compatibility of the cable properties with the telescope's pointing control system. No degradation in the performance of the drive system was found (DESI-0752).

\subsubsection{Fiber Connection}\label{sec:connect}

The focal plane spool box houses a fiber connection between the positioner fiber assembly and the remaining fiber cable. A connection is necessary to allow for positioner installation from the front of the focal plane and to facilitate the project schedule flow by isolating the positioner and focal plane fabrication, integration and test effort from the fiber cable development. (The connection also eases the process demands for AR coating of the positioner fiber assembly, as previously described.)

We have selected fusion splicing as the method of connection, following the comparative test of a variety of mechanical connector and splicing technologies. The key performance issue for connection schemes is their impact on throughput and FRD. Our demand of a low loss from these effects has proven difficult to demonstrate with commercial mechanical connectors, including single and multiple connector styles. Generally, significant connection losses occur at the mechanical interface due to lateral displacement and Fresnel losses at the junction. In addition, FRD performance is degraded by termination stress and angular misalignment. Our tests included the US Conec MTP$^\circledR$ connectors  (used by the SDSS-III APOGEE project) that are available in units that simultaneously connect 32 fibers. We also tested ceramic-ferrule military style single fiber connectors (Glenair) that can be ganged into multi-pin housings, and specialty single pin connectors by Diamond S.A. that promised precision lateral alignment.

\begin{figure}[!b]
\centering
\includegraphics[height=2.5in]{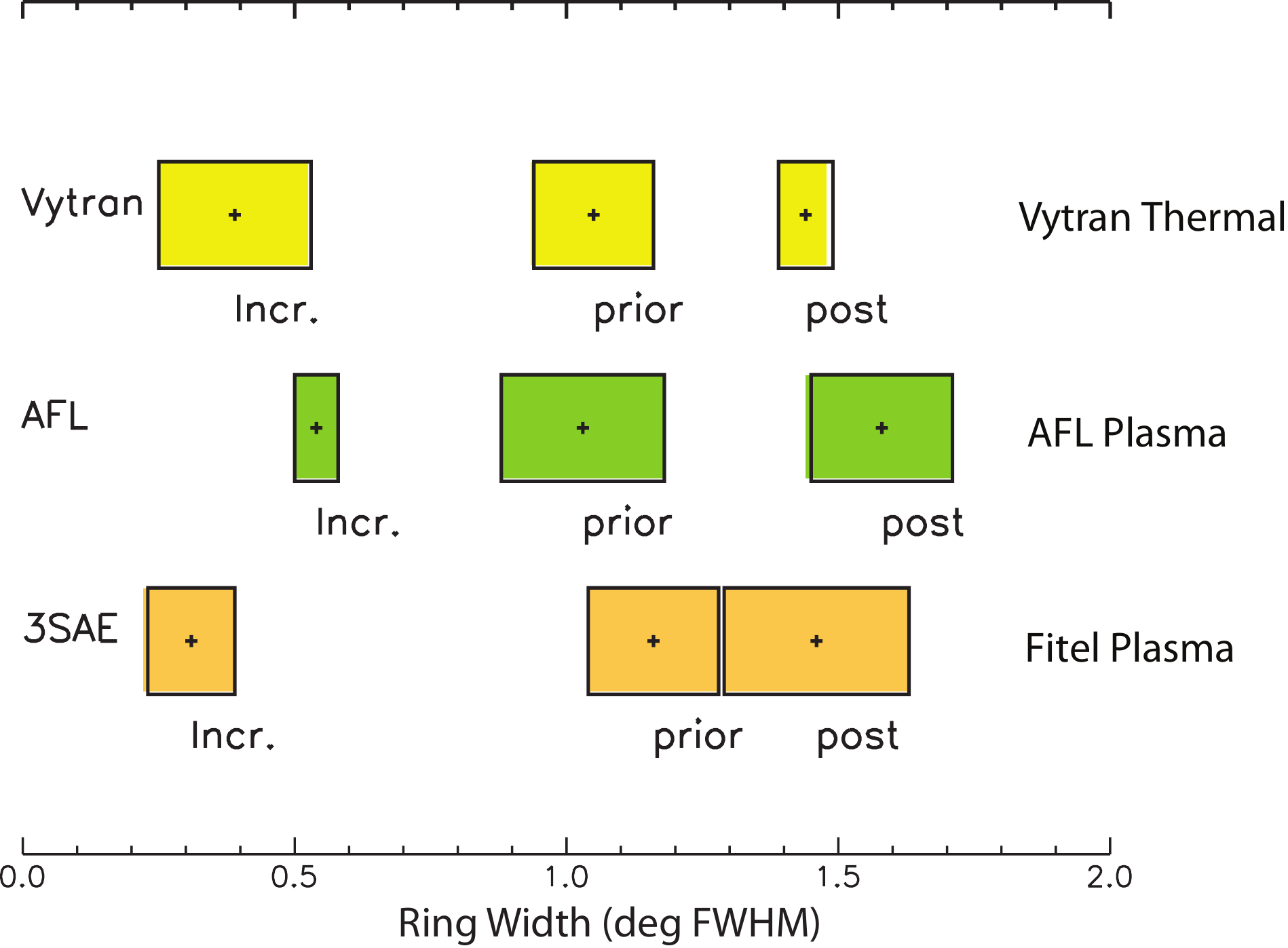}
\caption[FRD of fusion spliced fibers]{Summary of collimated FRD ring width increase for fusion spliced fibers by different commercial methods. A number of fibers were measured prior to splicing so that a baseline could be established. This range of baseline performances are shown as the width of the ``prior'' boxes; the range of post-slice performances are shown with the ``post'' boxes; and the range of incremental increases are shown with the ``Incr.'' (post minus prior) boxes. The 120~\micron core FBP fibers were illuminated at an angle corresponding to f/4.5.}
\label{fig:splice}
\end{figure}

Our investigation of the fusion splicing process as applied to the DESI optical fiber type has shown performance results far superior to mechanical connections. The high precision of fiber core alignment and resulting continuous glass interface yields bulk transmission losses of about 1\% and minor impacts on FRD ($\leq$0.2~deg increase in collimated FRD). These results were developed in a test campaign at the application laboratories of three major splice equipment vendors (Vytran, 3SAE, AFL). Competing methods for the key process steps of stripping, cleaving, fusing and re-coating, were tested and optimized. The FRD impact for fusion spliced fibers are shown in Figure~\ref{fig:splice}. 

\begin{figure}[htbp]
\centering
\includegraphics[height=2.5in]{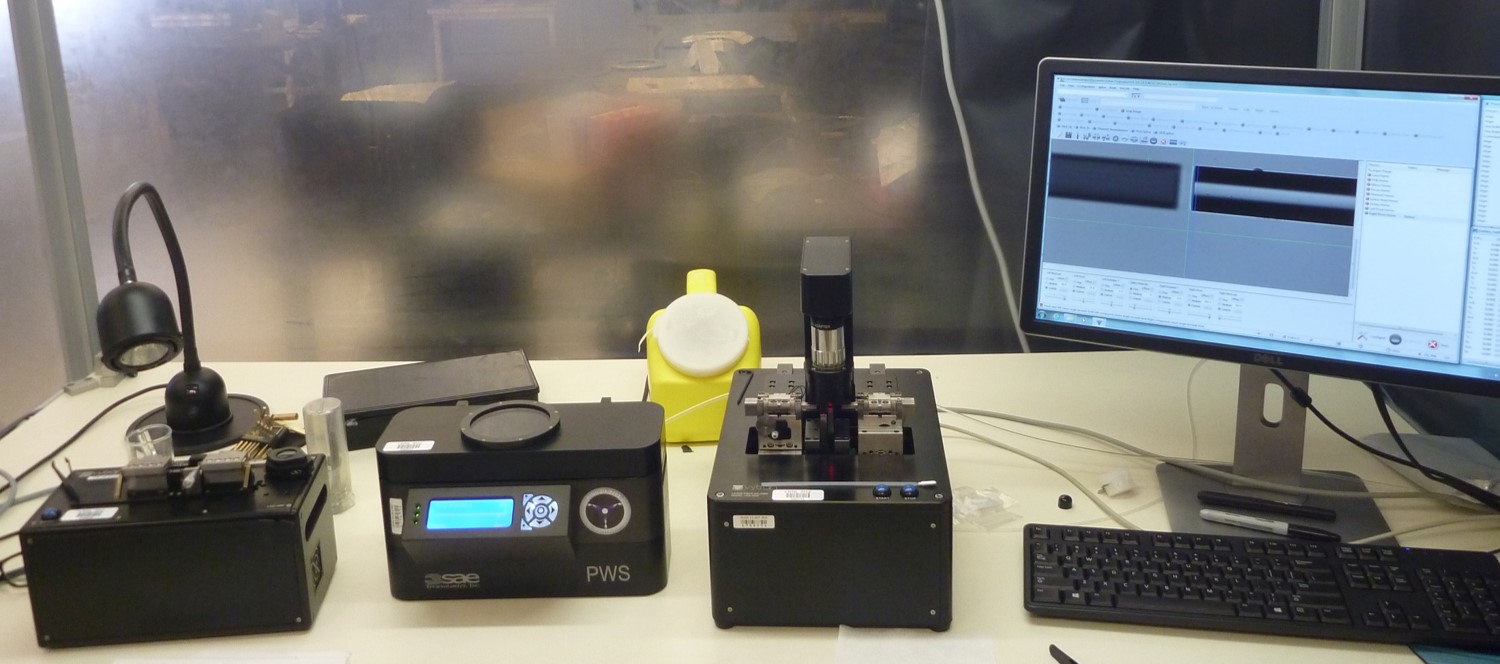}\\
\mbox{(a) splicing station showing fiber cleaver, buffer stripper, and splicer}\\
$\begin{array}{cc}
\includegraphics[height=1.04in]{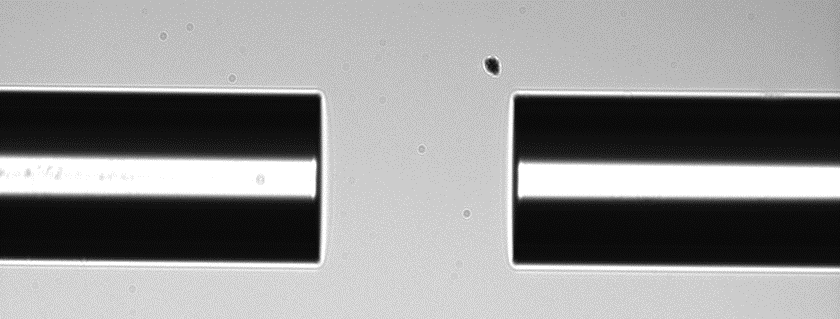}&\includegraphics[height=1.04in]{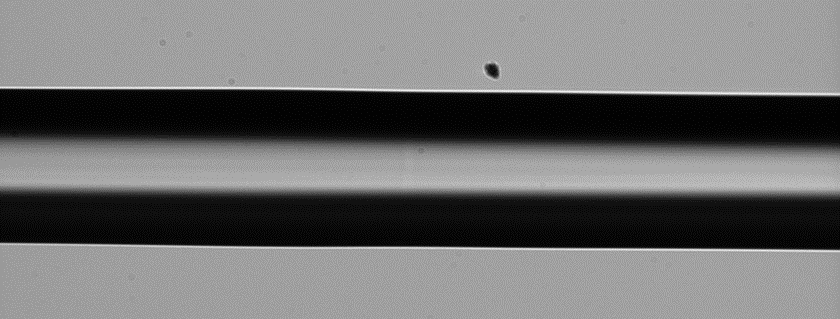}\\
\mbox{(b)  cleaved fibers ready to be spliced}&\mbox{(c) spliced fiber}\\
\end{array}$
\caption[Splicing Station]{Summary of collimated FRD ring width increase for fusion spliced fibers by different commercial methods. A number of fibers were measured prior to splicing so that a baseline could be established. This range of baseline performances are shown as the width of the ``prior'' boxes; the range of post-slice performances are shown with the ``post'' boxes; and the range of incremental increases are shown with the ``Incr.'' (post minus prior) boxes. The 120~\micron core FBP fibers were illuminated at an angle corresponding to f/4.5.}
\label{fig:splice_station}
\end{figure}

\begin{figure}[htbp]
\centering
\includegraphics[height=3in]{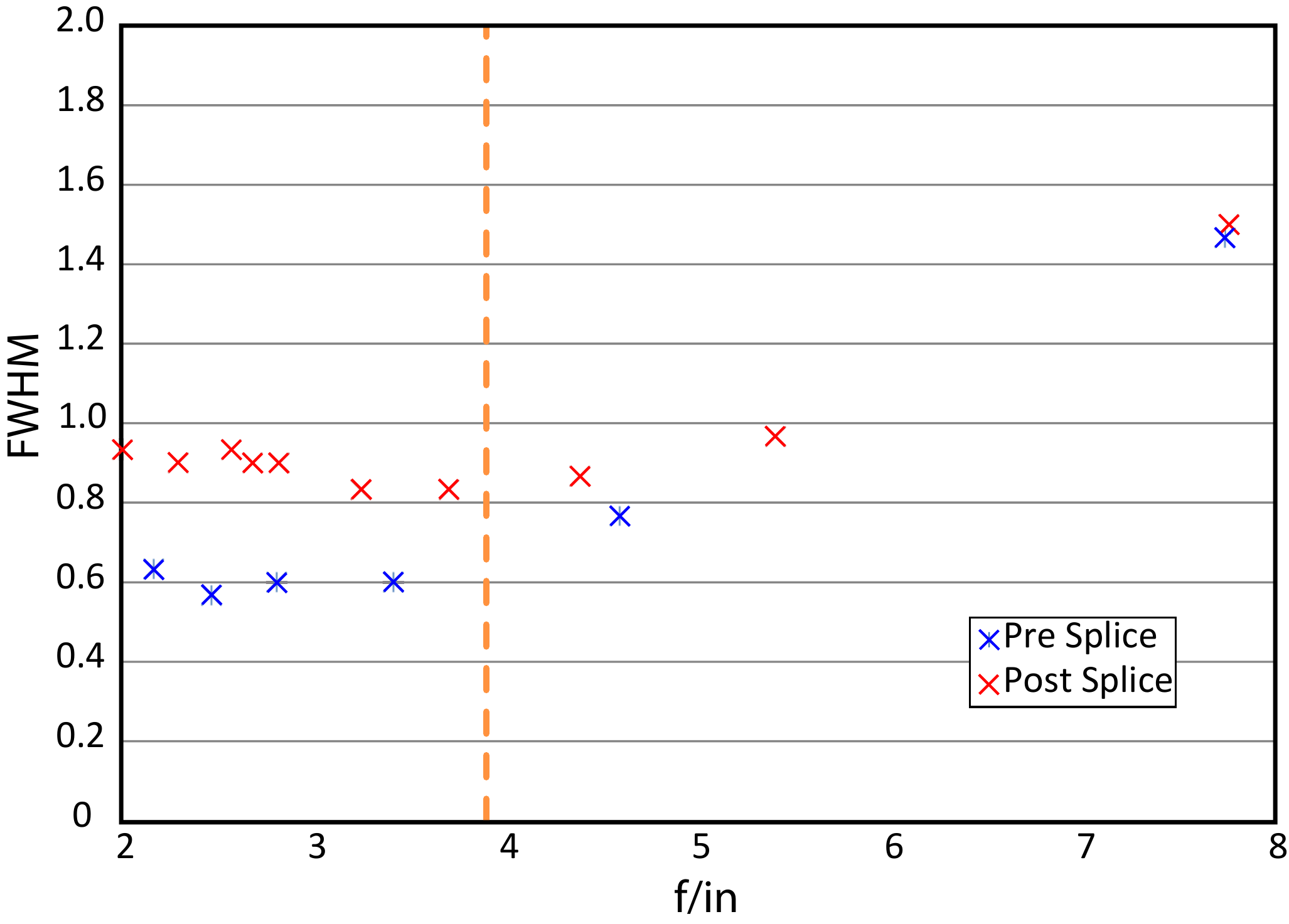}
\caption[Optimised splicing performance]{}
\label{fig:splice_performance}
\end{figure}

In 2015 Splicing equipment was purchased that enabled us to continue the FRD testing and optimization. Figure \ref{fig:splice_station} shows the splicing station setup in panel (a) and the images of the fiber pre- and post-splice in panel (b) and (c). Figure \ref{fig:splice_performance} shows the change in FRD between a single fiber that has been cleaved, and then the same fiber after it is cut in half and sliced back together. Prior testing has shown that an increase of 0.2 degrees at f/3.9 translates to a $\sim1$\% throughput loss to the spectrograph.

The fiber spool box has been designed to contain sufficient free fiber to fit splicing machinery for multiple fusion splice cycles. The box design includes segments of soft urethane foam (Poron$^\circledR$) for splice support and protection. The foam support has been proven to not induce measurable FRD while directly holding the optical fiber.

\subsection{Spectrograph Slit } 
\label{sec:slit}

There are 10 science slits, 2 test slits which are sparsely populated and one flat fielding slit. The science slits will 
be temporary removed to allow insertion of one of the other slits when necessary.  A slit is comprised of 20 blocks each containing 25 fibers. The naming convention is Block B0 through B19 from bottom (nearest to the bench) to top (furthest away from the bench. Within a block, fiber F0 (lowest in the block) to F24 (top of the block).

\subsubsection{Science Slits}

\begin{figure}[!t]
\centering
\includegraphics[height=2.5in]{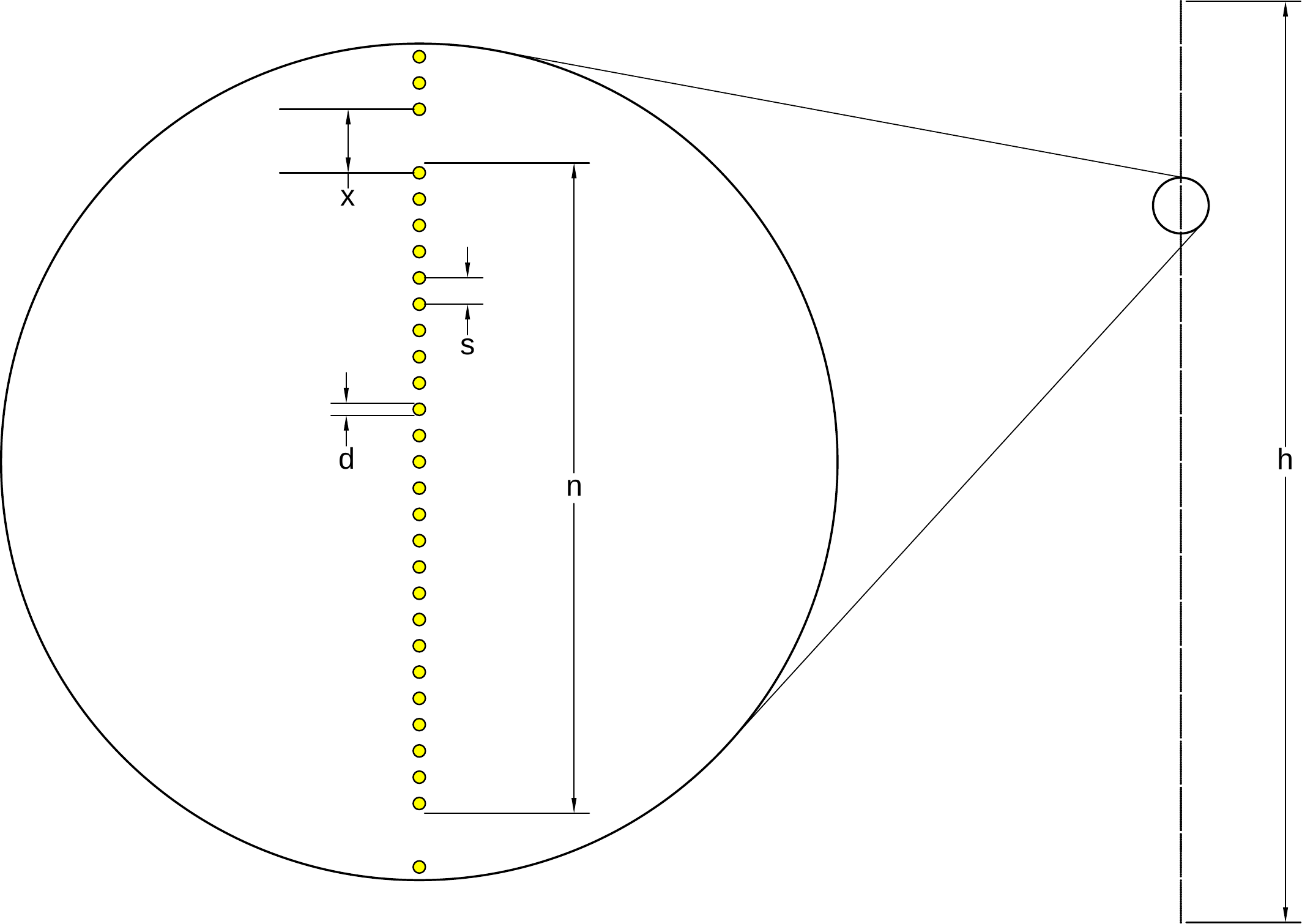}
\caption{Fiber Slit geometry.}
\label{fig:slit_req}
\end{figure}

Each spectrograph input is formed by a fiber slit assembly. The slit consists of a group of 500 fibers 
arranged in a linear array with a convex surface that satisfies the spectrograph axial focal prescription.
The fiber slit parameters are outlined in Table~\ref{tbl:slit_req} and the geometry is shown in 
Figure~\ref{fig:slit_req}. 

\begin{table}[!b]
\begin{center}
\caption{Fiber Slit Geometry Specifications for f/3.57.}
\label{tbl:slit_req} 
\begin{tabular}{ l l }
\hline
Specification & Requirement \\
\hline
 Fiber Diameter [$d$] & 107~\micron \\
Slit Height [$h$] &120.9~mm\\
Number of Fibers [$N$]&500\\
Fibers Per Group [$n$]&25\\
Number of Groups [$g$]&20\\
Fiber Spacing [$s$] & 230~\micron\\
Group Spacing [$x$] & 556~\micron\\
Radius of Curvature [convex]&468.3~mm\\
Fiber tip axial location& $\pm$25~\micron\\
\hline
\end{tabular}
\end{center}
\end{table}

Ideally, each fiber would be pointed toward the spectrograph entrance pupil with the optical axis slightly 
diverging from the axis of it's neighbors. This would mean a flared slit with all fibers coming from the radial 
center 468.3~mm behind them. In practice the 500 fibers are grouped to 20 blocks with 25 fibers each which are 
cemented into parallel v-grooves, with each block pointing into a slightly different direction. While this causes 
a linear shift in pupil illumination along one slit block, this effect is smaller than the natural pupil shift 
that happens anyhow along the whole slit even if each fiber axis would originates exactly from the same radius 
of slit curvature. ZEMAX simulations have shown that there is no additional light loss caused by this simplification,
 as the optical elements are oversized. 

The fiber on-center spacing is established by the spectrograph field size together with the desired unilluminated 
regions between the spectra on the sensor. Optical tolerances demand a precise location for the fiber tips with respect to focal distance. However lateral and fiber center tolerances are undemanding because the relative positions are set by the v-grooves which are manufactured to tight tolerances.

The slit array mechanical assembly can be compared to that used for the BOSS spectrograph slit~\cite{Smee13}. However, the 20 fiber blocks 
are bonded to a slit plate that is either ceramic or metal and includes machined structures acting as datum 
surfaces during assembly. The plate provides the mechanical interface to the spectrograph and is installed using 
a linear stage and a datum surface to allow reproducible installation inside the narrow gap of the spectrograph 
dichroic beam splitter. The assembly plate also supports and constrains each block's fiber bundle and terminates 
the protective sheaths of the bundles. The fiber blocks are the basic fabrication unit for the fiber system. 

There are many possible configurations for the mapping between the fiber cable and the slit so various studies were carried out to optimize the slit design with a view to making the assembly simple while conforming to the required tolerances. The preferred solutions are presented in documents DESI-0830, which detail the slit optical geometry and mapping between cable and slit components respectively. 

The adopted solution has 20 slit blocks mapped to 10 Miniflex tubes (diameter 4~mm) within the cable (plus an unpopulated tube to aid packing). Therefore, each tube contains 50 fibers, which maps to 2 slit blocks, each of which contains 25 fibers.

The blocks are essentially regular cuboids with v-shaped grooves cut in one surface parallel to one axis. The fibers are therefore parallel within each block despite the finite location of the virtual spectrograph input pupil which means that the focal surface containing the fiber slit is convex. Modeling by ray-tracing shows that there is no significant degradation of image quality using this simplifying configuration. The slit blocks, however, are angled and positioned along a curved surface.

The fiber blocks are fabricated by bonding the ends of the individual fibers between v-shaped grooves and a cover plate. The fiber ends are then cleaved and co-polished with the block surface. The v-grooves are machined in fused silica. 

This allows the fibers to be tested, without adding an additional termination, with light input into one slit block emerging from another slit block fed into test apparatus. After testing is complete, the fiber can be cut where it runs free at the top-end to separate the fibers into independent bundles of 25 where they can then be spliced with the top-end fibers. 

A prototype slit block with diverging grooves has been fabricated.
This is a greater challenge to fabrication than the proposed parallel-groove blocks so makes for a more stringent test than required. Metrology showed compliance with the mechanical tolerances for groove position and direction and FRD (DESI-0619).

\subsubsection{Calibration Slits}

For spectrograph calibration purposes it is planned to temporarily replace the science slits with replacements 
that allow to sample the point spread function along the slit extension and to allow a flat field calibration of 
the collimator-spectrograph train.

\paragraph{PSF calibration Slit}

\begin{figure}[!t]
\centering
\includegraphics[height=2.5in]{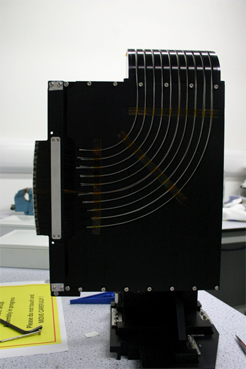}
\caption{Sparsely populated test slit to used for each spectrograph during initial integration and test.}
\label{fig:test_slit}
\end{figure}

This consists of a sparse array of single fiber outputs. Each block contains one fiber only to create 20 spectra widely separated for PSF and stray light analysis. The fiber population consists of the bottom blocks B0-B9 with the bottom fiber F0 installed only, the mid block B10 which is fully populated (F0-F24) and the top blocks with the upper fiber position F24 populated only. In total 44 fibers are installed which on the input side are densely packed in a stacked connector that is illuminated by an f/3.57 beam by the 12~mm round field of the Offer relay. This slit has been manufactured and is shown in figure \ref{fig:test_slit}.

\paragraph{Flat Field Slit}

This consists of a leaky fiber that emits light laterally. This creates a single, broad white light 
slit that can be use for flat field calibration of the collimator/spectrograph train. The specifications for this 
component is tabulated in Table \ref{tab:flatfieldslit}. A  candidate is the Corning Fibrance with an OD of 
230~$\pm$~10~\micron and an operational wavelength range between 405~nm and 1000~nm at least. Calculations show that an attenuation of 0.3 db/km results in 0.8\% variation along the slit, hence the large scale variation along the slit seems easy to be met. The continuum light source can be fed through the fiber itself. 

\begin{table}[tbh]
\begin{center}
\caption{Specifications for the flat fielding slit.}
\label{tab:flatfieldslit}
\begin{tabular}{|l|l|}
\hline
Uniformity along slit & 50\% \\
Uniformity along wavelength & 50\% \\
Variation along 300~\micron of slit & 1\% \\
Spectral variation & $<$ 1\% over $\lambda / 4000$ \\
Width of illumination & $<$ 300~\micron \\
Lateral slit location & $\pm$50~\micron \\
Piston tolerance & $\pm$30~\micron \\
Beam intensity & 4 $\times 10^{4}$ Hz$^{-1}$ mm$^{-2}$ nm$^{-1}$ \\
\hline 
\end{tabular}
\end{center}
\end{table}

 

\subsubsection{Mechanical Design}

The slit is located in a slot in the first spectrograph dichroic.  The mechanical design of the slit must be compatible with the space envelope defined in DESI-0543. The main challenges foreseen in mounting the slit blocks are that:

\begin{enumerate}
\item{The maximum overall slit assembly thickness is 3 mm (at the beam) to minimize optical path obscuration. Consequently a significant support structure which enforces a minimum curvature radii is required to route fibers to bundling points outside the beam area.}
\item{The slit is co-located with the front surface of the dichroic mirror and must be positioned through a narrow slit cut into that glass optic. A repeatable method of retracting the slit is required to allow the science slit to be exchanged with the flat field slit. The Piston location of the slit repositioning is expected to be of the order of $\pm$30~\micron. Lateral positioning of the slit is to be within $\pm$50~\micron.}
\end{enumerate}
 
\begin{figure}[!t]
\centering
\includegraphics[width=.9\textwidth]{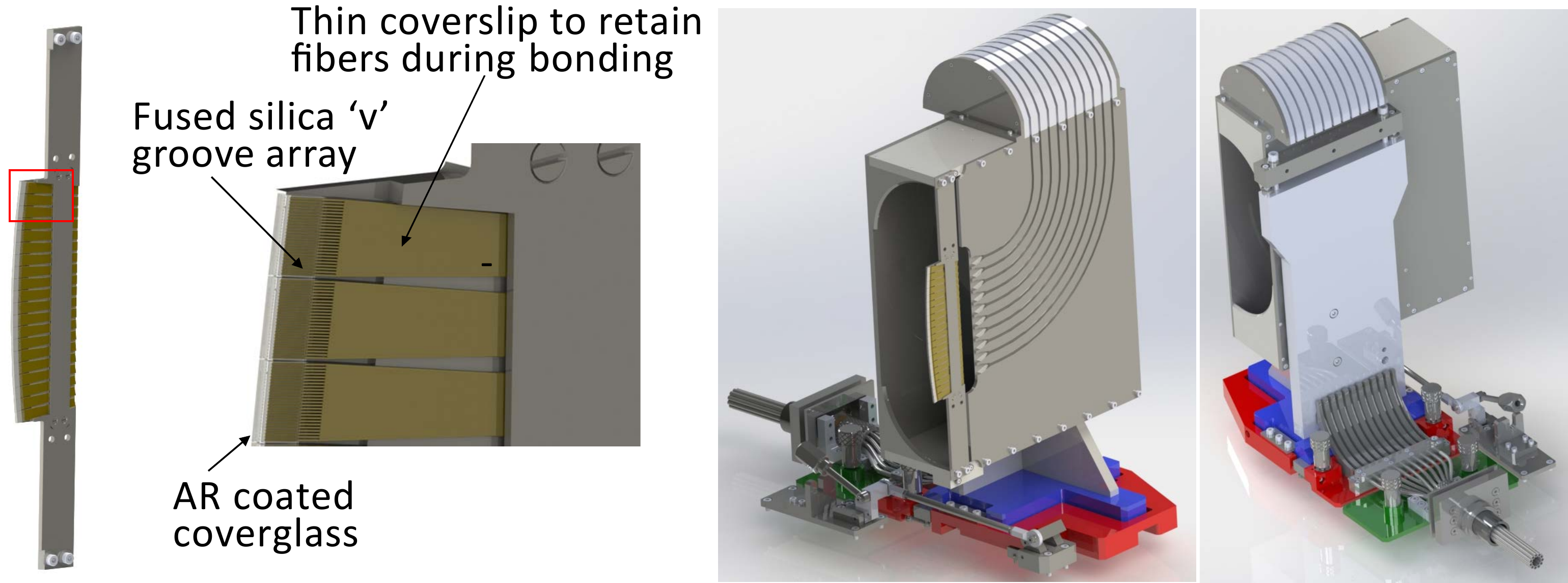}\\
\caption{The fiber slit block fabrication (left) and strawman slithead assembly (right)}
\label{fig:test_slit_block}
\end{figure}

A ``strawman'' mechanical design has been developed which is shown on the right hand side of Figure~\ref{fig:test_slit_block}. In this design the slit is mounted on preloaded linear rails which is translated using a leadscrew assembly. There are potential issues with access to the drive mechanism when the slit is mounted in the spectrograph, hence, alternative methods of retracting and positioning the slit are being considered.

To support the fiber ends with the required fiber spacing  a three-part assembly has been designed (see left panel of Fig. \ref{fig:test_slit_block}). The fibers will initially be placed in the grooves and bonded in place with a low-shrinkage epoxy with a thin coverglass on top. This assembly will provide sufficient lateral support to hold the fibers during polishing. After polishing, an AR-coated coverglass will be glued to the front of the array using a suitable optical adhesive (\eg, Norland NOA88). The final stage will be to bond the v-groove arrays onto a backing plate following the exact radius of curvature of the slit. The required tolerances will be achieved by using a bonding jig which will locate each array against two pins along the front of the AR window and one that contacts with a feature cut into the side of the v-groove arrays through slots cut into the backing plate.

After the finish polishing and fiber support bonding, the pairs of slit blocks that contain the same fibers will be tested with light input at one block and emerging from another. Tests will be for throughput, FRD, and optical alignment. An AR coated thin (250~\micron) coverglass will then be bonded over the fiber block face. The AR cover bonding uses optical index matched glue that will mitigate surface roughness or flaws in the fiber block polish. Using an AR cover plate avoids subjecting the bonded block assembly and its attached fiber and cable to coating chamber vacuum or thermal excursions. 

\subsection{Fiber System Assembly and Integration}

In order to ease integration flow  the fiber system will be produced in lots of 500 fibers. In addition to this, the PFA production will be decoupled from the cables and slit blocks. This  allows the PFAs to be manufactured while the cables are being built and the slits polished. The fiber cable and slits will be manufactured at Durham University. The PFA bonding and fiber system splicing must be performed in a clean tent.


\label{sec:fib_I_T}
Performance of the fiber system FRD and throughput are verified for each subsystem by testing prior to integration. Extensive fiber testing during the research and development phase has led to the development of optical testing setups which will be modified and automated. The testing happens at component, engineering model level, and production level.

\subsubsection{Component Level Testing}

Component level testing includes the acceptance of  components as they arrive from the manufacturer. FRD and transmission testing will be performed on the bulk fiber when it is received in order to ensure it has sufficient performance before it is used in any sub-systems.

\begin{itemize}

\item Optical fiber: FRD and throughput measurements on sample fibers of the received batch to verify the fiber meets the specification. The collimated laser FRD setup can be used as well as the throughput setup.

\item V-groove arrays: Visual inspection by microscope of chosen sample to verify compliance of groove number and dimensional properties.
\item V-groove lids: Visual inspection and dimensional metrology.

\end{itemize}

\subsubsection{Engineering Model Level Testing}

Engineering model level testing targets the first production arrays where it is particular important to monitor performance compliance to validate component selection and assembly methods.

\begin{itemize}
\item Slit block: FRD and throughput measurements of each fiber along a slit block. This will verify that the slit block architecture and assembly method is able to provide a compliant slit.
\item Block pointing accuracy: The parallelism can be quickly checked by illuminating all fiber inputs with the collimated laser beam of the FRD setup. If the superposition of the 25 annuli do not deviate from the theoretical expectation, the block is compliant. This test can be quickly done during FRD tests.
\item Slit pointing accuracy: Can be measured by laser metrology using the reflection of the polished front face and a screen. 

\end{itemize}

\subsubsection{Production Level Testing}

Production level testing includes, at a minimum, spot checks to flag  potential problems. Depending on  experiences during prototype testing it can be decided what sampling rate is required. If for example prototype testing revealed that there is no significant scatter in FRD performance along a slit block, one fiber per block can be measured to spot check compliance. In-situ verifications during assembly are possible at the slit assembly state.

\begin{itemize}

\item FRD and throughput: A sub-set of fibers along a block will be FRD and throughput measured. The size of the sub-set depends on the outcome of prototype testing.

\item Block pointing accuracy: Use of the annulus projection again while illuminating all fiber inputs with the collimated laser beam at the FRD setup. If the pattern follows the theoretical expectation, the block is compliant while deviations from the pattern indicate either a pointing direction or FRD problem that has to be investigated further.

\item Laser metrology at the polished and reflective v-groove front face: This can be done in-situ while cementing the arrays onto the slit plate, providing a second guide in addition to the mechanical datum. A screen with tolerance circles for each block position can be produced to streamline this process further.

\end{itemize}

As the fibers are terminated, all tests can be performed on double ended fibers of twice the final length before being cut in the middle. This assures the optical quality of the input surface is high enough without the need to polish input surfaces. After the fiber system has been spliced, a collimated FRD test will be performed before the system is released to telescope integration.

\clearpage

\section{Spectrographs}
\setcounter{equation}{0}\setcounter{figure}{0}\setcounter{table}{0}
\label{sec:Instr_Spectrographs}

\label{sec:Instr_Spectro}
The major spectrograph requirements are given in Table~\ref{tab:spectro-reqs} (extracted from  DESI-0613).  The bandpass, spectral resolution, point spread function (PSF) stability, and noise variance requirements flow from the science requirements document (DESI-0318). The throughput requirements flow from the system throughput budget (DESI-0347). The temperature requirements flow from the DESI environmental requirements document (DESI-0583). The fiber diameter was chosen to optimize the signal-to-noise ratio in emission line galaxies (DESI-0304). The focal ratio was chosen to ensure greater than 90\% throughput due to the focal ratio degradation of the fibers given the f/3.85 beam input into the fibers by the corrector (see Figure~\ref{fig:fib_FRD}).

\begin{table}[ht]
\small
  \centering
  \caption{Spectrograph Requirements.}
    \begin{tabular}{lrr}
    \hline
	 Item & Requirement & Current Design \\
    \hline
    Bandpass  &  360--980  nm  & complies  \\
	\noalign{\vskip 2mm}
               & $\ge 1,500$; 360 nm $< \lambda \le$ 555 nm & 2,000--3,200 \\
    Resolution ($\lambda/\Delta\lambda$) & $\ge 3,000 $; 555 nm $< \lambda \le$ 656 nm & 3,200--4,100 \\
                & $\ge 4,000$; 656 nm $< \lambda \le$ 980 nm & 4,100--5,100  \\
	\noalign{\vskip 2mm}
         & $\ge 21\%$; $\lambda =$ 360 nm & $39\%$    \\
         & $\ge 30\%$; $\lambda =$ 375 nm & $51\%$    \\
			& $\ge 43\%$; $\lambda =$ 400 nm & $60\%$    \\
			& $\ge 55\%$; $\lambda =$ 450 nm & $73\%$    \\
			& $\ge 57\%$; $\lambda =$ 500 nm & $69\%$    \\
			& $\ge 54\%$; $\lambda =$ 550 nm & $68\%$    \\
   End-to-end throughput & $\ge 50\%$;  $\lambda =$ 600 nm  & 69\%   \\
         & $\ge 56\%$; $\lambda =$ 650 nm & 72\%  \\
         & $\ge 58\%$; $\lambda =$ 700 nm & $70\%$     \\
         & $\ge 56\%$; $\lambda =$ 750 nm & $66\%$     \\
         & $\ge 63\%$; $\lambda =$ 800 nm & $78\%$     \\
         & $\ge 63\%$; $\lambda =$ 850 nm & $78\%$     \\
         & $\ge 62\%$; $\lambda =$ 900 nm & $73\%$     \\
         & $\ge 48\%$; $\lambda =$ 980 nm & $55\%$     \\
	\noalign{\vskip 2mm}
   Point Spread Function Stability & $\le 1\%$ & TBD$^{\dag}$ \\
   Number of Fibers & 5,000 & 5,000 \\
    Fiber diameter &      107 \micron & 107 \micron   \\
    Collimator f/\# &  $\le 3.57$ & 3.57 \\
    Noise Variance & $\le 25\%$ of sky & complies \\
    Operational Temperature & $-10^{\circ}$C to $+30^{\circ}$C & complies \\
    Survival Temperature & $-20^{\circ}$C to $+40^{\circ}$C & complies \\
	\hline
	$^\dag$ Will be verified with the prototype \\
    \end{tabular}
  \label{tab:spectro-reqs}
\end{table}

The spectrograph specifications are derived from the requirements and the baseline design and are given in Table~\ref{tab:spectro-specs}. These values lead to the optical design. Ten spectrographs must be used to support the 5,000 fibers of the DESI instrument. To simplify the optical design the spectrographs will be placed in a temperature controlled environment. This isolates the optics from the operational temperature swings of the Coud\'{e} room. The spectrographs will be mounted in five stacks of two in a temperature controlled enclosure in the Coud\'{e} room at the Mayall telescope. 

\begin{table}[ht]
\small
  \centering
  \caption{Spectrograph Baseline Design Specifications.}
    \begin{tabular}{lrr}
    \hline
	 Item & Specification & Current Design \\
    \hline
    Number of spectrographs &       10        &   complies  \\
    Detector pixel pitch &       15 \micron &   15 \micron \\
    Spectral detector elements &       4,096 pixels & complies\\
    Spatial detector elements &      4,096 pixels & complies \\
    Minimum resolution elements &       $\ge 3$ pixels & 3.4 pixels\\
    Fiber spacing (slit plane) &       230 \micron &   230 \micron \\
    Slit Height                &       120.9 mm & 120.9 mm\\
    Number of fibers (spatial) &       500 &       500\\
   Maximum rms Radius &  $\le 13$ \micron & 11.5 \micron \\
    Fiber - Fiber Crosstalk & $\le 0.5\%$ & 0.05--0.46\% \\
    $95\%$ Encircled energy diameter &     $\le 110 $      \micron & 61--82 \micron \\
    $50\%$ Encircled energy diameter &    $\le 50 $         \micron & 37--45 \micron \\
    Operational Temperature &  $20^{\circ}$C $\pm$2$^{\circ}$C& complies\\
    \hline
    \end{tabular}
  \label{tab:spectro-specs}
\end{table}

\subsection{Spectrograph Optical Design}

Once the specifications are defined a literature search (DESI-0755) of astronomical spectrographs was done to help limit the design space of the spectrograph. A trade study was performed on whether the optics should be refractive or reflective (DESI-0757). More discussion of the results of this study will be in the collimator and camera sections. For more detail on the spectrograph optical design see DESI-0334.

\begin{figure}[!ht]
\centering
\includegraphics[height=2.75in]{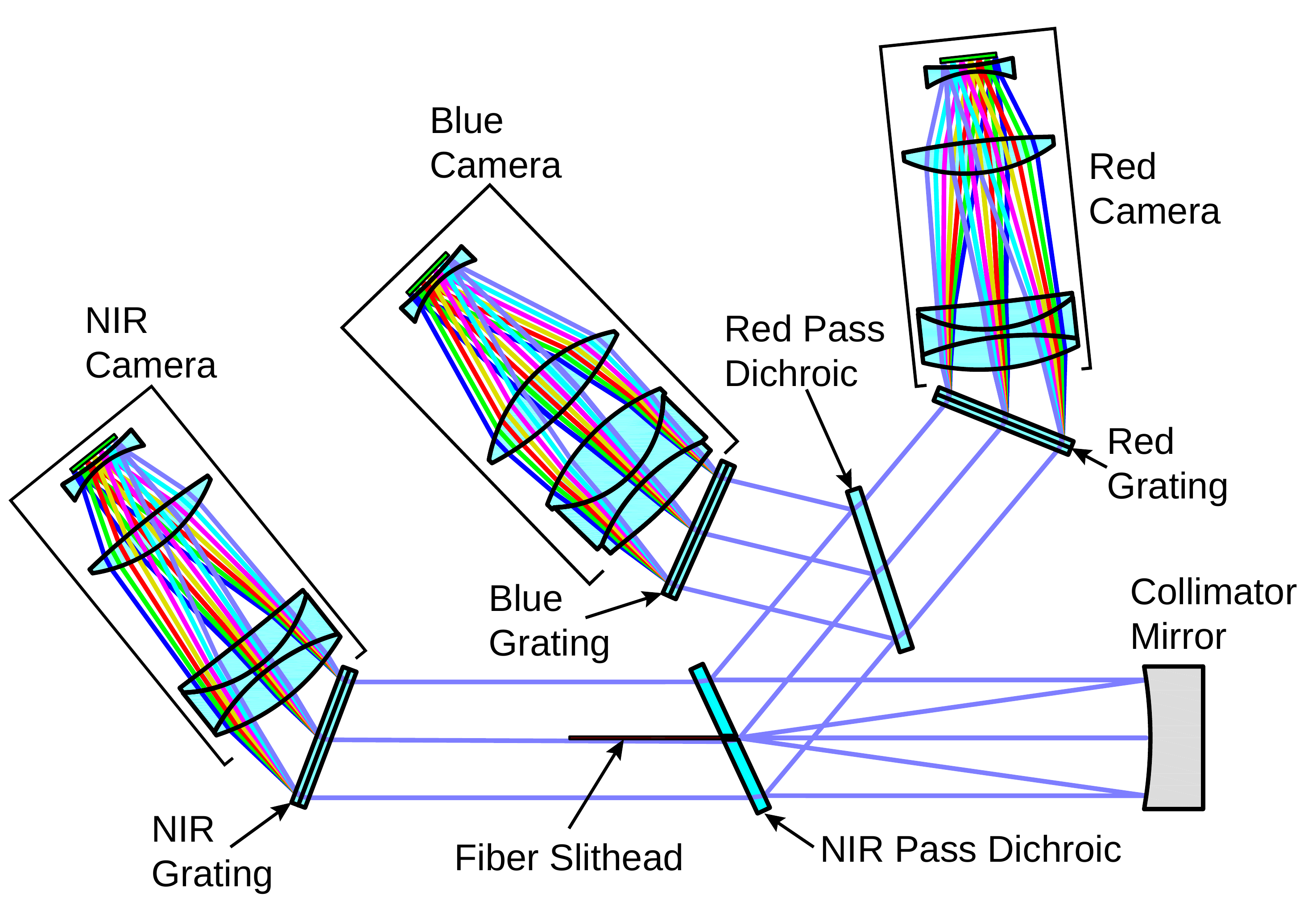}
\caption{The spectrograph schematic.}
\label{fig:specschematic}
\end{figure}

Since the spectral range is more than a factor of two, the spectrograph bandpass needs to be split into at least two channels, using dichroic filters, to eliminate second order contamination. A trade study was done to find the optimal number of channels (DESI-0756). Three channels had the highest throughput and could use standard 4096 x 4096 detectors. The schematic of the optical layout of the spectrograph is shown in Figure~\ref{fig:specschematic}. The light from the fiber slit is collimated by a spherical mirror. The light is then split into three spectral band passes using two dichroics. Each of the three channels disperses the light with a volume phase holographic (VPH) grating. The light is then focused onto the detector using a 5 lens camera.
  
The magnification of the spectrograph is the ratio of the camera to collimator focal ratios. The collimator accepts the f/3.57 light from the fibers. The fiber diameter is 107~\micron and it must project to a minimum of 45~\micron (3 pixels) on the detector. This gives the minimum spectrograph magnification of 0.42.  Given the slit height of 120.9~mm, which must map to less than 60.02 mm on the detector, the maximum spectrograph magnification is 0.496.

If the optics of the spectrograph have no aberrations, the projected fiber diameter on the detector is just $m\cdot{d}$, where $m$ is the magnification of the spectrograph and $d$ is the diameter of the fiber. The spectral resolution of an arm of the spectrograph can then be estimated from equation~\ref{eq:specres}.
\begin{equation}
R(\lambda) \approx \frac{2 \lambda D}{\sqrt{3} m d (\lambda_{max} - \lambda_{min})}.\label{eq:specres}
\end{equation}
$R$ is the FWHM spectral resolution, $\lambda$ is the wavelength, $D$ is the size of the detector, $\lambda_{max}$ is the maximum wavelength of the channel and $\lambda_{min}$ is the minimum wavelength of the channel. Since the diameter of the fiber and the detector width are fixed the only variables in the spectrograph design are the magnification and the minimum and maximum wavelengths of the channel. Initial spectrograph design parameters were derived using equation~\ref{eq:specres} and the spectral resolution requirements from Table~\ref{tab:spectro-reqs}. There is no unique solution; the design chosen has the maximum magnification possible that met the resolution requirements. This results in the slowest possible camera, which is easier to manufacture. The bands chosen that meet the requirements are: 360--593~nm, 566--772~nm, and 747--980~nm for the blue, red and near infrared (NIR) channels respectively. 

\subsubsection {Fiber Slit}
\label{par:fiber-slit} 

The slithead is very similar to the BOSS design \cite{Smee13} and consists of a thin, stiff plate with a curved edge of radius 468~mm having 500 fibers.
The 107 $\micron$ diameter fibers are mounted in groups of
twenty five in v-groove blocks, with 20 v-groove blocks being bonded to each slithead. The center-to-center spacing between fibers on adjacent v-groove blocks is 556 \micron, compared to 230 $\micron$ between fibers within a
v-groove block. The larger gap imaged on the CCDs is used to understand the PSF wings in the spectral extraction process (Section~\ref{sec:extract-sky}).  The total length of the arc is 120.9~mm
from outside edge to outside edge of the first and last
fibers, 18~mm shorter than the BOSS design. The slit is inserted into a slot milled into the first dichroic with the center fiber tangent to the reflective surface of the dichroic.

\subsubsection{Collimator}

\begin{figure}[!htb]
\centering
\includegraphics[height=2.25in]{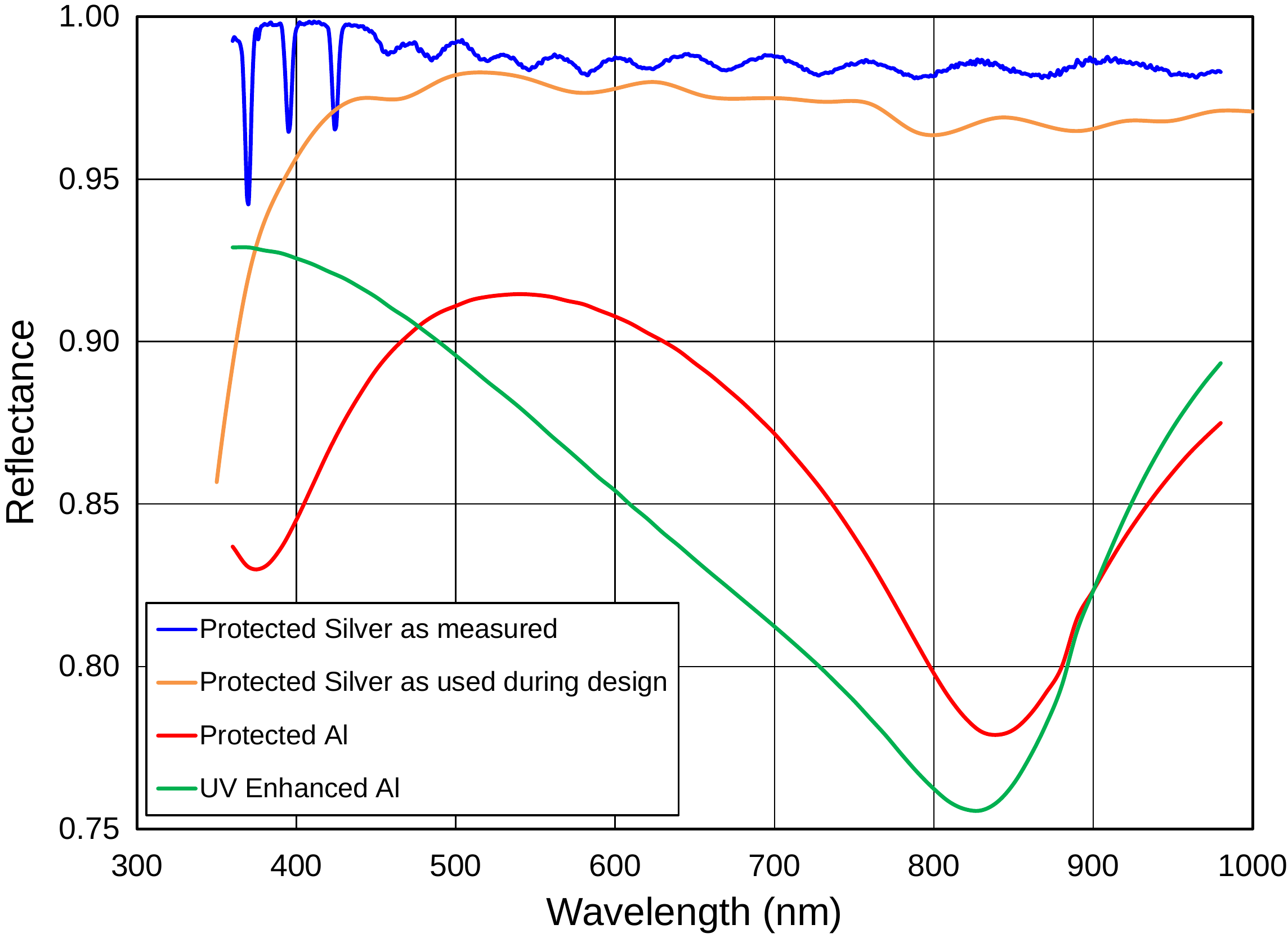}
\caption{Reflectance of silver versus aluminum. The orange curve is for silver has reported by BOSS and used during the spectrograph design. The blue curve is a DESI measurement from the first collimator.  Two types of Al are also shown as the red and green curves.}
\label{fig:agalref}
\end{figure}

The collimator can either use reflective or refractive optics. A trade was performed (DESI-0757) which showed that a reflective collimator was the best. It has fewer optical surfaces (only 1) and slightly higher throughput than a refractive collimator.  The spherical collimator mirror is similar to the BOSS design \cite{Smee13}. The collimator mirror is coated with an enhanced silver coating. A plot of the reflectivity of silver and aluminum is shown in Figure~\ref{fig:agalref}. The silver curve is from measurements of the BOSS collimator. The aluminum curves are from published vendor data. On average silver has about 10\% higher throughput than aluminum. The f/\# is 3.57 and the pupil is 126~mm.

\subsubsection{Dichroics}

Dichroic beam splitters are used to split the band pass into three. The dichroics can be placed between the fibers and the collimator or after the collimator. Dichroics in the diverging beam of the fibers will have a lower throughput than in the collimated beam. We have chosen to place them after the  collimator to increase the throughput of the spectrograph.  

\begin{figure}[!htb]
\centering
\includegraphics[height=2.25in]{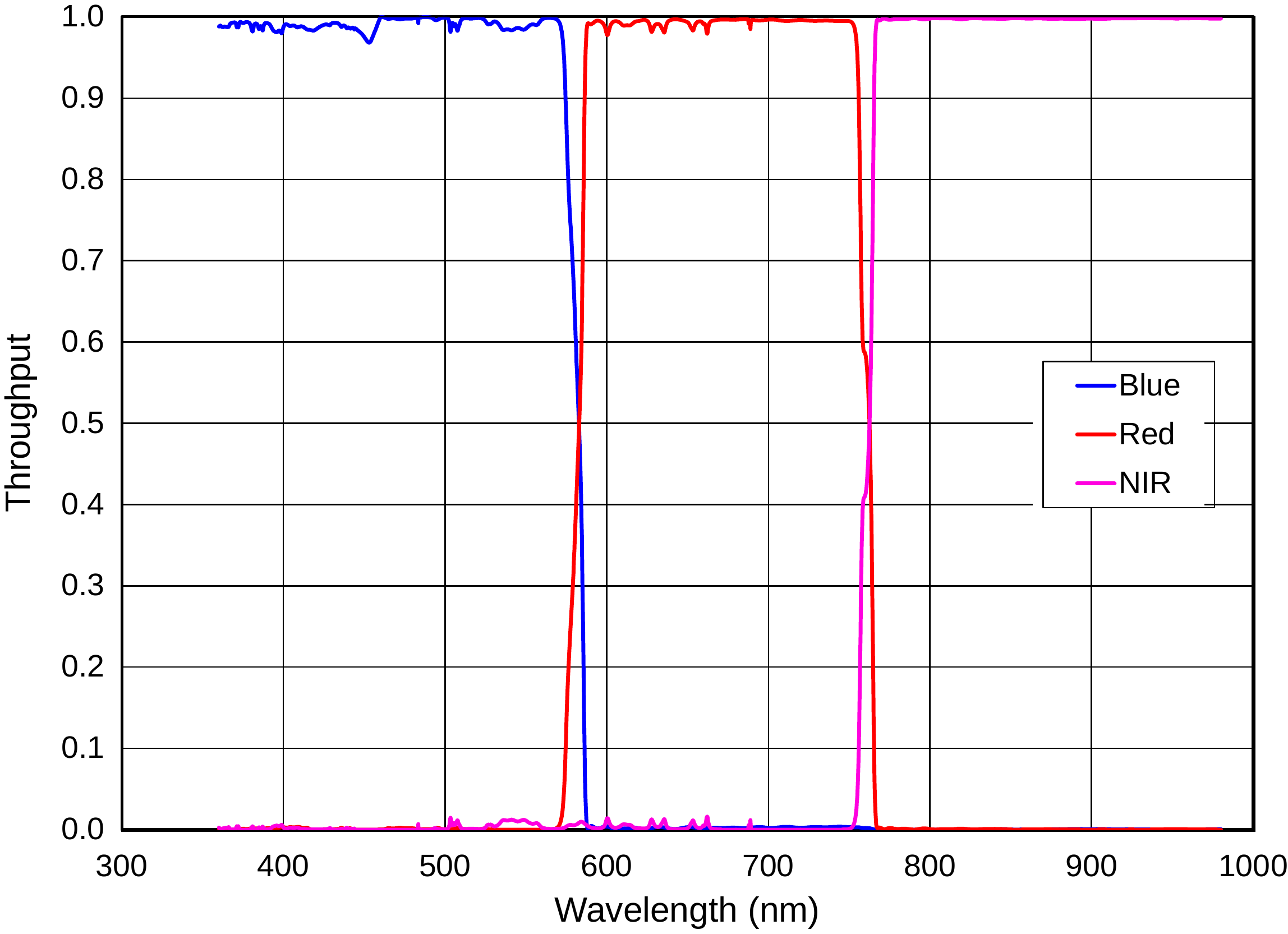}
\caption{The as measured  throughput for the first two DESI dichroics.}
\label{fig:dichroicthru}
\end{figure}

The dichroic beam splitters reflect the short band pass and transmit the longer band pass. They are fused silica optical flats coated with the dichroic on one side and an anti-reflection coating on the other. The specifications for the dichroic are shown in Table~\ref{tab:dichroic-reqs}. More detailed specifications are given in DESI-1058.  Figure~\ref{fig:dichroicthru} shows the measurements of the throughput for the first DESI dichroics.

\begin{table}[!htb]
\small
  \centering
  \caption{Dichroics Budgeted Requirements.}
    \begin{tabular}{lccl}
    \hline
         & NIR Pass Dichroic & Red Pass Dichroic & Throughput \\
    \hline
   Reflection Band (nm) &  360--747     & 360--566         & $>95\%$    \\
    Transmission Band (nm) &  772--980     & 593--747      & $>95\%$    \\
    Crossover Region R+T (nm) & 747--772  &  566--593 & $>90\%$ \\
   Crossover Width (nm) &  25     & 27         &   \\
    \hline
    \end{tabular}
  \label{tab:dichroic-reqs}
\end{table}

\subsubsection{Gratings}

The spectral resolution requirement, desired spectral overlap between the bands  and the detector define the three band passes. The band passes and grating parameters are given in Table~\ref{tab:grating-reqs}. Volume phase holographic (VPH) gratings have been chosen as the spectral dispersers. This type of grating delivers the best throughput over the relatively broad band passes of the three channels. The VPH grating is made by sandwiching a gelatin between two fused silica optical flat plates. Fringes are holographically recorded in the gelatin. $\Delta \beta$ is the dispersion angle, $\alpha$ is the incident angle, $1/\sigma$ is the line density, and $\phi$ is the tilt angle of the fringes. The fringes of the gratings are tilted to remove the Littrow ghost \cite{Burgh07}. Our studies showed that high efficiency is maintained at the band pass edges when the grating spectral dispersion $(\Delta \beta)$ was kept to less than 16\degree.  The clear aperture of the gratings are kept below 145~mm so smaller holographic recording tables can be used to reduce the cost.

\begin{table}[!htb]
\small
  \centering
  \caption{Spectrograph Grating Budgeted Requirements.}
    \begin{tabular}{lcccl}
    \hline
              & Blue Channel & Red Channel & NIR Channel & Driver\\
    \hline
    $\lambda_{min}$ (nm) &  360     & 566         & 747   & \\
    $\lambda_{max}$ (nm) &  593     & 772           & 980  &  \\
    $\Delta \beta~(\degree)$ & 15.74      & 15.5         & 15.42 & Throughput \\
   $\alpha~(\degree)$ &  10.44     & 18.12         & 20.76  & Littrow ghosts \\
   $1/\sigma~(l/mm)$ &   1101.9    & 1157.4     & 992.5 & Littrow ghosts \\
    $\phi~(\degree)$ &  -3.18     & -3.03          & -2.97     & Littrow ghosts\\
    Band Edge Efficiency & $>65\%$ & $>80\%$ & $>80\%$ & Throughput \\
    \hline
    \end{tabular}
  \label{tab:grating-reqs}
\end{table}

\begin{figure}[!ht]
\centering
\includegraphics[width=0.95\textwidth]{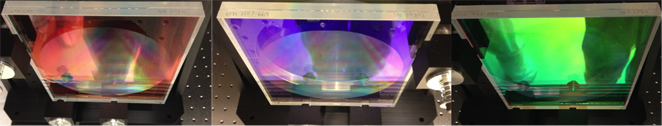}
\caption{Photographs of the blue, red and NIR gratings in test mounts, from left to right. 
}
\label{fig:gratphot_rgb}
\end{figure}

\begin{figure}[!ht]
\centering
\includegraphics[height=1.5in]{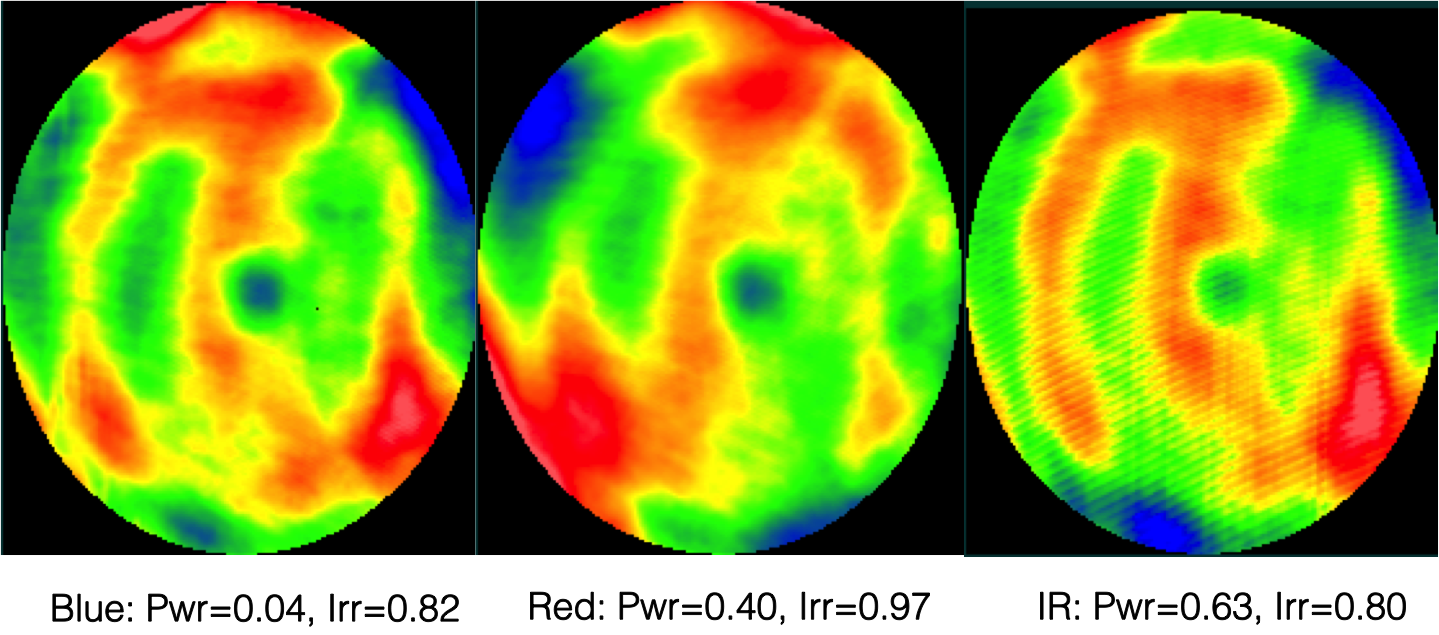}
\caption{Measured blue, red and NIR grating wavefront in -1 order, from left to right.
Power and Irregularity wavefront are indicated by Pwr and Irr, respectively.
}
\label{fig:Grating_Wavefronts}
\end{figure}

\begin{figure}[!ht]
\centering
\includegraphics[height=2.25in]{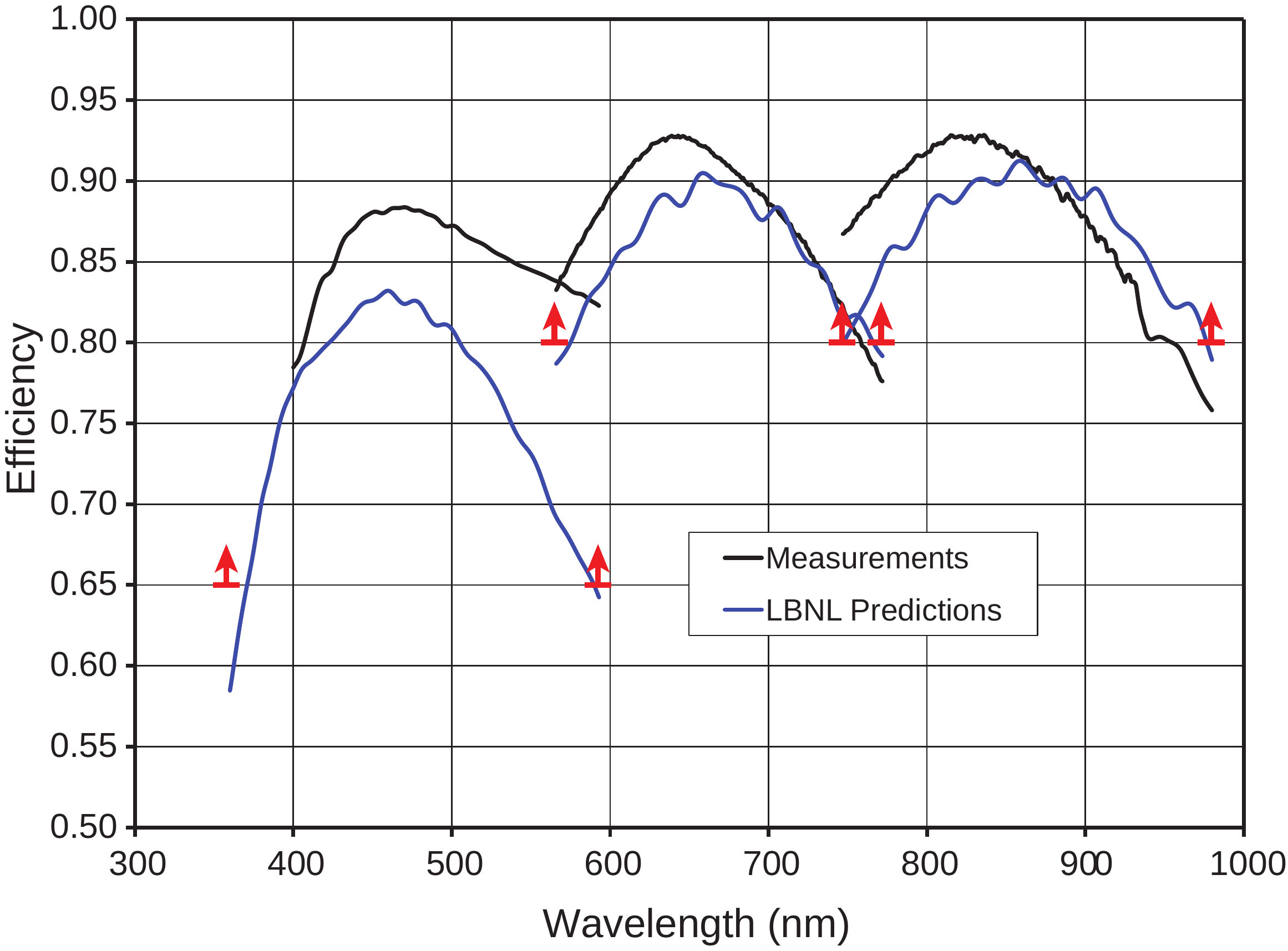}
\caption{The vendor measured efficiency of the gratings is shown in black. The predicted efficiency is shown in blue and the requirements at the band edges are shown in red. The vendor could not measure the efficiency of the blue grating below 400 nm.
}
\label{fig:grateff}
\end{figure}

\begin{figure}[!ht]
\centering
\includegraphics[height=2.25in]{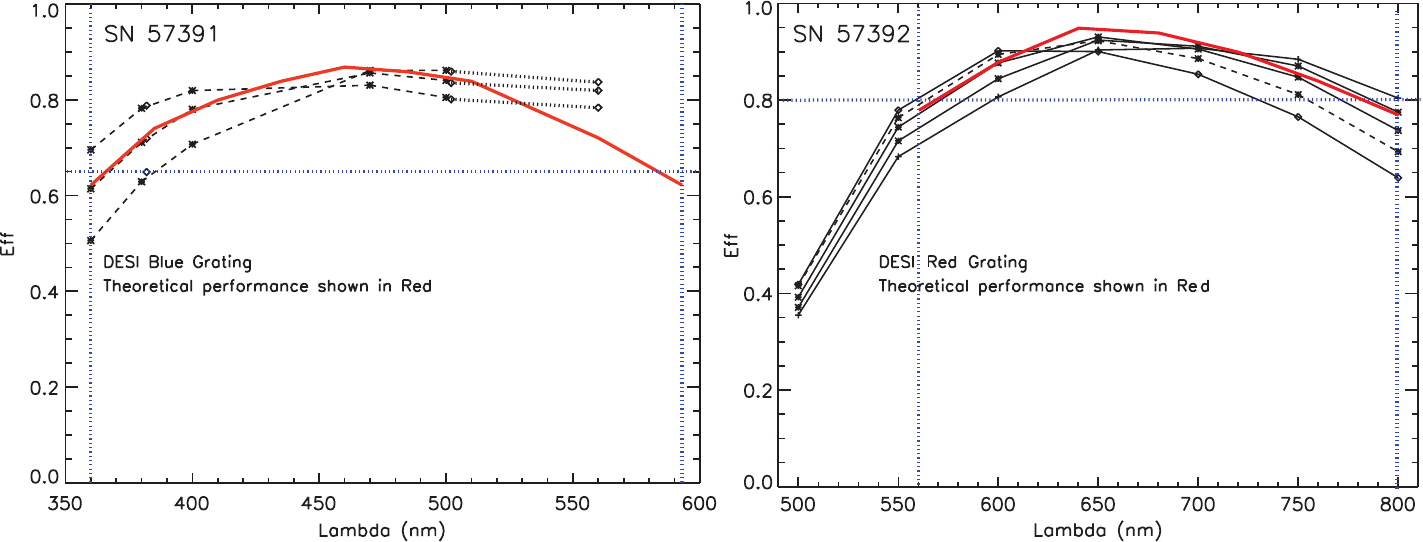}
\caption{The LBNL measured efficiency of the blue (left) and red (right) gratings are shown as a function of the incident angle, $\alpha$ on the left and right, respectively. The predicted efficiency is shown in red and the specification is in blue. Tilting the gratings incidence angle within $(<1\degree)$ of its specified value achieves efficiency optimization across the band, as desired. The vertical dashed lines are the bandpass edge locations and the horizontal dashed line are the required bandpass edge throughputs.}
\label{fig:lbnlgrateff}
\end{figure}

Full sized demonstration gratings for the three channels have been purchased and tested for their critical performance parameters of efficiency and wavefront error.  Figure~\ref{fig:gratphot_rgb} shows a photo of the three gratings in a test mount.  Figure~\ref{fig:Grating_Wavefronts} shows measurements of the wavefront error, from a Zygo interferometer. The peak to valley wavefront error for power and irregularity are below the budgeted 1.5 and 0.5 waves.
The efficiency of the gratings have been measured by the vendor and remeasured at LBNL including as a function of incident angle $\alpha$. The vendor results are shown in Figure~\ref{fig:grateff}. The LBNL results are shown in Figure~\ref{fig:lbnlgrateff}. The efficiency performance is good and performance optimization within $<1\degree$ of the specified incidence angle, ~$\alpha$, is achieved as required.

\subsubsection{Camera}

The cameras can use either reflective or refractive optics. The reflective Schmidt cameras use an aspheric corrector plate, spherical mirror and field flattener lenses. The detector obscures the beam with a loss of at least 10 - 15\%. Refractive cameras have no obscuration but require more optical surfaces. The BOSS camera \cite{Smee13} used 8 lenses, two with aspheric surfaces. The refractive design was chosen since it has higher throughput. See DESI-0757 for more details on the trade.

The focal ratio of the camera will determine the spectral resolution of the camera and the focal length will determine the spectral coverage. The optical requirements for the cameras are given in Table~\ref{tab:camera-reqs}. An area of the detector less than the full size was used to image the spectra in order to reduce the impact of edge effects and allow for alignment tolerances. The throughput requirements are budgeted from the requirements in Table~\ref{tab:spectro-reqs}.

\begin{table}[!ht]
\small
  \centering
  \caption{Camera Optical Budgeted Requirements.}
    \begin{tabular}{lccc}
    \hline
              & Blue & Red & NIR \\
    \hline
   Maximum rms spot radius (\micron) &  $<12$     & $<12$         & $<12$    \\
   Angular field (\degree) &  7.72     & 7.72           & 7.72    \\
   Spectral field (nm) & 360--593      & 566--772         & 747--980 \\
   Spectral field size (\degree) &  8.06     & 8.03     & 8.03 \\
   Focal length (mm) &  $212.8 \pm 1$   & $213.6 \pm 1$    & $214.0 \pm 1$ \\
   Focal Ratio & f/1.70 & f/1.70 & f/1.70 \\
   Spectral Detector Size (mm)	& 60.26 & 60.26 & 60.26 \\   
   Spatial Detector Size (mm)	& 60.02 & 60.02 & 60.02 \\
   \hline
     & $>86\%$    & $>92.5\%$           & $>92.4\%$   \\
     & @ $\lambda = 360$ nm & @ $\lambda = 560$ nm & @ $\lambda = 740$ nm \\
    \cline{2-4}
    Throughput & $>90\%$ & $>92\%$ & $>91\%$     \\    
     & @ $\lambda = 600$ nm  & @ $\lambda = 650$ nm    & @ $\lambda = 980$ nm \\
    \cline{2-4}
     &        & $>93\%$ &      \\
       &        & @ $\lambda = 780$ nm          &      \\
    \hline
    \end{tabular}
  \label{tab:camera-reqs}
\end{table}

The design considerations to reduce cost were: minimize the number of lenses, minimize the number of aspheric surfaces, minimize the volume of glass, use low cost glass with frequent melts. To maximize the throughput the number of lenses was minimized and only glasses with high transmission in the band pass were used. The last lens is fused silica since it is the vacuum seal to the cryostat. Fused silica is very strong and stiff and it is not radioactive which would increase the backgrounds in the detector.

The camera design was optimized using Zemax$^\circledR$.  The optical design was only allowed to have a maximum rms radius of $<$10 \micron to allow room for manufacturing and alignment tolerances. Physical constraints on the design were: window to CCD $>3$~mm, lenses 5~mm larger in radius than clear aperture for mounting, center thickness $>10$~mm, edge thickness $>5$~mm, total lens thickness $<80$~mm (but as small as possible), contact surfaces 0.25~mm center thickness, non contact surfaces $>5$~mm distance. The design started with variations on the BOSS design \cite{Smee13} which has 8 lenses, 2 aspheric surfaces and 7 glass types. An exhaustive study by three optical engineers on glass choice and design choice was done. The final design is a field flattened Petzval design with a contact triplet. All of the channels have 5 lenses with 2 aspheric surfaces (the front of the first lens and the front of the last lens).  The optical optimization requires that the detector cryostat window and CCD are tilted and offset from the optical axis.  The blue camera uses three glass types including CaF$_2$. The red and NIR cameras have 4 standard glass types. A schematic of the cameras is shown in Figure~\ref{fig:specschematic}.

\begin{figure}[!ht]
\centering
\includegraphics[width=5in]{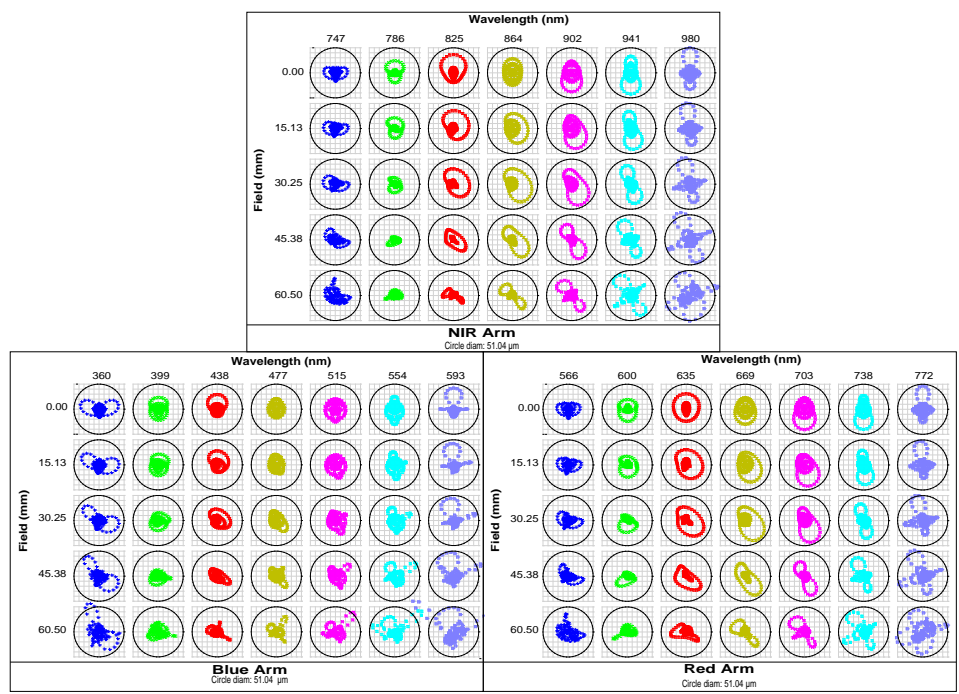}
\caption{The ray traced spot diagram for the spectrograph. The lower left figure is for the blue channel, the lower right figure is for the red channel and the upper figure is for the NIR channel. For each figure different fiber locations are shown from top to bottom (from 0.0 to 60.5 mm) and different wavelengths are shown from left to right. The circle is the size of the projected fiber onto the detector.}
\label{fig:spotdiagram}
\end{figure}

\begin{figure}[!ht]
\centering
\includegraphics[height=2.5in]{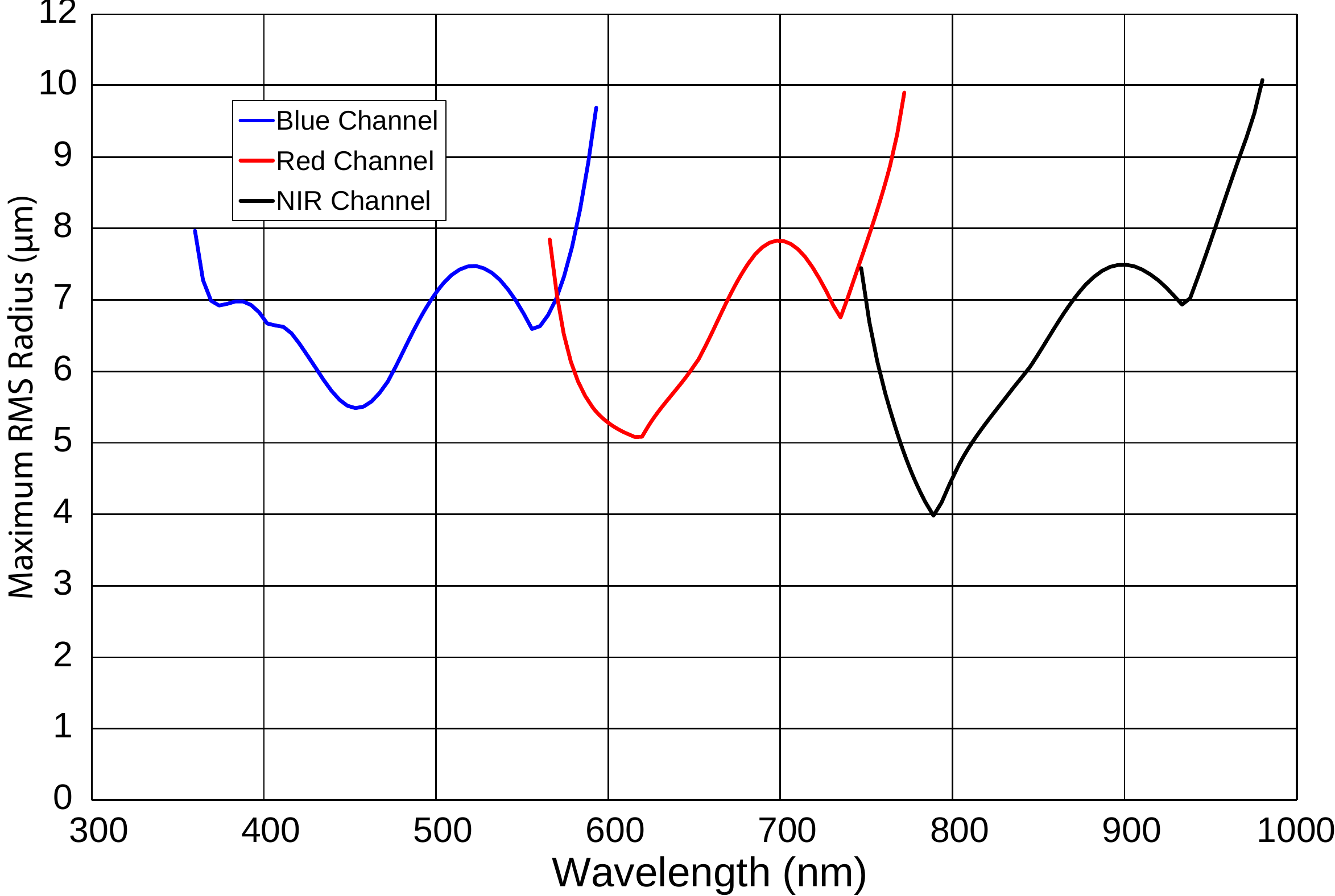}
\caption{The maximum rms radius versus wavelength for the optical design is plotted. The blue, red and black curves are for the blue, red and NIR channels of the spectrograph respectively. All wavelengths meet the $10~\micron$ rms requirement for the optical design.}
\label{fig:maxrmsradius}
\end{figure}

The optical design of the spectrograph was analyzed using Zemax$^\circledR$ to make sure that it meets requirements. The spot sizes for the spectrograph are shown in Figure~\ref{fig:spotdiagram}. The maximum rms radius versus wavelength is in Figure~\ref{fig:maxrmsradius}. The design meets the requirement of $<$10 \micron rms at all wavelengths.

The red and NIR triplets are bonded with UV curing glue. An FEA analysis of the triplet was done to determine the stress and strain on the glass and the glue.  Then a smaller doublet was made that would have twice the stress of the full size triplet. The doublet was manufactured and bonded. It was then put through three thermal cycles for the DESI survival temperature range of -20 to +40 $\degree$C (DESI-1107). The shape error was measured and a visual inspection of the lens were performed both before and after the thermal cycle. The doublet lens passed the thermal cycle test.

The blue triplet was initially going to be bonded with Silicon RTV. A full scale triplet was made and bonded. It failed the three cycle survival temperature test. Then the triplet was redesigned to use Cargille 1074 laser liquid to couple the lenses. The oil cell is very similar to the design used by the KOSMOS spectrograph \cite{KOSMOSoil}. The oil cell passed its three cycle survival temperature test. In addition, the 1074 oil was exposed to the silicone o-rings and diaphragm for 30 months and the transmission was remeasured. The transmission of the oil did not change as a result of exposure to the materials.

\begin{figure}[!ht]
\centering
\includegraphics[height=2.5in]{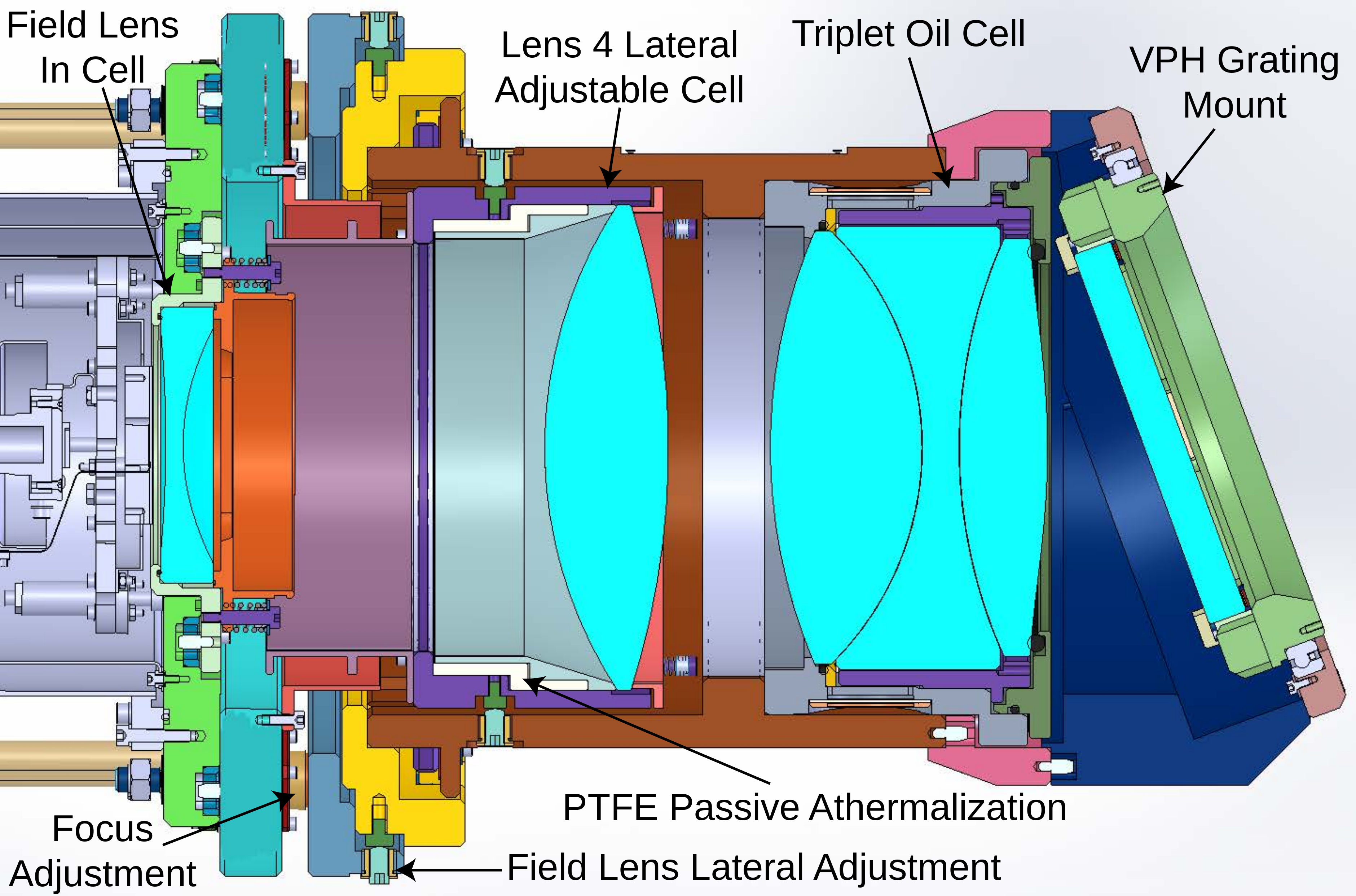}
\caption{Section of the mechanical design of the blue camera.}
\label{fig:bluesec}
\end{figure}

The mechanical design of the blue camera is shown in  Figure~\ref{fig:bluesec}. The VPH grating mount is shown in dark blue on the right. The triplet oil cell is shown next in gray and purple. The lateral adjustable cell for the fourth lens is shown next in pink and purple. Finally the field lens cell is shown in light green with a orange baffle. All of the lens cells are thermally compensated axially and radially.

An optical thermal analysis of the cameras was done. For the operation range 18--22$\celsius$ the red and the NIR cameras did not require passive athermalization. The blue camera did need passive athermalization. This was done by inserting PTFE plastic in the cell of the fourth lens. See the white piece in the cell in Figure~\ref{fig:bluesec}.

\begin{figure}[!ht]
\centering
\includegraphics[width=0.9\textwidth]{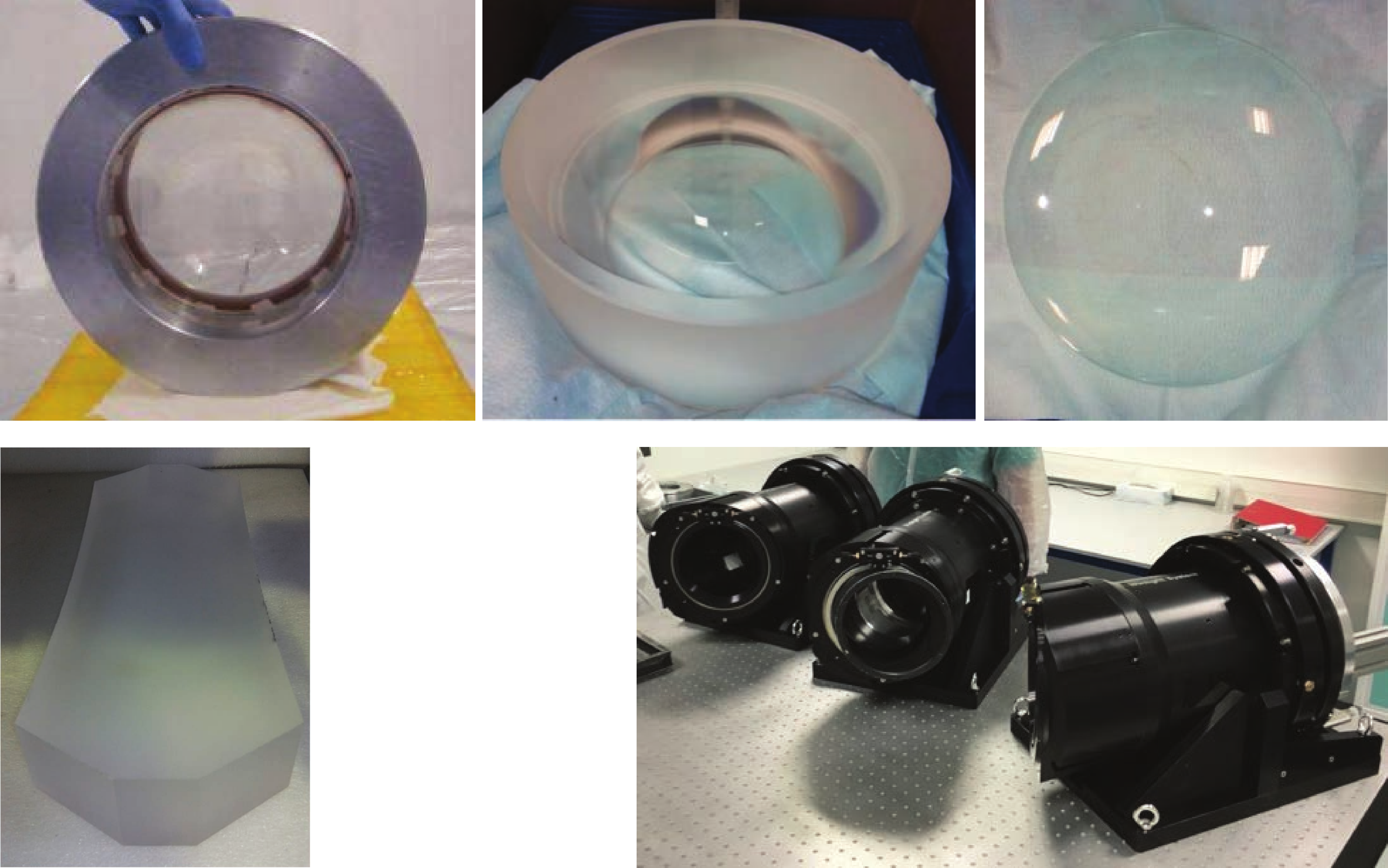}
\caption{Photographs of several prototype spectrograph components. : The test oil cell for the blue camera. C: 200 mm diameter L2 lens for the red camera. D: 180 mm diameter L1 lens for the red camera. E: Various camera barrel components. F: The three lens barrels for the red, NIR and blue cameras.}
\label{fig:specphot}
\end{figure}

The prototype spectrograph fabrication has started. Photographs of several of the parts under manufacture are shown in Figure~\ref{fig:specphot}. The collimator mirror, L1 and L2 lenses for the red camera, test oil cell for the blue camera, lens barrel parts and the three lens barrels are shown.
 
\subsubsection{Tolerancing and Simulation}

There are several effects that will change the imaging of the spectrograph from the optical design. The optics will not be manufactured or assembled perfectly. The illumination pattern from the fiber is not uniform. The detector will blur the images due to charge diffusion and depth of conversion. And diffraction will  add wings to the images. These effects were simulated to see if the as manufactured spectrograph (including detector) will meet the imaging requirements of the spectrograph.  

A preliminary tolerance study of the spectrographs was done to determine the as built performance of the spectrograph. The typical allowed tolerances for the cameras were 0.3\% in the radius of curvature, 0.2~mm for thickness, 0.05~mm for the decenter of the lenses, 1 arcminute for the tilts, 1 wave for transmitted power optical path difference (OPD), 1/2 wave transmitted irregular OPD. A few of the lenses for the red and NIR channels had to have slightly tighter requirements than listed above to meet the maximum rms radius of 12~\micron. Two of the radii of curvature in the blue channel have significantly tighter tolerances to meet the requirements. The detector tolerances were 0.015~mm P-V flatness, 0.125~mm piston and 2 arcminute tilt. The collimator mirror had a radius of curvature tolerance of $\pm$6~mm, an irregularity of 1/10 wave OPD a tilt of 1 arminute and piston of 0.125~mm for piston. Both dichroics had a power OPD of 1.5 wave and an irregularity OPD of 1 wave, 2 arcminutes of tilt and 0.125~mm of piston. The gratings had a power OPD of 1.5 waves, 1 wave of irregularity OPD, 2 arcminutes of tilt and 0.125~mm of piston. The compensator for the red and NIR channels was the distance and decenter of the window/detector assembly with respect to the other 4 lenses. The blue channel also added the spacing and decenter of the fourth lens with respect to the triplet as compensators.    

\begin{figure}[!t]
\centering
\includegraphics[height=2.5in]{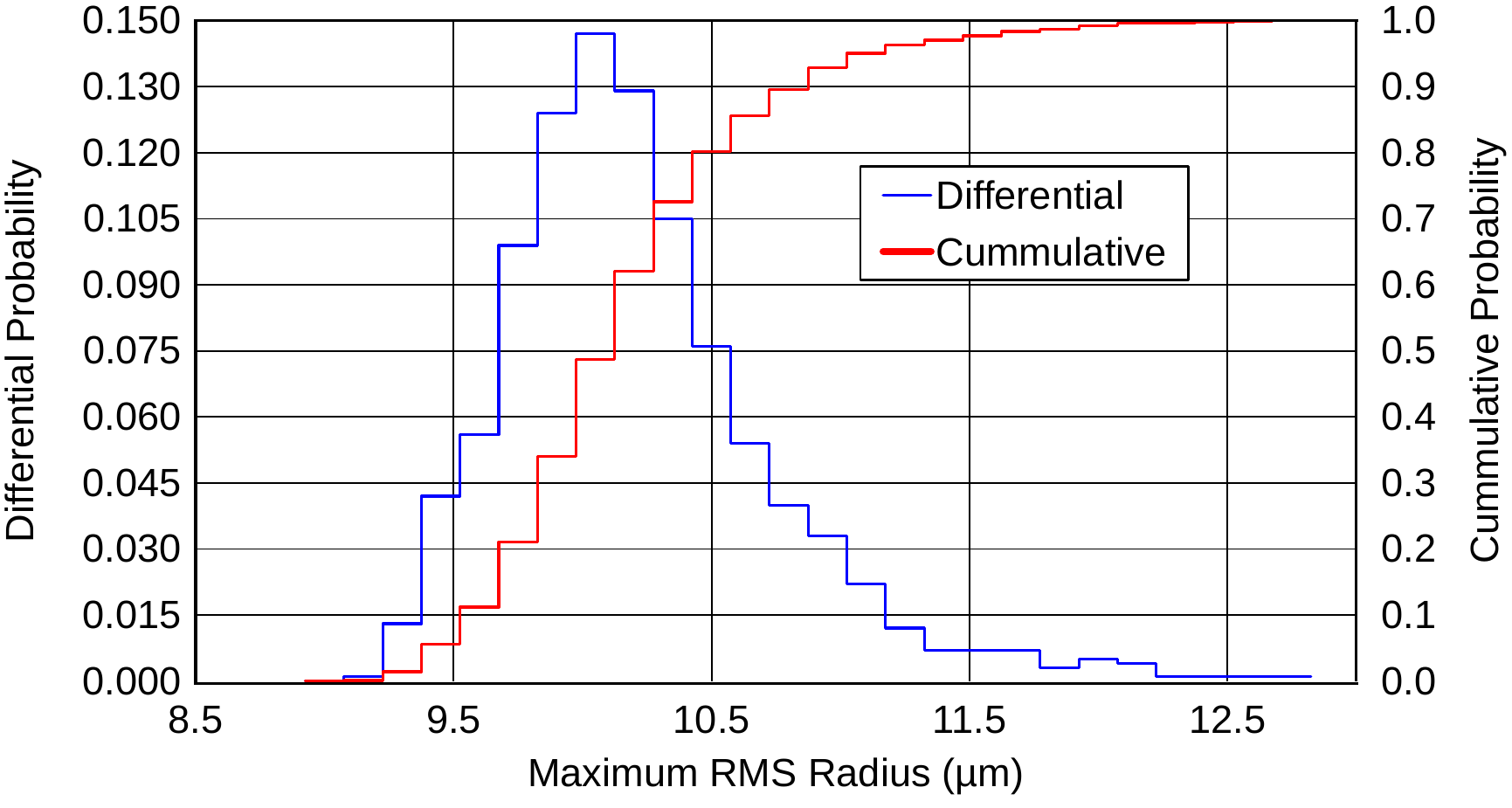}
\caption{Statistics from the Monte Carlo simulations for the tolerancing of the red channel of the spectrograph. The blue curve is the differential probability distribution and the red is the cumulative distribution.}
\label{fig:specmonte}
\end{figure}

A Monte Carlo simulation of 1,000 spectrographs was performed to determine if the tolerances meet the requirements. The results for the red channel are shown in Figure~\ref{fig:specmonte}. All the channels were similar to this.  The most likely spectrograph had a maximum rms radius of $\sim$10~\micron and 95\% of the spectrographs were better than 11~\micron. A representative description of the as built optics was chosen from the 1,000 Monte Carlos such that there is a 95\% chance that the as built camera would be better.  

A measured angular illumination pattern from a fiber was used as the input into the simulated as-built optics. Diffraction was then added to the output. This combined image was then input into a detector simulator to add the charge diffusion and depth of conversion effects. Two detector types were used in the simulation. For the blue channel a thin substrate detector with $5~\micron$ diffusion was used. For the red and NIR channels a $250~\micron$ thick fully depleted model was used with a $5~\micron$ diffusion. 

\begin{figure}[!h]
\centering
\includegraphics[height=2.3in]{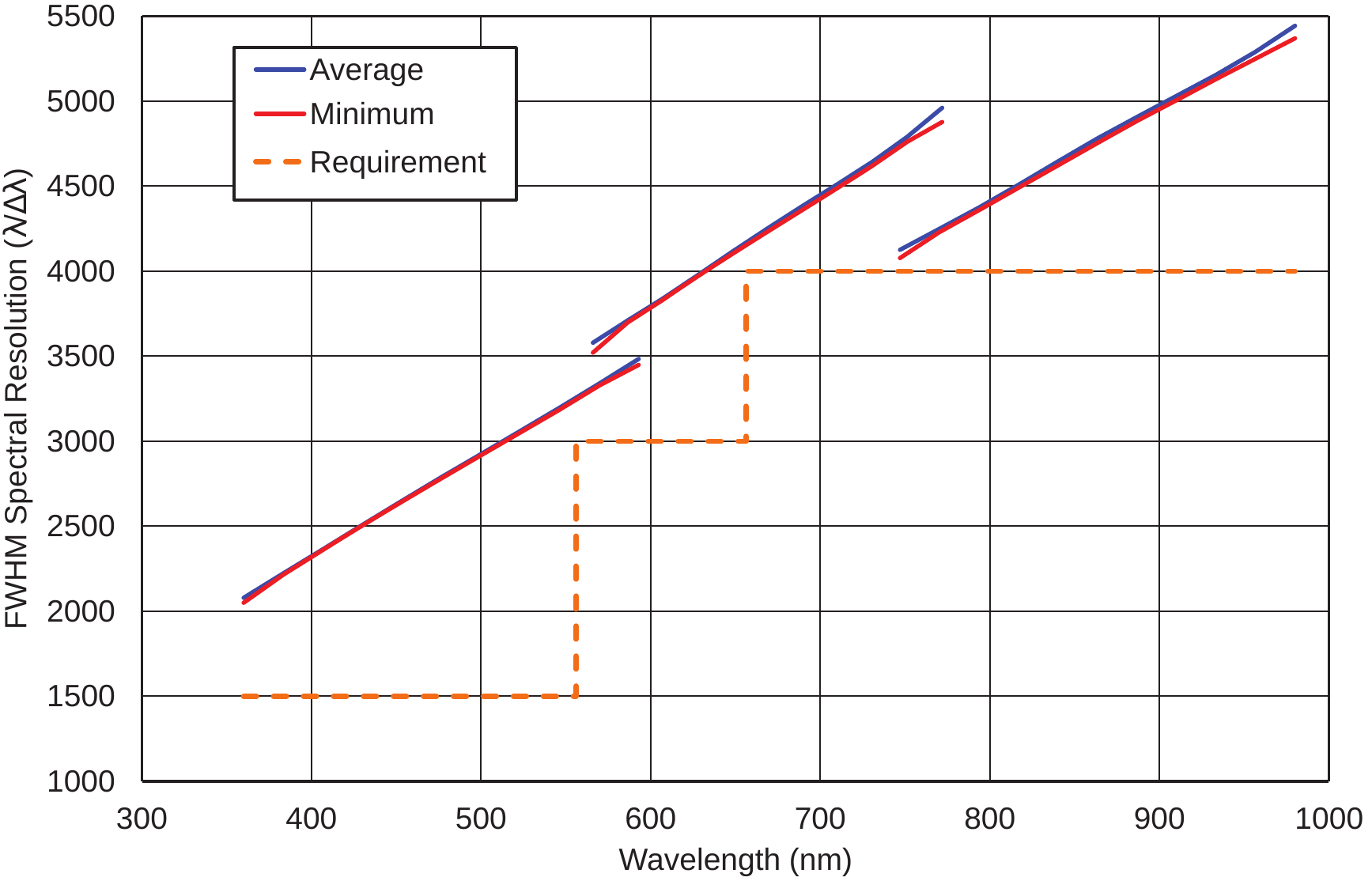}
\caption{The minimum and average FWHM  spectral resolution versus wavelength for the simulated spectrograph. The red curve is the minimum spectral resolution, the blue curve is the average resolution and the dashed orange curve is the requirement. }
\label{fig:spectralres}
\end{figure}

The FWHM spectral resolution($\lambda/\Delta\lambda$) for the simulated spectrograph is shown in Figure~\ref{fig:spectralres}. The red curve is the minimum spectral resolution for all fibers and the blue curve is the average. The requirement of Table~\ref{tab:spectro-reqs} is plotted as the dashed orange line. The spectral resolution requirement is met at all wavelengths. 

\begin{figure}[!h]
\centering
\includegraphics[height=2.4in]{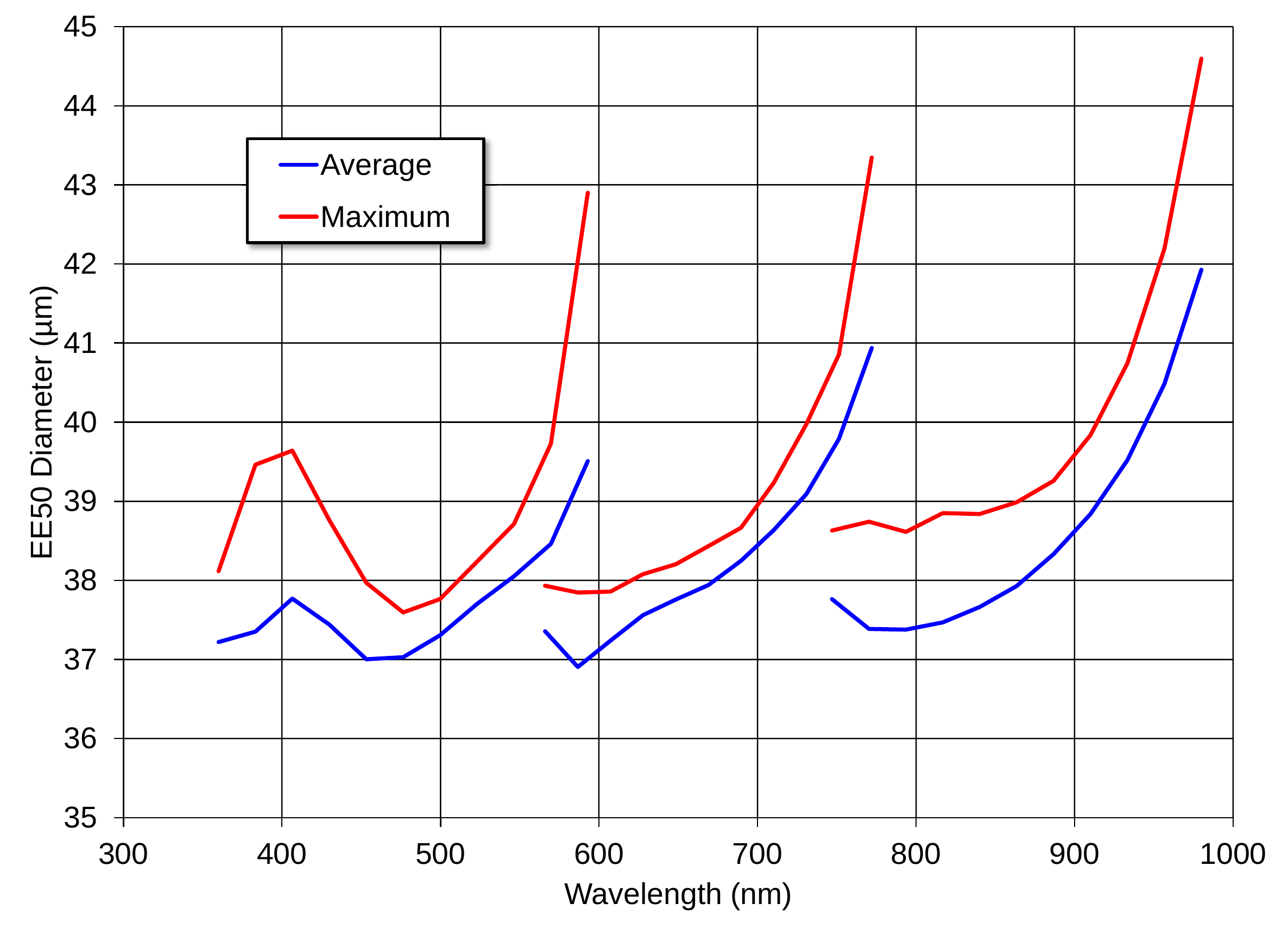}
\caption{The average (blue) and maximum (red) diameters of the $50\%$ encircled energy versus wavelength for the simulated spectrograph.  The requirement is $<50 \micron$.}
\label{fig:ee50diam}
\end{figure}

\begin{figure}[!h]
\centering
\includegraphics[height=2.4in]{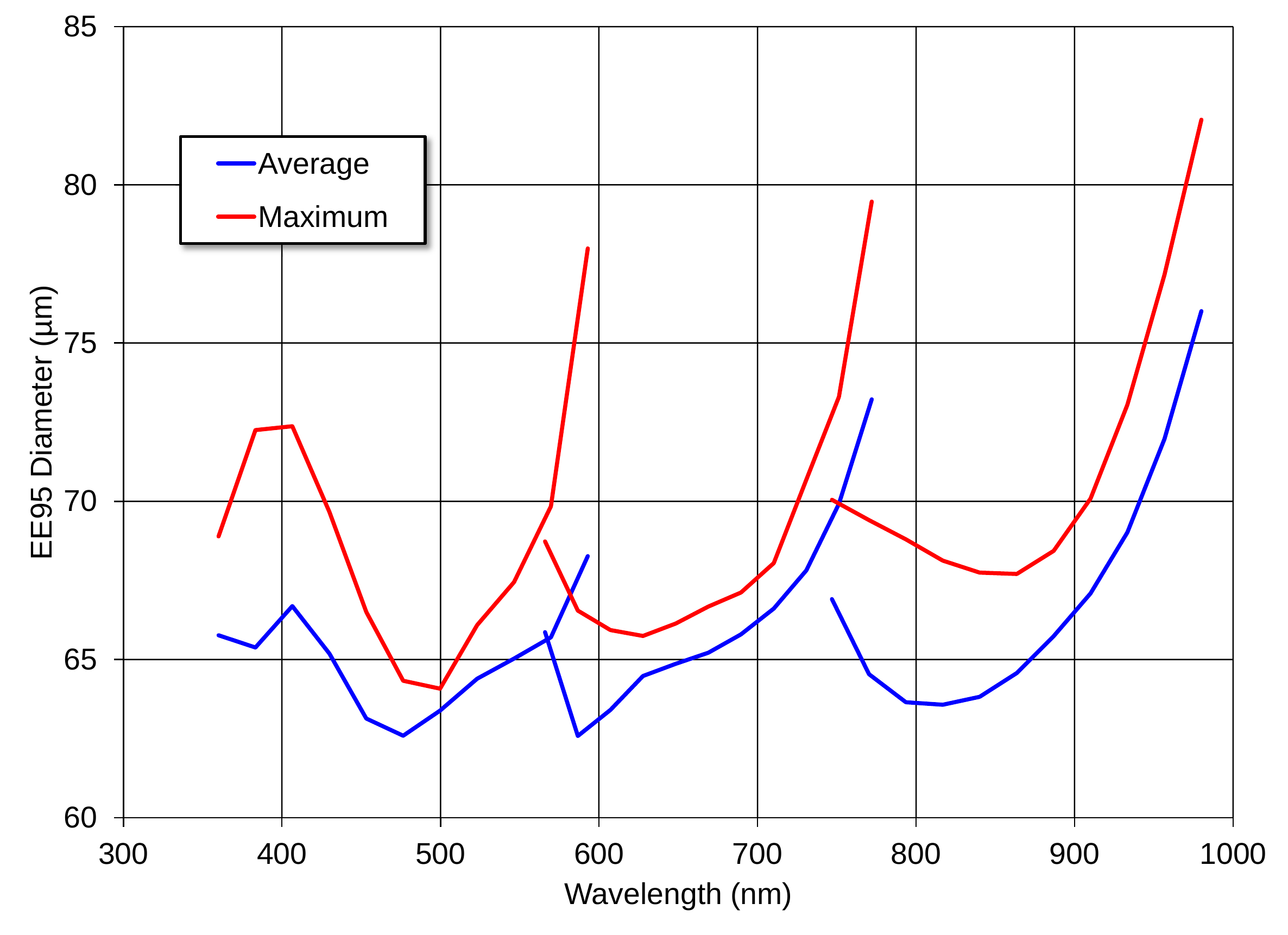}
\caption{The maximum diameter of the $95\%$ encircled energy versus wavelength for the simulated spectrograph.  The requirement is $<110 \micron$.}
\label{fig:ee95diam}
\end{figure}

The $50\%$ encircled energy diameter for the simulated spectrograph is shown in Figure~\ref{fig:ee50diam}. The red curve is the maximum for all fibers and the blue curve is the average. The simulated spectrograph meets the requirement of $<50~\micron$ from Table~\ref{tab:spectro-reqs} for all wavelengths and all channels. 

The $95\%$ encircled energy diameter for the simulated spectrograph is shown in Figure~\ref{fig:ee95diam}. The red curve is the maximum for all fibers and the blue curve is the average. The simulated spectrograph meets the requirement of $<110~\micron$ from Table~\ref{tab:spectro-reqs} for all wavelengths and all channels. 

\begin{figure}[tb]
\centering
\includegraphics[height=2.5in]{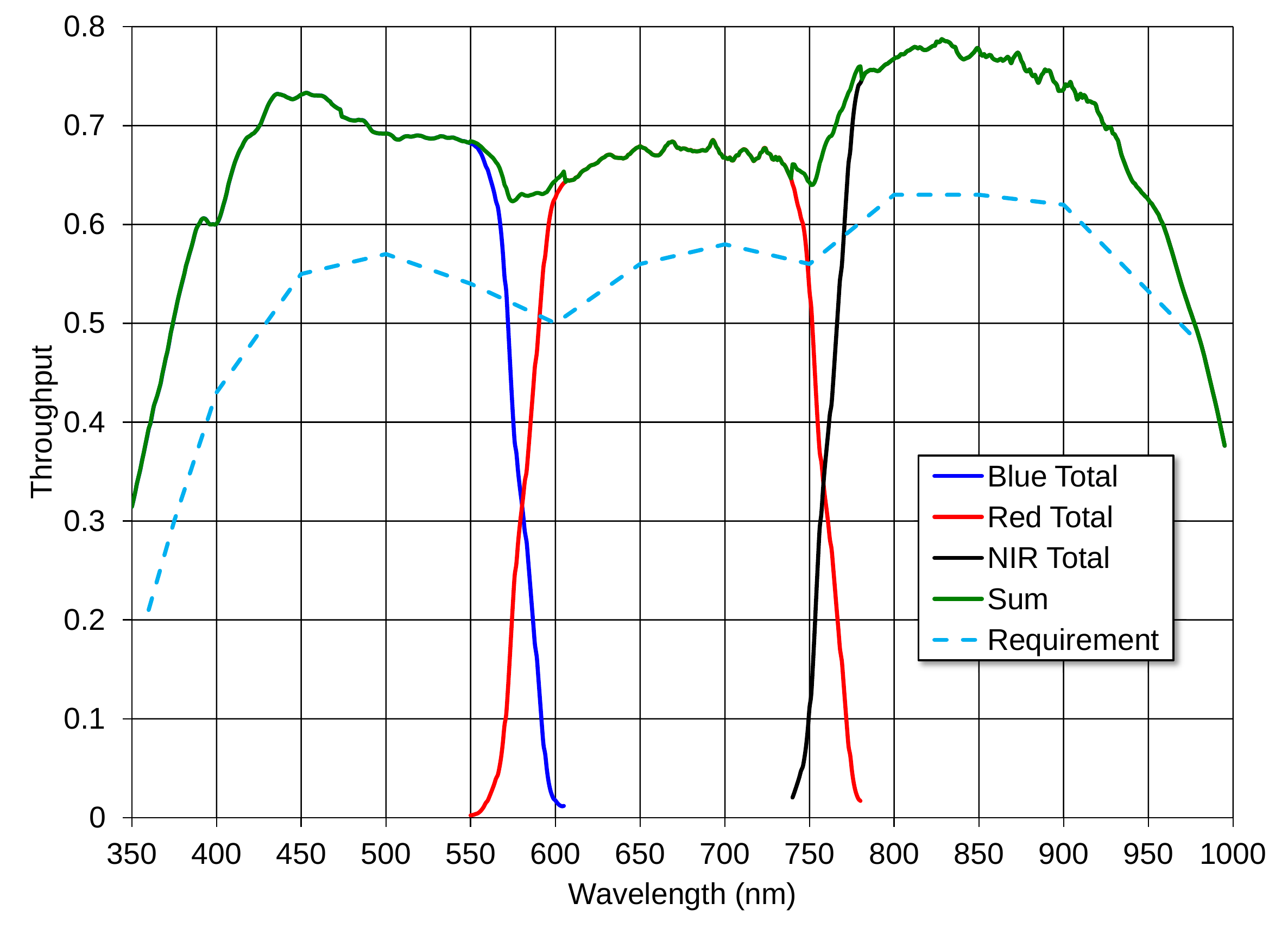}
\caption{The estimated throughput of the spectrograph. The blue, red and black curves are for the blue, red and NIR channels respectively. The green curve is the sum of the channel throughputs and the dashed blue curve is the requirement.}
\label{fig:specthroughput}
\end{figure}

The throughput of the spectrograph, from the fiber to the CCD, was also simulated. The transmission of the cameras were estimated using the published vendor data and Zemax. The 
antireflection coatings on the lenses were assumed to have a transmission of 99\% in blue channel and 99.5\% in the red and NIR channels. The collimator mirror reflectance is from the BOSS collimator measurements. The
dichroic performance was estimated by scaling measurements of the BOSS dichroic beam splitters. Grating efficiencies were the measurements from the vendor. The quantum efficiency of the CCD detectors was estimated using measurements of current detectors. The estimated throughput is shown in Figure~\ref{fig:specthroughput}. The blue, red and black curves are for the blue, red and NIR channels respectively. The green curve is the sum of the channel throughputs and the dashed blue curve is the requirement from Table~\ref{tab:spectro-reqs}. The throughput meets the requirement for all wavelengths. 

\subsubsection{Stray Light}

Two prelimiary ghost studies have been completed on the spectrograph. One by LBL (DESI-0372) and one by Winlight (DESI-1112). The results show that all ghosts are compliant with the specification of an irradiance of $< 10^{-5}$ of the incident irradiance. In addition, a preliminary stray light analysis was done (DESI-373). This study was used to define the surface roughness and particulate requirements for the optical surfaces. We are currently working with Photon Engineering (Tucson AZ) for a complete ghost and stray light analysis of the spectrograph.

\subsection{Spectrograph Mechanical and Electrical Design}
\label{sec:spectro_mech_elec_design}

\begin{figure}[!hbt]
\centering
\includegraphics[height=2.7in]{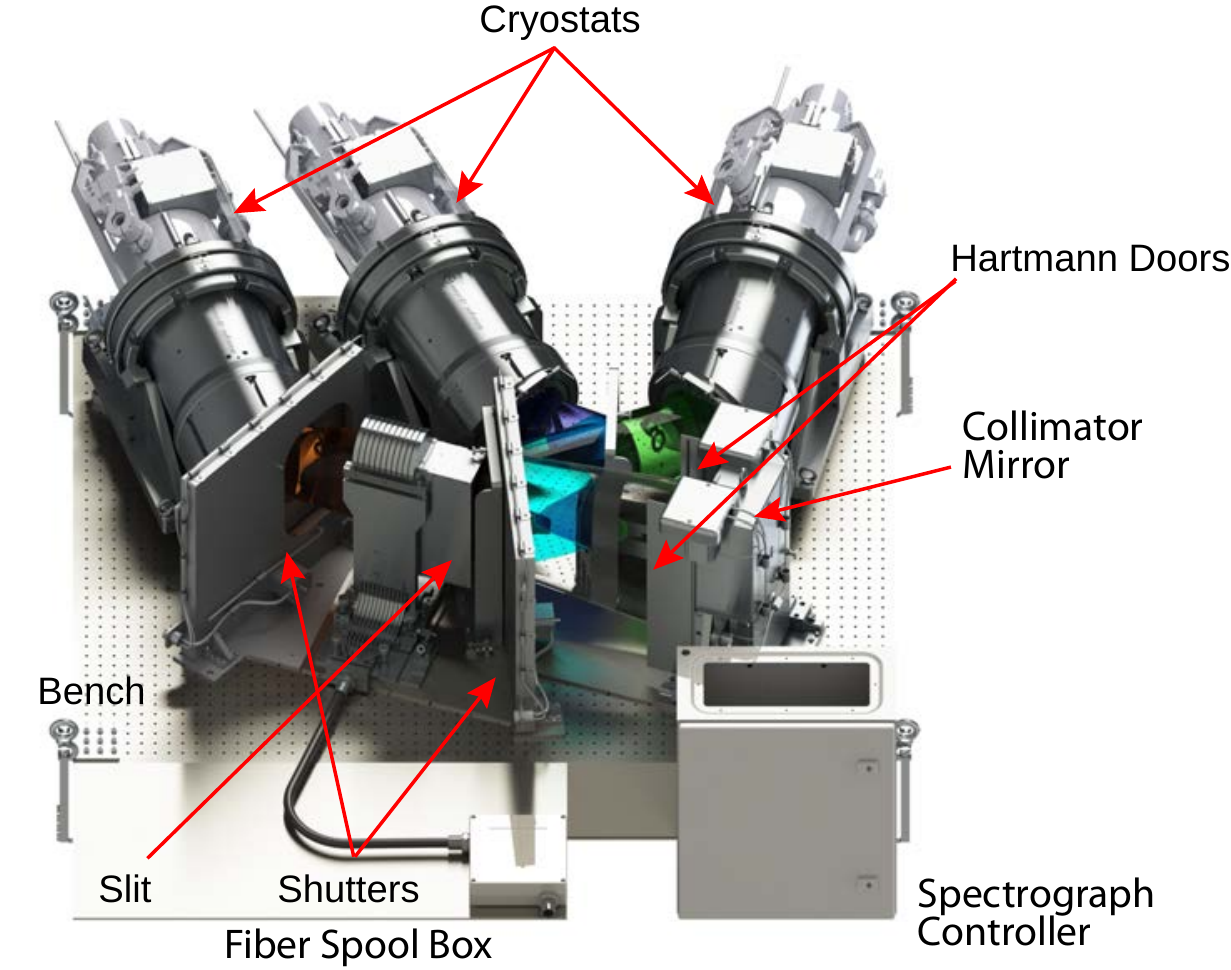}
\caption{A CAD rendering of the spectrograph. The colored cylinders are the light beams. The fiber illuminator, CCD flat field illuminator, thermal sensors and light enclosure are not shown. The spectrograph is 1.8 m wide~$\times$~1.4 m deep~$\times$~0.6 m high.}
\label{fig:specrender}
\end{figure}

A CAD rendering showing the major mechanical components of the spectrograph is shown in Figure~\ref{fig:specrender}. The major mechanical components of the spectrograph are the bench, shutters, Hartmann doors, fiber illuminator, CCD flat field illuminator, and thermal sensors.

\begin{figure}[!htb]
\centering
\includegraphics[height=3.5in]{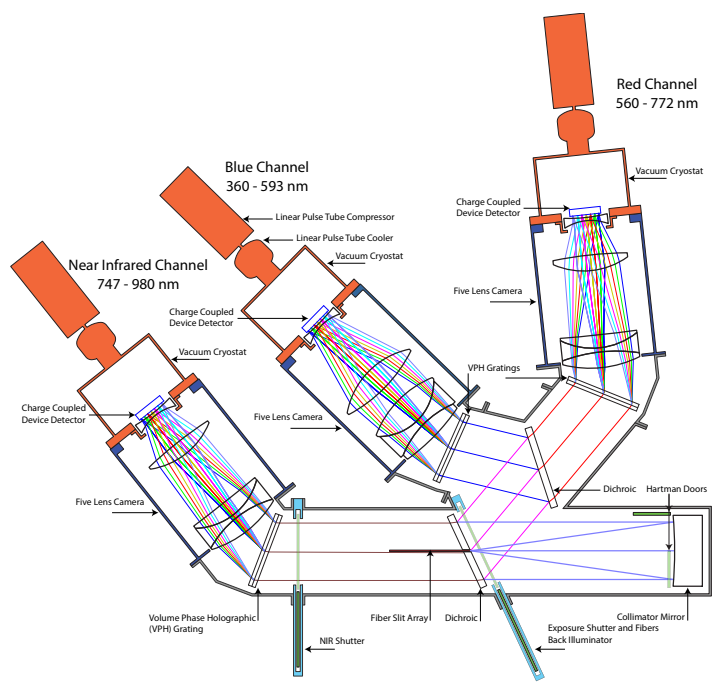}
\caption{Schematic of the mechanisms of the spectrograph.}
\label{fig:mechschem}
\end{figure}

The optical bench is an aluminum structure with precision machined interfaces to locate the seven opto-mechanical subassemblies: the fiber slit head, the collimator, the two dichroics, and the three camera/grating assemblies. At all of these opto-mechanical interfaces precision-machined reference datums and kinematic-locating features are employed to ensure accurate, repeatable alignment of the optical system. The optical components have passive thermal compensators in them to maintain the focus over the 18--22~\celsius~operational temperature range. There is an enclosure that also mounts to the bench. It blocks external light and is semi-hermetic so that a purge of dry air will prevent condensation on the cold cryostat window. A schematic of the mechanisms for the spectrograph is shown in Figure~\ref{fig:mechschem}.

\begin{figure}[!t]
\centering
\includegraphics[height=2.5in]{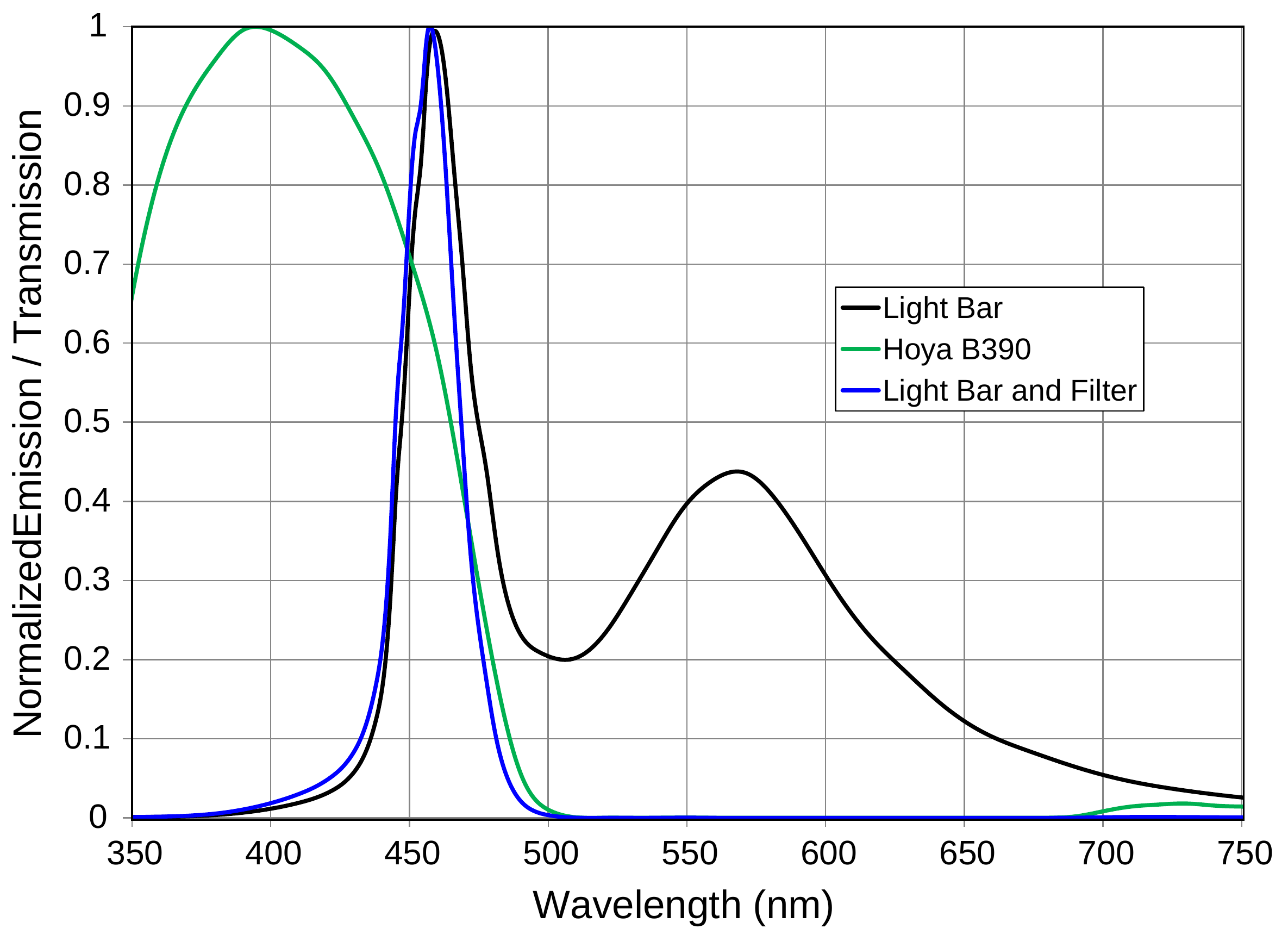} 
\caption{Spectral energy distribution of the Linear Chip On Board LED ({\it black}), transmission of the Hoya B390 filter ({\it green}), and output spectrum 
({\it blue}).} 
\label{fig:IllumSED}
\end{figure}

The fiber illuminator is an LED light source that is moved in front of the fiber slit head when it is in operation. It is used to back illuminate the fiber tips at the corrector focal plane so the Fiber View Camera (Section~\ref{sec:Instr_FVC}) can measure the location of the fibers. The fiber illuminator should produce at least $10^{10}$ photons/second into the $f/2.2$ acceptance cones of each fiber (DESI-0652, -0939). These photons should be in the 400--557~nm bandpass to be detectable by the fiber view camera, yet not interfere with the guiding CCD. A 3-Watt Linear Chip On Board LED, combined with a 3-mm thick Hoya B-390 filter, produces sufficient photon flux in the required bandpass (Figure~\ref{fig:IllumSED}). The LED will be mounted in a machined aluminum cell, covered with a thin film diffuser and the filter, and placed behind a narrow slit that will primarily direct light into the fiber ends. The fiber illuminator is mounted to the back of the exposure shutter and  positions the LEDs close to the fiber slit when closed.

\begin{figure}[!b]
\centering
\includegraphics[width=0.9\textwidth]{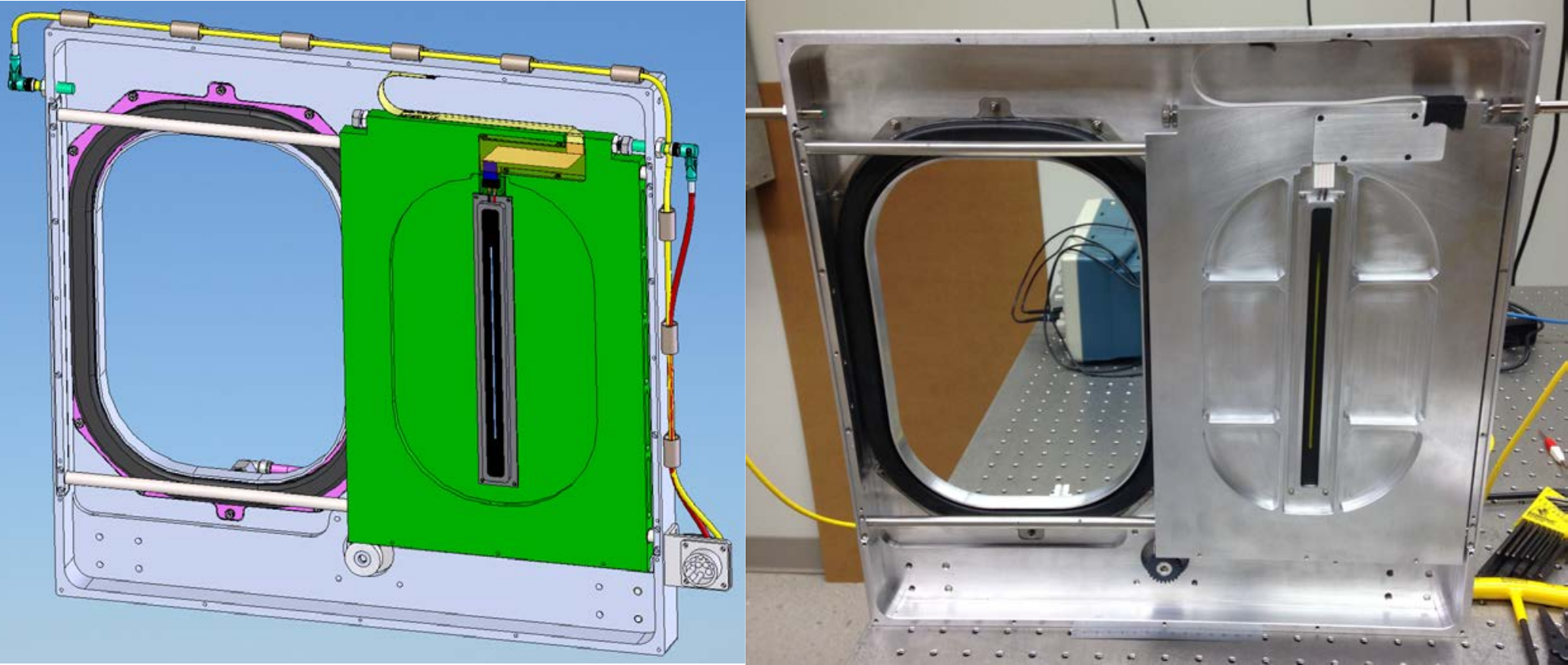} 
\caption{Left: Model for the Exposure Shutter shown in the open position and without one cover in place. This model shows the PneumaSeal inflateable seal around the aperture ({\it black}) and the fiber illuminator LED mounted in the middle of the shutter door ({\it green}). Right:  First fabricated unit.} 
\label{fig:ShutterConcept} 
\end{figure}

The shutters  (Figure~\ref{fig:ShutterConcept}) are used to block light from the CCDs during readout, including light from the fiber illuminator. The fiber illuminator will usually be operated during CCD readout, so the shutters are required to transmit less than $10^{-7}$ of the photons at the fiber tip (DESI-0652). The exposure shutter is mounted in front of the first dichroic. The NIR shutter is mounted in front of the NIR camera. The exposure shutter is used to control the exposure of the three channels simultaneously, while the NIR shutter is just needed to block the light from the fiber illuminator. Both shutters are equipped with a PneumaSeal inflateable seal that has been measured to provide $>10^{-9}$ attenuation of the LED flux. The seals operate at 10--15 psi and will only be inflated when the fiber illuminator is in operation.

All of the fibers in all channels will have the same exposure independent of the motion of the shutter because the shutter is near the pupil. The exposure shutter must open and close in less than 500 milliseconds, and must have a lifetime of at least 155,000 cycles (twice the expected number of exposures) (DESI-0584). The NIR shutter must open and close in less than 1 second, and have the same lifetime as the exposure shutter. The two shutters are identical mechanisms except the NIR shutter does not have a fiber illuminator. The shutters use a direct drive stepper motor with a rack and pinion gear, the blade slides on Frelon-coated aluminum bushings, and the inductive proximity sensors detect when the shutter is fully open and fully closed.

\begin{figure}[!t]
\centering
\includegraphics[height=3in]{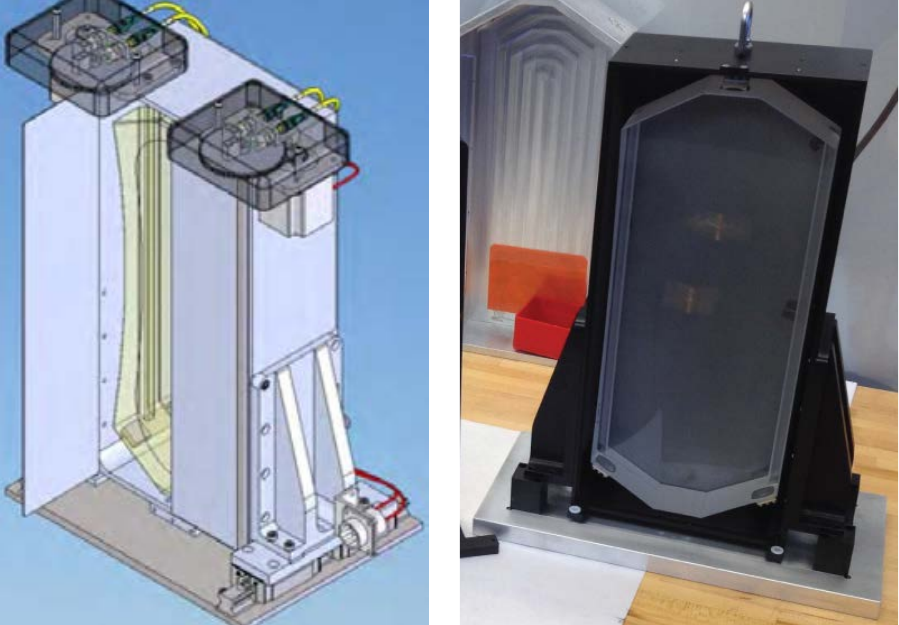}
\caption{Left: Model for the collimator mount and Hartmann doors with the left door open. Right: First collimator mount with an uncoated mirror installed.} 
\label{fig:Collimator}
\end{figure}

The Hartmann doors are in front of the collimator mirror and part of the collimator mount (Figure~\ref{fig:Collimator} left). The doors are a simple bi-fold design, and each operates as an independent mechanism. They are used to determine the focus by comparing the location of the centroid of spectral lines with the left or right door closed. The design is a smaller, modified version of the Hartmann doors used on BOSS \cite{Smee13}.

The collimator mount (Figure~\ref{fig:Collimator} right) has manual focus, tip, and tilt adjustments. The collimator is required to be insensitive to focus changes with less than 7.5~\micron/\celsius~shift. The collimator focus is athermalized with a re-entrant focus mechanism of high CTE material, which compensates for the change in distance between the slit and collimator with temperature. Analysis of the design indicates it should produce a residual sensitivity of $\sim$2~\micron/\celsius~due to uncertainties in CTE values. The collimator mirror is supported with three thin, stainless steel webs. These webs correct for the CTE differences between the aluminum mounting plate and the Zerodur mirror, yet minimize mirror distortion. FEA indicates that the mechanical mount produces less than 5~nm rms surface error in representative 128~mm diameter sub pupils (DESI-1077).

\begin{figure}[!t]
\centering
\includegraphics[height=3in]{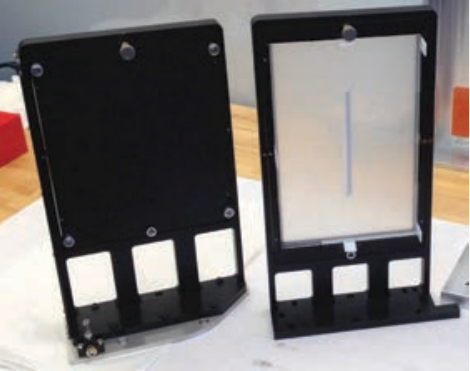}
\caption{The mounts for the two dichroics. The one on the right has an uncoated substrate with a machined slot for fiber optic slit head to project light onto the collimator mirror.} 
\label{fig:dichroic_mounts}
\end{figure}

There are two dichroics.  The first transmits the infrared to one spectrograph arm, and the second reflects the blue into another arm and transmits the red to the third arm.  The dichroic mounts are shown in Fig.~\ref{fig:dichroic_mounts}.  Note the machined slot in the left substrate.  This is where the slit project toward the collimator mirror.

The CCD flat field illuminator is a replacement slit head with a ``leaky'' fiber that produces a nearly uniform illumination along its length. It has the same radius of curvature and interface features of the fiber slit head. The leaky fiber is fed  from a continuum light source to produce continuous light both, spatially and spectrally, onto the CCD detectors.  This CCD flat field will be measured about once per year as was done for BOSS. 

Environmental sensors will be mounted at several locations on the spectrograph. Platinum RTDs will monitor the temperature of the spectrograph to ensure that it is within the operational range. There will also be a humidity sensor near each cryostat window to ensure that the air is above the dew point.

In addition to the mechanical system there is a spectrograph controller. The spectrograph controller is housed in an electronics box  on the optical bench. The controller includes motor controllers for the two shutters and two Hartmann doors, electronics for the fiber illuminator, air valves for the shutter seals, and connections for the temperature and humidity sensors. This controller is connected to the Instrument Control System (Section~\ref{sec:Instr_Control_System}) via Ethernet.  The spectrograph controller also contains a minicomputer that runs the code that operates the spectrograph and a router that isolates internal communication.

\begin{figure}[!htb]
\centering
\includegraphics[width=\textwidth]{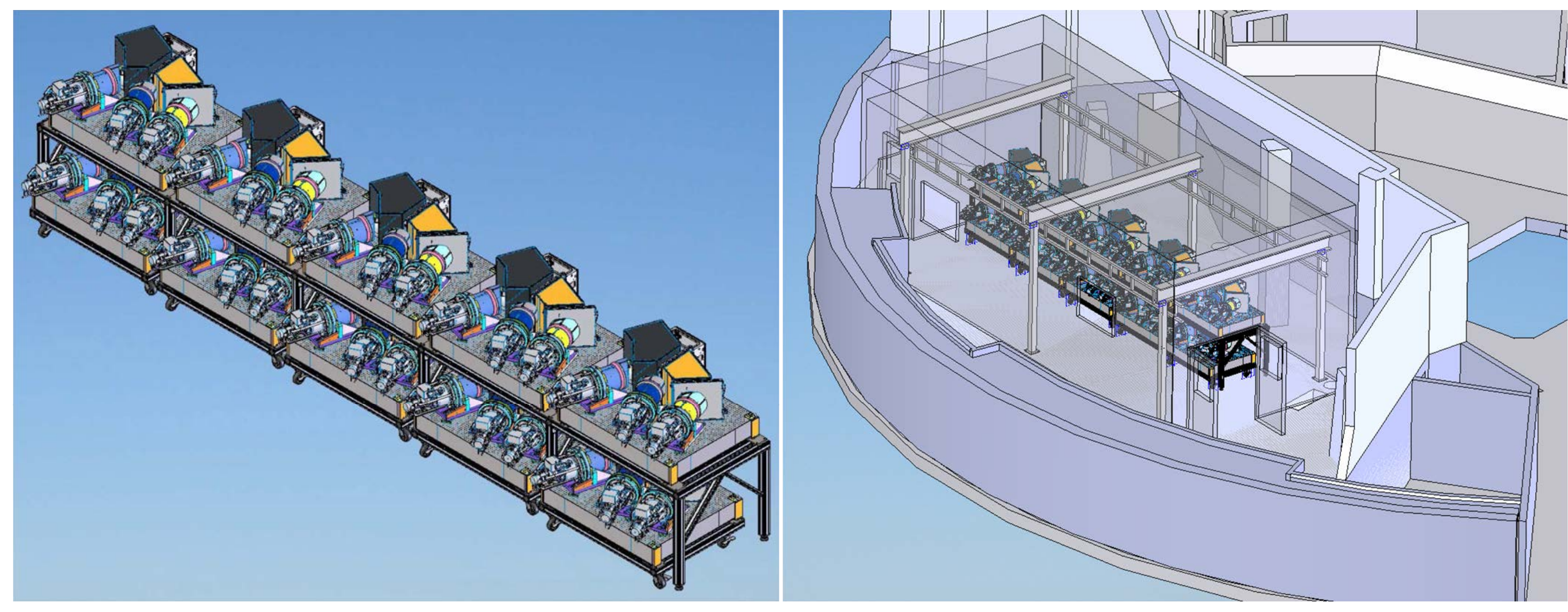} 
\caption{Left: The spectrographs are arranged in five racks, two tall. Right: A climate-controlled, insulated ``Shack'' is shown in the Coud\'{e} room that will contain the racks. An overhead gantry crane is inside the Shack for installation and removal of mechanisms, cameras, cryostats, or even the entire $\sim$500~kg spectrographs. 
}
\label{fig:ShackConcept}
\end{figure}

The ten spectrographs are mounted in five stacks of two in a climate controlled enclosure (Shack) in the Coud\'{e} room at the Mayall telescope (Figure~\ref{fig:ShackConcept}). The temperature in the Shack is maintained within 18--22~\celsius~and the relative humidity is kept below 85\%.  The cabling and access is designed to accommodate most maintenance operations without removing a spectrograph from the rack. The support rack will also include any vibration isolation that is needed.

The current concept for the Shack is to place the five racks in a row arrangement to allow access to the cryostat and slit sides of the spectrographs in their racks (Figure~\ref{fig:ShackConcept}). There is also sufficient space to include an overhead gantry crane, which offers a safe way to perform maintenance with a minimal number of personnel. The Shack includes a Shack controller, which measures the Shack temperature and humidity, as well as monitors glycol and other utilities.


\subsection{Spectrograph Integration and Test}

The integration of the spectrograph occurs in four phases. First  is the integration and alignment of the cameras. Second  is the integration of the optics onto the bench. Third  is the integration of the cryostats and mechanisms onto the spectrograph. Fourth is the integration of the spectrographs onto the racks at the Mayall telescope. 

The integration and alignment of the cameras will be done by the camera vendor. They will align the lenses and measure the point spread functions (PSF) using a translating microscope with camera. The throughput will also be measured. In addition, the camera will undergo thermal cycle testing (to the survival temperature limits, -20 to +40 \celsius  (DESI-0583)) to ensure that the alignment does not change and the camera survives. They will also measure the PSF over the operational range to ensure that it does not change.

The gratings, dichroics, collimator and camera will be installed onto the camera bench by the spectrograph integrator. The grating angle will be adjusted to maximize the throughput. The spectrograph will be aligned using a tooling slit and the PSF measured using a microscope and camera.  The PSF will be measured over the 18--22~\celsius~temperature range to make sure it remains within specification.

The thermal and humidity sensors will be installed on the cameras. The shutters, Hartmann doors and cryostat will be installed and aligned on the bench using a test fiber slit head. The throughput, PSF, and spectral dispersion of the spectrograph will then be measured. The PSF will be measured over the 18--22~\celsius~ temperature range to make sure it remains within specification. Finally scattered light will be quantified. 

The spectrograph will then be disassembled and  packed into its shipping container and shipped to NOAO. At Kitt Peak the spectrographs will be unpacked in the Fourier Transform Spectrograph (FTS) lab at the Mayall telescope.  The FTS lab is on the same floor as the large Coud\'{e} room. The spectrographs will be reassembled in the FTS lab and the functionality verified. They will then be rolled into the large Coud\'{e} room. In the  Coud\'{e} room is an insulated room where the spectrograph will be placed on 5 racks of two spectrogrpahs each. The Shack concept is shown in Figure~\ref{fig:ShackConcept}. After the spectrographs are placed in the rack, the last installation step is the addition of the fiber slit head.


\subsection{Cryostats}
\label{sec:Instr_Cryostats}
\begin{figure}[htb]
\centering
\includegraphics[height=3in]{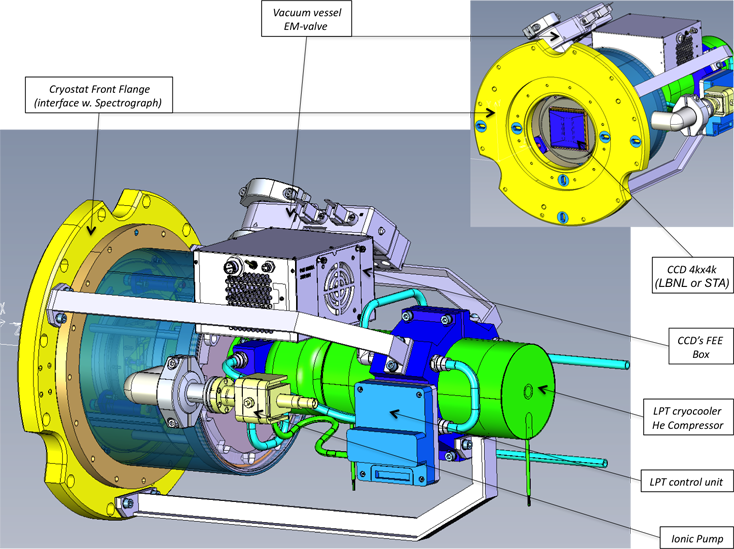}
\caption{3D model of the cryostat and view of the front side interface.}
\label{fig:cryo1}
\end{figure}

The ten DESI spectrographs each require three CCDs for the blue, red, and NIR channels. Each CCD is housed in is own cryostat, each in turn with its own refrigerator and CCD control electronics.
The main requirements for the design of the cryostat system are as follows:

\begin{itemize}
  \setlength{\itemsep}{1pt}
  \setlength{\parskip}{0pt}
  \setlength{\parsep}{0pt}
\item Cool CCDs to 145~K (LBNL CCDs) or 170~K (ITL CCDs) with a precision of 1~K and regulate their temperature to  $\pm$0.1~K.
\item Allow the CCDs to be aligned along the optical axis, in both transverse axes and tilt angles relative to the spectrograph optical axis.
\item Provide easy access to the instrument for ease of maintenance and fast replacement (typically, one cryostat to be replaced in less than 24 hours by 2 persons).
\item Monitor per cryostat the CCD temperature regulation, vacuum conditions, power supply, control and safety processes.
\end{itemize}

\subsubsection{Cryostat Design and Interfaces}

\paragraph{Mechanical Design}

One of the most important cryostat requirements is to have identical units in order to be able to react quickly in case of changes or failures. So we adopt the same mechanical design for all cryostats (except for the front optical lens) in order to have interchangeability for each spectrograph arm.

The detector focal plane is determined by the optical configuration of the spectrographs and will be slightly different in each channel. The last lens of each spectrograph arm has to be part of the cryostat due to its short distance to the CCD plane. It will act as the vacuum window of the cryostat vessel. Figure~\ref{fig:cryo1} shows the cryostat 3D model. One cryostat weighs 25~kg, is 310~mm in diameter and 480~mm long.

\begin{figure}[htb]
\centering
\includegraphics[width=0.9\textwidth]{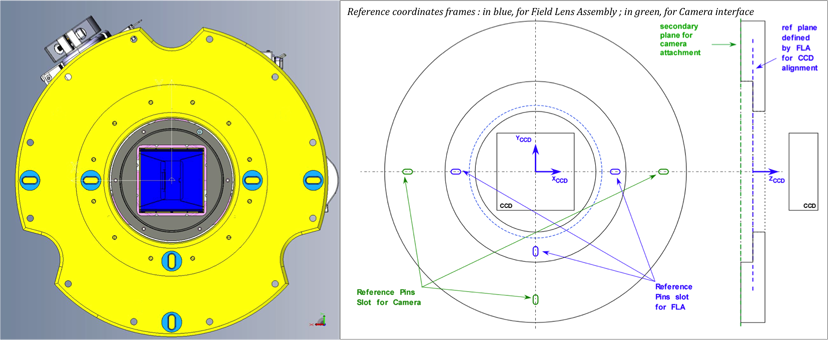}
\caption{ Reference frames system for the Cryostat Interface (blue for FLA, green for the spectrograph interface).}
\label{fig:cryo2}
\end{figure}

The reference plane of the interface between the spectrograph and the CCD is the front face of the cryostat flange. The reference frame for the alignment is defined by high precision pin-holes located on this flange (Figure~\ref{fig:cryo2} left).

The last lens will be assembled and aligned in its mechanical mount, the Field Lens Assembly (FLA).  Pin-holes will allow the camera and the FLA to be aligned by mechanical construction relative to the interface plane. Fine adjustment of the CCD plane position will rely on an XY$\theta$Z stage and an optical measurement table. First, the CCD plane will be positioned parallel to the reference plane by adding shims between the front flange rear face and the feet of the cold plate bearing the CCD, controlling the position with the optical table. This ensures that the CCD plane coincides with the XY plane as defined in Figure~\ref{fig:cryo2} right. Then, the CCD XY$\theta $Z position will be adjusted by moving the CCD with respect to the cold plate using the translation stage. The cold plate and its fixation elements were designed to ensure thermal expansion compensation between these pieces. The CCD is then fixed to the cold plate with bolts. Thus the alignment achieved at room temperature will be valid in cold conditions. The above procedure will allow us to align the CCD plane within 100--200~\micron along the optical Z axis and within 50~\micron in transverse XY plane, the spectrograph optical requirements. The final focusing precision, $\pm$15~\micron along Z,  will achieved by moving the camera relative to the cryostat, the camera vendor's responsibility.

\paragraph{Cryostat Vessels}

\begin{figure}[htb]
\centering
\includegraphics[height=3in]{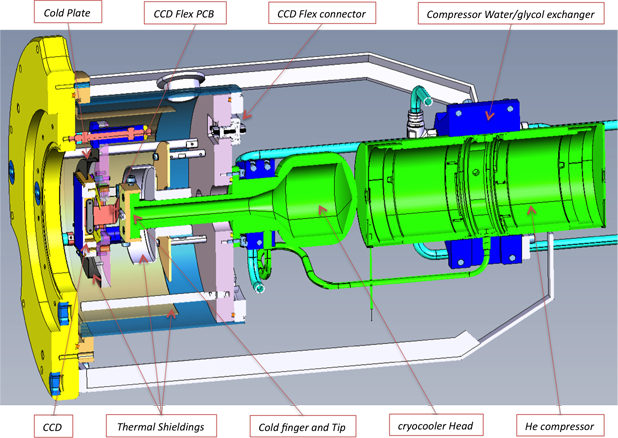}
\caption{Cryostat cross sectional view.}
\label{fig:cryo3}
\end{figure}

The cryostat vessel provides the mechanical connection to the spectrograph camera, supports the thermal and vacuum conditions, and interfaces with the control system and the CCD electronics. The cryostat is a metal cylinder which receives a front flange that integrates the FLA of the camera and  supports the CCD on the vacuuml side. A rear flange closes the vessel and supports the cryocooler (Figure~\ref{fig:cryo3}). It has several hermetic connectors, one for the CCD flex to the front-end electronics box and two for temperature regulation and the monitoring by the control system. The cylindrical side is equipped with 2 ports for vacuum pumping and pressure gauges.

Cooling power is supplied from the refrigerator to the CCD through a set of mechanical links; the CCD on its Invar package is mounted on an Invar cold plate connected with flexible copper braids to a copper cold base screwed to the cold tip of the cryocooler. The Invar cold plate provides the mounting and positioning of the CCD and supplies cooling with minimal thermal losses. The CCD package and cold plate are made with low thermal expansion materials to reduce differential thermal constraints.

The copper cold base is dimensioned to have a thermal capacitance suitable for the CCD temperature regulation. Pt100 resistors fixed on the CCD package, the cold plate, the Cold Tip and on the front flange close to the FLA ensure thermal control and monitoring. A resistive heater is glued on the cold plate to control warm-up operations.

Thermal shielding of the cryostat is provided in three pieces, one for the vessel sides, one for the rear flange and one for the front lens. The latter is different for the 3 channels of the spectrographs, which have lenses of different diameters. The shielding will be provided by polished Al plates. 

\subsubsection{Cryogenic System and Performance}

To produce cooling power for the 30 independent cryostats, a closed cycle cryocooler per cryostat is used, each with its own CCD temperature regulation. Pulse tube cooler technology is used in order to have a simple and robust system for the control of the 30 cameras, providing easy integration and maintenance.

To define an appropriate cryocooler model for DESI cryostats, we made a simple estimate of the power and temperature budget of the different elements of the cameras. The results are given in Table~\ref{tab:Cryo_power_budget}. The values are for a CCD temperature of 170~K and a maximum difference of 20~K with respect to the cold finger temperature. We would like the option to operate some CCDs as low as 145~K, which requires more than the 3~W. Thus we bought the Thales LPT9310 Cryocooler (4~W at 88~K) to evaluate in the prototype cryostat.

\begin{table}[!b]
\centering
\caption{Estimated radiative and conductive thermal losses
(at $10^{-5}$ mbar, convective heat transfer is negligible).}
\begin{tabular}{lcc}
\hline
Element                          & Loss (W) &  \% of total loss \\
\hline
Lens (radiative)                 & 1.9      & 62    \\
CCD dissipation                  & 0.12     & 4     \\
CCD electronic cables            & 0.23     & 7     \\
Cold plate / vessel (radiative)  & 0.16     & 5     \\
Cold plate supports (conductive) & 0.15	    & 6     \\
Cold base regulation capacity    & 0.5      & 16    \\
\hline
TOTAL                            & 3.06     & 100   \\
\hline
\end{tabular}
\label{tab:Cryo_power_budget}
\end{table}

\paragraph{LPT Cryocooler}

\begin{figure}[!t]
\centering
\includegraphics[height=2in]{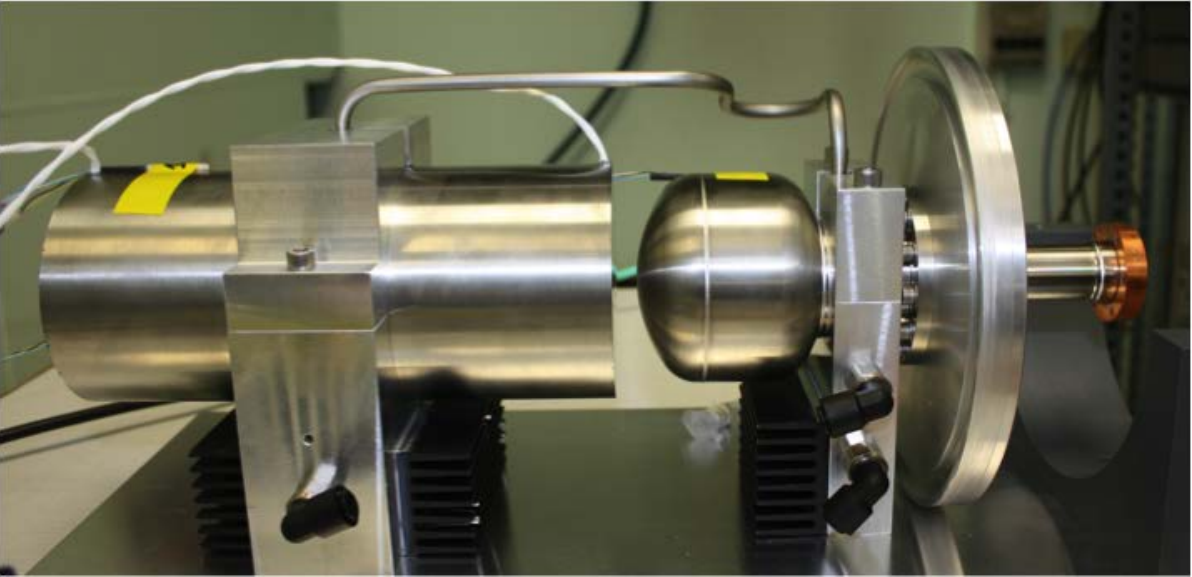}
\caption{Thales 4W LPT9310 in DESI test  stand.}
\label{fig:cryo4}
\end{figure}

Linear Pulse Tubes (LPT) were developed by the ``Service des Basses Temperatures'' (SBT) at CEA in Grenoble (France). The technology was transferred to Thales Cryogenics BV Company which provides several models of LPTs with different power and temperature ranges (Figure~\ref{fig:cryo4}). The LPT is a closed-cycle pulse tube cooler made of a compressor module and a cold finger, connected together by a metal tube. The compressor pistons are driven by integral linear electric motors and are gas-coupled to the pulse tube cold finger, which has no moving parts. This technology, combined with the proven design of the ultra-reliable flexure bearing compressors, results in extremely reliable and miniature cryocoolers with a minimum of vibrations. In addition, the compact magnetic circuit is optimized for motor efficiency and reduction of electromagnetic interferences. The compressor part is separated from the cold head by a He line which is bent in order to reduce vibration coupling.

\begin{figure}[!b]
\centering
\includegraphics[height=2.5in]{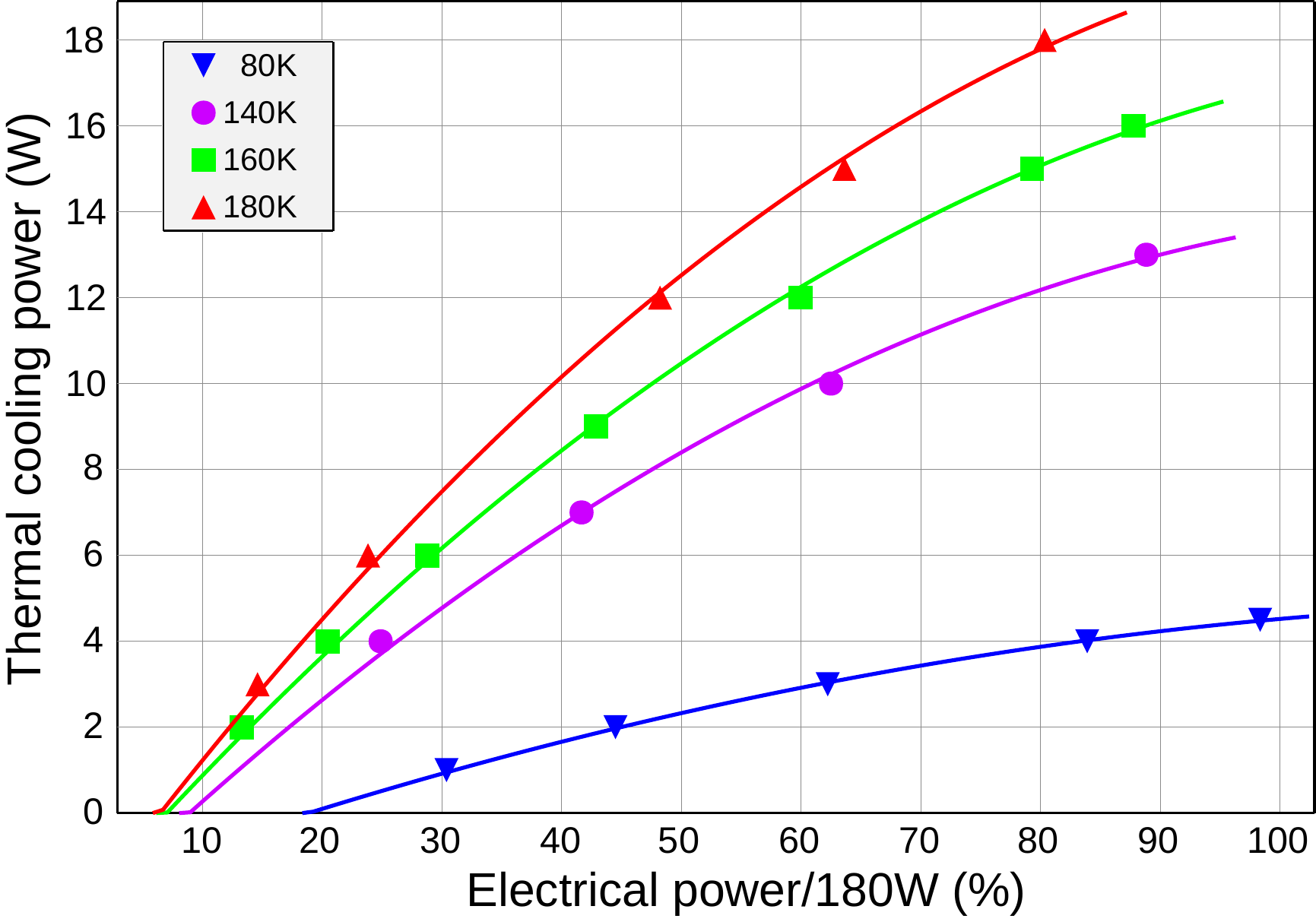}
\caption{Saclay measurements of the LPT9310 thermal response for various cold-tip temperatures in a test cryostat.}
\label{fig:cryo5}
\end{figure}

The LPTs have a mean time to failure of more than 90000 hours at full power. To guarantee this reliability over the lifetime of the instrument, we chose the LPT9310 model working at 25--30\% of its maximal power. One unit (Figure~\ref{fig:cryo4}) was integrated in a dedicated test cryostat at Saclay and its thermal performance validated. The results are shown in Figure~\ref{fig:cryo5}. To obtain the latter, the cold head and compressor were cooled by a liquid distribution system, Glycol or Water, at about 5--10~\celsius with a flow of about a liter per minute and per LPT. This will be provided for each LPT by an existing cooling distribution at the Mayall telescope.

\paragraph{Device Monitoring and Temperature Regulation}

\begin{figure}[!t]
\centering
\includegraphics[width=\textwidth]{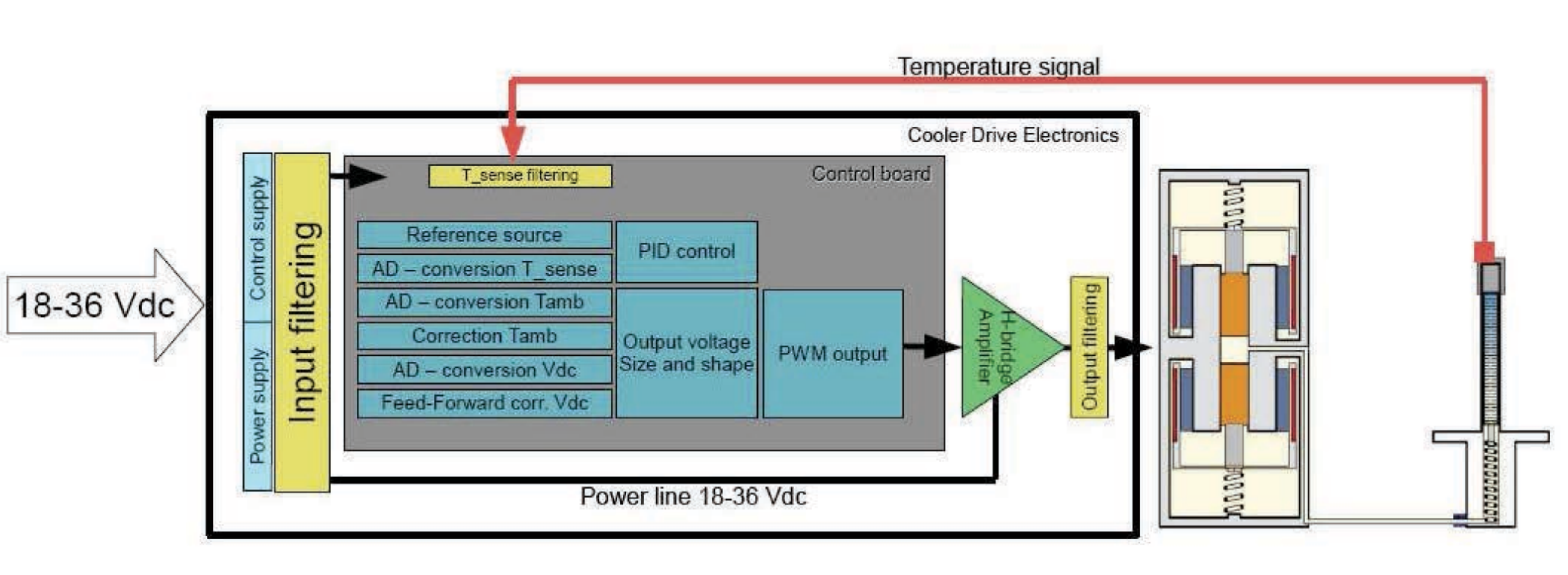}
\caption{LPT9310 cooler drive electronics block diagram.}
\label{fig:cryo6}
\end{figure}

The LPT compressor is powered with an AC voltage signal which sets the cold finger operating point in power and temperature. Changing this voltage allows the thermal performance to be tuned in a given range. The LPT is provided with an electrical interface called XPCDE (XP Cooler Drive Electronics) powered by an input DC signal  (Figure~\ref{fig:cryo6}). The CDE converts the input signal from DC to AC and adjusts the output voltage through a temperature regulation loop. This loop is based on an Proportional-Integral method, using a Pt100 temperature sensor. The XPCDE provides internal diagnoses about the thermal control process itself. Pre-tuning of the XPCDE is done by the manufacturer; fine tuning is done on each set of LPT unit in order to achieve extreme temperature stability. The tests (see Section 5.3) showed that the CDE is able to set and regulate the CCD temperature inside our cryostat provided that the regulation control loop P-I parameters are tuned.

\paragraph{Cryocoolers and Vacuum Maintenance}

The LPTs have a mean time to failure of more than 90000 hours at full power and do not require routine maintenance. Monitoring the response of the LPTs will be implemented in the cryostat control system. In case of a unit failure, the LPT will be replaced by a spare and the faulty unit returned to the manufacturer for examination and repair.

The design of the vacuum system takes into account the mechanical assembly of the 10 spectrographs mounted in five towers. There are two vacuum units to be shared between the 10 spectrographs. A vacuum unit is composed of a primary and secondary dry pumping machines (250 l/s) and a large conductance flexible line to be connected to a spectrograph (3 cryostats). With such a system, the time to reach $10{\pm}5$ mbar will be less than 5 hours. Each distribution line will be equipped with an automatic electro-magnetic valve to maintain the vacuum in the cryostats in case a pumping machine fails. Each cryostat pipe will be equipped with its own isolation and over-pressure valves.

As it is not intended to pump the cryostats during observation time, the vacuum will be maintained by an ion pump. This solution is currently being evaluated (performance, sizing and costing). The final design relies on the outgassing and leakage rate that will be quantified using the cryostats of the DESI spectrograph prototype.

Vacuum maintenance is planned between the observation periods. The cryostats will be warmed up and pumped. The actual periodicity of such operations will be defined when the tests on EM\#1 are completed.

\subsubsection{Prototype Cryostat Measurements and Model Validation}
\label{sec:cryoproto}

\begin{figure}[!t]
\centering
\includegraphics[height=3in]{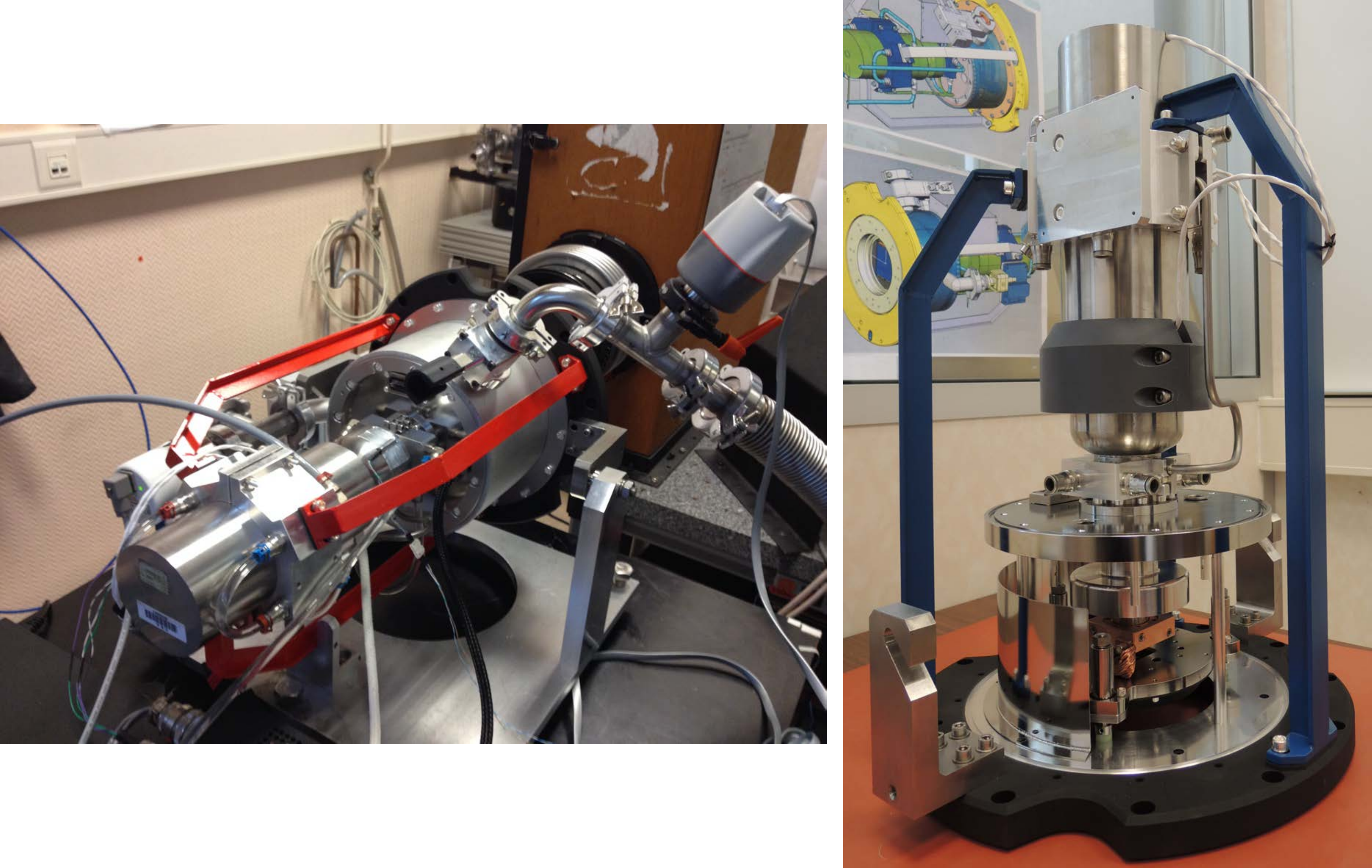}
\caption{Saclay first cryostat prototype.}
\label{fig:cryo7}
\end{figure}

A cryostat prototype (Figure~\ref{fig:cryo7}) was constructed including all the elements previously described. We used the LPT9310 cryocooler and its CDE to test the cooldown and temperature regulation of a mechanical grade LBNL CCD under real experimental conditions (DESI-0327):

\begin{itemize}
  \setlength{\itemsep}{1pt}
  \setlength{\parskip}{0pt}
  \setlength{\parsep}{0pt}
\item Thermal losses were measured between 2.8 W and 3.2 W for the requested CCD temperatures (145--180~K), in good agreement with our estimate.
\item LPT consumption was only 25--30\% of its maximum allowed power (Figure~\ref{fig:cryo5}).
\item Temperature precision of the CCD was about 1~K.
\item Temperature stability of the CCD was within 0.2~K peak to peak.
\item Cool down temperature rate was 1.5~K/min and could be adapted if needed.
\item cool down and temperature stability was reached in about 2.5 hours.
\item Vibration amplitude was less than 3~\micron (the instrument precision) at the detector in the directions parallel and perpendicular to the optical axis.
\end{itemize}

These show that we can easily reach all the requirements (CCD temperatures between 145--180~K, acceptable level of vibration, CCD temperature stability).

When extrapolating to 30 LPTs, the level of vibrations could be a concern. But, as each LPT is independently AC--powered by its own XPCDE and all XPCDE units are DC--powered, the vibrations of the 30 LPTs should not add in phase. We do not expect a higher vibration amplitude when operating with 30 LPTs, but this is carried as a risk. There are two mitigations if this proves to be a problem. The frequency of each LPT can be adjusted to different values. Isolation and damping of the compressor can be changed. It is important to avoid mechanical resonance of the structure supporting the spectrographs.

\subsubsection{ Cryostat Control System}

\begin{figure}[!htb]
\centering
\includegraphics[width=0.9\textwidth]{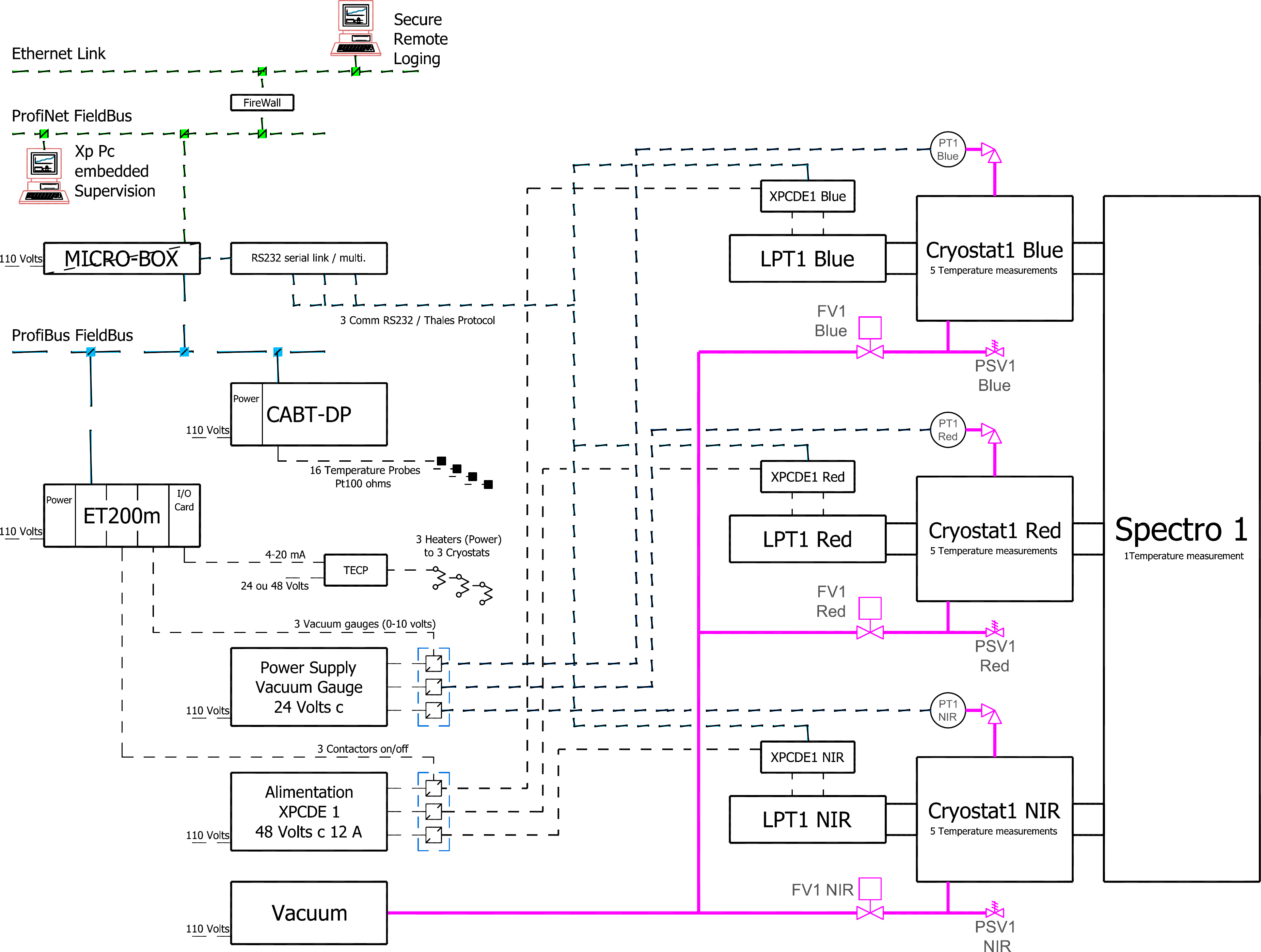}
\caption{ Architecture of the DESI cryostat Control System.}
\label{fig:cryo8}
\end{figure}

The control system will drive and maintain all 30 CCDs and cryostats at nominal operating conditions. A well-tested solution from prior projects has been adopted (MegaCam, Visir/VLT, LHC Atlas and CMS).
The main components of the control system (Figure~\ref{fig:cryo8}) are:

\begin{itemize}
  \setlength{\itemsep}{1pt}
  \setlength{\parskip}{0pt}
  \setlength{\parsep}{0pt}
\item Main unit MicroBox which integrate a Programmable Logic Controller (PLC) and a PC.
\item Multi-channel RS232 Bus link interface to communicate with the XPCDEs of each cryocooler.
\item Common group of Input/Output units to monitor the main parameters as pumping System, Chiller, vacuum distribution (pressure and valves).
\item Several (x10) groups of I/O analog and digital units for cryostats parameter. So, each group is associated to 3 cryostats of one spectrograph.
\item OPC server for the communication with the Instruments Control System. It allows the ICS to download the Slow Control database which contains all analog values or parameter's states of the cryostats. It also transmits all  commands from the ICS to the Slow Control.
\item Supervisory unit, ported on an external PC to ensure safe functioning. It allows to display the state of the cryostat system in a graphical mode. It also allows to control all the elements in an SuperUSer mode for maintenance operations or in case of failure.
\end{itemize}

The whole system is overseen by a Java module over Ethernet link. The general architecture of the system is based on the solution developed, implemented and tested on the Saclay prototype cryostat. 

The PLC is a Siemens Simatic S7-300 with a system core based on a UC319 mainframe. The program implemented in the PLC acquires in real time all variables corresponding to the monitoring and control of the instrument (vacuum and temperature monitoring), the control of the cold production unit (CCD cooling down and warming up), and the safety procedures on cryogenics, vacuum and electrical power.  A local network (based on an industrial bus, \eg, ProfiBus or ProfiNet) ensures communication with the remote plug-in I/O modules and with the PLC.

Temperature measurements are provided by the Pt100 temperature probes connected to temperature modules acquired by the PLC. The other analog sensors (heaters, vacuum gauges) are connected on a 4--20 mA or 0--10 V module. All measurement modules will be located in two cabinets.

Supervision software with user interface is implemented in a NISE industrial PC connected to the PLC. It ensures the control of the cryocoolers, of the vacuum system and the monitoring/recording of all variables, with possibly different levels of user access rights. This access will be available to the Instrument Control System via an OPC server (OPC\_UA communication Protocol).
This system will also allow the set-up to be remotely controlled via the Ethernet network that will be accessible from the internet through a secured interface.

The total electrical consumption for the cryostat system is estimated to be about 11~kW during cool-down. It is reduced to about 4~kW in
nominal operation conditions, assuming the LPTs are at 30\% of the maximal power and the pumping station is off (vacuum stationary state in cryostat during Science Observations). See details in Table~\ref{tab:cryo_power}.

\begin{table}[!t]
\centering
\caption{ Electrical power consumption of the Cryostat System.}
\includegraphics[width=\textwidth]{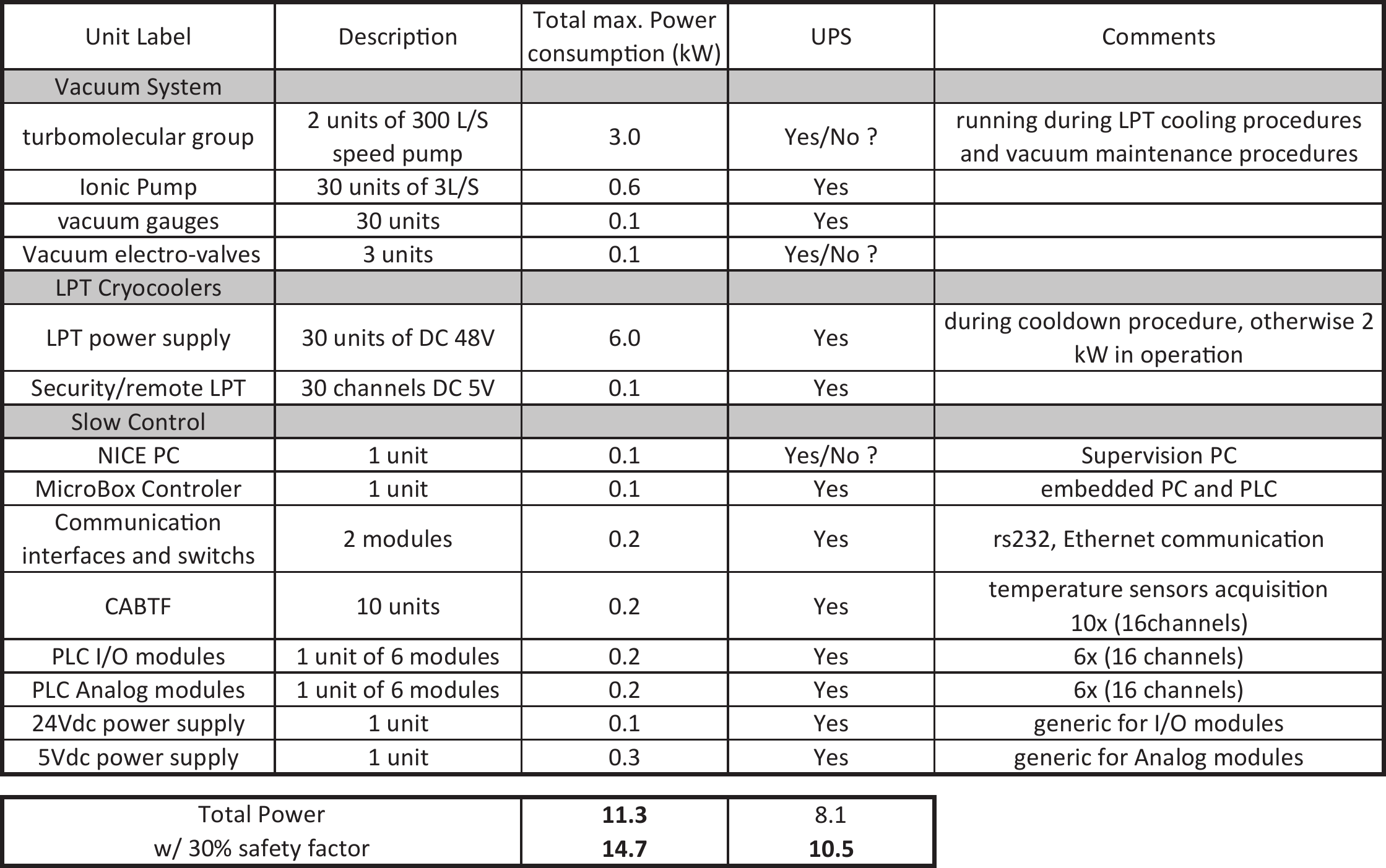}
\label{tab:cryo_power}
\end{table}

\subsubsection{Cryostats Integration and Tests}

The cryostat integration requires dedicated tools that will be built in Saclay.  In addition some elements will be provided by other members of the project:

\begin{itemize}
  \setlength{\itemsep}{1pt}
  \setlength{\parskip}{0pt}
  \setlength{\parsep}{0pt}
\item one special flat FLA to seal any cryostat during the test phase. This allows the CCD position to be optically controlled. The FLA will be provided by the spectrograph vendor.
\item one CCD per cryostat and one front-end electronics unit (with software) to allow us to perform functional tests of the CCDs once integrated and cooled.
\item one final spectrograph FLA per cryostat -- provided by the spectrograph vendor -- to close the cryostats before shipping.
\end{itemize}

\begin{figure}[!t]
\centering
\includegraphics[width=0.9\textwidth]{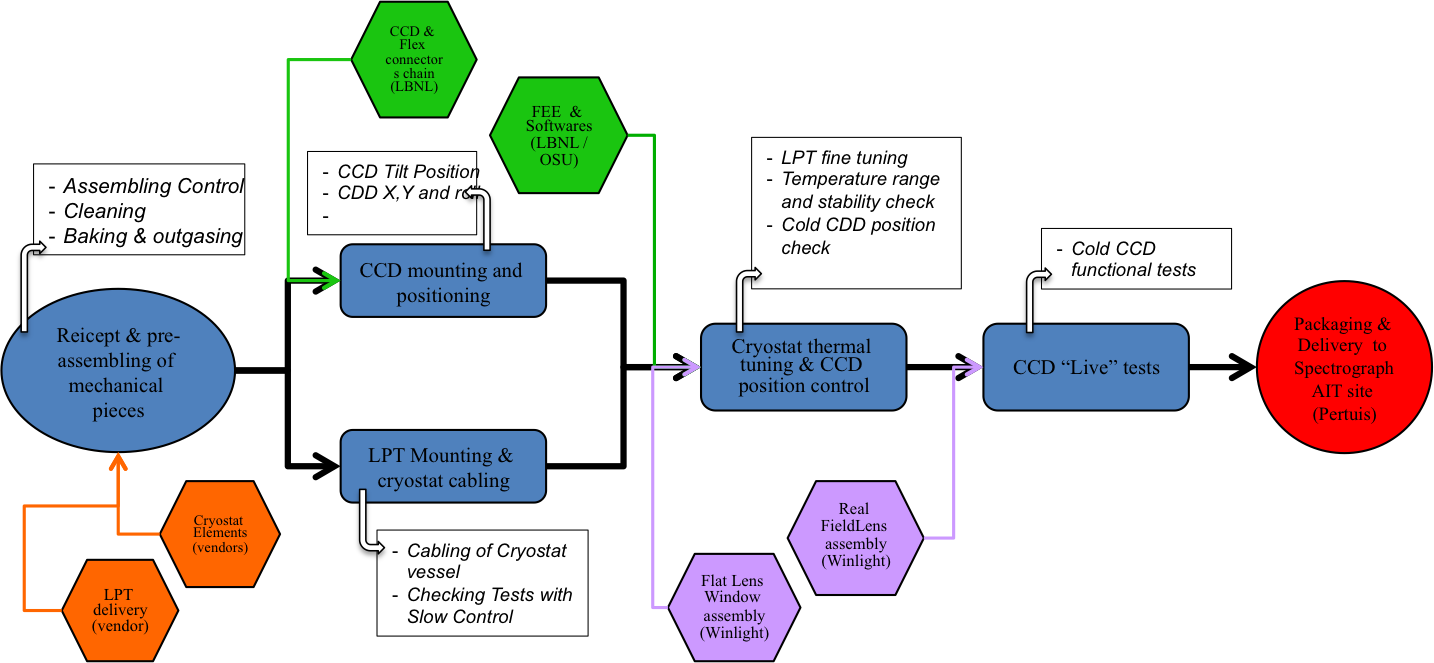}
\caption{Cryostat integration Flow.}
\label{fig:cryo9}
\end{figure}

The cryostat integration procedures and tests are composed of 5 steps, as shown in Figure~\ref{fig:cryo9}:

\begin{enumerate}
  \setlength{\itemsep}{1pt}
  \setlength{\parskip}{0pt}
  \setlength{\parsep}{0pt}
\item assembling control and cleaning-baking-outgasing operations of the mechanical elements of the cryostats.
\item CCD mounting and position tuning
\item LPT mounting and cabling/integration of all the sensors -- Check of vacuum leakages -- Tests with the Slow Control.
\item CCD integration in Cryostat  - Fine tuning of cryocooler regulation with CCD
\item functional tests of the CCD once closed with final FLA unit
\end{enumerate}

\begin{figure}[!ht]
\centering
\includegraphics[height=2.3in]{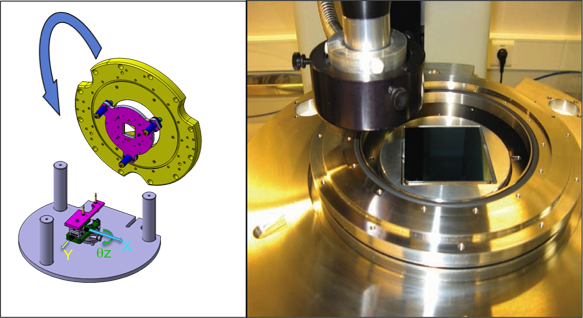}
\caption{Left: 3 axes stage tool for precise positioning of the CCD.  Right: optical machine used for CCD position measurements in the Saclay class 100 clean room.}
\label{fig:cryo11}
\end{figure}

The mechanical elements of the cryostat, including sensors and cryocooler, will be assembled and closed in a class 10000 clean room. The control system will then be connected to check data acquisition, vacuum stability and Security processes. 
The CCD will be integrated in a class 100 clean room. Its 3D position w.r.t the reference frame defined in the spectrograph interface control document (Figure~\ref{fig:cryo2}) will be measured by optical metrology. A specific 3 stages (XY$\theta$) tool supporting the front flange, CCD cold plate, and CCD assembly will allow the precise positioning of the CCD to be achieved (Figure~\ref{fig:cryo11}).

The cryostat will be closed with a ``flat lens'' FLA. This FLA will allow us to control the CCD position in the cryostat at low temperatures and for any orientation of the cryostat.  After all tests are completed, the flat lens FLA will be replaced by the spectrograph FLA and the fully integrated cryostat will be packaged and shipped.


\subsection{Sensors and Electronics}
\label{sec:Instr_Sensors}

The ten DESI spectrographs each require three CCDs for the blue, red, and NIR channels.
Each CCD is housed in is own cryostat, each in turn with its own refrigerator and CCD control electronics. 

\subsubsection{CCDs}

Each of the three channels of a spectrograph will use a 4k$\times$4k CCD with  15~\micron\ pixels.
The CCD specifications are defined in DESI-0741.  A subset of performance specifications are given in Table~\ref{tab:ccdspecs}.
 For the blue arm we baseline the ITL STA4150A provided by University of Arizona Imaging Technologies Lab. 
For the red and NIR channels we baseline the LBNL 4k$\times$4k,  250~\micron~thick CCD to optimize quantum efficiency.  The CCD format is the same as for BOSS, but has a lower read noise output.  

Both CCDs are known to meet the dark current, diffusion, and cosmetics requirements.  The QE requirements are within measured values. For readnoise,  Semiconductor Technology Associates report $<2.9$~e at 100~kilopixel/sec with 3.2~e achieved to date by ITL.  For the LBNL CCDs, lab measurements find a read noise of  1.8~e at 100~kilopixel/sec.

\begin{table}[htb]
\centering
\caption{CCD performance specifications.}
\begin{tabular}{lcc}
	\hline			
	Item	&	Specification	 &  \\
	\hline			
	Format	&	&	 \\
\hspace{10pt}	Pixel size	&	15~\micron	&  \\
\hspace{10pt}	Pixel count	&	$\geq (4096~\times~4096)$	&  \\
\hspace{10pt}	Readout channels	& 4	& \\
	\hline			
	Dark current	&	$<$ 10 e/pix/hr & \\
	Readnoise at 100 kpixel/sec	&	$<$ 3 e &	 \\
	Full well (3\% linearity deviation)	& $>$ 75,000 e- &  	\\
	Non-Linearity (from 200 e- to 75\% full well)& $<$ 1\%  &   \\
	\hline
	Charge transfer efficiency & & \\
\hspace{10pt}	Parallel	& $>$ 0.99999 &   \\
\hspace{10pt}	Serial 	& $>$ 0.99999 &  	   \\
	\hline
	Quantum efficiency	&		\\
\hspace{10pt}	360--400 nm		&	$>$ 75\% &	  \\
\hspace{10pt}	400--600 nm		&	$>$ 85\% &	 \\
\hspace{10pt}	600--900 nm		&	$>$ 85\% &	 \\
\hspace{10pt}	900--980 nm		&	$>$ 60\% &	 \\
	\hline			
	Lateral diffusion (rms at surface)	&	$<$ 5~\micron	\\
	\hline			
	Cosmetics	&		\\
\hspace{10pt}	Column defects - black or white 	&	$<$ 10 max \\
\hspace{10pt}	White spots 	&	$<$ 800 max	\\
\hspace{10pt}	Total (black \& white) spots 	&	$<$ 1500 max 	\\
\hspace{10pt}	Traps $>$ 200 e- 	&	$<$ 15 max	\\
	\hline			
	Flatness	&		\\
\hspace{10pt}	Blue channel  	&  $<$ 20~\micron~P-P	\\
\hspace{10pt}	Red and NIR channel & $<$ 15~\micron~P-P	\\
	\hline
	Corner pixels location knowledge	&	$\pm50~\micron$ \\
	\hline
\end{tabular}
\label{tab:ccdspecs}
\end{table}

\begin{figure}[!hb]
\centering
\includegraphics[width=0.9\textwidth]{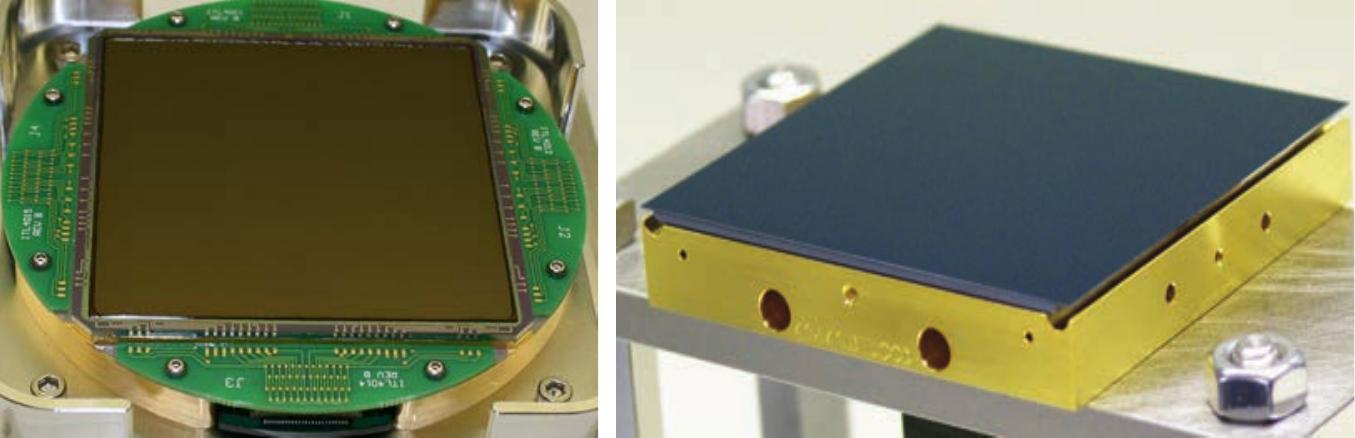} 
\caption{CCDs are 4k$\times$4k with 15~\micron\ pixels. Left: ITL package for STA 4150A CCD.  Right: LBNL 4k$\times$4k CCD package.}
\label{fig:CCD_pix}
\end{figure}

Figure~\ref{fig:CCD_pix} shows the packaging for LBNL and ITL CCDs that are compatible with the DESI cryostats.
The production of the LBNL CCDs will follow the DECam model where LBNL will produce and cold probe unpackaged CCD die that will in turn be packaged and characterized at FNAL.  FNAL has packaged seven devices to date using various quality die.  ITL  will be provided packaged and tested CCDs.

\begin{figure}[!htb]
\centering
\includegraphics[height=2.5in]{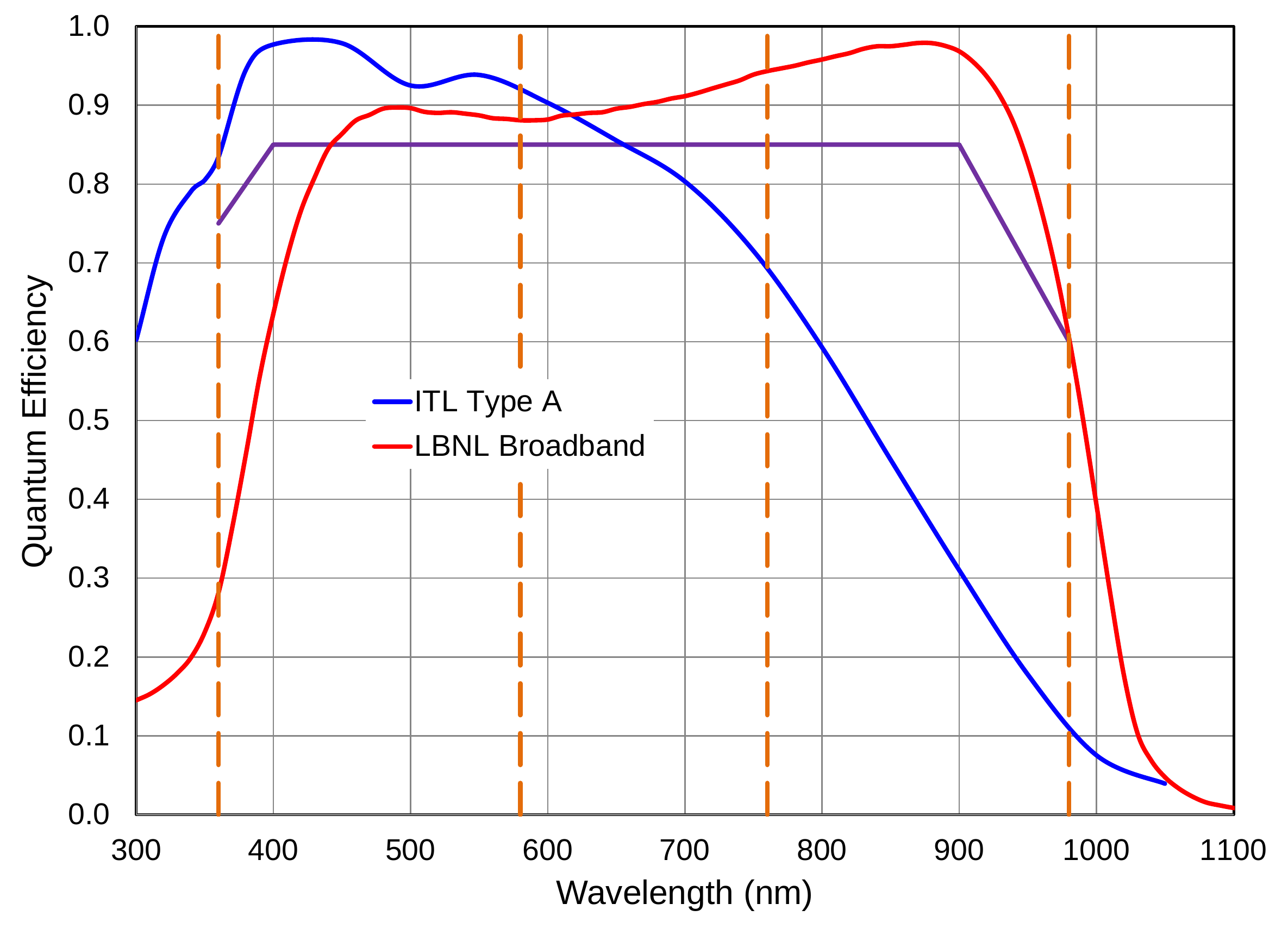} 
\caption{Quantum efficiency for  DESI : ITL STA5140A and  250~\micron\ thick LBNL CCDs.  Data are lab measurements. The purple line is the CCD specification. The vertical orange dashed lines are the locations of the band transitions for the three spectrograph channels.}
\label{fig:CCD_QEs}
\end{figure}

The quantum efficiency performance of both CCDs is well established. The red
curve in Figure~\ref{fig:CCD_QEs} shows measured QE for an LBNL CCD with a broadband antireflection coating. The high-side cutoff is determined by the CCD thickness.  We note that the QE in the red band for the CCD only meets the requirement on average at the moment.
Figure~\ref{fig:CCD_QEs} shows measured AR coating data from ITL (blue curve).  

\subsubsection{LBNL CCD Packaging}
\begin{figure}[!bt]
\centering
$\begin{array}{cc}
\includegraphics[height=1.9in]{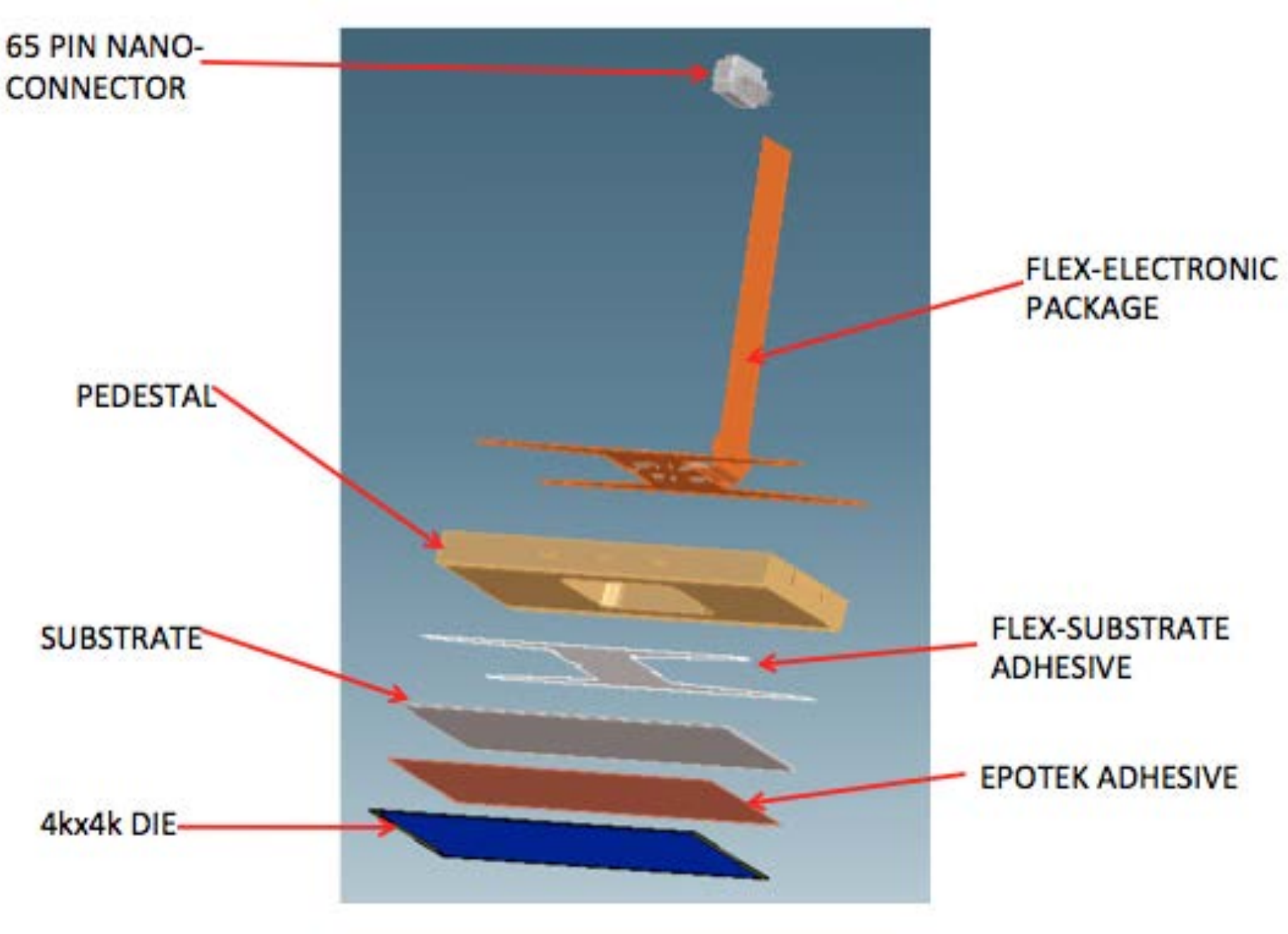} &
\includegraphics[height=1.9in]{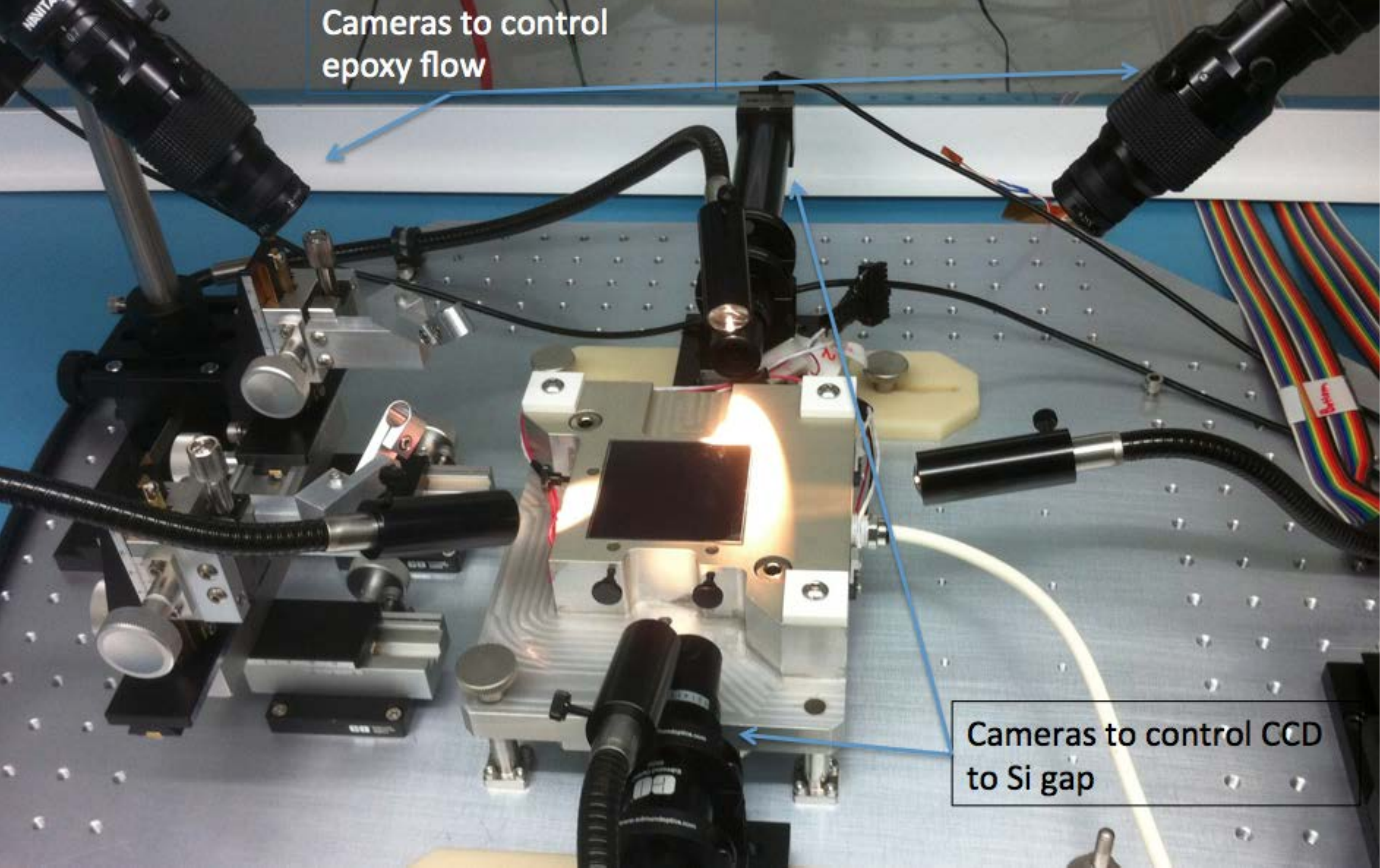}
\end{array}$
\caption{Left: Blowup of CCD package for the DESI 4kx4k LBNL CCDs.  Right: CCD packaging fixture showing four digital cameras to control the process. Two are used
to measure the gap between the CCD and the substrate, which can be adjusted with micrometers 
(not shown in the image). The other two cameras are used to monitor the flow of epoxy during the gluing process.
The fixture holding the substrate is not shown in this image. }
\label{fig:CCD_pack}
\end{figure}

The LBNL CCDs will be packaged at FNAL using the packaging technique developed by LBNL. 
An exploded view of the package for the 4k$\times$4k DESI sensors is shown in
Figure~\ref{fig:CCD_pack}. The detector is attached to a Si substrate with Epotek adhesive. The
Si substrate is attached to the custom readout flex circuit, which brings the signals to a 
64-pin connector. The flex circuit is wire-bonded to the CCD. Finally an Invar pedestal foot is attached to
the substrate to provide mechanical support for the detector.

The most critical step to produce a flat sensor is attaching the sensor to the Si substrate. 
The 80~\micron~gap between the sensor and the substrate needs to be precisely
controlled. This is achieved with an optical system permanently mounted on the
packaging fixture. This step is shown in Figure~\ref{fig:CCD_pack}.

\begin{figure}[!t]
\centering
\includegraphics[height=2in]{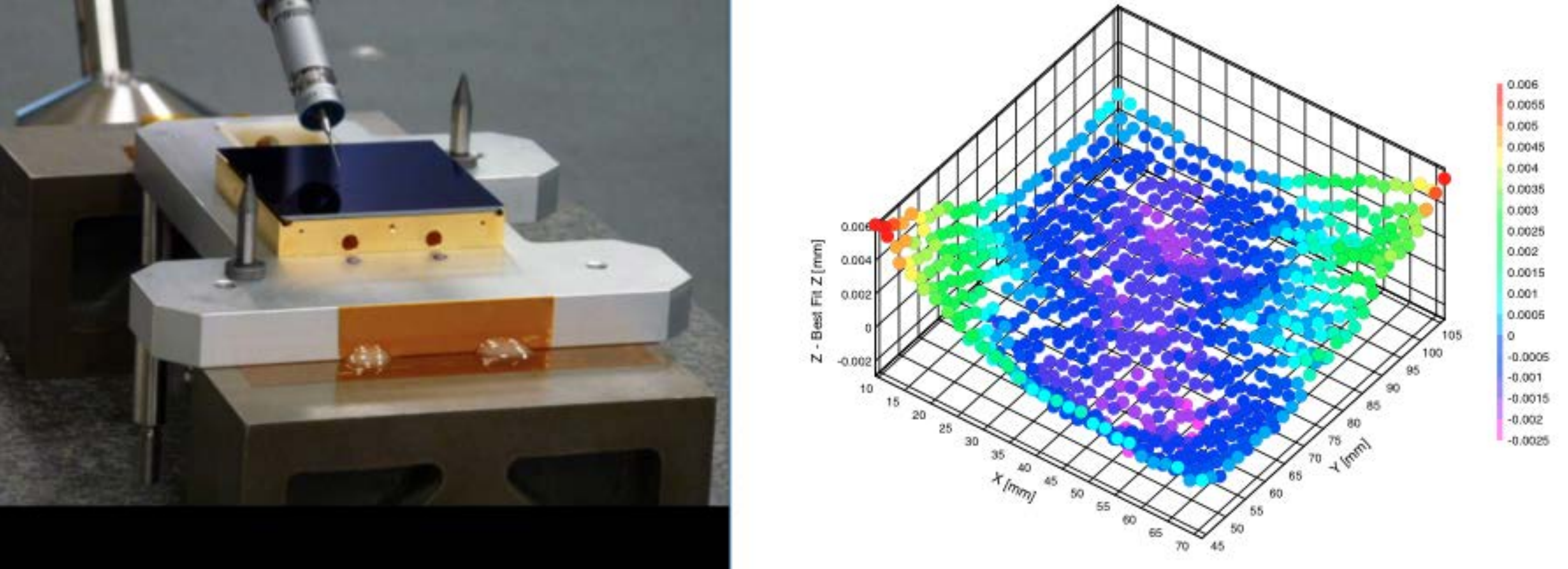}
\caption{Left: CMM touch probe measurements of the package flatness for a completed mechanical grade device. Right: Deviations from the best fit plane on the mechanical grade device. }
\label{fig:CCD_packQA}
\end{figure}

Seven mechanical and engineering grade detectors have been assembled using this fixture at FNAL. 
Room temperature and cryo flatness measurements have been performed.  An example is shown in Figure~\ref{fig:CCD_packQA} where the maximum
deviation from a flat surface is 6~\micron, this is acceptable for the instrument but is still
under study for improvement.

Given the two gluing steps on this process, four working days are required to complete a package. Taking
this into consideration we estimate a packaging production rate of one CCD per week.

\subsubsection{CCD Characterization}
Each LBNL detector packaged at FNAL will be tested against the technical requirements for the instrument. 
The detectors passing the requirements will be fully characterized in the CCD testing laboratory at FNAL. 
This facility was developed during the construction of DECam \cite{Kubik10}. The characterization will include: noise measurements,
linearity measurements for the complete dynamic range of the instrument, QE measurements in the range
between 360~nm and 1000~nm, and dark current. The CCDs will be
exposed to x-rays to characterize the diffusion of the detector and charge transfer efficiency.
The characterization will also include scanning clock and bias voltages around the
nominal operating parameters. 

\begin{figure}[!b]
\centering
\includegraphics[height=2in]{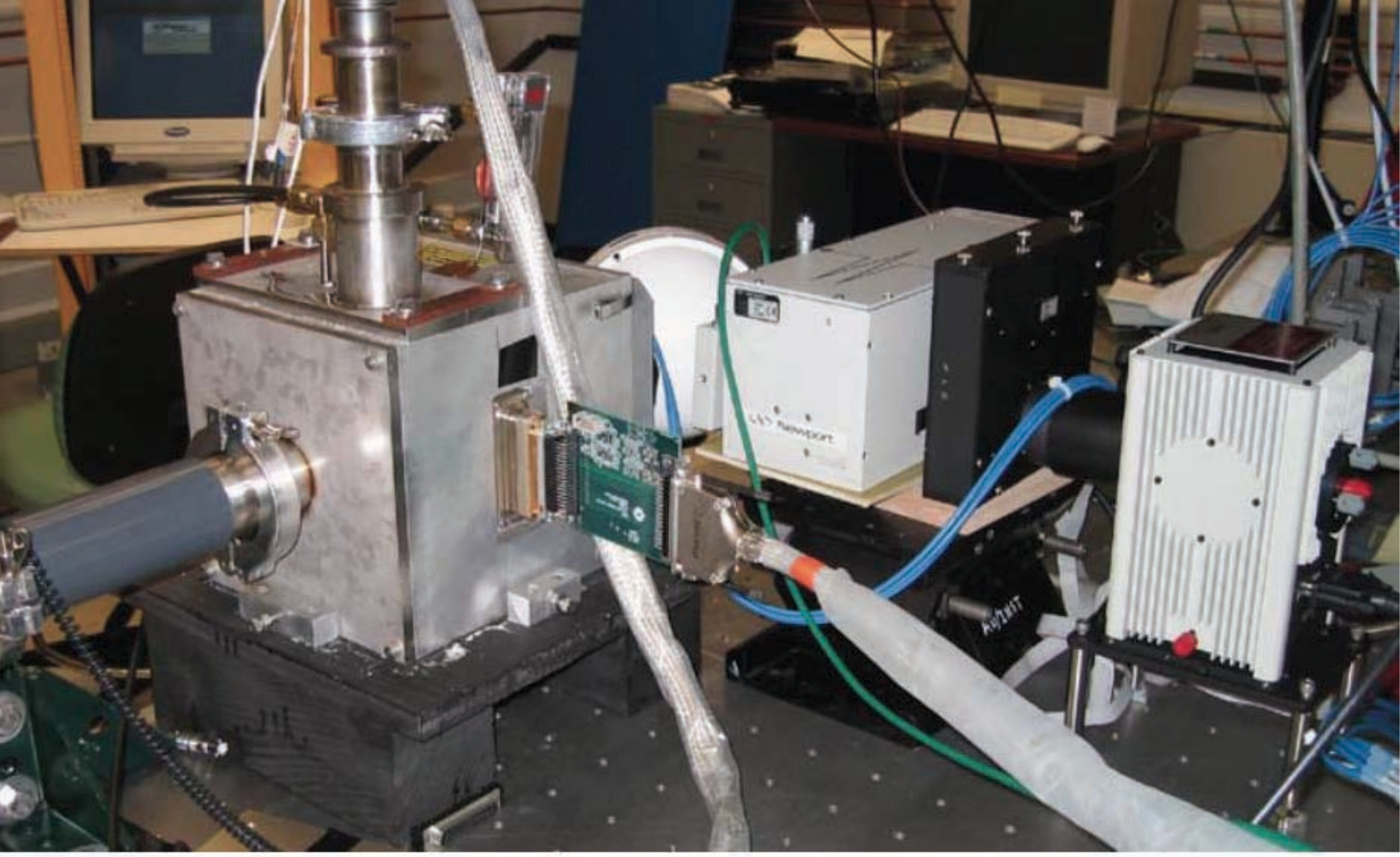} \hspace{0.25in}
\includegraphics[height=2in]{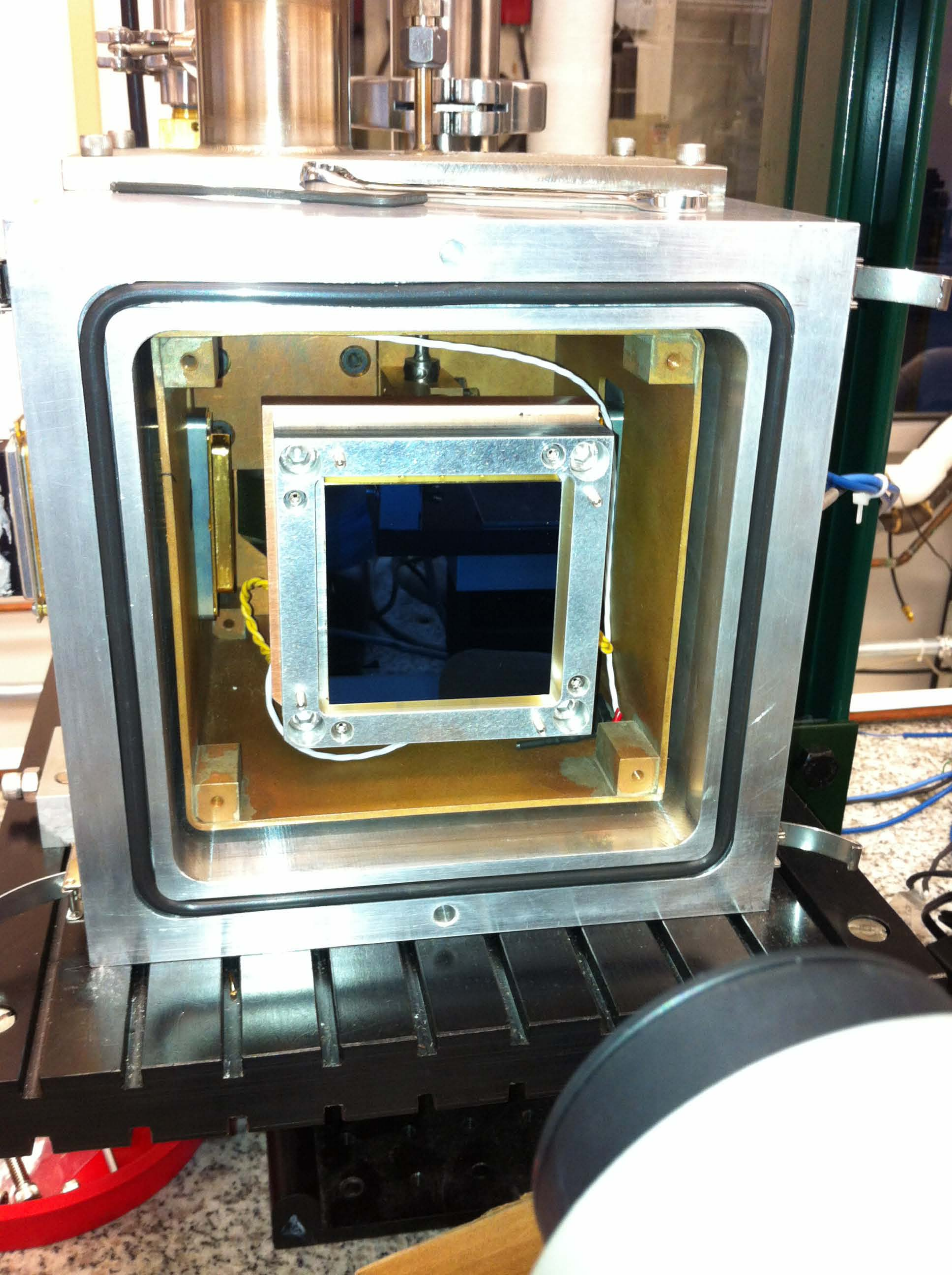}
\caption{Left: Test Station 1 (TS1). The illumination source consists of a power
regulated broad-band lamp, automated monochromator and filter wheel, shutter and integration sphere. An
x-ray source is located inside the vacuum vessel. Right: Test Station 2 (TS2).  TS2 with a mechanical grade 4k x 4k DESI CCD mounted inside the vacuum vessel. The
optical window has been removed to show the sensor inside. The front of the image shows the confocal light source.
}
\label{fig:CCD_TS}
\end{figure}

The baseline plan includes having two dedicated test stations for DESI detectors at FNAL.
Test Station 1 (TS1) will be on an optical bench with a stabilized illumination source, 
coupled to a monochromator with a wavelength range of 360~nm to 1000~nm.  TS1 can
operate with a stable temperature from room temperature to 145K. This test station is shown in
Figure~\ref{fig:CCD_TS} left.

Test Station 2 (TS2) shown in Figure~\ref{fig:CCD_TS} right will be used to measure the flatness of detectors at the operating temperature
of 145K. This measurement will be done using a confocal lamp focusing light on the cold CCD
surface, and capable of determining the reflecting plane with a precision of 1~\micron. This light source is mounted
on a precision X-Y stage to produce a topological map of the detector. 
This technique was successfully used by the FNAL team for flatness measurements in the DECam focal plane \cite{Diehl10}.

The testing cycle for each DESI detector will be about one week long divided in three
stages. Each detector will be installed in TS1 for the initial check against technical requirements. 
This initial test stage (Stage-1) will last about 1 day. After a detector passes Stage-1 testing, a full
characterization consisting of 2 days of data collection will be performed also using TS1, Stage-2.
Stage-3 testing consists of flatness measurements. For this purpose, the detector will be moved into 
TS2 for flatness measurements. These measurements are about 1 day long. 
Detectors completing Stage-1, Stage-2 and Stage-3 testing will be ready for installation in the
cryostats.

Automated software will control the testing of the detectors, in a manner similar to what was
done for DECam CCDs. The CCD testing facility at FNAL has four additional test stations  
with different capabilities typically used for other projects. The additional infrastructure
could be made available to the project if the schedule for DESI
demands a higher rate of testing, or new testing needs develop during the
fabrication process.

\subsubsection{Electronics} \label{sec:frontend_electronics}

The electronics module for each CCD is mounted on the warm side of the cryostat wall. This provides easy access
for replacement without disturbing the detector or spectrograph.
The LBNL-developed modules include local power generation from an isolated single DC voltage, CCD bias voltages generation,
programmable clock levels and pattern, 
CCD signal processing and digitization, and operating settings readback.  The module runs Linux supporting a suite of remotely callable routines for normal and special CCD operations.
Configuration and control of the electronics and delivery of science data
is over Ethernet. 

A level of complexity is introduced into this electronics because of the mixture of n-channel (ITL) and
p-channel (LBNL) CCDs. The CCD output structures required opposite sign DC biasing voltages and the electron-to-voltage
gains are of opposite sign.  The controller uses the same regulators and drive circuits for these with different power rail strapping.
Common clocking circuitry can work for both.  In addition, the LBNL devices 
require a high voltage depletion supply.

\begin{figure}[!htb]
\centering
\includegraphics[height=2.5in]{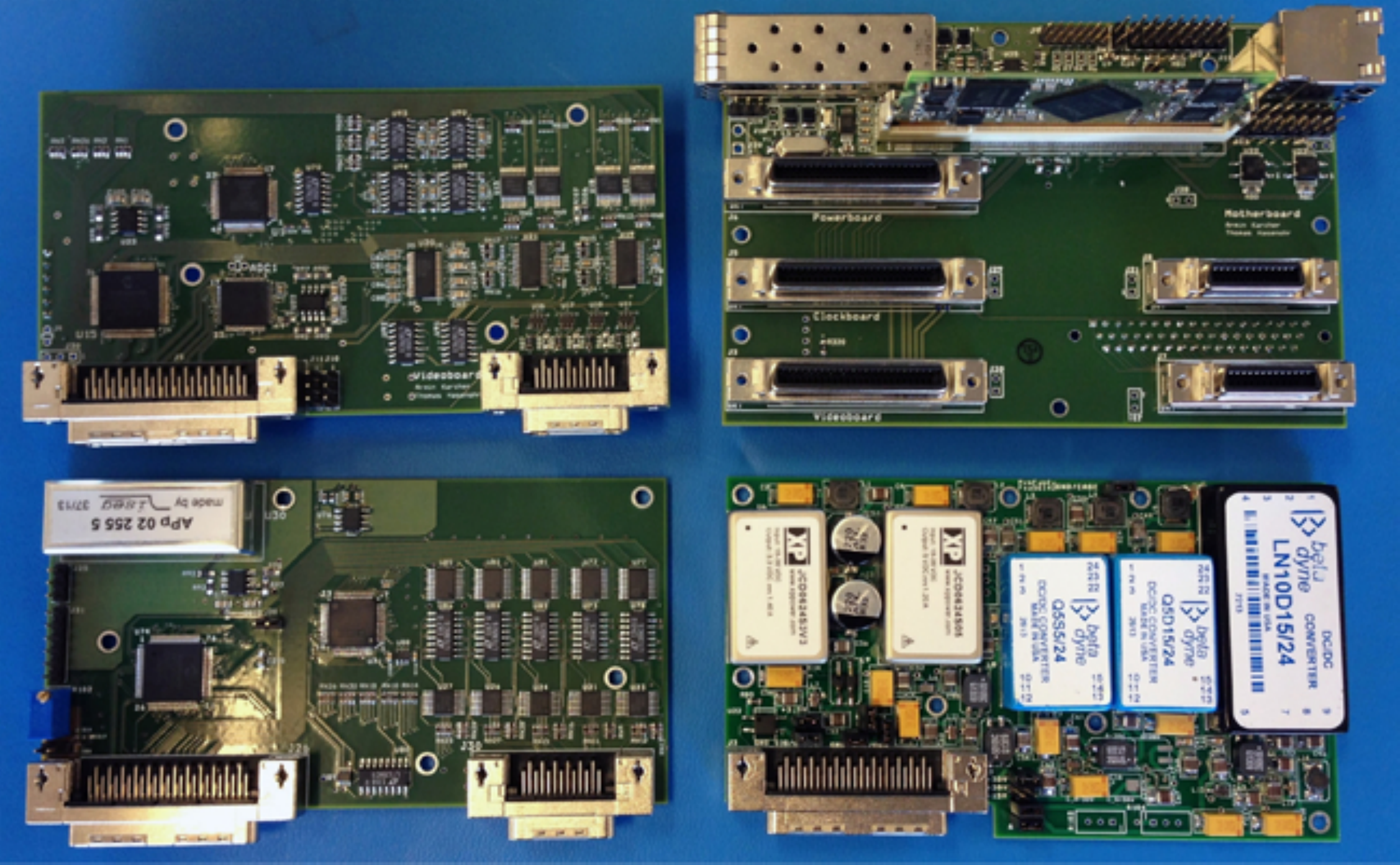}
\caption{CCD frontend electronics module components supporting both n- and p-channel CCDs. Boards clockwise from top right are: backplane with Xilinx FPGA card, network connectors, and connector to the CCD (on the backside);  controller power supplies; clock-level drivers and substrate HV voltage generator; and CCD DC bias voltages generators, preamps, and 4-channel ADC.}
\label{fig:CCD_FEE}
\end{figure}

LBNL has developed a generic four-channel CCD readout controller that supports both flavors of CCD. 
The analog signal processing and digitization is accomplished with an oversampling 100~MHz 16-bit ADC. Correlated double sampling using sums of samples of the pixel reset and video levels are done in an FPGA.  Images are transmitted as FITS files to the Instrument Control System. The production-ready hardware is shown in Figure~\ref{fig:CCD_FEE}.  

Several LBNL CCDs have been operated with system. It has shown equal or better  readnoise performance than a commercial dual-slope integrator controller from 50--300~kpixel/s.  At 100 kilopixel/sec, 1.8~e for the LBNL CCD has been attained.  Configured for ITL CCDs, the controller has  successfully operated  an ITL-provided STA510A $800 \times 1200$ small CCD, achieving the readnoise reported by ITL for the device.


\subsection{Calibration System}
\label{sec:Instr_Calibration_System}

The goal of the calibration system is two-fold,  
i) characterize the spectrograph optics geometrical response: location of fiber traces in the spectrograph CCDs, wavelength solution (wavelength as a function of CCD coordinates per fiber), spectrograph point spread function (PSF), and ii) characterize the response of the instrument: pixel level flat-fielding (correction for the relative variations of CCD quantum efficiency, per CCD pixel and wavelength), and fiber flat-fielding (correction of the relative variation of fiber throughput, per fiber and wavelength). 

For each spectrograph, the pixel-level calibration will be performed with the CCD flat field illuminator which is internal to the spectrograph (see \S\ref{sec:ccd_flat_field_illuminator}). Presently it is anticipated that this will be used once a year since it entails removal of the fiber slitheads. 

The rest of the calibration (spectrograph optics and fiber response) will be performed daily with dedicated calibration exposures using a set of calibration lamps (continuum and line lamps, see \S\ref{sec:calibration_lamps}) illuminating a screen placed on the dome of the Mayall telescope (see \S\ref{sec:calibration_dome_screen}).
Since the DESI spectrographs are mounted on stationary optical benches in a temperature controlled environment, their calibration should remain quite stable, and we do not expect to take calibration exposures during the night. 

In addition, the wavelength calibration accuracy and stability will be monitored with sky lines present in all target spectra (see \S\ref{sec:calibration_sky}). Also, within each exposure, fibers assigned to blank sky locations will be used for measuring the sky spectrum to subtract from object fibers, and  fibers assigned to standard stars will enable an approximate spectrophotometric flux calibration (see \S\ref{sec:calibration_stars}).

\subsubsection{Screen}
\label{sec:calibration_dome_screen}

\begin{figure}[!t]
\centering
\includegraphics[width=0.5\linewidth]{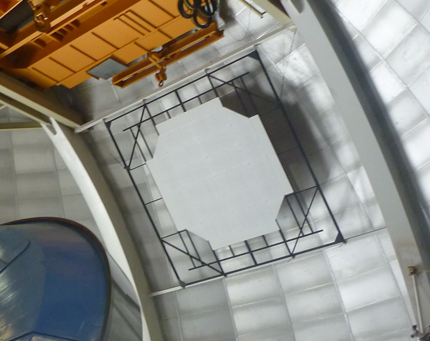}
\includegraphics[width=0.4\linewidth]{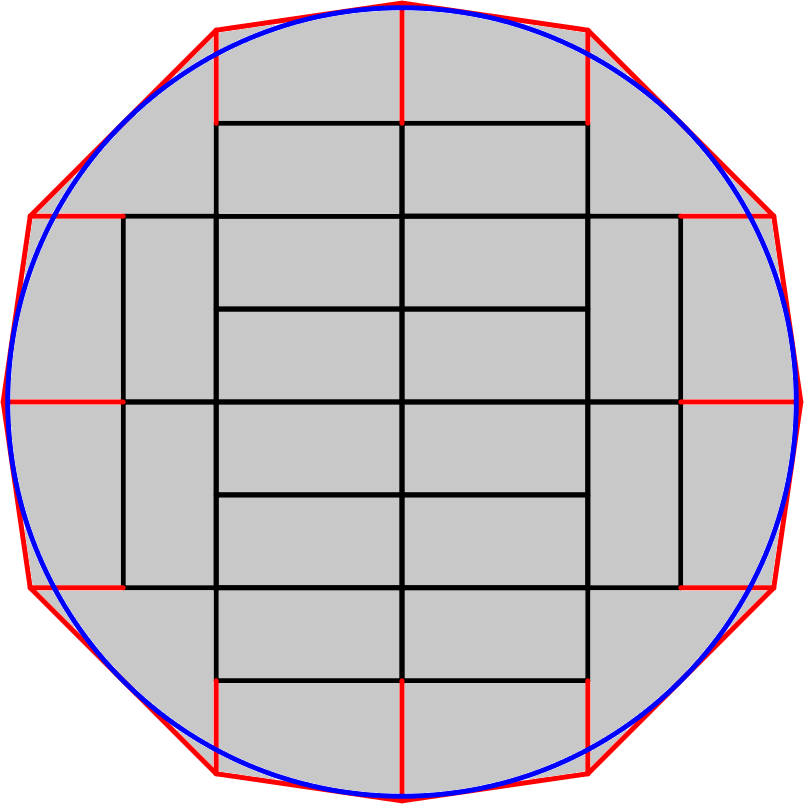}
\caption{Left: Picture of the current screen in the Mayall telescope dome. Right: schematic view of the new screen configuration.}
\label{fig:calibration_screen}
\end{figure}

The current Mayall screen mount will be reused (see Figure~\S\ref{fig:calibration_screen}, left panel) but the screen panels replaced to allow a full illumination of the telescope given the larger DESI field of view. The new design, with a useful diameter of 5173~mm (blue circle in Figure~\S\ref{fig:calibration_screen}) will provide a $\pm 230$~mm margin for lateral dome placement error. The screen panels will be coated with a {\it Permaflect} coating, which reflectance is better than 90\% from 350 to 1100~nm and the surface is nearly Lambertian (diffused intensity in direction $\theta$ proportional to $\cos \theta$); the reflectance is nearly independent of the incident angle\footnote{{\it Permaflect} properties: \url{https://www.labsphere.com/site/assets/files/2130/permaflect_94.pdf}}.

\subsubsection{Lamps}
\label{sec:calibration_lamps}

\begin{figure}[!t]
\centering
\includegraphics[width=0.4\textwidth]{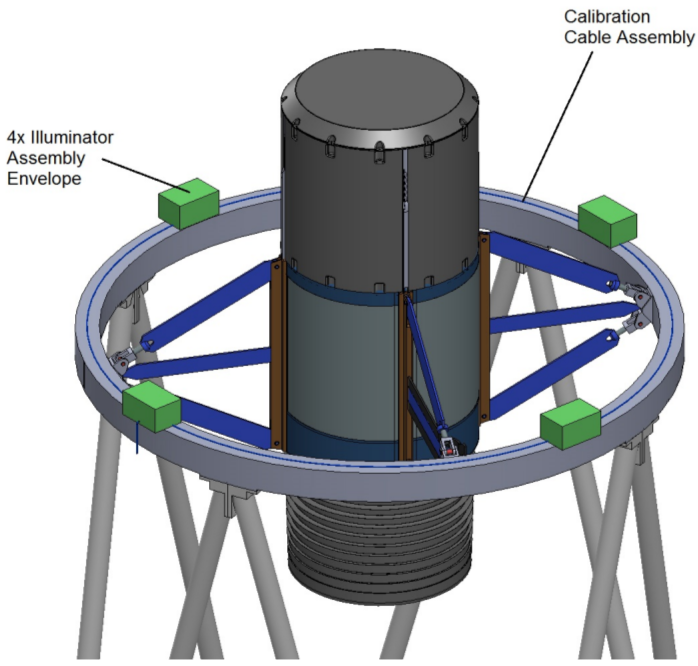}
\includegraphics[width=0.5\textwidth]{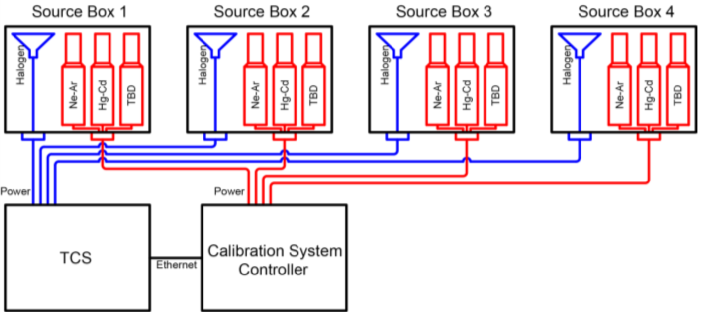}
\caption{Left: Location of the calibration lamp boxes on the upper ring. Right: Schematic view of the interface of the lamp controllers to the telescope and instrument control systems.}
\label{fig:calibration_lamps}
\end{figure}

See Figure~\S\ref{fig:calibration_lamps}

Calibration lamps are used to illuminate the pupil of the telescope. 
They require warm up and stabilization time that is significant compared to a science exposure time.  As such, these will be used during the day to establish a baseline calibration, but they will not be used throughout the night.  
Four boxes containing the same set of calibration lamps (broad-band and emission lines) will be placed on the upper ring of the telescope (see Figure~\ref{fig:calibration_lamps}). Four sets of lamps with isotropic emission give a field uniformity better than 1\% compared to the sky field of view variations (due to plate-scale and vignetting by the prime focus cage assembly), when assuming an ideal Lambertian reflectance screen. This meets the requirement of 5\% (The field uniformity affects the fiber flat-field and subsequent sky subtraction accuracy).

\begin{figure}[!h]
\centering
\includegraphics[width=0.9\textwidth]{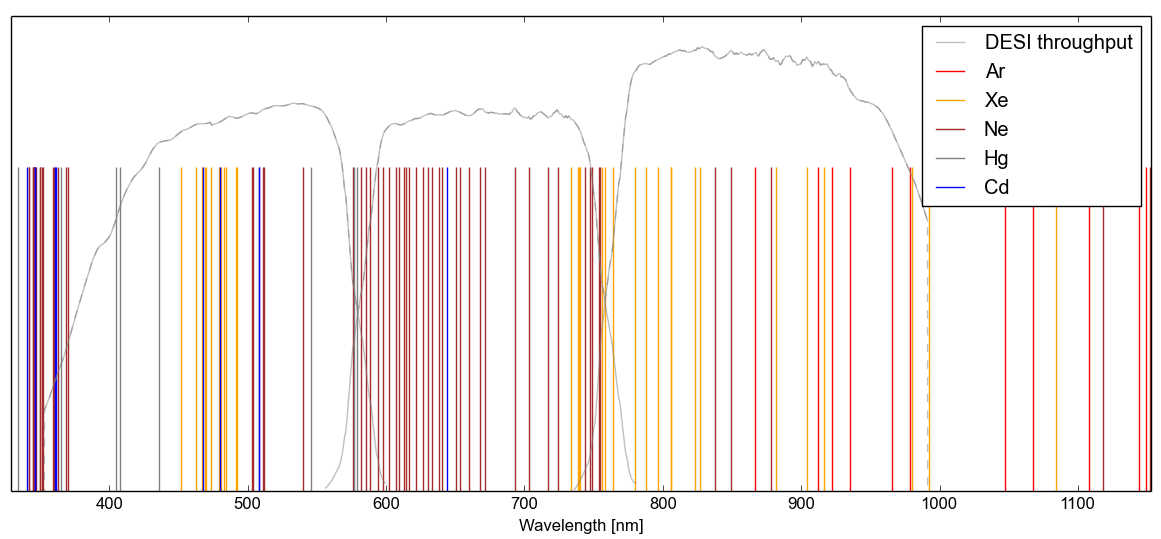}
\caption{Distribution of the main Ar, Cd, Hg, Ne and Xe lamp lines compared to DESI throughput.}
\label{fig:calibration_lamp_lines}
\end{figure}

Possible emission line lamps is currently being studied among a choice of low-pressure gas discharge lamps, with intensity adapted to DESI illumination system (with size, weight, heat dissipation and power consumption constraints). A set of lamps of Ar, Cd, Hg, Ne and Xe emission lines are needed to cover with a sufficient density the DESI wavelength range (see Figure~\ref{fig:calibration_lamp_lines}).
Each box will contain 3 or 4 line emission lamps and a continuum lamp (\eg, quartz iodine halogen) which will provide a mostly smooth spectrum that evenly illuminates all fibers, along with the required controllers interfaced to Telescope and Instrument Control Systems (see Figure~\ref{fig:calibration_lamps}, right panel).

\subsubsection{Sky Lines}
\label{sec:calibration_sky}

All spectra of all science exposures will unavoidably have multiple sky lines.
Fibers assigned to blank sky locations distributed throughout the focal plane
will be used to measure the sky on an exposure-by-exposure basis.
Each observation guarantees 40 sky spectra per spectrograph per exposure.
These are used for the detailed wavelength calibration and modeling
of the sky spectra to subtract from the science target spectra.

Since the sky lines are resolved and have known wavelengths, they will be
used to monitor the stability of the spectrographs (Section~\ref{sec:onlineQA}).

\subsubsection{Stars}
\label{sec:calibration_stars}

Standard stars will be selected as main-sequence F-stars based upon
color and magnitude cuts, with the detailed selection dependent upon
the available targeting datasets.  These stars have well understood and
mostly smooth continua that may be modeled using stellar templates with
temperature and surface gravity as tunable parameters.  By comparing the
observed standard star spectra with their modeled spectra and photometry 
(from the targeting data), a spectrophotometric solution is 
derived.  This solution is modeled across the focal plane to provide 
 spectrophotometric calibration to the spectra. This spectrophotometric solution describes the calibration of counts on
the detector to flux on the sky, which includes the instrument throughput and
atmospheric absorption bands.

\subsubsection{CCD Flat Field Illuminator}
\label{sec:ccd_flat_field_illuminator}

CCD pixel-level response will be measured using a flat field illuminator
as described in Section~\ref{sec:spectro_mech_elec_design}.  This mechanism
evenly illuminates the CCD in the spatial direction with a smooth spectrum such that each pixel
receives the same wavelength of light as it would from a science spectrum.
By comparing each pixel to the median of other pixels with the same
wavelength, a wavelength-dependent pixel level response is measured.
This is used by the raw data preprocessing to make pixel-level
response corrections.  BOSS employs a similar mechanism to calibrate its
spectrograph CCDs, taking measurements approximately once per year or
after any operation which breaks the CCD dewar vacuum.  DESI also plans
yearly CCD flat field calibrations since the illuminator mechanism requires
removing the slit heads.

\clearpage

\section{Instrument Readout and Control System}
\setcounter{equation}{0}\setcounter{figure}{0}\setcounter{table}{0}
\label{sec:Instr_Control_System}

The design of the DESI online system is based on the readout and control system architecture developed for the Dark Energy Camera \cite{Diehl2015}. This system was deployed on Cerro Tololo in 2012, and has been used successfully for both the DES survey and the community observing program. A detailed description of the DECam data acquisition system can be found in \cite{SPIEhonscheid2012} and \cite{SPIEhonscheid2014}. Important components of the DESI online system such as the dynamic exposure time calculator, the real-time data quality assessment and complex algorithms to convert on-sky target coordinates to fiber positions on the focal plane are based on the SDSS-III/BOSS online system. The DESI instrument control system (ICS) is tasked to
perform all control and monitor functions required to enable successful operation of the DESI instrument and completion of the
DESI science program.
The requirements for ICS are:

\begin{itemize}
  \setlength{\itemsep}{1pt}
  \setlength{\parskip}{0pt}
  \setlength{\parsep}{0pt}
\item Performance and Design
\begin{itemize}
\item In order to ensure maximum survey efficiency the ICS shall complete all activities between exposures on average in less than 120 seconds. A design goal shall be to achieve this in 60 seconds. A lower limit is set by the CCD readout time of 42 seconds.
\item  The ICS shall be robust and modular in the sense that the system shall meet operating requirements subject to total failure of any single computer or disk drive. Dedicated hardware interfaces are excluded from this requirement.
\item  The ICS hardware and software shall be designed so that system up-times of greater than 97\% can be achieved no later than the end of the first year of the 
survey.
\item The ICS architecture shall support small-scale test systems and sub-component integration tasks.
\end{itemize}
\item Monitor
\begin{itemize}
\item The ICS shall continuously monitor the operational parameters of the DESI instrument. The values of these parameters and other telemetry information shall be archived in a database for offline analysis.
\item The ICS shall provide both a web-based and a programmatic interface to the telemetry database.
\item The ICS shall provide a connection to the existing Mayall monitor systems to collect and archive environmental information. 
\item The ICS shall provide an alarm and error notification system. All alarm and error messages generated by the instrument shall be archived.
\item The ICS shall not be responsible for the hardware protection of DESI components.
\end{itemize}
\item Control
\begin{itemize}
\item The ICS shall provide all user interfaces required by the observer to operate the instrument. Secure remote access shall be supported.
\item The ICS shall interface with the Mayall telescope control system to automatically position the telescope to ensure high survey efficiency.
\item The ICS shall provide sufficiently fast and reliable communication paths between DESI sub-components. This includes the connection between the guider 
(GFA) and the telescope control system and between the alignment system (GFA) and the hexapod.
\item The ICS shall include a dynamic exposure time calculator.
\item The ICS shall provide the process to map the target coordinates received in RA/DEC format from data systems to focal plane coordinates that can be loaded into the fiber positioner.
\end{itemize}
\item Data Flow
\begin{itemize}
\item The ICS shall provide sufficient network bandwidth to retrieve the data from the front-end electronics without causing any additional deadtime and without 
extending the time between exposures.
\item The ICS shall format the data in multi-extension FITS format with a set of DESI specific header keywords. The images will be compressed with the Rice/tile algorithm before they are transferred off of the mountain. 
\end{itemize}
\item Online Hardware
\begin{itemize}
\item The ICS project shall be responsible for all DESI computer hardware at the Mayall. This shall include the online computers, disk storage, the internal network hardware and two multi-display workstations for the Mayall console room.
\item The ICS shall provide sufficient storage capacity to continue to operate DESI for at least three nights should the data transfer links off the mountain fail.
\end{itemize}
\end{itemize}

\begin{figure}[!b]
\centering
\includegraphics[width=0.75\textwidth]{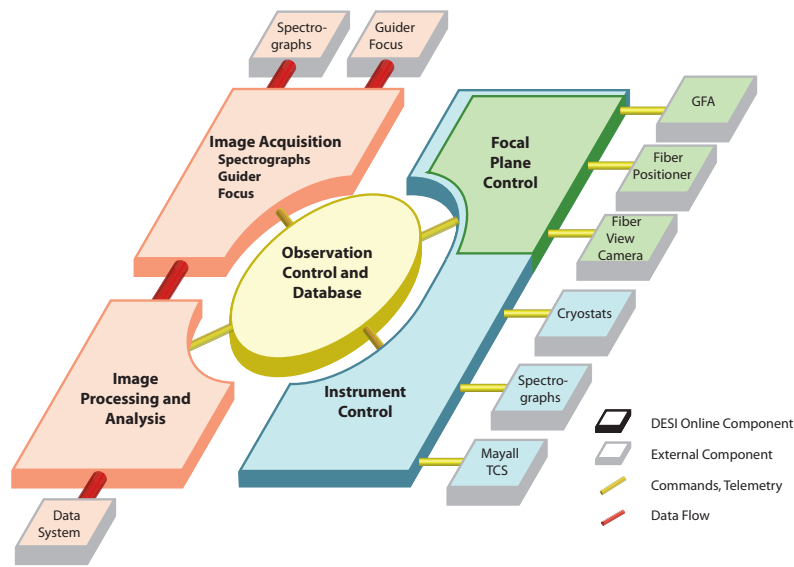}
\caption{Schematic view of the DESI readout and control system. The observer console and other user interfaces are not shown.}
\label{fig:daqoverview}
\end{figure}

The design of the DESI online system is based on these requirements. The overall architecture is shown schematically in Figure~\ref{fig:daqoverview}. At the core of the system sits the observation control system (OCS) that 
orchestrates the complex DESI exposure sequence. It interacts closely with the focal plane systems including the fiber positioner, the fiber view camera, and the guider, focus and alignment system. Data is acquired from the guider and the front end electronics in the spectrographs and flows through the image processing stage to a storage device from where the images are picked up by the DESI Data Systems group. The connection to the Mayall telescope control system (TCS) and the Mayall telemetry database is provided by the TCS Interface component also shown in Figure~\ref{fig:daqoverview}.
The readout system is tightly integrated with the DESI instrument monitoring and control system (IMCS). The IMCS monitors every component of the instrument  and detailed information about instrument status, operating conditions and performance will be archived in the DESI operations database. The monitoring and control system is designed to provide all components of the instrument access to the telemetry and alarm history databases even if the rest of the OCS and the readout system is offline.
The final component of the DESI online system (not shown) is a set of web-based user interfaces that includes the observer console, alarm history as well as telemetry and status displays.

Our detailed description of the DESI readout and control system begins in the next section with a discussion of the OCS followed by a review of the DESI exposure sequence and the activities that need to be coordinated by the online system between exposures. The  dataflow architecture, the operations database, the interface to the Mayall telescope, and the instrument control and monitoring system are covered in the next sections.
We conclude with a discussion of the online software we are developing for DESI, a description of the hardware configuration and the integration and testing plans.

\begin{figure}[!p]
\centering
\includegraphics[width=1.3\textwidth, angle=90]{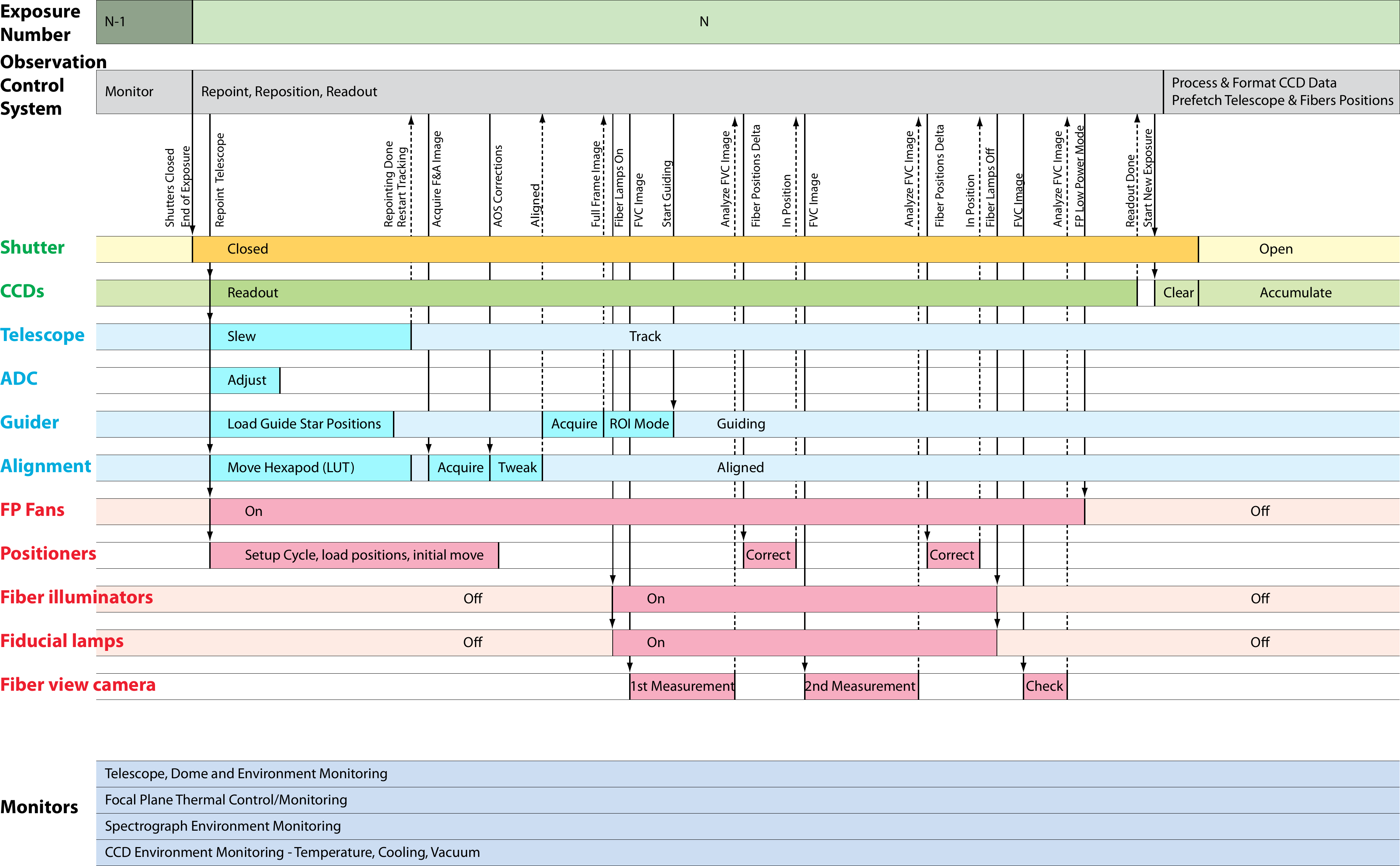}
\caption{ A typical DESI exposure sequence timeline. Lengths of the bars do not represent activity durations.}
\label{fig:timeline}
\end{figure}

\subsection{Observation Control}
\label{sec:OCS}

The OCS is the central 
component of the DESI readout system coordinating all aspects of the observation sequence. We have adopted a multi-threaded, pipelined architecture to maximize throughput. Exposure requests are handled by the OCS QueueManager and the Next-Field-Selector script that will be developed by the Data Systems group. The format for exposure requests has been defined (DESI-0560) and includes the 5,000 fiber coordinates, the pointing coordinates for the telescope, a list of guide stars, an expected exposure time and some book keeping information. The QueueManager can handle multiple requests, supports exposure scripts and provides a number of functions to manage the exposure request queue. The Exposure Sequencer is at the core of the OCS. It receives exposure requests from the QueueManager and orchestrates the complex DESI exposure sequence as described in the following section.
At the end of an exposure the OCS will initiate readout, digitization and image processing including quality assessment. Once the exposure has been written to a disk cache, the OCS notifies the data transfer 
system \cite{SPIEfitzpatrick2010} developed by the Data Systems group that image data are available to be transferred to the NOAO archive and the DESI processing site(s).

In addition, the OCS maintains the overall status of the DESI instrument and the online system. Software interlocks are used to monitor the readiness of the components that participate in the exposure sequence. An exposure can only be started when all required interlocks are set. When an interlock breaks or another error condition is detected, an alarm is issued to alert the observer.  

The architecture of the OCS is based on a well-tested design we developed for the Dark Energy Camera readout and control system. We have adopted this design to support the spectrograph tests at Winlight, France and a more advanced prototype is under development for the ProtoDESI test run later in 2016.

\subsection{Exposure Sequence}
\label{sec:expsequence}

A typical DESI exposure sequence is shown in Figure~\ref{fig:timeline}. The observation control system (OCS) is responsible for coordinating the different activities. In order to maximize survey throughput we will employ a pipelined architecture and set up for the next exposure while the previous image is being digitized and read out.

When the Dynamic Exposure Timer described in Section \ref{sec:etc} informs the OCS that the accumulation period of 
an exposure has ended, the OCS signals the spectrographs to close the shutters and instructs the frontend electronics to read out the CCDs. 

Information about the next pointing, target coordinates and guide star data has already been received from the Next-Field-Selector during the previous accumulation phase. Once the spectrograph shutters are closed, 
the OCS transmits the coordinates of the new field to the telescope. 
DESI uses a hexapod to control focus and alignment of the instrument. 
While the telescope slews to the new position, a lookup table is used to retrieve base values for all six hexapod parameters. These are combined with accumulated correction signals obtained by the active optics component of the GFA and sent to the hexapod to initiate the adjustment. The ADC is instructed to update for the new telescope pointing. The focal plane fans are turned on.
The PlateMaker application (see Section~\ref{sec:platemaker}) uses an optical model of the DESI corrector and corrections based on the previous exposure to convert the 5000 on-sky target coordinates to focal plane coordinates. These initial coordinates are sent to the fiber positioner to perform the initial move to the new configuration.

Once the telescope  and the hexapod are in position, the four focus and alignment cameras of the GFA system take a full-frame exposure. The out-of-focus star images (donuts) are analyzed to determine the wavefront and small adjustments are sent to the hexapod (``tweaks'') to complete telescope and focal plane alignment.

With the alignment complete, full-frame exposures are taken by the six guide cameras of the GFA system.
Using the astrometric solution derived from these images, the PlateMaker determines the offset between actual and requested telescope position. 
If a large offset is detected, a move command is sent to the Mayall telescope control system to correct the pointing and another set of full frame exposures is taken. This sequence is repeated until the telescope pointing is within 20 arcsec of the requested position. At this point in the exposure sequence, 
the guide CCDs are switched from full-frame to region-of-interest mode using pre-selected guide stars and the telescope tracking feedback loop is closed. Via the guide corrections the telescope position is tweaked to correct the residual offset and to match the desired target location. During an exposure, guider correction signals are sent at a rate of about 1~Hz which is well matched to the characteristics of the Mayall TCS.  

The PlateMaker refines the mapping of target coordinates to focal plane coordinates using telemetry information from the instrument control system and astrometry information obtained from the final set of  full frame guide camera exposures. When the fiber positioner signals completion of the initial move, the fiducial and fiber illuminators will be turned on and the fiber view camera acquires a first snapshot of actual fiber locations. The Fiber Positioner Control System, another component of the OCS, receives the updated target coordinates in the focal plane system from the PlateMaker, determines the corrections for each actuator and sends the new coordinates to the positioner system. One or two additional cycles will be required to complete the positioner setup. The maximum number of iterations is bound by the requirement to keep the time between exposures below 120 s (60 s). At the end of the fiber positioning cycles, the final fiber position is imaged,  recorded and then the focal plane fans as well as the fiber and fiducial illuminators are switched off. The fiber view camera takes a last image to verify that all light sources are off before the next exposure begins. 

Concurrent with these operations the OCS waits for CCD readout complete messages from all spectrograph cameras. With the readout complete and the fibers, telescope, hexapod and ADC in position 
the OCS sends an open command to the shutters and the next accumulation phase begins.
While the spectra are being acquired information about the next exposure, including telescope coordinates and target positions, is loaded into the OCS, and the cycle repeats. 
In the current baseline design we will not adjust the hexapod alignment or the ADC settings during an exposure but the OCS design provides sufficient flexibility to accommodate this should the need arise in the future.

The DESI DAQ system is designed to complete the entire sequence outlined above of activities between exposures in less than 120 seconds with a goal to accomplish these tasks in as little as 60 seconds.
Several of the key components that participate in the exposure sequence are discussed in following sections. 

\subsubsection{Active Optics System}
\label{sec:aos}
The Active Optics System (AOS) will adjust focus and alignment of the DESI corrector using out-of-focus stars in the four F\&A cameras of the GFA system. Each of these four cameras will contain a 30~mm by 15~mm CCD, arranged so that half of the sensor is intra-focal by 1.5~mm and half extra-focal by 1.5~mm. 
 The resulting out-of-focus stars (donuts) are images of the exit pupil distorted by optical aberrations.  The AOS will analyze an ensemble of donuts, with a forward fit of the images to a pupil-plane Zernike expansion. The measured aberrations are then used to infer deviations from optimal focus and alignment.  
The AOS will be based on the highly successful system developed for DECam \cite{SPIEroodman2010, SPIEroodman2012, SPIEroodman2014}, and will reuse almost all of the relevant online and offline code.  The DECam AOS is operating reliably, unsupervised and in closed-loop for all five degrees of freedom (focus, hexapod x and y decenter, tip and tilt - DECam does not use hexapod rotation).  Typical control of focus is maintained to better than 30~\micron, decenter to 300~\micron, and tip/tilt to $15$~arcsec. These values are likely limited by the repeatability of the mechanical alignment between the primary mirror and prime-focus camera, or by time-varying distortions caused by in-dome seeing, but not by the measurement error of the AOS itself.  As such, similar quality performance may be expected for DESI.  

There are some differences between the AOS and the donuts for DESI compared to DES, and each will require some small changes to the existing code.  First, the code which locates donuts in the GFA image, {\it dfdFinder}, has all hard-coded parameters and will need to be tuned for the GFA CCDs, to yield good donut finding efficiency. Next, the donut fitting code, {\it donutengine}, will need a detailed model of the DESI corrector's pupil function, accurately characterizing the vignetting for each of the four GFA cameras. The DESI pupil function is actually much simpler than the DES one, which must model the complicated inner obscuration.  The non-linear fitting code, {\it donutfit}, and the donut analysis code, {\it donutana}, will be reused unmodified.  The donut analysis code will, however, require the appropriate calibration data for DESI.

Several AOS calibration products will be required.  First, a series of images must be taken, dithered in a region with appropriate stellar density, to produce a map of the aberrations across the GFA cameras.  The DESI optical model predicts a significant change in astigmatism, coma and trefoil across the GFA cameras, and these variations must be characterized using measurements to account for differences between the as-built system and the idealized optical model. Second, the effect of focus and hexapod alignment on the astigmatism and coma must be measured, with images taken over a range of focal settings and hexapod decenters and tip/tilts. This data will be used to infer the 4x4 sensitivity matrix between astigmatism (or more accurately the focal plane slope of astigmatism) and coma and the hexapod decenter and tip/tilt.  Once these calibrations are complete a look-up table of the typical focus and hexapod alignment across the sky may be built.  Finally, a look-up table of the astigmatism of the Mayall 4-meter primary mirror can be constructed as well. 

The design of the DESI active optics system is complete and the required changes to the existing DECam code have been identified. Due to the lack of a hexapod, the AOS and donut software is not required for the ProtoDESI test later in 2016 and the bulk of the AOS coding and development work has been scheduled for 2017 and 2018.

\subsubsection{Plate Maker}
\label{sec:platemaker}
The function of the PlateMaker application is to map the target
coordinates received in RA/DEC format to focal plane coordinates that
can be sent to the fiber positioner.
This task must correct for mechanical and optical distortions using
a combination of guider images, Fiber View Camera (FVC) images, and a set
of fiducials at known location in the focal plane.  Ancillary information
from the ICS such as temperature, hexapod settings, and telescope pointing
information are also needed.

The PlateMaker task has six modules, which function as follows:

\begin{enumerate}

\item Target positions:  The target selection database will provide J2000
coordinates along with other information for 5000 targets and a list of
stars that overlap the guider CCD fields; additionally, the field
center, telescope coordinates, UTC, and focal plane temperature are
provided.  The PlateMaker receives this information via the Next-Field-Selector and the OCS. It is then used to apply precession, nutation,
aberration, refraction, and polar axis misalignment to produce apparent
coordinates for all objects relative to the apparent field center.

A nominal distortion pattern is applied to the target positions and sent
to the fiber positioner for the initial positioning.

\item Field Acquisition images:
Once the telescope has finished slewing and settled,
and the ADC has rotated to the desired position,
a set of field acquisition exposures made using the guide CCDs
is obtained.  These exposures
are analyzed to identify
stars, and the star instrumental coordinates (corrected for a nominal
distortion pattern) are matched to the star lists
from the target selection DB in order to determine the sky position,
scale factor, and any residual position angle offset for each guider
CCD.  This information is combined with knowledge of
the focal plane temperature in order to determine the absolute sky coordinates
of each pair of fiducials attached to each GFA assembly.

\item FVC image:  In parallel with obtaining the guider images and once
the fiber positioners are at the nominal target location, an image is
taken with the FVC of the back-illuminated target and fiducial fibers.
The image is analyzed to produce x, y spot positions of each fiber.  These
spot positions are matched with a pattern-matching algorithm to identify
which fiber corresponds to a particular spot.

\item FVC astrometric calibration: This small module takes the measured
spot positions of the GFA fiducials and matches them with the absolute sky
positions determined in step 2.  These matches are used to determine an
absolute astrometric calibration of the FVC focal plane.  If the FVC
camera lens and/or CCD introduce distortion,
it is presumed that this distortion has been calibrated in the laboratory,
since it cannot be determined at the telescope.

\item Distortion calculation: The spot positions of all the fiducials
are matched to their known positions (adjusted for focal plane temperature)
and used to determine coefficients for the distortion introduced by the
DESI corrector and ADC.  This mapping occurs from DESI focal plane to the
FVC focal plane and the reverse.

\item Calculate fiber positions:  At this point the apparent target positions
are mapped to the FVC focal plane and matched to the appropriate spot
position from the FVC image.  Errors are calculated, and corrected
positions are sent to the fiber positioner system.

The last few steps are iterated until all fibers are in position.

\end{enumerate}

\subsubsection{Guider}
\label{sec:guider}
The DESI guider system provides correction signals to the Mayall
telescope control system (TCS) to maintain a stable and accurate position
during science exposures. It consists of six guide cameras described in detail in Section~\ref{gfa_section} and code for star finding and centroiding algorithms that we call the guider.
The guider's task is to analyze the images taken by the GFA detectors,
to determine the position of the guide star centroids and to derive
corrections signals that are sent to the TCS. Tracking uncertainties
have to be less than 0.03~arcsec.

A detailed study of the GFA cameras showed that signal-to-noise ratios better than
10 can be achieved with stars brighter than magnitude 17.5 in R band. Six of the ten
DESI GFA modules are dedicated to guiding on up to 24 guide
stars, as the electronics will allow to configure up to 4
region-of-interests (ROI) per detector. The actual number of guide
stars used in an exposure will depend on the availability of enough
high-S/N stars. Combining all guider sensors we expect to have an average of 11 stars that meet the
target S/N even for the worst-case scenario near the galactic pole.

The DESI guider distinguishes two stages: the acquisition stage where guiding stars are selected and the tracking stage where guide stars are continuously monitored and corrections are sent to the telescope.
Several different operating modes have been implemented. All share the same code for the tracking stage but employ different acquisition strategies.
By default, the guider operates in catalog mode, where guide stars have been preselected from an external catalog. In self mode,  the guider analyzes the entire full-frame GFA images, and automatically selects the best guide stars, based on signal-to-noise, isolation, saturation,  and location with respect to the chip boundaries.

As part of the DESI exposure sequence, the  PlateMaker converts the guide star positions received as part of the exposure request to pixel coordinates for each of the 6 GFA sensors. This information is then submitted to the guider. The guider determines suitable regions of interest around the stars and uploads the configurations to the GFA cameras. With the acquisition phase complete, the GFA cameras start to produce guide star images at a configurable rate, typically around 1 Hz.  The guider enters tracking mode continuously analyzing the guide star images. Shifts in the overall centroid position are determined and error signals are sent to the Mayall TCS to correct telescope tracking.

The guider supports a shift function to change the reference centroid position. DESI will employ this functionality for the final adjustment of the telescope pointing to match the requested target coordinates. When the difference between the measured centroid position and the reference position goes below an adjustable threshold, the guider signals the OCS that the telescope is in position and that the next exposure can begin.

During the tracking phase, the first step in each iteration is to locate the guide stars in each of the ROI postage stamps. In order to avoid losing or mismatching a reference guide star, the guider algorithm always chooses the star with position and flux most similar to the reference guide star. For each of the guide stars found, the centroid is measured and the offset to the reference position is determined. Given the different orientations of the GFA cameras, the guider uses WCS information to convert the individual positions to world coordinates in the equatorial system.
As higher signal-to-noise stars provide more accurate centroids, the guider weighs each offset by the corresponding star's signal-to-noise before calculating the final offset that is sent as error signal to the telescope control system. In order to avoid combining spurious detections, the algorithm will analyze the agreement/discordance among the different guide star offsets (from 3 to 24) to discard the outliers and those stars that provide very inaccurate centroids. 

As guide star images are analyzed during the course of a science exposure, additional information such as atmospheric seeing, transparency and sky levels is extracted. This information is used by the exposure time calculator to determine the optimal exposure time that will maximize overall survey throughput.

\subsubsection{Exposure Time Calculator}
\label{sec:etc}
Most DESI targets will require only a single exposure. In order to maximize survey throughput, exposure times cannot be fixed in advance and must adapt to the changing conditions that determine the integrated signal-to-noise ratio. The online exposure-time calculator (ETC) is responsible for estimating the optimal exposure time during the open-shutter integration time, using real-time updates from the guide sensors.

The ETC is initialized for each exposure using the following inputs obtained via the DESI Online System:
\begin{itemize}
\item Target integrated signal-to-noise ratio (SNR) for an average spectroscopic target.
\item Configuration data for the Guide and Focus Array (GFA), including:
\begin{itemize}
\item Magnitudes and colors of guide stars that will be read out.
\item GFA camera readout cadence and exposure time.
\item Parameters for temperature dependence of GFA camera dark current.
\end{itemize}
\item Exposure $t_0$ when all targets should nominally be centered on their fiber and refraction effects should be completely removed by the Atmospheric Dispersion Corrector (ADC).
\item Telescope pointing in RA and DEC and corresponding nominal airmass.
\item Predicted sky spectrum based on recent calibrated sky-fiber observations. This last item is desirable but not strictly required.
\end{itemize}
The inner loop of the ETC consists of ingesting calibrated GFA images and sensor temperature measurements at about 1 Hz and deriving estimates of the corresponding incremental spectroscopic signal and background. Specifically, the ETC uses the following GFA outputs from each iteration of the guide loop:
\begin{itemize}
\item Up to 24 calibrated postage stamp images containing guide stars. These are used to estimate the focal plane PSF size and shape and to monitor relative changes in atmospheric transparency.
\item Pixel data for regions containing no stars or bright galaxies. These are used to estimate the combined sky level and dark current and could either come from the margins of sufficiently large postage stamps containing stars, or else from dedicated background stamps. We estimate that a total signal-free area of $\sim$1000 pixels per GFA exposure will be sufficient.
\item Measurements of each GFA's sensor temperature. These are used to separate the sky and dark-current contributions to the background estimate. We estimate that an absolute rms temperature measurement error of 1.0~\celsius is sufficient.
\end{itemize}

\begin{figure}[htb]
\begin{center}
\includegraphics[width=0.9\textwidth]{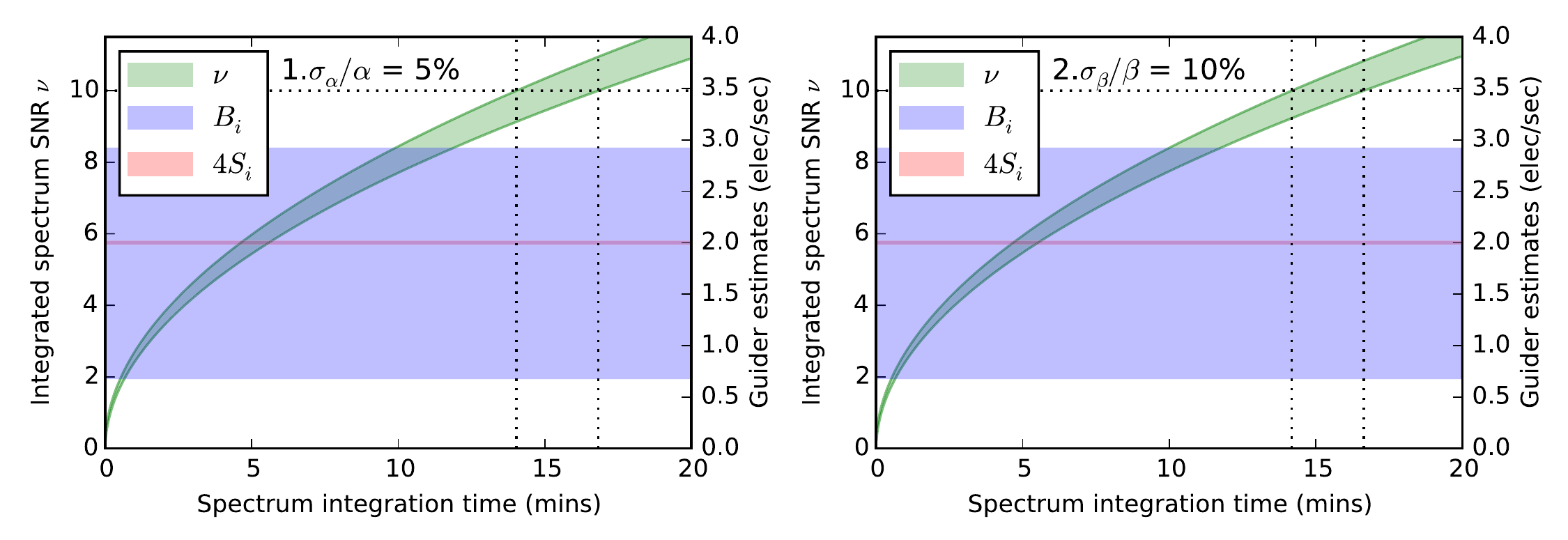}
\caption{Projected evolution of background (blue), signal (red) and integrated SNR (green) estimates during a 20-minute spectroscopic integration. The left-hand plot uses $\sigma_\alpha/\alpha = 0.05$ and the right-hand plot uses a larger error $\sigma_\beta/\beta = 0.10$ that gives a comparable error on $\sigma_\nu/\nu$. The width of each band shows estimated 1-$\sigma$ errors, including the (red) signal curve where the error is not visible.  Background (blue) and signal (red) estimates are per guide exposure and use the right-hand axis, in units of detected spectroscopic electrons per second.}
\label{fig:etc_forecast}
\end{center}
\end{figure}

DESI-1100 provides a detailed error analysis with forecasts of the expected ETC performance under various assumptions. The relevant quantities are the estimated integrated SNR $\nu$ and its error $\sigma_{\nu}$, with
\begin{equation}
\frac{\sigma_\nu^2}{\nu^2} \simeq \frac{1}{n}\left[
\frac{1}{4} \left(\frac{\sigma_B}{B}\right)^2 + \left(\frac{\sigma_S}{S}\right)^2\right] \; ,
\end{equation}
where $n$ is the number of accumulated guider exposures, and $S$ and $B$ are the independent estimates of the accumulated spectroscopic signal and background.  We find that the relative error on $S$ is negligible compared with the relative error on $B$, but that both are sufficiently small under reasonable assumptions that they are not likely to dominate the overall ETC error. Instead, the cross calibrations between the guider and spectrograph responses to signal ($\alpha$) and background ($\beta$) are expected to dominate. Figure~\ref{fig:etc_forecast} shows a sample forecast for the integrated SNR during a 20-minute exposure with different assumptions about these cross-calibration errors. Overall, we expect that the ETC will be able to determine the optimal time to end a typical exposure within a window of $\pm 90$ seconds.

\subsection{Readout and Dataflow}
\label{sec:readout}

\begin{figure}[htb]
\centering
\includegraphics[width=\textwidth]{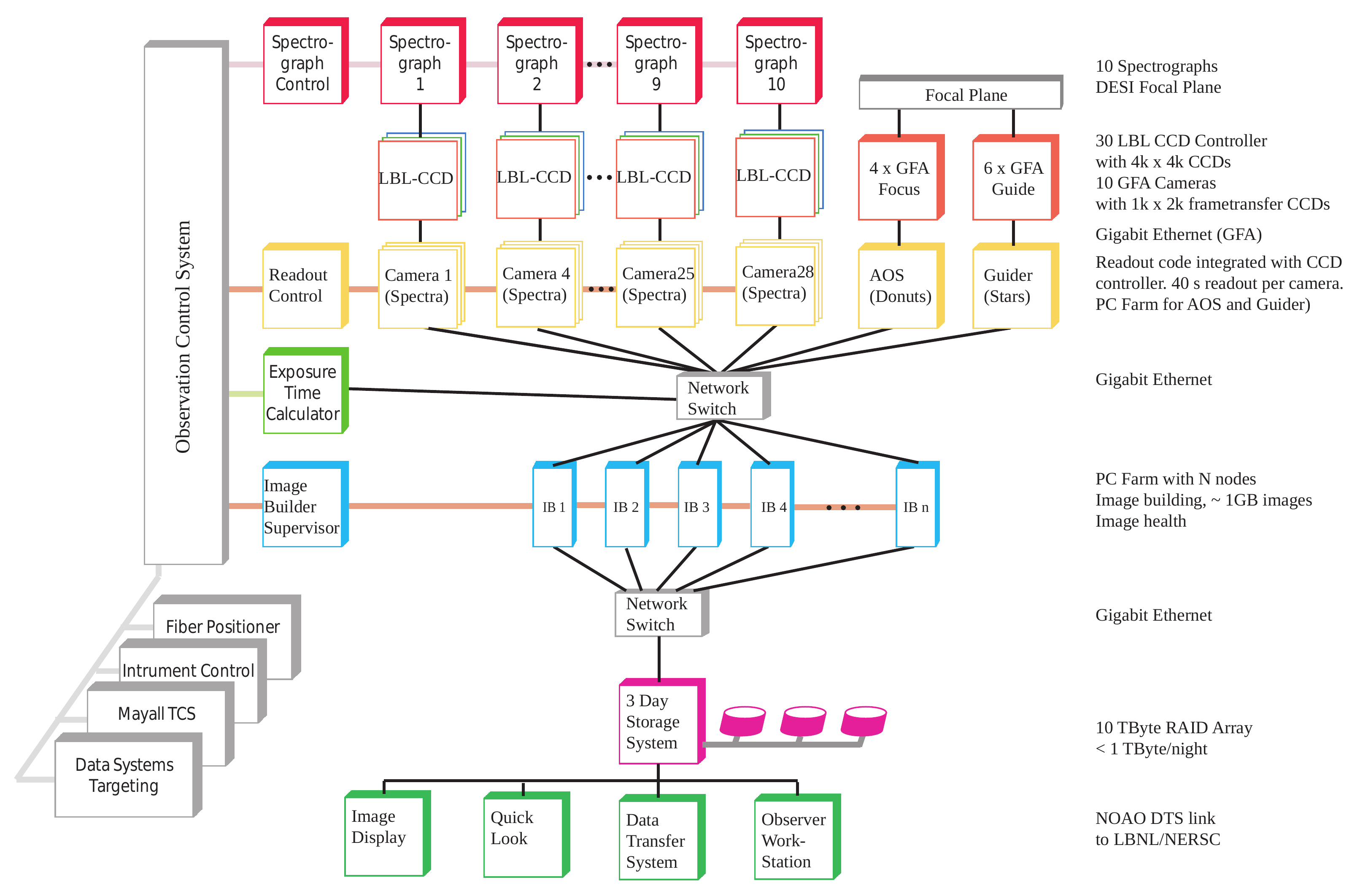}
\caption{Block diagram of the DESI data acquisition system.}
\label{fig:DESIDAQ}
\end{figure}

The DESI instrument consists of ten identical spectrographs each with three cameras covering different 
wavelength regions. Each camera uses a single 4k$\times$4k CCD with four readout amplifiers that operate 
in parallel. 
A default pixel clock of 100~kpixels/s results in a readout time of approximately 42 seconds. The charge 
contained in each pixel is reported as a 16-bit unsigned integer yielding a data volume of 34~MBytes per camera or 
about 1~GByte per exposure for the entire instrument. A schematic view of the DESI readout system is shown 
in Figure~\ref{fig:DESIDAQ}. We have adopted a modular approach consisting 
of 30 identical data sources, one for each camera. 
Data flows from the top starting 
with the CCDs and ending with the images stored as multi-extension FITS files on disk arrays in the computer room. Each 
CCD is connected to a CCD front\-end electronics module described in detail in Section \ref{sec:frontend_electronics}. 
Besides the digitizer, clock signals and voltage generators, the module includes a full frame 
buffer, a powerful ARM microprocessor, and a high speed network interface. The CCD controllers run regular Linux which enables us to fully integrate them with the DESI online system without the need for an additional interface. As shown in Figure~\ref{fig:DESIDAQ} a DESI Online System (DOS) application (Camera) runs on each CCD controller. Besides CCD readout and the data transfer to the Image Builder farm, this application handles the front end electronics configuration, collects telemetry information and monitors operating conditions. The integrated network connection combined with the modular 
design allows us to operate individual cameras with only a laptop computer, a network cable and of course 
the online software suite. We expect this to become a very valuable tool during construction, 
commissioning, and maintenance. The design of the CCD controller is complete and a software interface library has been developed. This software will be used and tested as part of the spectrograph tests at Winlight, France.

It is the responsibility of the Image Builder processes to collect data from all 30 CCD controllers, to combine the fragments to a single, multi-extension FITS file and to insert additional information about the current exposure into the FITS header. The images are compressed using the standard Rice/tile algorithm and saved to the 3-Day storage system. This disk array has sufficient capacity to record  several nights worth of images should the link off the mountain be unavailable for an extended period. In order to increase throughout and to allow for some online data processing for quality assurance purposes we will employ multiple Image Builder applications taking turns processing new exposures. Image Builder assignment and coordination will be managed by the Image Builder Supervisor application shown in Figure~\ref{fig:DESIDAQ}. 

When an image is complete, the Image Builder notifies the DTS to start the transfer off the mountain and the quicklook pipeline is alerted that the next exposure can be analyzed. A display in the console room shows the CCD images (or an observer selected subset) and summary information is shown on the observer console user interface. The images from the 10 GFA cameras are also recorded in the 3-Day storage system from where they are eventually transferred to the DESI processing facility at LBNL NERSC.

Data transfer from the CCD controllers will begin shortly after the start of digitization and will proceed concurrently with CCD readout. System throughput will be designed to 
match the CCD readout time of 42 seconds to avoid additional dead time between exposures. The required 
bandwidth of approximately 10~Mbits/s is easily achievable with current technology. 

The DESI readout and dataflow system is based on the same architecture we developed for the DECam online system. Functional prototypes of the camera and image builder applications have been developed and are operational. A single Image Builder system will be used for the spectrograph test system at Winlight, France and a more complex system with multiple Image Builder processes will be part of the ProtoDESI test run.

\subsubsection{Quality Assessment}
\label{sec:OnlineQA}

Continuous monitoring of both the hardware and the data quality is necessary to control systematic uncertainties at the level required to achieve the science goals of the project and to allow continuous, error-free operation of the DESI instrument. Several components of the DESI online system are designed to implement these quality assurance procedures. ImageHealth, a first check of the data quality of every exposure, runs during the image building stage. The image health algorithms determine mean and noise values for each CCD amplifier in both the overscan and data regions and perform additional simple tests at the pixel level to ensure that the spectrographs are functioning properly. 
During the night, the observer at the telescope will also require automated feedback on the data quality to gauge progress of the observations and to ensure that the instrument is performing as expected.
We will build on the experience of both the BOSS and DES teams which have  developed  quick reduction pipelines in operation at the Sloan telescope and the Blanco 4-m telescope, respectively.
A quicklook pipeline is a simplified version of a full reduction pipeline,
replacing the most expensive computational steps with simpler
(and in some cases more robust) algorithms.
Quicklook determines sky levels and calculates S/N estimates per exposure as a function of
wavelength and object magnitude.  This allows a robust, near-real-time
assessment of data quality as a cross check of the exposure length and
tile completeness as determined by the guider data and exposure time calculator.
Results from the quality assessment processes will be logged in the telemetry database that is part of the DESI operations database system. The DESI quality assessment system will be based on the quick reduce framework developed for DECam by our collaborators from Brazil. The algorithms will be based on the BOSS/eBOSS pipelines are a deliverable of the Data Systems group (WBS 1.8).

Additional quality assurance tools will be available to the observer including full data access. The DESI online system will automatically copy every exposure to an observer workstation where it can be accessed by standard tools such as ds9, idl or iraf for an in depth interactive analysis. The observer can choose to process the spectra and guider images with their private, customized algorithms. These operations are completely decoupled from the DESI image pipeline and do not affect the overall throughput.

\subsection{Operations Database}
\label{sec:opsDB}
At the heart of the DESI online system is the
database system that hosts the databases for DESI mountain-top operations. The system consists of a number of different components,  each having its own schema within the database. This structure keeps the different sets of tables organized while still permitting queries that join together tables in different schema. This would not be possible if we kept the different components in separate databases. However, for convenience in what follows we refer to the various components simply as databases. We have used the same architecture successfully for the DECam mountain-top database system.

\subsubsection{Telemetry Database}
These tables will be stored in the ``telemetry'' schema and will capture information obtained from the Instrument
Control System, the online QA software, the Mayall TCS and environmental systems, as well as from other DESI online components. The method of storing data in the telemetry database will be via the Shared Variable Engine (SVE). The data will be stored by the SVE client code. It uses the asynchronous call back feature of the SVE to trigger the data storage. The SVE client queries the database for the table schema so it does not need be provided with schema.  This make it easy to add new tables as required.  Making the data storage part of the SVE client allows us to easily log telemetry from devices even when a DOS instance is not running. There will be a Telemetry Viewer web-based application that allows users to make plots of quantities in the telemetry schema and retrieve data via SQL queries. 

\subsubsection{ConstantsDB}
The ConstantsDB is intended for data that is required to configure the ICS and  the instrument. This is data that changes infrequently but need to be version controlled. The tables are stored in the ``constants'' schema. A consistent set of constants is called a snapshot and is made up of groups that collect together related elements which contain the actual constants. 
We define the concept of a snapshot tag as a keyword that can be assigned to a different snapshot at different times. A tag like {\em CURRENT} allows the online system to load the current set of constants without having to update version numbers in the code every time a constant changes. There will be an API that allows the user to retrieve constants from the database in a variety of ways. A web-based tool will be provided for manipulating the contents of the ConstantsDB. Access control will be required in order to make changes but viewing will be open. The web tool will allow the contents of the ConstantsDB to be manipulated in a consistent way. 

\subsubsection{AlarmDB}
The AlarmDB is part of the DOS Alarm System. These tables are stored in the ``alarm'' schema. Both DOS and non-DOS applications can raise alarms by inserting them into the alarm database. The database sends out notifications that include information about the alarm level and the name of the DOS instance that raised the alarm. Any number of programs can connect to the database and listen for these notifications.  DOS provides a set of standard actions including support for auto-dialer/phone calls, emails and forwarding of alarms to the DOS shared variable system. The software will be designed so that it can be used by both DOS and non-DOS applications. The only requirement is that the alarm database is operational and accessible. There will also provide a web-based viewer that allows users to view the alarm table in the database. This is intended for archival use as the real-time display of alarms will go through the DOS GUI.

\subsubsection{ExposureDB}
The ExposureDB tracks the status of an exposure as it passes through the system. The tables are stored in the ``exposure'' schema. There is a python API that allows programs throughout the system to read and write the status of an exposure. The exposure table is responsible for automatically generating the exposure numbers of each exposure. There will be a web-based viewer that allows users to view the exposure table in the database. This is intended for archival use as the real-time display of exposures will go through the DOS GUI.

\subsubsection{Guider Image Data Management - GuiderImageDB}
The guider images need to be archived for use by the data management system. The guiding rate is expected to be about 1 Hz so using a nominal exposure time of 1800 seconds along with the 6 guide cameras and two postage stamps per camera there will be about 22000 guider postage stamps per science exposure. Taking the assumption of 1940 hours of observing per year yields 80 million stamps per year that have to be managed. Each stamp is assumed to be about 10KB which is 800 GB of data per year. To avoid managing large numbers of small files on disk, they will be aggregated so that all the postage stamps from a single science exposure are collected into a multi-extension FITS file (MEF). There will be an IMAGE extension which will be a FITS data cube containing the 22,000 postage stamps along with the appropriate header information (for example the exposure number that the guider images match with). There will also be a FITS binary table that contains one row per postage stamp to capture the appropriate information per stamp, for example magnitude of the star in the stamp. The file size will be of order a few hundred MB. The file will be written as part of the image building step in DOS and then the file can be transmitted to DTS for transport to NERSC. There will be an association in the database between a given guider MEF and the science exposure from the ExposureDB. We can also store the information in the FITS binary table in the database if required. 
The files on disk at KPNO will be managed in the same way that the image data will be managed, as space is needed they will be deleted. 

\subsection{ConditionsDB}
The fiber positioners will require regular calibration. Typically a set of measurements will be taken and then they need to be uploaded into the database and then the fiber positioner control code will need to download them from the database. The frequency and volume of the data is such that the ConstantsDB is not a good solution. However the need to store calibration data on a frequent basis is common to all High Energy Physics experiments so we plan to use the Conditions Database developed for the Nova experiment at FNAL. There is both a web interface to the Conditions Database (which uses the same technology as the other web-based viewers) as well as an python API so it will be straightforward to integrate it into the DOS software.

\subsection{Targeting Information Data Management - TargetInfoDB}
The information about the targeting for each pointing of the instrument will be captured in the database.  We will need to record the RA/DEC of the requested position, the focal plane coordinates and the actual position achieved. The information will also need to be associated to the relevant science exposure from the ExposureDB.
This will include recording the positions for all 5000 fiber positioners after each move (a move consists of moving all 5000 fibers to new positions). There are expected to be about 3 moves per target position and we estimate that each move will generate about 3 MB of data so there will be about 9 MB of data for each science exposure. We also need to store the centroid information of the fiber locations as recorded by the Fiber View Camera. This will be about 100 KBytes per move. If we store the centroid information for each move then that will be 300 KBytes per science exposure.  A python API will be provided that allows the data to be written to the database from the relevant DOS applications.

\subsubsection{Survey Strategy DB}
The survey strategy DB used by the Data Systems group will record the field centers of the full survey, the fibers assigned to each field, the priority of each field for observing,
the date each field was observed and the approximate date that each
field is expected to be observed according to the long-term simulations.

\subsubsection{Electronic Log Book}
There will be an Electronic Logbook provided for recording operational details during installation and commissioning and continuing on into survey operations. We are planning on using the ECL product from the Fermi National Accelerator Laboratory (FNAL). This is a web application with a PostgreSQL database backend. It is in widespread use by multiple experiments at FNAL as well as by the DECam instrument at CTIO.

\subsubsection{Technology choices}
The database server will be PostgreSQL\footnote{http://www.postgresql.org}. We will use the current stable production release at the time that the DOS system is installed at KPNO. The interface to the database will be developed using SQLAlchemy\footnote{http://www.sqlalchemy.org}. The underlying database driver is psycopg2\footnote{http://initd.org/psycopg} . The web viewers use mod\_wsgi to connect with the Apache server. The plotting in the Telemetry Viewer is done using ChartDirector\footnote{http://www.advsofteng.com}

\subsubsection{Reliability and recovery}
We will use the PostgreSQL streaming replication to make a standby copy of the database at KPNO. In the event of a disk failure on the master this will allow us to switch to the standby and continue operations with minimal downtime. This arrangement is use for the Dark Energy Camera and has already demonstrated its usefulness during a disk failure, only a couple of hours of downtime were incurred while the switch was made. 
We will also use the same streaming replication to have a query-able read-only copy of the database at one or more offsite locations, referred to as a mirror. This will allow access to all the telemetry and other data from outside the KPNO firewall. We will also run copies of the various web viewers against this mirror database(s). 
We will also make static backups of the database on the appropriate timescale. These will be stored on another machine in the online system but will also be transferred back to NERSC for storage. We will also develop tools to monitor the health and status of the database. 

\subsubsection{Development status}
Many of the key parts of the database system have operated reliably for the last four years for the DECam instrument at CTIO. These components will undergo some modest evolution as a result of the experience gained with DECam. This allows us to focus effort on the new parts needed for DESI such as the GuiderDB and the TargetInfoDB. The use of a hot standby for quick failover and the use of an offsite mirror to provide collaboration-wide access to the telemetry data have proved to be extremely beneficial for DECam and will be a key element of the strategy for DESI.

\subsection{Instrument Monitor and Control System}
\label{sec:imcs}
Hardware monitoring and control of the DESI instrument is the responsibility of the instrument monitor and control 
system (IMCS).  
We distinguish three sets of IMCS applications: The first set consists of critical 
systems such as the CCD cryostats and the monitor system for the frontend 
electronics have to operate at all times. {\bf Fail-safe systems and interlocks for critical 
and/or sensitive components will be implemented in hardware and are the responsibility of the device 
designer.} Control loops and monitor functions for these  applications will use PLCs or other programmable 
automation controllers that can operate stand-alone without requiring the rest of the DESI IMCS to be 
online. Sensor information and application status, alarms, and error messages produced by these components will be archived in 
the DESI telemetry database on the mountain.  The DESI online system includes the tools necessary to access this information for viewing and data mining purposes.

The second set of applications is made up by all systems that have to operate around the clock but are not directly part of the instrument protection system. This includes the Mayall environmental, dome and telescope telemetry systems, the spectrographs and the spectrograph shack,  focal plane systems as well as monitoring of the computer and networking hardware. All telemetry and alarm information generated by these system will be archived in the operations database. 
The connection between DOS and these devices/controllers is typically provided by a DOS device application (see Section \ref{sec:framework}) that includes the code necessary to communicate with the device hardware and uses the standard DOS communication protocol to interact with the rest of the online system. The 
DOS group provides the higher level software for these applications, while the device specific code will be developed by the groups responsible for the respective components. 

The third set of instrument control applications consists of components that participate more actively 
in the image acquisition process such as the PlateMaker, the guider and the focus and alignment system. Similar to the first 
two sets, these applications will also use the operations database to archive the information. Applications in the third group will be started and stopped with the rest of the readout system and will not operate continuously.

DESI control applications can be classified by the systems they are supporting. Each of the 10 DESI spectrographs is managed by its own DOS framework application. This {\bf Spectrograph} process controls the spectrograph hardware using a dedicated {\bf Spectrograph Controller}  together with three {\bf CCD Controller} processes - one for each of the spectrograph's cameras. The {\bf Spectrograph Controller} runs on an 
embedded Linux computer in the spectrograph electronics box that controls all spectrograph mechanisms such as the shutters, the fiber back illumination and the Hartmann doors. The {\bf CCD Controllers} run directly on the LBNL CCD controller hardware described in Section 5.
These controller processes are implemented as DOS device applications providing them with full time access to the operations database and the DOS communication system.
A separate computer/process is used to control the spectrograph shack and to monitor environmental and operational parameters.
The cryostat system for the spectrograph CCDs is part of the instrument protection system and has an independent control system. The cryostat control system includes an OPC-UA server\footnote{http://www.opcfoundation.org} that allows DOS readonly access to monitor operating conditions and react to error and alarm conditions. All telemetry information and alarms collected from the spectrograph controllers, the cryostat control system and the spectrograph shack is archived in the operations database.

DESI will employ a software based {\bf Time Distribution System} using a multicast-based protocol to distribute exposure related information to the spectrographs. The same mechanism is used to synchronize the fiber illumination system. Several software solutions were evaluated. The multicast option we adopted for DESI  achieved a typical jitter between the different spectrographs of less than 10 ms. This is less than the variability in the spectrograph mechanisms itself and meets DESI requirements. While some jitter in exposure start time among the spectrographs can be tolerated, the actual exposure time must be controlled to higher accuracy. This is handled by a dedicated timer integrated with the spectrograph electronics.

The complex DESI focal plane and the prime focus cage assembly require a number of instrument control applications including the {\bf Fiber Positioner Supervisor}, {\bf the FVC interface}, the alignment system with the {\bf Hexapod Controller} and the {\bf ACD Controller}, and the {\bf GFA System} already described in Section~\ref{sec:aos}. Most of these applications are joint developments between DOS and the corrector barrel/cage system or the Focal Plane Systems with the responsibilities and deliverables defined in a series of Interface Control Documents (ICDs). 

\subsection{Telescope Operations Interface}
\label{sec:TCSInterface}

\begin{figure}[!hbt]
\centering
\includegraphics[width=0.7\textwidth]{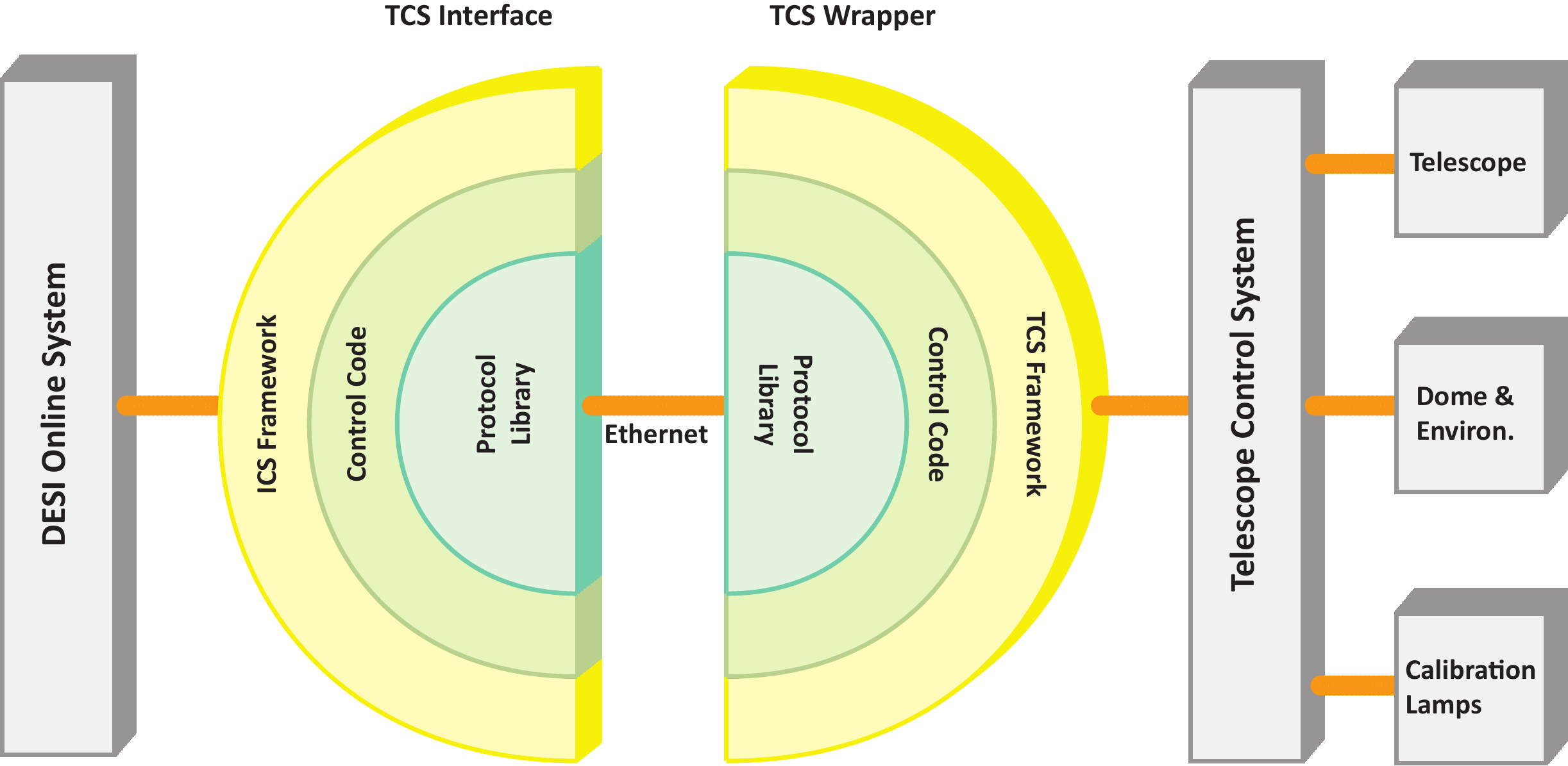}
\caption{Block diagram of the Interface between the DESI Online System and the Mayall Telescope Control System.}
\label{fig:tcsinterface}
\end{figure}

The DESI Online System connects with the existing Mayall telescope control system (TCS) to communicate new pointing coordinates and to send correction signals derived from the guider. In return DESI will 
receive telescope position and status information from the Mayall TCS. On the DOS side the data exchange is handled by the {\bf TCS Interface} application. Since the dome environmental 
system and most of the observatory instrumentation for weather and seeing conditions are already connected 
to the TCS, we will not access these devices directly but monitor them through the {\bf TCS Interface}. The DESI flat-field and calibration lamps are part of the telescope system and will therefore be controlled through the {\bf TCS Interface}.
Similar to the design developed by our group for the Dark Energy Camera and the Blanco telescope, the DESI 
online system will include a TCS interface process that acts as conduit and protocol translator between 
the instrument and the telescope control systems. A block diagram of the associated software modules is shown in Figure~\ref{fig:tcsinterface}. The message passing protocol and commands are detailed in DESI-0473 in the document database.

During an exposure, the DESI guider and the telescope servo systems form a closed feedback loop 
to allow the telescope to track a fixed position on the sky. For an imaging survey it is sufficient 
to have a stable position. DESI, however, requires a precise absolute position so that the fibers are correctly 
positioned on their targets. Given a pointing request, the Mayall slews into position with a typical accuracy of
3 arcsec. 
Using the guide CCDs in the focal plane we will then locate the current position to 0.03 arcsec accuracy. As previously discussed,  we will send realtime pointing corrections to the TCS to adjust the telescope position, if the offset between requested and actual position is larger than a certain fraction of the fiber positioner motion. To improve telescope pointing accuracy, DESI will provide the current hexapod settings to the TCS for every exposure.

The {\bf TCS Interface} application and the connection between DOS and the Mayall TCS were tested during a 2 day/night test run at KPNO in January 2016. We successfully move the telescope under DOS control, read back dome, telescope and environmental data and observed the telescope react to guider correction signals generated on the DOS side by guider emulator. This system is ready for the ProtoDESI test run later this year.

\subsection{Online Software}
\label{sec:OnlineSoftware}

The DOS application processes described in the previous sections are built upon a layer of 
infrastructure software that facilitates message passing and information sharing in a distributed 
environment. Other system level components of the DESI online system include the configuration system and support for user interfaces and observer consoles. In this section we will focus on the infrastructure layer of the DESI online software and the design on the DOS system components.

\subsubsection{Infrastructure Software}	

The DESI readout and control system is implemented as a distributed multi-processor system.
Due to the nature of this architecture, inter-process communication takes a central place in the design of the DESI infrastructure software. Copying the design of the very successful DECam online system, our solution is based on the Python Remote Objects (PYRO) software package. PYRO is a free, advanced and powerful distributed object technology system written entirely in Python. It allows objects to interact just like normal Python objects even when they are spread over different computers on the network. PYRO handles the network communication transparently. PYRO provides an object-oriented form of remote procedure calls similar to Java's Remote Method Invocation (RMI). A name server supports dynamic object location, which renders (network) configuration files obsolete. Using the name server, DESI processes can locate and establish communication links with each other irrespective of the underlying hardware architecture; an important feature for any multi-processor communication system. In a test system, for example, most of the applications might share one or two computers whereas on the mountain everything is spread out over many computers to maximize performance.

The DOS communication software distinguishes between command messages and telemetry data. Commands are used to request information from a remote application or to activate a remote action. The command or message passing system is implemented using a Client-Server design pattern with a thin custom software layer (PML) on top of PYRO. PML introduces the concepts of roles and devices to provide a uniform naming scheme for all DESI online applications. 
The telemetry system is based on the publish-subscribe design pattern using ideas similar to data distribution services, an emerging standard for publish-subscribe middleware for distributed systems. Again built on the core functionality provided by PYRO, we developed a concept called Shared Variables that we have adapted to the DESI online system. Consisting of a client stub library and a central server (Shared Variable Engine or SVE), this system allows user applications to publish information such as temperature readings or readout status to a virtual data space. Other applications can subscribe to information placed in this virtual data space and will receive updates whenever a publisher submits a new value. Publishing and subscribing to a shared variable are completely decoupled. A publisher needs no knowledge of who will subscribe to this variable and vice versa. The shared variable system supports asynchronous callbacks, guaranteed delivery, multiple publishers of the same shared variable and group subscriptions.

In order to ensure long term support for DESI, we have ported the DECam software to Python~3 and upgraded all underlying software packages such as Pyro, PostgreSQL and the GUI toolkit to the latest versions. The default operating system is RedHat Enterprise Linux 7/Centos 7 but since the software is written entirely in Python other systems like Mac OS and various Linux distributions used by embedded controllers are easily supported as well. All code resides in an svn repository hosted at LBNL. Our release management tools are based on eUPS originally developed for SDSS and now maintained by the LSST project. To support test stands and code development we use virtual machines to distribute ready-to-use systems that include not only the DOS software but also copies of the operations database and the web-based viewing tools.

\subsubsection{Configuration}

Initialization and configuration of a complex distributed system such as the DESI online system is a multi-step process.  
The complete system will have about 60 nodes ranging from embedded controllers to server class computers and an even larger number of applications. Some device application will already be running and need to be integrated with the newly started processes. In order to start the online processes in the correct order, on the correct computer, and with the correct arguments with a single command, we have developed a flexible startup system called the {\bf Architect}. 
The {\bf Architect} is responsible for connecting to multiple hosts and starting DOS system services such as the PYRO name server, the SVE and the logging system as well as all user applications. Initialization files in standard Windows .ini notation are used to describe the configuration. The .ini files selects the participating nodes, specifies which applications should be started on each node and provides calling arguments and other configuration information for each process. Once all applications and services are running, the {\bf Architect} monitors the processes at the operating system level. The {\bf Architect} implements a management interface to provide full process control so that individual processes can be stopped and restarted without having to restart the entire system.

\subsubsection{Application Framework}
\label{sec:framework}

\begin{figure}[!htb]
\centering
\includegraphics[height=3in, angle=-90]{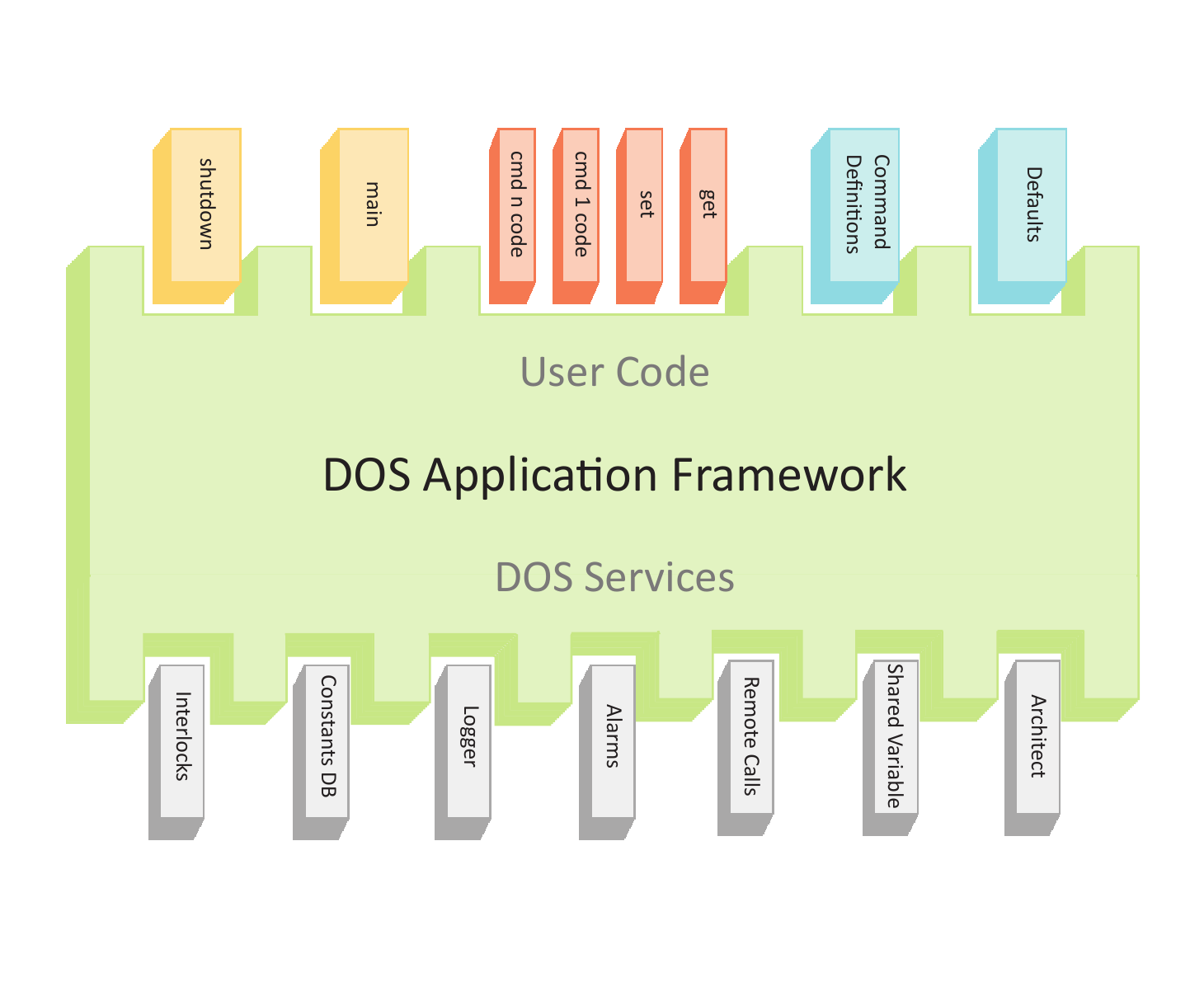}
\caption{Schematic view of the DOS Application Framework. The Framepwork provides standardized access to DOS services such as the Architect configuration system, the DOS cloud (shared variables) and the messaging system (PML). The user provided pieces are shown on the top.}
\label{fig:framework}
\end{figure}

All DESI online applications will be based on a common software model. This Application Framework (Figure~\ref{fig:framework}) serves as a base class and gives the same basic structure to all applications. It provides a unified interface to all DESI  services such as the configuration system (Architect), the DOS data cloud (Shared Variables), the message passing and remote procedure system (PML), alarms, logging, the constants database and the interlock and monitoring system. The Framework manages all resources centrally which allows it to unsubscribe all shared variables and to close all PML
connections when the application is about to exit. Proper exit handling is critical for the ability to stop and restart individual processes without the need to end the entire instance. Additional functionality provided by the Application Framework include a heartbeat that can be used to monitor the overall state of the system and a standardized management interface with process control functions and access to an application state variable.  The user code inherits the application framework class and provides default configuration settings and the list of commands that will be exposed to the online system. Configuration settings can be updated at run time by the Architect. The code to execute the commands has to be written by the user. If the application requires access to the runloop, the user can overwrite the main() method and custom exit handlers can be added by overwritting the shutdown() method. The DOS Application Framework significantly reduces the complexity of application development. It is lightweight and since it is written in pure Python code it is very easy to port to different host environments. For DESI, the application framework will be available not only on workstations and servers but also on most of the embedded controllers of the instrument control system including the CCD controller module, the spectrograph controller and elements of the focal plane control system. The application framework supports stand alone operation for devices that operate all the time and not just when the rest of the online system is running. The user code is identical to that for regular application and full access to the mountain database is available. Software support is included to connect devices operating in this mode to the rest of the DESI online system when this is started.

\subsubsection{User Interfaces and Observer Console}

\begin{figure}[!hbt]
\centering
\includegraphics[width=0.9\textwidth]{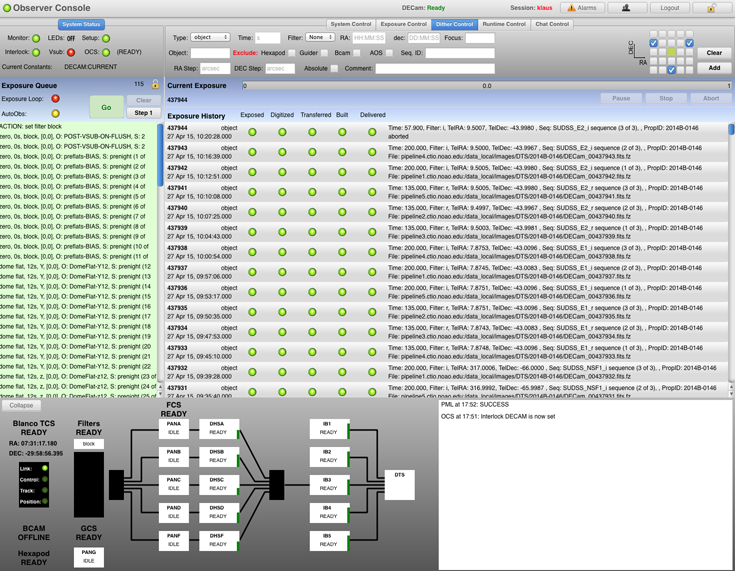}
\caption{DESI will use web-based user interface similar to the DECam observer console shown here.}
\label{fig:gui}
\end{figure}

The DESI observer console acts as the primary observer interface for day to day operation. This set of graphical user interfaces includes elements for system and exposure control, alarm displays and telemetry monitors. Combined, they present the user with the most commonly used information and everything that is needed to operate the DESI instrument. Near real-time performance is needed to provide quick access to more specific information and to achieve the responsiveness expected from a modern system. The DESI GUI architecture follows the Model-View-Controller (MVC) pattern first developed for Smalltalk but now in common use for large applications. The MVC pattern is based on the realization that all applications are, essentially, interfaces that manipulate data. Following our successful implementation of this approach for DECam, we are implementing the controller component as a web server so that standard web browsers can be used to render the views. Using web browsers provides many desirable features such as platform independence, remote access, a large number of 3rd party tools, and a certain level of security. Early concerns regarding usability and performance in a real-time environment have been addressed by recent advances in browser technology. Thanks to faster rendering and JavaScript execution as well as new standards such as HTML5 and the Web-Socket API that allow for advanced functionality such as bidirectional socket communication and 2D drawing contexts, web browsers are now on a near equal footing with desktop GUI toolkits. 
As an example, Figure~\ref{fig:gui} shows the DECam observer console GUI. A similar design will be developed for DESI.

The client side (web browser) code for DESI user interface is based on HTML/CSS, JavaScript and SproutCore,an HTML5 application framework.  Websockets, defined as part of the new HTML5 standard and now supported by all major browsers, are used to provide the most responsive user experience. Conceptually similar to traditional TCP/IP sockets, websockets allow the DESI DAQ system to push information updates and state changes directly to the GUI. We have successfully ported the graphical user interface system from DECam to the DESI environment and in the process updated many of the underlying software packages. Specialized observer consoles for the spectrograph tests at Winlight, France, and the ProtoDESI tes are currently under development. 
 
\begin{table} [htb]
\centering
\caption{DESI online computer needs. Networking and infrastructure equipment is not listed.}\label{tab:hardware}
\begin{tabular}[width=0.95\textwidth]{|l|l|l|}
\hline
Function & Configuration & Comment\\
\hline
Database Server & Rackmount Server & DESI Operations Database\\
 & (16 cores, 64GB RAM, 24 TB) & \\
Backup Database Server & Rackmount Server & Database will be mirrored \\
User Server & Rackmount Server & Repository, home areas\\
Backup User Server & Rackmount Server &  Hot-Standby backup server\\
GFA System & Rackmount Workstation  & Guider, Wavefront analysis \\
 & (12 cores, 64 GB, 2x2TB) & \\
Focus and Alignment & Rackmount Workstation (5 units) & Donut analysis is CPU intensive\\
 & &  (Based on DECam experience) \\
Image Pipeline & Rackmount Workstation (5 units) & Incl. acquisition, processing \\
Quicklook Pipeline & Rackmount Workstation (2 units) & Based on BOSS experience\\
Fiber Positioner Interface & Rackmount Workstation & \\
Fiber View Camera & Rackmount Workstation & Based on performance tests \\
DAQ Core Processes & Rackmount Workstation (2 units) & Incl. webserver, PlateMaker\\
ICS & Rackmount Workstation  & Includes TCSInterface \\
DTS & Rackmount Workstation & Interface to DTS system\\
Observer Workstation & Desktop Workstation & Includes 6 displays\\
 & (6 cores, 32 GB, 6 TB) & \\
Observer Console & Desktop Workstation & Includes 6 displays\\
Image Display & Desktop Workstation & Includes 1 display\\
\hline
\end{tabular}
\end{table}

\subsection{Readout and Control System Hardware}
\label{sec:DAQhardware}

The ICS project is responsible for all DESI computer hardware at the Mayall telescope. This includes the online computers, disk storage, the internal network infrastructure and two multi-display observer workstations for the Mayall control room. The backbone network infrastructure in the Mayall building will be provided by the observatory. The observatory is also responsible for computer and network security and will provide a firewall and other access control solutions. The main user work place, the observer console, is modeled after the solution we developed for DECam and the Blanco console room. The DESI and DECam computer systems are similar in scale and we use our experience with DECam to design the DESI system detailed in Table \ref{tab:hardware}. In order to maximize system availability and reliability we operate the key server systems with hot-standby backup servers. This configuration provides synchronized backup copies of the filesystem and the DESI database and allows for rapid switch over should the primary server fail. We had one such incident with DECam. We were able to quickly switch over to the backup server minimizing the loss of observing time. 
The computer equipment will be housed in the computer room at the Mayall telescope. Clean power and cooling will be provided by the observatory. The system will be configured to allow remote monitoring and system management as much as possible. The number of different workstation configurations will be minimized to facilitate maintenance.   
Additional smaller test systems are required for development and to support ProtoDESI.

\begin{figure}[htb]
\centering
\includegraphics[width=0.8\textwidth]{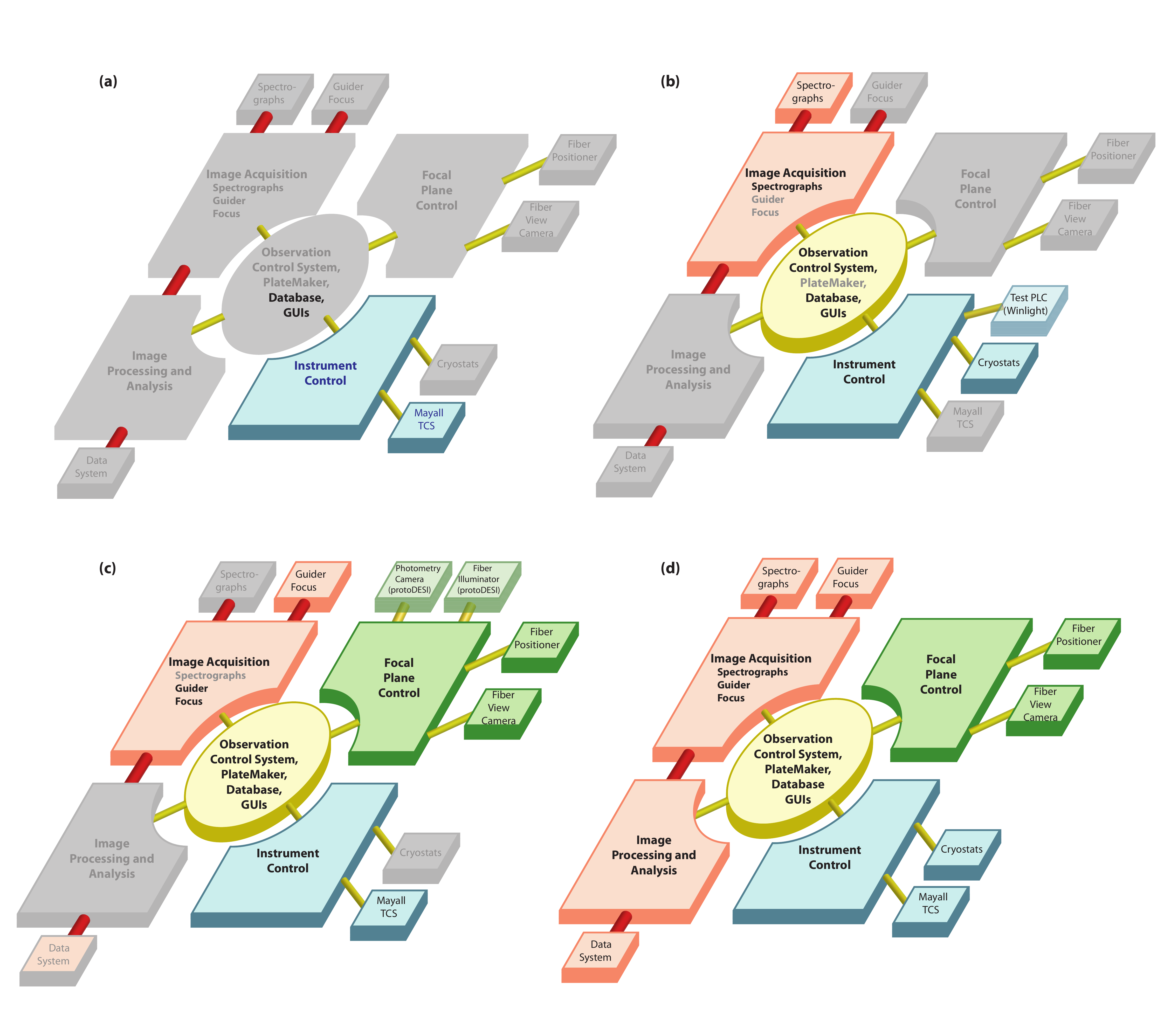}
\caption{DESI DAQ development cycles: (a) Mayall TCS integration (b) Spectrograph test stand at Winlight (c) ProtoDESI system (d) Full system test of the DESI ICS using hardware emulators.}
\label{fig:daqdevelopment}
\end{figure}

\subsection{Integration and Testing}
\label{sec:DAQintegration}

The DESI instrument control system is connected to most major components of the instrument as well as the Mayall telescope. Given the large number of interfaces, we have designed a project plan that makes integration and testing an integral part of the ICS project. For each DESI component that interfaces with the online system, we developed interface control documents (ICD) that detail the hardware connection and define deliverables and responsibilities. These documents have been reviewed and will be placed under change control. The ICDs are accompanied by implementation documents that specify details of the API such as data formats, methods, telemetry and alarms. While the hardware components are under development, the ICS team creates software simulators for each interface implementing the protocol as defined in the ICDs and the implementation documents. Adopting a spiral development approach with increasingly complex systems, we have scheduled a number of tests throughout the construction project. Some of these tests are coordinated with the project to support other activities, for example spectrograph testing and evaluation. The four key DOS system tests are: ICS-Mayall TCS integration, single spectrograph support, ProtoDESI, and full system tests using emulated hardware components. Each of these tests requires a different set of DOS applications and system services as shown in Figure~\ref{fig:daqdevelopment}. The Mayall TCS test was completed successfully in Januar 2016. A complete single spectrograph data acquisition system is being finalized and will be shipped to Winlight spring 2016. The ProtoDESI  test in August and September 2016 will feature complete ICS prototypes for the focal plane and data flow systems together with a number of prototype hardware controllers. While the project does not include a standard test-stand facility for all hardware and software components, we can take advantage of the modular architecture of the ICS and run the prototype system at the different productions sites. During these tests we will integrate with the available hardware controllers and use our software emulators for the remaining components. Similar integration tests with the complete production system will be repeated at the end of the construction project.

\clearpage

\section{Data Systems}\label{s7:DataMgt}
\setcounter{equation}{0}\setcounter{figure}{0}\setcounter{table}{0}
Data Systems covers target selection, survey planning, data processing, and data management.
Although separate topics, these are closely related in the design and
execution of the survey and thus they are jointly planned and managed.
The DESI construction project deliverables of Data Systems are:
\begin{itemize}
\item \textbf{Target Selection Pipeline}: Code to generate a catalog of
    targets (ELG, LRG, QSO, BGS, MWS, standard stars, and blank sky locations)
    given multi-band photometric catalogs and selection criteria as input.
\item \textbf{Survey Planning and Integration with Operations} \\
    (fiber assignment, field selection, quicklook):
    \begin{itemize}
    \item A defined set of telescope pointings (tiles) on the sky
        that cover the survey footprint;
    \item Code to assign input targets to focal plane fibers for each tile,
        prioritizing
        targets of different classes to maximize the overall survey impact
        (``fiber assignment'');
    \item Code to prioritize how those tiles should be observed to maximize
        survey efficiency (``field selection''); and
    \item Code to assess spectroscopic data quality shortly after
        observations complete (``quicklook'').
    \end{itemize}
\item \textbf{Spectroscopic Pipeline} code to convert raw data into:
    \begin{itemize}
    \item wavelength-calibrated, sky-subtracted, flux-calibrated spectra;
    \item spectroscopic classifications and redshifts; and
    \item measurements of the 3D survey efficiency for generating
        large scale structure catalogs.
    \end{itemize}
\item \textbf{Data Transfer, Archive, and Distribution}: Tools to automatically
    transfer, back-up, and
    distribute raw and processed data.  This system integrates with the
    Spectroscopic Pipeline to process new data as they arrive,
    and distribute processed data when they are available.  This activity
    also has oversight of documentation, coding standards, and tagged
    releases of both reduced data and code.
\item \textbf{Collaboration Tools}: Tools to facilitate the efficient operations
    of the collaboration including a wiki, code repository, bug tracking
    system, mailing lists, document repository, and website.
\end{itemize}

From the perspective of Data Systems, DESI is a modest expansion of what
is already done with SDSS-III/BOSS: 10 spectrographs with 3 channels each instead of
2 spectrographs with 2 channels each --- a factor of 7.5$\times$ more data
per exposure.   Wherever possible, we have made
technology and requirements choices based upon the experience of the highly
successful BOSS spectroscopic pipeline and SDSS-III data management systems.
The spirit of these systems also influences our design choices,
for example:
\begin{itemize}
    \item the primary interface between pipeline steps is defined
        by a ``data model'' for those files rather than function call
        signatures or object-oriented class interfaces;
    \item simplicity and pragmatism are key design considerations;
    \item files on disk are the authoritative data product rather than
        a centralized database (though databases will be used);
    \item software should be runnable on a laptop without an
        internet connection for debugging and development.
\end{itemize}
At the same
time, we are taking the opportunity to modernize significant portions of the
system for updated algorithms and long term maintainability.

\begin{figure}[tb]
\centering
\includegraphics[width=5in]{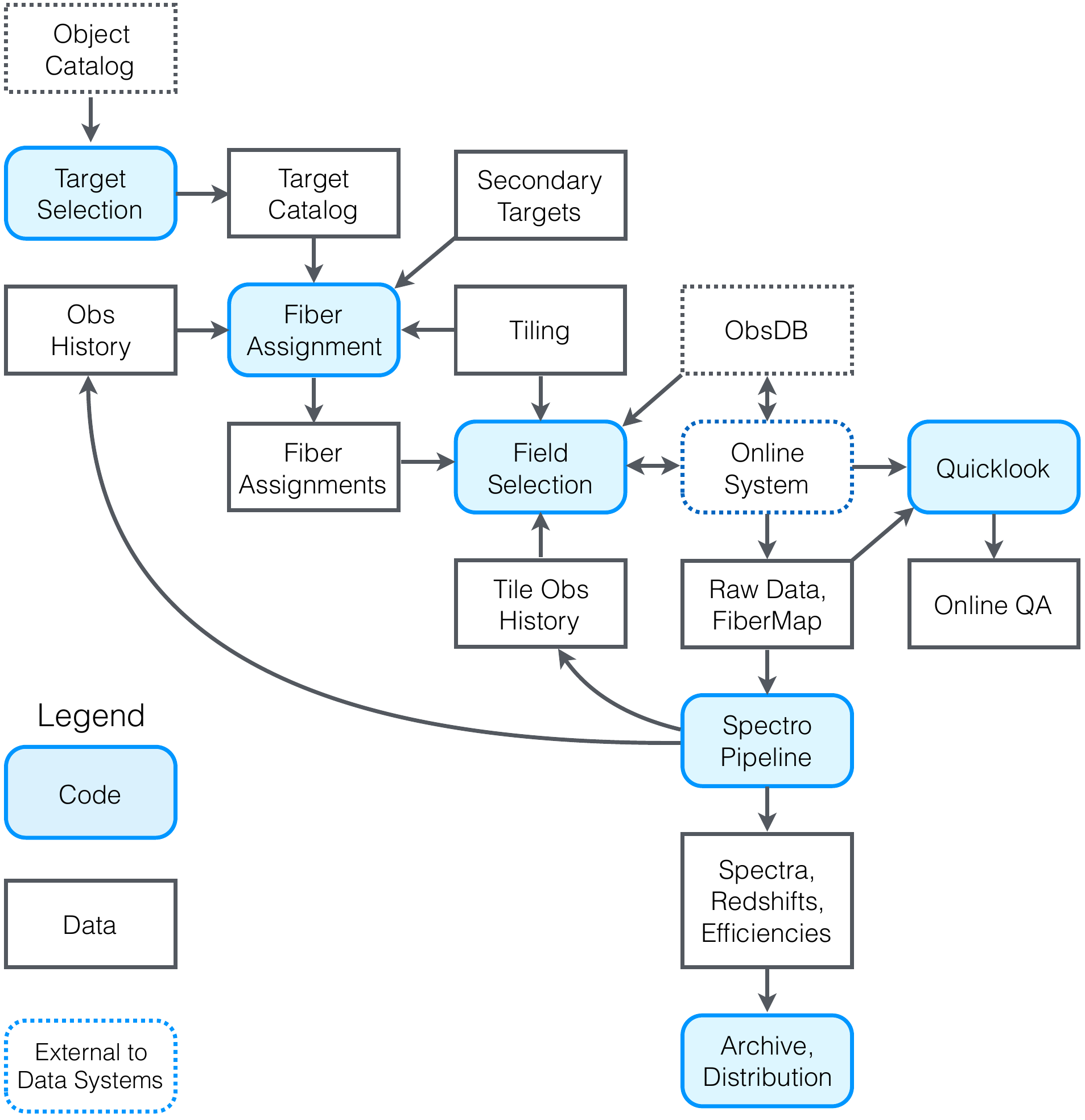}
\caption{
\label{fig:datablock}
Block diagram of DESI Data Systems data flow.
}
\end{figure}

Figure~\ref{fig:datablock} shows a block diagram of the DESI data flow,
including interfaces between Data Systems and the online Instrument Control
System (ICS).

\begin{itemize}
    \item {\bf Target selection} takes an input catalog of detected objects
    and selects which ones DESI could observe.  Advanced version exists.
    \item {\bf Fiber assignment} chooses which targets should be observed for
    each tile on the sky, using priority ranking as necessary to choose
    between multiple targets visible by a single fiber.  Advanced version exists.
    \item {\bf Field selection} chooses which tiles should be observed
    in which order.
    This includes a daily update of the tile priorities and optimal observing
    hour angles (``afternoon planning'') and a real-time decision of
    what tile to observe next (``next field selector'').
    It passes this choice to the DESI Online
    System, which performs the observations to produce raw data.  In progress.
    \item The {\bf spectroscopic pipeline} processes these raw data to produce
    spectra, redshifts, and efficiencies for generating a large scale structure catalog.
    Advanced algorithms exist; automated pipeline in-progress.
    \item {\bf Quicklook} software performs a fast simplified version of
    the spectroscopic pipeline during observations to provide rapid
    turnaround quality assurance plots and metrics.  Started.
    \item Data are {\bf archived} to tape and at a geographically separate
    backup site; all stages of the data processing are available to the
    DESI collaboration via {\bf data distribution} tools.
    In progress, based upon previously existing tools.
    \item {\bf Simulations} and {\bf data challenges} (not shown) are
    used for development and testing of these items and to prepare science analyses.
    Advanced versions exist; operations simulations planned.
\end{itemize}

\subsection{Target Selection Pipeline}

Target selection is described in detail in
the Target Selection section of the Science Final Design Report.
This includes the astrophysical question of selecting targets such as galaxies
and QSOs from the imaging datasets,
as well as selecting regions of blank sky for sky subtraction and
standard stars for flux calibration.
The data management team is responsible for archiving the targeting
datasets and coordinating the data flow from the selected targets through
fiber assignment, observation, and downstream to analyses. In particular,
large scale structure analyses rely upon information both about the targets that were
observed \emph{and} targets that were selected but never observed due to
limited fiber resources.

Data Systems will apply versioned selection algorithms to all of the
versioned imaging datasets and will generate catalogs of targets to observe.
These catalogs will provide a unique ID for each target that is traceable back
to the input images, the processing version of the input catalogs,
and the version of the
target selection algorithm used.  This target ID is passed forward through
the data processing and provides traceable provenance of why each
target was selected and which input catalogs contributed to that selection.
The target catalog will contain approximately as many targets as
are observable over the course of the full DESI survey.
The target selection pipeline itself does not prioritize targets ---
that is the responsibility of the survey planning team in coordination
with fiber assignment (see Sec.\ \ref{sec:fiberpriorities}).


The target selection pipeline is based upon python selection functions applied
to imaging catalog files provided by the Legacy Surveys (DECaLS, BASS, MzLS)
data releases.  These result in a target catalog file containing the variables
used for the selection (e.g. magnitudes), variables needed for downstream
processing (e.g. RA, dec), and variables needed for data provenance traceability
(e.g. the unique target ID).  The input imaging catalog format is documented at
\url{http://legacysurvey.org/dr2/catalogs/} and the output target selection
catalog format is documented at
\url{http://desidatamodel.readthedocs.org/en/latest/DESI_TARGET/targets.html}.

\begin{figure}[tb]
\centering
\includegraphics[width=\textwidth]{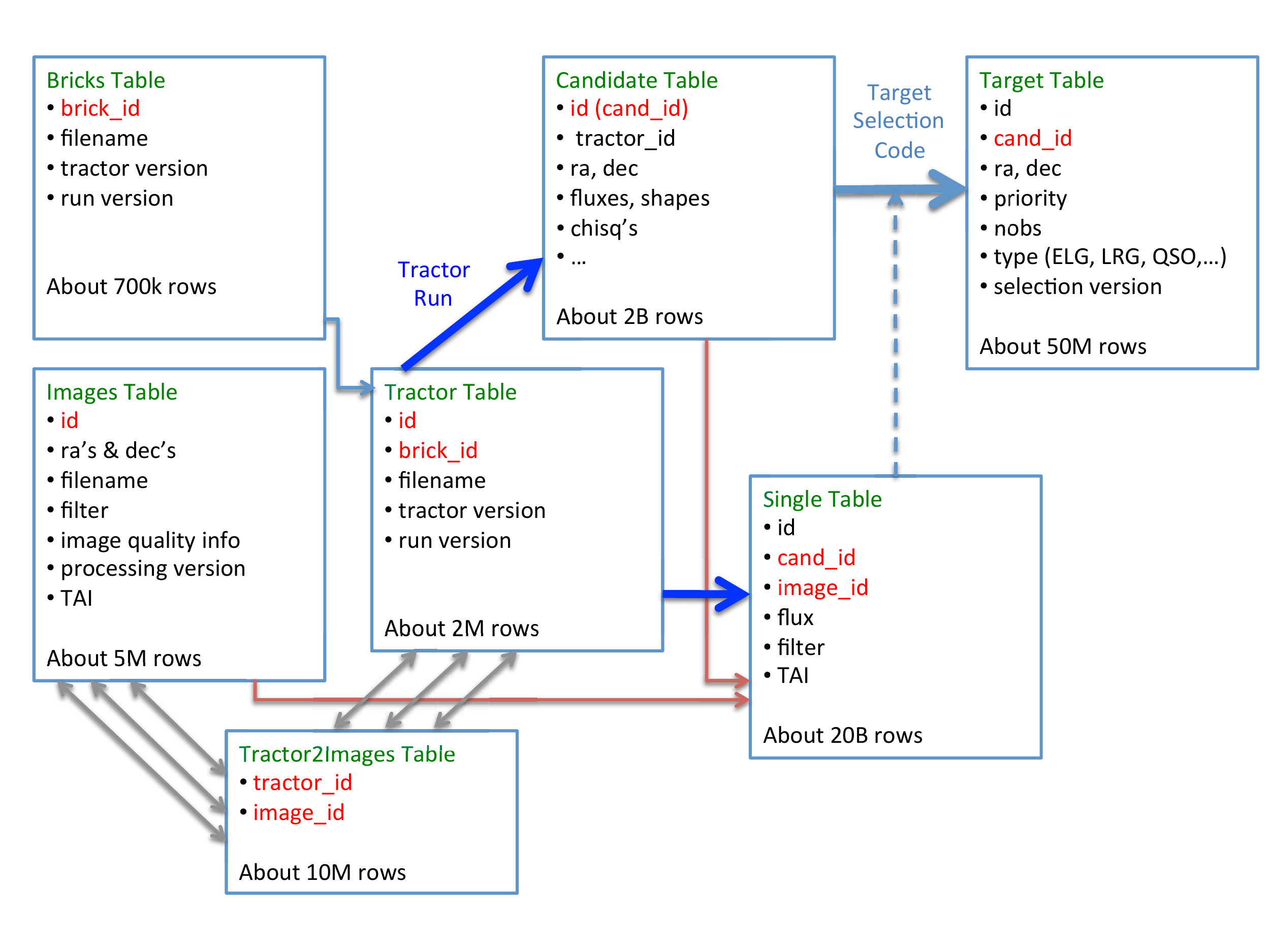}
\caption{
\label{fig:targetdb}
Block diagram of target selection database schema.
}
\end{figure}

These data are also loaded into a Postgres database for higher level queries
and investigations.
Figure \ref{fig:targetdb} shows a block diagram of the target selection
database which tracks these catalogs and their provenance back to the
input imaging data.  The Images table tracks the original input images.
The Bricks, Tractor, Tractor2Images, and Single tables track how those
images are processed to identify objects for the Candidates Table.
Target selection then applies cuts to the Candidates table to produce
the Targets table with the list of targets that DESI should observe.
The target selection will employ a PostgreSQL database hosted
at NERSC.

In order to better coordinate with scientists working on optimizing target selection,
the database will be extended to include tools to track target density as a function
of sky position. 
To better characterize varying sky densities,
functionality will also be extended to include masking of
problematic regions (as also described in Sec.\ \ref{sec:ang}).
In addition to the PostgreSQL database, versioned command line tools will be developed 
in Python that allow on-the-fly calculation of
targets and target densities from flat files of sources, without the need to load database tables.
Ultimately, all of the command line tools written to characterize targets on-the-fly 
will be incorporated as functions into the PostgreSQL database.

Note that the processed images used by the
target selection pipeline are a deliverable of the DESI science collaboration,
not the DESI project itself. The data management team will, however, 
make periodic releases
of targets, tied to releases of the imaging data used for targeting (c.f.\ the Baseline Imaging Datasets
subsection 
of the Science Final Design Report).
The target selection pipeline may be refined and an updated version released in tandem with
these target releases. Upon the commencement of DESI, the final target selection pipeline 
will be released to the community in an ``as is" fashion, without continued maintenance
or support. 

The computing requirements for target selection are presented
in Section~\ref{sec:ts}.

\subsection{Survey Planning, Fiber Assignment, and Operations}
\label{sec:surveyplan}

The survey footprint, field centers, overall survey plan and 
survey validation
plan are described in the Science FDR.  As described before, the baseline survey
is designed to cover 14k sq.\ deg.  For the unlikely case of a descoped and reduced
focal plane with only 3000 fibers, we also model a program covering 9k sq.\ deg.
Data Systems is responsible for developing
and implementing this content with fiber assignment, the next field selector,
and quick-look, as described in the following sections.

\subsubsection{Fiber Assignment}
\label{sec:fiber_assignment}

Fiber assignment is the process of selecting which targets are
assigned to which fibers on which tile, and thus also which targets
are {\em not} assigned to any fiber in the regions where there are
more targets than available fibers.  Fiber assignment depends upon the
details of the fiber positioner layout, the survey tiling pattern
on the sky, the input target catalogs, and spectroscopic pipeline
feedback for targets that have been previously observed.
For each tile, Fiber Assignment will provide the
Right Ascension and declination (RA, dec), and target ID
selected for each fiber.
These outputs are passed by the Next Field Selector to the
Instrument Control System.

\paragraph{Merged Target List}
\label{sec:MTL}

A precursor step of fiber assignment is to generate a Merged Target List (MTL),
which combines the information from multiple sources: the DESI Key Project (\ie,
the primary DESI survey) target catalog; results from any previous observations
of these targets; catalogs of standard stars (for flux calibration) and blank
sky locations (for modeling sky subtraction); and possibly secondary target
catalogs.  Results from the spectroscopic pipeline will be used to inform which
targets have been fully observed and which require additional observations.  The
code that generates the merged target list also assigns target priorities based
upon the target selection bits and past observing history of the target.  For
the DESI main survey, these will be fixed to provide the desired ranking among
LRGs, ELGs, and QSOs.  Secondary targets will be provided priorities, assigned
in a way that will assure that the Key Project is not significantly impacted.
Each collection of secondary targets will be assigned a new target bit,
analogous to the LRG, ELG, QSO, BGS, MWS bits. Each secondary target will also
have a unique identifier, just as the Key Project targets.

The MTL provides flexibility in assigning target priorities for the Key Project,
the Bright Time Survey, and secondary target programs.  The priorities will not
be built into the final Fiber Assignment code but rather the MTL will define the
various priorities, as encoded in the targetmask.yaml file in the desitarget
code repository.\footnote{
  \url{https://github.com/desihub/desitarget/blob/master/py/desitarget/targetmask.yaml}}

\paragraph{Key Project Target Fiber Assignment}
\label{sec:fiberpriorities}

Roughly 25,000 Key Project targets will fall on a single plate and
each spot will be revisited, on average, a bit over five times.  So with
5,000 fibers we have a near match of fibers and targets.  However,
fluctuations either due to Poisson statistics or to clustering will
result in domains where not all targets can be measured (and other
domains where there will be fibers with no unobserved targets within
reach).  Thus choices need to be made.

Approximately 20\% of the time allocated to LRG, ELG, and QSO observations will occur when the
moon is above the horizon.  Because the identifying features in ELG spectra will be found
preferentially at the longest wavelengths covered by the DESI spectrographs, spectroscopic
classification of ELG targets is less vulnerable to moon contamination than is classification
of LRG and QSO targets.  For this reason, we will assign fibers only to ELG targets in a subset of
the fields that cover the DESI footprint.  These fields will be observed when the moon is above the horizon.

As discussed in the Variations of Forecasts with Survey Parameters subsection of the Science FDR, the marginal Figure of
Merit is higher for each \lyaf\ QSO, QSO, and LRG than it is for each ELG.
This relative FoM for each target class and required exposure times
leads to the following baseline prioritization scheme for DESI targets for the 80\% of allocated
time in which the moon is below the horizon:
\begin{enumerate}
  \setlength{\itemsep}{1pt}
  \setlength{\parskip}{0pt}
  \setlength{\parsep}{0pt}
\item confirmed \lyaf\ QSOs with known redshift $z>2.1$, except during grey time, during which only ELG targets will be observed.  The grey-time program will extend over the course of
five years, using the fraction of time with the highest illumination from the moon
as described in the survey section of the Science FDR.
\item QSOs targets not yet observed.
\item LRGs that have been observed but not to their requisite depth
(each LRG requires 1--3 observations depending on their magnitude).
\item LRGs targets not yet observed.
\item ELGs targets.
\end{enumerate}

Every exposure also requires standard stars and sky fibers for
calibration.  For each tile we require that every one of the ten
petals has at least 10 standard stars and 40 sky fibers.  If the initial
assignments don't permit this, some ELG targets are sacrificed for the
purpose.

Because the ranges of the positioners overlap, the assignments must avoid
physical collisions to maintain high fiber assignment efficiency.
The Fiber Assignment code tests whether an assignment would lead to a collision with a previously assigned fiber, using the specific shape and dimensions of the positioner.  All potential assignments leading to a collision are rejected and an alternative assignment is sought.
The final check that there are no collisions will be the responsibility of the Instrument Control System, and the positioners are designed to
be robust even in the case of accidental collisions.  However, the highest
efficiency is obtained by not requesting assignments that would collide in the
first place.

As will be shown in the following section and Table~\ref{tab:fullfocalplaneyield},
this prioritization strategy results in nearly 100\% rate of fiber assignment to the QSO
and LRG samples, with roughly 78\% yield for ELG targets.

\paragraph{Performance of Tiling and Fiber Assignment}
The performance of the tiling and fiber assignment has been studied with mock
catalogs from N-body simulations.  Those simulations assign ELGs, LRGs and QSOs to dark matter
halos at the areal densities and redshift distributions presented in
the Target Selection section of the Science FDR.
The completeness for each target class is presented in Table \ref{tab:fullfocalplaneyield}
for the full focal plane and in Table \ref{tab:pacmanyield} for the reduced focal plane
with 3000 fibers.
The QSO targets are separately listed as those at $z<2.1$ (QSO-tracer), which are used as
tracers only, and as those at $z>2.1$ which are used for the \lyaf.
A fraction of the LRG and QSO targets will be targeting failures, and have
been listed separately as ``bad'' targets.
The \lyaf\ QSOs are not assigned to the fifth-pass grey time observations,
but some of these targets still receive more than four observations.

\begin{table}\begin{center}
\caption{Performance of the baseline tiling simulated on 14k sq.\ deg.\ of the mock
galaxy sample for the full focal plane.
All densities are shown in (sq. deg.)$^{-1}$.
}\label{tab:fullfocalplaneyield}
\begin{tabular}{lrrrrrrrrrr}\hline
&\multicolumn{6}{c}{Times Observed}&           Initial         &Fibers&Number&Fraction\\
 &  0  &      1  &      2   &     3  &      4   &      5 &Number.& Used &  Observed&Obtained \\ \hline
QSO \lya\   &     0 &     1 &   4 & 15 & 26 & 1 &    49 &   169 &    49&0.989 \\
 QSO Tracer   &     1 &   118 &   0 &  0 &  0 & 0 &   119 &   118 &   118 &0.989\\
        LRG   &    15 &    35 & 247 &  0 &  0 & 0 &   298 &   530 &   283 &0.948\\
        ELG   &   531 & 1,879 &   0 &  0 &  0 & 0 & 2,411 & 1,879 & 1,879 &0.780\\
   Fake QSO   &     0 &    89 &   0 &  0 &  0 & 0 &    90 &    89 &    89 &-\\
   Fake LRG   &     2 &    47 &   0 &  0 &  0 & 0 &    50 &    47 &    47 &-\\
 \hline
\end{tabular}\end{center}
\end{table}

\begin{table}\begin{center}
\caption{Performance of the tiling simulated on 9k sq.\ deg.\ of the mock
galaxy sample for the reduced ``Pacman'' focal plane.
All densities are in (sq. deg.)$^{-1}$. 
}\label{tab:pacmanyield}
\begin{tabular}{lrrrrrrrrrr}\hline
&\multicolumn{6}{c}{Times Observed}&           Initial         &Fibers&Number&Fraction\\
 &  0  &      1  &      2   &     3  &      4   &      5 & Number& Used &  Observed&Obtained \\ \hline
   QSO Ly-a   &     1 &     1 &   4 & 14 & 28 & 0 &    49 &   169 &    48 &0.979\\
 QSO Tracer   &     2 &   117 &   0 &  0 &  0 & 0 &   119 &   117 &   117 &0.979\\
        LRG   &    16 &    31 & 250 &  0 &  0 & 0 &   298 &   532 &   282 &0.944\\
        ELG   &   487 & 1,924 &   0 &  0 &  0 & 0 & 2,411 & 1,924 & 1,924 &0.798\\
   Fake QSO   &     1 &    88 &   0 &  0 &  0 & 0 &    90 &    88 &    88 &-\\
   Fake LRG   &     2 &    47 &   0 &  0 &  0 & 0 &    50 &    47 &    47 &-\\
\hline
\end{tabular}\end{center}
\end{table}

The fiber assignment algorithm will have as input the Merged Target List (MTL) .
The true nature of the QSO and LRG targets is unknown until they
are first observed.  The fakes are not observed again.  The QSO tracers and
ELGs are not observed again, since a single exposure is sufficient
to measure their redshift.
The \lya\ QSOs continue to be observed to a maximum of five observations.  The LRGs are observed to a maximum of
two exposures.  This may be modified in the future to allow LRGs to be judged as requiring one, two, or three observations, depending on
their brightness.

\paragraph{Calibration Target Fiber Assignment}
\label{sec:stdstar}

In addition to key project targets (ELGs, LRGs, QSOs), every exposure must
include standard stars for flux calibration and fibers assigned to blank
sky for modeling and subtracting the night sky spectrum from science fibers.
These calibration targets will be assigned to remaining fibers after the key
project target fiber assignment has been completed.
When necessary, fibers assigned to ELG targets may be
exchanged for standard star targets if density fluctuations result in
an insufficient number of standard stars available on the spare fibers.

Based upon BOSS sky and calibration star studies, 40
sky fibers and 10 standard stars are required
for each spectrograph, representing
10\% of the total fiber budget.
Locations of blank sky and standard star targets are provided by the target
selection pipeline and included in the MTL.

Many more ``blank sky'' locations than the required density will be provided.
These will be selected as random sky positions for which there are
no detected objects within a sufficient distance to ensure that
flux from extended objects is excluded.
From this abundance of possible ``sky'' locations, the fiber assignment
will select 40 per spectrograph per exposure.

\paragraph{Secondary Target Fiber Assignment}

The final and optional step in fiber assignment is for secondary target
requests.  These could be assigned to unused fibers, and may also override
standard star and blank sky fiber assignment within the constraints of
calibration target uniformity and minimum numbers required.  However, in most
cases, key project targets will be assigned first, then calibration targets, and
then (optionally) secondary targets.  Performing the fiber assignment in these
steps allows the survey to pre-plan the key project and calibration fiber
assignments, while allowing flexibility for last minute target-of-opportunity
style secondary target requests.  This decouples the key project targeting from
external program secondary target requests, allowing both greater implementation
freedom.  All this will be managed through the MTL.

In exceptional cases, it may be possible for secondary targets to override
key project targets, with the necessary bookkeeping of this fiber assignment
override passed forward to final analyses.  We do not anticipate this to
be a standard mode of operation but structurally this will be a collaboration
management decision rather than an artificial constraint from the design of
the fiber assignment and data management systems.
The default operation will be to assign unused fibers to additional
``sky'' targets and spectrophotometric standard stars.

\paragraph{Target Assignment Bookkeeping}

Large-scale structure clustering analysis depends upon a detailed
knowledge of the efficiency for successfully observing each target.
Upon the completion of fiber assignments for a suite of tiles,
the fiber assignment status must therefore be tracked for each target.
As described in Section~\ref{sec:opsDB}, the status for each target will be updated
following observations to reflect observing conditions and spectroscopic classification.  This will be done by the MTL.

The interface with instrument control system (ICS) decouples the details of tiling
and fiber assignment from the mechanics of actually observing.  For each
tile, Data Systems will provide a file defining the field center,
the mid-point observation time, the
sky coordinates (RA, DEC) of each assigned target, and the design
wavelength for each target.
The ICS will use this information to translate the sky coordinates
to physical coordinates on the focal plane (in mm) for each fiber
positioner (PlateMaker; Section~\ref{sec:expsequence}).
Although there is an atmospheric dispersion corrector (ADC),
there are still small-term offsets due to the atmosphere that
effect the translation to physical coordinates.
In principle, each target location could be optimized to a
different design wavelength, although these effects are small
within the ADC design range of $1<$ Airmass $<2.0$.
The default operation will be to use the same design wavelength for
all targets.

The well-defined interface between target lists and PlateMaker allows
flexibility for creating custom tile definitions for commissioning, debugging,
and potential non-DESI observing programs.  A verification program will
allow pre-testing that fiber assignments are physically possible.

\paragraph{Bright Time Survey Fiber Assignment}

The above description of fiber assignment has focused on the dark time
key project survey, but the same code is directly applicable to the bright
time fiber assignment as well.  The basic structure fits both programs:
each has a pre-defined tiling of the sky and input catalogs of various target
classes with priorities and requested number of observations.  They will
differ only in the inputs, while using the same fiber assignment code.


\subsubsection{Afternoon Planning}\label{subsec:afternoonplan}

While fiber assignment occurs offline, afternoon planning, next field selector, and quicklook are run
at the mountain in real-time.  In the afternoon planning stage, information on the status of each field
in the footprint will be consolidated, hour angles assigned to all fields that remain to be observed, and potential fields
for the night's observing prioritized according to the long-term survey plan and moon phase.
The Instrument Control System (ICS) will run the afternoon planning software as a Python module.
The afternoon planning software will identify a subsample of fields from the full footprint to observe
and pass the information to the next field selector for real time observations.
The afternoon planning should work naturally for
dark and bright time programs and require the same basic formats.

The afternoon planning software will communicate with fiber assignment to determine which fields
are available to observe.  The software will communicate with the abbreviated online data reductions (quicklook)
to gauge the depth of exposures for each field.  The afternoon planning software will communicate
with the full data reduction pipeline to confirm or override the status implied by the online reductions.
Fields will be prioritized for observation each night based on the moon position and long term strategy.

The global planning with hour angle assignments can be updated as often as
every afternoon.
These modifications to the design of each field will be recorded
before the night begins, thus appearing seamless to the software that governs mountain operations.
To optimize survey progress, the results of survey simulations will be used to establish the
algorithm to prioritize fields each afternoon.
This algorithm will form the basis of the afternoon planning software
that will be run in full operations during the five-year survey.

\subsubsection{Next Field Selector}\label{subsec:fieldselector}

The next field selector will implement the survey plan in real-time based upon current
observing conditions, past observing history, and what part of the sky is currently visible.
The Instrument Control System (ICS) will run the next field selector as a Python module.
The next field selector will choose the next field to observe
and pass the information from fiber assignment to ICS for plate mapping and acquisition of that field.
At a minimum, the information passed to ICS will include the field center, RA and DEC of each fiber assignment,
guide star information, and expected exposure time.  The next field selector should work naturally for
dark and bright time programs and require the same basic formats.

To optimize survey progress, the results of survey simulations will be used to establish the
algorithm to prioritize field selection every time the survey moves to a new
location on the sky.  This algorithm
will be a function of sky conditions, LST, and moon position.
This algorithm will form the basis of the software for the next field selector
which will be run in full operations during the five-year survey.

\subsubsection{Quicklook Extraction Pipeline}
\label{sec:onlineQA}

Unlike BOSS, where observations are accumulated in exposures of fixed duration,
DESI will require more precise tuning of exposure times to minimize time lost to
over-exposures.
DESI will determine exposure depth during open-shutter time
using information from the guide camera as described in Section~\ref{sec:etc}.
The guider images will be used to determine sky conditions such as atmospheric
transparency, seeing, and sky background.
The exposure time calculator will use this information to
predict exposure times at the beginning of a field and update approximate
exposure depths during each exposure to determine when the field meets
a threshold depth to meet survey goals.
These exposure depths will be tuned to a fixed S/N ratio for a fiducial target
at a fiducial redshift.
For example, spectral templates representing the ELG sample will be fed to the exposure time calculator
to estimate the required open shutter time to achieve S/N$=7$ per pixel integrated over the [OII] emission
line doublet at the DESI flux limit.

While the real-time exposure progress will be judged by the exposure time calculator,
there will be quick-look spectral reductions run on all science exposures to ensure that
the data meet the quality predicted at the end of the exposure.
These reductions will be used as a diagnostic to ensure that the instrument is performing correctly,
allowing the observing staff to address problems quickly should they arise.
Under normal operations all observing decisions are automated, while
reserving the ability for operator override for testing and exceptional
circumstances.

This quick-look code will be a simplified version of the full data reduction
pipeline described in the following section, replacing the most
computationally expensive steps with faster approximations so that results
are available within two minutes of readout.
Since it is based upon the full data reduction pipeline,
the data group will provide the software for this analysis, which will be
integrated into mountain operations software.  The results of these
quick-look reductions will be archived and distributed as a data product.
Similarly, guide camera data, temperature measurements, and any
other meteorological information used for making observing decisions
will be archived.

\subsubsection{Survey Simulations}
\label{sec:surveysim}

Software simulations of the full survey will be used to model the order in which tiles are observed
and under what conditions each exposure is acquired.  Survey simulations will use the same data
flow and algorithm code as ICS will use for managing the actual observations.
Field centers, target classes associated with each field, hour angles, and long term strategy will
all be input to the survey simulations.  Weather and moon conditions will be simulated for each night of the year.
The afternoon planning software and next field selector will be used by the simulator to prioritize fields each
night and select each field in real time for observing.

Survey simulations confirm the algorithms developed for the afternoon planning, next field selector and exposure time calculator
subsystems and also test the interface to ICS.  These simulations will inform the survey coverage and uniformity
of clustering in the end-to-end data challenge.
The survey simulations will address the following specific questions regarding Survey Planning:

\begin{itemize}

\item  What is the final algorithm for afternoon planning?

\item  What is the final algorithm for the next field selector?

\item  What is the possible variation in areal coverage due to weather?
This will help assess how robust the afternoon planning, next-field selector and long-term
survey plan will be to seasonal bad weather.

\item  What is the algorithm for determining real-time exposure times?
This would require estimates of seeing, atmospheric transparency, and
sky background as part of the survey simulator.  These will be used in combination
with simulated spectra to test the algorithm for the exposure time calculator and
the total exposure time required to complete each field.

\item  What is the area and target completeness we expect as a function of time?
The clustering group will need semi-completed footprints to test
issues like reconstruction, fiber collisions, and incomplete target
samples.

\item  What is the effect on redshift efficiency of varying sky background
and atmospheric transparency as a function of time?  Science analyses
must correct for these redshift efficiency variations when constructing
Large Scale Structure catalogs.

\end{itemize}

\subsubsection{Survey Quality Assurance Interface}
\label{sec:qainterface}

The quicklook software described in Section~\ref{sec:onlineQA} will be used to inform the observer on the quality of
each image immediately following the completed exposure.  However, the quicklook routines will not provide a high-level
overview of survey progress or global data quality.  The Survey Quality Assurance Interface will be an online tool
accessible to observers and members of the collaboration to track the progress of the survey, planned observations,
and data quality.

The quality assurance interface will record information from the operations database,
the quicklook extractions, the full pipeline, and the catalogued output of the full data reductions pipeline.
It will be used to ensure consistency between the survey progress recorded at the mountain and downstream
in the analysis.

It will provide visualization tools to access this information.  The tools will interface the Operations database and offline
processing results.  For example, the interface should display all of the fields in a map that is color-coded by completeness,
with an option to show individual layers of fields.  It should be able to display histograms of various
metrics averaged over each exposure such as sky background and fiducial S/N.  The interface should have the ability
to show observation progress as a function of time, including projections for future observations.
The quality assurance interface will require coordination with ICS.

\subsubsection{Planning for Commissioning and Survey Validation}
\label{sec:SV}

DESI will conduct a phase of commissioning and 
survey validation
to establish general operational procedures, evaluate instrument performance,
and validate selection algorithms applied to each class of target.
The commissioning strategy is outlined in Section~\ref{s6:Commish}.
Members of
the Data team will participate in commissioning to ensure that requirements needs are met.
The Data team must confirm that the data model is complete for downstream
analysis.  Likewise, the Data team must test the operations software and iterate with developers
to ensure that it is well-documented and observations are easily performed by those who are not
familiar with the full system.

Once calibration routines, operational procedures, and instrument performance are understood,
time before the main survey should be dedicated to science observations to the extent possible.
These 
survey validation
observations will broaden the main sample to inform the final target selection and analysis.
The exact scope of each of these observations is not yet defined, but one should expect samples on the
order of 1\% of the final cosmology sample.  In other words, each program will likely take one month
to complete with a complete spectrograph and should produce hundreds of thousands of spectra.
There is overlap between some of these programs, so they should be performed in parallel whenever possible.

Equally important, the observations conducted during 
survey validation
should produce samples that are
intrinsically interesting and can lead to immediate analysis and results.  There may be an early data release
to facilitate dedicated analysis of the 
survey validation program.
A list of 
survey validation
programs is included in the Science FDR and the associated documents.

%

\subsection{Data Processing}
\label{sec:dataprocessing}

\subsubsection{Spectroscopic Data Reduction Pipeline}

The DESI spectroscopic data reduction pipeline (DRP) is a software
system that will calibrate and transform raw CCD data from the DESI spectrograph detectors
into extracted spectra with classifications, redshifts,
and appropriate error estimates and warning flags.
The requirements for the subcomponents of this pipeline are
derived entirely from the top-level requirements of
redshift success rate for survey-quality data.
In order to produce redshift catalogs at the required level of
accuracy and precision, the DRP will also generate
high-quality instrument calibration files and one-dimensional
target spectra.

\begin{figure}[tbh]
\centering
\includegraphics[height=3.5in]{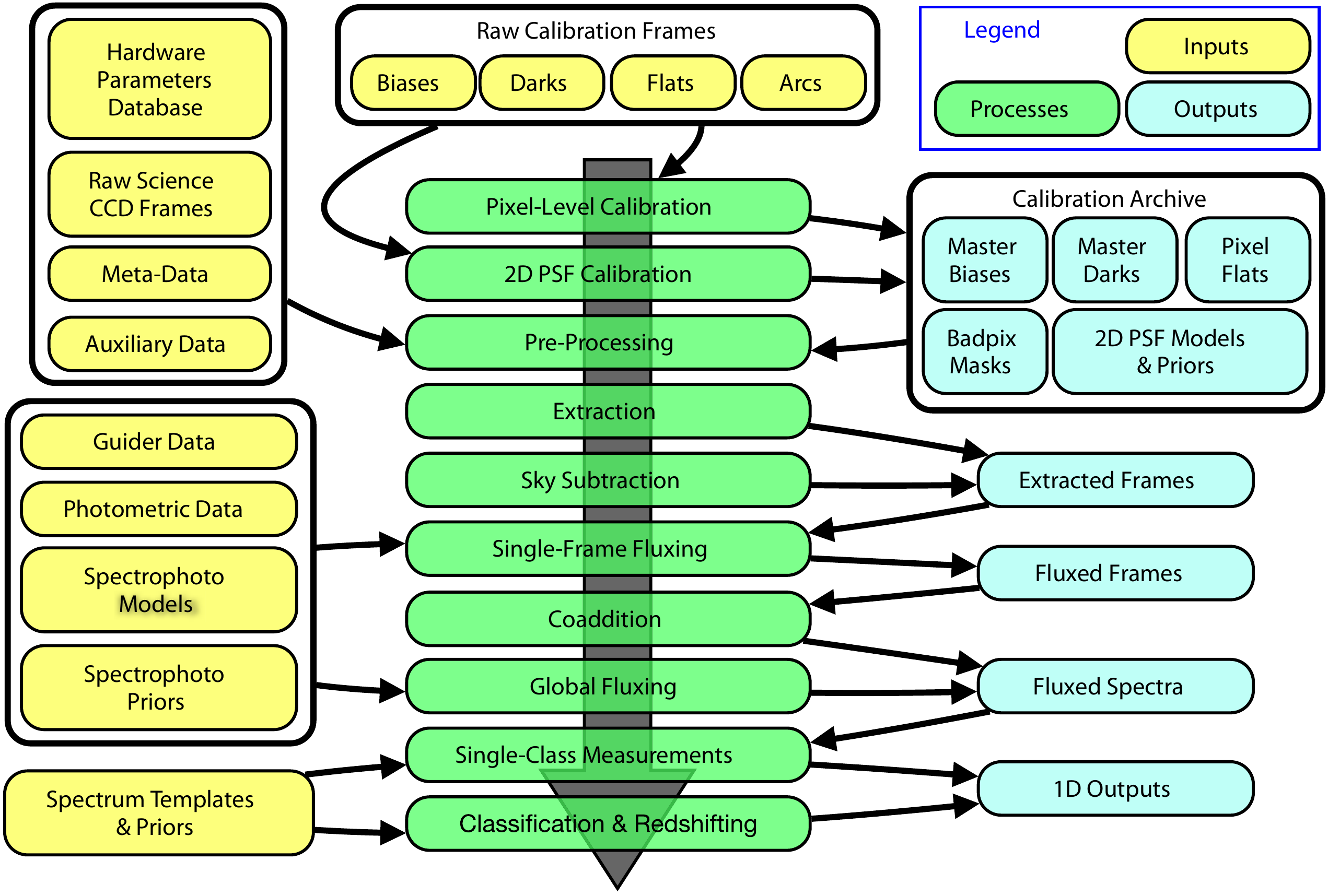}
\caption{\label{fig:DRPfigure}
Schematic processing flow of the DESI DRP conceptual design.
Successive steps receive data from previous steps either
through intermediate data products or through direct
API/function calls.}
\end{figure}

The design of the DESI DRP is based on the mature
and successful model of the SDSS and BOSS spectroscopic pipelines
\cite{Bolton2012}, with improvements and
modifications as necessary to reflect technical and
scientific differences between the DESI and SDSS/BOSS projects.
The schematic processing flow of the DESI DRP design is illustrated
in Figure~\ref{fig:DRPfigure}.  In the following subsections, we briefly
describe the major steps in this pipeline, highlighting in particular the
aspects that are new for DESI, as well as points at which multiple software
solutions are still under consideration as a hedge against unanticipated
developments in hardware design or survey operations.

\paragraph{Pixel-level Calibration and Preprocessing}

The first stage of the DESI DRP will derive and apply the necessary
calibrations to the pixels of the DESI science CCD detectors.
One suite of software tools and calibration data-taking
procedures will be developed to enable estimation of the gains, dark
currents, bias patterns, bad-pixel maps, pixel-to-pixel
quantum-efficiency (QE) variations, nonlinearities, and electronic
crosstalk signatures of all amplifiers on all DESI science detectors.
These tools will be applied to the calibration frames, in order to develop
a pixel-level calibration library that is updated with each lunation
and also with every adjustment of the detector hardware itself.
The details of coordination between hardware adjustment,
calibration data-taking, and pixel-level calibration processing
are a matter for the operations phase, but they can be greatly facilitated
in the construction phase through the adoption of a calibration-product
data model using a simple date-based file-naming scheme such as is currently
used for BOSS\@.

A second set of routines will be developed to apply these bias, dark, gain,
QE, crosstalk, and non-linearity
corrections to individual frames, to estimate the read noise in
each amplifier, to estimate inverse-variance images based on read
noise and photon-count noise, and to flag bad pixels and cosmic-ray hits.
The resulting corrected raw data images and their associated
inverse-variance images will then passed to the subsequent steps in the DRP\@.
Sufficient algorithmic implementations of all aspects of pixel-level calibration are
found within the BOSS pipeline, and thus the construction of this subcomponent
of the DESI DRP will largely be a re-coding of existing software within a more
modular Python framework.  This preprocessing step is fast enough ($<1$~sec)
that it can always be applied directly to raw data and passed forward; it is
not necessary to re-write the preprocessed data back to disk.

The lossy fiber system described in Section~\ref{sec:spectro_mech_elec_design}
will be used to measure the individual pixel response to generate
``pixel flats'' for raw data pre-calibration prior to spectral extraction.
These calibration data and code for analyzing them are
archived and distributed with the science
data as part of the data management system,
as described in Section~\ref{sec:transfer}.

\paragraph{2D PSF Calibration}

The ``2D PSF calibration'' stage of the DRP will derive and
record the calibration of the optical path between the telescope
focal plane and the science CCD planes that is necessary to enable
spectral extraction.
This includes determination of the location of the traces of the spectra
on the CCDs, the characterization of the shape of the two-dimensional
PSFs of the spectrographs as they vary over their camera fields of view,
the measurement of the throughput of fibers relative
to one another (``flat-fielding''), and the determination of the
shift of PSF centroid as a function of wavelength in each fiber.
Taken together, these calibration aspects define a transfer matrix
that maps input counts at the focal plane as a function of wavelength
and fiber number into output illuminating counts on the plane
of the DESI science CCDs,
along with any charge-transfer inefficiency effects
that further modify the effective PSF\@.

The determination of these calibrations
will be made on the basis of flat (incandescent) and arc
(gas discharge) lamps uniformly illuminating the pupil and the
focal plane of the telescope.

The spectrograph optics are very stable since they are mounted on
unmoving optical benches in a temperature controlled room\footnote{in
contrast to BOSS where the spectrograph optical benches are attached to
a moving telescope at ambient temperature}.
These will be periodically calibrated using the
continuum and arc lamp calibration system described in detail in Section~\ref{sec:Instr_Calibration_System}.
These exposures will measure the
spectral trace positions, wavelength solution, and PSF shape
as projected onto the CCDs.

Due to the anticipated stability of the DESI
systems relative to the SDSS/BOSS systems, we plan to develop a library
of 2D PSF calibration data to serve as a prior-knowledge source
for routine re-calibration, and
to identify the dominant modes of variation in the spectrograph transfer functions.
We expect furthermore to verify and tune our 2D PSF calibrations against the
optical design model of the DESI spectrographs.  During the DRP construction
development phase, we will implement multiple
mathematical basis sets for the representation of 2D PSFs in order to have
flexibility in choosing an implementation best suited to the data delivered
by the as-built optical system.

A first empirical model of the 2D PSF has been developed and tested
successfully against simulations of DESI arc lamp images. 
It uses a Gauss-Hermite basis to describe the core of the PSF, with a Lorenztian to describe
 its tails. The image simulations are based on a PSF model derived from the engineering design of
the spectrograph, including optical aberrations, diffraction, the mean free path of NIR photons at
 large incidence angle, and charge diffusion in the CCD.

The 2D PSF calibration parameters determined in this stage
are used to perform the extraction of science spectra from the DESI
CCDs in the next stage of the DRP\@.
We expect that our calibration regime will determine these
parameters with statistical errors that are small enough relative
to noise levels in the science exposures that they can
be neglected, but we will implement tracking of 2D PSF
calibration-parameter noise if this proves necessary.

Pairs of Hartmann exposures on arc lamps
will be used to monitor focus of the spectrographs, which is significantly
more sensitive and precise than using the widths of the arc lines alone.
Daily bias and dark exposures will also monitor the CCDs.
These will be used by the daily spectral extraction pipeline and to
monitor the overall stability and health of the spectrographs.

The primary moving part affecting DESI calibrations is the fiber system.
Flexure in the fibers could affect the focal ratio degradation (FRD) losses
or the near-field image at the fiber output,
and thus the effective fiber throughput and PSF as projected onto the CCD.
These in turn affect the PSF model and the flux calibration, and thus the ability to
accurately model and subtract the sky from the science fibers.
If necessary, these effects could be partially ameliorated in software
by normalizing sky lines in science fibers according to the ratio
of modeled flux in the model for sky background or by explicitly absorbing sky subtraction residuals with a PCA basis while
fitting the galaxy and QSO templates.  Alternately, if fiber flexure effects
are reproducible, PSF and throughput calibrations could be performed each
morning using the exact fiber positions of the exposures taken throughout
the night.  Ongoing lab tests and simulations seek to identify the scale
of this issue and whether there are any impacts on Data Systems software.
The first spectrograph testbench data will provide the definitive study of
the spectrograph PSF stability under changes of temperature and fiber flexure.

\paragraph{Spectral Extraction and Sky Subtraction}
\label{sec:extract-sky}

This component of the DRP will extract one-dimensional
background-subtracted spectra and associated noise
vectors from individual preprocessed
two-dimensional science CCD exposures and their associated 2D PSF calibration
solutions.  The spectra delivered will have units of flat-fielded counts.

The red channels of the DESI spectrographs are designed to operate at sufficiently
high resolution that the free spectral range between night-sky lines is large
enough to cover the necessary cosmological volume even if isolated wavelengths are
compromised by sky-subtraction systematics.
Realizing the fullest possible cosmological signal from the
faint ELG targets of the DESI experiment nevertheless requires
sky subtraction of uniformly high quality, and the DRP plan is therefore to implement

spectral extraction using the ``spectroperfectionism'' (or 2D-PSF) algorithm of \cite{Bolton10}.
This algorithm offers the advantage of extracting the data using the
actual 2D PSF of the spectrograph, avoiding the biases associated with
non-separability of this PSF that can translate into sky-subtraction
residuals when applying the row-by-row cross-section-based
``optimal extraction'' algorithm of \cite{Horne86}.
The output spectra,
noise vectors, and resolution estimates from the 2D-PSF extractions are furthermore
desirable in that they deliver a closer approximation to lossless compression
of the raw data \cite{Bolton2012b}, and can be tuned to a convenient and common wavelength baseline
immediately upon extraction.  Furthermore, the extracted flux samples from the 2D-PSF
algorithm are free of covariance with one another.

In its most straightforward
implementation, the 2D-PSF method is more
computationally expensive than the row-by-row method.  Work done during
the R\&D phase has demonstrated that 2D-PSF extraction
can be implemented using an iterative extraction procedure
that operates on sub-regions of the data, rendering the problem tractable at
DESI scale even with today's computing resources.  For example, extraction of a single
frame can be performed in about one minute on 120 nodes of a modern supercomputer, and
can scale to many more nodes as needed.
This implementation requires a periodic separation
between adjacent fiber spectra on the detector (``bundle gaps''),
and this requirement is satisfied
by the spectrograph design (see Section~\ref{par:fiber-slit}).

Two implementations of the spectroperfectionism algorithm in Python and C++ using MPI have been
implemented and tested on pixel level simulations. 
Both provide the same output which is, for each fiber, an array of uncorrelated flux density, their variance, and a resolution matrix to convert any spectral model to the resolution of the spectrograph. Both codes allow the extraction on an arbitrary wavelength grid.

The resolution matrix is used for all subsequent reduction steps. For instance, when fitting for the average illumination spectrum for the fiber flat-fielding or when fitting for the sky spectrum, we consider a spectral model at infinite resolution, and degrade it to the resolution of each fiber individually. This is an important improvement over the existing SDSS/BOSS pipeline.

\paragraph{Flux Calibration and Coaddition}

The DRP will perform flux calibration of the extracted and
sky-subtracted DESI count spectra
by matching physical stellar models to the count spectra
of spectrophotometric standard stars observed during the same exposures
as the primary survey targets.  These matches will be used to derive
the conversion vector from counts to specific flux units.  Data on seeing and guiding
from the guide camera may furthermore be used to correct for the differential
effects of seeing and fiber sampling on extended galaxies as compared to point-like
calibration stars.  Flux-calibrated spectra from individual cameras and
exposures will then be combined into single output spectra.
The final coadded spectra will then be adjusted with low-order correction vectors to bring
the synthetic photometry of standard stars and galaxies into agreement
on average with available broadband photometry.
This procedure is substantially similar to the methods used in BOSS currently
to achieve the $\sim$5--10\% spectrophotometric accuracy that is sufficient to
enable the application of flux-calibrated templates for classification
and redshift measurement,
as well as the measurement of clustering
in the \lyaf~when combined with additional quasar continuum-normalization techniques.
More advanced approaches to spectrophotometric calibration
for further optimization of the DESI \lyaf~cosmology analysis
will be pursued by the pipeline and \lyaf~science working groups
in collaboration during the operations phase.

A first version of the calibration and coaddition algorithms have been implemented
and applied to simulated spectra of all DESI targets in 2015 (starting for CCD-level simulations). We were able to recover the input spectra at the expected statistical precision. More detailed simulations including CCD defects will be used to further test and improve the algorithms.

\paragraph{Classification and Redshift Measurement}\label{sec:spec1d}


The DRP will perform physical classification and redshift measurement
of targets observed by DESI.  The redshift and classification reported
for each object will the best fit redshifted solution among the
various target models. This work inherits from the experience of the
BOSS spectroscopic survey analysis \cite{Bolton2012}.

The development of the redshift code is organized in two phases.  In a
first phase, a variety of approaches to redshift fitting will be
investigated by several groups within the DESI collaboration. This
work is organized with a series of data challenges based on
simulations.  In a second phase, the best algorithms will be
considered for the DESI redshift fitting code and further refined. It
is likely that different algorithms will be better suited for
different target classes.

The approaches currently pursued span a large range of possibilities
 
\begin{itemize}

\item $\chi^2$ template-fitting of a large library of archetype
  spectra (``Redmonster'' software: T. Hutchinson et al., in preparation).
\item $\chi^2$ fitting using a linear combination of a restricted set of
  templates.
\item Non-linear models.
\item Dedicated emission line fit for ELGs.
\item Bayesian estimators.

\end{itemize}

One of the challenges of these developments is to make use of the
accurate resolution provided by the pipeline (wavelength dependent,
one per fiber for each camera arm), using individual exposures rather
than co-added spectra, and accounting for the spectroscopic data
reduction imperfections, within a finite CPU-time budget.  Another is
to build accurate templates, either from theoretical models
(population synthesis codes for ELG and LRG, theoretical stellar
models), or empirical models derived from existing data sets (for
instance PCA models for quasars).  Photometric information will also
be used to weight the possible best fit solutions.

The developments, tests and comparisons of those approaches are
organized within a series of redshift data challenges. For each
challenge, two sets of simulated spectral data are made available to
the collaboration, a test sample with the input target classes,
redshifts and spectra, and a validation sample with the truth hidden
to the competitors until the end of the challenge.
 
Preliminary results from the first redshift challenge are very
encouraging. They are however probably too optimistic as only
statistical noise was considered in the simulations, without any
pipeline artifacts.  We foresee to increase the level of realism of
the simulations for the next challenges, including unmasked cosmic ray
signal, spectro-photometric calibration errors, systematic sky
residuals, Galactic dust extinction, and a range of observation
conditions.  The second challenge as the first one will be based on
fast simulations of spectra, emulating the statistical noise and DRP
systematic errors, while the last ones will make use of the full
pipeline simulations, starting from raw CCD images processed as the
DESI data. In parallel, the newly developed algorithms will be tested
with existing public data of the BOSS survey.

The output of the successive redshift data challenges will also be
used for a fast simulation of the redshift efficiency and precision
needed for the development of the galaxy clustering analysis.

\paragraph{Progress and Quality Assurance Tracking}

The DRP will automatically generate files summarizing the survey progress
and quality metrics for each processed field.  These will be available as
plots, binary tables, and web pages for a quick assessment of the status.

\paragraph{Prototype Spectroscopic Pipeline Results}

\begin{figure}[tb]
\centering
\includegraphics[width=\textwidth]{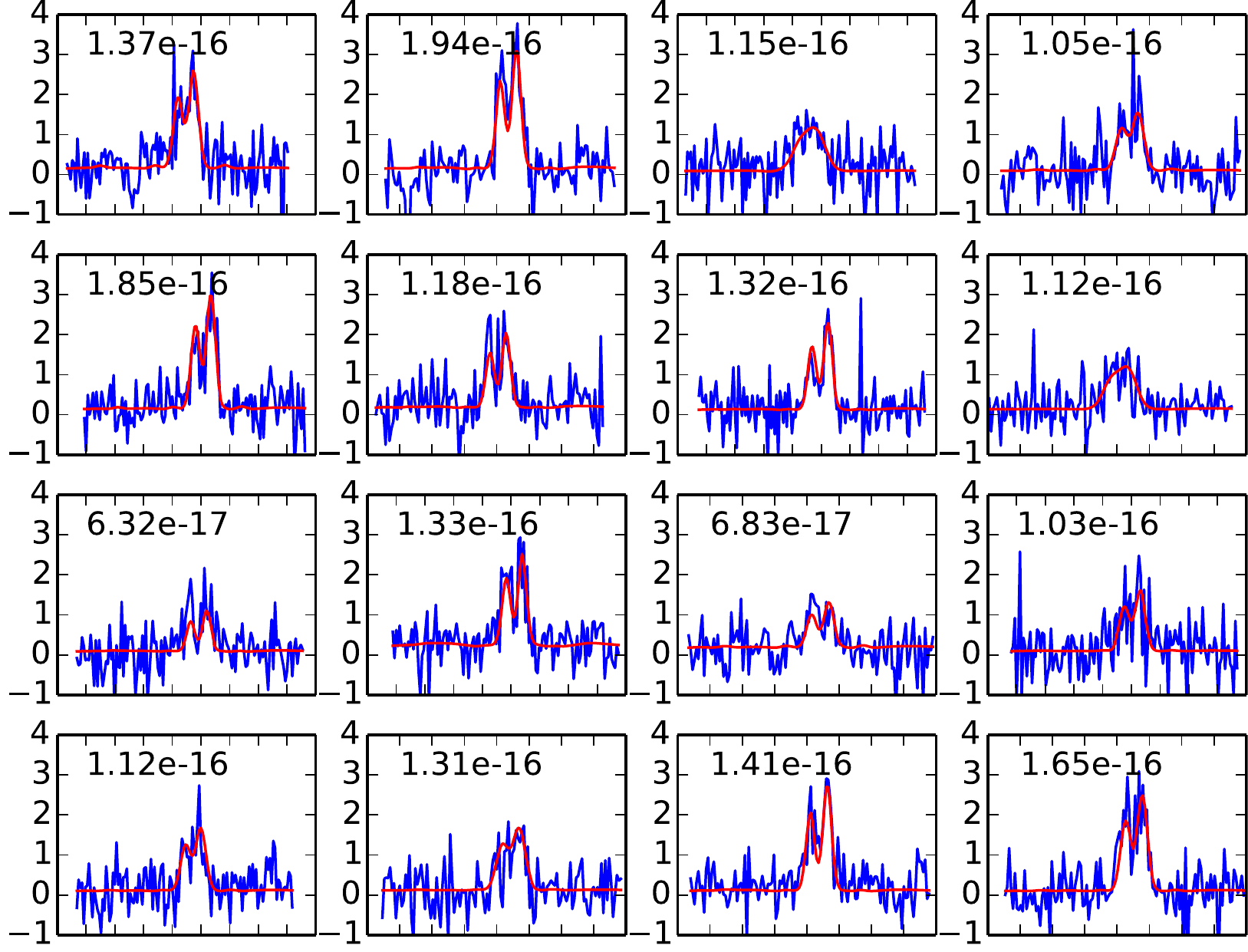}
\caption{
\label{fig:elg_gallery}
Example ELG spectra extracted by the DESI pipeline starting from pixel-level
raw data simulations.  The inset number is the [OII] flux in erg/s/cm$^2$.
Blue shows the extracted calibrated flux [$10^{-17}$~erg/s/cm$^2$/\AA]
{\it vs}.\ wavelength;
red shows the input truth.
}
\end{figure}

\begin{figure}[tb]
\centering
\includegraphics[width=\textwidth]{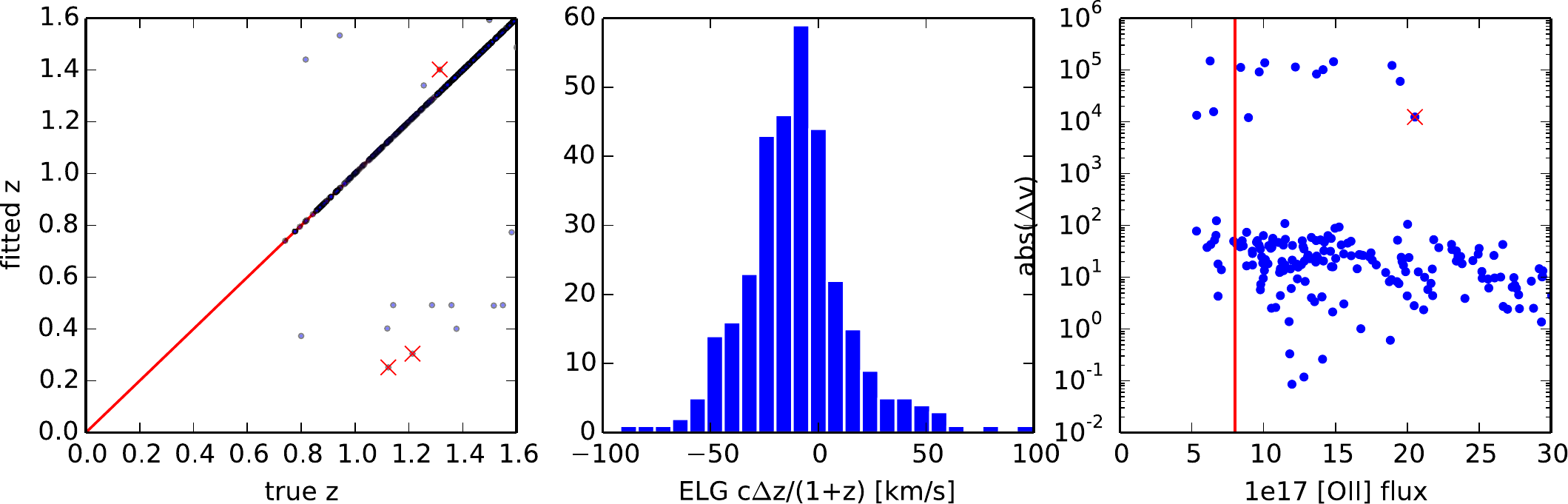}
\caption{
\label{fig:zelg}
Redshift fitting results from pixel-level raw data simulations of ELGs
processed with the prototype DESI spectroscopic pipeline.
}
\end{figure}

Figure~\ref{fig:elg_gallery} shows example ELG spectra extracted and
calibrated by the prototype DESI spectroscopic pipeline, starting from
pixel-level raw data simulations (see Section~\ref{sec:pixsim}) and proceeding
through extraction, sky subtraction, flux calibration, and redshift fitting.
The inset numbers are the [OII] flux;
the nominal DESI flux limit is $8\times10^{-17}$ erg/s/cm$^2$ and 
a more typical case is $1.5\times10^{-16}$  erg/s/cm$^2$.
Figure \ref{fig:zelg} shows results of fitting redshifts to these and
additional ELG spectra.  The left
plot shows the fitted redshift $z$ versus~the true redshift.
1\% are ``catastrophic outliers'' flagged with red $\times$ symbols where
the pipeline was wrong but raised no flags; this is well below the 5\%
requirement.
The other outliers were flagged by the pipeline as uncertain measurements,
and count against the overall efficiency.
The middle plot shows the redshift accuracy.  The bias
(10~km/s) and scatter (30~km/s) are well below the requirements
(60~km/s bias and 150~km/s scatter).  The right panel shows the
redshift accuracy in km/s versus ELG [OII] line flux; the red vertical
bar is the nominal DESI flux limit.  Although these results are
encouraging, we recognize that
the current simulations are too clean and these metrics will get worse.

\subsubsection{Software Development}

The DESI DRP will be a Python-based
system, supplemented by calls to C/C++ code as appropriate for
computationally intensive steps.
Python is ideal because of its free and open-source nature, its
wide and increasing adoption within physics and astronomy,
the availability of mature and full-featured numerical libraries,
its support for linking to compiled modules from lower-level
languages, its ease of maintenance,
and its facility for combining multiple coding styles (procedural,
object-oriented, scripting) within a single flexible software system.
All current DESI pipeline developers are already working
in Python or Python/C/C++.

The primary mode of parallelization for the DESI DRP will be through
an ``embarrassingly parallel'' mode where individual exposures or
pointings are processed by individual computer cores independently
of other data.  Some pipeline steps are parallelized using OpenMP (C++) or
multiprocessing (Python) to parallelize within a single node.
The computationally intensive low-level extraction code is parallelized
across multiple nodes using MPI.

The schedule for DESI DRP development, implementation, and testing
consists of: an initial technical design
specification phase,
a core library development phase, a prototype system integration and testing phase,
a full-scale testing and refinement phase,
a commissioning phase coincident with DESI hardware commissioning,
and an operations and maintenance phase during the course of routine
survey operations.  Throughout the pipeline construction process, the functioning
of individual software components and integrated systems will be
verified through unit and functional tests, including an automated
nightly raw data through redshifts integration test for a few fibers.
Higher level integration, functionality, and performance will be tested
through periodic ``data challenges''
based on data from eBOSS, DESI simulations, and DESI spectrograph teststand data.
Development and testing will be done on a range of
small-scale (personal) and large-scale (\eg, NERSC) computing systems to
ensure that the DRP scales and executes correctly across a range of architectures
and operating systems.

All intermediate and final data products of the DESI DRP will be archived in
FITS files with a documented data model.  The intermediate files are anticipated
to be of interest both for pipeline debugging and development, as well as for
novel scientific analysis methods that need to access the data and calibrations
at a lower level.



All software for DESI data processing will be managed in a version-controlled
code repository.  All collaborators will have access to this repository.
Both internal and public data releases will be based upon tagged versions
of the code, available at the same time as the data.

\subsection{Data Transfer, Archive, and Distribution}
\label{sec:transfer}

\subsubsection{Raw Data Transfer}

Raw data will be transferred daily from KPNO via NOAO
to the central data repository at NERSC.
There it will be backed up to the HPSS tape storage
system\footnote{\url{http://www.nersc.gov/systems/hpss-data-archive/}}.
By retaining files at NOAO, we can maintain an additional, geographically
remote copy of the raw data.

Before transfer, raw data will be compressed, and directories containing nightly
data will be permission-locked so that no further data can be written.
Checksum of files will be computed before transfer to insure data integrity
at every stage of transfer and backup.  As part of this process,
the operations database will also be backed up, both in a flat file form
and on a remote clone database.

Failures of the checksums at any point in this chain of steps will
trigger human investigation of the problem, and recovery from the original
files at KPNO.

The NOAO Data Transport System (DTS)~\cite{2010ASPC..434..260F,doi:10.1117/12.857329}
is currently serving as the data transfer pipeline for the Dark Energy Survey,
and we are planning to adapt it to DESI, which has much more modest transfer
requirements.
See Section~\ref{sec:spec} for further details of data bandwidth requirements.

\subsubsection{Operational Data Archiving}

In addition to the raw spectroscopic data, Data Systems will also archive
operational data such as the Guide-Focus-Alignment camera output,
temperature and pressure measurements, the history of commands sent to the
fiber positioners, their recorded position, engineering monitoring data,
and the images and analyzed data from the fiber view camera.

\subsubsection{Processed Data Distribution}

The DESI processing pipeline will deliver results as flat files that will be
placed in a simple, transparent hierarchy.  To the extent that databases are
used for intermediate or final results, the database will be interchangeable
with flat files, that is, one can be reconstructed from the other.  Standard
file types, such as FITS, will be used throughout.  The directory hierarchy
(which could, in principle, be spread across multiple file systems)
will be visible to the collaboration through a variety of
services including http, rsync and
Globus~Online\footnote{\url{https://www.globus.org/}}.  These data access
methods are already provided by NERSC.

Catalog data generated by the processing pipeline will be collected and placed
into a catalog database.  Catalog data will include, but will not be limited to,
celestial coordinates, redshift, spectral classification (galaxy, QSO, \etc),
spectral metadata (exposure date, fiber number, \etc) and targeting data.
This database may also contain the full
spectral data (flux versus wavelength) and associated metadata.

This database will be accessible to the collaboration through a web-based
API.  The specific details of this access method are still being developed.
We would use the interfaces provided by the SDSS-III
SAS\footnote{\url{http://data.sdss3.org}} and the SDSS-III
API\footnote{\url{http://api.sdss3.org}} as a guide to developing a
web-based data access method for DESI.

DESI plans to backup all reduced data to remote locations,
potentially including NOAO as
described above, in addition to HPSS tape storage at NERSC.  We are developing
tools that help automate the HPSS backup process.
The databases described above will be periodically
dumped to files that can be backed up to HPSS tape storage.

Following the software development practices of SDSS/BOSS,
all raw and reduced data files will have a data model that describes the
content of the file.  This documentation is the primary interface definition
between different steps of the data processing.
Automation will ensure that all raw and reduced
data files conform to their data models and will also identify files that
either fail to conform or that lack a data model.  The data model will
be both human and machine readable.  SDSS currently uses HTML format;
DESI is currently developing a data model based on
reStructuredText\footnote{\url{http://docutils.sourceforge.net/rst.html}}
which is more human-friendly than raw HTML but still easily
processed by standard Python-based tools.

All code used for processing data will be documented using standard techniques
such as Sphinx\footnote{\url{http://sphinx-doc.org}} (for Python code) or
Doxygen\footnote{\url{http://doxygen.org}} (for C/C++ code).
Files will also contain metadata that points back to documentation relevant
to the generation of that file, such as code
name and version, the data model URL, \etc\   High-level documentation such as
algorithm descriptions, user tutorials and technical term definitions will
be managed with an off-the-shelf content management system to ensure that
documenters do not get bogged down with formatting of their documentation.

\subsubsection{Data Releases}

Tagged versions of the processed data and associated code will be released to
the DESI science collaboration approximately yearly.  These will serve as
the standard datasets for publications for consistency across analyses and
future traceability.  Additionally, the most recent
reductions of the data will always be available to the entire collaboration
as soon as they are processed, typically within 24 hours of observation.

After a proprietary period, these tagged data and software will also be
publicly released, with the intention of producing high-quality science
in a timely manner.
Data releases will include software and additional documentation as needed.
DESI's management policy for digital data is provided in the Project
Execution Plan (DESI-0381).

Once vetted for internal (and eventually public) release,
the tagged data will not be writable by the DESI collaboration.
These data will be provided with checksums to verify the integrity of the
data.  Access to public data may involve the
same methods described above, but with no password required and with
access limited to public data only.

\subsection{Large Scale Structure Catalog}

The spectroscopic data reduction pipeline outputs classifications, redshifts,
and uncertainties for each observed target. The likelihood of successfully
measuring the redshift of desired targets will depend on the redshift and
angular position of targets. These variations are induced by effects such as
target selection under different photometric data conditions,
efficiencies of the fiber assignment algorithm, and from the pipeline performance
under different observation conditions. If not corrected, they will induce an
artificial pattern in the distribution of observed targets and may
significantly bias clustering measurements.

The Large Scale Structure Catalog code (LSScat) handles the bookkeeping for
these efficiency measurements. It begins with a complete list of targets along
with the results of the fiber assignment algorithm and redshift measurements.
Its output is an estimate of the galaxy overdensity field as a function of
angular and redshift coordinates. The galaxy overdensity field is typically
reported as a list of 3D galaxy positions and a set of weights assigned to each
galaxy, as well as the \emph{selection function} that quantifies the expected
completeness of the survey as a function of both angular and radial coordinates,
target class, and potentially other target properties. For most practical
purposes, the selection function will be given as a set of ``random'' galaxies,
where the relative density of ``random'' galaxies in different parts of the
survey is proportional to the relative completeness of those regions.

Correlation functions or power spectra can then be straightforwardly computed
through weighted pair counts of the galaxy and random catalogs. Below we detail
three intermediate steps for developing the final weights, 3D selection
function, and random galaxy catalogs.

\subsubsection{Angular Selection Function and Angular Systematics} 
\label{sec:ang}

The angular selection function, or mask, specifies the expected number density
of targets as a function of angular coordinates.   This function describes the
geometrical structure of the imaging and spectroscopic footprints, and requires
knowledge of pertinent variations in, \eg, number of tilings,
completeness, depth, and seeing across the survey. Large scale structure
catalogs from many galaxy redshift surveys (including BOSS) have used the mangle
software package\footnote{\url{http://space.mit.edu/~molly/mangle/}} for the
geometrical description and related computations.  Mangle divides the celestial
sphere into regions in which conditions are relatively uniform.  For instance,
in BOSS these are regions within the imaging footprint covered by a unique set
of spectroscopic tiles.  Each region is defined by an arbitrary number of
``edges'' that must each be part of some circle on the sphere; see \cite{Ham04}
for more details. In DESI (unlike BOSS) the fibers can not be put anywhere on a
plate, which implies that the mangle regions will have to be defined at the
level of fiber patrol radii. Given that DESI survey will consist of more than 50
million unique fiber locations, the number of mangle regions will be extremely
large which makes this approach impractical. 

Significant nonuniformity in the angular selection function will be present
already in the targeting catalog. The nonuniformity will be generated by, \eg, variations in galactic extinction and/or stellar density, observing
conditions when the imaging data was taken, or photometric calibration
variations. These variations are expected to be smooth (compared to the fiber
scale) and can be mitigated by a weighting function that will apply relative
weights to objects in ``random'' catalog based on their angular positions.

Additional, purely angular, inefficiencies will be introduced at a later stage,
\eg, a broken fiber or a bright star will result in an area that is
effectively unobservable. These effects can be described by a veto mask that
removes all targets and random galaxies in specific areas. 

\subsubsection{Radial Selection Function and Radial Systematics}\label{sec:rad}

For DESI targets, particularly ELGs, the probability of obtaining a successful
redshift will depend on the observing conditions at the time the spectrum as
taken (\eg, sky background), the \otwo flux of the target, and whether
the redshift of the object puts the \otwo doublet inside a sky line. Redshift
success rate may also be a function of location on the focal plane during
observation. This information must be used to define at each point on the sky
the \emph{radial survey selection function}, which is the actual redshift
distribution of the targets times the probability of obtaining a successful
redshift as a function of \eg target \otwo flux, line width, and
profile, as well as the actual redshift.

\subsubsection{Effects of Fiber Assignment}
\label{sec:undoing_non_random}

The fiber assignment algorithm will induce additional non-uniformity into the LSS
catalogs. There are three main reasons behind this. The main source of
non-uniformity is that the tiling scheme will unavoidably result in some areas
of the footprint being covered by more plates than the others. The redshift
measurement efficiency will be lower in the parts of the sky that are reachable
by a lower number of fibers. This will induce spurious correlations on large
scales (order of a DESI field).  The next leading source of non-uniformity is
related to DESI having multiple target types with different priority. QSOs and
LRGs need multiple observations and are assigned higher priority compared to ELG
targets. This implies that ELG targets are less likely to be observed in the
vicinity of higher priority targets. The last source of non-uniformity is the
variation in the density of targets. Since DESI fibers positions are restricted
within a field, in regions with high target density some targets will not be
assigned a fiber. All these sources of non-uniformity impact lower priority
targets (ELGs) more than the higher priority objects (\lya QSOs, target QSOs, LRGs).

To mitigate the spurious signal induced by first two sources of non-uniformity we
process ``random'' catalogs by the same fiber assignment pipeline as the
target catalog. Below is a brief description of how this is done for ELG
targets. We create a large number of ``random'' catalogs with the same average
density of targets as in targeting catalog. Targets other than the ELGs are
placed at exactly the same positions as in the targeting catalog. The ELGs
themselves are put in such a way that they follow the observed N(z) but are
assigned random angular positions. These ``random'' catalogs are then
``observed'' using the same fiber assignment pipeline as for real targets.
Successfully ``observed'' ELGs from all ``random'' catalogs are then
concatenated into the final ``random'' catalog for ELG sample. This procedure
naturally results in a ``random'' catalog in which the density of
``observed'' targets is higher in areas with higher density of fibers and lower
in areas around higher priority targets. In the limit of large number of
``random'' input catalogs, this procedure is expected to completely remove
non-uniformities generated by first two sources.

\begin{figure*} 
\centering                                                                
{\includegraphics[width=0.45\textwidth]{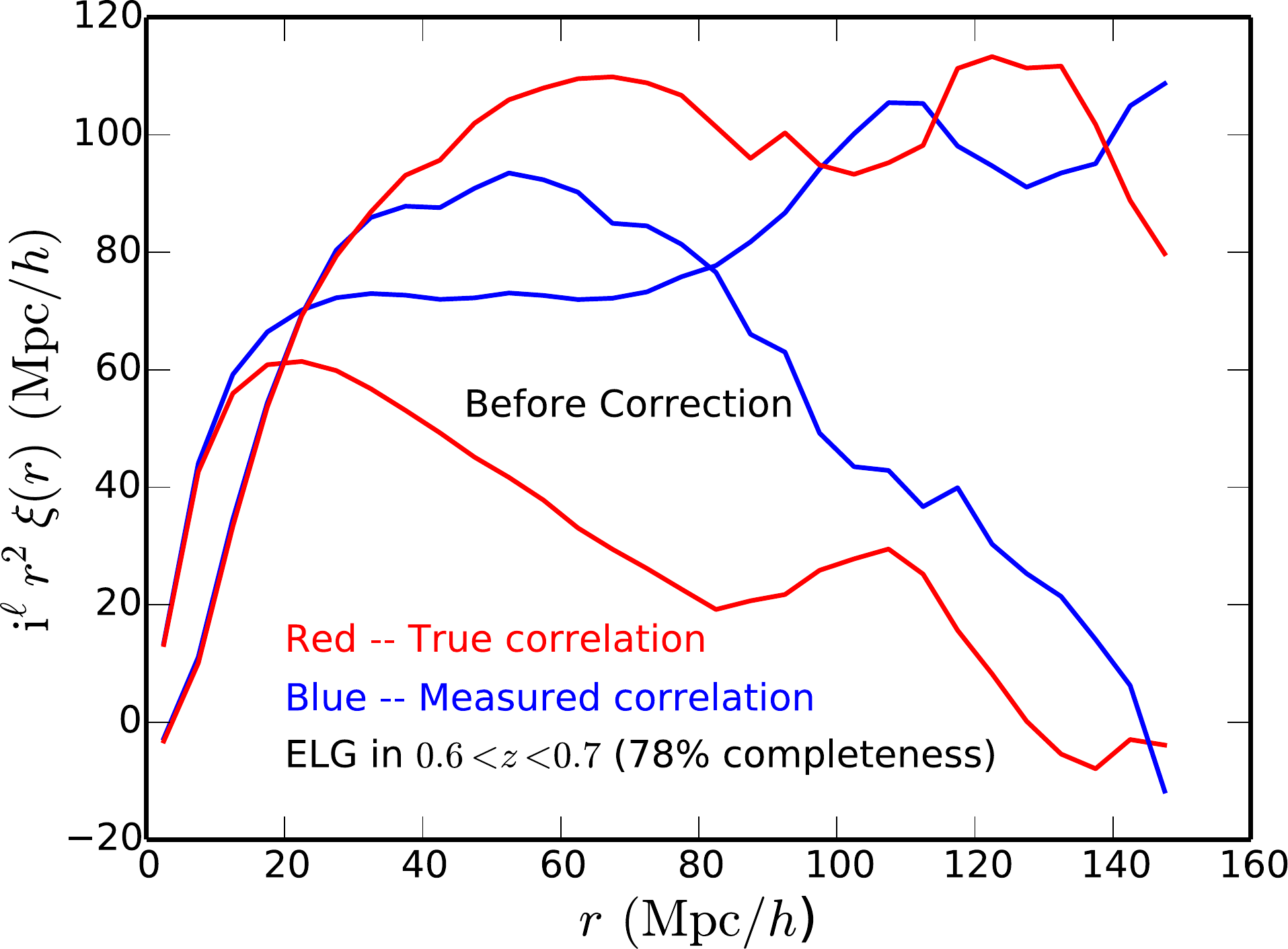}}                  
{\includegraphics[width=0.45\textwidth]{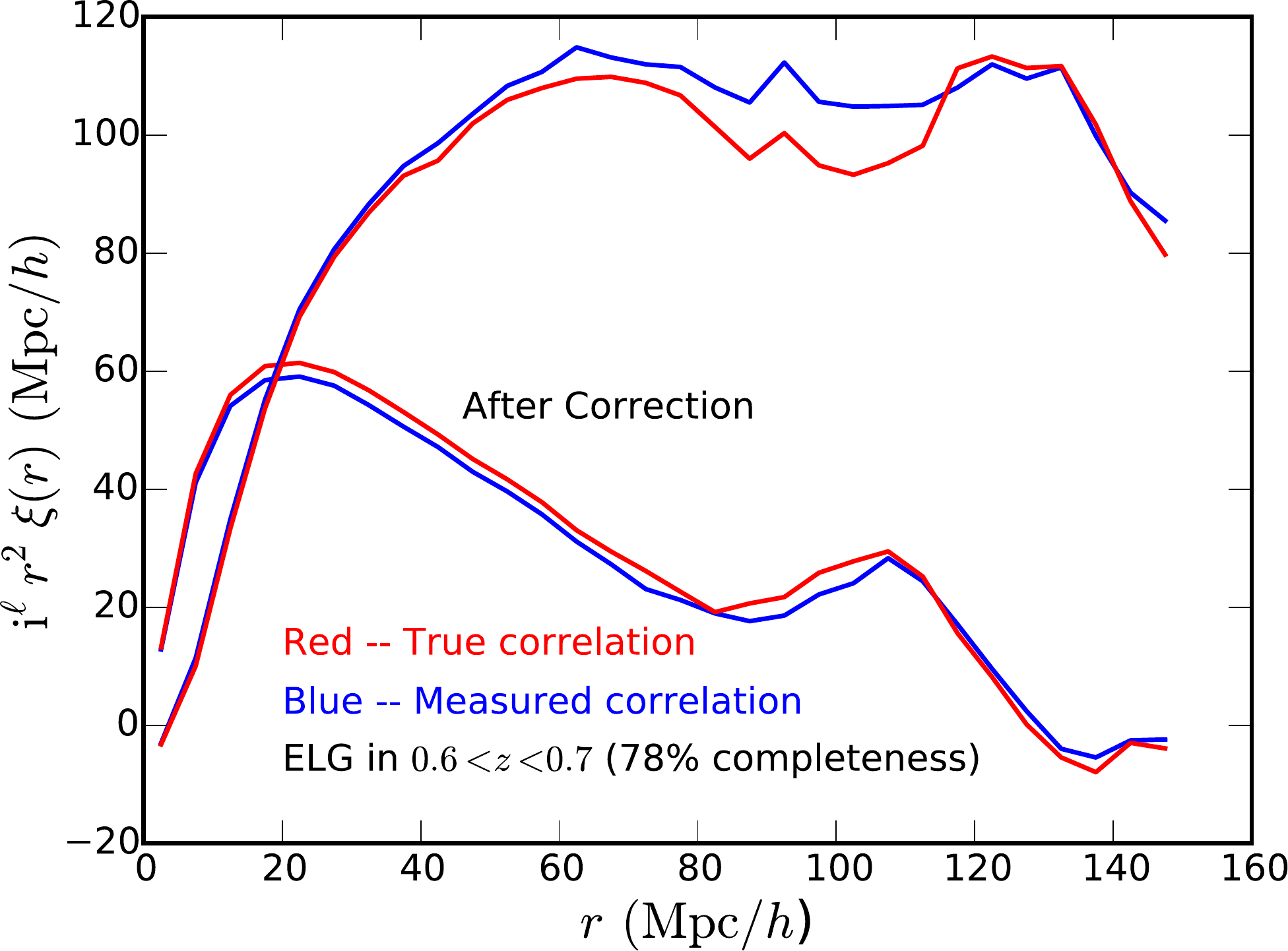}}                  
\caption{Red lines show the input correlation function (first two Legendre
moments) of ELGs from N-body simulations. The blue lines on the left panel show the
measured correlation function with no correction for the fiber assignment. The
blue lines on the right panel show the measured correlation function after correcting
for the fiber assignment as described in Sec.~\ref{sec:undoing_non_random}.}                                                                      
\label{fig:TPCF}                                                              
\end{figure*}

Figure~\ref{fig:TPCF} demonstrates how this procedure works on simulated target
catalog. The blue lines show input correlation function of the simulations
(first two angular multipoles). The red lines on the left panel show the
correlation function computed without correcting for the fiber assignment
effects. The induced spurious signal significantly distorts the signal on all
scales. Red lines on the right panel show the correlation function computed
after correcting for the fiber assignment effects as described above. After
corrections most of the bias is removed and the measured correlation function is
close to the input. 

As successively more realistic models of tiling and fiber assignment are
developed (see Section~\ref{sec:fiber_assignment}), these will be tested
by running full analyses on the catalogs of ``observed'' galaxies generated by
implementing the fiber assignment code on mock catalogs. We expect the success
of this approach to be insensitive to the details of fiber assignment algorithm
and tiling strategy.

A small bias remains in the correlation function even after using the
``observed'' randoms. This bias is likely to be coming from the fact that the
efficiency of fiber assignment anti-correlates with the density of targets. This
bias can be mitigated by weighting targets while calculating the correlation
function, \eg\ by taking all targets that haven't been
assigned fibers and up-weighting their nearest neighbors that were assigned fibers.

\subsection{Simulations}
\label{sec:sims}

Simulations are needed throughout the design, development, and operations
of DESI.  Some of these focus on individual aspects, \eg, to optimize a
particular piece of the hardware design or to tune an individual algorithm.
Other simulations provide mock data for the spectral extraction pipeline
development prior to obtaining real data.  End-to-end simulations will
ensure system level integration and scaling of the software, while also
validating the design performance as a whole.
Cosmological simulations are needed as input for particular pieces
that require realistic cosmology to test their performance. The
following sections describe DESI simulations, tracing through the
order of data flow.

\subsubsection{Cosmological Simulations}

Cosmological simulations are critical inputs for the development, testing and
validation of the full DESI pipeline.  In particular they are essential to
\begin{itemize}
  \setlength{\itemsep}{1pt}
  \setlength{\parskip}{0pt}
  \setlength{\parsep}{0pt}
\item understand the effect of the tiling and fiber assignment strategies
\item model the radial selection function of the different targets for the
    development of the large scale structure catalog
\item provide a realistically clustered mock input catalog for
    algorithm development
\item serve as input for the end-to-end simulation pipeline for system
    level testing
\end{itemize}
The cosmological simulations needed to test the pipeline are a subset of
the simulations needed for final science analysis, and thus the
Data Systems team will coordinate with the Cosmological Simulations
Working Group (and other science working groups) to
obtain the simulations needed.

The cosmological simulations as needed by DESI are
constructed with a multi-step procedure.
First, given a cosmological model
an N-body simulation is run using either baryons plus dark matter or
solely dark matter.  Second, galaxies and quasars are populated within
this N-body simulation:  Gas-hydrodynamical simulations need a way of
identifying the objects, while solely dark matter simulations
require a scheme such as a statistical halo occupation distribution (HOD)
model or a physical semi-analytic galaxy formation model.
The last step, which also depends upon how the N-body simulations are run,
is the creation of a lightcone from the simulation outputs (unless created
already on the fly).

The differing methods have various strengths and weaknesses.
For example, HOD models can be tuned to real data distributions to accurately
reproduce the clustering signal of a given population at a given cosmic
epoch.  On the other hand, the final DESI targeting data do not yet
exist and thus the data to tune against is not yet available.
In the meantime, semi-analytic models can provide approximate predictions
for samples that are either not yet defined or for which no data are
available yet. Both approaches are needed for a
project which aims to sample different galaxy populations at a fixed
cosmic epoch.

A selection of requirements for the mocks are:
\begin{itemize}

\item[a.] the redshift space clustering of the main targets (LRGs,
ELGs and QSOs) needs to be accurately recovered to enable detailed
tests of the impact of the tiling and fiber assignment algorithms.
In a first instance,
the correct angular clustering is necessary, while the redshift space
clustering as function of redshift will be important to understand any
remaining biases.

\item[b.] it will be necessary to create mocks with all three target
classes
covering the full redshift range of interest (\ie, $0.6<z<3.5$)
to ensure any effects related to the tiling are properly
understood, leading to a well defined angular selection function.

\item[c.] the radial selection function of the mocks needs to match
that of the targets, to ensure one can incorporate the instrumental
limitations in recovering the correct redshift given the source
redshift and flux.

\item[d.] the mock galaxies need to have similar properties to the
ones used to select the targets in the first place to ensure one can
run the targeting selection algorithm on the mocks, like on the real
data, ensuring that one can model its limitations and understand the
impact photometric selections have on the sample definition and its
associated clustering signal.

\end{itemize}

We envisage the need for deep spectroscopic data covering the
parameter space used in the target selection to ensure we can
properly model and constrain the properties of mock galaxies to be as
similar as possible to the real DESI targets. The combination of
galaxy formation mocks, which have a certain level of predictive
power, and statistical mocks, requiring larger datasets to be trained
on, is key in this process.

The final requirement on the size of the cosmological
simulations is not yet set.
Mass resolution of a few 10$^9$~M$_\odot$ (and possibly as low as a
few 10$^8$~M$_\odot$) is necessary to model the
LRGs, ELGs and QSOs within the same volume, while the mass resolution
needed for the \lyaf~QSO has to be several orders of
magnitude higher.  Creating a suite of mock catalogs of
several hundred of square degrees each (\ie, much less than the
full DESI footprint of $\sim$14,000~deg$^2$)
ought to be sufficient to test the end-to-end
simulation pipeline, including the tiling algorithm, the instrument
simulations, and the target selection. Even though DESI targets are
selected to be in the range $0.6<z<3.5$, it is imperative to have mock
catalogs to $z=0$ to ensure the target selection can be modeled in
full.

Finally there is a need for a few full scale DESI specific simulations and
mocks to be produced to properly test the whole pipeline, as only with the
large amount of data this requires can one truly test the reliability of the
end-to-end pipeline: any order of magnitude scaling cannot be assumed to work
without being tested.

The cosmological simulations needed for the final science analyses
will likely be of similar resolution as those described
here, although of much greater size to fully cover the DESI footprint,
and many more realizations will be needed to enable error estimates and
cosmological parameter searches.
In addition to N-body cosmological simulations focused on the galaxy broadband
power spectrum, final analysis of the \lyaf\ broadband power will
require large hydrodynamic simulations and
quasar mock catalogs with astrophysical features such as complex quasar
continua, high column density absorbers or metal absorption lines, and
instrumental features such as noise, sky subtraction residuals, and errors in
the spectrophotometric calibration.
These final science analysis simulations
will be provided by the Cosmological Simulations Working Group in coordination
with the other science working groups.

\subsubsection{Operations Simulations}
\label{sec:opssim}

Section~\ref{sec:surveysim} describes
an operations simulator for survey footprint and long term observation
planning.  The prototype version of this simulator currently guides survey
design and will be further developed to connect input simulated targeting
catalogs to simulated
``as-observed'' data for instrument and pipeline simulations.

\subsubsection{Target Spectral Templates}

Spectral templates are needed for two purposes: fitting models to the data
to measure classifications and redshifts (see Section~\ref{sec:spec1d}),
and generating simulated data for the spectral extraction pipeline algorithm
development and end-to-end tests of the DESI design.

We have developed a set of empirical ELG templates based on spectral
synthesis models fitted to broad- and medium-band spectrophotometry of
$\sim20,000$ galaxies with $0.5<z<2$ and $r\lesssim24$ in the COSMOS
field.  We have incorporated realistic variations in the emission-line
ratios in the templates using spectroscopy of faint galaxies from the
zCOSMOS survey.  In particular, the model spectra account for the
measured redshift evolution in the \otwo doublet ratio due to the
differing physical conditions in the interstellar medium of
intermediate-redshift galaxies.  Using these 1D templates as a
baseline, we are using measured rotation curves of emission-line
galaxies at $z\sim1$ observed by TKRS/DEEP2 to construct 3D spectral
datacubes (2D sky position versus wavelength), in order to account for
realistic variations in the emission-line center (due to galactic
rotation) and line-width (\ie, velocity dispersion).  With 3D
datacubes in hand, we will simulate the effective 1D spectrum that
each spectroscopic fiber will obtain based on the position of the
target galaxy in the focal plane.

QSO templates have been developed from SDSS and BOSS observations, as well
as LRG templates from BOSS ancillary data taken with DESI-like targeting
requirements.  These ELG, LRG, and QSO templates will be combined with
the cosmological simulations described in the previous section to generate
input datasets for DESI instrument and pipeline simulations.

Current studies use sky spectra measured with UVES \citep{Haunschik03};
future studies could add variations by calibrating the high-resolution
UVES sky spectra to observed spectra from KPNO or alternately BOSS.
Atmospheric extinction uses the widely available {\tt knpoextinct.dat} curve
from 3400~\AA\ to 9000~\AA, augmented with high resolution water
absorption bands as measured at KPNO~\cite{ADey13}.

Arc and flat lamp spectra are based upon BOSS calibration lamps,
deconvolved to model the arc lines as delta-functions.  These allow
simulation of DESI calibration data, \eg, for measuring the PSF
shape and fiber-to-fiber throughput variations of simulated data.

\subsubsection{Instrument Simulations}
\label{sec:InstrumentSimulations}

Instrument simulations guide the hardware design process and provide
inputs for the spectral pipeline development.
This is optimized both at the high-level to ensure that DESI meets
the science requirements, and at the low-level for where the spectral
extraction pipeline places requirements on the hardware design, \eg, for
fiber-to-fiber cross talk and PSF stability between calibration and science
exposures.

\begin{figure}[tb]
\centering
\includegraphics[width=3in]{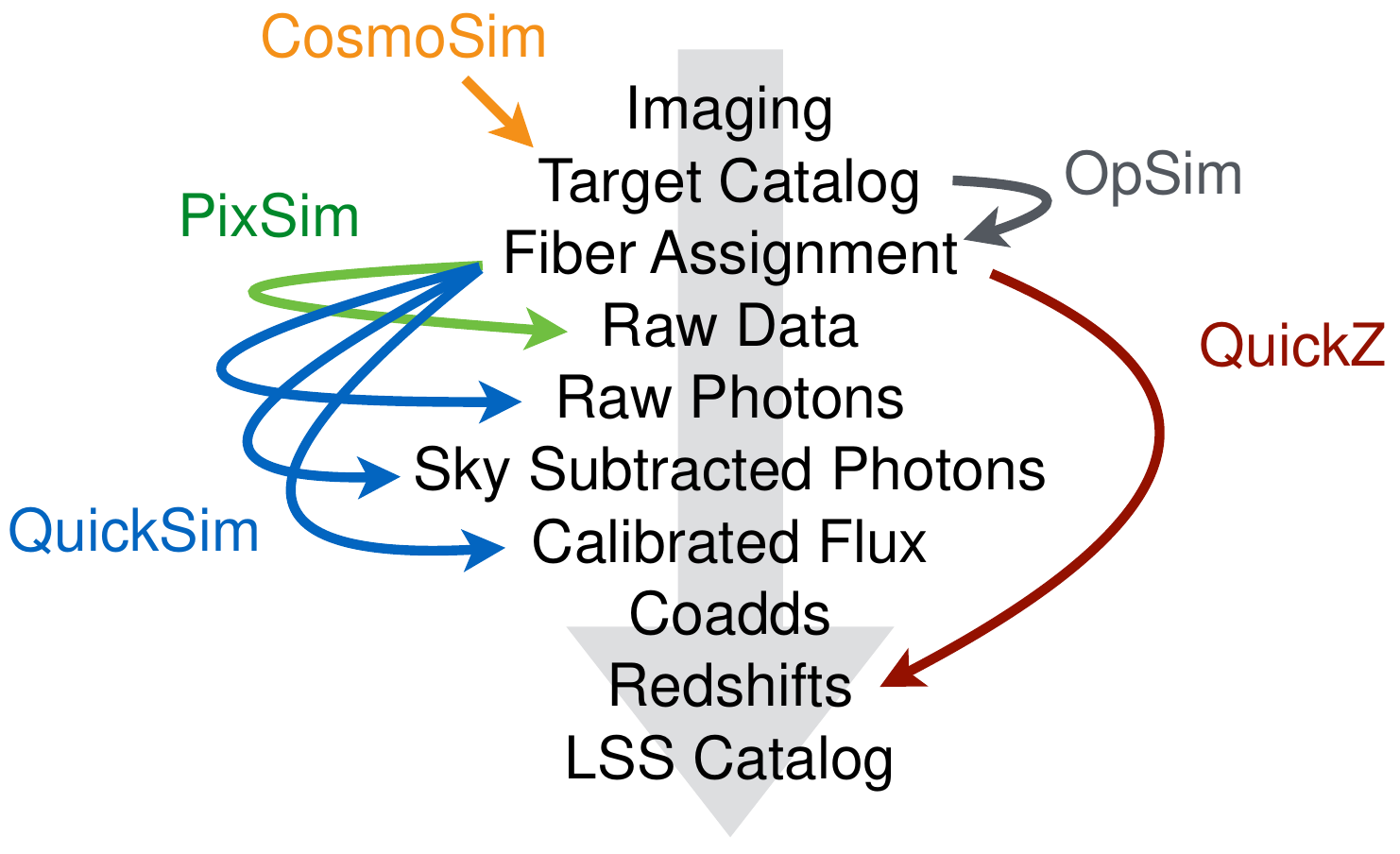}
\caption{
\label{fig:datasims}
Diagram of different types of simulations under development for DESI.
}
\end{figure}

Data Systems is developing multiple levels of simulations for code
development, testing, and verification; these are shown in
Figure~\ref{fig:datasims}.
\begin{itemize}
    \item CosmoSim: input cosmology simulations to produce mock target
        catalogs.  Exists.
    \item PixSim: pixel-level simulations of raw data to be used for
        spectroscopic pipeline development and study of detailed systematics.
        Exists.
    \item QuickSim: a fast spectral simulator that emulates the DESI
        spectroscopic resolution and throughput.  It propagates signal,
        sky, and detector noise but bypasses the computationally expensive
        raw data pixel simulations and extractions.  Exists.
    \item OpSim: a survey simulator that traces what tiles would be observed
        in what order based upon the real next field selector and a Monte
        Carlo realization of weather and observing conditions.  Planned.
    \item QuickZ: an ultra fast simulator that applies DESI efficiencies
        to an input target catalog to generate an output redshift catalog.
        Planned in conjunction with the science collaboration using tools
        provided by Data Systems.
\end{itemize}

CosmoSim has been used for fiber assignment algorithm development.
QuickSim was used for the majority of the performance studies in this report.
PixSim has been used for spectroscopic pipeline development and
Figures~\ref{fig:elg_gallery} and \ref{fig:zelg}; we are in
progress with calibrating/verifying QuickSim versus PixSim.

\begin{figure}[tb]
\centering
\includegraphics[height=3in]{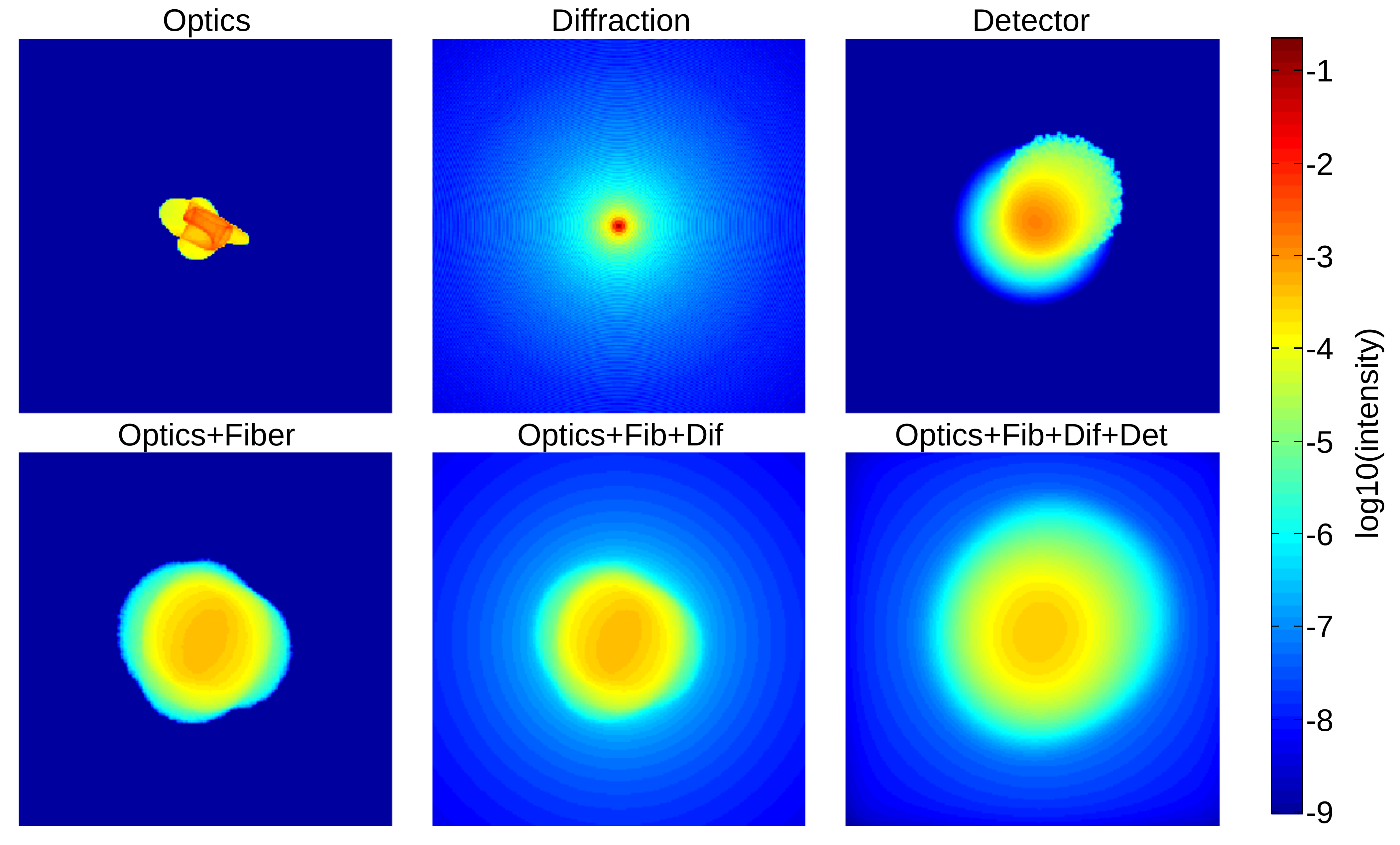}
\caption{
\label{fig:psfspots}
Simulated spectrograph PSF spots including various effects.
Top row left to right: infinitesimally small fiber for optics only, diffraction only, detector only.
Bottom row left to right: 107~\micron fiber for optics only,
optics and diffraction, and optics and diffraction and detector.
Each spot covers $15 \times 15$ CCD pixels.
Note that these are on a log color scale.
}
\end{figure}

Hardware simulations are based upon a throughput budget provided by
systems engineering and a spectrograph point-spread-function (PSF) model
provided by the spectrograph engineering team.
The PSF is simulated using Zemax
ray traces of the input fiber illumination pattern at the slit,
plus diffraction and CCD detector absorption effects.
These are modeled on a grid of wavelengths and fiber slit positions
and then interpolated to arbitrary wavelengths and fibers.
Figure~\ref{fig:psfspots} show these spots on a log color scale for a 980~nm
monochromatic illumination of fiber 0 on a 500~\micron thick CCD,
which is the worst case for PSF distortion.
The various panels show different contributions to the PSF
covering a size of $15 \times 15$ CCD pixels;
the bottom right spot is the final spot used for simulations.
Equivalent spots are generated for 10 other fiber positions and 32 other
wavelengths\footnote{Finer grids are possible if needed.}.

In addition to these simulations, the DESI algorithms are written in a
modular fashion so that they can also be applied to real BOSS data.
This provides a valuable cross check on the performance of these algorithms
in the presence of data quality issues that arise in real data.  That also
guides areas where the DESI simulations need to improve to be more realistic.

\paragraph{desimodel}
\label{sec:desimodel}

DESI hardware and operations parameters used by the simulations are
are stored in machine readable formats in the ``desimodel'' product
of the DESI code repository.
Tagged versions of desimodel provide unambiguous descriptions of
each reference DESI design.

The hardware parameters include the
throughput, PSF model, Mayall mirror geometric area, CCD dimensions,
focal plane positioner layout, plate scale, and fiber size.
These have documented traceability back to DocDB entries for details.
In addition to hardware parameters (Section~\ref{sec:performance}),
desimodel includes survey parameters
such as median seeing, sky emission and extinction spectra, default
exposure time, reference spectra for each target class, and $dn/dz$
distributions expected from targeting.  All parameters necessary
to simulate the reference DESI mission are contained in desimodel.

Ownership of each class of parameters is coordinated: systems engineering
controls the hardware parameters; data systems controls the
astrophysical quantities such as the impact of seeing, night sky brightness,
and extinction; and the science collaboration provides reference spectra
and targeting $dn/dz$ distributions.

%

\paragraph{Quicksim}

Quicksim code included in desimodel simulates the $S/N$ properties of
DESI spectra assuming error propagation of photon shot noise plus CCD
readout noise.  It goes directly from an input spectrum to
a resolution-convolved output spectrum plus noise.  It treats the PSF
dispersion as Gaussian in the wavelength direction, and parameterizes
the effective number of pixels affected by CCD readout noise using the
true cross-dispersion shape of the PSF for fiber 100 (which is neither the
best nor the worst PSF fiber).  Quicksim is available in both IDL and
Python versions for end-user flexibility.

\paragraph{Specter Pixel-Level Simulations}
\label{sec:pixsim}

\begin{figure}[tb]
\begin{centering}
\includegraphics[height=2.5in]{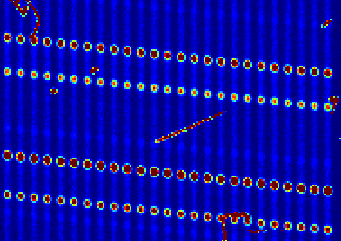}
\caption{
\label{fig:specter-img}
Example CCD pixel-level simulation of 25 sky spectra, including CCD readout
noise and cosmic rays superimposed from real data.  This is approximately
$1/1500$ of the data from a single DESI exposure.
}
\end{centering}
\end{figure}

Detailed pixel-level simulations use the
Specter toolkit\footnote{\url{https://github.com/desihub/specter/}},
with a DESI-specific wrapper in desisim\footnote{\url{https://github.com/desihub/desisim/}}.
It simulates all 500 fibers per CCD and generates raw data frames mimicking
DESI raw data, including fiber-to-fiber cross talk, PSF distortions at
the edges of the CCDs, and cosmic rays.
A subset of a single simulated frame is shown in
Figure~\ref{fig:specter-img}.
Specter provides inputs for spectral pipeline development and enables studies
of systematics not addressed by QuickSim, e.g.\ fiber-to-fiber cross talk, PSF
stability, multi-fiber sky subtraction, and calibration requirements.
Specter will also be used to generate
a full survey of simulated raw data to perform end-to-end data
challenges.  Since Specter is much slower than QuickSim, it will be used
only for studies that require pixel-level detail.

\subsubsection{DESI Performance Studies}\label{sec:desiperformance}

\paragraph{ELG Signal-to-Noise}
\label{sec:elgcase}


\begin{figure}[!tb]
\center
\includegraphics[width=0.9\textwidth]{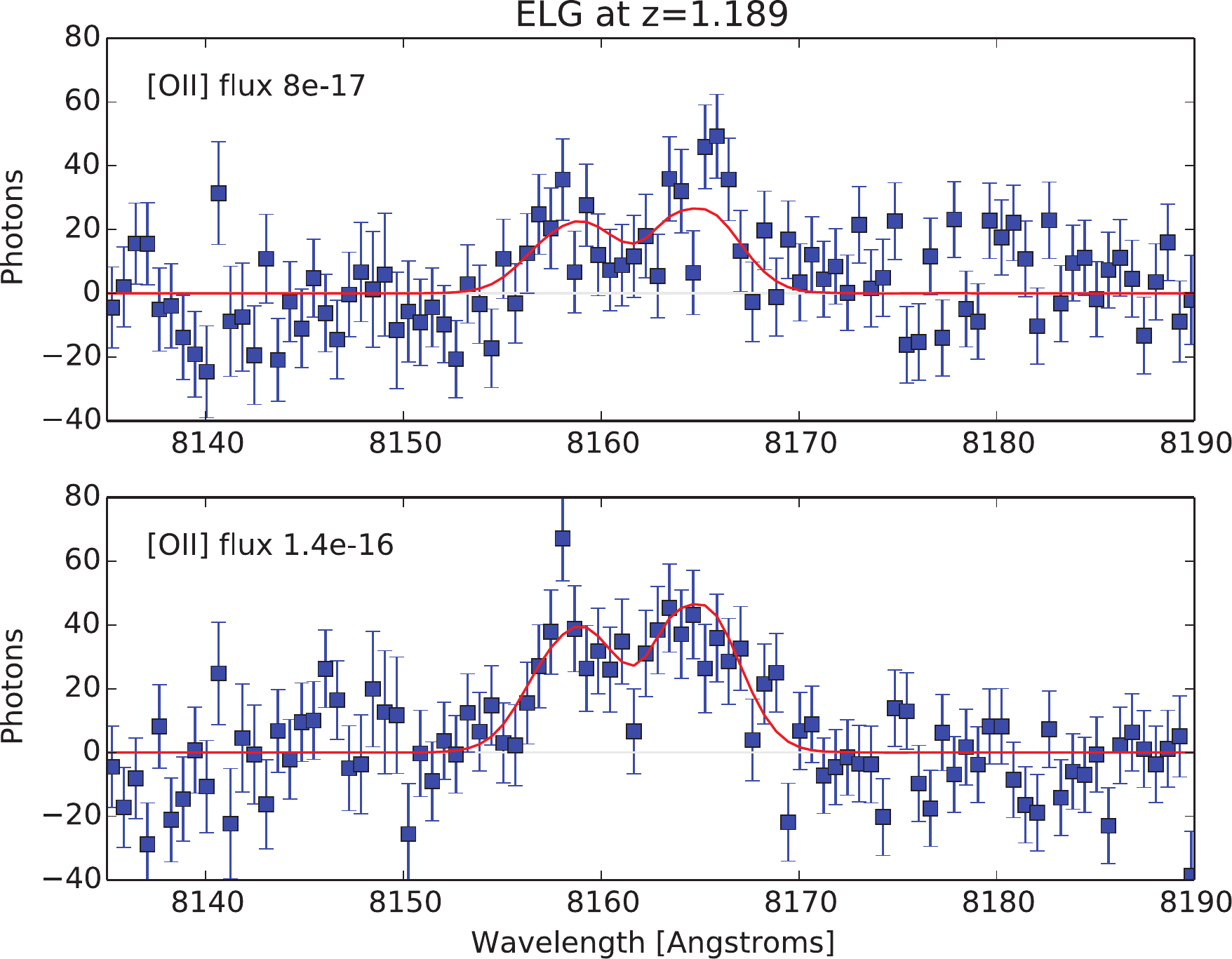}
\caption{
A quicksim simulation of the \otwo emission line doublet at
a limiting flux of
F(\otwo)=$0.8\times10^{-16}$ ergs s$^{-1}$ cm$^{-2}$ (top)
and the median case of
F(\otwo)=$1.4\times10^{-16}$ ergs s$^{-1}$ cm$^{-2}$ (bottom)
for a reference 1000 second exposure.
The simulated emission lines have a velocity width of 70 km/s and
a ratio of $1:1.3$. The red curves represent the input spectra at the resolution of the instrument (expected number of collected photons per pixel row), and the blue squares a random realization of the data with noise.}
\label{fig:ELGquicksim}
\end{figure}

Figure~\ref{fig:ELGquicksim} shows the quicksim output of a 1000~second
exposure at zenith with 1.1~arcsec seeing and no galactic dust
extinction.
These
conditions define a ``reference'' exposure S/N.  Actual observations will
be longer or shorter to achieve this same S/N given the current conditions
(see Section~\ref{sec:onlineQA}).
This ELG is at $z=1.189$ where the \otwo doublet is
between sky lines;
The nearest sky line is a faint line at 8105~\AA.

The top plot shows a limiting
flux case of F(\otwo)=$0.8\times10^{-16}$ ergs s$^{-1}$ cm$^{-2}$
with a binning of 0.6~\AA\ corresponding to the typical CCD pixel size
in the NIR arm of the spectrograph.  Although the spectrum is noisy,
this is a $7\sigma$ detection, as defined by the significance of the \otwo
doublet amplitude derived from an optimal fit using known line profiles and
flux ratio.
The bottom plot of Figure~\ref{fig:ELGquicksim} shows the median flux case of
F(\otwo)=$1.4\times10^{-16}$ ergs s$^{-1}$ cm$^{-2}$ with a $11.8~\sigma$
detection.

\begin{figure}[!bt]
\center
\includegraphics[height=3in]{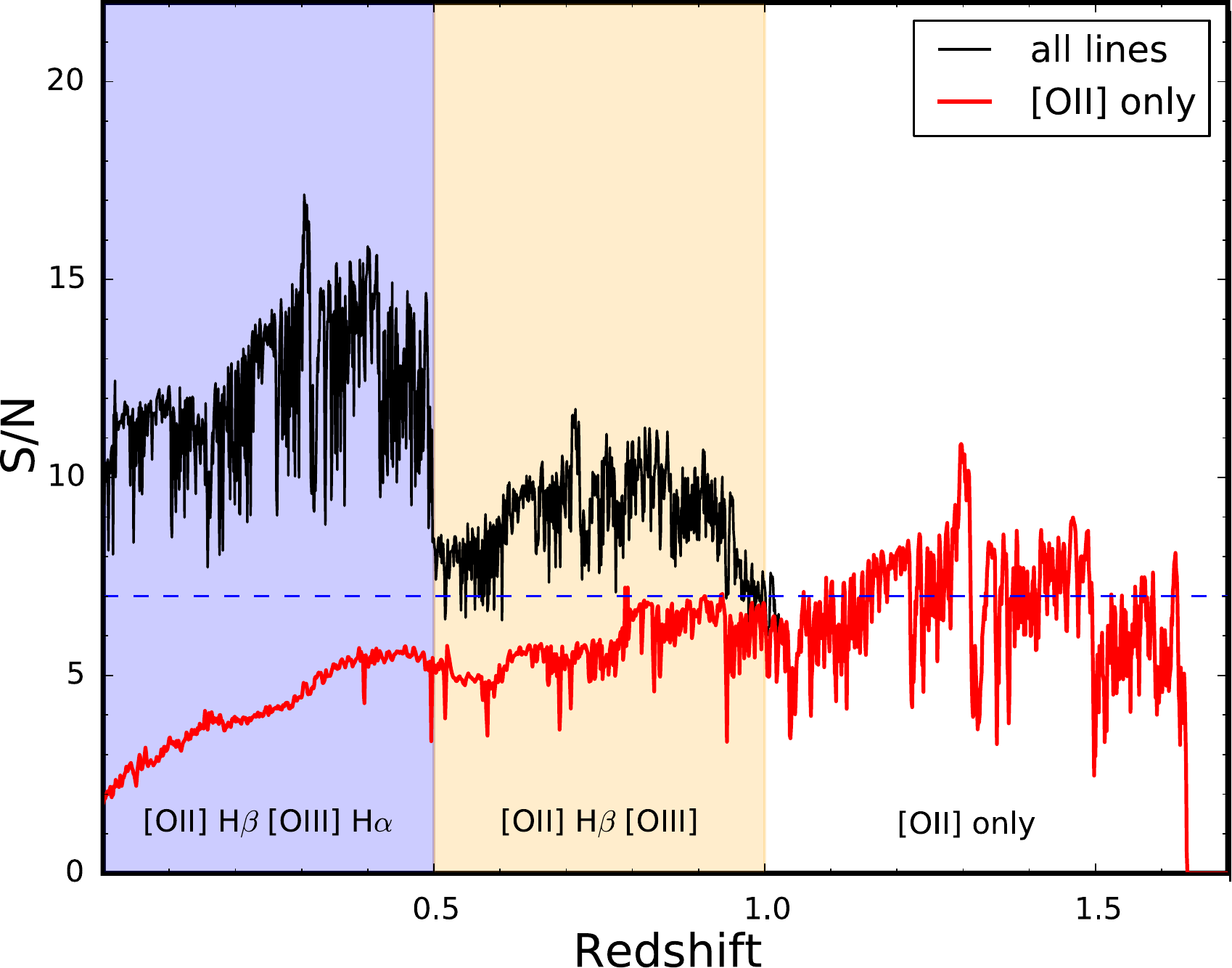}
\caption{The emission line S/N assuming constant line ratios for
F(\otwo)=$8\times10^{-17}$ ergs s$^{-1}$ cm$^{-2}$
across the full DESI redshift range for a reference 1000 second exposure.
The black line shows the combined S/N from all emission lines observable
in the waveband, while the red line shows the S/N only from \otwo.
The horizontal blue dashed line shows
S/N$>7$ where DESI achieves an unambiguous \otwo\ detection.}
\label{fig:ELGSNz}
\end{figure}

To accurately predict the fraction of targeted ELGs that will have high-confidence redshifts, we must
understand the DESI redshift detection efficiency versus redshift and
\otwo flux.
Figure~\ref{fig:ELGSNz}
shows the emission line S/N for a 1000 second reference exposure
of an F(\otwo)=$8\times10^{-17}$ ergs s$^{-1}$ cm$^{-2}$ ELG template spectrum.
This spectrum is redshifted in $\Delta z$=0.001 increments
over the full DESI spectral range.
The S/N for other \otwo\ fluxes is scaled from this example as described
in the following section.

The template spectrum uses constant emission line flux ratios
relative to F(\otwo), with a velocity dispersion of 70 km s$^{-1}$.
The template \otwo doublet line ratio is 1:1.37,
and the other line ratios are \otwo/H$\beta$=2.5,
\othree(5007~\AA)/H$\beta$=1.6,
\othree(5007~\AA)/\othree(4959~\AA)=2.984 and H$\alpha$/H$\beta$=3.

For $z<0.5$, the combined S/N from all emission
lines is strongly enhanced by the presence of H$\alpha$. Between
$0.5< z <1.0$, H$\alpha$\ is redshifted beyond $\lambda>9800$ \AA\ and a
redshift must be measured from a combination of H$\beta$, \othree, and \otwo.
The \otwo doublet will be the only
emission line available for redshift measurement beyond $z>1$, therefore
requiring that the doublet is
resolved for an unambiguous two-line identification in this redshift range.
S/N$>7$, shown in the blue dashed line, ensures that the line is
unambiguously detected as an \otwo\ doublet and not a mis-identified
single line.
Note that the change of sensitivity to the \otwo doublet with redshift for
$z<0.3$ is principally due to the change of the spectrograph throughput with
wavelength below $\lambda < 5000$\AA (see Figure~\ref{fig:specthroughput}).


\paragraph{Redshift Success Rates}
\label{sec:zsuccess}


\begin{figure}[!bt]
\center
\includegraphics[height=3in]{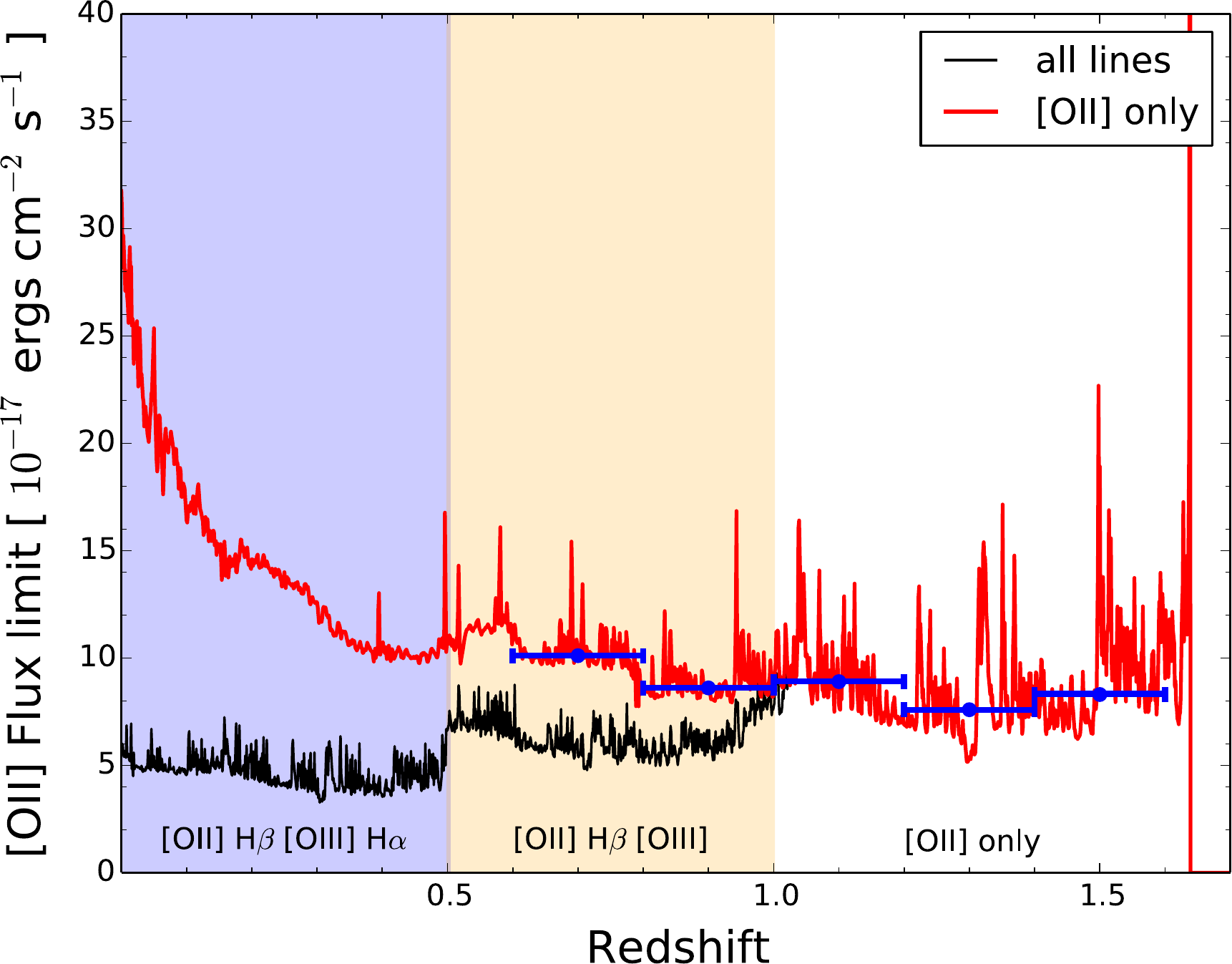}
\caption{\otwo flux limit for a $S/N>7$ detection of the \otwo doublet only (red curve) or a combination of emission lines (black curve) vs
redshift for a reference 1000 second exposure (a single exposure), at zenith, without Milky Way dust extinction. We have assumed here a fixed line ratio between the \otwo doublet and the other lines.  The blue points show the
binned median of the red curve.
\label{fig:FluxLimit}}
\end{figure}

A robust estimate of the success rate versus~redshift and \otwo\ flux
can be derived from Figure~\ref{fig:ELGSNz} because the noise is largely
dominated by the sky Poisson noise (with a small contribution of readout noise).
Requiring $S/N>7$ translates
into a \otwo flux limit shown in Figure~\ref{fig:FluxLimit}. For galaxies above this limit, we can safely assume
a high redshift success rate of $95\%$. The conservative $5\%$ inefficiency accounting for CCD defects and cosmic
ray hits at the CCD location of the emission lines. This flux limit is used in Target Selection section of the Science FDR to
derive a redshift efficiency given an input ELG target selection catalog,
and is the basis of the $dN/dz$ distributions used for the cosmology
projections in the Science FDR.


\paragraph{Trade Studies}
\label{sec:trades}

\begin{figure}[!hbt]
\center
\includegraphics[width=.95\textwidth]{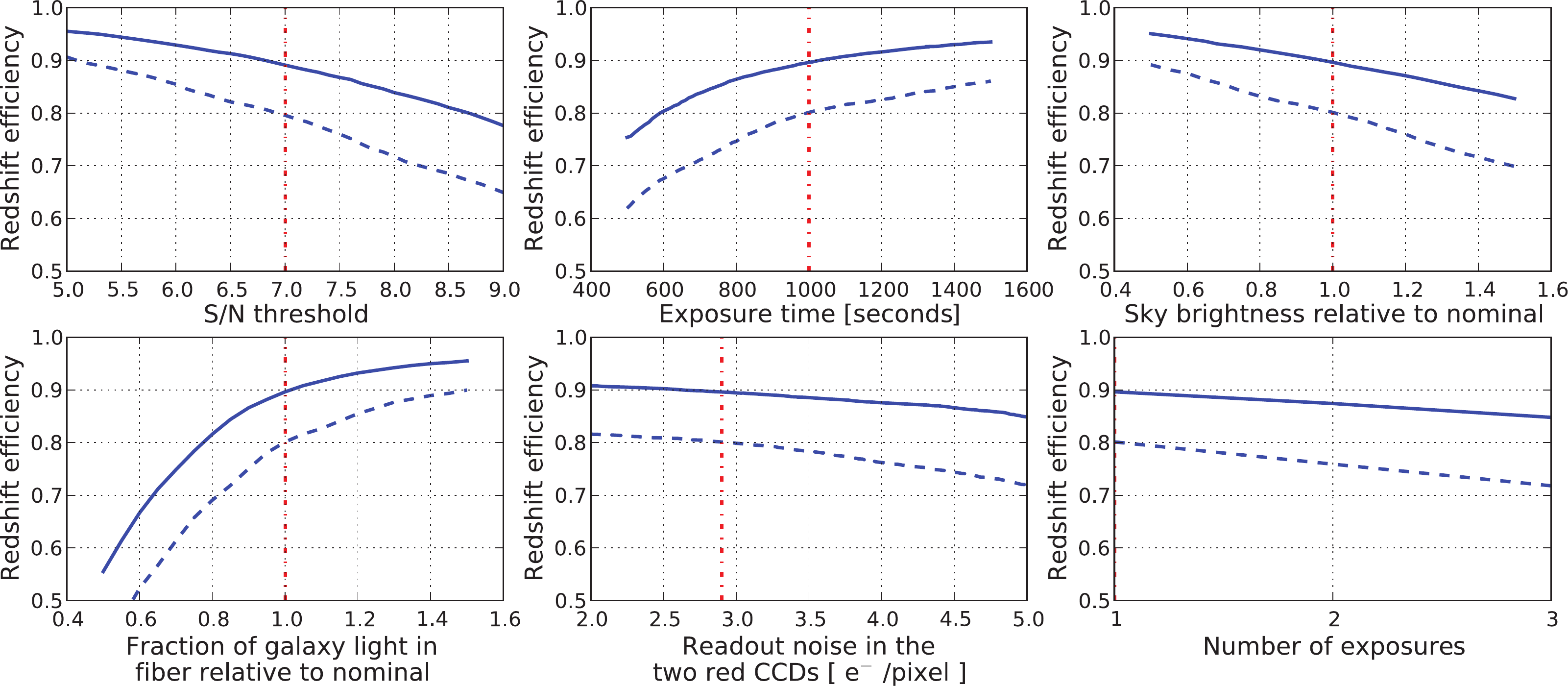}
\caption{Redshift success rate for ELGs as a function of S/N threshold, exposure time, sky brightness, fraction of galaxy light collected in the fibers, readout noise and number of exposures. The distribution of ELGs as a function of redshift and \otwo flux follows that described in the Target Selection section of the Science FDR. The solid curve is the efficiency when all emission lines are used, and the dashed curve when only the \otwo doublet is considered. The vertical red dashed-dotted lines are the nominal values. In this plot, we do not include the 5\% provisional inefficiency related to CCD defects and cosmic rays mentioned in the text.\label{fig:nztrade}}
\end{figure}

In developing a conceptual design for the DESI experiment, it is necessary to perform trade studies around
the baseline instrument design and observation plan during the R\&D phase.   Ultimately, the scientific impact of
these design decisions must be quantified in a useful metric such as the FoM described by the Dark Energy Task Force~\cite{DETF06}.
In the case of DESI ELGs, we can use the above simulations to evaluate the change of redshift success rate with S/N threshold, exposure time, sky brightness, the fraction of light from the target galaxy collected in the fiber, CCD readout noise and number of exposures (which increases the contribution of readout noise). The results are shown in Figure~\ref{fig:nztrade}.
 This study is based on a distribution of ELG \otwo fluxes as a function of redshift consistent with the selection presented in the Target Selection section of the Science FDR. Assuming uncorrelated uncertainties of 10\% on the system throughput (equivalent to a change of exposure time), the sky brightness, and the fraction of galaxy light collected in the fibers results in an uncertainty of only 4\% on the redshift success rate for ELGs. This is a consequence of the fact that most targets are significantly brighter than the flux limit defined above.

\subsubsection{DESI End-to-End Survey Simulations}
\label{sec:end2endsims}

End-to-end simulations provide a system-level integration test of the
survey plan and data reduction software, as well as a method to test for
system-level systematics \citep{spokes:2013, Refregier:2013un}.
Examples of such systematics include coupling between tiling,
fiber allocation, and source clustering; or between
targeting color selection, photometric calibration,
and final source populations.  Simulating and testing these effects
ahead of time enables optimization of the algorithms and verification
of the science goals given the ``as-designed'' hardware and ``as-built''
software.  End-to-end simulations have also been used by other
cosmology experiments such as LSST \citep{LSSTSciBookv2}
and CMB experiments.

We expect end-to-end simulations to be used throughout the life cycle of DESI.
At early stages, they will be used to optimize the instrument and survey
strategy. On intermediate time scales they will be used to demonstrate the
feasibility of the baseline design, to shape the development of the DESI
data management system, and to test its scaling performance prior to
first light.
During the operations phase, they will be important for the science
exploitation of the data as a tool to control systematics.

Final end-to-end simulations will use the actual DESI data systems components,
inserting simulated data as if they were real data.  A cosmology
simulation will provide a simulated targeting catalog, upon which the real
fiber assignment code is run using the same file formats and interfaces as
the final real data.  The operations simulator will simulate observing
conditions and select tiles using the real next field selector.  We will
use these to generate pixel-level raw data simulations that the data
reduction pipeline will process as if they were real raw data.
This tests both the scaling
and integrity of the data flow, with the final BAO analysis testing the
software ``as-built'' performance.
A subset of these data will be used to test the data transfer system and
its connection to automatically launching data processing as data arrive
at NERSC.

A data challenge approximately a year before first light will focus on
end-to-end integration.  This provides time to address any integration issues
that arise during the end-to-end tests.

\subsection{Computing Requirements}
\label{sec:compreq}


There are several distinct activities within the DESI project that
have substantial computing requirements, notably target selection in
the near term, full survey simulations in the mid-term, and the data
reduction pipeline during survey operations.  Tables \ref{tab:store},
\ref{tab:trans}, and \ref{tab:comp} provide an overview of these by
year, split by disk storage, data transfer, and computation
requirements.  These include computing requirements beyond the
timescale of DESI construction since one requirement of DESI Data
Systems is that it will be viable to run on available NERSC resources
during the years of operation.




\afterpage{

\begin{table}[!t]
\centering
\small

\caption{Cumulative data volume in terabytes used by the DESI
project by fiscal year.  \textbf{Note:  these fiscal years assume
current schedule predictions.}  All of this data is archived to tape (HPSS)
and kept on disk.  These numbers do not include science analyses.
}

\begin{tabular}[width=0.95\textwdith]{lrrrrrrrrrr}

\hline
Activity & FY15 & FY16 & FY17 & FY18 & FY19 & FY20 & FY21 & FY22 & FY23 \\
\hline

Targeting -- WISE                   &      68 &      95 &     121 &     121 &     121 &     121 &     121 &     121 &    121 \\
Targeting -- DECam\footnotemark[1] 	&     132 &     197 &     263 &     263 &     263 &     263 &     263 &     263 &    263 \\
Targeting -- Mosaic\footnotemark[1] &       0 &      47 &      94 &      94 &      94 &      94 &      94 &      94 &     94 \\
Targeting -- Bok\footnotemark[1] 	&       0 &      23 &      47 &      70 &      70 &      70 &      70 &      70 &     70 \\
Targeting -- ZTF\footnotemark[1] 	&       0 &       0 &      47 &      47 &      47 &      47 &      47 &      47 &     47 \\
Instrument Sims\footnotemark[2] 	&       0 &      15 &     150 &     150 &     150 &     150 &     150 &     150 &    150 \\
Cosmology Sims\footnotemark[3] 		&      10 &      10 &      10 &      10 &      10 &      10 &      10 &      10 &     10 \\
DESI Data\footnotemark[4] 			&      38 &      77 &     115 &     232 &     349 &     465 &     582 &     698 &    815 \\
Science Collaboration\footnotemark[5] &   100 &     300 &     500 &    1000 &    2000 &    3000 &    4000 &    5000 &   5000 \\
Total 			                	&     348 &     764 &    1374 &    1987 &    3104 &    4220 &    5337 &    6454 &   6570 \\

\hline
\end{tabular}
\begin{flushleft}
{\footnotesize
(1) Assumes processed data is factor of 4 larger than raw data, and includes
additional 1/6 of total for reprocessing workspace.\\
(2) 2017 is the full, 5-year survey simulation and processing.\\
(3) This is the space for one cosmology simulation needed for instrument
simulations.\\
(4) This is raw data plus two copies of processed data at any given time.
Includes data from early spectrograph testing.\\
(5) Needed for science analysis, and budgeted separately.\\
}
\end{flushleft}
\vspace{-0.25em}

\label{tab:store}
\end{table}


\begin{table}[!ht]
\centering
\small

\caption{Data transfer requirements in terabytes for the DESI project by
fiscal year.  \textbf{Note:  these fiscal years assume current
schedule predictions.}  These numbers are the total volume of data moved in
each year.
The instrument simulations and target catalogs will be
generated at NERSC and do not need to be moved.  The cosmology simulations
may be generated at the DOE Leadership Class Facilities, and this is included
below.  The numbers for data processing represent the raw data transfer from
KPNO.  Data distribution from NERSC to the collaboration assumes one major
data release per year and is based on network statistics from BOSS.}

\begin{tabular}[width=0.95\textwidth]{lrrrrrrrrrr}

\hline
Activity & FY15 & FY16 & FY17 & FY18 & FY19 & FY20 & FY21 & FY22 & FY23 \\
\hline

Targeting -- WISE 	& 	 80 &	  0 & 	  0 &	  0 & 	  0 & 	  0 &	  0 &	  0 & 0 \\
Targeting -- DECam 	&    55 &    55 &    55 &    55 &     0 &     0 &     0 &     0 & 0 \\
Targeting -- Mosaic 	&     0 &     8 &     8 &     0 &     0 &     0 &     0 &     0 & 0 \\
Targeting -- Bok 	&     4 &     4 &     4 &     0 &     0 &     0 &     0 &     0 & 0 \\
Targeting -- ZTF 	&     0 &     0 &    27 &     0 &     0 &     0 &     0 &     0 & 0 \\
Cosmology Sims 		&    30 &    30 &    30 &    30 &    30 &    30 &    30 &    30 & 30 \\
DESI Data to NERSC 	&     0 &     0 &     0 &     6 &     6 &     6 &     6 &     6 & 6 \\
DESI Data serving 	&     0 &     0 &     0 &   100 &   100 &   100 &   100 &   100 & 100 \\
Total 				&   169 &    97 &   124 &   191 &   136 &   136 &   136 &   136 & 136 \\

\hline
\end{tabular}

\label{tab:trans}
\end{table}


\begin{table}[!ht]
\centering
\small

\caption{Computational requirements of project activities by fiscal year,
given in millions of 2012 CPU-hours.  \textbf{Note:  these fiscal years
assume current schedule predictions.}}

\begin{tabular}{lrrrrrrrrrr}

\hline
Activity & FY15 & FY16 & FY17 & FY18 & FY19 & FY20 & FY21 & FY22 & FY23 \\
\hline

Targeting\footnotemark[1] 			& 	3.6 &	7.2 & 	9.6 &	 12 & 	 10 & 	 10 &	 10 &	 10 &  10 \\
Instrument Sims\footnotemark[2] 	&   2.3 &   2.3 &    45 &   2.3 &   2.3 &   2.3 &   2.3 &   2.3 & 2.3 \\
Cosmology Sims\footnotemark[3] 		&     2 &     2 &     2 &     4 &     4 &     4 &     4 &     4 &   4 \\
Data Processing\footnotemark[4] 	&     0 &     0 &     0 &    18 &    36 &    54 &    72 &    90 & 108 \\
Science Collaboration\footnotemark[5] &   5 &     5 &     5 &    10 &    20 &    20 &    30 &    40 &  50 \\
Total 			                	&   12.9 &  16.5 &  61.6 &  46.3 &  72.3 &  90.3 &  118.3 &   146 & 174 \\

\hline
\end{tabular}
\begin{flushleft}
{\footnotesize
(1) Assumes running Tractor on all targeting data 4 times per year until
survey start, and then performing Monte Carlo analyses on the target
selection.\\
(2) In 2017 we simulate and analyze a full 5 year survey.\\
(3) This is only the cosmology simulation to support the project needs, not
science analysis.  One per year until survey start, then two per year.\\
(4) Assumes 3 reprocessings per year.\\
(5) Needed for science analysis.\\
}
\end{flushleft}
\vspace{-0.25em}

\label{tab:comp}
\end{table}

} 

CPU usage for data processing will be provided through NERSC allocations,
\ie, with no direct cost to the DESI project.  Similarly, input
simulations will use allocations at NERSC, other DOE leadership class
facilities, or at computing centers provided by collaborating institutions.
Disk space, however, will be purchased at NERSC to support these activities.
The projected disk space requirements and input assumptions are detailed in a
spreadsheet in DocDB (DESI-0738).

Here we describe the requirements by activity.
All processing requirements are described in circa 2012 CPU core hours.
In the next 1--2 years, NERSC will transition to a many-core Intel
architecture, with $\geq$60 cores per node.
Although such systems will have an ever-increasing theoretical peak number
of flops, actually leveraging those new architectures will require some work.
Fortunately, processing of the full DESI dataset can be parallelized both at
the task level (extracting the spectra from many exposures) and within a task
(extracting smaller groups of fibers over a reduced wavelength range for a
single exposure).
In our development of software tools and algorithms we will strive to build
this parallelism into our design, in order to use these systems effectively.


\subsubsection{Target Selection}
\label{sec:ts}

See the Target Selection section of the Science FDR for a discussion of the
process of selecting targets for spectroscopy by DESI.
In order to perform the target selection in the years leading up to the
beginning of observations, we must transfer substantial amounts of data
from other locations to NERSC.
At NERSC, we use simultaneous astrometric and photometric measurements across
multiple targeting datasets to identify and select objects.
This analysis will dominate the disk space usage prior to the start of the
survey, and will use moderate levels of CPU time as detailed below.

The primary instruments that will be used for DESI target selection are the
Wide-field Infrared Survey Explorer (WISE), the Dark Energy Camera (DECam)
on the Blanco telescope, and the Bok telescope and Mosaic imager on Kitt Peak.
In addition, data from other facilities such as SDSS, the Canada France Hawaii
Telescope (CFHT), and the Zwicky Transient Factory (ZTF)
may be used to supplement target selection.
Network transfers of the raw data from these instruments to NERSC is well
within current capabilities.  In addition to the uncompressed raw data, we
must also compute and store derived products (processed images, image stacks,
catalogs, \etc) needed for target selection.
We store the raw and processed data on disk as well as archiving it to tape
storage.  We additionally use a small amount of disk space for testing the
data reduction techniques.  Detailed yearly totals for both network use and
disk space are given in Tables \ref{tab:store} and \ref{tab:trans}.

The dominant computational cost of target selection is the process of
extracting catalogs from combined datasets (see the Tractor section of the
Science FDR). We estimate that a full target selection run (using all data)
will be performed four times a year as part of the development for the
data releases 1--2 times per year.
Based on production Tractor runs with
imaging data covering 12\% of the DESI footprint, we can extrapolate the
future CPU usage in Table~\ref{tab:comp}.


\subsubsection{Cosmological Simulations}
\label{sec:cossim}



N-body and hydrodynamical simulations are needed as inputs for developing
the algorithms for tiling, fiber assignment, and large scale structure catalog
generation since each of these steps impacts the systematics of the measured
cosmology.
The numbers given below are for a single realization of the simulated sky;
many tests could be performed with a much smaller simulated area.
Final science analyses will require many such realizations, but that is under
the purview of the science collaborations and is not scoped here.

For both N-body and hydrodynamic simulations, the low-level (``level-1'')
data outputs are stored on the computer where they were generated
(for example, the DOE Leadership Class Facilities or the machine-specific
scratch disk at NERSC).
We do not transfer the level-1 data, and instead produce ``level-2''
data products which can be transferred to the NERSC global filesystem for
further work.
This level-2 data consists of the mock DESI galaxies and approximately
40 snapshots of halo data needed to construct those mocks.
The total size of these items is approximately 10~TB.
Performing one N-body simulation and constructing the mocks is estimated
to require 2M CPU hours.
We anticipate creating the first such simulation early in the project, to
enable rapid use for pipeline development.
The CPU requirements in Table \ref{tab:comp} reflect the likely need to
generate one such simulation per year (to incorporate new features or fix
bugs) in the years leading up to the full survey data challenge.
Only one such simulation is required to be stored on project disk at any
given time; previous simulations may be staged from tape to scratch disk
as needed.


\subsubsection{Instrument Simulations}
\label{sec:instsim}

Pixel level instrument simulations are used to develop the hardware design
and generate simulated data for the spectroscopic data reduction pipeline;
see Sections~\ref{sec:InstrumentSimulations} through \ref{sec:end2endsims}.
Current full CCD pixel-level simulations require 450 CPU hours per DESI exposure
to generate 30~CCD frames with 5,000~spectra.
Generating a full exposure requires constructing a PSF from a realization of
the instrument model (using a nominal model plus fluctuations within the design
tolerance), using that PSF to project a set of simulated spectra onto the CCD,
and adding realistic noise and other artifacts.
If we chose a fixed PSF across many exposures, we could reduce the
computational cost of the simulation in exchange for less realism.
In addition to the full survey data challenge (see below), we anticipate
needing a couple million (circa 2012) CPU hours per year for smaller testing
campaigns of specific features of the simulation and extraction tools.  This
estimate is based on current experiences testing these tools on BOSS and
simulated DESI data.  These test campaigns will generate 10--50~TB of data.
Because instrument simulations are generated and analyzed on the machines
at NERSC (which all share a common filesystem), no data transfers are needed
for this activity.

A full survey data challenge in late 2017 will verify the full Data Systems
chain and guide the development work for the final year before first light.
This data challenge will begin with mock targeting catalogs and proceed
through simulations of the observing strategy, field selection, and fiber
assignment.  These will be used to generate a full survey of simulated
pixel-level data that will flow through the data transfer and archive system
and be analyzed by the data reduction pipeline as if it were real data.
A large scale structure catalog will be produced from these simulations and
the power spectrum / correlation function will be analyzed, thus ensuring the
complete bookkeeping chain from targeting through science analysis.

Since this is a full-sized dataset, processing it will require the
computational resources outlined in Section \ref{sec:spec}, as well as the
resources required to simulate images from the input catalogs and fiber
allocation methods.


\subsubsection{Spectroscopic Data Processing}
\label{sec:spec}

Raw data are transferred daily from KPNO to NERSC via NOAO in Tucson.
Copies of the raw data will be archived at both NOAO and NERSC
to provide geographically separate backups.  At NERSC the data is
preprocessed, calibrated and made available to software which performs
spectral extraction and redshift determination.
The output data products are hosted and served to the science collaboration
through the NERSC science gateways.

When estimating the size of the Raw data, there are several possible ways of
computing the data volume per year.  Three such calculations can be found
in the spreadsheet documenting computing resources by year (DESI-0738).
The final estimate we use is based on the scaling of compressed BOSS raw data
volumes (by 30 versus 4 CCD frames per exposure), plus a margin to allow for
uncertainty in the number of daily calibration images and the data compression
factor.  This gives us 10~TB/year of raw data, plus an additional 0.6~TB/year of
auxiliary data from acquisition and guiding.

BOSS spectral extractions result in 2.5 times more extracted data
than raw data volume.  DESI spectra will be similar, with the addition
of the ``resolution matrix'' to encode the spectral resolution.  A
naive encoding of this matrix would be quite large, but we anticipate
that a sufficiently accurate parameterization will result in extracted data
being 5 times larger than raw data, \ie, 50~TB per year.  While
modestly large by current standards, this will be a trivial amount of disk
space by the start of the DESI survey.


The network link from KPNO is large enough to enable data transfers in real
time as soon as the data are taken.  The computing systems at KPNO will have
sufficient disk storage to cache at least one week of raw data in the case of
significant network outages or delays in data transfer.
After processing, calibrated images and spectral domain products are served
from NERSC to the rest of the collaboration.  This processed data is also
backed up to NOAO.  Based on experience serving BOSS processed data, we
anticipate serving $\mathcal{O}$(100~TB) of data per year to the DESI science
collaboration.

Constructing the PSF from calibration images will require about two
CPU-hours per CCD frame.
Assuming three re-calibrations per day, this amounts to less than 100,000
CPU-hours per year total.
Our baseline spectral extraction software splits the spectra by
independent ``bundles'' of fibers on the CCD and overlapping wavelength
regions which are then reassembled to avoid edge effects when extracting
sub-regions.  Current development code for spectral extraction can extract a
single DESI frame with about 15 (circa 2012) CPU hours.  If we assume
that we have 20 minute exposures and that we have 221 nights of observing with
an average of 10 hours per night, this gives us about 6630 exposures per year.
At 30 frames per exposure, this is approximately 200,000 frames per year of
science data.  We assume an additional factor of two for calibration images,
which gives us 400,000 frames and a total of six million CPU hours per year of
survey data to do the extractions.  These estimates include both the
dark time and the bright time programs, and the possibility of taking multiple
exposures per tile due to high extinction or poor conditions.
Based upon BOSS redshift fitting code, we expect
classifications and redshifts to take less than 100k CPU-hours per year
of DESI data.


Based on BOSS operational experience, we allocate 3 full reprocessings
of all data per year, obtaining the CPU
hour estimates summarized in Table \ref{tab:comp}.
While fairly large by current computing standards, these should be quite
manageable by the start of DESI.

\subsubsection{Science Analysis for Cosmology}

The DESI science collaboration will coordinate a broad research program in
Cosmology and this is described in detail in Part 1 of the Final Design
Report.  This effort will require substantial computing resources in the form
of CPU hours and disk space.  The CPU hours will be provided by standard
NERSC allocations, and the disk space is budgeted separately from the DESI
project costs.  We include these required resources in tables \ref{tab:store}
and \ref{tab:comp} in order to give a more complete picture of our
requirements.

The needed disk and computing resources are dominated by the
generation, storage, and analysis of mock catalogs and their parent
simulations, more so than the actual DESI data.  These mock catalogs
are critical for the control of systematic errors at the level of
DESI precision, and analyses such as for neutrino masses and modified
gravity require connection to the underlying dark matter and gas
distributions.  Typically, one uses hundreds to thousands of mock
catalogs, for dozens of choices of control parameters, and
then repeated several times through the life of the survey.  While
the underlying cosmological simulations will be run with other
computing allocations, storing suitable data products and then
analyzing them is the responsibility of the DESI collaboration.
Even retaining simplified and subsampled data products for hundreds
to thousands of simulations adds up to PBs of data.

\subsubsection{Computing Requirements Summary}

Data transfer volume is well within the current capabilities of
ESNET and is not expected to be a problem.
CPU requirements are large but not intractable, dominated by processing
the full survey data challenge in 2017 and the final reprocessings of the
entire dataset at the end of the survey.
Disk space requirements for targeting data will likely be the most expensive
piece of hardware computing since it is both large and needed prior to
commencing the survey.  Thus overall, the computing requirements are large but
entirely manageable.  Tables \ref{tab:store}, \ref{tab:trans},
and \ref{tab:comp}
are a summary of the spreadsheet in DESI-0738.

\subsection{Collaboration Tools}
\label{sec:collabtools}


We have developed an Integrated Collaboration Environment system based on
Trac\footnote{\url{http://trac.edgewall.org}} software.  Trac provides a wiki,
source code browser, bug tracking system, timeline and milestone tracking,
and basic search facilities.  We have further integrated Trac with a set
of mailing lists, based on GNU Mailman\footnote{\url{http://www.gnu.org/software/mailman/}}.
This integration allows mailing lists to be searched from within the Trac
environment.

For the last year, we have been developing the open-source portions of the DESI
pipeline in the \texttt{desihub} group on GitHub\footnote{\url{https://github.com/desihub}}.
In addition to the features of GitHub itself, like issue tracking, GitHub also
allows us to plug in to several other services:
\begin{itemize}
    \item Travis Continuous Integration (CI) service\footnote{\url{https://travis-ci.org}}.
    This service runs unit tests on the code every time a change is checked in to
    the repository.  It also allows verification of code added to branches before
    it is reintegrated with the master code base.
    \item Coveralls test coverage service\footnote{\url{https://coveralls.io}}.
    This service runs after a Travis test is complete.  It displays what
    parts of the code were actually tested, and what parts of the code were
    bypassed, and which might still contain hidden bugs.  The goal is to have
    perfect (100\%) coverage, meaning every line of code is tested in some
    way, though practical concerns (the code might create large data files)
    may make some tests difficult to write.
    \item Read the Docs documentation service\footnote{\url{https://readthedocs.org}}.
    This service automatically process the documentation associated with a
    software product, and produces a public web page.  For example, the
    ``desispec'' product, the core of the DESI spectral pipeline,
    is documented at \url{http://desispec.readthedocs.org}.
    It will automatically post documentation for any new code that is checked in,
    but stable versions (usually the latest tag) can also be designated for
    long-term use.
\end{itemize}

In addition, we have code repositories based on
Subversion\footnote{\url{http://subversion.apache.org/}}, for code related
to DESI hardware, and for internal documents.
Membership in the DESI Trac also grants access to our Subversion repositories.
This feature allows DESI collaborators to immediately access DESI code without
additional sign-ups or requests.

Document Control is based on DocDB\footnote{\url{http://docdb.fnal.gov/doc/}},
which we have already begun to integrate with Trac.  Because Trac
is highly extensible through plugins, we expect to make further developments
in DocDB-Trac integration.  Similar to how we have integrated Trac and Subversion
access, DocDB access is also connected to a Trac account.

The Integrated Collaboration Environment is backed up daily to a server at the
University of Utah. This backup can be activated quickly
in the case of a temporary or catastrophic loss of the primary site.


We are developing a public website for DESI based on
WordPress\footnote{\url{http://wordpress.org}}.  This can serve as the basis
for public outreach and documentation of our public data releases (see also
Section~\ref{sec:transfer}).

We will provide a system for announcing projects and publications similar to
a system used by SDSS.  This allows the entire collaboration to be informed
of projects and working groups that may be of interest.  The announcement of
publications facilitates collaboration-wide review of papers written using
DESI data.

\subsection{Coding Standards}

The majority of the Data Systems code is written in python and open source
on \url{https://github.com/desihub} (see \S\ref{sec:collabtools}).
Code revisions are tracked using git following a basic workflow of:
\begin{itemize}
    \item Development plans are discussed on telecons, the desi-data email list,
        and/or github issues.
    \item Implementation is done in a git branch and then the branch is pushed
        to github.
    \item The author opens a pull request which is reviewed by someone else
        prior to merging it into the master branch.
    \item Unit tests are included with each package and are run by the author
        on their development machines, and by Travis in a standardized test
        environment.  All tests must pass prior to merging new code into the
        master branch.
    \item Automated coverage tests also flag when testing coverage has
        decreased.  Whether this is acceptable or not prior to merging is
        handled on a case-by-case basis.  In most cases, new code is required
        to include unit tests that will increase the overall coverage.
    \item A nightly integration test on two different NERSC machines also
        checks cross repository consistency from pixel-level data simulations
        through redshifts, using the latest master version from each repository.
\end{itemize}

Coding standards are documented on the DESI wiki and are periodically
reviewed, discussed, and updated.  These cover python coding style, unit tests,
documentation, dependencies, file formats, code organization, and tagging
procedures.

When possible, functionality is split into a DESI-agnostic package that is
wrapped by a DESI-specific package.  For example,
specsim\footnote{\url{https://github.com/desihub/specsim}} provides a
general toolkit for rapid simulations of fiber fed spectrographs, while
desisim\footnote{\url{https://github.com/desihub/desisim}} uses that to
simulate DESI itself.  This allows cross project development coordination
(most specifically with SDSS/eBOSS), and allows DESI to cross check the
performance of its code using real world data.

\clearpage

\section{Integration and Test}\label{s6:IandT}
\setcounter{equation}{0}\setcounter{figure}{0}\setcounter{table}{0}

The DESI instrument will consist of numerous separate elements supplied by distinct and distant collaborators, all finally coming together at the Mayall Telescope.  Comprehensive planning and management of integration and testing activities, of interfaces between elements, and of preparation of the Mayall Telescope for the installation of DESI will be crucial to project success.  The Level 2 Integration and Test Managers are primarily responsible for these activities.  In addition, the DESI Project Engineer will be engaged in integration and test planning and strategy.  

Periodic workshops dedicated to integration and test planning will be held with the collaborating institutions to involve them in the planning and execution of the integration, interfacing, and testing tasks they will be party to.  The first of these workshops was the initial meeting of the Installation and Commissioning Task Force (ICTF) during April 2014 at Kitt Peak and NOAO Headquarters in Tucson.  That meeting raised previously unidentified issues and yielded a number of Action Items (all identified, tracked and responses referenced in DESI-0781), each of which addresses an integration question or issue.  An ICTF Report (DESI-0776), referencing those issues and actions, has been published.

This section describes Integration and Test all the way through installation and functional verification of DESI on the Mayall telescope.  Commissioning of DESI is described in the next section.

\subsection{Instrument Integration and Test}

The DESI Instrument consists of a Corrector, Telescope Fixed Top End Structure, Hexapod, Corrector Alignment System, Focal Plane Assembly, Fiber System, Spectrograph System, Fiber View Camera, and Instrument Control System, as shown in Figure~\ref{fig:block-diagram}.  Each of these major elements is the responsibility of distinct collaborators or groups of collaborators.  Most of these elements will first be brought together at the Mayall Telescope during the Installation Campaign.  To ensure successful integration of these elements, a comprehensive Integration and Test plan is required, along with rigorous Interface Control.

\begin{figure}[tbh]
\centering
\includegraphics[width=\textwidth]{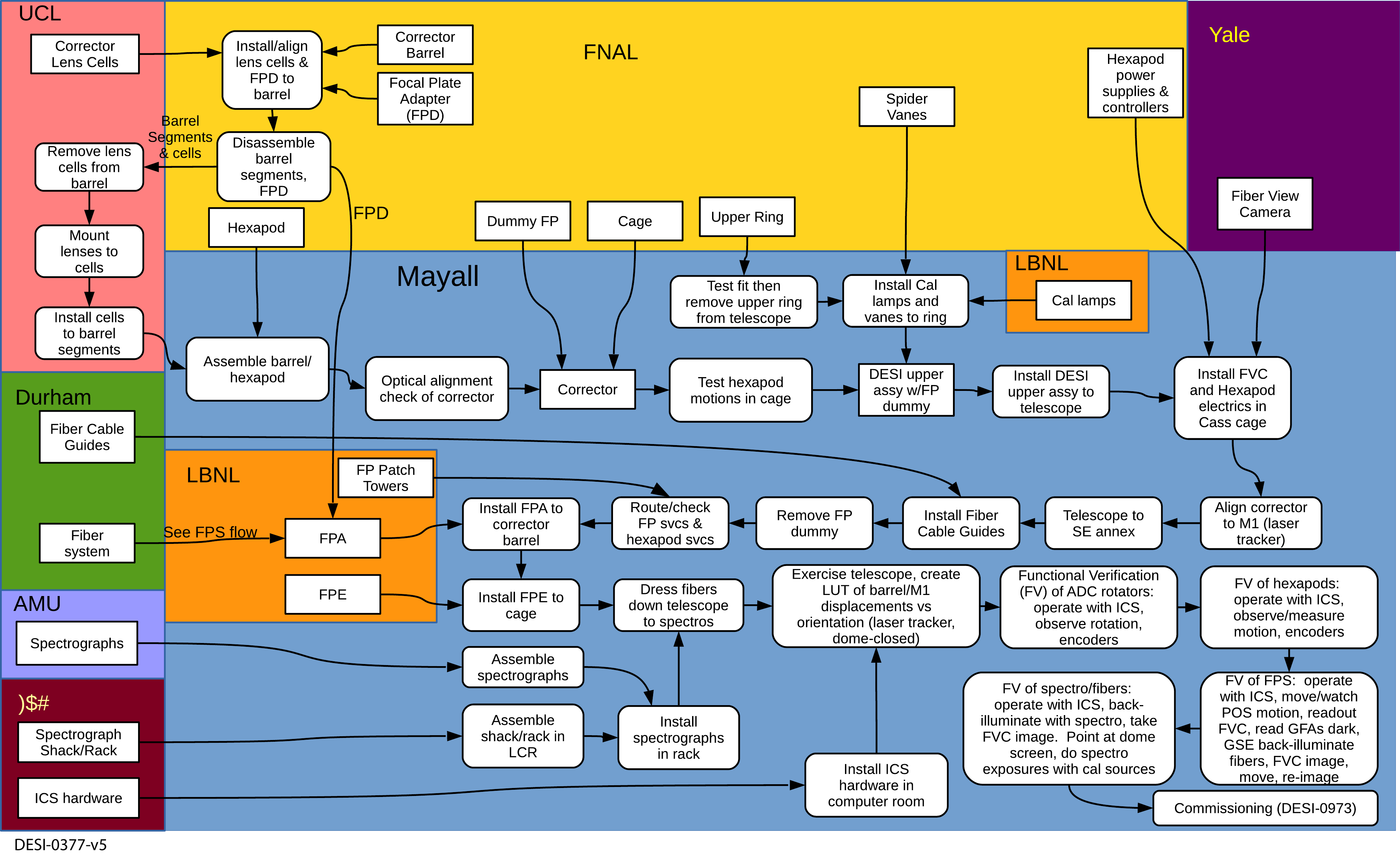}
\caption{High-level DESI Integration and Test flow}
\label{fig:IandTflow}
\end{figure}

An Integration and Test Flow has been developed (DESI-0377).  Figure \ref{fig:IandTflow} illustrates a high-level synopsis of this Integration and Test Flow.  To minimize problems due to faulty elements, components and subsystems are to be verified before integration into higher level of assemblies.  Depending on the component or subsystem, verification may be as simple as a visual inspection or as complex as a complete functional test under environmental extremes.  
System-level integration and test steps are the ultimate responsibility of and require approval by the Level 2 I\&T Managers.

%
%

This subsection describes system-level I\&T activities up until the major DESI instrument elements are handed off to NOAO on-site.  These activities include bringing major subsystems together prior to shipment to NOAO, along with post-delivery verification at NOAO.  Integration of DESI to the Mayall will primarily be performed by NOAO and their contractors, with help from DESI personnel.

Post-delivery verification will be carried out primarily by the instrument team responsible for the system level I\&T with support from DESI and NOAO personnel.  The DESI components will be delivered to NOAO in various stages of integration.  Upon receipt, the shipping crates will be inspected for visible signs of damage, and any environmental monitors used to track relevant conditions during transit (including dynamic loading, temperature, humidity, and air pressure) will be checked.  After unpacking, the hardware will be inspected visually and tested to verify functionality.  Depending on the environmental and testing requirements, this verification will take place in the NOAO-Tucson high-bay facility, the Instrument Handling Facility on Kitt Peak (which includes a class-10000 clean room), or the Mayall Telescope building on Kitt Peak.

After passing these post-shipping tests, elements will be handed over to NOAO, who will execute the integration with the Mayall Telescope as described in Section~\ref{sec:installationintro}.
%

\subsubsection{Corrector Barrel/Corrector Optics/ADC Rotators}

Integration, alignment, and test of the Corrector Assembly prior to delivery to NOAO is described in Section~\ref{sec:alignment}.

The three segments of the Corrector assembly will be shipped to NOAO in three individual shock-isolated crates.  Unpacking, verification, and integration of the Corrector will occur in a tented area on the ground floor of the Mayall building.  This area will be outfitted with a portable gantry crane, along with the UCL turntable/pencil beam setup.  Reassembling in the Mayall building will limit transportation loads on the fully assembled corrector.

The ADC rotators will be run post-delivery at least one full revolution in each direction.  Motor current will be monitored and verified to be within specification.

Post-delivery to NOAO, laser pencil beam/rotary table testing (as described in Section~\ref{sec:alignment}),  will be done after the three segments are integrated together.  This mainly verifies that the lenses have not shifted significantly since pre-ship testing.

\subsubsection{Telescope Fixed Top End Structure/Hexapod}

The hexapod assembly will be delivered to the supplier of the Cage, Spider, and DESI Upper Ring.  The cage and hexapod will be integrated together, and a corrector/focal plane mass dummy will be integrated to the hexapod.  Functional testing of the hexapod will include running hexapod algorithms and measuring associated motions of the corrector/focal plane mass dummy.  In addition, range of motion, power consumption, and temperatures of the hexapod motors will be measured.

Following hexapod verification, the mass dummy, hexapod assembly, and cage all will be separated from each other.  The hexapod will be shipped to NOAO in the crate it arrived in.  The spider vanes will be shipped together in a crate.  The cage will be shipped in a crate by itself.

The upper ring will be a monolithic weldment 5.4~m in diameter.  It will necessarily be an oversize load for on-highway transportation.  It will be plastic-wrapped and mounted on a flat-bed truck for shipping to NOAO. 

After delivery to NOAO, the hexapod system will be run on the bench through a test sequence of motions.  The direction of the hexapod motions will also be measured. Motor current will be monitored and verified to be within specification.

Post-delivery verification of the elements of the telescope fixed top end structure include unpacking and visual inspection.  In addition, prior to integrating the spider vanes and cage to the new upper ring, the new upper ring will be test fit to the Mayall Serrurier trusses as described in Section~\ref{sec:initialtopend}.

\subsubsection{Focal Plane/Fiber System}

\paragraph{Pre-Shipment}

The Focal Plane Assembly and Fiber System are closely interwoven.  Each fiber positioner will be integrated with its fiber ferrule and associated positioner fiber assembly with a cleaved end.  Then each of the ten focal plate segments will be populated with 500 fiber positioners with their fiber pigtails and electrical wires routed through the focal plane fiber management system to that segment's fiber junction box, containing 500 fiber ends.  As groups of twelve positioners are integrated, those twelve positioners and fibers will be tested to verify basic mechanical, electronic, and optical operation.  This is an opportunity to address issues without having to de-integrate positioners after they are fully surrounded by neighbors.  Once the segment is fully populated, basic operation of all of the fiber positioners, GFA modules, and field fiducial illuminators will be verified using a test sequence driven by a PC. Integration and test of the focal plane segments to this point (including field fiducial illuminators and GFA modules) will be the responsibility of WBS 1.4.

Each petal assembly will be surveyed using a coordinate measuring machine (CMM) with a touch probe to map the relative locations of each field fiducial illuminator, the GFA fiducial illuminators, and the reference tooling balls installed in the focal plate.  With this set of known relations, fiber view camera measurements of the illuminated field fiducials, GFA fiducials, and back-illuminated science fibers can map accurately to
physical locations, establishing the focal plate scale and the relation of the fiber tips to the guide sensor pixels.

In parallel with the assembly of the focal plane, ten $\sim$35~m 500-fiber cable assemblies from slithead to free, bare fiber ends will be built and tested as described in Section~\ref{sec:fib_I_T}. Each of these fiber cable assemblies will be integrated to its respective focal plane segment assembly by threading the fiber ends into the cable side of the fiber junction box, strain-relieving the mechanical cable end to the fiber junction box, fusion bonding the cable fiber ends to the positioner fiber pigtail ends by extending them out from the junction box to a fusion splicer, dressing the now-continuous fiber loops into the fiber junction box, then installing covers on the junction box.  Fiber integrity will be tested by illuminating the slit block and looking for dark fiber tips at the positioners.  Diagnosis and repair of discontinuous fibers will take place at this stage.  Basic operational tests of the positioners, GFA modules, and field fiducials will be performed.

Once integrated with its fiber cable, each focal plate segment will be integrated one at a time with the focal plate interface ring/focal plate adapter assembly.  Once all ten segments are integrated and the center connection plates installed, a CMM with a touch probe will be used to verify alignment of the petals to each other and the integration ring.  Fiber integrity and basic operational tests of the positioners, GFA modules, and field fiducials will be repeated.  

The complete focal plane assembly with integrated fiber cables will be crated together and shipped to NOAO as a single unit. The Focal Plane rear closeout and thermal management system will be crated and shipped separately.

\paragraph{Post-Delivery}

To avoid contamination, unpacking, inspection, and testing of the Focal Plane/Fiber System assembly will be done in a clean tent.

The integrity of the fibers will be verified by illuminating each slithead one at a time and observing light from each fiber tip of the corresponding focal plane segment.  A record of this test will be a digital photograph of each segment with lighted fiber tips.

The liquid cooling loops in the heat exchangers will be plumbed and leak-checked.

The cooling fans will be run and current monitored and verified to be within specification.

The field fiducial illuminators will be powered and visually confirmed to function.

The GFA sensors will be turned on and tested for functionality, sensitivity, and contrast.

A fiber positioner drive computer will be connected to the fiber communication system, and all 5000 fiber positioners will be operated without using a fiber view camera.  The positioners may be operated individually or in groups.  This test is designed only to verify the operation of the positioners, not their accuracy.

To avoid the need for collision avoidance, the positioners will be commanded to move only within the 10.4~mm envelope that avoids any contact with neighbors.  Operation will be verified visually and recorded on video to aid in identifying and diagnosing problem positioners.

\subsubsection{Spectrographs/Cryostats}

CCDs and their front end electronics will be installed and aligned in the cryostats by the cryostat supplier.  Non-contact metrology will be used to accurately locate and orient the active area of the detector relative to precision interfacing features on the cryostat.  The CCD location will be verified cold, under vacuum through a dummy Field Lens Assembly with a flat window.  This will ensure that the Field Lens Assembly is correctly located in relation to the CCD, and that each cryostat is interchangeable with every other cryostat of the same channel.  Interchangeability of cryostats of the same channel is required so that if a cryostat fails, it can be replaced with a spare stored at the Mayall.  The failed cryostat will be refurbished and become the spare, which must be suitable for any spectrograph.

The Field Lens Assembly will be supplied by the spectrograph camera supplier.  Because it also acts as the vacuum and contamination barrier to the cryostat vacuum vessel, it will be integrated to the cryostat by the cryostat supplier after the dummy Field Lens Assembly is removed.  The integrated and sealed cryostats will be shipped to the spectrograph integrator under vacuum.  One complete, tested spare cryostat for each channel (three spare cryostats in total) will be shipped to NOAO to minimize downtime in the event of a cryostat failure in-service.

As described in Section~\ref{sec:Instr_Spectrographs}, the spectrograph integrator will assemble and align the spectrograph without the cryostats but with optics, gratings, shutters, illuminators.  Testing of throughput and PSF  will be done at that integration level.

The cryostats with integrated detectors and field lens assemblies will then be integrated to each spectrograph assembly by the spectrograph integrator.  A test slithead will be used to do complete calibration, resolution, and stray light testing.  Each tested and characterized spectrograph will be shipped to NOAO as an assembly including its three cryostats.

Post-delivery verification of the spectrographs will include operation of essentially all functions, but not verification of optical performance (\eg, no PSF or throughput measurements).

After external and internal inspection, the spectrograph purge will be connected, the cryostats will be plumbed to cooling water and leak-checked, the vacuum in the vacuum vessel will be regenerated as needed, and detector cool down using the LPTs will be started and allowed to stabilize overnight.

A test slithead with illuminator will be installed.

The exposure shutter and NIR shutter will be verified by opening and closing individually.

The Hartmann shutters will be verified by opening and closing one half at a time.

The fiber back-illuminator will be operated.

CCD/FEE operation will be verified using the test slithead illuminator.

\subsubsection{Instrument Control System}

Integration and test of the Instrument Control System is described in Section~\ref{sec:DAQintegration}.

\paragraph{Fiber View Camera} 

After arrival at the Mayall, the fiber view camera will be connected to a computer and its function verified by imaging a test pattern.

\subsection{Mayall Preparation}

\begin{figure}[!h]
\centering
\includegraphics[height=3in]{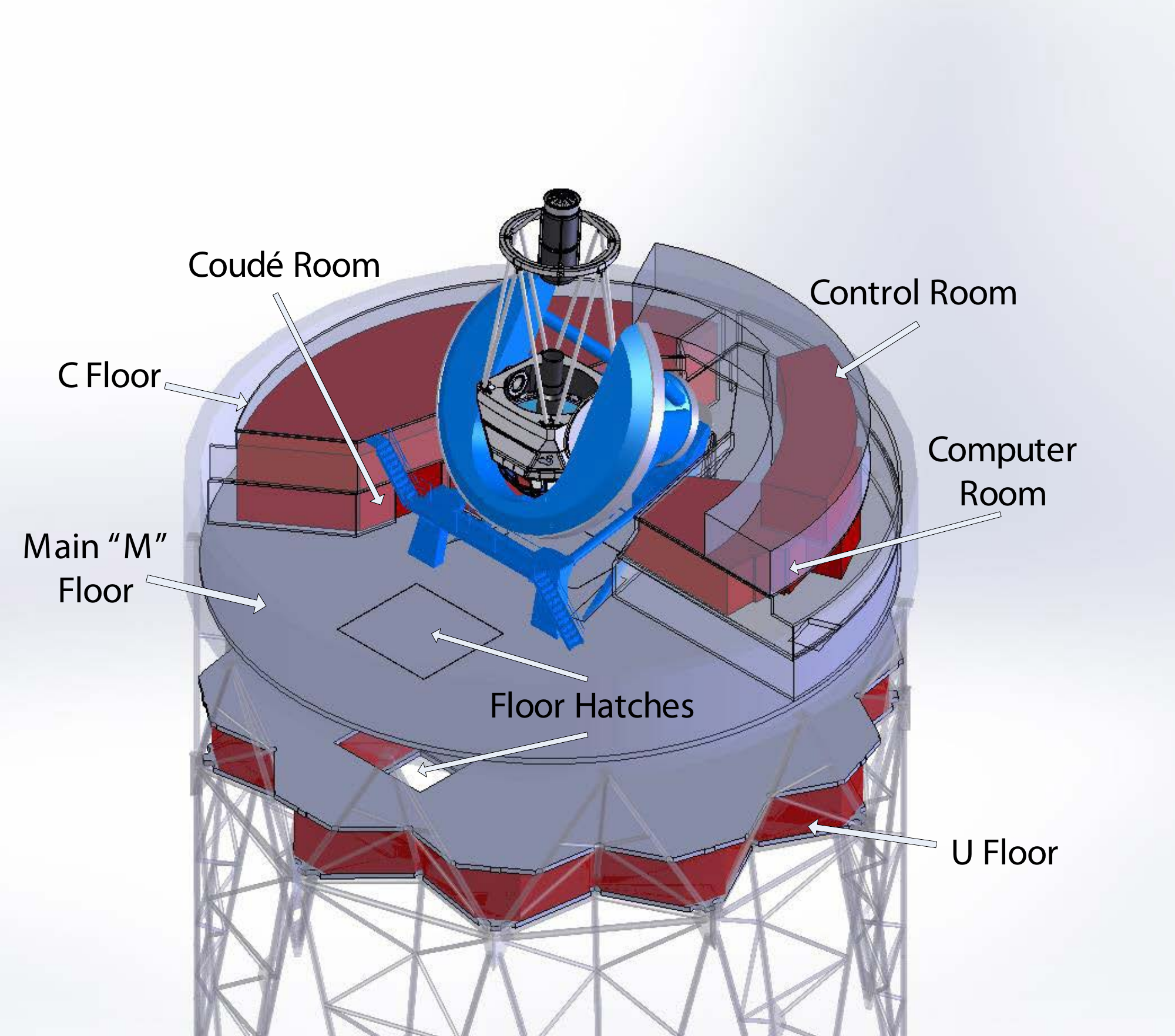}
\caption{Iso view of the Mayall telescope at zenith pointing, showing the general layout of the major supporting facilities referred to throughout the text.  The old FTS lab is on the same floor level as the Coud\'{e} room, below and behind the Computer Room from this viewing angle.}
\label{fig:IsoViewTelescopeInChamberAnnotated}
\end{figure}

This subsection describes the work needed to prepare for the installation of the DESI instrument on the Mayall telescope.  Much of this work will begin while the telescope is still in use for regular scientific observing with its existing instruments.  Figure \ref{fig:IsoViewTelescopeInChamberAnnotated} illustrates the general layout of the telescope and the surrounding rooms and facilities that will be mentioned frequently throughout this subsection and the next.

\subsubsection{Measurements}

Measurements of the Mayall M1 surface and environmental monitoring of key locations at the Mayall will be used to help define requirements and guide the development of the DESI instrument.

\paragraph{Laser Tracker Measurement of M1 Surface} 

\begin{figure}[!h]
\centering
\includegraphics[height=2.5in]{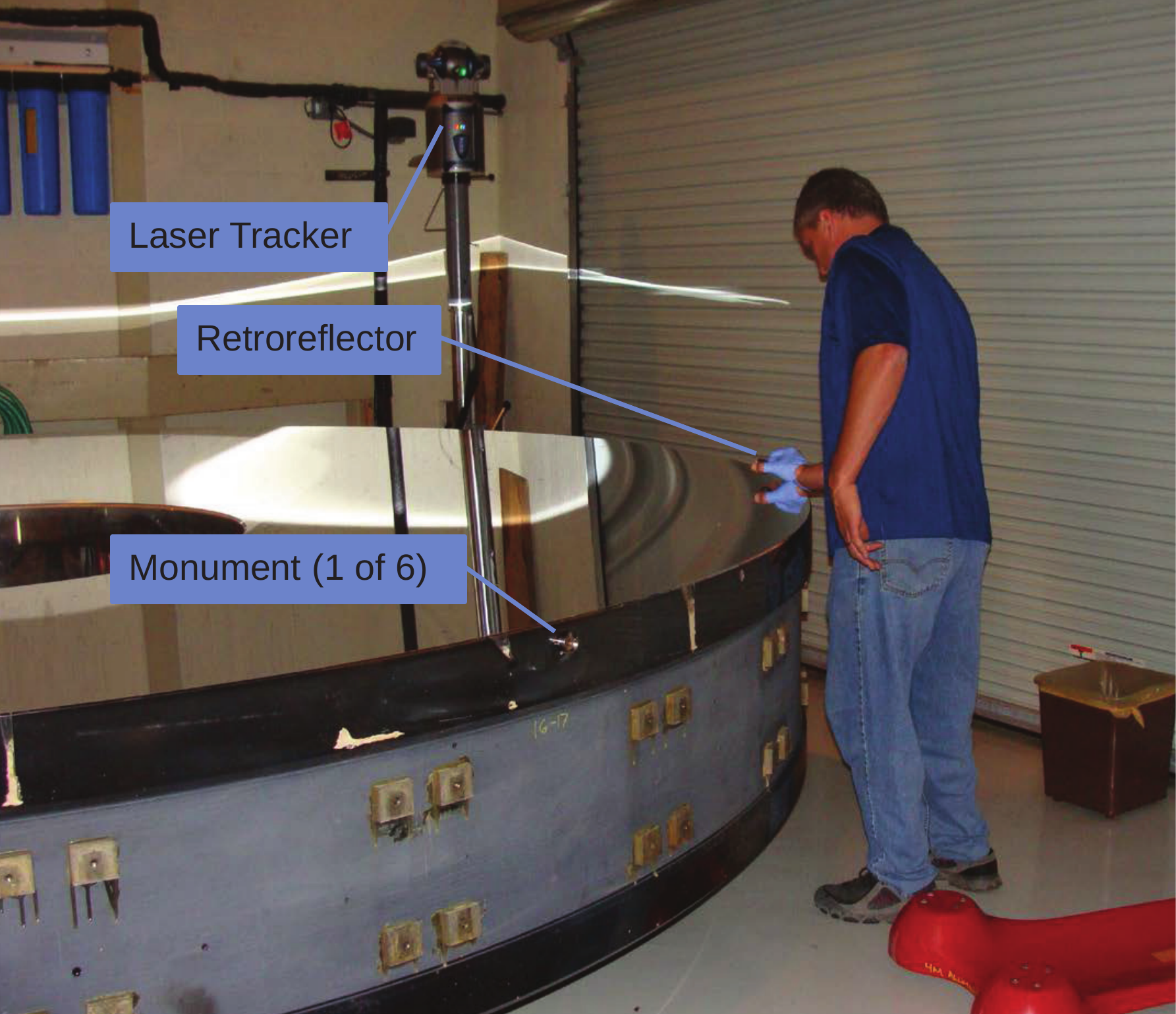}
\caption{Laser tracker surveying of the Mayall primary mirror in July 2014.  Just before re-aluminization, the mirror was washed and its optical surface, outer diameter, and inner hole diameter were surveyed relative to the six permanent monuments on the mirror outer diameter.  This survey yielded the relation between the mirror optical axis and the fixed monuments, to help align DESI to the primary mirror.}
\label{fig:M1Survey}
\end{figure}

Just prior to re-aluminizing the Mayall primary mirror in July 2014, its front surface, inner hole diameter, and outer surface diameter were surveyed using a laser tracker.  These measurements were made relative to six magnetic laser tracker retro-reflector monuments that were permanently bonded to the outer diameter of the mirror in 2013.  Figure~\ref{fig:M1Survey} shows the mirror being surveyed on the ground floor of the Mayall building.

The aspheric profile of the mirror surface was fitted to a solid model of the mirror to identify the location and orientation of the mirror optical axis relative to the fixed monuments.  The survey was done with the laser tracker located in three different azimuthal positions around the mirror, and the results of the three surveys bundled to give a best-fit relation of the surfaces to the monuments.  These measurements will be used to align the DESI cage and corrector barrel to the Mayall primary mirror, using a laser tracker and tracker monuments on the DESI hardware.

The individual measurements of the monument locations relative to the mirror axis deviate from the bundled values by an average of 25 microns, and the locations of the centroids of the six monuments deviate from survey to survey by 13 microns. The inferred orientations of the optical axis (the Y-axis in the model) vary by less than 4 arcseconds from survey to survey.   The mirror optical axis was found to be offset 9~mm from the mirror hole and outer diameter, indicating that those cylindrical surfaces are not good references for optical alignment.

\paragraph{Environmental Monitoring} 

Data on the thermal and vibrational environment of the telescope and immediately associated facilities are needed for the detailed design.  The most critical parameters are the thermal and vibration environment at prime focus and in the Coud\'{e} Room.  Work has already begun in several areas.  

Accelerometers were installed temporarily and used in measurement campaigns in the prime focus cage and in the large Coud\'{e} room.  Accelerometers in the prime focus cage were used to measure the background vibration spectrum during normal operation and to assess the additional vibration from different types of cooling fans.  These data will provide critical input to the design of any cooling system needed for the DESI prime focus hardware. Accelerometers on the floor of the Coud\'{e} room were used to measure the spectra of vibrations transmitted through the floor from different pieces of telescope and facility equipment that are currently used during normal observing.  Similar measurements with the accelerometers on the cart of a large instrument stored nearby were used to assess the vibration mitigation provided by the wire-rope shock isolators in the cart. These data will be used in modeling the mechanical structures of the spectrographs and their racks, and for incorporating any needed damping devices into the racks and their mountings.  

Temperature monitoring in the originally planned DESI spectrograph location (the old Fourier Transform Spectrograph or FTS lab, on the main dome floor West of the telescope mount) has been done with a combination of the existing building monitoring system and portable USB temperature loggers.  Similar monitoring has begun for the Coud\'{e} room (East of the telescope mount on the same floor) using monitors connected to the building environmental control system.  The Coud\'{e} room is the preferred location for the spectrographs versus the FTS Lab as it allows a shorter fiber run and it offers quite a bit more space for a thermally controlled enclosure and other support equipment. The Coud\'{e} room monitoring program found that the room temperature follows longer-term temperature changes in the dome and outside, but with a time lag of 24--36 hours, and roughly a factor of 2 dampening in response amplitude.  The Coud\'{e} room also experiences seasonal variations in temperature, typically ranging between 10~\celsius and 20~\celsius, with recorded extremes during the monitoring period of 0 and 25~\celsius.  These data will help define the requirements for the thermally controlled spectrograph enclosure.  Parallel monitoring of temperatures in the dome is helping to define the operating range required of the prime focus components.  The interior of the dome saw temperature variations typically between 0 and 25~\celsius, with recorded extremes of during the recording period of -10 and 28~\celsius.  An extreme low temperature of -18.1~\celsius was recorded in the dome in February 2011.   

\subsubsection{Observatory Facility Modifications}

\paragraph{Cooling System Modifications} 

Heat removal from equipment and temperature control of the telescope chamber and support equipment is currently provided by the Mayall facility chiller.  The  chiller, a Carrier Aqua Snap Model 30RAN045, was installed in 2009 and has a nominal cooling capacity of 45 tons at sea level.  It can deliver about 43 tons of cooling after de-rating for the working altitude of 6875 feet.  Under the typical load conditions currently encountered, the Mayall facility uses no more than 50\% of the capacity of the chiller.  Thus there is about 20 tons (~72 kW) of cooling capacity available presently.  In addition, during DESI operations there will be no need to cool the two He compressors currently used on the NEWFIRM cryogenic system; these compressors impose about 5 kW of heat load each during NEWFIRM operations.  In the DESI era, this 10 kW of heat removal will also be available for DESI equipment, making a total of about 82 kW of heat removal available for DESI equipment.  
  
Detailed heat budgets for the DESI equipment are under development, but information about the expected system loads is contained in both the Observatory Facility Cooling Requirements document, DESI-0484, and the respective ICDs: DESI-0389 for the Focal Plate Assembly, DESI-0621 (Hexapod) and DESI-0388 (Fiber View Camera) for the Cassegrain cage, DESI-0470 for the Coud\'{e} room, and DESI-0473 for the computer system.  Where requirements are found in the relevant ICD, that document controls and is referred to in DESI-0484.  Where no cooling requirements appear in the ICD, DESI-0484 governs.  Summaries of these loads are reflected in Table~\ref{tab:powerbudget}.

\begin{table}
\centering
\caption{DESI heat sources and locations (Estimated average power during science operations, reference document DESI-0484).}
\begin{tabular}{lllll}
\hline
Subsystem & ICD Document & Location & Mean Heat Load (W) \\
\hline
Focal Plane & 389 & Prime Focus & 723 \\
Hexapod & 621 & Cassegrain Cage & 155  \\
Fiber View Camera & 388 & Cassegrain Cage & 50 \\
Spectrograph Shack & 470 & Coud\'{e} Room & 5,000 \\
Spectrograph Equipment & 470 & Coud\'{e} Room & 1,000 \\
Computer Equipment & 473 & Computer Room & 23,500 \\
\hline
Total & & & 28,046 \\
\hline
\end{tabular}
\label{tab:powerbudget}
\end{table}

Of the six items listed in Table~\ref{tab:powerbudget}, only the last three will depend on the Facility chilled glycol system for heat removal.  The first three will simply dissipate their heat into the dome air, directly or indirectly.  The Hexapod and Fiber View Camera are very small average heat loads; due to their location in the Cassegrain cage, direct dissipation into ambient air of these small loads will not adversely affect telescope image quality.  For the Focal Plane system, heat will be actively removed from the elements inside the Prime Focus Enclosure shroud:  fiber positioners with electronics, Guide-Focus-Alignment sensors (GFAs) with electronics, focal plane cooling fans, field fiducial illuminators, and power supplies.  The fiber positioners are by far the largest peak heat load, but are operated only intermittently.  The relatively high thermal time constant of the focal plane system allows the cooling system to be sized for mean heat load rather than peak.  Heat will be removed from the top end by means of a liquid loop running from a liquid-air heat exchanger located on the main floor of the dome, powered by a pump to account for the height gain to the telescope top end.  The requirement is to maintain the temperature within the focal plate thermal shroud at ambient or slightly above.  It is important not to cool the focal plate assembly below ambient in order to prevent condensation from forming within the shroud under humid conditions.  Therefore this system will simply extract the modest average heat load from the ambient air within the shroud and deposit it away from the telescope beam; it will return ambient-temperature coolant to the heat exchangers in the shroud. Within the thermal shroud, the cooling liquid will be plumbed through heat sinks in direct contact with heat sources other than the positioners, and through one or more air-liquid heat exchangers in a gentle forced-convection air loop with the fiber positioners.  The hexapod and ADC motors (not listed in Table~\ref{tab:powerbudget})are expected to operate with such low duty cycle and power levels they will produce negligible effects on image quality if they are allowed to cool by conduction to the structure and by natural convection to the air.  The calibration lamps on the top ring will not operate during observations and will not have active heat removal.

In the Coud\'{e} room, there will be a chiller to supply cooling water to the thirty Linear Pulse Tube (LPT) cryocoolers that cool the science CCDs.  This chiller and the spectrograph shack environmental control system will reject their waste heat into the building glycol system.  The Coud\'{e} room is not currently served by the building glycol system, but the facility preparations will include extending the glycol system to the Coud\'{e} room and providing glycol distribution where needed within the Coud\'{e} room.  A contract has been let to a consulting engineering firm to design the additional glycol distribution needed to serve the Coud\'{e} room, and the design was received in May 2015.  The design identified the glycol circuit on the immediate "outside" (dome side) of the Coud\'{e} room wall as having both sufficient excess flow capacity (beyond current needs) and appropriate coolant temperature for the DESI equipment.  The minor additional glygol distribution plumbing will be added during the 2016 Mayall Summer Shutdown, as installation requires a temporary shutdown of the Primary Mirror Temperature Control system which is served by this glycol circuit.  Other DESI equipment in the Coud\'{e} room, the cryostat control electronics and possibly an air dryer (if the Mayall facility air dryer is deemed insufficient), will vent their modest heat production into the room air. It is not expected that active cooling of the Coud\'{e} room will be necessary; if this changes, the room air can easily be kept near dome-ambient by a small air-handling unit (AHU) that includes a heat exchanger for chilling the air using building glycol.

In the Cassegrain cage, the DESI heat sources will be the fiber view camera system and the hexapod power supplies and drive electronics.  Their small heat output will be satisfactorily removed by convection into the ambient air and by conduction into their mounting hardware.

DESI will use two full-height racks of computer and network equipment, to be located in the computer room.  There are now three half-height racks (on casters) of computers serving the current science instruments.  These computers will no longer be used in the DESI era and will be removed prior to or at the start of DESI equipment installation.  The net increase in heat loading resulting from DESI installation will therefore be quite modest.  The air conditioning unit in the computer room is currently loaded fairly lightly, and it has more than adequate capacity to extract the additional heat load.  

\paragraph{Electrical System Modifications} 

DESI will be locating a great deal of new equipment distributed among various locations at the Mayall telescope, including prime focus; the Cassegrain cage; the spectrograph room (large Coud\'{e} room); the Mayall computer room; and the Mayall control room annex.  Table~\ref{tab:elecbudget} shows predicted DESI electrical power loads at their respective locations at the Mayall.  

\begin{table}
\centering
\caption{Estimated DESI equipment peak electrical loads and locations (reference DESI-0483).}
\begin{tabular}{lllll}
\hline
Subsystem & ICD Document & Location & Peak Power (W) \\
\hline
Focal Plate & 389 & Prime Focus & 16,196 \\
ADC Rotator & n/a & Prime Focus & 100 \\
Calibration Lamps & 470 & Top Ring & 610 \\
Hexapod & 621 & Cassegrain Cage & 1,300 \\
Fiber View Camera & 388 & Cassegrain Cage & 50 \\
Spectrograph Shack & 470 & Coud\'{e} Room & 27,400 \\
Spectrograph Equipment & 470 & Coud\'{e} Room & 2,500 \\
Computer Equipment & 473 & Computer Room & 23,500 \\
\hline
Total & & & 61,602 \\
\end{tabular}
\label{tab:elecbudget}
\end{table}

Table~\ref{tab:elecbudget} and Table~\ref{tab:powerbudget} are not directly comparable.  The electrical service must be sized to handle the maximum peak load possible, plus a suitable margin, so peak loads are shown in Table~\ref{tab:elecbudget}.  However, during DESI operations only average heat loads must be extracted, so average loads are shown in Table~\ref{tab:powerbudget}. Also, the peak electrical loads in Table~\ref{tab:elecbudget} include some equipment that will not be running at all except during installation or major maintenance periods; one example is the turbo-molecular vacuum pump for the spectrograph cryostats in the Coud\'{e} room.  The electrical service must be sized to provide for these loads, but they will add no heat to the room during operations. On the other hand, some items will either operate continuously ({it e.g.,} the computers) or at duty factors approaching 100\% during DESI operations ({\it e.g.,} the Fiber View Camera), so these items have the same values in both tables.  

There are currently one clean 15 amp 110~VAC circuit and one dirty 15 amp 110~VAC circuit running to prime focus.  It will not be possible to run new electrical wiring all the way there before DESI installation, because DESI installation will involve replacing the entire top ring and top end assembly of the telescope.  Also, there is currently not enough capacity available in the clean power distribution system on the main dome floor to serve DESI's loads on the telescope. 
It is clear that some expansion of electrical services will also be required to and within the Coud\'{e} room.  The current service to that room is quite minimal, and the new loads there will be large.  The computer room may require some additional power capacity or outlets, although likely quite minor and available without expanding the capacity of the electrical panels serving those areas.  NOAO has supplied the anticipated DESI loads to its consulting engineering firm and obtained a design of the system expansions needed, including design of sufficient modern UPS capacity for the expected clean power loads.  NOAO will begin work on implementing the expansions in April 2016 via an outside contractor, and the work is required to be completed by the end of the 2016 Mayall Summer Shutdown.
  
It is assumed that equipment supplied by international collaborators will be specified to run on U.S.-standard AC frequency and voltages.

\paragraph{Telescope Modifications} 

Aside from the upper ring and spider, DESI requires two mechanical changes to the existing hardware on the telescope.  The first is at the baffle extending up from the center hole in the primary mirror.  The upper portion of the baffle and the heavy weldment designed to hold an M3 mirror for the Coud\'{e} light path (no longer used) vignette the wide DESI field of view.  They will be removed during the process of DESI installation, but removal is not necessary until then.  This assembly also provides the contact point for the petals of the primary mirror cover.  With the upper baffle gone, the cover petals will require extensions to cover the circular area left open by removal of the existing baffle.  NOAO currently plans to design and fabricate new cover petals that are long enough to cover the entire primary mirror including the central hole (formerly filled by the upper baffle).  These new cover petals will be installed during the DESI installation process, but they will require design and fabrication before DESI installation begins.

DESI also requires a new mechanical subsystem on the telescope, the e-chain cable wraps for carrying the DESI optical fibers.  One set of e-chains goes around the outside of the East Declination bearing, at a large enough radius to put it completely outside of the Declination hard-stop blocks mounted on the East side of the center section.  The other set of e-chains goes in front of the South bearing journal to accommodate motion about the Hour Angle axis.  The e-chains themselves are commercial off-the-shelf parts (IGUS part number 2828.01.200).  NOAO will design and fabricate the hardware for attaching these e-chains to the telescope structure along with the other supports or sliding surfaces that will be needed, beginning in October 2015.  Fixed support brackets for the optical fibers will be designed to support the fiber cables along the upper truss en route to the Declination cable wrap.  Fixed brackets will also be designed and built for the run along the wide elliptical tube between the ``horseshoe'' and the South bearing journal to support the cables en route to the Hour Angle cable wrap.  All of the cable wraps and support hardware are planned for installation immediately following the ProtoDESI testing campaign.  Installing these cable wraps about a year prior to the full DESI installation will allow plenty of time for testing and adjusting the wraps to enable smooth operation of the telescope with these new mechanical loads.
  
In addition to the mechanical changes, the Mayall Telescope Control System (TCS) software will require an interface to the comprehensive DESI Instrument Control System (ICS) that will control the DESI hardware and sequence the many complex operations required during observing.  An Interface Control Document (ICD) defining this interface has been drafted and approved by both NOAO and OSU, which is responsible for writing the ICS software. This ICD, DESI-0473, defines the protocols that will be used for passing commands and queried from the ICS, and for the responses from the TCS. It also includes examples of the commands and queries, with properly formatted responses.  NOAO's preparations for DESI include coding the TCS modules required to implement this interface, and testing them with simulation packages that represent the ICS.  This TCS interface development will in fact be required for the operation of the ProtoDESI test hardware, so the software work was completed by the end of calendar year 2015.  It was tested extensively and successfully during two nights of telescope engineering time January 21 and 22, 2016.  The interface is now ready for the arrival of ProtoDESI, expected in August 2016.

\paragraph{Spectrograph Room Modifications} 
\label{parag:spectroroom}


The large Coud\'{e} room will require substantial modifications to turn it into a space suitable for the DESI hardware.  First and most obviously, an environmental enclosure (the spectrograph shack) will have to be built to enclose the spectrograph stacks.  Figure~\ref{fig:ShackConcept} shows the approximate location of the spectrograph enclosure (``shack'') in the Coud\'{e} room.  The design of the shack is the responsibility of OSU under DESI WBS 1.6, but the services it will need must be supplied by NOAO.  This enclosure must maintain the spectrographs at $20\pm{2}\degree${C} in the presence of much larger diurnal and seasonal swings in the Coud\'{e} room as a whole.  In addition, there will be requirements on humidity, both within the spectrographs and (probably) inside the environmental enclosure.  The environmental system will need to address both of these.  It must be large enough to enclose the spectrograph stacks in their racks as well as the fiber cables and junction boxes after they come through the wall, and all the wiring, data cables, gas piping and cold liquid distribution needed for the cryostats.  inside the spectrograph shack, the working space must be large enough for minor work {\it in situ} on the spectrographs and for access with  suitable carts for removal of a cryostat or spectrograph for major servicing.  Access to the shack will be via doors or removable panels on the shack.

Second, the large Coud\'{e} room will also be the best place to locate any other operational equipment that must be near the spectrographs, the telescope or the dome chamber, but not out in the dome environment directly.  This category includes a water chiller for cooling the cryostat closed-cycle refrigerators (the LPT coolers), and the air dryer for supplying dry compressed air to the spectrographs.  It may also include the pump for the cooling loop that circulates dome-ambient temperature liquid to the Focal Plane Assembly.  In addition, there will be considerable support equipment -- including one or more vacuum pumps, a laminar flow bench, work benches, electronic test equipment, and computers -- that will be used in the Coud\'{e} room during spectrograph installation and maintenance periods.  All this operational and support equipment will need adequate electric power and network connections, and the operational equipment will also require cooling; these services must be planned carefully into the early stages of the process of upgrading the Coud\'{e} room.

\paragraph{Mayall Control Room Annex} 

The present Mayall control room cannot accommodate more than 5 people comfortably.  Given the complexity of DESI and its numerous interactions with the telescope and dome environment, some multiple of that number of people will be needed at the telescope during the commissioning and 
survey validation
phases.  Even after regular operations commence, periodic engineering nights will be needed, which again will likely require a larger number of people than can work comfortably in the existing space.  Also, provision needs to be made for participation by remote collaborators throughout the commissioning, 
survey validation,
and operations phases.  This means providing for convenient videoconferencing capability into the control room space, with screens, cameras, and microphones positioned so that remote collaborators can be fully engaged in the control room work.  The layout of the present control room does not permit placement of these items in locations where easy communication with everyone is practical.

The best option for providing additional working space is the conversion of a little-used lounge area on the U Floor, four floors below the control room.  If properly arranged and equipped, this room can easily provide working space for 15 or more people, as well as remote participants via dedicated video (\eg, skype) screens and cameras.  It is already within the climate-controlled part of the building, and it is also not far from the computer room, so providing the necessary network access would be straightforward.  It would require only repainting, provision of adequate electrical and network services, modernized lighting, and appropriate furniture and computer equipment.  During the final design phase, a layout for the room was developed with an eye toward promoting successful nighttime interactions as well as making the space suitable for as many night workers as possible, as well as collaborators at remote locations.  The lounge area furnishings can easily be relocated to another part of the building.  

The existing control room will not be taken completely out of service, but during the DESI era it will no longer be used for nighttime operations.  Having the telescope operator co-located with the DESI commissioning and science teams will be ideal for efficient nighttime work, encouraging rapid and efficient communication.  The TCS software and guide cameras will all be operated over Ethernet, so other than provision of an emergency stop switch there will be no hard-wired connections required between the telescope systems and the U Floor area.  The old control panel with its mechanical Selsyn dials will remain in place on the C Floor and continue to be used for movement of the telescope required by daytime engineering activities.
 
\paragraph{Calibration System} 

\begin{figure}
\centering
\includegraphics[height=2.5in]{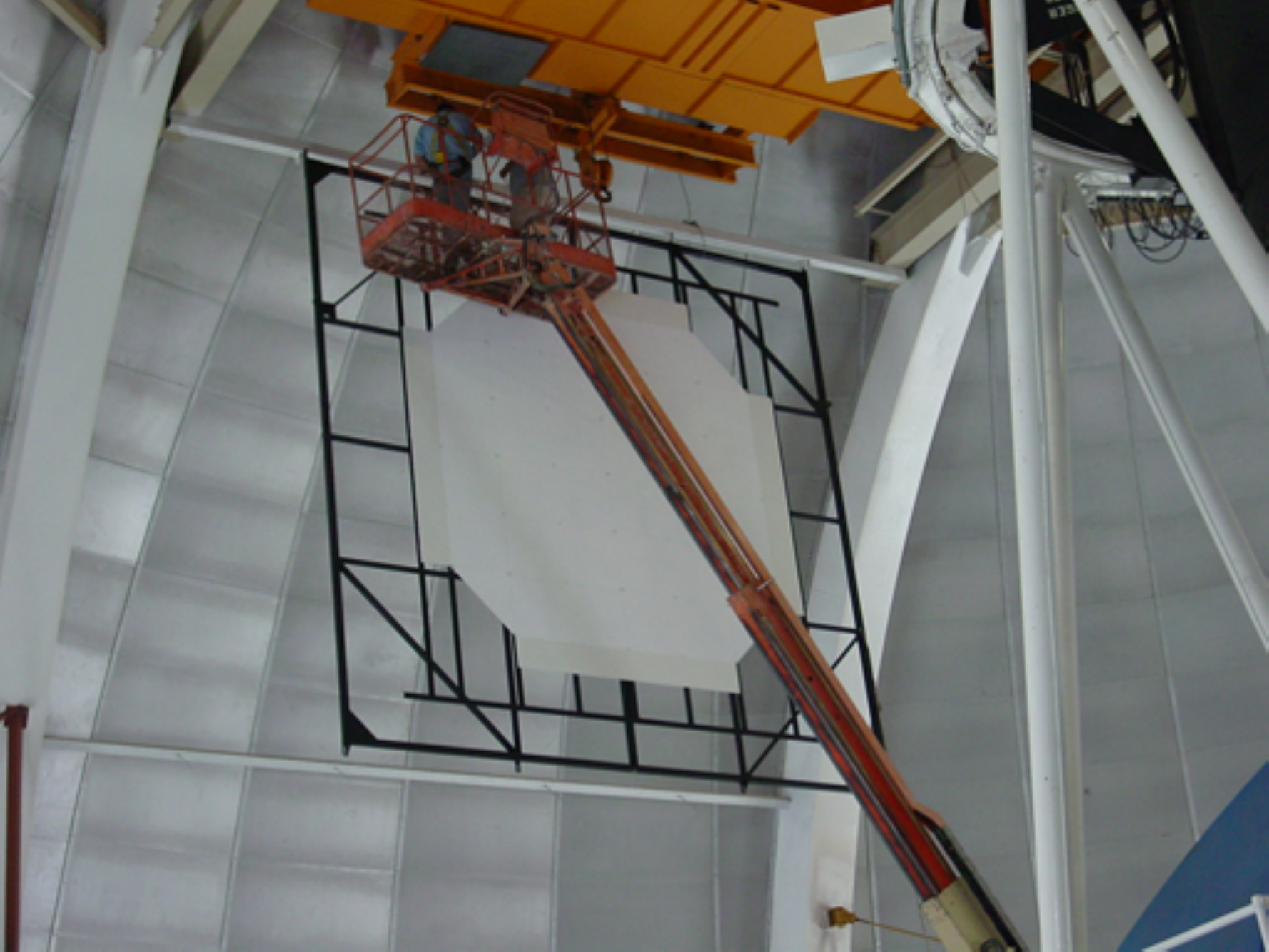}
\caption{Installing the existing reflection screen used for daytime flat fielding.  The screen is being mounted to the inside of the dome, between the arch girders and just below the attachment point for the dome crane rails.  Note the use of a long-reach boomlift. For safety reasons, the telescope drives must be locked out and tagged out during the entire operation.}
\label{fig:MayallInstallationMetalWhiteSpot}
\end{figure}

DESI will require systems for calibrating detector response (\ie, flat fielding), wavelength mapping, and line spread function determination.  These systems will require light sources of significant power, projection systems inside the dome, and reflecting screens for uniformly scattering the projected light into the telescope beam.  Dome-mounted equipment must be robust enough to withstand wind buffeting, making installation correspondingly complex and time-consuming.  As with DECam, the projectors per se must be mounted on the telescope top ring, so those items will be installed as part of the overall DESI installation.  The current design of the calibration system calls for the light sources to be co-located on the top ring with the projection optics.  The existing reflecting screen mounted on the inside of the Mayall dome is not large enough to cover the DESI field of view.  However, the frame (see  Figure~\ref{fig:MayallInstallationMetalWhiteSpot}) is large enough. NOAO will add an additional ring of panels, similar to those seen in the outer ring in Figure~\ref{fig:MayallInstallationMetalWhiteSpot}, to provide sufficient area.  Installation of these panels will be require specialized equipment (\eg, a rented boomlift) and while underway will prevent all other work on the telescope or use of the dome cranes.   

\paragraph{Computer Room} 

DESI will come with a large number of computer systems for instrument control, observation planning, image processing, and data buffering (among other uses).  A careful survey of the Mayall's current computer room was made during the final design phase.  Some of the computers currently there (three half-height racks, all on casters) can be retired and removed, as they support instruments that will not be used during the DESI era; such removal of course cannot begin until the telescope shuts down for DESI installation.  With those racks out, there will be ample floor space for the anticipated two full-height racks coming with DESI.  As noted above, substantial changes to the electric power system will not be needed, but some additional outlets may be required.  Such changes will be complete by the end of the 2016 Mayall Summer shutdown.  

\paragraph{Ground Floor Garage} 

A protected space needs to be prepared within the high-bay area of the Mayall garage for re-assembling the DESI corrector, checking its alignment, and assembling the corrector and hexapod into the new Prime Focus cage frame.  This space requires a clean tent with about 20 square meters of floor area and about 5 meters of working height and the tent must be slightly positively pressurized to keep out dust.  Also, the garage door will be locked out of operation and sealed against dust infiltration during the period of corrector assembly.  This tent will be constructed in place, much like a similar facility at the Blanco telescope used to re-assemble the DECam corrector.  The working space also requires a crane with a hook height of at least 6 meters and a working capacity of at least 7 metric tons.  This crane may be an A-frame crane purchased commercially, or it may be provided by the smaller of KPNO's two mobile cranes.  This mobile crane is routinely used within the Mayall garage area and has more than the required load capacity and hook height.  For use in this application, provisions will need to be made for providing outside air to the crane engine and venting its exhaust to the outside of the building. Such ``breathing'' can easily be provided by flexible metal ductwork running through temporary penetrations in the outside wall.  The use of the mobile crane is substantially cheaper and easier to implement, so that option has been selected as the baseline plan, pending final approval by the Systems Engineer and the Corrector team at UCL.


\subsubsection{Installation Planning}

Detailed planning of tasks must be carried out before DESI installation begins.  Tasks need to be in the proper order and carefully coordinated to avoid conflicts over personnel resources, physical resources such as cranes, and working space.  All resources required for each task need to be clearly identified and their availability confirmed.  Work should not begin until this plan has been completed and thoroughly reviewed.

The first draft of such a plan has been done, although it is still very clearly a work in progress.  First, there is a widely varying level of detail possible at this stage of the DESI project.  While there is reasonable detail on steps that Mayall personnel are familiar with (\eg, removing the primary mirror), there is considerably less detail on other steps where designs of DESI equipment and its handling features are less advanced.  Of necessity, time requirements and resource needs are still purely notional for installing things that are still in design development.  The draft plan is currently undergoing almost continuous revision and refinement as designs develop, and as various alternative approaches are considered and refined. This plan will continue to evolve throughout the design and fabrication phases of the DESI project, in order to keep up with the developing project design details.  

At the same time, it will be mandatory to develop safety plans for the entire installation process.  These will include procedures for every task identified in the plan; every procedure will be accompanied by a Job Hazard Analysis (JHA).  These procedures and the associated JHAs will develop along with the plan, seeing revisions as needed to keep them current with the state of the DESI equipment designs.  Safety planning will also encompass identification of all critical lifts to be carried out during installation, and writing of Critical Lift Plans for every critical lift.  Finally, the safety planning will cover the training required for all personnel, ranging from training in general KPNO conditions and policies to specific training required for individual tasks.  Like the installation plan itself, the basic safety documents have been drafted, but they are expected to be revised throughout the DESI design and construction period.  Task-specific safety documents and lift plans will continue to be written and revised up to the point of readiness to begin installation.  

NOAO intends to call a formal installation readiness review prior to the shutdown of the telescope.  This external review will cover the final detailed installation plan and the safety plans, as well as the status of the preparation work described above.

\subsubsection{Installation Equipment Preparation}

\begin{figure}[!tb]
\centering
\includegraphics[height=3.0in]{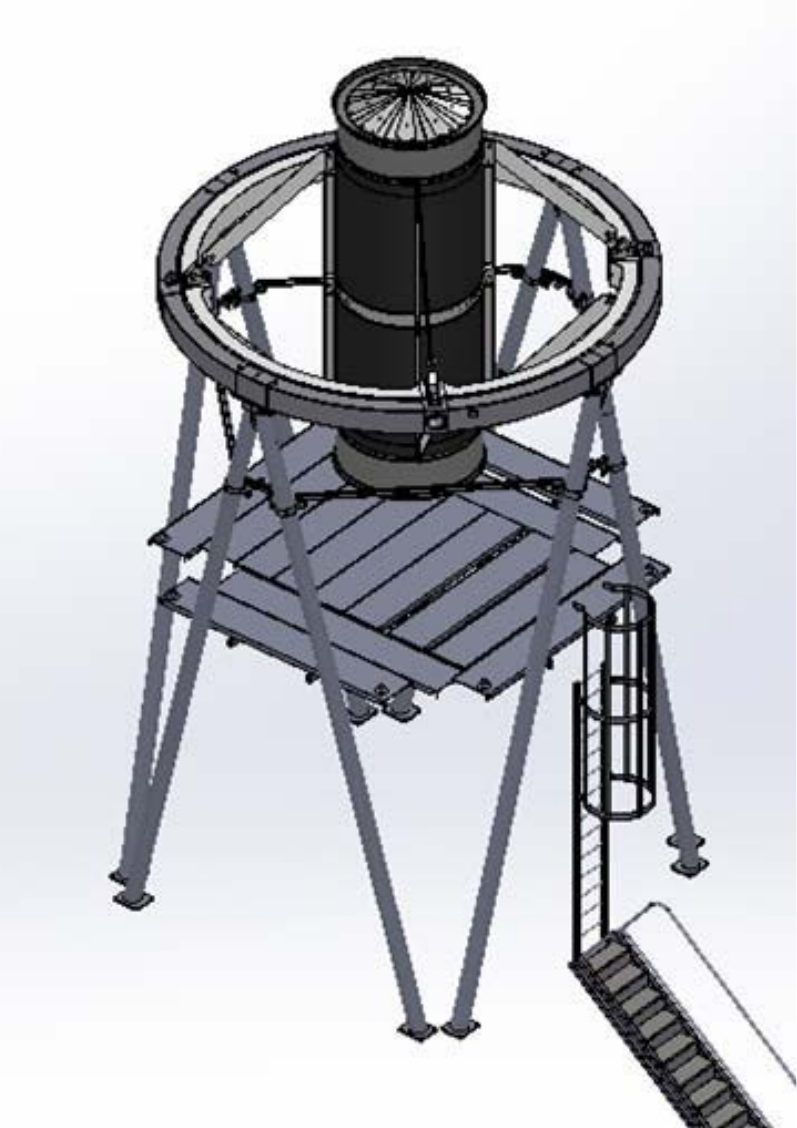}
\caption{Extract from preliminary design for the lower of two Zenith work platforms, showing the stairs-plus-ladder access arrangement.  Also shown, above the lower platform, are the braces for securing the upper Serrurier truss members when the top ring system is removed.}
\label{fig:LowerPlatformWithLadder}
\end{figure}

\begin{figure}[!bt]
\centering
\includegraphics[height=3.0in]{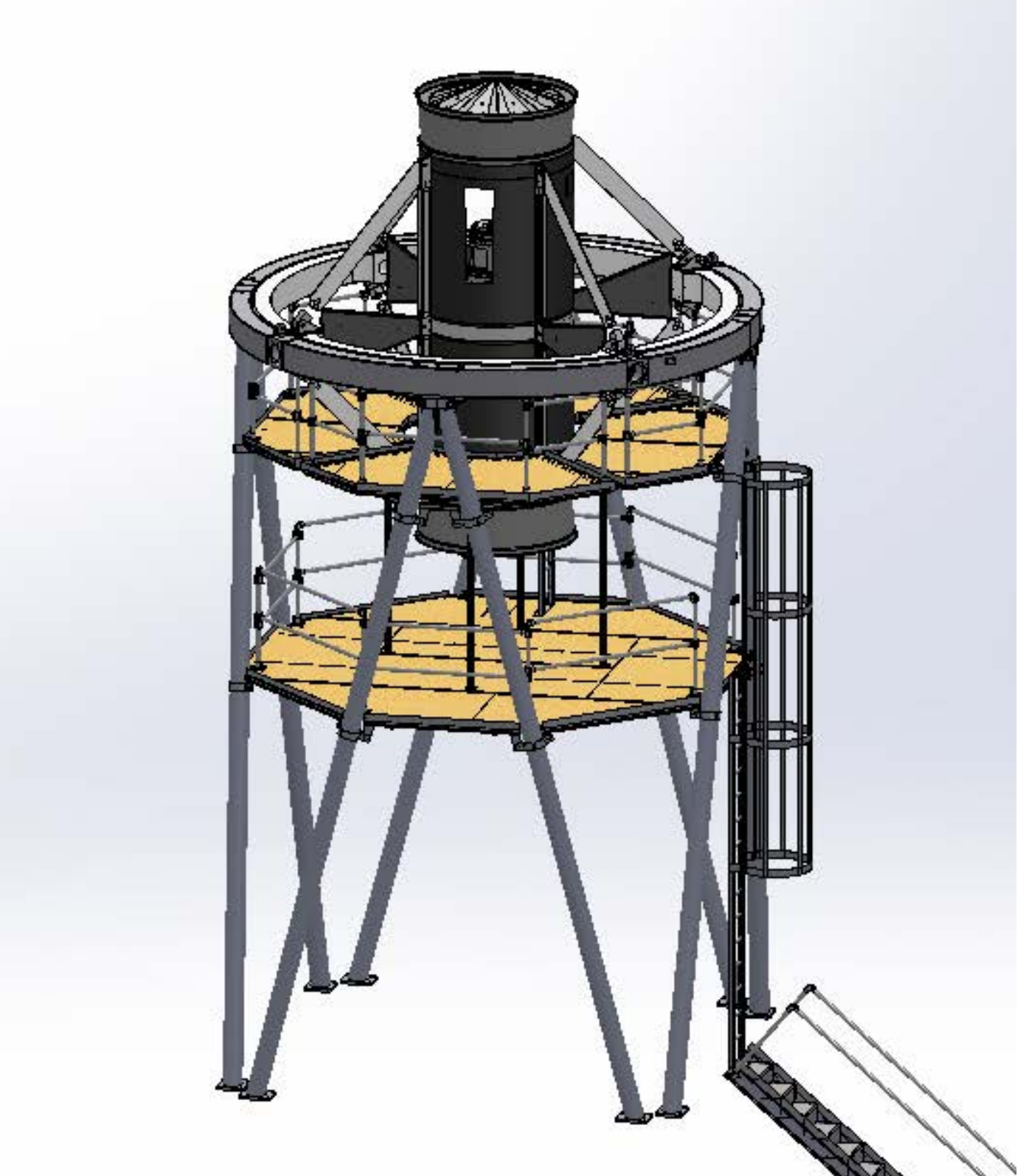}
\caption{Preliminary design for both upper and lower top end platforms, showing access ladder and stairs, and showing upper platform supported by upper truss braces.}
\label{fig:CompleteTopEndPlatforms}
\end{figure}

In connection with the installation planning, NOAO is also designing and will construct two items of specialized equipment needed during DESI installation.  The first item is the set of work platforms needed for the removal of the existing top and flip rings and the installation of the new DESI top end.  These platforms will allow safe work spaces for personnel while the telescope is locked in a zenith-pointing orientation. Their design is based on that for the similar platforms used at CTIO during DECam installation, although some changes are planned such as allowing access via a combination of stairs and a ladder (see Figure~\ref{fig:LowerPlatformWithLadder}).  The second item is a set of braces for securing the upper Serrurier truss members in position prior to removal of the existing top end.  These braces (also depicted in Figure~\ref{fig:LowerPlatformWithLadder}) must be secure enough to hold these truss members in position with nothing attached to the truss tops, and yet must be easy enough to install and remove that those steps can be carried out by personnel working either on the zenith work platforms or in a mobile boomlift.  The completed preliminary design incorporates these braces into the upper Zenith work platform, using the braces as part of the support for the upper work platform.  Figure~\ref{fig:CompleteTopEndPlatforms} shows the complete design with all features visible. 

\subsection{Installation}
\label{sec:installationintro}

DESI Installation will require that the telescope be inoperable for more than a year.  Obviously, once this process begins, the telescope will no longer be usable for nighttime observing.  The launch of this phase thus represents a major milestone for the project, as well as a major change in the scientific capabilities available to the US astronomical community.  The schedule and sign-off procedures for beginning the DESI installation will be worked out by the funding agencies.  The next subsections describe the current understanding of the DESI installation steps.  These will be further developed during the detailed design phase of the project.

\begin{figure}[!bt]
\centering
\includegraphics[height=2.5in]{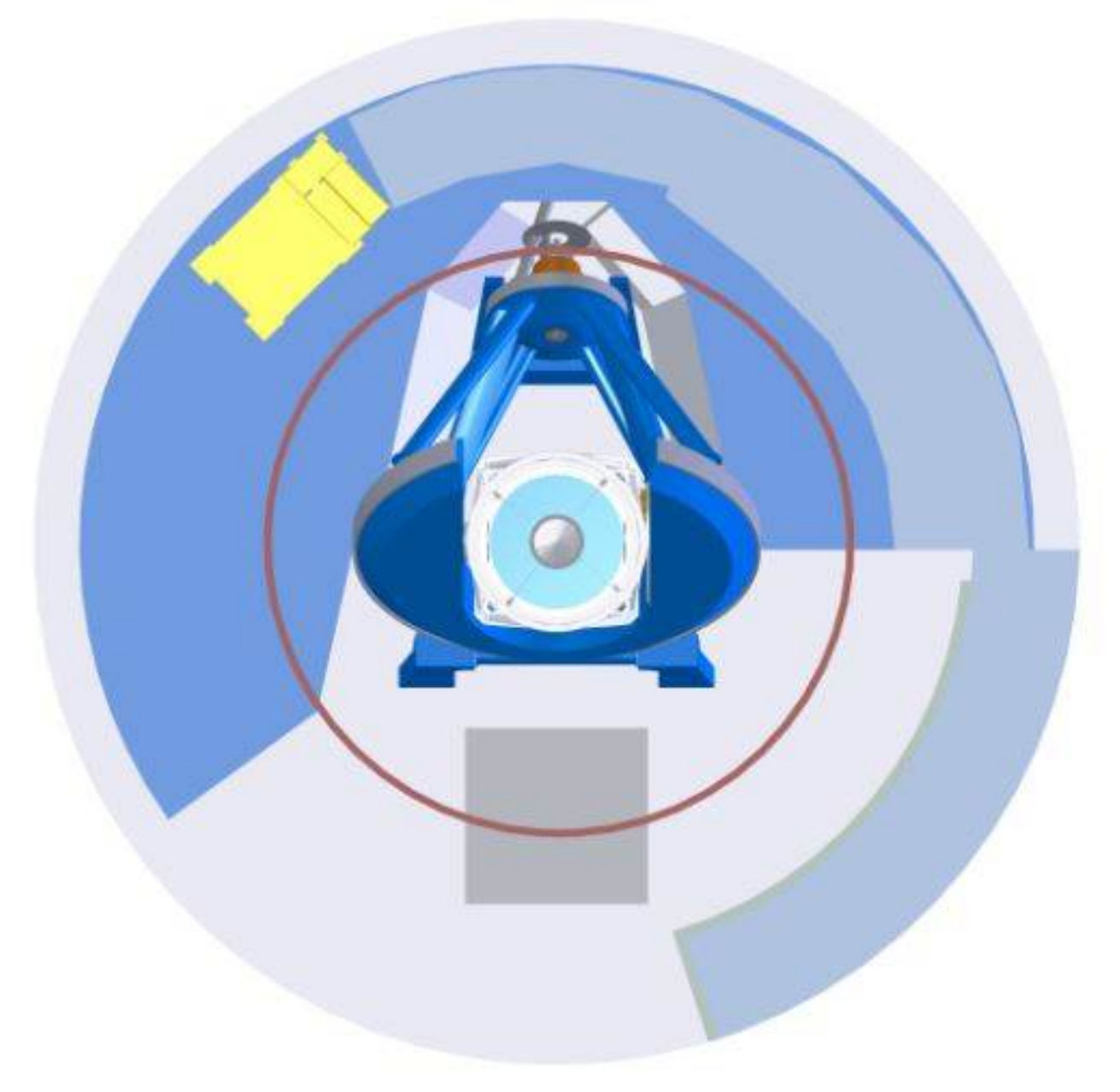}
\caption{Plan view of telescope and dome with the crane limits marked (orange circle).  North is down in this view.}
\label{fig:PlanViewTelescopeWithCraneLimits}
\end{figure}

Throughout this discussion, it is important to remember some of the physical parameters of the dome and the dome working environment that will drive much of the installation process.  First, the 5-ton and 50-ton dome cranes have a limited reach from the dome center.  Figure~\ref{fig:PlanViewTelescopeWithCraneLimits} shows the range of motion of those cranes; the outer edge of the orange circle represents the limit of reach for the 5-ton crane, and the inner edge represents the limit of reach for the 50-ton crane.  

Second, the dome cranes travel radially within the circle shown in Figure~\ref{fig:PlanViewTelescopeWithCraneLimits} along their rails, but they are moved azimuthally only by rotating the dome itself.  The dome rotation drive system has no brakes, depending rather on rolling resistance and mass to bring the dome to a stop.  It can be positioned reliably to within about a foot by an experienced operator, but the dome cranes cannot be relied upon for precise horizontal positioning of loads.  The lack of brakes also means that the dome, along with the cranes and any loads suspended from them, can be moved by the wind under strong wind conditions.  It will thus be necessary to specify and observe wind speed limits in planning and carrying out all critical lifts involving the dome cranes.  

Third, the dome hatch (the gray square North of the telescope in Figure~\ref{fig:PlanViewTelescopeWithCraneLimits}) has a cover that is divided roughly in thirds.  The center hatch section can be removed without disturbing the outer two sections, but removing either of the outer sections requires first removing the center section.  All three sections are heavy steel-and-concrete constructions designed to support the weight of the primary mirror and its cell.  Removing or replacing them requires use of the 50-ton dome crane itself, and so the hatch sections must be stored within crane range, usually adjacent to the hatch on one side, whenever hatch opening is required.

\subsubsection{Removal of Existing Telescope Hardware}

Just prior to the start of DESI installation, any instruments not currently mounted on the telescope, along with their handling carts or other support equipment, will be removed from the Mayall building and stored elsewhere on Kitt Peak or in Tucson.  As soon as the telescope is taken out of service, all computers that will not be needed during DESI operations will be shut down and removed from the computer room.  The computer removal does not interfere in any way with the telescope mechanical work underway in the dome, so it can proceed independently.    

\paragraph{Preparation of Top End for Removal} 

\begin{figure}[htb]
\centering
\includegraphics[height=2.5in]{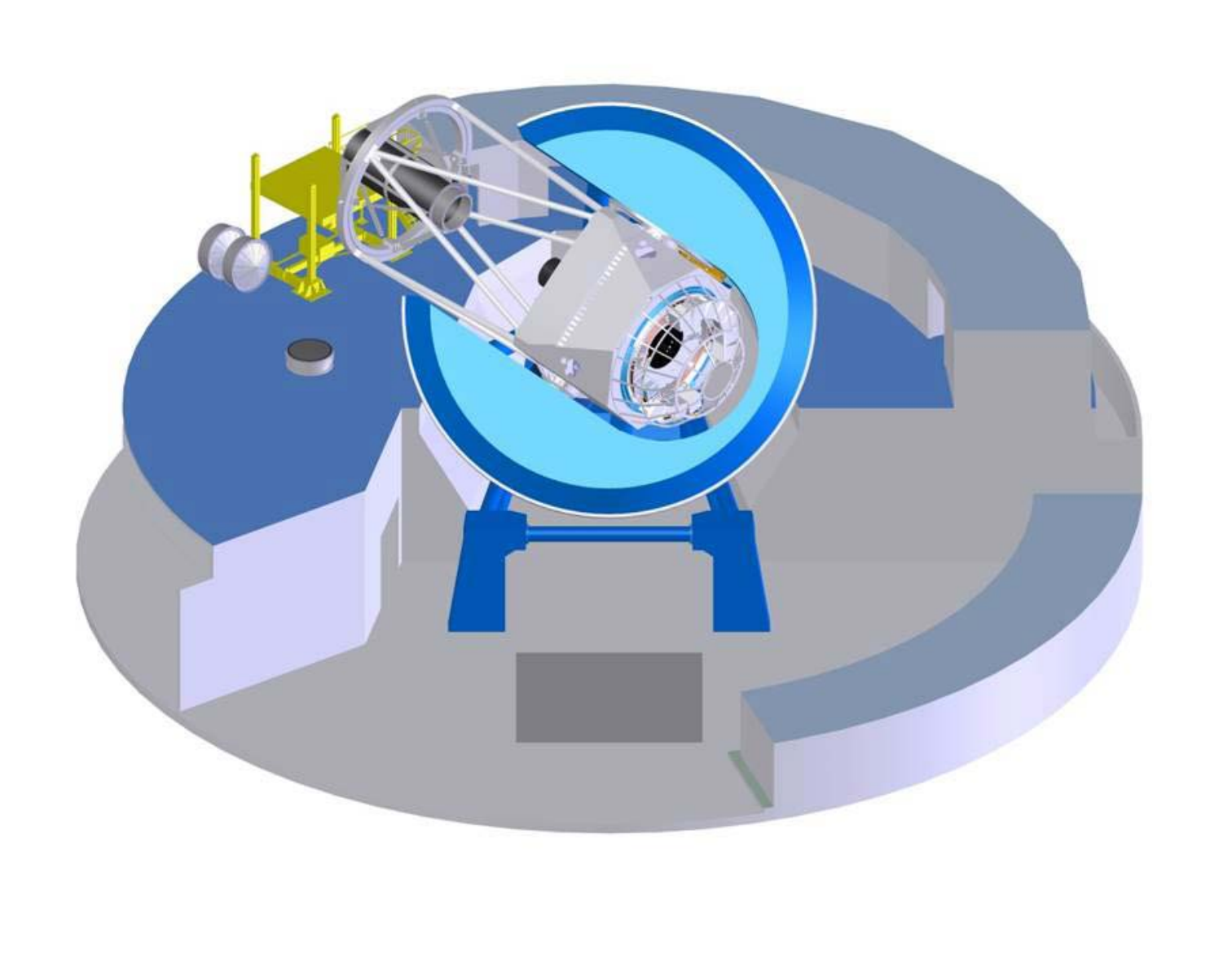}
\caption{Telescope positioned at the Southeast Annex.  The secondary mirror and its covers have just been removed and are laid off to the side.}
\label{fig:GobleSBTelescopeStrippingAtSEAnnex}
\end{figure}

DESI requires the complete removal and replacement of the existing telescope top end, including the upper ring, flip ring, spider vanes, and prime focus assembly.  In preparation for this removal, first the telescope will be positioned at the South limit.  The Mosaic prime focus imager will be taken off, boxed, and removed from the building for storage.  The prime focus corrector optics will be removed as well.  Suitable counterweights will be installed in place.  The telescope will then be moved to the Southeast Annex and physically restrained.  The \begin{math} f/8 \end{math} M2 will be removed, placed in a protective crate, and then removed from the building to external storage.  The M2 covers will be removed, and the M2 counterweight installed to maintain telescope balance (Figure~\ref{fig:GobleSBTelescopeStrippingAtSEAnnex}). All existing service connections -- power, data, air, vacuum, \etc\ -- will be removed from the prime focus cage and stripped back to points below the top rings.  Clamps and mounting hardware for the Zenith work platforms and for the upper truss bracing will be installed on the truss.      

\paragraph{Bottom End Disassembly} 

\begin{figure}[htb]
\centering
\includegraphics[height=2.5in]{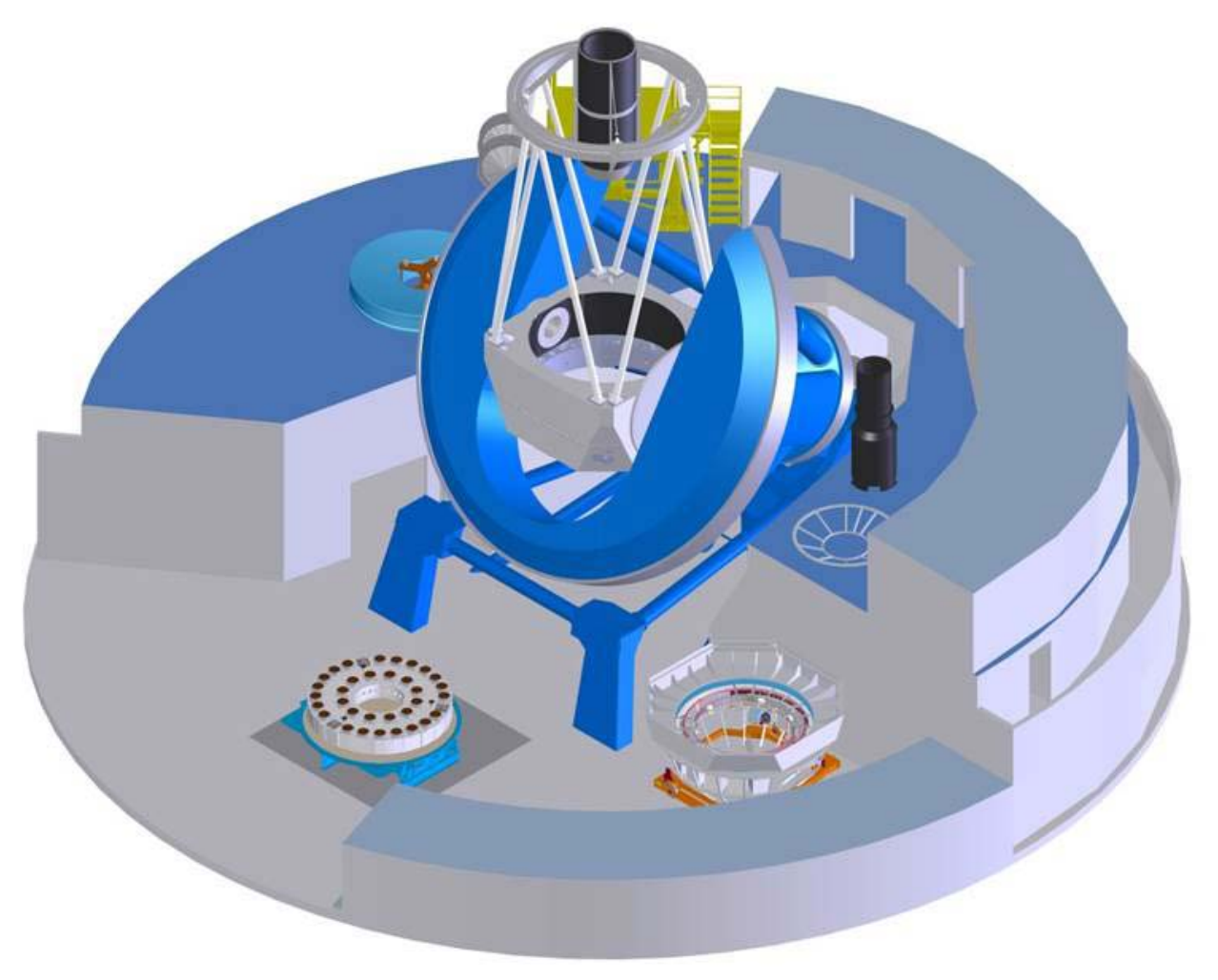}
\caption{The Mayall telescope with the bottom end fully disassembled and the components staged for removal from the dome.}
\label{fig:GobleSBBottomEndDisassembly3}
\end{figure}

The telescope will be rebalanced and returned to the zenith-pointing position, locked out and tagged out, and physically restrained.  All needed handling equipment will be staged and prepared for M1 removal.  The currently mounted Cassegrain instrument will be removed, placed in its handling cart, and removed from the building for long-term storage on Kitt Peak or in Tucson.  The Cassegrain equipment cage and lower shell will all be removed from the telescope bottom end.  The existing central baffle assembly will be removed and lifted off the telescope.  All removed items will be staged on the main dome floor (the M floor) or the console room level (the C floor).  Next the M1 and its cell will be removed from the telescope, and the M1 removed from the cell.  The M1 will be lowered to the ground floor and stored in the aluminizing chamber for cleanliness and safety. The M1 cell, the lower shell and the Cassegrain cage will then be re-installed on the telescope to make space in the dome for DESI installation.  The central baffle assembly will be removed from the building along with any other items still remaining.  Figure \ref{fig:GobleSBBottomEndDisassembly3} shows the telescope with the bottom end completely disassembled and all components staged for removal from the dome.  The primary mirror is on the C floor, above the Coude room.  The central baffle assembly is on the C floor above the old FTS lab.  The mirror cell is on the hatch.  Staging these items in these locations is part of the normal process used for primary mirror re-aluminizations throughout the service life of the Mayall.  

\paragraph{Top End Removal} 

\begin{figure}[!t]
\centering

\begin{minipage}[t]{0.45\textwidth}
\centering
\includegraphics[width=\textwidth]{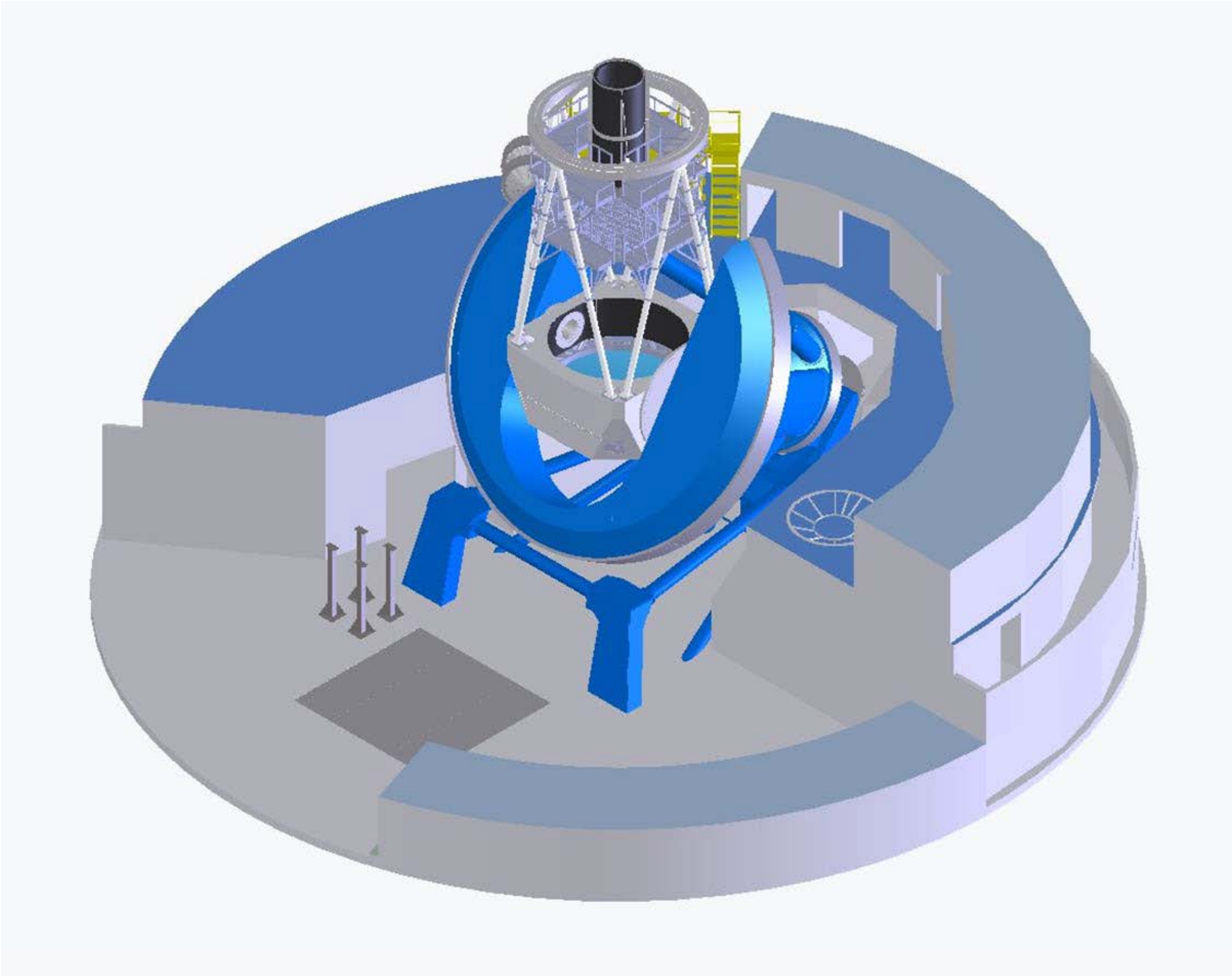}
\caption{Bracing and work platforms in place for removal of top end.}
\label{fig:WorkPlatformsAtZenith}
\end{minipage}

\end{figure}

Equipment needed for disassembling the top end will be brought up to the M floor and placed in position. The Zenith work platforms and the incorporated bracing for the Upper Serrurier Truss will be securely installed on the upper truss using a rented articulated boomlift.  Figure~\ref{fig:WorkPlatformsAtZenith} shows the bracing and platforms in place.  The bolts and pins securing the outer ring to the upper truss ends will be loosened slightly to ensure that they can be removed.  Then, a rented mobile crane will be parked outside the dome, the dome aperture (slit) opened, and the crane rigged to the upper rings.  The bolts and pins will be completely removed, and the old top end will be lifted off the truss and out through the open dome slit, to be lowered directly onto a waiting flat-bed truck trailer.  The top end structure can then be driven away to another location on Kitt Peak (near the McMath-Pierce Solar Telescope) where it will be wrapped weather tight and stored.  The telescope is now ready for DESI installation.

%
%


\subsubsection{Installation of DESI Hardware}

The task of installing DESI includes work both on and off the telescope.  One of the major off-telescope tasks (assembly of the prime focus cage with its major components) must be scheduled critically to be ready in time for installation on the telescope.  Other off-telescope tasks, such as installation of the spectrographs, may (depending on the readiness dates of the spectrographs) be paced by the on-telescope work in that there are very limited opportunities to bring spectrographs into the building while the on-telescope work is going on.  Completing the installation as quickly as practical will require very careful scheduling and detailed work coordination to ensure safety, efficiency, and quality of work.

\paragraph{Initial Assembly and Installation of New Telescope Top End}
\label{sec:initialtopend}

\begin{figure}[!t]
\centering

\begin{minipage}[t]{0.45\textwidth}
\centering
\includegraphics[width=\textwidth]{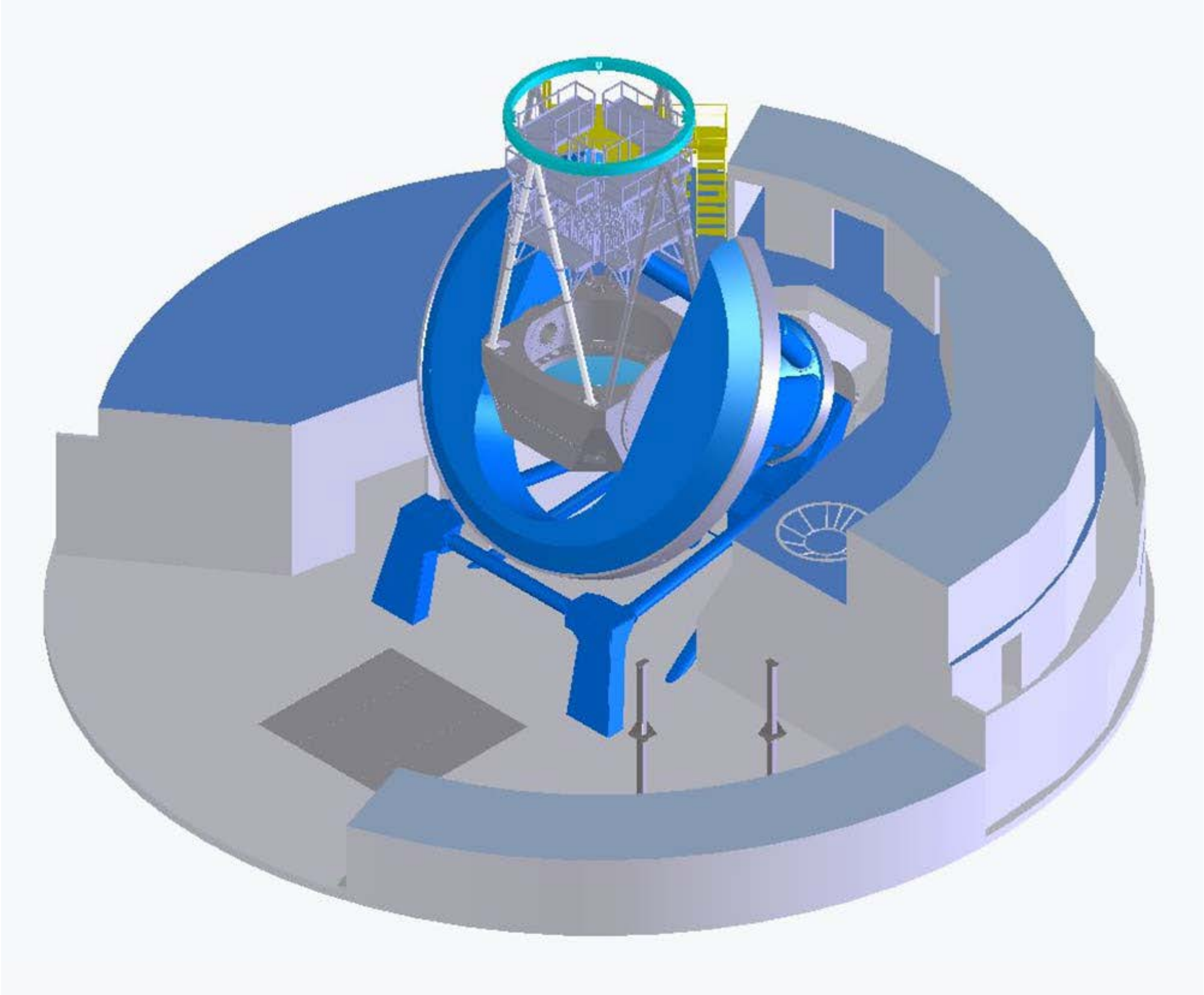}
\caption{Test fit of the DESI top ring to the truss.}
\label{fig:TestFitDESITopRing}
\end{minipage}
\begin{minipage}[t]{0.45\textwidth}
\centering
\includegraphics[width=\textwidth]{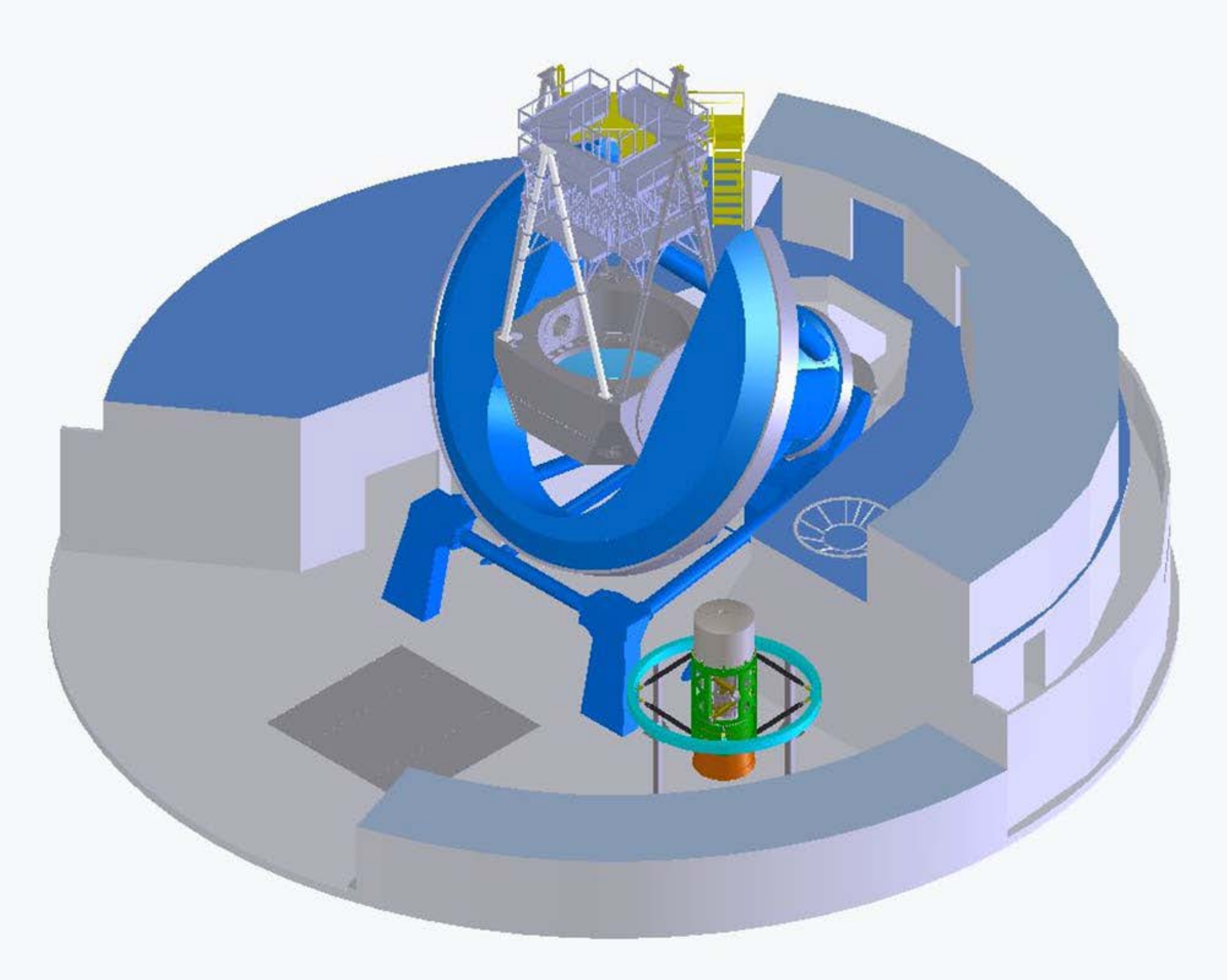}
\caption{DESI prime focus cage installed in top ring on dome floor.}
\label{fig:AssembleDESITopEndFloor}
\end{minipage}

\end{figure}

First, prior to any on-telescope assembly work, the DESI prime focus cage assembly, consisting of the cage structure, the corrector barrel, the hexapod and a dummy weight in place of the focal plate assembly (the focal plate adapter, the focal plate and the fiber bundle), will be assembled in the garage high-bay on the ground floor of the Mayall building.   Assembling the cage inside the Mayall building will avoid the need for truck transport of the assembly, but the assembly work there will interfere with the ability to remove old Mayall components from the building or bring DESI components into the building.  Ingress or egress for components too large to fit in the personnel elevators will have to be scheduled around the cage assembly work.  

Inside the dome, the new DESI top ring will be lifted to the top of the telescope truss and test-fit to the truss members (Figure~\ref{fig:TestFitDESITopRing}).  After a satisfactory mechanical alignment of the ring to the upper truss is obtained, the ring will be punch-marked in place, for later drilling and tapping of the screw and pin holes for attaching the ring to the upper truss.  Then, the empty top ring will be lowered back to the M floor and placed on work stands near but not on top of the main hatch in the floor.  The ring will then be drilled for the bolts and pins needed for truss attachment, and the bolt holes will be tapped.

\begin{figure}[!t]
\centering
\includegraphics[height=2.5in]{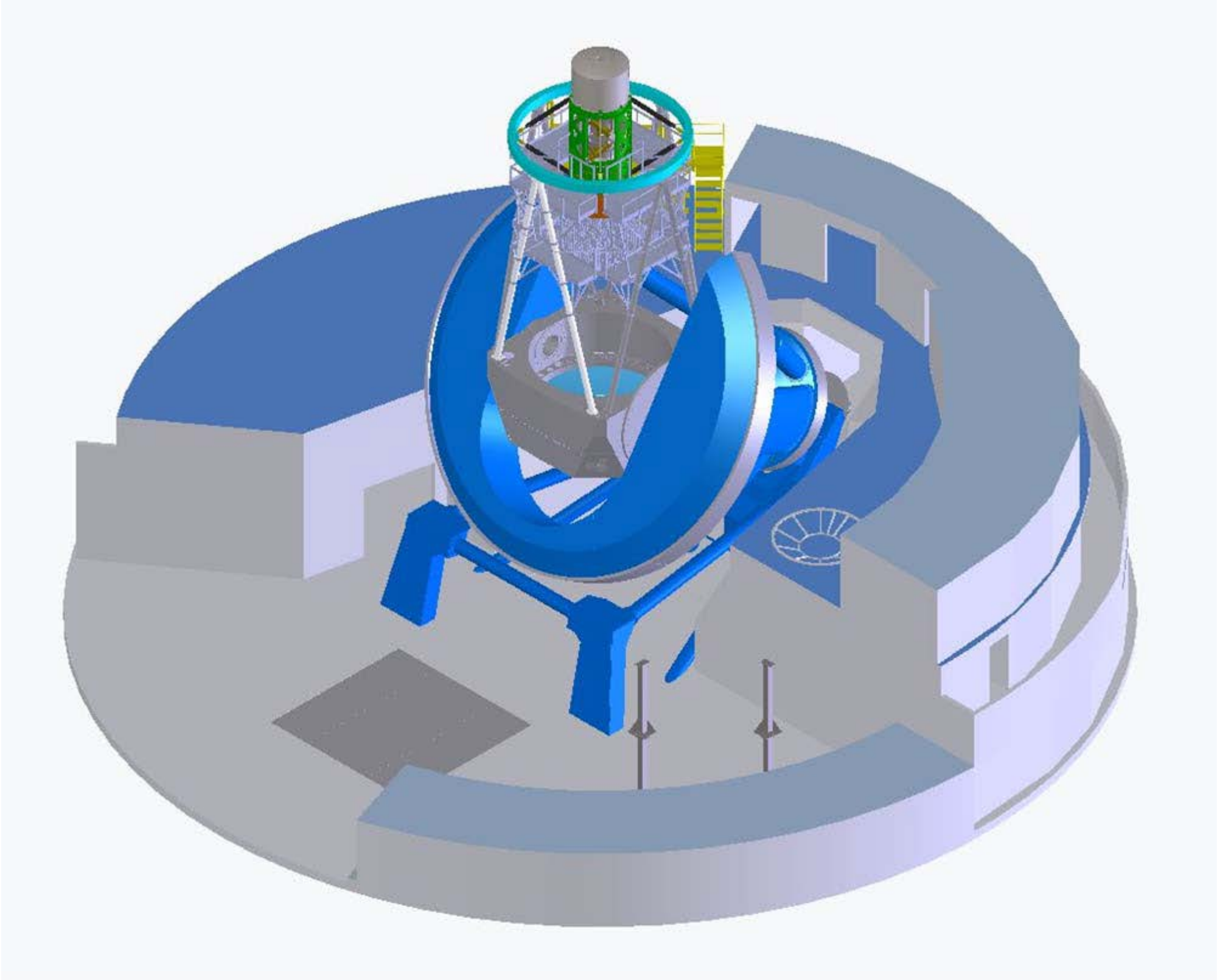}
\caption{DESI top end assembly installed on telescope.}
\label{fig:InstallDESITopEnd}
\end{figure}

The prime focus cage assembly will then be lifted into the dome and placed in position at the center of the top ring, and the spider vanes will be installed to connect the two (Figure~\ref{fig:AssembleDESITopEndFloor}).  Then, the cables and other connections for the ADC and the Focal Plate services will be extended from the cage to the top ring and dressed in place for later connection.  The lamps and projection optics for the calibration system will also be installed at this time, and their wiring dressed in place to the ring for later connection.  Finally, the completed top end assembly will be lifted to the top of the telescope truss and bolted in place atop the truss members (Figure~\ref{fig:InstallDESITopEnd}).  All connections for the Focal Plate, the ADC, the calibration lamps, and the hexapod will then be completed between the unterminated service lines at the top of the truss and the dressed connections from the hardware on the top end assembly.  Basic ``on-off'' tests will be performed at this point to verify operability of both the ADC and calibration lamps.  The hexapod cannot be tested until its power supplies are installed (next section).  The top end assembly fixtures will be removed from the main floor to make room for the telescope bottom end re-assembly.

\paragraph{Re-Assembly of Telescope Bottom End}
\label{sec:rebuildbottom}

Fine alignment of the Prime Focus cage requires the laser tracker monuments (retro-reflector target mounts) that are permanently bonded to the edge of the primary mirror.  Re-installation of the primary mirror must therefore take place before fine alignment.  

With the upper portion of the central baffle assembly removed, the primary mirror cover petals will be replaced with revised versions that extend to the center, fully covering the entire mirror and Fiber View Camera.  The next steps are to recoat the primary mirror with aluminum, remove the lower shell and mirror cell, lift the mirror up to the M floor, re-assemble the mirror in its cell, and re-install the cell and lower shell on the telescope.  Then the Cassegrain equipment cage will be re-installed along with the mirror support control equipment it contains. The hexapod power supplies will be installed in the Cassegrain cage, and the power cables from the supplies to the hexapod will be dressed along the outside of the center section and up one of the upper truss members along the West side of the truss where they will be out of the way of the optical fiber installation later.  The hexapod can now be tested for operability; the hexapod specialists (from FNAL) can carry out these tests in parallel with the following steps  in the Cassegrain cage.  Following that, the Fiber View Camera will be installed at the back of the mirror cell in place of the now-removed Cassegrain rotator, connected and tested, and aligned with visible targets on the corrector barrel or prime focus cage.  The counterweights on the bottom end of the telescope will be adjusted as necessary to provide a proper balance with the weight of the new DESI top end.  The new DESI top end is required to weigh no more than the weight of the existing top end that is being removed.  However, the bottom end of the telescope will be substantially lighter (more equipment is being removed from the bottom end than is being added) so additional counterweights will be required at Cassegrain.  This stage ends with a coarsely balanced telescope that is ready to be moved but still physically restrained for safety.

\paragraph{Top End Mechanical Alignment}
\label{sec:measurealign}

Next, with the hexapod adjusted so the barrel is at the midpoint of its range of motion relative to the cage, detailed measurements will be taken of the position the Prime Focus corrector barrel relative to the laser tracker target monuments on M1.  These monuments were previously used to locate the optical axis of M1 relative to its outer circumference.  Using a laser tracker mounted on a temporary fixture above the center section, it will be possible to survey simultaneously four of the six M1 retro-reflector locations and four retro-reflectors fixed to the corrector barrel.  These measurements will be used to adjust the tension on the spider vanes, if necessary, to position the new cage assembly so as to align the corrector barrel as closely as possible to the M1 optical axis with the hexapod at the middle of its travel range.

\paragraph{DESI Focal Plate and Fiber Cable Installation}
\label{sec:fibercable}

\begin{figure}[!b]
\centering
\includegraphics[height=2.5in]{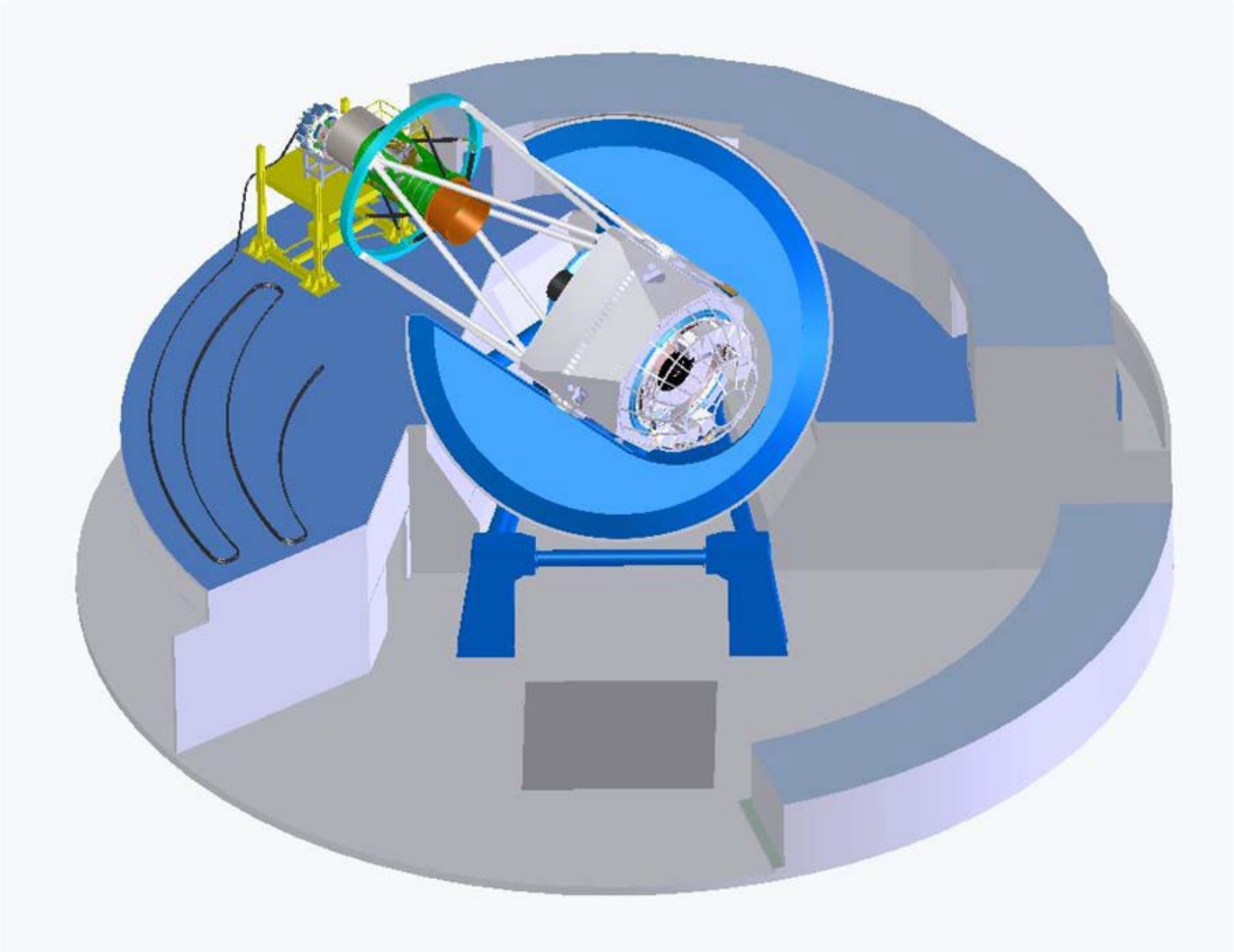}
\caption{Installing the DESI focal plate assembly at the Southeast Annex.  Note fiber cables laid out on C floor to North of SE Annex platform.}
\label{fig:InstallDESIFocalPlane}
\end{figure}

After the bottom end equipment is re-installed, the physical restraints on the telescope will be removed, and the lock-outs and tag-outs removed from the telescope drive equipment.  The telescope will then be moved over to position the top end above the Southeast Annex work platform.   See Figure~\ref{fig:InstallDESIFocalPlane}. The telescope will be locked-out and tagged-out in this position, and the routine physical restraints applied to prevent movement.   The dummy weight representing the focal plate assembly will be removed from the top of the corrector barrel, and the real focal plate installed in its place.  Doing this work at the Southeast Annex allows the roughly 49.5-m long fiber cables attached to the focal plate to be laid out in an orderly fashion along the floor while the focal plate is installed.  It also enables installation of the delicate focal plate at a location where use of a custom lifting fixture becomes straightforward, and it provides easy access for the personnel carrying out the task. After the focal plate and its adapter are secured to the barrel, the connections to the required services will be secured to the focal plate using the cables and fluid connections previously installed on the top end.  Then the optical fiber cables will be mounted along the spider vanes to the top ring using the mounting hardware previously installed during the top end assembly.  Finally the focal plate enclosure will be secured over the entire assembly.

\begin{figure}[!t]
\centering
\includegraphics[height=2.5in]{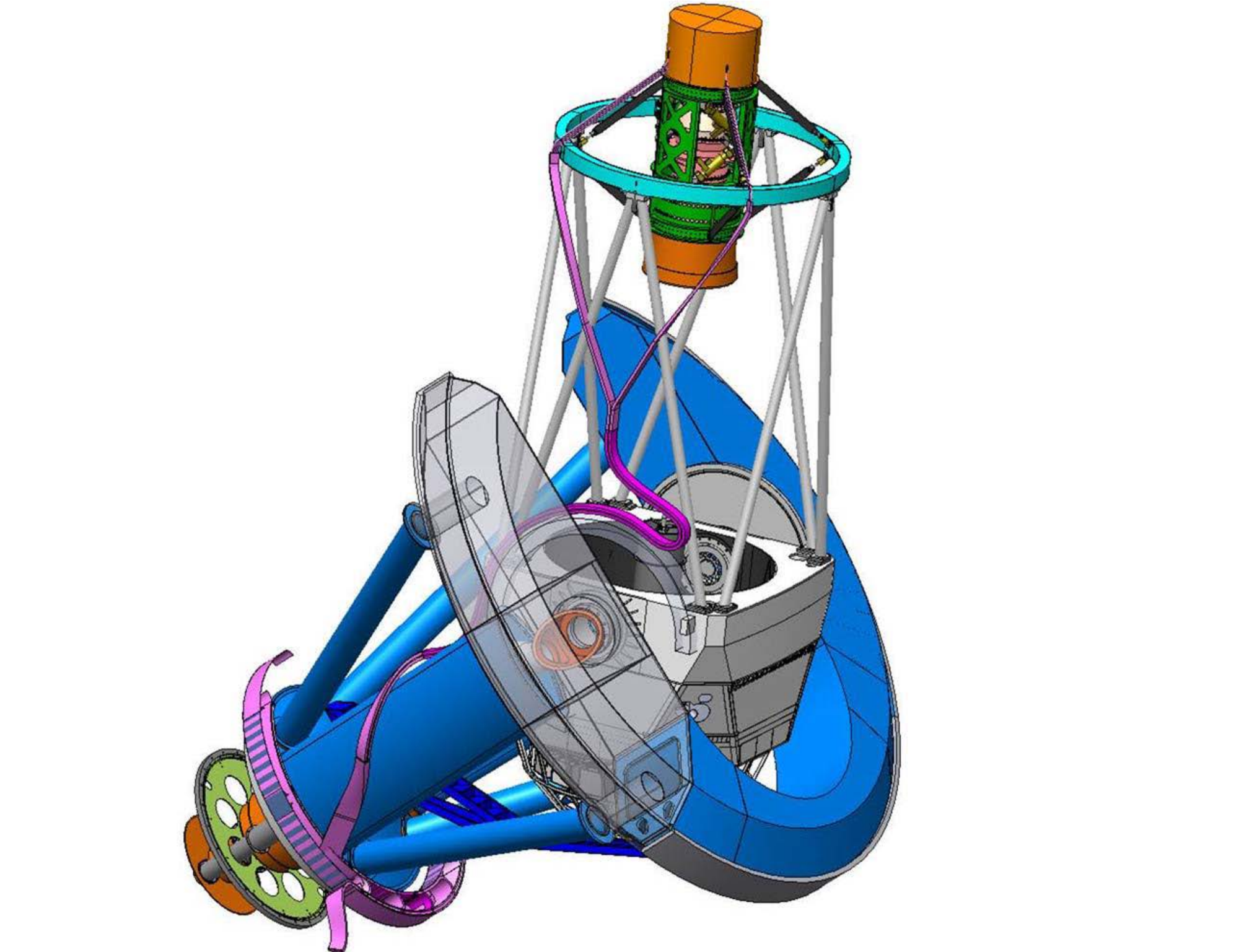}
\caption{Fiber routing on telescope, illustrating locations and space requirements for the Dec and HA cable wraps.}
\label{fig:FiberRoutingOnTelescope}
\end{figure}

\begin{figure}[!h]
\centering
\includegraphics[height=2.5in]{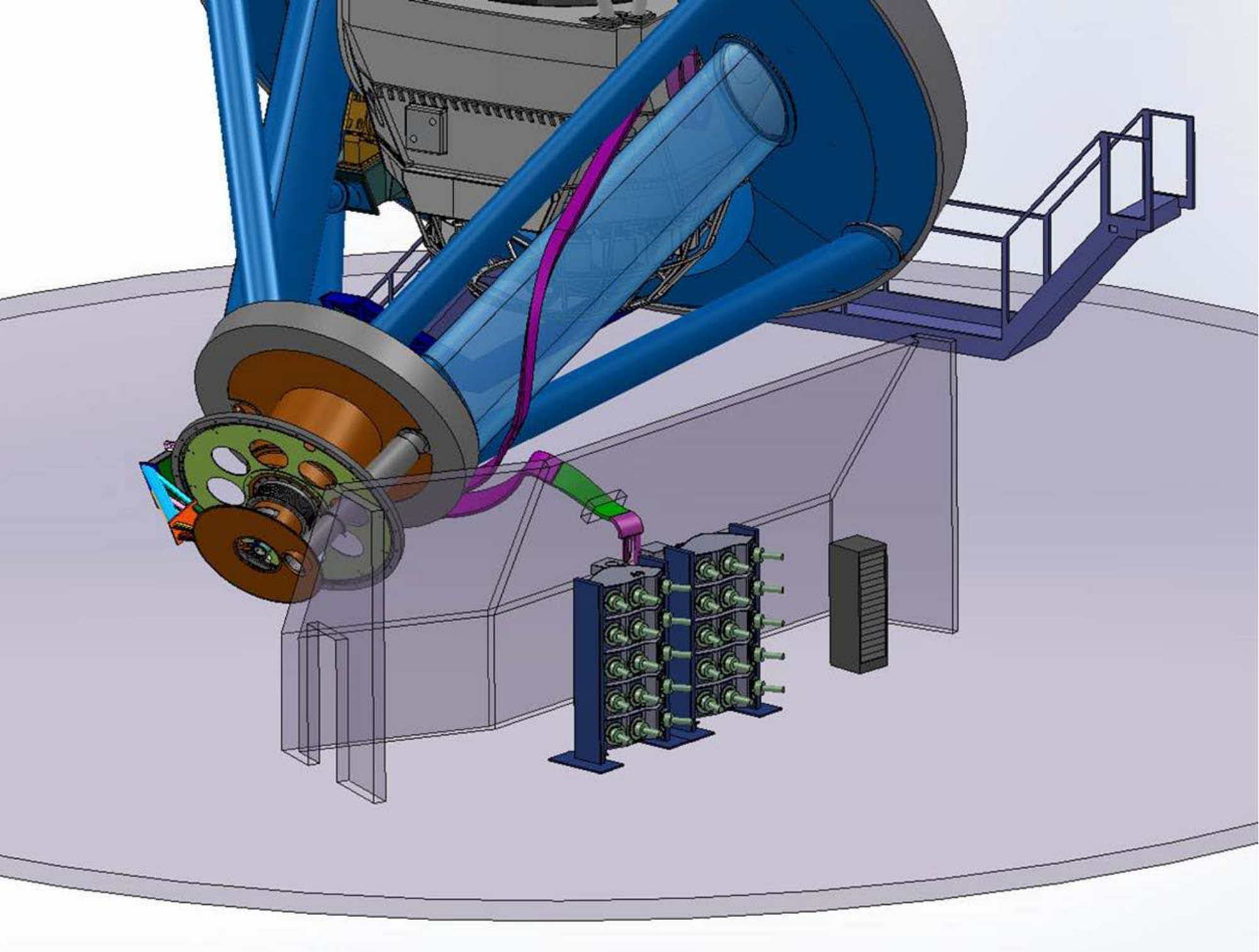}
\caption{Fiber routing through wall of Coud\'{e} room to spectrographs.  Note that spectrographs will be arranged in five short stacks and not the two tall stacks shown here.}
\label{fig:FiberRoutingThroughWall}
\end{figure}

The telescope will now be released from the physical restraint at the Southeast Annex, and slowly returned to a Zenith-pointing orientation.  The optical fiber cables will be carefully handled and paid out during this movement to protect them and position them appropriately for dressing along the outside of the upper truss.  Once the telescope is at Zenith, locked out, tagged out, and restrained in place, the fiber cables can be secured in previously installed brackets to bring them down the outside of the truss and centered over the East Declination bearing.  The cables will then be laid, one at a time, into the Declination cable wraps.  The next step sees the cables secured in brackets along the large elliptical tube on the East side of the moving Polar Axis structure, down to the Hour Angle bearing cable wrap just North of the South Bearing journal.  The cables will then be laid, one at a time, into the Hour Angle bearing cable wrap.  They will exit that cable wrap and be passed, again one at a time, through an opening in the adjacent wall separating the telescope chamber from the Coud\'{e} room.  Figure~\ref{fig:FiberRoutingOnTelescope} shows the routing of the fiber cables on the telescope with the telescope pointing roughly to zenith for ease of illustration.  Figure~\ref{fig:FiberRoutingThroughWall} shows the routing through the wall of the Coud\'{e} room.

\paragraph{Telescope Motion Tests and Final Alignment Check}
\label{sec:movetest}

With the telescope ready to move, the telescope balance will be checked and bottom-end counterweights changed as needed, but it is expected that only small adjustments will be needed at this point.  The telescope will then be put through a series of motion tests to confirm that the new loads in the cable wraps can be handled within nominal drive current limits.  The telescope motion servo systems will be re-tuned to optimize performance with the new weight, mechanical moments, and cable wrap loads on the telescope.  If there are any problems with meeting slew and settle time requirements, they will be resolved before proceeding further.  After that, there will be a final alignment check performed with the laser tracker system at zenith and various other orientations to ensure that movement of the top end position with changing position is within the expected limits.   This data may also be used to build a preliminary look-up table for positioning the hexapod following telescope slews.   

\paragraph{Spectrographs}
\label{sec:spectrographs}

The ten DESI spectrographs will be installed on their racks in an environmentally controlled chamber (the ``spectrograph shack'') in the Coud\'{e} room.  The work of installing and testing the spectrographs thus does not directly interfere with the flow of work in the main dome.  It is expected that the spectrograph components will fit within the personnel elevators, obviating the need for use of the dome crane to lift them, but requiring more extensive re-assembly on the main floor.  If so, this will reduce the conflict with in-dome work.  The re-assembly will take place in a temporary clean tent set up for this purpose in the old FTS lab on the main floor, to the West of the telescope (see Figure \ref{fig:IsoViewTelescopeInChamberAnnotated}).  After assembly, they can be rolled on carts across the main floor to the Coud\'{e} room, requiring only a brief pause in the on-telescope work.  Once the spectrographs are in the Coud\'{e} room, the rest of the work on them can proceed without interfering with the work on the telescope, provided that sufficient personnel are available for both tasks, and provided that both groups observe safety restrictions on access to the others' work areas.  

However, if the spectrographs must be lifted up from the ground floor through the M floor hatch using the 5-ton dome crane, then the work in the telescope dome must pause for about one day each time a batch of spectrographs is to be delivered to allow time for opening the center section of the main hatch, carrying out the lift, and closing the main hatch.  This longer interruption will require very careful coordination with the on-telescope work.    

Once the spectrographs are in the Coud\'{e} room, they must be installed in the racks, and connected to power, cooling, and data services.  The slit heads will arrive attached to the long fiber cables, so they will be available only after the fiber cables are installed on the telescope and through the cable wraps, but initial testing can be done with a test slit head prior to fiber installation.  Once the slit heads have reached the spectrograph rack, they can be installed on as many spectrographs as are ready to receive them at the time.  The remaining slit heads will be secured temporarily for their protection to keep them out of the way while the rest of the spectrographs are installed and connected.  

Timing of the spectrograph installation process will largely depend on when the spectrographs are completed and delivered from the integrator to Kitt Peak.  Any units that arrive before work starts on the telescope could be installed immediately as long as their environmental enclosure is ready, as work in the Coud\'{e} room will not interfere with night-time science observing prior to the telescope shutdown.  Similarly, spectrographs delivered after the in-dome work is complete can be installed immediately as well.  Timing of spectrograph receipt is not critical to the installation schedule, as on-sky testing can begin with less than all spectrographs (one is sufficient for initial testing).

\paragraph{Computer Equipment}
\label{sec:computers}

The Mayall computer room is located on the mezzanine level one floor above the M floor.  The computer room has its own elevator access, and work in that room can proceed independently of the on-telescope installation.  The computer racks and equipment can be brought up in the elevators without tying up the dome crane or main floor hatch, further separating the computer installation from the main installation.  The only requirement on this work is that it be fully completed in time to support the initial testing of other hardware as the other hardware is installed.  This of course assumes that any clearing of unneeded equipment, rearrangement of the room and expansion of the power or cooling services are all finished before the first DESI computers arrive.

\paragraph{Support to Other DESI Subsystem Tests}
\label{sec:supportothersubsystemtests}

After installation is complete, NOAO staff will be needed to support functional checkout of DESI subsystems.  First, the mechanisms in the prime focus cage -- the ADC drives and hexapods -- will need to be tested, their performance measured, and any required adjustments or repairs carried out.  In the case of the hexapod, the confirming measurements will require use of the laser tracker system mounted temporarily above the top surface of the telescope center section.  Next, the focal plate system will require testing to make sure that the connections to the cooling fluid system are leakproof, to verify that the movable fibers and fixed fiducials illuminate properly and can be imaged with the Fiber View Camera, and finally that the movable fibers can be positioned accurately.  The GFA sensors will also need to be tested at this time to verify that they are sensitive to light and communicating properly with the control software.  Third, the slit heads will need to be connected to the spectrographs.  Finally, the fiber system and spectrographs will need testing together, both by back-illuminating the fibers from the spectrographs and by taking flat and arc-line calibration images using the dome screen.

\paragraph{Telescope Re-Verification}
\label{sec:telescopereverify}

As soon as the motion tests following fiber cable installation are complete, the telescope will be ready for re-verification through night-time observations.  These will consist primarily of gathering pointing data to update the telescope pointing model, checking the slew and settle times on-sky to verify the performance of the motion control servo loops, verifying pointing, measuring open-loop tracking performance, verifying guiding under ICS control, and gathering data on top end movement due to flexure to build any look-up tables for control of the DESI hexapod.  Tests will also be conducted to verify the proper performance of the interaction between the telescope, the DESI hexapod and the DESI guide-focus-alignment (GFA) sensors.  All these on-sky tests require use of the DESI GFA sensors for finding stars, centering stars, and making low-order measurements of the incoming wavefronts to build and refine the look-up tables for hexapod positioning as a function of telescope pointing.  

Some of these checks constitute verification that performance under ICS control post-installation matches pre-installation performance. A coarse grid of positions that spans the (HA, Dec) range of DESI Survey operations will be used. Slew and settle times will be verified for large (30 degree) and small (5 degree) slews. Pointing, open-loop tracking, and guiding will be checked using stars and the GFA sensors. The same tests can be performed using the GFA sensor with ProtoDESI to establish pre-installation performance. Pointing accuracy around the sky shall be verified after the pointing model is updated. Open loop tracking will be checked for random jitter and cumulative errors over a ~5 minute interval, using (i) sequences of short exposures and (ii) single long exposures with small nonsidereal drift rates to produce star trails in the axis normal to that under check. Guiding performance will be characterized by image growth of stellar PSFs for 1-20 minute exposures (the longest time matching the average Survey exposure time), and by the power spectrum of guider correction signals to capture small random jitter. Image quality around the sky will be checked with the GFAs, measuring stellar PSFs on a night of good seeing. This will be done without using the hexapod for tip-tilt correction, in accordance with the pre-installation test condition.

Once these steps are complete, the telescope is ready for system-level testing on-sky with the rest of the DESI instrument.

\subsection{Safety}

This safety description focuses on personnel and equipment safety throughout the installation, commissioning and operational phases of the Dark Energy Spectroscopic Instrument (DESI) project. The Plan addresses working conditions and procedures at the Kitt Peak site, as well as the management structure that positively impacts safety.

The DESI project is centrally managed but executed by several teams in various locations and with different funding sponsors. The DESI Installation and Commissioning Safety Plan covers these project phases while recognizing and relying on existing Environmental, Health and Safety Policies at Kitt Peak.

This section provides the plan description and references other project documents for additional details.

\subsubsection{Environmental, Health and Safety Plans}
\label{sec:safetyplans}

The detailed guidelines for safe working conditions are found in site specific Environmental, Health and Safety (EHS) Plans. Each DESI collaborating organization and contractor has established Safety Plans that will govern the specific activities at their own location. The draft safety plan for the Kitt Peak site can be found under DESI-1004.  NOAO expects and indeed requires that all staff, permanent to the location or visiting, will follow these local requirements during all work at Kitt Peak.  

\paragraph{Compliance with Regulations, Codes and Standards}

NOAO EHS plans comply with all applicable EHS regulations and requirements in Title 29 of U.S. Code of Federal Regulations (CFR).  This includes part 1910, ``OSHA Safety and Health Standards for General Industry'', and part 1926, ``Safety and Health Regulations for Construction'', 49 CFR Federal Motor Carrier Safety Administration, and 40 CFR Protection of Environment and others that may apply.  NOAO considers the above CFRs as a minimum standards. 

\paragraph{Environmental, Health and Safety Plan Responsibilities}

Work conducted on NOAO sites, by participating organizations, shall follow NOAO Safety Plans. These plans include identification of the local Safety Manager/Coordinator and lines of authority to which all personnel working on NOAO activities at that location shall adhere. NOAO Safety plans take the precedence over home institution regulations. 

\paragraph{EHS Plan Expectations}

NOAO has a set of guidelines, expected to be followed, from the local EHS plans while working at Kitt Peak or other NOAO sites. The Kitt Peak Safety Manager and the project management will work with the collaborating institutions to insure that local plans meet Project expectations. The Kitt Peak Safety Manager is responsible for the review of the local plans to see that expectations are met, that all elements (documentation) are available to the staff, and that issues of nomenclature and document titles are reconciled where the differences are identified. The following topics are expected to be addressed in local EHS plans:

\begin{itemize}
\item Safety Manager/Coordinator
\item Visitors and Access Policy
\item Personal Protection Equipment
\item Hazardous Material Management
\item Emergency Contacts and Procedures-Kitt Peak Emergency Manual
\item Utility Use and Servicing
\item Lock-out Processes
\item Housekeeping
\item Reporting and Records
\end{itemize}

\paragraph{Contractor Safety and Health Plans}

All NOAO Contractors are required to develop, submit, and implement a project and site-specific safety and health management plan. The plan must be extended to its own employees, project employees, other workers, and members of the public. The contractor is also required to ensure compliance with the plan. The details of the contractor requirements are subject to specific policies from NOAO.  Contractor requirements are expected to address the same information as NOAO EHS plans, as laid out in the preceding section.

For work contracted at NOAO specific locations the contractor's safety record and ongoing Safety Plan will be an important element in the selection criteria. All contractors are required to complete a Safety and Health Questionnaire for Contractors that will be used to evaluate potential contract employers. All general contractors shall require subcontractors to comply with the Contractors Safety and Health Specification. The Safety and Health Questionnaire for contractors will be provided to potential Contractors during the proposal documents to be completed. The questionnaire will be used by NOAO to assess contractor's safety performance and Plans and will be used during the selection process.

\paragraph{EHS Plan Availability to Personnel}

NOAO will maintain a Safety Plan Website that will include links to all applicable NOAO EHS plans and policies to facilitate convenient access to the information to everyone working on NOAO tasks. This is a particularly important for visiting staff who work at other institutions where access to the information is not routine.

\subsubsection{Safety Reviews and Hazard Analysis}
\label{sec:reviewanalysis}

NOAO will conduct periodic design and procedure reviews focused on compliance with safety regulations, this Plan and good practices. The frequency of the reviews will be commensurate with the stage of the project as defined in the Project Management Plan. Reviews will be conducted by safety professionals, the Project Management team and front line supervisors and will focus on specific subsystems of the observatory, the observatory as a whole, and the Safety Plan itself. Included in the safety reviews will be a periodic Hazard Assessment and Hazard Analysis of each subsystem.

\paragraph{Safety Reviews}

NOAO will conduct periodic design and procedure reviews focused on compliance to the Safety Plan and applicable standards. The frequency of the reviews will correspond with the stage of the project as defined in the Project Management Plan. The Kitt Peak Safety Manager with the support of other safety professionals and project leaders will lead the reviews. Reviews will focus on specific subsystems of the observatory, contracted efforts, the observatory as a whole, and the Safety Plan itself.  Included in the reviews will be the periodic Hazard Assessment and Hazard Analysis of each subsystem. The Contractor's hazard analysis will also be reviewed to confirm that the deliverables established in the contract documents are consistent.  The Project Manager(s) and the Kitt Peak Safety Manager will determine other safety reviews and their scope during the project period.

\paragraph{Job Hazard Analysis}

NOAO will utilize a formal job hazard analysis to identify, analyze, and determine the solution for the hazards throughout the project work being done at NOAO. This hazard resolution method for NOAO consists of the series of analytic steps summarized below:

\begin{itemize}
\item Define the physical and functional characteristics of the proposed project by employing the information available, and relating the interaction between people, procedures, equipment and the environment.
\item Identify known hazards related to all aspects of the DESI project work being done and determine their causes.
\item Assess the hazards to determine the severity and probability, and to recommend means for their elimination or control.
\item Implement corrective measures to eliminate or control the individual hazards, or accept the corresponding hazards.
\item Conduct follow-up analysis to determine the effectiveness of preventive measures, address new or unexpected hazards, and issue additional recommendations if necessary.
\end{itemize}

One example is already available.  NOAO has drafted and posted a procedure for installing ProtoDESI at the existing Mayall prime focus corrector, along with the Job Hazard Analysis for this procedure.  These documents can be found on the DESI DocDB system at DESI-1587 and DESI-1588, respectively.  

\subsubsection{Safety Management}
\label{sec:safetymanagement}

The NOAO Project Management team is responsible and held accountable for incorporating the Safety Plan policies, standards, rules, and principles into the project work. To emphasize the commitment to safety, a single NOAO-Kitt Peak Safety Manager will be named to manage, execute, and verify compliance to the Safety Plan. This safety professional reports directly to NOAO-Kitt Peak Director and is accountable for the implementation of the plan. Other team members from NOAO collaborating institutions and outside affiliations will assist the Safety Manager. For appropriate phases of the DESI installation and commissioning, Safety advisors/Project leads will be employed to provide front line review of work in progress. NOAO will also maintain a safety web site to communicate with the large distributed NOAO team and facilitate the dissemination of key information. 

\paragraph{Kitt Peak Safety Manager}

The Kitt Peak Safety Manager is responsible for the NOAO/Kitt Peak Safety Plan, and its implementation, and helps management ensure compliance. The Safety Manager reports directly to the Kitt Peak Director and works closely with NOAO subsystems managers, systems engineers and the safety managers/coordinators at each participating organization responsible for the implementation of local EHS Plans. The Safety Manager has the authority and responsibility to report EHS issues and to make recommendations to the Project Managers. The Kitt Peak Safety Manager encourages the project team to value safety and manage all hazards in their area of responsibility.

The Safety Manager, working with the institutional Safety Managers/Coordinators, will be actively involved in many of the aspects of the project. This group is responsible for
\begin{itemize}
\item Preparing the Safety Plan deliverable documents;
\item Supporting the Plan level interface with NOAO and others;
\item Developing and establishing safety design criteria and safety design requirements as needed;
\item Participating in hazard analysis if required;
\item Evaluating design changes for their impact to the safety, health or the environment;
\item Coordinate and verify adequate emergency response systems and procedures that will be used during the project execution;
\item Coordinate EHS activities needed during the project.
\end{itemize}

\paragraph{NOAO Associate Director for Kitt Peak}

NOAO's commitment to safety extends through all levels of management. The NOAO Associate Director for Kitt Peak, also known as the Kitt Peak director, and then NOAO-DESI project manager will implement the NOAO Installation and Commissioning Safety Plan and EHS objectives by:

\begin{itemize}
\item Confirming that the Safety Plan is established and integrated throughout NOAO.
\item Confirming that known hazards are identified, eliminated, or controlled within established Plan hazard acceptability parameters.
\item Reinforcing that NOAO operations are performed in accordance with applicable project safety requirements, and applicable governmental safety regulations.
\end{itemize}

\paragraph{NOAO DESI Project Manager}

The NOAO DESI Project Manager has the day-to-day responsibility for ensuring that EHS Management practices are incorporated into local activities of the DESI Project. The NOAO DESI Project Manager will implement the NOAO Installation and Commissioning Safety Plan in close coordination with the Kitt Peak Safety Manager to ensure all subsystem managers to keep EHS issues a priority in all aspects of the project. Specifically the project manager and subsystem manager will:

\begin{itemize}
\item Make EHS considerations a part of all planning processes by identifying known hazards, determining what standards apply, implementing control, determining the competencies required to do the work safely, and finally assuring that each of these elements are in place before work is authorized to proceed.
\item Focus on safe accomplishment of the operation, understanding assignments and carrying out the core safety management functions correctly and efficiently. These principles are dependent upon both management commitment and employees/individual involvement and accountability.
\item Ensure the Safety Plan is established and integrated throughout the DESI project.
\item Ensure that known hazards are identified, eliminated or controlled within established Plan hazard acceptability parameters.
\item Review and approve safety analysis and Safety Plan documents.
\item Ensuring that necessary documents are submitted to agencies that require them.
\end{itemize}

\paragraph{NOAO Safety Team}

A NOAO project safety team will be established and will consist of representatives from each of the project groups and critical subsystems as determined by the Project Manager(s). The purpose of this team is to consult on EHS issues, provide policy advice, evaluate Plan effectiveness, and make recommendations to the project team. The project safety team will meet at appropriate intervals as required and determined by the team and safety manager. 

\paragraph{Team Members}

All members of NOAO team as well as DESI team members working at Kitt Peak are responsible for integrating safety into their work and supporting the EHS plan as established for the work location. NOAO and DESI staff are encouraged and empowered to understand the work environment and identify the signs, procedures and conditions that they consider unsafe, whether specifically addressed in EHS plans or otherwise occurring. Every team member working at Kitt Peak, regardless of who employs them, can stop work and seek technical assistance from the safety manager for guidance or resolution of safety issues.

\subsubsection{Design for Safe Installation and Operation}
\label{sec:safedesign}

NOAO will follow the best installation and commissioning practices and comply with the appropriate laws and standards in all aspects of the DESI project.  The NOAO team, the safety manager, project managers and engineers will work closely to apply the correct safety standard to the different elements. In all cases, they will be consistent with applicable standards from US Federal Government and other institutions, including: 

\begin{itemize}
\item U.S. Occupational Safety and Health Administration (OSHA)
\item Environmental Protection Agency (EPA)
\item National Fire Protection Association (NFPA)
\item National Optical Astronomy Observatory 
\item Department of Energy National Laboratories
\end{itemize}

Cooperation between the Safety Manager, Engineering staff and the safety team will be required to keep everyone involved with the review and hazard analysis processes described in this Plan and the project hazard management processes. As operating procedures are developed, the job hazard analysis will be included to protect people, equipment, and processes from known hazards.  

As discussed in the preliminary safety plan and hazard analysis documents (DESI-1004), the major hazards during the Kitt Peak work include (but are not limited to) working at height, and overhead lifts of heavy, high-value equipment.  The procedure for every task involving such hazards will be accompanied by Job Hazard Analyses and Critical Lift Plans which will cover all the hazard mitigations required for the task.  As an example, in the case of at-height work, mitigations will include: designing the at-height work platforms to allow two means of ingress and egress; ensuring that all personnel have appropriate Personal Protective Equipment (PPE) for the job -- fall restraint harnesses with attachment gear, hard hats, etc -- and are fully trained in how to use it; and ensuring that an adequate number of personnel fully trained in at-height rescue are available at all times.

\clearpage

\section{Commissioning}\label{s6:Commish}
\setcounter{equation}{0}\setcounter{figure}{0}\setcounter{table}{0}
\subsection{Overview}

This section describes the plan for commissioning of the DESI instrument. A comprehensive commissioning plan is being developed and is available  in DESI-0973.  As we gain experience with the DESI components and ProtoDESI we will update the commissioning plan.   To provide context this section also briefly describes the tests that will occur before the commissioning begins (Functional Verification) and the period of 
Survey Validation
 that occurs after commissioning but prior to the start of the 5 year survey.  

The purposes of Commissioning and 
Survey Validation
are to test and optimize the full DESI system and to ensure that it meets the requirements.   Commissioning begins at the completion of functional tests of the telescope and DESI systems following installation of the focal plane/fiber positioner system in the prime focus cage on the Mayall telescope.  Commissioning is currently scheduled to take place from January to June 2019.  Completion of commissioning marks the conclusion of the DESI DOE construction project and readiness for CD-4.  A 
Survey Validation
 period follows commissioning, and at its conclusion DESI should be ready for the 5 year DESI survey to begin.  
Technical support of DESI will be reduced, but will not end with the completion of the project.   An operations budget will provide funding for the effort needed for operation of the survey after completion of the construction project. 

Commissioning is largely a science-driven activity; by contrast, the phases of installation and functional verification are primarily engineering activities.   Commissioning starts with on-sky testing.   The activities to be carried out are systems tests to evaluate the performance and to identify causes where performance specifications are not met.  The ability to respond to the unexpected with fast decisions is important, such as  being able to adjust the observing schedule in near real-time depending on weather and other circumstances, while still trying to keep within the bounds of the time provided.   The DESI project key performance parameters,, KPPs, (see DESI Project Execution Plan) require that the DESI system has at least 6 spectrographs and that it is has demonstrated the ability to obtain simultaneous spectra of hundreds of galaxies prior to completion of the commissioning period and the DOE construction project.


\subsection{Functional Verification}

Functional Verification of all the DESI systems will happen after each have been installed. Nearly all of them require the use of the DESI Online System (DOS).  To define the state of the DESI systems at the start of commissioning we include a list of some of these tests here. 
\begin{itemize}
\item	Laser tracker: measure primary mirror to PF cage flexure using the laser tracker, build information for the hexapod look up table.  The laser tracker will be removed from the telescope after that use.
\item	Spectrograph system: Operate spectrograph climate control system, pump down and operate cryostats and LPTs, dark exposures with spectrograph CCDs (involves data system also), operate spectrograph accessories
\item	Focal Plane System: Leak check FP cooling services, check cooling system aliveness (fans, temperature sensors), move positioners one petal at a time, look at them through porthole, turn on and read out Fiber View Camera (FVC), turn on and read out Guide and Focus Arrays (GFAs) (darks), illuminate field fiducials and image with FVC.
\item	Barrel: Rotate Atmospheric Dispersion Correctors (ADCs), visually monitor and or monitor encoder, operate Hexapod, visually monitor and/or measure barrel/cage relative motion directly.
\item	Telescope (and fiber wraps): Use ICS/TCS connection to point telescope with dome closed.  Operate at slew and tracking speeds.  Observe fiber wraps and telescope drive motor torques.
\item	Spectrograph/fiber system: Point at dome screen, do spectrograph exposures with each of the various calibration lamps (involves data system also).
\item	Fiber/FP: Back-illuminate fibers with GSE back-illuminator, take and readout FVC image, move all positioners 1~mm, repeat FVC.
\item	Spectrograph/fiber: Back-illuminate fibers with spectrograph back-illuminator, take and readout FVC image.
\item	The DESI Data Transport System (DDTS) is a DESI project deliverable.  The DDTS will have data paths set up and be thoroughly tested prior to commissioning.  We will exercise it during commissioning by relying on it as the primary data transport mechanism.

\item	Quicklook pipelines on the mountain debugged and tested.
\item	Full data reduction pipeline at LBNL run and tested on calibration frames (flats and arcs)
\item	Confirm that appropriate metadata is recorded in fits files and observation database for each exposure
\end{itemize}

\subsection{Commissioning}

Commissioning is when these separate sub-systems are operated as a system and on-sky tests begin. Broadly, commissioning includes:
\begin{itemize}
\item	Tests that are designed to prove the correct functioning of the DESI system. 
\begin{itemize}
\item	Verify that installation on the telescope has not degraded any performance characteristics (overlaps with functional verification tests).
\item	Evaluate the optical performance, end-to-end.
\item	Test and calibrate telescope-camera systems such as the fiber view camera, the automatic focus system, the guide system, the mirror support system (air pads), and the alignment system.
\item	Test the ability of the telescope to point and track such that fibers stay located on targets for the required exposure times.
\item	Collect astronomical and calibration data that will allow testing of the Pipelines. 
\item	Test the robustness (fault-tolerance), as a system.
\item	Determine initial operational efficiency (will be optimized later). 
\end{itemize}
\item	The science pipelines will be used during the commissioning period to provide feedback to the commissioning team.  They will be optimized during the 
survey validation
period. Quick-look pipelines will be needed on the mountain We will determine at a later date if the full set of pipelines will need to run on the mountain as well as at LBNL (NERSC).
\end{itemize}

Commissioning will require broad participation from the collaboration.  Some activities will happen on Kitt Peak but many require attention from people off-site.  Excellent connectivity is essential.  Activities on Kitt Peak include:
\begin{itemize}
\item	Calibration, characterization of CCDs and spectrograph response using lamps (the dome has been demonstrated to be dark enough to do this).
\item	Monitor spectrograph shack climate controls
\item	Operating positioners and imaging backlit fibers using Fiber View Camera (may require a special filter for daytime testing).
\item	Operations training
\item	Final safety signoff
\end{itemize}

Examples of offsite activities include:
\begin{itemize}
\item	Exercise the reduction pipelines, with real data.   
\item	Evaluate the data quality
\item	Monitor the data transport system (DTS) performance and reliability. 
\item	These groups will provide quick feedback to the on-site commissioning team. 
\end{itemize}

Formally the Commissioning period can be declared complete when the DESI system has demonstrated that it can routinely obtain simultaneous spectra of hundreds of galaxies.

\subsection{Survey Validation}

Survey Validation
continues the transition to a system capable of executing the DESI survey.  It will be led by the scientific team but will likely need periodic technical support from the commissioning team. SV tests the end-to-end system by demonstrating that a substantive sample of data collected with the DESI protocols can be reduced and meets the science requirements.   It begins the process of measuring and improving operational efficiency. DESI-0943 contains the detailed planning.

At the end of commissioning and 
survey validation
DESI will not be finished, in the sense that optimization of the active optics control system, provision of better calibrations and characteristics, tweaking and bug-fixing of software, \etc are all activities that are expected to continue for some time.    Given successful commissioning, all these activities can take place while DESI is being used for science, normally with updates being implemented and tested during scheduled telescope maintenance periods and engineering runs.

\subsection{Commissioning Strategy}

It is expected that all 10 spectrographs will be available at the start of the commissioning.  Even if not, the DESI online system will be able to deal with different numbers of spectrographs and petals.  In addition we will construct a back illumination system for the fibers that are not in a spectrograph so that they may be included in the tests of the fiber view camera and fiber positioning tests.  We note that project Key Performance Parameters can be met with as few as 6 spectrographs.

The Commissioning plan (DESI-0973) contains a summary of the expected weather data during the February to July time period, based on a study of historical weather patterns (DESI-0993).  Averages and extremes of usable hours per month and useful nights are tabulated using data from 2005-2014.  The bottom line is that for average weather we can expect the commissioning period to have roughly 50\% clear nights and about 75\% of nights of useful data taking.

The commissioning plan details the commissioning tasks and divides them into four levels of increasing complexity (see Chapter 4 of DESI-0973), gradually approaching readiness for 
survey validation.
Tuning and improvement of performance and reliability will continue after commissioning is complete.

\subsection{Commissioning Roles}

 Scientists and engineers from the DESI Collaboration and NOAO will participate in the commissioning activities.   They will be organized as a day crew, a night crew, and a remote crew.   The night crew will run the nighttime tests.   The day crew will respond to nighttime problems and run scheduled daytime tests.   
 The night crew in particular will likely be populated as teams, working approximately on a weekly schedule.  Sub-system experts will be made available, many remotely, depending on the schedule.   The remote crew will be geographically dispersed (Tucson, LBL and other DESI institutions) and will generally be monitoring pipeline outputs and conducting more deep analyses of data, where appropriate using codes developed for DESI.   A daily``1pm meeting'' will be used to communicate progress and problems between the daytime and nighttime crews.
 This model was used in the successful commissioning of DECam.

Specific Roles are listed below:

\begin{enumerate}
  \setlength{\itemsep}{1pt}
  \setlength{\parskip}{0pt}
  \setlength{\parsep}{0pt}
\item The DESI Project Scientists head the commissioning of the DESI System.  Responsibilities include preparing the day-to-day plan, assigning staff, seeing that the plan is executed, monitoring progress, signing off on planning changes, reporting to senior management.  

\item NOAO Telescope  Scientist - The scientist present at the telescope who is the expert on telescope operations and performance.   This may be staffed by NOAO.

\item On-duty scientists and staff - Other scientists and technical staff present at Kitt Peak during commissioning. This will be a mixture of NOAO staff and DESI collaboration members.  These may work day shift, night shift, or swing shift.   One will be the NOAO Telescope Scientist. Some may work directly on executing the commissioning programs, others on analysis, troubleshooting, \etc\    There will be one person specifically assigned to be responsible for maintaining a Punch List.  
\end{enumerate}

\subsection{Summary}

In many ways the DESI system is similar to the DECam system that was installed on the Blanco telescope in 2012.  In particular re-commissioning of the telescope control system, tracking, pointing and commissioning of the communications between the TCS and the DESI guide and focus systems will follow similar steps.  An initial version of the commissioning plan has been prepared.  It builds on the experience of DECam and will continue to be developed over the course of the project as we gain experience with the DESI specific systems.  Experience with ProtoDESI 
will provide critical input into the commissioning plans and will inform future updates to the commissioning plan.

\clearpage

\section*{Acknowledgements}
\addcontentsline{toc}{section}{Acknowledgements}
This research is supported by the Director, Office of Science, Office of High Energy Physics of the U.S. Department of Energy under Contract No. 
DE–AC02–05CH1123, and by the National Energy Research Scientific Computing Center, a DOE Office of Science User Facility under the same 
contract; additional support for DESI is provided by the U.S. National Science Foundation, Division of Astronomical Sciences under Contract No. 
AST-0950945 to the National Optical Astronomy Observatory; the Science and Technologies Facilities Council of the United Kingdom; the Gordon 
and Betty Moore Foundation; the Heising-Simons Foundation; the National Council of Science and Technology of Mexico, and by the DESI 
Member Institutions: Aix-Marseille University;  Argonne National Laboratory; Barcelona Regional Participation Group; Brookhaven National Laboratory; 
Boston University; Carnegie Mellon University; CEA-IRFU, Saclay; China Participation Group; Cornell University; Durham University;  École Polytechnique 
Fédérale de Lausanne; Eidgenössische Technische Hochschule, Zürich;  Fermi National Accelerator Laboratory;  Granada-Madrid-Tenerife Regional 
Participation Group; Harvard University; Korea Astronomy and Space Science Institute; Korea Institute for Advanced Study; Institute of Cosmological  
Sciences, University of Barcelona; Lawrence Berkeley National Laboratory; Laboratoire de Physique Nucléaire et de Hautes Energies; Mexico Regional 
Participation Group; National Optical Astronomy Observatory; Siena College; SLAC National Accelerator Laboratory;  Southern Methodist University; 
Swinburne University; The Ohio State University; Universidad de los Andes; University of Arizona; University of California, Berkeley; University of California, 
rvine; University of California, Santa Cruz; University College London; University of Michigan at Ann Arbor; University of Pennsylvania; University of Pittsburgh; 
University of Portsmouth; University of Queensland; University of Toronto; University of Utah; UK Regional Participation Group; Yale University. T
he authors are honored to be permitted to conduct astronomical research on Iolkam Du’ag (Kitt Peak), a mountain with particular significance to the 
Tohono O’odham Nation.  For more information, visit desi.lbl.gov.

%
%

\clearpage

\addcontentsline{toc}{section}{References}
\printbibliography

\noindent$\rule{4in}{0.15mm}$\\
Author Institutions
\begin{small}
\begin{enumerate}
  \setlength{\itemsep}{1pt}
  \setlength{\parskip}{0pt}
  \setlength{\parsep}{0pt}

\item [$^{1}$]2137 Frederick Reines Hall, Irvine, CA 92697, USA
\item [$^{2}$]Aix Marseille Univ, CNRS, LAM,  13388 Marseille, France
\item [$^{3}$]Aix Marseille Univ, CNRS, OHP, 04870 Saint-Michel-l'Observatoire, France
\item [$^{4}$]Aix Marseille Universit\'{e}, CNRS/IN2P3, CPPM UMR 7346, 13288, Marseille, France
\item [$^{5}$]Alphabet Inc., 1650 Charleston Rd. Mountain View, CA 94043, USA
\item [$^{6}$]AMNH, Department of Astrophysics, American Museum of Natural History, New York, NY 10024, USA
\item [$^{7}$]APC, Universit\'{e} Paris Diderot-Paris 7, CNRS/IN2P3, CEA, Observatoire de Paris, 10, rue Alice Domon \& Léonie Duquet, Paris, France
\item [$^{8}$]Argonne National Laboratory, High-Energy Physics Division, 9700 S. Cass Avenue, Argonne, IL 60439, USA
\item [$^{9}$]Astronomy Department, Yale University, P.O. Box 208101 New Haven, CT 06520-8101, USA
\item [$^{10}$]Brookhaven National Laboratory, Upton NY 11973, USA
\item [$^{11}$]Carreterra M\'{e}xico-Toluca S/N, La Marquesa, Ocoyoacac, Edo. de M\'{e}xico C.P. 52750,  M\'{e}xico
\item [$^{12}$]CEA Saclay, IRFU  F-91191 Gif-sur-Yvette, France
\item [$^{13}$]Center for Cosmology and AstroParticle Physics, The Ohio State University, 191 West Woodruff Avenue, Columbus, OH 43210, USA
\item [$^{14}$]Centre for Advanced Instrumentation, Department of Physics, Durham University, South Road, Durham, DH1 3LE, UK
\item [$^{15}$]Centre for Astrophysics \& Supercomputing, Swinburne University of Technology, P.O. Box 218, Hawthorn, VIC 3122, Australia
\item [$^{16}$]Centre for Extragalactic Astronomy, Department of Physics, Durham University, South Road, Durham, DH1 3LE, UK
\item [$^{17}$]Centre for Theoretical Cosmology, Department of Applied Mathematics and Theoretical Physics, Wilberforce Road, Cambridge CB3 0WA, UK
\item [$^{18}$]Cerro Tololo Inter-American Observatory (CTIO), Colina El Pino s/n, Casilla 603, La Serena, Chile
\item [$^{19}$]CIEMAT, Avenida Complutense 40, E-28040 Madrid, Spain
\item [$^{20}$]Clippinger Laboratories, Room 333, Ohio University, Athens, OH 45701, USA
\item [$^{21}$]Departamento de F\'{i}sica, Universidad de Guanajuato - DCI, C.P. 37150, Leon, Guanajuato, M\'{e}xico
\item [$^{22}$]Departamento de F\'isica, Universidad de los Andes, Cra. 1 No. 18A-10, Edificio Ip, Bogot\'{a}, Colombia 
\item [$^{23}$]Department of Astronomy \& Astrophysics, University of Toronto, 50 St.~George Street, Toronto, ON, Canada M5S 3H4
\item [$^{24}$]Department of Astronomy and Astrophysics, University of California, Santa Cruz, 1156 High Street, Santa Cruz, CA 95065, USA
\item [$^{25}$]Department of Astronomy and Space Science, Sejong University, Seoul 143-747, Republic of Korea
\item [$^{26}$]Department of Astronomy, The Ohio State University, 4055 McPherson Laboratory, 140 W 18th Avenue, Columbus, OH 43210, USA
\item [$^{27}$]Department of Astronomy, University of California, Berkeley, CA 94720-3411, USA
\item [$^{28}$]Department of Astronomy, University of Michigan, 1085 S. University Avenue, Ann Arbor, MI 48109-1107, USA
\item [$^{29}$]Department of Astronomy, Yale University, Steinbach Hall, 52 Hillhouse Avenue, New Haven, CT 06511, USA
\item [$^{30}$]Department of Physics \& Astronomy and Pittsburgh Particle Physics, Astrophysics, and Cosmology Center (PITT PACC), University of Pittsburgh, Pittsburgh, PA 15260, USA
\item [$^{31}$]Department of Physics \& Astronomy, Ohio University, Athens, OH 45701, USA
\item [$^{32}$]Department of Physics \& Astronomy, University  of Wyoming, 1000 E. University, Dept.~3905, Laramie, WY 82071, USA
\item [$^{33}$]Department of Physics \& Astronomy, University College London, Gower Street, London, WC1E 6BT, UK
\item [$^{34}$]Department of Physics and Astronomy, Siena College, 515 Loudon Road, Loudonville, NY 12211, USA
\item [$^{35}$]Department of Physics and Astronomy, The University of Utah, 115 South 1400 East, Salt Lake City, UT 84112, USA
\item [$^{36}$]Department of Physics and Astronomy, University College London, 3rd Floor, 132 Hampstead Road, London, NW1 2PS, UK
\item [$^{37}$]Department of Physics and Astronomy, University of California, 4129 Frederick Reines Hall, Irvine, CA 92697, USA
\item [$^{38}$]Department of Physics and Center for Cosmology and Particle Physics, New York University, New York, NY 10003, USA
\item [$^{39}$]Department of Physics and JINA Center for the Evolution of the Elements, University of Notre Dame, Notre Dame, IN 46556, USA
\item [$^{40}$]Department of Physics and Michigan Center for Theoretical Physics, University of Michigan, Ann Arbor, MI 48109, USA
\item [$^{41}$]Department of Physics, Carnegie Mellon University, 5000 Forbes Avenue, Pittsburgh, PA 15213, USA
\item [$^{42}$]Department of Physics, Harvard University, 17 Oxford Street, Cambridge, MA 02138, USA
\item [$^{43}$]Department of Physics, Kansas State University, 116 Cardwell Hall, Manhattan, KS 66506, USA
\item [$^{44}$]Department of Physics, Southern Methodist University, 3215 Daniel Avenue, Dallas, TX 75275, USA
\item [$^{45}$]Department of Physics, The Ohio State University, 191 West Woodruff Avenue, Columbus, OH 43210, USA
\item [$^{46}$]Department of Physics, University of Arizona, 1118 E. Fourth Street, PO Box 210081, Tucson, AZ 85721, USA
\item [$^{47}$]Department of Physics, University of California, Berkeley, 366 LeConte Hall MC 7300, Berkeley, CA 94720-7300, USA
\item [$^{48}$]Department of Physics, University of Michigan, 450 Church St., Ann Arbor, MI 48109, USA
\item [$^{49}$]Department of Physics, University of Warwick, Gibbet Hill Road, Coventry, CV4 7AL, UK
\item [$^{50}$]Ecole Polytechnique F\'{e}d\'{e}rale de Lausanne, CH-1015 Lausanne, Switzerland
\item [$^{51}$]European Space Astronomy Centre (ESAC), 38205 Villanueva de la Ca\~{n}ada, Madrid, Spain
\item [$^{52}$]Fermi National Accelerator Laboratory, PO Box 500, Batavia, IL 60510, USA
\item [$^{53}$]Harvard-Smithsonian Center for Astrophysics, Harvard University, 60 Garden Street, Cambridge, MA 02138, USA
\item [$^{54}$]HCTLab Research Group, Escuela Politecnica Superior, Universidad Aut\'{o}noma de Madrid, C/Francisco Tomas y Valiente 11, 38049, Spain
\item [$^{55}$]Instituci\'{o} Catalana de Recerca i Estudis Avan\c{c}ats (ICREA), Pg.~de Llu\'{i}s Companys 23, 08010 Barcelona, Spain
\item [$^{56}$]Institut de C\`{i}encies de l'Espai, IEEC-CSIC, Campus UAB, Carrer de Can Magrans s/n, 08913 Bellaterra, Barcelona, Spain
\item [$^{57}$]Institut de Fisica d’Altes Energies (IFAE), The Barcelona Institute of Science and Technology, Campus UAB, 08193 Bellaterra Barcelona, Spain
\item [$^{58}$]Institute for Astronomy, ETH Z\"{u}rich, Wolfgang-Pauli-Strasse 27, CH-8093 Z\"{u}rich, Switzerland
\item [$^{59}$]Institute for Astronomy, University of Edinburgh, Royal Observatory, Edinburgh EH9 3HJ, UK
\item [$^{60}$]Institute for Computational Cosmology, Department of Physics, Durham University, South Road, Durham DH1 3LE, UK
\item [$^{61}$]Institute of Astronomy, University of Cambridge, Madingley Road, Cambridge, CB3 0HA, UK
\item [$^{62}$]Institute of Cosmology \& Gravitation, University of Portsmouth, Dennis Sciama Building, Portsmouth PO1 3FX, UK
\item [$^{63}$]Instituto de Astrofisica de Andaluc\'{i}a, Glorieta de la Astronom\'{i}a, s/n, E-18008 Granada, Spain
\item [$^{64}$]Instituto de Astrof\'{i}sica de Canarias, C/ Vía L\'{a}ctea, s/n, 38205 San Crist\'{o}bal de La Laguna, Santa Cruz de Tenerife, Spain
\item [$^{65}$]Instituto de Astronomia, Universidad Nacional Aut\'{o}noma de M\'{e}xico, Apartado Postal 70–264, 04510 M\'{e}xico D.F., M\'{e}xico
\item [$^{66}$]Instituto de C\`{i}encias del Cosmoc, (ICCUB) Universidad de Barcelona (IEEC-UB), Mart\'{i} i Franqu\`{e}s 1, E08028 Barcelona
\item [$^{67}$]Instituto de F\'{i}sica Te\'{o}rica (IFT) UAM/CSIC, Universidad Aut\'{o}noma de Madrid, Cantoblanco, E-28049, Madrid, Spain
\item [$^{68}$]Instituto de F\'{i}isica, Universidad Nacional Aut\'{o}noma de M\'{e}xico, Cd. M\'{e}xico C.P. 04510
\item [$^{69}$]Kavli Institute for Astronomy and Astrophysics at Peking University, PKU, 5 Yiheyuan Road, Haidian District, Beijing 100871, P.R. China
\item [$^{70}$]Kavli Institute for Cosmology, Cambridge, University of Cambridge, Madingley Road, Cambridge CB3 0HA, UK
\item [$^{71}$]Kavli Institute for Particle Astrophysics and Cosmology and SLAC National Accelerator Laboratory, Menlo Park, CA 94305, USA
\item [$^{72}$]Key Laboratory of Optical Astronomy, National Astronomical Observatories, Chinese Academy of Sciences, Beijing 100012, P.R. China
\item [$^{73}$]Korea Astronomy and Space Science Institute, 776, Daedeokdae-ro, Yuseong-gu, Daejeon 34055, Republic of Korea
\item [$^{74}$]Laboratoire d’Astrophysique, Ecole Polytechnique F\'{e}d\'{e}rale de Lausanne (EPFL), Observatoire de Sauverny, CH-1290 Versoix, Switzerland
\item [$^{75}$]Laborat\'{o}rio Interinstitucional de e-Astronomia, Rua Gal. Jose Cristino 77, Rio de Janeiro, RJ 20921-400, Brazil
\item [$^{76}$]Lawrence Berkeley National Laboratory, 1 Cyclotron Road, Berkeley, CA 94720, USA
\item [$^{77}$]Lawrence Livermore National Laboratory, P.O. Box 808 L-211, Livermore, CA 94551, USA
\item [$^{78}$]Ludwig-Maximilians University Munich, University Observatory, Scheinerstr.~1, 81679 Munich, Germany
\item [$^{79}$]McWilliams Center for Cosmology, Carnegie Mellon University, 5000 Forbes Avenue, Pittsburgh, PA 15213, USA
\item [$^{80}$]National Astronomical Observatories, Chinese Academy of Sciences, A20 Datun Rd. 100012, Beijing, P.R. China
\item [$^{81}$]National Optical Astronomy Observatory, 950 N. Cherry Avenue, Tucson, AZ 85719, USA
\item [$^{82}$]Observatorio Nacional, R. Gal. Jose Cristino 77, Rio de Janeiro, RJ 20921-400, Brazil
\item [$^{83}$]Physics Department, Stanford University, Stanford, CA 93405, USA
\item [$^{84}$]Physics Department, Yale University, P.O. Box 208120, New Haven, CT 06511, USA
\item [$^{85}$]Physics Dept., Boston University, 590 Commonwealth Avenue, Boston, MA 02215, USA
\item [$^{86}$]School of Mathematics and Physics, University of Queensland, 4101, Australia
\item [$^{87}$]School of Physics, Korea Institute for Advanced Study, 85 Hoegiro, Dongdaemun-Gu, Seoul 02455, Republic of Korea
\item [$^{88}$]Sorbonne Universit\'{e}s, UPMC Université Paris 06, Universit\'{e} Paris-Diderot, CNRS-IN2P3 LPNHE 4 Place Jussieu, F-75252, Paris Cedex 05, France
\item [$^{89}$]Space Sciences Laboratory, University of California, Berkeley, 7 Gauss Way, Berkeley, CA  94720, USA
\item [$^{90}$]Steward Observatory, University of Arizona, 933 N. Cherry Avenue, Tucson, AZ 85721, USA
\item [$^{91}$]SUPA, School of Physics and Astronomy, University of St Andrews, St Andrews, KY16 9SS, UK
\item [$^{92}$]University of California Observatories, 1156 High Street, Sana Cruz, CA 95065, USA
\item [$^{93}$]University of Science and Technology, Daejeon 34113, Republic of Korea

\end{enumerate}
\end{small}

\end{document}